\documentclass[nofootinbib,aps,eqsecnum,titlepage,preprint]{revtex4}%
\usepackage{amsfonts}
\usepackage{amsmath}
\usepackage{amssymb}
\usepackage{graphicx}
\usepackage[stable]{footmisc}%
\providecommand{\U}[1]{\protect\rule{.1in}{.1in}}
\begin{document}
\preprint{5$^{th}$ edition}
\title[Fr\"{o}hlich Polarons]{Fr\"{o}hlich Polarons\\Lecture course including detailed theoretical derivations}
\thanks{The printed version of these Lectures is copyrighted by TQC -- Departement
Fysica -- Universiteit Antwerpen / Jozef T. L. Devreese}
\thanks{$^{\ast\ast}$Electronic address: \textsl{jozef.devreese@uantwerpen.be}}
\author{Jozef T. L. Devreese$^{\ast\ast}$}
\affiliation{Theory of Quantum and Complex Systems (TQC), Universiteit Antwerpen, CDE,
Universiteitsplein, 1, B-2610 Antwerpen, Belgium}

\begin{abstract}
Based on a course presented by the author at the International School of
Physics Enrico Fermi, CLXI Course,."Polarons in Bulk Materials and Systems
with Reduced Dimensionality", Varenna, Italy, 21.6. - 1.7.2005, including
further developments since 2005.

In the present course, an overview is presented of the fundamentals of
continuum-polaron physics, which provide the basis of the analysis of polaron
effects in ionic crystals and polar semiconductors. These Lecture Notes deal
with \textquotedblleft large\textquotedblright, or \textquotedblleft
continuum\textquotedblright, polarons, as described by the Fr\"{o}hlich
Hamiltonian. The emphasis is on the polaron optical absorption, with detailed
mathematical derivations.

Appendix A treats optical conductivity of a strong-coupling polaron.

Appendix B considers Feynman's path-integral polaron treatment approached
using time-ordered operator calculus.

Appendix C is devoted to the many-body large polaron optical conductivity in
Nb doped strontium titanate.

Appendix D contains summary of the present state of the problem of the polaron
mobility. \textbf{It is remarkable that the theory of the polaron mobility
developed by Kadanoff \cite{Kadanoff}, which was recognized during a long
time, needs a correction factor as found in Ref. \cite{Los1984} and
independently confirmed in the recent work \cite{SB2014}.}

\textbf{Appendix E lists recent publications on Fr\"{o}hlich polarons in
Nature, Science and Physical Review Letters appeared from 2005 to 2015.}

\end{abstract}
\date{\today}
\startpage{1}
\endpage{1000}
\maketitle

\newpage

\section*{\emph{Preface to the 5th edition}}

Since 2005, when the first edition of the present Lecture Course was prepared,
polaron physics continued to intensely develop, involving new areas and
testing new powerful methods. In subsequent editions, these new developments
are included in order to emphasize which of them we consider important.

Renewed interest in large (Fr\"{o}hlich) polarons has been inspired by recent
experimental advances in the determination of the band structure of highly
polar oxides \cite{Meevasana}. The optical response of complex oxides clearly
reveals the polaron features and can shed light on the band structure of a
crystal and its polaron characteristics. The interpretation of the measured
data is essential to achieve a comprehensive understanding and to optimize
practical application of functional materials. In particular, the question
whether the polarons are large or small is often a subject of intense
discussions, for example, in the case of SrTiO$_{3}$ and TiO$_{2}$, key
materials in many technological sectors.

In the recent ARPES measurements \cite{Meevasana} no clear signatures of
small-polaron phenomena in $n$-doped strontium titanate were found, and the
conclusion was reached that small polarons are not formed in strontium
titanate. The electron-phonon coupling strength in strontium titanate
$\alpha\approx3.6$ obtained in Ref. \cite{pVDM-PRL2008} is typical for a
rather moderate coupling that makes the formation of small polarons in the
conduction band of SrTiO$_{3}$ hardly possible. On the contrary, recent
density functional calculations \cite{Franchini1,Franchini2015} show that
excess electrons form small polarons if the density of electronic carriers is
sufficiently high. This opens the interesting possibility to study an
interplay of small and large polarons in SrTiO$_{3}$ and other oxides. In Ref.
\cite{DKMM2010}, the many-large-polaron model gives then a convincing
interpretation of the experimentally observed mid-infrared band of
SrTi$_{1-x}$Nb$_{x}$O$_{3}$.

The polaron theory is a testing field for new powerful theoretical quantum
field methods, such as the Diagrammatic Quantum Monte Carlo (DQMC) method.
Applied first to the calculation of the ground-state energy of a Fr\"{o}hlich
polaron \cite{Msch1}, DQMC has been successful in the calculation of the
optical conductivity of the Fr\"{o}hlich polaron \cite{Msch2}. This inspired
attempts to develop analytical methods for the polaron optical response. The
recent work on the strong-coupling large-polaron optical conductivity
\cite{SC} shows a good agreement with DQMC in the strong-coupling limit. In
Ref. \cite{Berciu}, the momentum average approximation is applied to derive an
analytic expression for the optical conductivity of a small polaron, that very
well matches the DQMC data.

The polaron theory has found recently several new interesting applications.
One of them is the theoretical interpretation of the physics of an impurity
immersed in an atomic Bose-Einstein condensate. In Refs.
\cite{BECpol1,BECpol2}, the ground-state energy of the BEC polaron has been
studied on the basis of a Fr\"{o}hlich type Hamiltonian using the Feynman
variational technique and the DQMC method. In Ref. \cite{Grusdt}, the problem
of the BEC polaron has been treated using the renormalization group method. It
successfully retrieves the DQMC results in the whole (available for the
comparison) range of the particle-phonon coupling strength.

In summary, polaron physics recently demonstrated new fascinating
developments, that makes the present lecture course timely and relevant.

\newpage
\tableofcontents
\newpage

\part{Single polaron}

\section{Introduction. The "standard" theories}

\subsection{The polaron concept}

\begin{figure}[h]
\begin{center}
\includegraphics[height=.3\textheight]{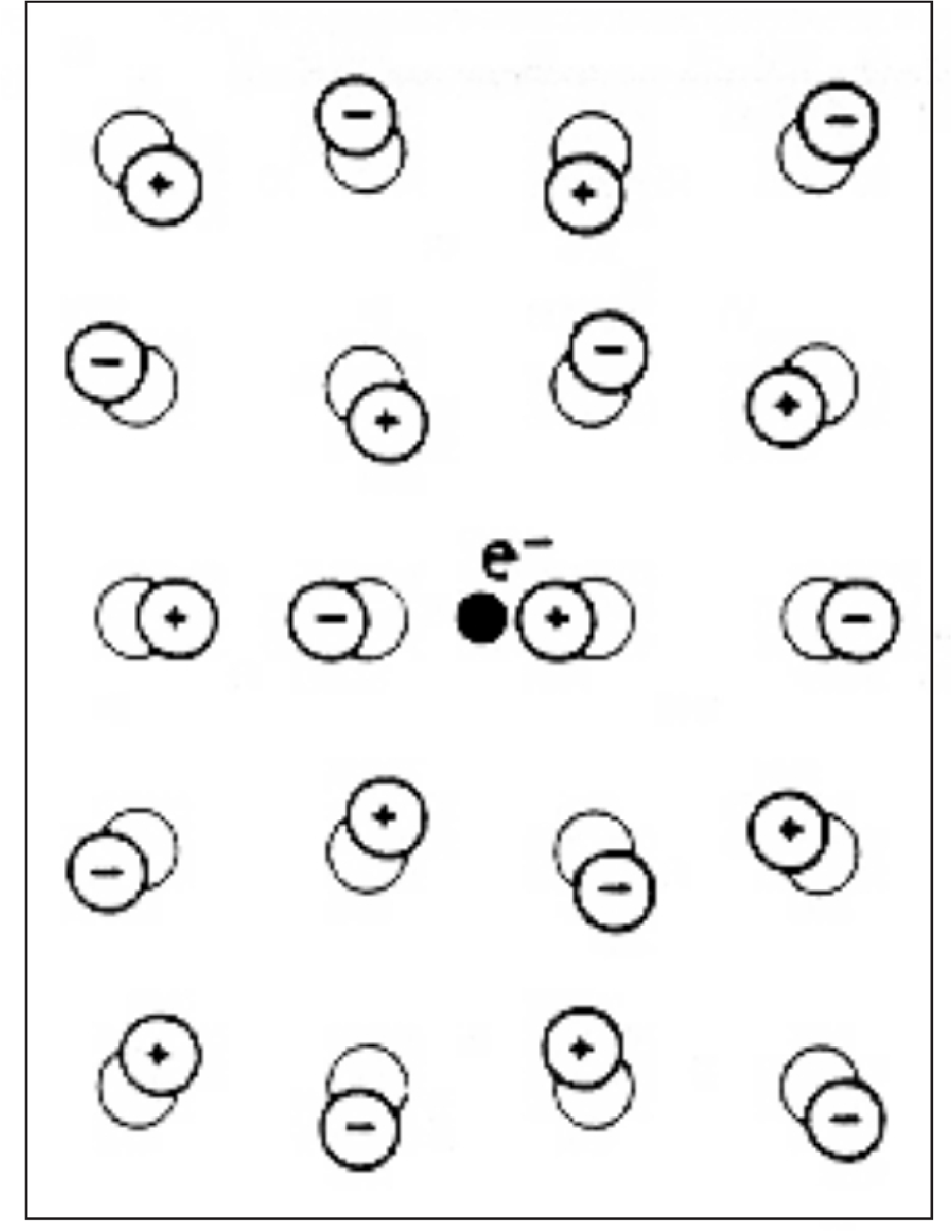}
\end{center}
\caption{Artist view of a polaron. A conduction electron in an ionic crystal
or a polar semiconductor repels the negative ions and attracts the positive
ions. A self-induced potential arises, which acts back on the electron and
modifies its physical properties. (From \cite{Devreese03}.)}%
\label{fig_scheme1}%
\end{figure}

A charge placed in a polarizable medium is screened. Dielectric theory
describes the phenomenon\ by the induction of a polarization around the charge
carrier. The idea of the autolocalization of an electron due to the induced
lattice polarization was first proposed by L. D. Landau \cite{Landau}. In the
further development of this concept, the induced polarization can follow the
charge carrier when it is moving through the medium. The carrier together with
the induced polarization is considered as one entity (see Fig.\thinspace
\ref{fig_scheme1}). It was called a \textit{polaron} by S. I. Pekar
\cite{Pekar1946,Pekar1946a}. The physical properties of a polaron differ from
those of a band-carrier. A polaron is characterized by its \textit{binding (or
self-) energy} $E_{0}$, an \textit{effective mass} $m^{\ast}$ and by its
characteristic \textit{response} to external electric and magnetic fields
(e.~g. dc mobility and optical absorption coefficient).

%\newpage

%\newpage

If the spatial extension of a polaron is large compared to the lattice
parameter of the solid, the latter can be treated as a polarizable continuum.
This is the case of a \textit{large (Fr\"{o}hlich)} polaron. When the
self-induced polarization caused by an electron or hole becomes of the order
of the lattice parameter, a \textit{small (Holstein)} polaron can arise
\cite{Fehske}. As distinct from large polarons, small polarons are governed by
short-range interactions.

\subsection{Intuitive concepts}

\paragraph{The polaron radius. Large polarons vs small polarons}

Consider the LO\ phonon field with frequency $\omega_{\text{\textrm{LO}}}$
interacting with an electron. Denote by $\Delta\nu$ the quadratic mean square
deviation of the electron velocity. In the electron-phonon interaction is
weak, the electron can travel a distance%

\begin{equation}
\Delta x\approx\frac{\Delta\nu}{\omega_{\text{LO}}} \label{I1}%
\end{equation}
during a time $\omega_{\text{\textrm{LO}}}^{-1},$characteristic for the
lattice period,because it is the distance within which the electron can be
localized using the phonon field as measuring device. From the uncertainty
relations it follows%

\begin{align}
\Delta p\Delta x  &  =\frac{m}{\omega_{\text{LO}}}\left(  \Delta\nu\right)
^{2}\approx\hbar\nonumber\\
\Delta\nu &  \sim\sqrt{\frac{\hbar\omega_{\text{LO}}}{m}},\nonumber\\
\Delta x  &  \sim\sqrt{\frac{\hbar}{m\omega_{\text{LO}}}}. \label{I2}%
\end{align}

At weak coupling $\Delta x$ is a measure of the polaron radius $r_{p}$. To be
consistent, the polaron radius $r_{p}$ must be considerably larger than the
lattice parameter $a.$(this is a criterion of a \textquotedblleft large
polaron\textquotedblright). Experimental evaluation of the polaron radius
leads to the follwing typical values: $r_{p}\approx10\mathring{A}$ for alkali
halides, $r_{p}\approx20\mathring{A}$ for silver halides, $r_{p}%
\approx100\mathring{A}$ for II-VI, II-V semiconductors. The continuum
approximation is not satisfied for transition metal oxides (NiO, CaO, MnO), in
other oxides (UO$_{2},$NbO$_{2}...$). For those solids the \textquotedblleft
small polaron\textquotedblright\ concept is used. In some substances (e.g.
perovskites) some intermediate region between large and small polarons is realized.

\paragraph{The coupling constant \cite{Fr54}}

Consider the case of \textit{strong electron-phonon interaction} in a polar
crystal. The electron of mass $m$ is then localized and can - to a first
approximation - be considered as a static charge distribution within a sphere
with radius $l_{1}.$ The medium is characterized by an average dielectric
constant $\bar{\varepsilon},$which will be defined below.

The potential energy of a sphere of radius $l_{1}$uniformly charged with the
charge $e$ in a vacuum is (see Eq. (8.6) of Ref. \cite{FLP-1})%
\begin{equation}
U_{vac}=\frac{3}{5}\frac{e^{2}}{l_{1}}. \label{U0}%
\end{equation}
The potential energy of a uniformly charged sphere in a medium with the
high-frequency dielectric constant $\varepsilon_{\infty}$ is%
\begin{equation}
U_{1}=\frac{3}{5}\frac{e^{2}}{\varepsilon_{\infty}l_{1}}. \label{U1}%
\end{equation}
This is the potential energy of the \textit{self-interaction} of the charge
$e$ uniformly spread over the sphere of radius $l_{1}$ in a medium with the
dielectric constant $\varepsilon_{\infty}$. In a medium with an inertial
polarization field (due to LO phonons), the potential energy of the uniformly
charged sphere is%
\begin{equation}
U_{2}=\frac{3}{5}\frac{e^{2}}{\varepsilon_{0}l_{1}}, \label{U2}%
\end{equation}
where $\varepsilon_{0}$ is the static dielectric constant. The polaron effect
is then related to the change of the potential energy of the interaction of
the charged sphere due to the inertial polarization field. This change is the
potential energy $U_{2}$ of the uniformly charged sphere in the presence of
the inertial polarization field minus the potential energy of the
\textit{self-interaction }$U_{1}$of the charge $e$ uniformly spread over the
sphere in a medium without the inertial polarization:%
\begin{equation}
U_{pol}\equiv U_{2}-U_{1}=\frac{3}{5}\frac{e^{2}}{l_{1}}\left(  \frac
{1}{\varepsilon_{0}}-\frac{1}{\varepsilon_{\infty}}\right)  =-\frac{3}{5}%
\frac{e^{2}}{\bar{\varepsilon}l_{1}}, \label{Upol}%
\end{equation}
with%
\[
\frac{1}{\bar{\varepsilon}}=\frac{1}{\varepsilon_{\infty}}-\frac
{1}{\varepsilon_{0}}.
\]

The electron distribution in a sphere may be non-uniform, what may influence
the numerical coefficient in Eqs. (\ref{U1}) to (\ref{Upol}). In this
connection one can use the estimate \cite{Fr54}%
\begin{equation}
U_{pol}\sim-\frac{e^{2}}{\bar{\varepsilon}l_{1}}. \label{U3}%
\end{equation}

The restriction of the electron in space requires its de Broglie wave length
to be of the order $l_{1},$so that its kinetic energy is of the order
$4\pi^{2}\hbar^{2}/2ml_{1}^{2}.$Minimizing the total energy with respect to
$l_{1}$leads to%

\[
\frac{\partial}{\partial l_{1}}\left(  -\frac{e^{2}}{l_{1}\bar{\varepsilon}%
}+\frac{4\pi^{2}\hbar^{2}}{2ml_{1}^{2}}\right)  =0\Longrightarrow\frac
{1}{l_{1}}=\frac{e^{2}m}{4\pi^{2}\hbar^{2}\bar{\varepsilon}},
\]
wherefrom the binding energy is%

\begin{equation}
U_{1}=-\frac{e^{4}m}{8\pi^{2}\hbar^{2}\bar{\varepsilon}^{2}}. \label{I4}%
\end{equation}

For \textit{weak coupling}, one can neglect the kinetic energy of the
electron. Taking the polaron radius according to (\ref{I2}), $r_{p}%
=\sqrt{2\hbar/m\omega_{\text{LO}}},$ the binding energy is%
\begin{equation}
U_{2}=-\frac{e^{2}}{r_{p}\bar{\varepsilon}}=-\frac{e^{2}}{\bar{\varepsilon}%
}\sqrt{\frac{m\omega_{\text{LO}}}{2\hbar}}. \label{I5}%
\end{equation}
We note that%

\begin{equation}
\frac{U_{1}}{\hbar\omega_{\text{LO}}}=-\frac{e^{4}m}{8\pi^{2}\hbar^{3}%
\bar{\varepsilon}^{2}\omega_{\text{LO}}}=-\frac{1}{4\pi^{2}}\left(
\frac{U_{2}}{\hbar\omega_{\text{LO}}}\right)  ^{2}. \label{I6}%
\end{equation}
Following the conventions of the field theory, the self energy at weak
coupling is written as%

\[
U_{2}=-\alpha\hbar\omega_{\text{LO}}.
\]
Therefore the so-called Fr\"{o}hlich polaron coupling constant is%

\begin{align}
\alpha &  =\frac{e^{2}}{\bar{\varepsilon}}\sqrt{\frac{m}{2\hbar^{3}%
\omega_{\text{LO}}}}\nonumber\\
&  \equiv\frac{e^{2}}{\hbar c}\sqrt{\frac{mc^{2}}{2\hbar\omega_{\text{LO}}}%
}\frac{1}{\bar{\varepsilon}}. \label{I7}%
\end{align}
For the average dielectric constant one shows that%

\[
\frac{1}{\bar{\varepsilon}}=\frac{1}{\varepsilon_{\infty}}-\frac
{1}{\varepsilon_{0}},
\]
where $\varepsilon_{\infty}$ and $\varepsilon_{0}$ are, respectively, the
electronic and the static dielectric constant of the polar crystal. The
difference $1/\varepsilon_{\infty}-1/\varepsilon_{0}$ arises because the ionic
vibrations occur in the infrared spectrum and the electrons in the shells can
follow the conduction electron adiabatically.

\paragraph{Polaron mobility \footnote{\textbf{See also Appendix D
\textquotedblleft Notes on the polaron mobility\textquotedblright.}}}

Here we give a simple derivation leading to the gross features of the mobility
behaviuor, especially its temperature dependence. The key idea is that the
mobility will change because the number of phonons in the lattice, with which
the polaron interacts, is changing with temperature.

The phonon density is given by%

\[
n=\frac{1}{e^{\frac{\hbar\omega_{\text{LO}}}{kT}}-1}.
\]

The mobility for large polaron is proportional to the inverse of the number of phonons:%

\[
\mu\approx\frac{1}{n}=e^{\frac{\hbar\omega_{\text{LO}}}{kT}}-1
\]
and for low temperatures $kT\ll\hbar\omega_{\text{LO}}$%

\begin{equation}
\mu\approx e^{\frac{\hbar\omega_{\text{LO}}}{kT}}. \label{I8}%
\end{equation}
The mobility of continuum poarons decreases with increasing temperature
following an exponential law. The slope of the straight line in $\ln\mu$ vs
$1/T$ is characterized by the LO phonon frequesncy. Systematic study
performed, in particular, by Fr\"{o}hlich and Kadanoff, gives%

\begin{equation}
\mu=\frac{e}{2m\omega_{\text{LO}}}e^{\frac{\hbar\omega_{\text{LO}}}{kT}}.
\end{equation}

The small polaron will jump from ion to ion under the influence of optical
phonons. The lerger the numver of phonons, the lerger the mobility. The
behaviuor of the small polaron is the opposite of that of the large polaron.
One expects:%

\[
\mu\approx n=\frac{1}{e^{\frac{\hbar\omega_{\text{LO}}}{kT}}-1}.
\]

For low temperatures $kT\ll\hbar\omega_{\text{LO}}$ one has:%

\begin{equation}
\mu\approx e^{-\frac{\hbar\omega_{\text{LO}}}{kT}}: \label{I9}%
\end{equation}
the mobility of small polaron is thermally activated. Systematic analysis
within the small-polaron theory shows that%

\begin{equation}
\mu\approx e^{-\gamma\frac{\hbar\omega_{\text{LO}}}{kT}}:
\end{equation}
with $\gamma\sim5.$

\subsection{The Fr\"{o}hlich Hamiltonian}

Fr\"{o}hlich proposed a model Hamiltonian for the \textquotedblleft
large\textquotedblright\ polaron through which {its} dynamics is treated
quantum mechanically {(\textquotedblleft Fr\"{o}hlich
Hamiltonian\textquotedblright)}. The polarization, carried by the longitudinal
optical {(LO)} phonons, is represented by a set of quantum oscillators with
frequency $\omega_{\mathrm{LO}}$, {the long-wavelength LO-phonon frequency,}
and the interaction between the charge and the polarization field is linear in
the field \cite{Fr54}:%
\begin{equation}
H=\frac{\mathbf{p}^{2}}{2m_{b}}+\sum_{\mathbf{k}}\hbar\omega_{\mathrm{LO}%
}a_{\mathbf{k}}^{+}a_{\mathbf{k}}+\sum_{\mathbf{k}}(V_{k}a_{\mathbf{k}%
}e^{i\mathbf{k\cdot r}}+V_{k}^{\ast}a_{\mathbf{k}}^{\dag}e^{-i\mathbf{k\cdot
r}}), \label{eq_1a}%
\end{equation}
where $\mathbf{r}$ is the position coordinate operator of the electron with
band mass $m_{b}$, $\mathbf{p}$ is its canonically conjugate momentum
operator; $a_{\mathbf{k}}^{\dagger}$ and $a_{\mathbf{k}}$ are the creation
(and annihilation) operators for longitudinal optical phonons of wave vector
$\mathbf{k}$ and energy $\hbar\omega_{\mathrm{LO}}$. The $V_{k}$ are Fourier
components of the electron-phonon interaction
\begin{equation}
V_{k}=-i\frac{\hbar\omega_{\mathrm{LO}}}{k}\left(  \frac{4\pi\alpha}%
{V}\right)  ^{\frac{1}{2}}\left(  \frac{\hbar}{2m_{b}\omega_{\mathrm{LO}}%
}\right)  ^{\frac{1}{4}}. \label{eq_1b}%
\end{equation}
The strength of the electron--phonon interaction is {expressed by} a
dimensionless coupling constant $\alpha$, which is defined as:
\begin{equation}
\alpha=\frac{e^{2}}{\hbar}\sqrt{\frac{m_{b}}{2\hbar\omega_{\mathrm{LO}}}%
}\left(  \frac{1}{\varepsilon_{\infty}}-\frac{1}{\varepsilon_{0}}\right)  .
\label{eq_1c}%
\end{equation}
In this definition, $\varepsilon_{\infty}$ and $\varepsilon_{0}$ are,
respectively, the electronic and the static dielectric constant of the polar crystal.

In Table~\ref{Table1} the Fr\"{o}hlich coupling constant is given for a few
solids\footnote{In some cases, due to lack of reliable experimental data to
determine the electron band mass, the values of $\alpha$ are not well
established.}.

\begin{table}[pbh]
\caption{\textit{Electron-phonon coupling constants} (After Ref.
\cite{Devreese03}) }%
\label{Table1}
\begin{center}
\bigskip%
\begin{tabular}
[c]{@{}llllll}\hline
Material & $\alpha$ & Ref. & Material & $\alpha$ & Ref.\\\hline
InSb & {0.023} & \cite{KartheuserGreenbook}\phantom{coupling} & AgCl & 1.84 &
\cite{Hodby}\\
InAs & 0.052 & \cite{KartheuserGreenbook} & KI & 2.5 &
\cite{KartheuserGreenbook}\\
GaAs & 0.068 & \cite{KartheuserGreenbook} & TlBr & 2.55 &
\cite{KartheuserGreenbook}\\
GaP & {0.20} & \cite{KartheuserGreenbook} & KBr & 3.05 &
\cite{KartheuserGreenbook}\\
CdTe & {0.29} & \cite{Grynberg} & Bi$_{12}$SiO$_{20}$ & 3.18 & \cite{Biaggio}%
\\
ZnSe & {0.43} & \cite{KartheuserGreenbook} & CdF$_{2}$ & 3.2 &
\cite{KartheuserGreenbook}\\
CdS & {0.53} & \cite{KartheuserGreenbook} & KCl & 3.44 &
\cite{KartheuserGreenbook}\\
$\alpha$-Al$_{2}$O$_{3}$ & 1.25 & \cite{sapphire} & CsI & 3.67 &
\cite{KartheuserGreenbook}\\
AgBr & 1.53 & \cite{Hodby} & SrTiO$_{3}$ & 3.77 & \cite{Ferroelectrics1992}\\
$\alpha$-SiO$_{2}$ & 1.59 & \cite{corundum} & RbCl & 3.81 &
\cite{KartheuserGreenbook}\\\hline
\end{tabular}
\end{center}
\end{table}\bigskip

{In deriving the form of $V_{k}$, expressions (\ref{eq_1b}) and (\ref{eq_1c}),
it was assumed that (i) the spatial extension of the polaron is large compared
to the lattice parameter of the solid (\textquotedblleft
continuum\textquotedblright\ approximation), (ii) spin and relativistic
effects can be neglected, (iii) the band-electron has parabolic dispersion,
(iv) in line with the first approximation it is also assumed that the
LO-phonons of interest for the interaction, are the long-wavelength phonons
with constant frequency $\omega_{\mathrm{LO}}$. }

The model, represented by the Hamiltonian (\ref{eq_1a}) (which up to now could
not been solved exactly) has been the subject of extensive investigations,
see, e.~g., Refs.
\cite{Pekar,KW63,A68,Devreese72,Mitra,Devreese96,AM96,Mishchenko2000}. In what
follows the key approaches of the Fr\"{o}hlich-polaron theory are briefly
reviewed with indication of their relevance for the polaron problems in nanostructures.

\subsection{Infinite mass model [\textquotedblleft shift\textquotedblright%
--operators]\label{sec-shift}}

Here some insight will be given in the type of transformation that might be
useful to study the Fr\"{o}hlich Hamiltonian (\ref{eq_1a}). For this purpose
the Hamiltonian will be treated for a particle with infinite mass
$m_{b}\rightarrow\infty,$ (which is at $\mathbf{r}=0):$%
\begin{equation}
H^{\infty}=\sum_{\mathbf{k}}\hbar\omega_{\mathrm{LO}}a_{\mathbf{k}}%
^{+}a_{\mathbf{k}}+\sum_{\mathbf{k}}(V_{k}a_{\mathbf{k}}+V_{k}^{\ast
}a_{\mathbf{k}}^{\dag}),
\end{equation}
which can be transformed into the following expression with \textquotedblleft
shifted\textquotedblright\ phonon operators:%
\begin{equation}
H^{\infty}=\sum_{\mathbf{k}}\hbar\omega_{\mathrm{LO}}\left(  a_{\mathbf{k}%
}^{\dag}+\frac{V_{k}}{\hbar\omega_{\mathrm{LO}}}\right)  \left(
a_{\mathbf{k}}+\frac{V_{k}^{\ast}}{\hbar\omega_{\mathrm{LO}}}\right)
-\sum_{\mathbf{k}}\frac{\left\vert V_{k}\right\vert ^{2}}{\hbar\omega
_{\mathrm{LO}}}.
\end{equation}

To determine the eigenstates of this Hamiltonian, one can perform a unitary
transformation which produces the following \textquotedblleft
shift\textquotedblright\ of the phonon operators:%
\[
a_{\mathbf{k}}\rightarrow b_{\mathbf{k}}=a_{\mathbf{k}}+\frac{V_{k}^{\ast}%
}{\hbar\omega_{\mathrm{LO}}},a_{\mathbf{k}}^{\dag}\rightarrow b_{\mathbf{k}%
}^{\dag}=a_{\mathbf{k}}^{\dag}+\frac{V_{k}}{\hbar\omega_{\mathrm{LO}}}.
\]
The transformation%
\begin{equation}
S=\exp\left[  -\sum_{\mathbf{k}}a_{\mathbf{k}}^{\dag}\frac{V_{k}^{\ast}}%
{\hbar\omega_{\mathrm{LO}}}+\sum_{\mathbf{k}}\frac{V_{k}}{\hbar\omega
_{\mathrm{LO}}}a_{\mathbf{k}}\right]  \label{S}%
\end{equation}
is canonical:
\[
S^{\dag}=\exp\left[  -\sum_{\mathbf{k}}a_{\mathbf{k}}\frac{V_{k}}{\hbar
\omega_{\mathrm{LO}}}+\sum_{\mathbf{k}}\frac{V_{k}^{\ast}}{\hbar
\omega_{\mathrm{LO}}}a_{\mathbf{k}}^{\dag}\right]  =S^{-1}%
\]
and has the desired property:%
\[
S^{-1}a_{\mathbf{k}}S=a_{\mathbf{k}}-\frac{V_{k}^{\ast}}{\hbar\omega
_{\mathrm{LO}}},S^{-1}a_{\mathbf{k}}^{\dag}S=a_{\mathbf{k}}^{\dag}-\frac
{V_{k}}{\hbar\omega_{\mathrm{LO}}}.
\]
The transformed Hamiltonian is now:%
\[
S^{-1}H^{\infty}S=\sum_{\mathbf{k}}\hbar\omega_{\mathrm{LO}}a_{\mathbf{k}%
}^{\dag}a_{\mathbf{k}}-\sum_{\mathbf{k}}\frac{\left\vert V_{k}\right\vert
^{2}}{\hbar\omega_{\mathrm{LO}}}.
\]
The eigenstates of the Hamiltonian contain an integer number of phonons
$\left(  \left\vert n_{\mathbf{k}}\right\rangle \right)  .$The eigenenergies
are evidently:%
\[
E=\sum_{\mathbf{k}}n_{\mathbf{k}}\hbar\omega_{\mathrm{LO}}-\sum_{\mathbf{k}%
}\frac{\left\vert V_{k}\right\vert ^{2}}{\hbar\omega_{\mathrm{LO}}}.
\]
This expression is divergent at it is often the case in field theory of point
charges are considered. A transformation of the type $S$ has been of great
interest in developing weak coupling theory as shown below.

\subsection{The "standard" theories}

\subsubsection{Weak coupling via a perturbation theory}

{For actual crystals $\alpha$-values typically range from $\alpha=0.02$ (InSb)
up to $\alpha\sim3$ to $4$ (alkali halides, some oxides), see Table 1. A
weak-coupling theory of the polaron was developed originally by Fr\"{o}hlich
\cite{Fr54}. He derived the first weak-coupling perturbation-theory results:
\begin{equation}
E_{0}=-\alpha\hbar\omega_{\mathrm{LO}} \label{eq_5a}%
\end{equation}
and }%
\begin{equation}
m^{\ast}=\frac{m_{b}}{1-\alpha/6}. \label{eq_5b}%
\end{equation}
Expressions (\ref{eq_5a}) and (\ref{eq_5b}) are rigorous to order $\alpha$.

\subsubsection{Weak coupling via a canonical transformation [\textquotedblleft
shift\textquotedblright-operators]}

Inspired by the work of Tomonaga on quantum electrodynamics (Q. E. D.), Lee,
Low and Pines (LLP) \cite{LLP} analyzed the properties of a weak-coupling
polaron starting from a formulation based on canonical transformations (cp.
the results of the subsection \ref{sec-shift}).As hown by them, the unitary
transformation%
\begin{equation}
U=\exp\left\{  \frac{i}{\hbar}\left(  \mathbf{P}-\sum_{\mathbf{k}}%
\hbar\mathbf{k}a_{\mathbf{k}}^{\dag}a_{\mathbf{k}}\right)  \cdot
\mathbf{r}\right\}  ,
\end{equation}
where $\mathbf{P}$ is a "c"-number representing the \textit{total system
momentum }allows to eliminate the electron co-ordinates from the system.
Intuitively one might guess this transformation by writing the exact wave
function in the form%
\[
\Psi_{\text{total }H}=\exp\left(  \frac{i}{\hbar}\mathbf{p}\cdot
\mathbf{r}\right)  \left\vert \Phi\right\rangle .
\]
It is plausible that the \textquotedblleft Bloch\textquotedblright\ factor
$\exp\left(  i/\hbar\mathbf{p}\cdot\mathbf{r}\right)  $ attaches the system to
the electron as origin of the co-cordinates. After this transformation the
Hamiltonian (\ref{eq_1a}) becomes:%
\begin{equation}
\mathcal{H}=U^{-1}HU=\frac{\left(  \mathbf{P}-\sum_{\mathbf{k}}\hbar
\mathbf{k}a_{\mathbf{k}}^{\dag}a_{\mathbf{k}}\right)  ^{2}}{2m_{b}}%
+\sum_{\mathbf{k}}\hbar\omega_{\mathrm{LO}}a_{\mathbf{k}}^{\dag}a_{\mathbf{k}%
}+\sum_{\mathbf{k}}(V_{k}a_{\mathbf{k}}+V_{k}^{\ast}a_{\mathbf{k}}^{\dag}).
\end{equation}
If, for the sake of simplicity, the case of total momentum equal to zero is
considered, this expression becomes:%
\begin{equation}
\mathcal{H}=\sum_{\mathbf{k,k}^{\prime}}\frac{\hbar^{2}\mathbf{k\cdot
k}^{\prime}a_{\mathbf{k}}^{\dag}a_{\mathbf{k}^{\prime}}^{\dag}a_{\mathbf{k}%
}a_{\mathbf{k}^{\prime}}}{2m_{b}}+\sum_{\mathbf{k}}\left(  \hbar
\omega_{\mathrm{LO}}+\frac{\hbar^{2}k^{2}}{2m_{b}}\right)  a_{\mathbf{k}%
}^{\dag}a_{\mathbf{k}}+\sum_{\mathbf{k}}(V_{k}a_{\mathbf{k}}+V_{k}^{\ast
}a_{\mathbf{k}}^{\dag}). \label{HCORR}%
\end{equation}
The first term of this Hamiltonian is the correlation energy term involving
different values for $\mathbf{k}$ and $\mathbf{k}^{\prime}.$If one
diagonalizes the second and the trird term of the Hamiltonian (\ref{HCORR})
(this can be done exactly by means of the "shifted-oscillator canonical
transformation"\ $S$ (\ref{S})), the result of LLP is found. The expectation
value of the first term is zero for the wave function $S\left\vert
0\right\rangle .$ Therefore one is sure to obtain a variational result. It is
remarkable that merely extracting the $\mathbf{k}=\mathbf{k}^{\prime}$ term
from the expression%
\begin{equation}
\sum_{\mathbf{k,k}^{\prime}}\frac{\hbar^{2}\mathbf{k\cdot k}^{\prime
}a_{\mathbf{k}}^{\dag}a_{\mathbf{k}^{\prime}}^{\dag}a_{\mathbf{k}%
}a_{\mathbf{k}^{\prime}}}{2m_{b}} \label{CORR}%
\end{equation}
eliminates the divergency from the problem (cp. with the case $m_{b}%
\rightarrow\infty$) and is equivalent to the sophisticated theory by Lee, Low
and Pines (LLP), which corresponds thus to neglect of the term (\ref{CORR}).
The details of the LLP theory are given in Appendix 1. The explicit form for
the energy is now%
\[
E=-\sum_{\mathbf{k}}\frac{\left\vert V_{k}\right\vert ^{2}}{\hbar
\omega_{\mathrm{LO}}+\frac{\hbar^{2}k^{2}}{2m_{b}}}=-\alpha\hbar\omega.
\]
This self energy is no longer divergent. The divergence is elmininated by the
quantum cut-off occurring at $k=\sqrt{2m_{b}\omega_{\mathrm{LO}}}/\hbar.$

For the self energy the LLP result is equivalent to the perturbation result.
The effective mass however is now given by%
\[
m^{\ast}=m_{b}\left(  1+\frac{\alpha}{6}\right)  ,
\]
a result, which follows if one considers the case $\mathbf{P\neq}0$ and which
is also exact for $\alpha\rightarrow0$. However, the LLP effective mass is
different from the perturbation result if $\alpha$ insreases.

{The LLP approximation has often been called \textquotedblleft
intermediate-coupling approximation\textquotedblright. However its range of
validity is the same as that of perturbation theory to order $\alpha$. The
significance of the LLP approximation consists of the flexibility of the
canonical transformations together with the fact that it puts the Fr\"{o}hlich
result on a variational basis. }

To order $\alpha^{2}$, the analytical expressions for the coefficients are
$\alpha^{2}$: $2\ln(\sqrt{2}+1)-{\frac{3}{2}}\ln2-{\frac{\sqrt{2}}{2}}%
\approx-0.01591962$ for the energy and ${\frac{4}{3}}\ln(\sqrt{2}+1)-{\frac
{2}{3}}\ln2-{\frac{5\sqrt{2}}{8}}+{\frac{7}{36}}\approx0.02362763$ for the
polaron mass \cite{R68}.

At present the following weak-coupling expansions are known: for the energy
\cite{S86,SS89}
\begin{equation}
\frac{E_{0}}{\hbar\omega_{\mathrm{LO}}}=-\alpha-0.0159196220\alpha
^{2}-0.000806070048\alpha^{3}-\ldots, \label{eq_7c}%
\end{equation}
and for the polaron mass \cite{R68}
\begin{equation}
\frac{m^{\ast}}{m_{b}}=1+\frac{\alpha}{6}+0.02362763\alpha^{2}+\ldots
\label{eq_7d}%
\end{equation}

\subsubsection{Strong coupling via a canonical transformation
[\textquotedblleft shift\textquotedblright-operators]\label{Strong}}

Historically, the strong coupling limit was studied before all other
treatments (Landau, Pekar {\cite{Pekar,LP48}}). Although it is only a formal
case because the actual crystals seems to have $\alpha$ values smaller than 5,
it is very interesting \ because it contains some indication of the
intermediate coupling too: approach the excitations from the strong coupling
limit and extrapolate to intermediate coupling is interesting because it is
expected that some specific strong coupling properties \textquotedblleft
survive\textquotedblright\ at intermediate coupling. In what follows, a
treatment, equivalent to that of Pekar, but in second quantization and written
with as much analogy to the LLP treatment as possible is given.

We start from the Fr\"{o}hlich Hamiltonian (\ref{eq_1a}). At strong coupling
one makes the assumption ({a \textquotedblleft
Produkt-Ansatz\textquotedblright)\ for the polaron wave-function%
\begin{equation}
|\Phi\rangle=|\varphi\rangle|f\rangle\label{eq_2a}%
\end{equation}
where $|\varphi\mathbf{\rangle}$ is the \textquotedblleft
electron-component\textquotedblright\ of the wave function (}$\left\langle
\varphi|\varphi\right\rangle =1).${The \textquotedblleft
field-component\textquotedblright\ of the wave function }$|f\rangle$
($\left\langle f|f\right\rangle =1)$ {parametrically depends on $|\varphi
\mathbf{\rangle}$. The Produkt-Ansatz (\ref{eq_2a}) --- or Born-Oppenheimer
approximation --- implies that the electron adiabatically follows the motion
of the atoms, while the field cannot follow the instantaneous motion of the
electron. Fr\"{o}hlich showed that the approximation (\ref{eq_2a}) leads to
results, which are only valid for sufficiently large $\alpha\rightarrow\infty
$, i.~e. in the strong-coupling regime. A more systematic analysis of
strong-coupling polarons based on canonical transformations applied to the
Hamiltonian (\ref{eq_1a}) was performed in Refs. \cite{BT49,B50,T51}. }

The expectation value for the energy is now:%
\[
\left\langle H\right\rangle =\left\langle \varphi\right\vert \frac
{\mathbf{p}^{2}}{2m_{b}}\left\vert \varphi\right\rangle +\left\langle
f\right\vert \left[  \sum_{\mathbf{k}}\hbar\omega_{\mathrm{LO}}a_{\mathbf{k}%
}^{+}a_{\mathbf{k}}+\sum_{\mathbf{k}}(V_{k}a_{\mathbf{k}}\rho_{\mathbf{k}%
}e^{i\mathbf{k\cdot r}}+V_{k}^{\ast}a_{\mathbf{k}}^{\dag}\rho_{\mathbf{k}%
}^{\ast})\right]  \left\vert f\right\rangle
\]
with
\[
\rho_{\mathbf{k}}=\left\langle \varphi\right\vert e^{i\mathbf{k\cdot r}%
}\left\vert \varphi\right\rangle .
\]
We wish to minimize $\left\langle H\right\rangle ,$ but also
\[
\left\langle f\right\vert \left[  \sum_{\mathbf{k}}\hbar\omega_{\mathrm{LO}%
}a_{\mathbf{k}}^{+}a_{\mathbf{k}}+\sum_{\mathbf{k}}(V_{k}a_{\mathbf{k}}%
\rho_{\mathbf{k}}e^{i\mathbf{k\cdot r}}+V_{k}^{\ast}a_{\mathbf{k}}^{\dag}%
\rho_{\mathbf{k}}^{\ast})\right]  \left\vert f\right\rangle
\]
has to be minimized. This expression will be minimized if $\left\vert
f\right\rangle $ is the ground state wave function of the \textquotedblleft
shifted\textquotedblright\ oscullator-type Hamiltonian. As we can diagonalize
this Hamiltonian exactly:%
\begin{align}
&  \sum_{\mathbf{k}}\hbar\omega_{\mathrm{LO}}a_{\mathbf{k}}^{+}a_{\mathbf{k}%
}+\sum_{\mathbf{k}}(V_{k}a_{\mathbf{k}}\rho_{\mathbf{k}}e^{i\mathbf{k\cdot r}%
}+V_{k}^{\ast}a_{\mathbf{k}}^{\dag}\rho_{\mathbf{k}}^{\ast})\label{SC1}\\
&  =\sum_{\mathbf{k}}\hbar\omega_{\mathrm{LO}}\left(  a_{\mathbf{k}}^{\dag
}+\frac{V_{k}\rho_{\mathbf{k}}}{\hbar\omega_{\mathrm{LO}}}\right)  \left(
a_{\mathbf{k}}+\frac{V_{k}^{\ast}\rho_{\mathbf{k}}^{\ast}}{\hbar
\omega_{\mathrm{LO}}}\right)  -\sum_{\mathbf{k}}\frac{\left\vert
V_{k}\right\vert ^{2}\left\vert \rho_{\mathbf{k}}\right\vert ^{2}}{\hbar
\omega_{\mathrm{LO}}},
\end{align}
we can apply a canonical transformation similar to (\ref{S}):%
\begin{equation}
S=\exp\left[  \sum_{\mathbf{k}}\left(  \frac{V_{k}\rho_{\mathbf{k}}}%
{\hbar\omega_{\mathrm{LO}}}a_{\mathbf{k}}-\frac{V_{k}^{\ast}\rho_{\mathbf{k}%
}^{\ast}}{\hbar\omega_{\mathrm{LO}}}a_{\mathbf{k}}^{\dag}\right)  \right]  ,
\end{equation}
which has the property:%

\[
S^{-1}a_{\mathbf{k}}S=a_{\mathbf{k}}-\frac{V_{k}^{\ast}\rho_{\mathbf{k}}%
}{\hbar\omega_{\mathrm{LO}}},S^{-1}a_{\mathbf{k}}^{\dag}S=a_{\mathbf{k}}%
^{\dag}-\frac{V_{k}\rho_{\mathbf{k}}^{\ast}}{\hbar\omega_{\mathrm{LO}}}.
\]
The transformed Hamiltonian is now:%

\begin{align*}
&  S^{-1}\left[  \sum_{\mathbf{k}}\hbar\omega_{\mathrm{LO}}a_{\mathbf{k}}%
^{+}a_{\mathbf{k}}+\sum_{\mathbf{k}}(V_{k}a_{\mathbf{k}}\rho_{\mathbf{k}%
}e^{i\mathbf{k\cdot r}}+V_{k}^{\ast}a_{\mathbf{k}}^{\dag}\rho_{\mathbf{k}%
}^{\ast})\right]  S\\
&  =\sum_{\mathbf{k}}\hbar\omega_{\mathrm{LO}}a_{\mathbf{k}}^{\dag
}a_{\mathbf{k}}-\sum_{\mathbf{k}}\frac{\left\vert V_{k}\right\vert
^{2}\left\vert \rho_{\mathbf{k}}\right\vert ^{2}}{\hbar\omega_{\mathrm{LO}}}.
\end{align*}
The phonon vacuum $\left\vert 0\right\rangle $ provides a minimum:%
\[
\left\langle 0\right\vert S^{-1}\left[  \sum_{\mathbf{k}}\hbar\omega
_{\mathrm{LO}}a_{\mathbf{k}}^{+}a_{\mathbf{k}}+\sum_{\mathbf{k}}%
(V_{k}a_{\mathbf{k}}\rho_{\mathbf{k}}e^{i\mathbf{k\cdot r}}+V_{k}^{\ast
}a_{\mathbf{k}}^{\dag}\rho_{\mathbf{k}}^{\ast})\right]  S\left\vert
0\right\rangle =-\sum_{\mathbf{k}}\frac{\left\vert V_{k}\right\vert
^{2}\left\vert \rho_{\mathbf{k}}\right\vert ^{2}}{\hbar\omega_{\mathrm{LO}}}.
\]
Hence, the Hamiltonian (\ref{SC1}) is minimized by the ground state wave
function%
\begin{equation}
S\left\vert 0\right\rangle =\exp\left[  \sum_{\mathbf{k}}\left(  \frac
{V_{k}\rho_{\mathbf{k}}}{\hbar\omega_{\mathrm{LO}}}a_{\mathbf{k}}-\frac
{V_{k}^{\ast}\rho_{\mathbf{k}}^{\ast}}{\hbar\omega_{\mathrm{LO}}}%
a_{\mathbf{k}}^{\dag}\right)  \right]  \left\vert 0\right\rangle .
\label{SCWF}%
\end{equation}
It gives the ground state energy%
\begin{equation}
E_{0}=\left\langle \varphi\right\vert \frac{\mathbf{p}^{2}}{2m_{b}}\left\vert
\varphi\right\rangle -\sum_{\mathbf{k}}\frac{\left\vert V_{k}\right\vert
^{2}\left\vert \rho_{\mathbf{k}}\right\vert ^{2}}{\hbar\omega_{\mathrm{LO}}},
\label{SC2}%
\end{equation}
which is still a functional of $\left\vert \varphi\right\rangle $. The
functionals $\rho_{\mathbf{k}}$ are different for differerent excitations.

\paragraph{Ground state of strong-coupling polarons}

For the ground state one considers a Gaussian wave function:%

\[
\left\vert \varphi_{"1s"}\right\rangle =C\exp\left(  -\frac{m_{b}\Omega_{0}%
}{2\hbar}r^{2}\right)
\]
with a variational parameter $\Omega_{0}$.%
\begin{align*}
\left\langle \varphi_{"1s"}|\varphi_{"1s"}\right\rangle  &  =C^{2}\int
d^{3}r\exp\left(  -\frac{m_{b}\Omega_{0}}{\hbar}r^{2}\right)  =C^{2}\left[
\int_{-\infty}^{\infty}dx\exp\left(  -\frac{m_{b}\Omega_{0}}{\hbar}%
x^{2}\right)  \right]  ^{3}\\
&  =C^{2}\left(  \frac{\sqrt{\pi}}{\frac{m_{b}\Omega_{0}}{\hbar}}\right)
^{3}=C^{2}\left(  \frac{\pi\hbar}{m_{b}\Omega_{0}}\right)  ^{3/2}=1\Rightarrow
C^{2}=\left(  \frac{m_{b}\Omega_{0}}{\pi\hbar}\right)  ^{3/2}%
\end{align*}%
\[
\left\vert \varphi_{"1s"}\right\rangle =\left(  \frac{m_{b}\Omega_{0}}%
{\pi\hbar}\right)  ^{3/4}\exp\left(  -\frac{m_{b}\Omega_{0}}{2\hbar}%
r^{2}\right)  .
\]
For the further use, we introduce a notation $C_{1}^{2}=\left(  \frac
{m_{b}\Omega_{0}}{\pi\hbar}\right)  ^{1/2}.$ Such a wave function is
consistent with the localization of the electron, which we expect for large
$\alpha$. The kinetic energy in (\ref{SC2}) for this function is calculated
using the representation of the operator $\mathbf{p}^{2}=-\hbar^{2}\nabla
^{2}=$ $-\hbar^{2}\left(  \nabla_{x}^{2}+\nabla_{x}^{2}+\nabla_{x}^{2}\right)
$:%
\begin{align*}
\left\langle \varphi_{"1s"}\right\vert \frac{\mathbf{p}^{2}}{2m_{b}}\left\vert
\varphi_{"1s"}\right\rangle  &  =-\frac{\hbar^{2}}{2m_{b}}C^{2}\int d^{3}%
r\exp\left(  -\frac{m_{b}\Omega_{0}}{2\hbar}r^{2}\right) \\
&  \times\left(  \nabla_{x}^{2}+\nabla_{x}^{2}+\nabla_{x}^{2}\right)
\exp\left(  -\frac{m_{b}\Omega_{0}}{2\hbar}r^{2}\right) \\
&  =-3\frac{\hbar^{2}}{2m_{b}}C^{2}\int_{-\infty}^{\infty}dx\exp\left(
-\frac{m_{b}\Omega_{0}}{2\hbar}x^{2}\right)  \nabla_{x}^{2}\exp\left(
-\frac{m_{b}\Omega_{0}}{2\hbar}x^{2}\right) \\
&  \times\int_{-\infty}^{\infty}dy\exp\left(  -\frac{m_{b}\Omega_{0}}{\hbar
}y^{2}\right)  \int_{-\infty}^{\infty}dz\exp\left(  -\frac{m_{b}\Omega_{0}%
}{\hbar}z^{2}\right)
\end{align*}%
\begin{align*}
&  =-3\frac{\hbar^{2}}{2m_{b}}C_{1}^{2}\int_{-\infty}^{\infty}dx\exp\left(
-\frac{m_{b}\Omega_{0}}{2\hbar}x^{2}\right) \\
&  \times\nabla_{x}\left[  -\frac{m_{b}\Omega_{0}}{\hbar}x\exp\left(
-\frac{m_{b}\Omega_{0}}{2\hbar}x^{2}\right)  \right] \\
&  =3\frac{\hbar^{2}}{2m_{b}}C_{1}^{2}\frac{m_{b}\Omega_{0}}{\hbar}%
\int_{-\infty}^{\infty}dx\left(  1-\frac{m_{b}\Omega_{0}}{\hbar}x^{2}\right)
\exp\left(  -\frac{m_{b}\Omega_{0}}{\hbar}x^{2}\right) \\
&  ==3\frac{\hbar^{2}}{m_{b}}C_{1}^{2}\frac{m_{b}\Omega_{0}}{\hbar}\int
_{0}^{\infty}dx\left(  1-\frac{m_{b}\Omega_{0}}{\hbar}x^{2}\right)
\exp\left(  -\frac{m_{b}\Omega_{0}}{\hbar}x^{2}\right) \\
&  =3\frac{\hbar^{2}}{m_{b}}C_{1}^{2}\frac{m_{b}\Omega_{0}}{\hbar}\left[
\frac{\sqrt{\pi}}{2\sqrt{\frac{m_{b}\Omega_{0}}{\hbar}}}-\frac{m_{b}\Omega
_{0}}{\hbar}\frac{\sqrt{\pi}}{4\sqrt{\left(  \frac{m_{b}\Omega_{0}}{\hbar
}\right)  ^{3}}}\right] \\
&  =3\frac{\hbar\Omega_{0}}{4}C_{1}^{2}\sqrt{\frac{\pi\hbar}{m_{b}\Omega_{0}}%
}=3\frac{\hbar\Omega_{0}}{4}\sqrt{\frac{m_{b}\Omega_{0}}{\pi\hbar}}\sqrt
{\frac{\pi\hbar}{m_{b}\Omega_{0}}}=\frac{3}{4}\hbar\Omega_{0}.
\end{align*}

The functional%
\begin{align}
\rho_{\mathbf{k}"1s"}  &  =\left\langle \varphi_{"1s"}\right\vert
e^{i\mathbf{k\cdot r}}\left\vert \varphi_{"1s"}\right\rangle =C^{2}\int
d^{3}r\exp\left(  -\frac{m_{b}\Omega_{0}}{\hbar}r^{2}+i\mathbf{k\cdot
r}\right) \nonumber\\
&  =C^{2}\int d^{3}r\exp\left(  -\frac{m_{b}\Omega_{0}}{\hbar}\left[
r^{2}+i\frac{\hbar}{m_{b}\Omega_{0}}\mathbf{k\cdot r-}\frac{\hbar^{2}k^{2}%
}{4m_{b}^{2}\Omega_{0}^{2}}+\frac{\hbar^{2}k^{2}}{4m_{b}^{2}\Omega_{0}^{2}%
}\right]  \right) \nonumber\\
&  =C^{2}\exp\left(  -\frac{m_{b}\Omega_{0}}{\hbar}\frac{\hbar^{2}k^{2}%
}{4m_{b}^{2}\Omega_{0}^{2}}\right)  \int d^{3}r\exp\left(  -\frac{m_{b}%
\Omega_{0}}{\hbar}\left[  \mathbf{r}+i\frac{\hbar}{2m_{b}\Omega_{0}}%
\mathbf{k}\right]  ^{2}\right)  \Rightarrow\nonumber\\
\rho_{\mathbf{k}"1s"}  &  =\exp\left(  -\frac{\hbar k^{2}}{4m_{b}\Omega_{0}%
}\right)  . \label{SCRHO}%
\end{align}

The second term in (\ref{SC2}) is then
\begin{align}
-\sum_{\mathbf{k}}\frac{\left\vert V_{k}\right\vert ^{2}\left\vert
\rho_{\mathbf{k}"1s"}\right\vert ^{2}}{\hbar\omega_{\mathrm{LO}}}  &
=-\frac{V}{\left(  2\pi\right)  ^{3}}%
%TCIMACRO{\dint }%
%BeginExpansion
{\displaystyle\int}
%EndExpansion
d^{3}k\frac{\hbar\omega_{\mathrm{LO}}}{k^{2}}\frac{4\pi\alpha}{V}\left(
\frac{\hbar}{2m_{b}\omega_{\mathrm{LO}}}\right)  ^{\frac{1}{2}}\exp\left(
-\frac{\hbar k^{2}}{2m_{b}\Omega_{0}}\right) \nonumber\\
&  =-\frac{\alpha\hbar\omega_{\mathrm{LO}}}{2\pi^{2}}.4\pi\left(  \frac{\hbar
}{2m_{b}\omega_{\mathrm{LO}}}\right)  ^{\frac{1}{2}}%
%TCIMACRO{\dint \nolimits_{0}^{\infty}}%
%BeginExpansion
{\displaystyle\int\nolimits_{0}^{\infty}}
%EndExpansion
dk\exp\left(  -\frac{\hbar k^{2}}{2m_{b}\Omega_{0}}\right) \nonumber\\
&  =-\frac{2\alpha\hbar\omega_{\mathrm{LO}}}{\pi}\left(  \frac{\hbar}%
{2m_{b}\omega_{\mathrm{LO}}}\right)  ^{\frac{1}{2}}\frac{\sqrt{\pi}}{2}%
\sqrt{\frac{2m_{b}\Omega_{0}}{\hbar}} =-\frac{\alpha\hbar}{\sqrt{\pi}}%
\sqrt{\Omega_{0}\omega_{\mathrm{LO}}}. \label{SCFIELD}%
\end{align}
The variational energy (\ref{SC2}) thus becomes%
\begin{equation}
E_{0}=\frac{3}{4}\hbar\Omega_{0}-\frac{\hbar\omega_{\mathrm{LO}}\alpha}%
{\sqrt{\pi}}\sqrt{\frac{\Omega_{0}}{\omega_{\mathrm{LO}}}}. \label{SC3}%
\end{equation}
Putting
\[
\frac{\partial E_{0}}{\partial\Omega_{0}}=0,
\]
one obtains%
\[
\frac{3}{4}=\frac{\alpha}{2\sqrt{\pi}}\sqrt{\frac{\omega_{\mathrm{LO}}}%
{\Omega_{0}}}\Longrightarrow\sqrt{\frac{\Omega_{0}}{\omega_{\mathrm{LO}}}%
}=\frac{2\alpha}{3\sqrt{\pi}}\Longrightarrow\frac{\Omega_{0}}{\omega
_{\mathrm{LO}}}=\frac{4}{9}\frac{\alpha^{2}}{\pi}\Longrightarrow
\]%
\begin{equation}
\Omega_{0}=\frac{4}{9}\frac{\alpha^{2}}{\pi}\omega_{\mathrm{LO}}. \label{SC4}%
\end{equation}
Substituting (\ref{SC4}) in (\ref{SC3}), we find t{he ground state energy of
the polaron $E_{0}$ (calculated with the energy of the uncoupled
electron-phonon system as zero energy):}
\[
E_{0}=\frac{3}{4}\hbar\frac{4}{9}\frac{\alpha^{2}}{\pi}\omega_{\mathrm{LO}%
}-\frac{\hbar\omega_{\mathrm{LO}}\alpha}{\sqrt{\pi}}\frac{2\alpha}{3\sqrt{\pi
}}=\left(  \frac{1}{3}-\frac{2}{3}\right)  \frac{\alpha^{2}}{\pi}\hbar
\omega_{\mathrm{LO}}\Longrightarrow
\]%
\begin{equation}
E_{0}=-\frac{1}{3}\frac{\alpha^{2}}{\pi}\hbar\omega_{\mathrm{LO}}%
=-0.106\alpha^{2}\hbar\omega_{\mathrm{LO}}. \label{SC5}%
\end{equation}
{The strong-coupling mass of the polaron, resulting again from the
approximation (\ref{eq_2a}), is given \cite{E65} as:
\begin{equation}
m_{0}^{\ast}=0.0200\alpha^{4}m_{b}. \label{eq_3d}%
\end{equation}
More rigorous strong-coupling expansions for $E_{0}$ and $m^{\ast}$ have been
presented in the literature \cite{M75}:
\begin{equation}
\frac{E_{0}}{\hbar\omega_{\mathrm{LO}}}=-0.108513\alpha^{2}-2.836,
\label{eq_4a}%
\end{equation}%
\begin{equation}
\frac{m_{0}^{\ast}}{m_{b}}=1+0.0227019\alpha^{4}. \label{eq_4b}%
\end{equation}
}The strong-coupling ground state energy (\ref{SC5}) is lower than the LLP
ground state energy for $\alpha>10.$

\paragraph{The excited states of the polaron: SS, FC, RES}

In principle, excited states of the polaron exist at all coupling. In the
general case, and for simplicity for\textbf{ }$\mathbf{P}=0$, a continuum of
states starts at $\hbar\omega_{\mathrm{LO}}$ above the ground state of the
polaron. This continuum physically corresponds to the scattering of free
phonons on the polaron. Those \textquotedblleft
scattering.states\textquotedblright\ (SS) were studied in \cite{DE64} anf for
the first time more generally in \cite{E65} are not the only excitations of
the polaron. There are also \textit{internal} excitation states corresponding
to the excitations of the electron in the potential it created itself. By
analogy with the excited states of colour centers, the following terminology
is used.

(i) The states where the electron is excited in the potential belonging to the
ground state configuration of the lattice are called {\textit{Franck-Condon}
(FC) states }

(ii) E{xcitations of the electron in which the lattice polarization is adapted
to the electronic configuration of the excited electron (which itself then
adapts its wave function to the new potential, etc. \ldots leading to a
self-consistent final state), are called \textit{relaxed excited state} (RES)
\cite{Pekar}.}

\paragraph{Calculation of the lowest FC state}

The formalism used until now is well adapted to treat the polaron excitations
at strong coupling. The field dependence of the wave function is (\ref{SCWF}).
For the FC\ state the $\rho_{\mathbf{k}}$ are the same as for the ground state
(\ref{SCRHO}). Physically $\rho_{\mathbf{k}}$ tells us, to what electronic
distribution the field is adapted. The electronic part of the excited wave
function is $2p$-like:%
\begin{equation}
\left\vert \varphi_{"2p"}\right\rangle =C_{"2p"}z\exp\left(  -\frac
{m_{b}\Omega_{p}}{2\hbar}r^{2}\right)  \label{SCWF2P}%
\end{equation}
with a parameter $\Omega_{p}$, which is equal to $\Omega_{0}:$%
\begin{align*}
\left\langle \varphi_{"2p"}|\varphi_{"2p"}\right\rangle  &  =C_{"2p"}^{2}%
\int_{-\infty}^{\infty}dzz^{2}\exp\left(  -\frac{m_{b}\Omega_{p}}{\hbar}%
z^{2}\right)  \left[  \int_{-\infty}^{\infty}dx\exp\left(  -\frac{m_{b}%
\Omega_{p}}{\hbar}x^{2}\right)  \right]  ^{2}\\
&  =C_{"2p"}^{2}\left(  \frac{\sqrt{\pi}}{\frac{m_{b}\Omega_{p}}{\hbar}%
}\right)  ^{2}\frac{\sqrt{\pi}}{2\left(  \frac{m_{b}\Omega_{p}}{\hbar}\right)
^{3/2}}=C_{"2p"}^{2}\left(  \frac{\pi\hbar}{m_{b}\Omega_{p}}\right)
^{3/2}\frac{\hbar}{2m_{b}\Omega_{p}}=1\Rightarrow\\
C_{"2p"}^{2}  &  =\left(  \frac{m_{b}\Omega_{p}}{\pi\hbar}\right)  ^{3/2}%
\frac{2m_{b}\Omega_{p}}{\hbar}%
\end{align*}%
\[
\left\vert \varphi_{"2p"}\right\rangle =\left(  \frac{m_{b}\Omega_{0}}%
{\pi\hbar}\right)  ^{3/4}\left(  \frac{2m_{b}\Omega_{p}}{\hbar}\right)
^{1/2}z\exp\left(  -\frac{m_{b}\Omega_{0}}{2\hbar}r^{2}\right)  .
\]
We introduce still a notation
\[
C_{2}^{2}=\frac{2}{\sqrt{\pi}}\left(  \frac{m_{b}\Omega_{p}}{\hbar}\right)
^{3/2}%
\]
The FC state energy is, similarly to (\ref{SC2}),%
\begin{equation}
E_{FC}=\left\langle \varphi_{"2p"}\right\vert \frac{\mathbf{p}^{2}}{2m_{b}%
}\left\vert \varphi_{"2p"}\right\rangle -\sum_{\mathbf{k}}\frac{\left\vert
V_{k}\right\vert ^{2}\left\vert \rho_{\mathbf{k}"1s"}\right\vert ^{2}}%
{\hbar\omega_{\mathrm{LO}}}. \label{SC6}%
\end{equation}
The kinetic energy term is
\begin{align}
\left\langle \varphi_{_{"2p"}}\right\vert \frac{\mathbf{p}^{2}}{2m_{b}%
}\left\vert \varphi_{_{"2p"}}\right\rangle  &  =-\frac{\hbar^{2}}{2m_{b}%
}C_{_{"2p"}}^{2}\int d^{3}rz\exp\left(  -\frac{m_{b}\Omega_{p}}{2\hbar}%
r^{2}\right) \nonumber\\
&  \times\left(  \nabla_{x}^{2}+\nabla_{x}^{2}+\nabla_{x}^{2}\right)  \left[
z\exp\left(  -\frac{m_{b}\Omega_{p}}{2\hbar}r^{2}\right)  \right] \nonumber\\
&  =-\frac{\hbar^{2}}{2m_{b}}C_{_{"2p"}}^{2}\int_{-\infty}^{\infty}%
dzz\exp\left(  -\frac{m_{b}\Omega_{p}}{2\hbar}z^{2}\right)  \nabla_{z}%
^{2}\left[  z\exp\left(  -\frac{m_{b}\Omega_{p}}{2\hbar}z^{2}\right)  \right]
\nonumber\\
&  \times\int_{-\infty}^{\infty}dy\exp\left(  -\frac{m_{b}\Omega_{p}}{\hbar
}y^{2}\right)  \int_{-\infty}^{\infty}dz\exp\left(  -\frac{m_{b}\Omega_{p}%
}{\hbar}z^{2}\right)  +\frac{2}{4}\hbar\Omega_{p}\nonumber\\
&  =-\frac{\hbar^{2}}{2m_{b}}C_{2}^{2}\int_{-\infty}^{\infty}dzz\exp\left(
-\frac{m_{b}\Omega_{p}}{2\hbar}z^{2}\right) \nonumber\\
&  \times\nabla_{z}\left[  \left(  1-\frac{m_{b}\Omega_{p}}{\hbar}%
z^{2}\right)  \exp\left(  -\frac{m_{b}\Omega_{p}}{2\hbar}z^{2}\right)
\right]  +\frac{1}{2}\hbar\Omega_{p}\nonumber\\
&  =\frac{\hbar^{2}}{2m_{b}}C_{2}^{2}\int_{-\infty}^{\infty}dzz\exp\left(
-\frac{m_{b}\Omega_{p}}{2\hbar}z^{2}\right) \nonumber\\
&  \times\left[  \frac{2m_{b}\Omega_{p}}{\hbar}z+\left(  1-\frac{m_{b}%
\Omega_{p}}{\hbar}z^{2}\right)  \frac{m_{b}\Omega_{p}}{\hbar}z\right]
\exp\left(  -\frac{m_{b}\Omega_{p}}{2\hbar}z^{2}\right) \nonumber\\
+\frac{1}{2}\hbar\Omega_{p}  &  =\frac{\hbar^{2}}{2m_{b}}C_{2}^{2}%
\int_{-\infty}^{\infty}dz\left(  3-\frac{m_{b}\Omega_{p}}{\hbar}z^{2}\right)
\frac{m_{b}\Omega_{p}}{\hbar}z^{2}\exp\left(  -\frac{m_{b}\Omega_{p}}{\hbar
}z^{2}\right) \nonumber\\
+\frac{1}{2}\hbar\Omega_{p}  &  =\frac{\hbar\Omega_{p}}{2}C_{2}^{2}\left[
\frac{3\sqrt{\pi}}{2\left(  \frac{m_{b}\Omega_{p}}{\hbar}\right)  ^{3/2}%
}-\frac{m_{b}\Omega_{p}}{\hbar}\frac{3\sqrt{\pi}}{4\left(  \frac{m_{b}%
\Omega_{p}}{\hbar}\right)  ^{5/2}}\right]  +\frac{1}{2}\hbar\Omega
_{p}\nonumber\\
&  =\frac{\hbar\Omega_{p}}{2}C_{2}^{2}\frac{3\sqrt{\pi}}{4\left(  \frac
{m_{b}\Omega_{p}}{\hbar}\right)  ^{3/2}}+\frac{1}{2}\hbar\Omega_{p}\nonumber\\
&  =\frac{\hbar\Omega_{p}}{2}\frac{2}{\sqrt{\pi}}\left(  \frac{m_{b}\Omega
_{p}}{\hbar}\right)  ^{3/2}\frac{3\sqrt{\pi}}{4\left(  \frac{m_{b}\Omega_{p}%
}{\hbar}\right)  ^{3/2}}+\frac{1}{2}\hbar\Omega_{p}\nonumber\\
&  =\frac{3}{4}\hbar\Omega_{p}+\frac{1}{2}\hbar\Omega_{p}=\frac{5}{4}%
\hbar\Omega_{p}. \label{SCKIN}%
\end{align}
For the FC state, $\Omega_{p}=\Omega_{0}.$ The second term in (\ref{SC6}) is
precisely (\ref{SCFIELD}),
\[
-\sum_{\mathbf{k}}\frac{\left\vert V_{k}\right\vert ^{2}\left\vert
\rho_{\mathbf{k}"1s"}\right\vert ^{2}}{\hbar\omega_{\mathrm{LO}}}=-\frac
{\hbar}{\sqrt{\pi}}\sqrt{\Omega_{0}\omega_{\mathrm{LO}}},
\]
and the FC energy (\ref{SC6}) becomes%
\begin{align*}
E_{FC}  &  =\frac{5}{4}\hbar\Omega_{0}-\frac{\hbar\omega_{\mathrm{LO}}\alpha
}{\sqrt{\pi}}\sqrt{\frac{\Omega_{0}}{\omega_{\mathrm{LO}}}}\\
&  =\frac{5}{4}\hbar\frac{4}{9}\frac{\alpha^{2}}{\pi}\omega_{\mathrm{LO}%
}-\frac{2}{3}\frac{\alpha^{2}}{\pi}\hbar\omega_{\mathrm{LO}}=\left(  \frac
{5}{9}-\frac{2}{3}\right)  \frac{\alpha^{2}}{\pi}\hbar\omega_{\mathrm{LO}%
}\Longrightarrow
\end{align*}
{The energy of the lowest FC state is, within the Produkt-Ansatz
\cite{KED-FC}:
\begin{equation}
E_{\mathrm{FC}}=\frac{\alpha^{2}}{9\pi}\hbar\omega_{\mathrm{LO}}%
=0.0354\alpha^{2}\hbar\omega_{\mathrm{LO}}. \label{SC7}%
\end{equation}
The fact that this energy is positive, is presumably due to the choice of a
harmonic potential. The real potential the electron sees is anharmonic, and a
bound state may be expected.}

\paragraph{Calculation of RES}

The electronic part of the excited wave function is (\ref{SCWF2P}) with a
variational parameter $\Omega_{p},$which is determined below. The variational
RES energy is, similarly to (\ref{SC2}),
\begin{equation}
E_{RES}=\left\langle \varphi_{"2p"}\right\vert \frac{\mathbf{p}^{2}}{2m_{b}%
}\left\vert \varphi_{"2p"}\right\rangle -\sum_{\mathbf{k}}\frac{\left\vert
V_{k}\right\vert ^{2}\left\vert \rho_{\mathbf{k}"2p"}\right\vert ^{2}}%
{\hbar\omega_{\mathrm{LO}}}. \label{SC7a}%
\end{equation}
Here the kinetic energy\ term is given by Eq. (\ref{SCKIN}). The functional,
which is now needed, is
\begin{align*}
\rho_{\mathbf{k}"2p"}  &  =\left\langle \varphi_{"2p"}\right\vert
e^{i\mathbf{k\cdot r}}\left\vert \varphi_{"2p"}\right\rangle =C_{"2p"}^{2}\int
d^{3}rz^{2}\exp\left(  -\frac{m_{b}\Omega_{p}}{\hbar}r^{2}+i\mathbf{k\cdot
r}\right) \\
&  =C_{"2p"}^{2}\int d^{3}rz^{2}\exp\left(  -\frac{m_{b}\Omega_{p}}{\hbar
}\left[  r^{2}+i\frac{\hbar}{m_{b}\Omega_{p}}\mathbf{k\cdot r-}\frac{\hbar
^{2}k^{2}}{4m_{b}^{2}\Omega_{p}^{2}}+\frac{\hbar^{2}k^{2}}{4m_{b}^{2}%
\Omega_{p}^{2}}\right]  \right) \\
&  =C_{"2p"}^{2}\ exp\left(  -\frac{m_{b}\Omega_{p}}{\hbar}\frac{\hbar
^{2}k^{2}}{4m_{b}^{2}\Omega_{p}^{2}}\right)  \int d^{3}rz^{2}\exp\left(
-\frac{m_{b}\Omega_{p}}{\hbar}\left[  \mathbf{r}+i\frac{\hbar}{2m_{b}%
\Omega_{p}}\mathbf{k}\right]  ^{2}\right)  \Rightarrow\
\end{align*}%
\begin{align}
\rho_{\mathbf{k}"2p"}  &  =\exp\left(  -\frac{\hbar k^{2}}{4m_{b}\Omega_{p}%
}\right)  C_{2}^{2}\int_{-\infty}^{\infty}dzz^{2}\exp\left(  -\frac
{m_{b}\Omega_{p}}{\hbar}\left[  z+i\frac{\hbar}{2m_{b}\Omega_{p}}k_{z}\right]
^{2}\right) \nonumber\\
&  =.\exp\left(  -\frac{\hbar k^{2}}{4m_{b}\Omega_{p}}\right)  C_{2}^{2}%
\int_{-\infty}^{\infty}dz\left(  z-i\frac{\hbar}{2m_{b}\Omega_{p}}%
k_{z}\right)  ^{2}\exp\left(  -\frac{m_{b}\Omega_{p}}{\hbar}z^{2}\right)
\nonumber\\
&  =\exp\left(  -\frac{\hbar k^{2}}{4m_{b}\Omega_{p}}\right)  C_{2}^{2}%
\int_{-\infty}^{\infty}dz\left[  z^{2}-\left(  \frac{\hbar}{2m_{b}\Omega_{p}%
}k_{z}\right)  ^{2}\right]  ^{2}\exp\left(  -\frac{m_{b}\Omega_{p}}{\hbar
}z^{2}\right) \nonumber\\
&  =\exp\left(  -\frac{\hbar k^{2}}{4m_{b}\Omega_{p}}\right)  C_{2}%
^{2}\left\{  \frac{\sqrt{\pi}}{2\left(  \frac{m_{b}\Omega_{p}}{\hbar}\right)
^{3/2}}-\left(  \frac{\hbar}{2m_{b}\Omega_{p}}k_{z}\right)  ^{2}\frac
{\sqrt{\pi}}{\left(  \frac{m_{b}\Omega_{p}}{\hbar}\right)  ^{1/2}}\right\}
\nonumber\\
&  =\left(  1-\frac{\hbar k_{z}^{2}}{2m_{b}\Omega_{p}}\right)  \exp\left(
-\frac{\hbar k^{2}}{4m_{b}\Omega_{p}}\right)  . \label{SC8}%
\end{align}

Further, we substitute (\ref{SC8}) in the second term in the r.h.s. of Eq.
(\ref{SC7a}):%
\begin{align*}
-\sum_{\mathbf{k}}\frac{\left\vert V_{k}\right\vert ^{2}\left\vert
\rho_{\mathbf{k}"1s"}\right\vert ^{2}}{\hbar\omega_{\mathrm{LO}}}  &
=-\frac{V}{\left(  2\pi\right)  ^{3}}%
%TCIMACRO{\dint }%
%BeginExpansion
{\displaystyle\int}
%EndExpansion
d^{3}k\frac{\hbar\omega_{\mathrm{LO}}}{k^{2}}\frac{4\pi\alpha}{V}\left(
\frac{\hbar}{2m_{b}\omega_{\mathrm{LO}}}\right)  ^{\frac{1}{2}}\\
&  \times\left(  1-\frac{\hbar k_{z}^{2}}{2m_{b}\Omega_{p}}\right)  ^{2}%
\exp\left(  -\frac{\hbar k^{2}}{2m_{b}\Omega_{p}}\right) \\
&  =-\frac{\alpha\hbar\omega_{\mathrm{LO}}}{2\pi^{2}}\left(  \frac{\hbar
}{2m_{b}\omega_{\mathrm{LO}}}\right)  ^{\frac{1}{2}}2\pi%
%TCIMACRO{\dint \nolimits_{-1}^{1}}%
%BeginExpansion
{\displaystyle\int\nolimits_{-1}^{1}}
%EndExpansion
dx\\
&  \times%
%TCIMACRO{\dint \nolimits_{0}^{\infty}}%
%BeginExpansion
{\displaystyle\int\nolimits_{0}^{\infty}}
%EndExpansion
dk\left(  1-\frac{\hbar k^{2}x^{2}}{m_{b}\Omega_{p}}+\frac{\hbar^{2}k^{4}%
x^{4}}{4m_{b}^{2}\Omega_{p}^{2}}\right)  \exp\left(  -\frac{\hbar k^{2}%
}{2m_{b}\Omega_{p}}\right) \\
&  =-\frac{\alpha\hbar\omega_{\mathrm{LO}}}{\pi}\left(  \frac{\hbar}%
{2m_{b}\omega_{\mathrm{LO}}}\right)  ^{\frac{1}{2}}\\
&  \times%
%TCIMACRO{\dint \nolimits_{0}^{\infty}}%
%BeginExpansion
{\displaystyle\int\nolimits_{0}^{\infty}}
%EndExpansion
dk\left(  2-\frac{2}{3}\frac{\hbar k^{2}}{m_{b}\Omega_{p}}+\frac{2}{5}%
\frac{\hbar^{2}k^{4}}{4m_{b}^{2}\Omega_{p}^{2}}\right)  \exp\left(
-\frac{\hbar k^{2}}{2m_{b}\Omega_{p}}\right) \\
&  =-\frac{\alpha\hbar\omega_{\mathrm{LO}}}{\pi}\left(  \frac{\hbar}%
{2m_{b}\omega_{\mathrm{LO}}}\right)  ^{\frac{1}{2}}\\
&  \times\left[  \frac{\sqrt{\pi}}{\left(  \frac{\hbar}{2m_{b}\Omega_{p}%
}\right)  ^{1/2}}-\frac{1}{3}\frac{\hbar}{m_{b}\Omega_{p}}\frac{\sqrt{\pi}%
}{2\left(  \frac{\hbar}{2m_{b}\Omega_{p}}\right)  ^{3/2}}+\frac{1}{5}%
\frac{\hbar^{2}}{4m_{b}^{2}\Omega_{p}^{2}}\frac{3\sqrt{\pi}}{4\left(
\frac{\hbar}{2m_{b}\Omega_{p}}\right)  ^{5/2}}\right]
\end{align*}
\begin{align*}
&  =-\frac{\alpha\hbar}{\sqrt{\pi}}\sqrt{\omega_{\mathrm{LO}}\Omega_{p}%
}\left[  1-\frac{1}{3}+\frac{3}{20}\right] \\
&  =-\frac{\alpha\hbar}{\sqrt{\pi}}\sqrt{\omega_{\mathrm{LO}}\Omega_{p}}%
\frac{60-20+9}{60}=-\frac{49}{60}\frac{\alpha\hbar}{\sqrt{\pi}}\sqrt
{\omega_{\mathrm{LO}}\Omega_{p}}%
\end{align*}

The variational energy (\ref{SC7a}) becomes%
\[
E_{RES}=\frac{5}{4}\hbar\Omega_{p}-\frac{49}{60}\frac{\alpha\hbar}{\sqrt{\pi}%
}\sqrt{\omega_{\mathrm{LO}}\Omega_{p}}.
\]

Putting
\[
\frac{\partial E_{RES}}{\partial\Omega_{p}}=0,
\]
one obtains%
\[
\frac{5}{4}=\frac{49\alpha}{120\sqrt{\pi}}\sqrt{\frac{\omega_{\mathrm{LO}}%
}{\Omega_{p}}}\Longrightarrow\sqrt{\frac{\Omega_{p}}{\omega_{\mathrm{LO}}}%
}=\frac{49\alpha}{150\sqrt{\pi}}\Longrightarrow\frac{\Omega_{p}}%
{\omega_{\mathrm{LO}}}=\left(  \frac{49}{150}\right)  ^{2}\frac{\alpha^{2}%
}{\pi}\Longrightarrow
\]

\begin{align*}
E_{RES}  &  =\frac{5}{4}\hbar\left(  \frac{49}{150}\right)  ^{2}\frac
{\alpha^{2}}{\pi}\omega_{\mathrm{LO}}-\frac{\hbar\omega_{\mathrm{LO}}\alpha
}{\sqrt{\pi}}\frac{49}{60}\frac{49\alpha}{150\sqrt{\pi}}=\left(  \frac{5}%
{4}-\frac{5}{2}\right)  \left(  \frac{49}{150}\right)  ^{2}\frac{\alpha^{2}%
}{\pi}\hbar\omega_{\mathrm{LO}}\\
&  =-\frac{5}{4}\left(  \frac{49}{150}\right)  ^{2}\frac{\alpha^{2}}{\pi}%
\hbar\omega_{\mathrm{LO}}=-\frac{49^{2}}{120\times150}\frac{\alpha^{2}}{\pi
}\hbar\omega_{\mathrm{LO}}\Longrightarrow
\end{align*}
{The energy of the RES is (see Refs. \cite{DE64,E65}):
\begin{equation}
E_{\mathrm{RES}}=-0.042\alpha^{2}\hbar\omega_{\mathrm{LO}}. \label{eq_3c}%
\end{equation}
The effective mass of the polaron in the RES is given \cite{E65} as:
\begin{equation}
m_{RES}^{\ast}=0.621\frac{\alpha^{4}}{81\pi^{2}}m_{b}=0.0200\alpha^{4}m_{b}.
\end{equation}
}

\begin{figure}[h]
\begin{center}
\includegraphics[height=.3\textheight]{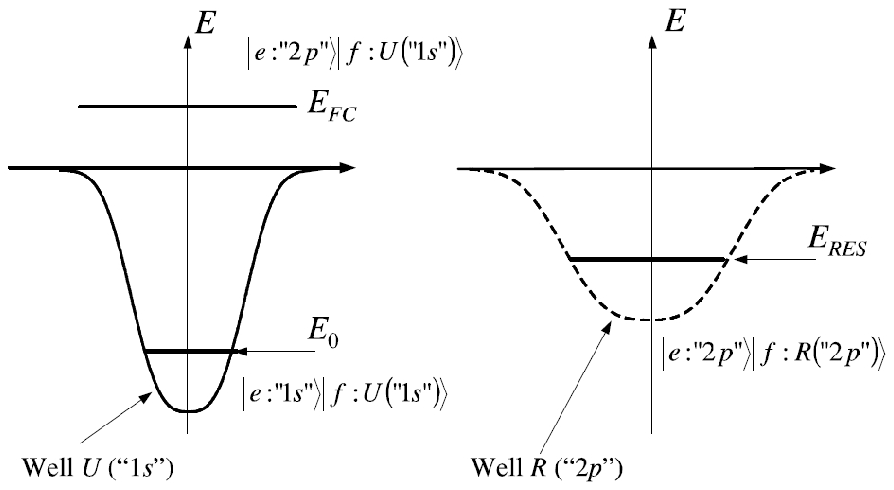}
\end{center}
\caption{Structure of the energy spectrum of a polaron at strong coupling:
$E_{0}$ --- the ground state, $E_{\mathrm{RES}}$ --- the (first) relaxed
excited state; the Franck-Condon states ($E_{\mathrm{FC}}$). In fact, both the
Franck-Condon states and the relaxed excited states lie in the continuum and,
strictly speaking, are resonances.}%
\label{fig_statesA}%
\end{figure}

{The structure of the energy spectrum of the strong-coupling polaron is shown
in Fig.\thinspace\ref{fig_statesA}.}

%\newpage

%\newpage

{The significance of the strong-coupling large polaron theory is formal only:
it allows to test \textquotedblleft all-coupling\textquotedblright\ theories
in the limit $\alpha\rightarrow\infty$. }Remarkably, the effective
electron-phonon coupling strength significantly increases in systems of low
dimension and low dimensionality.

\subsubsection{All-coupling theory. The Feynman path integral}

Feynman developed a superior all-coupling polaron theory using his
path-integral formalism\thinspace\cite{Feynman}. He studied first the
self-energy $E_{0}$ and the effective mass $m^{\ast}$ of polarons\thinspace
\cite{Feynman}.

Feynman got the idea to formulate the polaron problem into the Lagrangian form
of quantum mechanics and then eliminate the field oscillators,
\textquotedblleft\ldots in exact analogy to Q.~E.~D. \ldots(resulting in)
\ldots a sum over all trajectories \ldots\textquotedblright. The resulting
path integral (here limited to the ground-state properties) is of the form
(Ref. \cite{Feynman}):
\begin{equation}
\langle0,\beta|0,0\rangle\!=\! \int\! \mathcal{D}\mathbf{r}(\tau)\exp\left[
-\frac{1}{2}\int_{0}^{\beta}\mathbf{{\dot{r}}}^{2}d\tau\!+\! \frac{\alpha
}{2^{\frac{3}{2}}}\int_{0}^{\beta}\! \int_{0}^{\beta}\frac{e^{-|\tau-\sigma|}%
}{|\mathbf{r}(\tau)-\mathbf{r}(\sigma)|}d\tau d\sigma\right]  , \label{eq_8a}%
\end{equation}
where $\beta=1/(k_{B}T)$. (\ref{eq_8a}) gives the amplitude that an electron
found at a point in space at time zero will appear at the same point at the
(imaginary) time $\beta$. This path integral (\ref{eq_8a}) has a great
intuitive appeal: it shows the polaron problem as an equivalent one-particle
problem in which the interaction, non-local in time or \textquotedblleft
retarded\textquotedblright, occurs between the electron and itself.
Subsequently Feynman showed how the variational principle of quantum mechanics
could be adapted to the path-integral formalism and he introduced a quadratic
trial action (non-local in time) to simulate (\ref{eq_8a}).

Applying the variational principle for path integrals then results in an upper
bound for the polaron self-energy at all $\alpha$, which at weak and strong
coupling gives accurate expressions. Feynman obtained smooth interpolation
between a weak and strong coupling (for the ground state energy). The
weak-coupling expansions of Feynman for the ground-state energy and the
effective mass of the polaron are:
\begin{equation}
\frac{E_{0}}{\hbar\omega_{\mathrm{LO}}}=-\alpha-0.0123\alpha^{2}%
-0.00064\alpha^{3}-\ldots\>(\alpha\rightarrow0), \label{eq_7a}%
\end{equation}%
\begin{equation}
\frac{m^{\ast}}{m_{b}}=1+\frac{\alpha}{6}+0.025\alpha^{2}+\ldots
\>(\alpha\rightarrow0). \label{eq_7b}%
\end{equation}
In the strong-coupling limit Feynman found for the ground-state energy
energy:
\begin{equation}
{\frac{E_{0}}{\hbar\omega_{\mathrm{LO}}}\equiv\frac{E_{3D}(\alpha)}%
{\hbar\omega_{\mathrm{LO}}}=-0.106\alpha^{2}-2.83-\ldots\hspace*{0.3cm}%
(\alpha\rightarrow\infty)} \label{eq_9a}%
\end{equation}
and for the polaron mass:
\begin{equation}
\frac{m^{\ast}}{m_{b}}\equiv\frac{m_{3D}^{\ast}(\alpha)}{m_{b}}=0.0202\alpha
^{4}+\ldots\hspace*{0.3cm}(\alpha\rightarrow\infty). \label{eq_9b}%
\end{equation}

Over the years the Feynman model for the polaron has remained the most
successful approach to this problem. The analysis of an exactly solvable
(\textquotedblleft symmetrical\textquotedblright) 1D-polaron model
\cite{DE64,DE1968}, Monte Carlo schemes \cite{Mishchenko2000,Ciuchi} and other
numerical schemes \cite{DeFilippis2003} demonstrate the remarkable accuracy of
Feynman's path-integral approach to the polaron ground-state energy.
Experimentally more directly accessible properties of the polaron, such as its
mobility and optical absorption, have been investigated subsequently. Within
the path-integral approach, Feynman et al. studied later the mobility of
polarons\thinspace\cite{FHIP,TF70}. Subsequently the path-integral approach to
the polaron problem was generalized and developed to become a powerful tool to
study optical absorption, magnetophonon resonance and cyclotron resonance
\cite{KED1969,DSG1972,PD86,catau2,TDPRB01}.

In Ref. \cite{DEK1975}, a self-consistent treatment for the polaron problem at
all $\alpha$ was presented, which is based on the Heisenberg equations of
motion starting from a trial expression for the electron position. It was used
to derive the effective mass and the optical properties of the polaron at
arbitrary coupling. A variational justification of the approximation used
in\ Ref. \cite{DEK1975} (through a Stiltjes continuous fraction) is reproduced
in Appendix 2.

\subsubsection{On Monte Carlo calculations of the polaron free energy}

In Ref. \cite{bgsprb285735}, using a Monte Carlo calculation, the ground-state
energy of a polaron was derived as $E_{0}=\lim_{\beta\rightarrow\infty}\Delta
F,$ where $\Delta F=F_{\beta}-F_{\beta}^{0}$ with $F_{\beta}$ the free energy
per polaron and $F_{\beta}^{0}=\left[  3/\left(  2\beta\right)  \right]
\ln\left(  2\pi\beta\right)  $ the free energy per electron. The value
$\beta\hbar\omega_{\text{LO}}=25$, used for the actual computation in Ref.
\cite{bgsprb285735}, corresponds to $T/T_{D}=0.04$ ($T_{D}=\hbar
\omega_{\text{LO}}/k_{B};$ $\hbar\omega_{\text{LO}}$ is the LO phonon energy).
So, as pointed out in Ref. \cite{pdprb316826}, the authors of Ref.
\cite{bgsprb285735} actually calculated the \textit{free energy} $\Delta F$,
rather than the polaron \textit{ground-state energy}.

To investigate the importance of temperature effects on $\Delta F,$ the
authors of Ref. \cite{pdprb316826} considered the polaron energy as obtained
by Osaka \cite{osaka1959}, who generalized the Feynman \cite{Feynman} polaron
theory to nonzero temperatures:%
\begin{align}
\frac{\Delta F}{\hbar\omega}  &  =\frac{3}{\beta}\ln\left(  \frac{w}{v}%
\frac{\sinh\frac{\beta_{0}v}{2}}{\sinh\frac{\beta_{0}w}{2}}\right)  -\frac
{3}{4}\frac{v^{2}-w^{2}}{v}\left(  \coth\frac{\beta_{0}v}{2}-\frac{2}%
{\beta_{0}v}\right) \nonumber\\
&  -\frac{\alpha}{\sqrt{2\pi}}\left[  1+n\left(  \omega_{\text{LO}}\right)
\right]  \int_{0}^{\beta_{0}}du\frac{e^{-u}}{\sqrt{D\left(  u\right)  }},
\label{e1}%
\end{align}
where $\beta_{0}=\beta\hbar\omega_{\text{LO}},$ $n\left(  \omega\right)
=1/\left(  e^{\beta\hbar\omega}-1\right)  ,$ and%
\begin{equation}
D\left(  u\right)  =\frac{w^{2}}{v^{2}}\frac{u}{2}\left(  1-\frac{u}{\beta
_{0}}\right)  +\frac{v^{2}-w^{2}}{2v^{3}}\left(  1-e^{-vu}-4n\left(  v\right)
\sinh^{2}\frac{vu}{2}\right)  . \label{e2}%
\end{equation}
This result is variational, with variational parameters $v$ and $w,$ and gives
an upper bound to the exact polaron free energy.

\begin{figure}
[h]
\begin{center}
\includegraphics[
height=10.2121cm,
width=9.2346cm
]%
{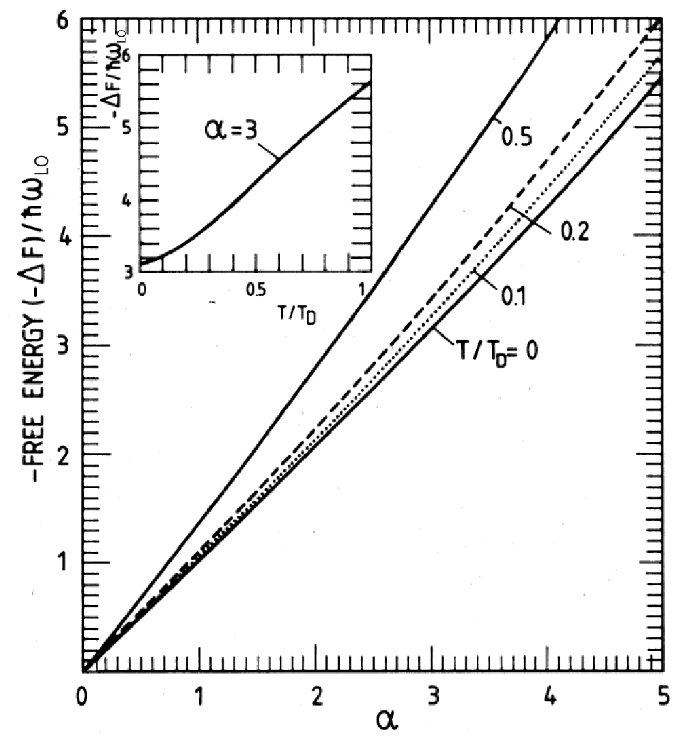}%
\caption{Contribution of the electron-phonon interaction to the free energy of
the Feynman polaron as a function of the electron-phonon coupling constant
$\alpha$ for different values of the lattice temperature. Inset: temperature
dependence of the free energy for $\alpha=3$. (From Ref. \cite{pdprb316826}.)}%
\label{pd316826}%
\end{center}
\end{figure}

The results of a numerical-variational calculation of Eq. (\ref{e1}) are shown
in Fig. \ref{pd316826}, where the free energy $-\Delta F$ is plotted (in units
of $\hbar\omega_{\text{LO}}$) as a function of $\alpha$ for different values
of the lattice temperature. As seen from Fig. \ref{pd316826}, (i) $-\Delta F$
increases with increasing temperature and (ii) the effect of temperature on
$\Delta F$ increases with increasing $\alpha$.

%\newpage%
%

%TCIMACRO{\FRAME{fhFU}{9.2346cm}{10.2121cm}{0pt}{\Qcb{Contribution of the
%electron-phonon interaction to the free energy of the Feynman polaron as a
%function of the electron-phonon coupling constant $\alpha$ for different
%values of the lattice temperature. Inset: temperature dependence of the free
%energy for $\alpha=3$. (From Ref. \cite{pdprb316826}.)}}{\Qlb{pd316826}%
%}{part1fig3.eps}{\special{ language "Scientific Word";  type "GRAPHIC";
%maintain-aspect-ratio TRUE;  display "ICON";  valid_file "F";
%width 9.2346cm;  height 10.2121cm;  depth 0pt;  original-width 11.3126cm;
%original-height 12.5208cm;  cropleft "0";  croptop "1";  cropright "1";
%cropbottom "0";  filename 'Part1Fig3.eps';file-properties "XNPEU";}} }%
%BeginExpansion
%EndExpansion

%\newpage

In Table \ref{tablePD316826}, the Monte Carlo results \cite{bgsprb285735},
$\left(  \Delta F\right)  _{\text{MC}}$, are compared with the free energy of
the Feynman polaron, $\left(  \Delta F\right)  _{\text{F}},$ calculated in
\cite{pdprb316826}. The values for the free energy obtained from the Feynman
polaron model are\textit{\ lower }than the MC results for $\alpha\lesssim2$
and $\alpha\geq4$ (but lie within the 1\% error of the Monte Carlo results).
Since the Feynman result for the polaron free energy is an upper bound to the
exact result, we conclude that for $\alpha\lesssim2$ and $\alpha\geq4$ the
results of the Feynman model are closer to the exact result than the MC
results of \cite{bgsprb285735}.%

%TCIMACRO{\TeXButton{B}{\begin{table}[htbp] \centering}}%
%BeginExpansion
\begin{table}[htbp] \centering
%EndExpansion
\caption{Comparison between the free energy of the Feynman polaron theory,
$-(\Delta F)_{\rm F}$, and the Monte Carlo results of
Ref.~\cite{bgsprb285735}, $-(\Delta F)_{\rm MC}$, for $T/T_D=0.04$. The
relative difference is defined as $\Delta=100\times[(\Delta F)_{\rm
F}-(\Delta F)_{\rm MC}]/(\Delta F)_{\rm MC}$. (From Ref.~\cite{pdprb316826})
\label{tablePD316826}}%
\begin{tabular}
[t]{|c|c|c|c|}\hline\hline
$\alpha$ & $-\left(  \Delta F\right)  _{\text{F}}$ & $-\left(  \Delta
F\right)  _{\text{MC}}$ & $\Delta$ (\%)\\\hline
$0.5$ & $0.50860$ & $0.505$ & $0.71$\\\hline
$1.0$ & $1.02429$ & $1.020$ & $0.42$\\\hline
$1.5$ & $1.54776$ & $1.545$ & $0.18$\\\hline
$2.0$ & $2.07979$ & $2.080$ & $-0.010$\\\hline
$2.5$ & $2.62137$ & $2.627$ & $-0.21$\\\hline
$3.0$ & $3.17365$ & $3.184$ & $-0.32$\\\hline
$3.5$ & $3.73814$ & $3.747$ & $-0.24$\\\hline
$4.0$ & $4.31670$ & $4.314$ & $0.063$\\\hline\hline
\end{tabular}%
%TCIMACRO{\TeXButton{E}{\end{table}}}%
%BeginExpansion
\end{table}%
%EndExpansion

\subsubsection{On the contributions of the $N$-phonon states to the polaron
ground state}

\begin{figure}
[h]
\begin{center}
\includegraphics[
height=15.6817cm,
width=10.4779cm
]%
{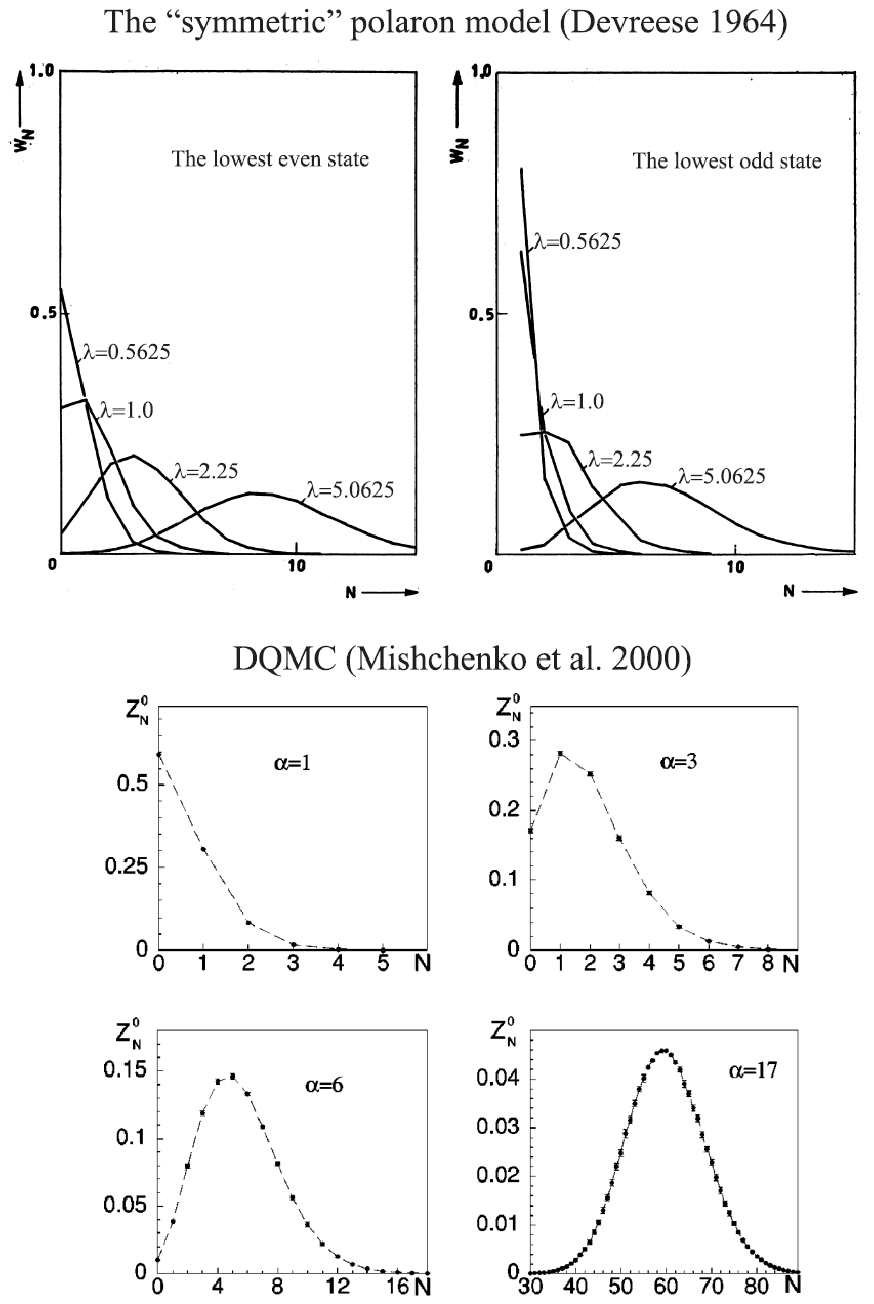}%
\caption{\textit{Upper panel}: The phonon distribution functions $W_{N}$ in
the \textquotedblleft symmetric\textquotedblright\ polaron model for various
values of the effective coupling constant $\lambda$ at $\kappa=1,\mathbf{P}=0$
(from \cite{DThesis}, Fig. 23). \textit{Lower panel}: Distribution of
multiphonon states in the polaron cloud within DQMC method for various values
of $\alpha$ (from \cite{Mishchenko2000}, Fig. 7). }%
\label{NPolarons}%
\end{center}
\end{figure}

The analysis of an exactly solvable (\textquotedblleft
symmetric\textquotedblright) 1D-polaron model was performed in Refs.
\cite{DThesis,DE64,DE68}. The model consists of an electron interacting with
two oscillators possessing the opposite wave vectors:\textbf{\ }$\mathbf{k}$
and {\normalsize -}$\mathbf{k.}$The parity operator, which changes
$a_{\mathbf{k}}$ and $a_{-\mathbf{k}}$ (and also $a_{\mathbf{k}}^{\dag}$ and
$a_{-\mathbf{k}}^{\dag}$), commutes with the Hamiltonian of the system. Hence,
the polaron states are classified into the even and odd ones with the
eigenvalues of the parity operator $+$1 and $-$1, respectively. For the lowest
even and odd states, the phonon distribution functions $W_{N}$ are plotted in
Fig. \ref{NPolarons}, upper panel, at some values of the effective coupling
constant $\lambda$ of the \textquotedblleft symmetric\textquotedblright%
\ model. The value of the parameter%
\[
\varkappa=\sqrt{\frac{\left(  \hbar k\right)  ^{2}}{m_{b}\hbar\omega
_{\mathrm{LO}}}}%
\]
for these graphs was taken 1, while the total polaron momentum $\mathbf{P}=0$.
In the weak-coupling case ($\lambda\approx0.6$) $W_{N}$ is a decaying function
of $N$. When increasing $\lambda$, $W_{N}$ acquires a maximum, e.g. at $N=8$
for the lowest even state at $\lambda\approx5.1$. The phonon distribution
function $W_{N\text{ }}$has the same character for the lowest even and the
lowest odd states at all values of the number of the virtual phonons in the
ground state. (as distinct from the higher states).\ This led to the
conclusion that the lowest odd state is an internal excited state of the polaron.

In Ref \cite{Mishchenko2000}, the structure of the polaron cloud was
investigated using the diagrammatic quantum Monte Carlo (DQMC) method. In
particular, partial contributions of $N$-phonon states to the polaron ground
state were found as a function of $N$ for a few values of the coupling
constant $\alpha,$ see Fig. \ref{NPolarons}, lower panel. It was shown to
gradually evolve from the weak-coupling case ($\alpha=1$) into the
strong-coupling regime ($\alpha=17$). Comparion of the lower panel to the
upper panel in Fig. \ref{NPolarons} clearly shows that the evolution of the
shape and the scale of the distribution of the $N$-phonon states with
increasing $\alpha$ as derived for a large polaron within DQMC method
\cite{Mishchenko2000} is \textit{in remarkable agreement }with the results
obtained within the "symmetric" 1D polaron model \cite{DThesis,DE64,DE68}.

\newpage%

%TCIMACRO{\FRAME{fhFU}{10.4779cm}{15.6817cm}{0pt}{\Qcb{\QTR{it}{Upper panel}:
%The phonon distribution functions $W_{N}$ in the \textquotedblleft
%symmetric\textquotedblright\ polaron model for various values of the effective
%coupling constant $\lambda$ at $\kappa=1,\QTR{bf}{P}=0$ (from \cite{DThesis},
%Fig. 23). \QTR{it}{Lower panel}: Distribution of multiphonon states in the
%polaron cloud within DQMC method for various values of $\alpha$ (from
%\cite{Mishchenko2000}, Fig. 7). }}{\Qlb{NPolarons}}{part1fig4.eps}%
%maintain-aspect-ratio TRUE;  display "ICON";  valid_file "F";
%width 10.4779cm;  height 15.6817cm;  depth 0pt;  original-width 14.6603cm;
%original-height 21.986cm;  cropleft "0";  croptop "1";  cropright "1";
%cropbottom "0";  filename 'Part1Fig4.eps';file-properties "XNPEU";}} }%
%BeginExpansion
%EndExpansion

\newpage

\subsection{Polaron mobility \footnote{\textbf{See also Appendix D
\textquotedblleft Notes on the polaron mobility\textquotedblright.}}}

The mobility of large polarons was studied within various theoretical
approaches(see Ref. \cite{PD1984} for the detailed references). Fr\"{o}hlich
\cite{Frohlich1937} pointed out the typical behavior of the large-polaron
mobility
\begin{equation}
\mu\propto\exp(\hbar\omega_{\mathrm{LO}}\beta),
\end{equation}
which is characteristic for weak coupling. Here, $\beta=1/k_{B}T$, $T$ is the
temperature. Within the weak-coupling regime, the mobility of the polaron was
then derived, e. g., using the Boltzmann equation in Refs.
\cite{HS1953,Osaka1961} and starting from the LLP-transformation in Ref.
\cite{LP1955}.

A nonperturbative analysis was embodied in the Feynman polaron theory, where
the mobility $\mu$ of the polaron using the path-integral formalism was
derived by Feynman et al. (usually referred to as FHIP) as a static limit
starting from a frequency-dependent impedance function. For sufficiently low
temperature $T$ the mobility then takes the form \cite{FHIP}
\begin{equation}
\mu=\left(  \frac{w}{v}\right)  ^{3}\frac{3e}{4m_{b}\hbar\omega_{\mathrm{LO}%
}^{2}\alpha\beta}\mbox{e}^{\hbar\omega_{\mathrm{LO}}\beta}\exp\{(v^{2}%
-w^{2})/w^{2}v\} \ , \label{eq:P24-1}%
\end{equation}
where $v$ and $w$ are (variational) functions of $\alpha$ obtained from the
Feynman polaron model.

Using the Boltzmann equation for the Feynman polaron model, Kadanoff
\cite{Kadanoff} found the mobility, which for low temperatures can be
represented as follows:
\begin{equation}
\mu=\left(  \frac{w}{v}\right)  ^{3}\frac{e}{2m_{b}\omega_{\mathrm{LO}}\alpha
}\mbox{e}^{\hbar\omega_{\mathrm{LO}}\beta}\exp\{(v^{2}-w^{2})/w^{2}v\} \ ,
\label{eq:P24-1-kadanoff}%
\end{equation}
The weak-coupling perturbation expansion of the low-temperature polaron
mobility as found using the Green's function technique \cite{LK1964} has
confirmed that the mobility derived from the Boltzmann equation is exceedingly
exact for weak coupling ($\alpha/6\ll1$) and at low temperatures ($k_{B}%
T\ll\hbar\omega_{\mathrm{LO}}$). As shown in Ref. \cite{Kadanoff}, the
mobility of Eq. (\ref{eq:P24-1}) differs by the factor of $3/(2\beta
\hbar\omega_{\mathrm{LO}})$ from that derived using the polaron Boltzmann
equation as given by Eq. (\ref{eq:P24-1-kadanoff}).

In the limit of weak electron-phonon coupling and low temperature, the FHIP
polaron mobility of Eq. (\ref{eq:P24-1}) differs by the factor of
$3/(2\beta\hbar\omega_{\mathrm{LO}})$ from the previous
result~\cite{HS1953,LP1955,Osaka1961}, which, as pointed out in Ref.
\cite{FHIP} and in later publications (see, e.g.,
Refs.\cite{Kadanoff,LK1964,PD1984}), is correct for $\beta\gg1$. As follows
from this comparison, the result of Ref.~\cite{FHIP} is not valid when
$T\rightarrow0$. As argued in Ref.~\cite{FHIP} and later confirmed, in
particular, in Ref.~\cite{PDpss1983} the above discrepancy can be attributed
to an interchange of two limits in calculating the impedance. In FHIP, for
weak electron-phonon coupling, one takes $\lim_{\Omega\rightarrow0}%
\lim_{\alpha\rightarrow0}$, whereas the correct order is $\lim_{\alpha
\rightarrow0}\lim_{\Omega\rightarrow0}$ ($\Omega$ is the frequency of the
applied electric field). It turns out that for the correct result the mobility
at low temperatures is predominantly limited by the absorption of phonons,
while in the theory of FHIP it is the emission of phonons which gives the
dominant contribution as $T$ goes to zero~\cite{PDpss1983}.

The analysis based on the Boltzmann equation takes into account the phonon
emission processes whenever the energy of the polaron is above the emission
threshold. The independent-collision model, which underlies the
Boltzmann-equation approach, however, fails in the \textquotedblleft strong
coupling regime\textquotedblright\ of the large polaron, when the thermal mean
free path becomes less than the de Broglie wavelength; in this case, the
Boltzmann equation cannot be expected to be adequate \cite{FHIP,HB1999}%
.\footnote{\textbf{For the polaron mobility in the weak- and strong-coupling
regimes, see also Appendix D \textquotedblleft Notes on the polaron
mobility\textquotedblright.}}

In fact, the expression (\ref{eq:P24-1}) for the polaron mobility was reported
to adequately describe the experimental data in several polar materials (see,
e.g., Refs. \cite{Brown1972,HB1999,Hendry2004}). Experimental work on alkali
halides and silver halides indicates that the mobility obtained from
Eq.~(\ref{eq:P24-1}) describes the experimental results quite accurately
\cite{Brown1972}. Measurements of mobility as a function of temperature for
photoexcited electrons in cubic $n$-type Bi$_{12}$SiO$_{20}$ are explained
well in terms of large polarons within the Feynman approach \cite{HB1999}. The
experimental findings on electron transport in crystalline TiO$_{2}$ (rutile
phase) probed by THz time-domain spectroscopy are quantitatively interpreted
within the Feynman model \cite{Hendry2004}. One of the reasons for the
agreement between theory based on Eq. (\ref{eq:P24-1}) and experiment is that
in the path-integral approximation to the polaron mobility, a Maxwellian
distribution for the electron velocities is assumed, when applying the
adiabatic switching on of the Fr\"{o}hlich interaction. Although such a
distribution is not inherent in the Fr\"{o}hlich interaction, its
incorporation tends to favor agreement with experiment because other
mechanisms (interaction with acoustic phonons etc.) cause a Gaussian distribution.

\section{Optical Absorption. Weak coupling}

\subsection{Optical absorption at weak coupling [within the perturbation
theory]}

\begin{figure}[h]
\begin{center}
\includegraphics[height=.2\textheight]{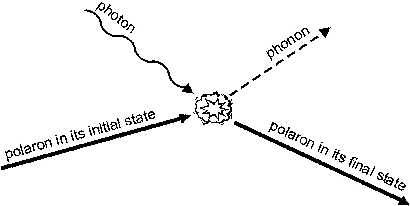}
\end{center}
\caption{Elementary polaron scattering process describing the absorption of an
incoming photon and the generation of an outgoing phonon.}%
\label{fig_scheme}%
\end{figure}

At zero temperature and in the weak-coupling limit, the optical absorption is
due to the elementary polaron scattering process, schematically shown in
Fig.\thinspace\ref{fig_scheme}.

%\newpage

%\newpage

In the weak-coupling limit ($\alpha\ll1$) the polaron absorption coefficient
was first obtained by V. Gurevich, I. Lang and Yu. Firsov \cite{GLF62}, who
started from the Kubo formula. Their optical-absorption coefficient is
equivalent to a particular case of the result of J. Tempere and J. T. Devreese
(Ref. \cite{TDPRB01}), with the dynamic srtucture factor $S(\mathbf{q}%
,\omega)$ corresponding to the Hartree-Fock approximation (see also
\cite{Mahan}, p. 585). At zero temperature, the absorption coefficient for
absorption of light with frequency $\Omega$ can be expressed in terms of
elementary functions in two limiting cases: in the region of comparatively
high polaron densities ($\hbar(\Omega-\omega_{\mathrm{LO}})/\zeta\ll1$)
\begin{equation}
\Gamma(\omega)=\frac{1}{\epsilon_{0}nc}\frac{2^{1/2}N^{2/3}\alpha}{(3\pi
^{2})^{1/3}}\frac{e^{2}}{(\hbar m_{b}\omega_{\mathrm{LO}})^{1/2}}{\frac
{\omega-1}{\omega^{3}}}\Theta(\omega-1), \label{weak}%
\end{equation}
and in the low-concentration region ($\hbar(\Omega-\omega_{\mathrm{LO}}%
)/\zeta\gg1$)
\begin{equation}
\Gamma(\omega)=\frac{1}{\epsilon_{0}nc}\frac{2Ne^{2}\alpha}{3m_{b}%
\omega_{\mathrm{LO}}}{\frac{(\omega-1)^{1/2}}{\omega^{3}}}\Theta(\omega-1),
\label{weak1}%
\end{equation}
where $\omega=\Omega/\omega_{\mathrm{LO}}$, $\epsilon_{0}$ is the dielectric
permittivity of the vacuum, $n$ is the refractive index of the medium, $N$ is
the concentration of polarons and $\zeta$ is the Fermi level for the
electrons. A step function
\[
\Theta(\omega-1)=\left\{
\begin{array}
[c]{lll}%
1 & \mbox{if} & \omega>1,\\
0 & \mbox{if} & \omega<1
\end{array}
\right.
\]
reflects the fact that at zero temperature the absorption of light accompanied
by the emission of a phonon can occur only if the energy of the incident
photon is larger than that of a phonon ($\omega>1$). In the weak-coupling
limit, according to Eqs.\thinspace(\ref{weak}), \ (\ref{weak1}), the
absorption spectrum consists of a \textquotedblleft one-phonon
line\textquotedblright.

At nonzero temperature, the absorption of a photon can be accompanied not only
by emission, but also by absorption of one or more phonons.

A simple derivation in Ref.\thinspace\cite{DHL1971} using a canonical
transformation method gives the absorption coefficient of free polarons, which
coincides with the result (\ref{weak1}) of Ref. \cite{GLF62}.

\subsection{Optical absorption at weak coupling [within the
canonical-transformation method \cite{DHL1971} (DHL)]}

The optical absorption of large polarons as a function of the frequency of the
incident light is calculated using the canonical-transformation formalism by
Devreese, Huybrechts and Lemmens (DHL) Ref. \cite{DHL1971}. A simple
calculation, which is developed below in full detail, gives a result for the
absorption coefficient, which is exact to order $\alpha.$

We start from the Hamiltonian of the electron-phonon system interacting with
light is written down using the vector potential of an electromagnetic field
$\mathbf{A}\left(  t\right)  $:%
\begin{equation}
H=\frac{1}{2m_{b}}\left(  \mathbf{p}+\frac{e}{c}\mathbf{A}\left(  t\right)
\right)  ^{2}+\sum_{\mathbf{k}}\omega_{\mathrm{LO}}a_{\mathbf{k}}%
^{+}a_{\mathbf{k}}+\sum_{\mathbf{k}}\left(  V_{\mathbf{k}}a_{\mathbf{k}%
}e^{i\mathbf{k\cdot r}}+V_{\mathbf{k}}^{\ast}a_{\mathbf{k}}^{+}%
e^{-i\mathbf{k\cdot r}}\right)  . \label{dhl-1}%
\end{equation}
The electric field is related to the vector potential as%
\begin{equation}
\mathbf{E}\left(  t\right)  =-\frac{1}{c}\frac{\partial\mathbf{A}\left(
t\right)  }{\partial t}. \label{dhl-2}%
\end{equation}
Within the electric dipole interaction the electric field with frequency
$\Omega$ is%

\begin{equation}
\mathbf{E}\left(  t\right)  =\mathbf{E}\cos\left(  \Omega t\right)
\Rightarrow\label{dhl-3}%
\end{equation}%
\begin{equation}
\mathbf{A}=-\frac{c}{\Omega}\mathbf{E}\sin\left(  \Omega t\right)  .
\label{dhl-4}%
\end{equation}
When expanding $\frac{1}{2m_{b}}\left(  \mathbf{p}+\frac{e}{c}\mathbf{A}%
\left(  t\right)  \right)  ^{2}$ in the Hamiltonian, we find%
\begin{equation}
\frac{1}{2m_{b}}\left(  \mathbf{p}+\frac{e}{c}\mathbf{A}\left(  t\right)
\right)  ^{2}=\frac{\mathbf{p}^{2}}{2m_{b}}+\frac{e}{m_{b}c}\mathbf{A}\left(
t\right)  \cdot\mathbf{p+}\frac{e^{2}}{2m_{b}c^{2}}A^{2}\left(  t\right)
\label{dhl-4a}%
\end{equation}
where the first term is the kinetic energy of the electron, and the second
term describes the interaction of the electron-phonon system with light%
\begin{equation}
V_{t}=\frac{e}{m_{b}c}\mathbf{A}\left(  t\right)  \cdot\mathbf{p=}-\frac
{e}{m_{b}\Omega}\mathbf{E\cdot p}\sin\Omega t \label{dhl-5}%
\end{equation}%
\begin{align}
V_{t}  &  \equiv V\sin\Omega t,\label{dhl-6}\\
V  &  \equiv-\frac{e}{m_{b}\Omega}\mathbf{E\cdot p}. \label{dhl-7}%
\end{align}
Since $\mathbf{A}\left(  t\right)  $ does not depend on the electron
coordinates, the term $\frac{e^{2}}{2m_{b}c^{2}}A^{2}\left(  t\right)  $ in
(\ref{dhl-4a}) does not play a role in our description of the optical
absorption. The total Hamiltonian for the system of a continuum polaron
interacting with light is thus%
\[
H_{tot}=H_{pol}+V_{t},
\]
where $H_{pol}$ is Fr\"{o}hlich's Hamiltonian (\ref{eq_1a}).

The absorption coefficient for absorption of light with frequency $\Omega$ by
free polarons is proportional to the probability $P(\Omega)$ that a photon is
absorbed by these polarons in their ground state,%
\begin{equation}
\Gamma_{p}(\Omega)=\frac{N}{\varepsilon_{0}cn2E^{2}}\Omega P(\Omega).
\label{DHL1}%
\end{equation}
Here $N$ is number of polarons, which are considered as independent from each
other, $\varepsilon_{0}$ is the permittivity of vacuum, $c$ is the velocity of
light, $n$ is the refractive index of the medium in which the polarons move,
$E$ is the modulus of the electric field vector of the incident photon. If the
incident light can be treated as a perturbation, the transition probability
$P(\Omega)$ is given by the Golden Rule of Fermi:%
\begin{equation}
P(\Omega)=2\pi%
%TCIMACRO{\dsum \nolimits_{f}}%
%BeginExpansion
{\displaystyle\sum\nolimits_{f}}
%EndExpansion
\left\langle \Phi_{0}\left\vert V\right\vert f\right\rangle \left\langle
f\left\vert V\right\vert \Phi_{0}\right\rangle \delta(E_{0}+\Omega-E_{f}).
\label{DHL2a}%
\end{equation}
$V$ is the amplitude of the time-dependent perturbation given by
(\ref{dhl-7}). \textbf{ }The ground state wave function of a free polaron is
$\left\vert \Phi_{0}\right\rangle $ and its energy is $E_{0}.$ The wave
functions of all possible final states are $\left\vert f\right\rangle $ and
the corresponding energies are $E_{f}$. The possible final states are all the
excited states of the polaron. The main idea of the present calculation is to
avoid the explicit summation over the final polaron states, which are poorly
known, by eliminating all the excited state wave functions $\left\vert
f\right\rangle $ from the expression (\ref{DHL2a}).

With this aim, the representation of the $\delta$-function is used:%
\[
\delta(x)=\frac{1}{\pi}\operatorname{Re}%
%TCIMACRO{\dint \nolimits_{-\infty}^{0}}%
%BeginExpansion
{\displaystyle\int\nolimits_{-\infty}^{0}}
%EndExpansion
dt\exp\left[  -i\left(  x+i\varepsilon\right)  t\right]  .
\]
This leads to:%
\begin{align*}
P(\Omega)  &  =2\operatorname{Re}%
%TCIMACRO{\dsum \nolimits_{f}}%
%BeginExpansion
{\displaystyle\sum\nolimits_{f}}
%EndExpansion%
%TCIMACRO{\dint \nolimits_{-\infty}^{0}}%
%BeginExpansion
{\displaystyle\int\nolimits_{-\infty}^{0}}
%EndExpansion
dt\left\langle \Phi_{0}\left\vert V\right\vert f\right\rangle \left\langle
f\left\vert V\right\vert \Phi_{0}\right\rangle \exp\left[  -i(\Omega
+i\varepsilon+E_{0}-E_{f})t\right] \\
&  =2\operatorname{Re}%
%TCIMACRO{\dsum \nolimits_{f}}%
%BeginExpansion
{\displaystyle\sum\nolimits_{f}}
%EndExpansion%
%TCIMACRO{\dint \nolimits_{-\infty}^{0}}%
%BeginExpansion
{\displaystyle\int\nolimits_{-\infty}^{0}}
%EndExpansion
dt\exp\left[  -i(\Omega+i\varepsilon)t\right]  \left\langle \Phi_{0}\left\vert
V\right\vert f\right\rangle \left\langle f\left\vert e^{iHt}Ve^{-iHt}%
\right\vert \Phi_{0}\right\rangle .
\end{align*}
Using the fact that%
\[%
%TCIMACRO{\dsum \nolimits_{f}}%
%BeginExpansion
{\displaystyle\sum\nolimits_{f}}
%EndExpansion
\left\vert f\right\rangle \left\langle f\right\vert =1
\]
and the notation%
\begin{align*}
e^{iHt}V(0)e^{-iHt}  &  =V(t),\\
\frac{dV(t)}{dt}  &  =i\left[  H,V(t)\right]
\end{align*}
we find%
\begin{equation}
P(\Omega)=2\operatorname{Re}%
%TCIMACRO{\dint \nolimits_{-\infty}^{0}}%
%BeginExpansion
{\displaystyle\int\nolimits_{-\infty}^{0}}
%EndExpansion
dt\exp\left[  -i(\Omega+i\varepsilon)t\right]  \left\langle \Phi_{0}\left\vert
V(0)V(t)\right\vert \Phi_{0}\right\rangle . \label{DHL8}%
\end{equation}
Defining%
\begin{equation}
R(\Omega)=%
%TCIMACRO{\dint \nolimits_{-\infty}^{0}}%
%BeginExpansion
{\displaystyle\int\nolimits_{-\infty}^{0}}
%EndExpansion
dt\exp\left[  -i(\Omega+i\varepsilon)t\right]  \left\langle \Phi_{0}\left\vert
V(0)V(t)\right\vert \Phi_{0}\right\rangle , \label{dhl-10}%
\end{equation}
one has%
\begin{equation}
P(\Omega)=2\operatorname{Re}R(\Omega). \label{DHL10}%
\end{equation}
Substituting (\ref{dhl-7}) to (\ref{dhl-10}), we find that%
\begin{equation}
R\left(  \Omega\right)  =\left(  \frac{e}{m_{b}\Omega}\right)  ^{2}%
\int_{-\infty}^{0}dte^{-i\left(  \Omega+i\varepsilon\right)  t}\left\langle
\Phi_{0}\left\vert \left(  \mathbf{E\cdot p}\left(  0\right)  \right)  \left(
\mathbf{E\cdot p}\left(  t\right)  \right)  \right\vert \Phi_{0}\right\rangle
\label{dhl-11}%
\end{equation}
and hence%
\begin{equation}
P\left(  \Omega\right)  =2\left(  \frac{e}{m_{b}\Omega}\right)  ^{2}%
\operatorname{Re}\int_{-\infty}^{0}dte^{-i\left(  \Omega+i\varepsilon\right)
t}\left\langle \Phi_{0}\left\vert \left(  \mathbf{E\cdot p}\left(  0\right)
\right)  \left(  \mathbf{E\cdot p}\left(  t\right)  \right)  \right\vert
\Phi_{0}\right\rangle . \label{dhl-12}%
\end{equation}

It is convenient to apply the first LLP transformation $S_{1}$(\ref{eq_6a}),
which eliminates the electron operators from the polaron Hamiltonian:%
\begin{align*}
H  &  \longrightarrow\mathcal{H}=S_{1}^{-1}H_{pol}S_{1}=\mathcal{H}%
_{0}+\mathcal{H}_{1}:\\
\mathcal{H}_{0}  &  =\frac{\mathbf{P}^{2}}{2m_{b}}+\sum_{\mathbf{k}}\left(
\omega_{\mathrm{LO}}+\frac{k^{2}}{2m_{b}}-\frac{\mathbf{k}\cdot\mathbf{P}%
}{m_{b}}\right)  a_{\mathbf{k}}^{\dag}a_{\mathbf{k}}+\sum_{\mathbf{k}}%
(V_{k}a_{\mathbf{k}}+V_{k}^{\ast}a_{\mathbf{k}}^{\dag}),\\
\mathcal{H}_{1}  &  =\frac{1}{2m_{b}}\sum_{\mathbf{k}}\mathbf{k\cdot
k}^{\prime}a_{\mathbf{k}}^{\dag}a_{\mathbf{k}^{\prime}}^{\dag}a_{\mathbf{k}%
}a_{\mathbf{k}^{\prime}},
\end{align*}
where $\mathcal{H}_{0}$ can be diagonalized exactly and gives rise to the
self-energy $E=-\alpha\omega_{\mathrm{LO}},$and $\mathcal{H}_{1}$ contains the
correlation effects between the phonons. The optical absorption will be
calculated here for the total momentum of the system $\mathbf{P=0.}$

In the LLP approximation the explicit form of the matrix element in
(\ref{dhl-11}) is%
\begin{equation}
\left\langle \Phi_{0}\left\vert \left(  \mathbf{E\cdot p}\left(  0\right)
\right)  \left(  \mathbf{E\cdot p}\left(  t\right)  \right)  \right\vert
\Phi_{0}\right\rangle =\left\langle 0\left\vert S_{2}^{-1}S_{1}^{-1}%
\mathbf{E}\cdot\mathbf{p}S_{1}S_{1}^{-1}\mathbf{E}\cdot\mathbf{p}(t)S_{1}%
S_{2}\right\vert 0\right\rangle , \label{DHL15}%
\end{equation}
where $S_{1}$and $S_{2}$ are the first (\ref{eq_6a}) and the second
(\ref{eq_6b}) LLP transformations. The application of $S_{1}$ gives:%
\[
S_{1}^{-1}\mathbf{p}(t)S_{1}=S_{1}^{-1}e^{iHt}\mathbf{p}e^{-iHt}S_{1}%
=S_{1}^{-1}e^{iHt}S_{1}S_{1}^{-1}\mathbf{p}S_{1}S_{1}^{-1}e^{-iHt}%
S_{1}\mathbf{.}%
\]

Using $\mathcal{H}=S_{1}^{-1}HS_{1},$ we arrive at $S_{1}^{-1}e^{iHt}%
S_{1}=e^{i\mathcal{H}t}.$Further we recall $S_{1}^{-1}\mathbf{p}%
S_{1}=\mathbf{P}-\sum_{\mathbf{k}}\hbar\mathbf{k}a_{\mathbf{k}}^{\dagger
}a_{\mathbf{k}}+\mathbf{p,}$ where $\mathbf{P=0}$ and \ $\mathbf{p}$ is set 0
(see Appendix 1). This results in%

\[
S_{1}^{-1}\mathbf{p}(t)S_{1}=e^{i\mathcal{H}t}\mathbf{p}e^{-i\mathcal{H}%
t}=-\sum_{\mathbf{k}}\hbar\mathbf{k}e^{i\mathcal{H}t}a_{\mathbf{k}}^{\dagger
}a_{\mathbf{k}}e^{-i\mathcal{H}t}=-\sum_{\mathbf{k}}\hbar\mathbf{k}%
a_{\mathbf{k}}^{\dagger}(t)a_{\mathbf{k}}(t).
\]
Then (\ref{DHL15}) takes the form%
\begin{equation}
\left\langle \Phi_{0}\left\vert \left(  \mathbf{E\cdot p}\left(  0\right)
\right)  \left(  \mathbf{E\cdot p}\left(  t\right)  \right)  \right\vert
\Phi_{0}\right\rangle =\left\langle 0\left\vert S_{2}^{-1}\left(
\sum_{\mathbf{k}}\mathbf{E}\cdot\mathbf{k}a_{\mathbf{k}}^{\dagger
}a_{\mathbf{k}}\right)  \left(  \sum_{\mathbf{k}}\mathbf{E}\cdot
\mathbf{k}a_{\mathbf{k}}^{\dagger}(t)a_{\mathbf{k}}(t)\right)  S_{2}%
\right\vert 0\right\rangle . \label{DHL17a}%
\end{equation}
Here the second LLP transformation is given by (\ref{eq_6b}) with%
\begin{equation}
f_{k}=-\frac{V_{k}^{\ast}}{\omega_{\mathrm{LO}}+\frac{k^{2}}{2m_{b}}}
\label{DHL17c}%
\end{equation}
and the vacuum is defined by $a_{\mathbf{k}}\left\vert 0\right\rangle =0.$ The
calculation of the matrix element (\ref{DHL17a}) proceeds as follows:%
\begin{align}
&  \left\langle 0\left\vert S_{2}^{-1}\left(  \sum_{\mathbf{k}}\mathbf{E}%
\cdot\mathbf{k}a_{\mathbf{k}}^{\dagger}a_{\mathbf{k}}\right)  \left(
\sum_{\mathbf{k}}\mathbf{E}\cdot\mathbf{k}a_{\mathbf{k}}^{\dagger
}(t)a_{\mathbf{k}}(t)\right)  S_{2}\right\vert 0\right\rangle \nonumber\\
&  =\left\langle 0\left\vert S_{2}^{-1}\left(  \sum_{\mathbf{k}}%
\mathbf{E}\cdot\mathbf{k}a_{\mathbf{k}}^{\dagger}a_{\mathbf{k}}\right)
S_{2}S_{2}^{-1}e^{i\mathcal{H}t}S_{2}S_{2}^{-1}\left(  \sum_{\mathbf{k}%
}\mathbf{E}\cdot\mathbf{k}a_{\mathbf{k}}^{\dagger}a_{\mathbf{k}}\right)
S_{2}S_{2}^{-1}e^{-i\mathcal{H}t}S_{2}\right\vert 0\right\rangle \nonumber\\
&  =\left\langle 0\left\vert S_{2}^{-1}\left(  \sum_{\mathbf{k}}%
\mathbf{E}\cdot\mathbf{k}a_{\mathbf{k}}^{\dagger}a_{\mathbf{k}}\right)
S_{2}e^{iS_{2}^{-1}\mathcal{H}S_{2}t}S_{2}^{-1}\left(  \sum_{\mathbf{k}%
}\mathbf{E}\cdot\mathbf{k}a_{\mathbf{k}}^{\dagger}a_{\mathbf{k}}\right)
S_{2}e^{-iS_{2}^{-1}\mathcal{H}S_{2}t}\right\vert 0\right\rangle .
\label{DHL18a}%
\end{align}

Further on, we calculate%
\[
S_{2}^{-1}\mathcal{H}S_{2}=H_{0}+H_{1}%
\]

where%
\[
H_{0}=S_{2}^{-1}\mathcal{H}_{0}S_{2}=S_{2}^{-1}\left[  \sum_{\mathbf{k}%
}\left(  \omega_{\mathrm{LO}}+\frac{k^{2}}{2m_{b}}\right)  a_{\mathbf{k}%
}^{\dag}a_{\mathbf{k}}+\sum_{\mathbf{k}}(V_{k}a_{\mathbf{k}}+V_{k}^{\ast
}a_{\mathbf{k}}^{\dag})\right]  S_{2}%
\]

Further we use $S_{2}^{-1}a_{\mathbf{k}}S_{2}=a_{\mathbf{k}}+f_{\mathbf{k}}$:%
\begin{align*}
H_{0}  &  =\sum_{\mathbf{k}}\left(  \omega_{\mathrm{LO}}+\frac{k^{2}}{2m_{b}%
}\right)  a_{\mathbf{k}}^{\dag}a_{\mathbf{k}}+\sum_{\mathbf{k}}\left(
\omega_{\mathrm{LO}}+\frac{k^{2}}{2m_{b}}\right)  \left\vert f_{\mathbf{k}%
}\right\vert ^{2}\\
&  +\sum_{\mathbf{k}}\left(  \omega_{\mathrm{LO}}+\frac{k^{2}}{2m_{b}}\right)
\left(  a_{\mathbf{k}}^{\dag}f_{\mathbf{k}}+a_{\mathbf{k}}f_{\mathbf{k}}%
^{\ast}\right)  +\sum_{\mathbf{k}}\left[  V_{k}\left(  a_{\mathbf{k}%
}+f_{\mathbf{k}}\right)  +V_{k}^{\ast}\left(  a_{\mathbf{k}}^{\dag
}+f_{\mathbf{k}}^{\ast}\right)  \right] \\
&  =\sum_{\mathbf{k}}\left(  \omega_{\mathrm{LO}}+\frac{k^{2}}{2m_{b}}\right)
a_{\mathbf{k}}^{\dag}a_{\mathbf{k}}+\sum_{\mathbf{k}}\frac{\left\vert
V_{k}\right\vert ^{2}}{\left(  \omega_{\mathrm{LO}}+\frac{k^{2}}{2m_{b}%
}\right)  }-2\sum_{\mathbf{k}}\frac{\left\vert V_{k}\right\vert ^{2}}{\left(
\omega_{\mathrm{LO}}+\frac{k^{2}}{2m_{b}}\right)  }\\
&  =\sum_{\mathbf{k}}\left(  \omega_{\mathrm{LO}}+\frac{k^{2}}{2m_{b}}\right)
a_{\mathbf{k}}^{\dag}a_{\mathbf{k}}-\sum_{\mathbf{k}}\frac{\left\vert
V_{k}\right\vert ^{2}}{\left(  \omega_{\mathrm{LO}}+\frac{k^{2}}{2m_{b}%
}\right)  }.
\end{align*}

The last term can be calculated analytically:%
\begin{align*}
\sum_{\mathbf{k}}\frac{\left\vert V_{k}\right\vert ^{2}}{\omega_{\mathrm{LO}%
}+\frac{k^{2}}{2m_{b}}}  &  =\frac{V}{\left(  2\pi\right)  ^{3}}\int
d^{3}k\left(  \frac{\omega_{\mathrm{LO}}}{k}\right)  ^{2}\frac{4\pi\alpha}%
{V}\left(  \frac{1}{2m_{b}\omega_{\mathrm{LO}}}\right)  ^{\frac{1}{2}}%
.\frac{1}{\omega_{\mathrm{LO}}+\frac{k^{2}}{2m_{b}}}\\
&  =\frac{\alpha\omega_{\mathrm{LO}}}{2\pi^{2}}4\pi%
%TCIMACRO{\dint \nolimits_{0}^{\infty}}%
%BeginExpansion
{\displaystyle\int\nolimits_{0}^{\infty}}
%EndExpansion
dk\left(  \frac{1}{2m_{b}\omega_{\mathrm{LO}}}\right)  ^{\frac{1}{2}}\frac
{1}{1+\frac{k^{2}}{2m_{b}\omega_{\mathrm{LO}}}}\\
&  =\frac{2\alpha\omega_{\mathrm{LO}}}{\pi}%
%TCIMACRO{\dint \nolimits_{0}^{\infty}}%
%BeginExpansion
{\displaystyle\int\nolimits_{0}^{\infty}}
%EndExpansion
d\kappa\frac{1}{1+\varkappa^{2}}=\frac{2\alpha\omega_{\mathrm{LO}}}{\pi
}\left.  \arctan\varkappa\right\vert _{0}^{\infty}=\alpha\omega_{\mathrm{LO}},
\end{align*}

\[
H_{0}=-\alpha\omega_{\mathrm{LO}}+\sum_{\mathbf{k}}\left(  \omega
_{\mathrm{LO}}+\frac{k^{2}}{2m_{b}}\right)  a_{\mathbf{k}}^{\dag}%
a_{\mathbf{k}}.
\]
The term%
\[
H_{1}=S_{2}^{-1}\mathcal{H}_{1}S_{2}%
\]
will be neglected:%
\[
e^{iS_{2}^{-1}\mathcal{H}S_{2}t}\approx e^{iH_{0}t}.
\]

Neglecting $H_{1}$, consistent with the LLP description, introduces no error
in order $\alpha$. Therefore (\ref{DHL18a}) becomes%
\begin{equation}
\left\langle 0\left\vert
\begin{array}
[c]{c}%
\sum_{\mathbf{k}}\mathbf{E}\cdot\mathbf{k}\left(  a_{\mathbf{k}}^{\dagger
}a_{\mathbf{k}}+f_{\mathbf{k}}a_{\mathbf{k}}^{\dagger}+f_{\mathbf{k}}^{\ast
}a_{\mathbf{k}}+f_{\mathbf{k}}f_{\mathbf{k}}^{\ast}\right)  e^{iH_{0}t}\\
\times\sum_{\mathbf{k}}\mathbf{E}\cdot\mathbf{k}\left(  a_{\mathbf{k}%
}^{\dagger}a_{\mathbf{k}}+f_{\mathbf{k}}a_{\mathbf{k}}^{\dagger}%
+f_{\mathbf{k}}^{\ast}a_{\mathbf{k}}+f_{\mathbf{k}}f_{\mathbf{k}}^{\ast
}\right)  e^{-iH_{0}t}%
\end{array}
\right\vert 0\right\rangle . \label{DHL19}%
\end{equation}
For\textbf{ }$\mathbf{P}=0$ there is no privileged direction and
$\sum_{\mathbf{k}}\mathbf{E}\cdot\mathbf{k}f_{\mathbf{k}}f_{\mathbf{k}}^{\ast
}=0,$ (\ref{DHL19}) reduces to:%
\[
\left\langle 0\left\vert \sum_{\mathbf{k}}\mathbf{E}\cdot\mathbf{k}%
f_{\mathbf{k}}^{\ast}a_{\mathbf{k}}e^{iH_{0}t}\sum_{\mathbf{k}}\mathbf{E}%
\cdot\mathbf{k}f_{\mathbf{k}}a_{\mathbf{k}}^{\dagger}e^{-iH_{0}t}\right\vert
0\right\rangle .
\]
From the equation of motion for $a_{\mathbf{k}}^{\dagger}:$%
\[
\frac{da_{\mathbf{k}}^{\dagger}(t)}{dt}=i\left[  H_{0},a_{\mathbf{k}}%
^{\dagger}\right]  =i\left(  \omega_{\mathrm{LO}}+\frac{k^{2}}{2m_{b}}\right)
a_{\mathbf{k}}^{\dag},
\]
it is easy now to calculate%
\[
e^{iH_{0}t}a_{\mathbf{k}}^{\dagger}e^{-iH_{0}t}=a_{\mathbf{k}}^{\dagger}%
\exp\left[  i\left(  \omega_{\mathrm{LO}}+\frac{k^{2}}{2m_{b}}\right)
t\right]  .
\]
The matrix element (\ref{DHL17a}) now becomes%
\[
\left\langle \Phi_{0}\left\vert \left(  \mathbf{E\cdot p}\left(  0\right)
\right)  \left(  \mathbf{E\cdot p}\left(  t\right)  \right)  \right\vert
\Phi_{0}\right\rangle =\sum_{\mathbf{k}}\left(  \mathbf{E}\cdot\mathbf{k}%
\right)  ^{2}f_{\mathbf{k}}^{\ast}f_{\mathbf{k}}\exp\left[  i\left(
\omega_{\mathrm{LO}}+\frac{k^{2}}{2m_{b}}\right)  t\right]  +O(\alpha^{2}).
\]
The transition probability (\ref{DHL8}) is then given by the expression%
\begin{align*}
P(\Omega)  &  =2\operatorname{Re}\frac{e^{2}}{m_{b}^{2}\Omega^{2}}%
\sum_{\mathbf{k}}\left(  \mathbf{E}\cdot\mathbf{k}\right)  ^{2}f_{\mathbf{k}%
}^{\ast}f_{\mathbf{k}}%
%TCIMACRO{\dint \nolimits_{-\infty}^{0}}%
%BeginExpansion
{\displaystyle\int\nolimits_{-\infty}^{0}}
%EndExpansion
dt\exp\left[  -i\left(  \Omega+i\varepsilon-\omega_{\mathrm{LO}}-\frac{k^{2}%
}{2m_{b}}\right)  t\right] \\
&  =2\pi\frac{e^{2}}{m_{b}^{2}\Omega^{2}}\sum_{\mathbf{k}}\left(
\mathbf{E}\cdot\mathbf{k}\right)  ^{2}\left\vert f_{\mathbf{k}}\right\vert
^{2}\delta\left(  \Omega-\omega_{\mathrm{LO}}-\frac{k^{2}}{2m_{b}}\right)  .
\end{align*}
Using (\ref{DHL17c}), we obtain%
\begin{align*}
P(\Omega)  &  =\frac{2\pi e^{2}}{m_{b}^{2}\Omega^{2}}\sum_{\mathbf{k}}%
\frac{\left(  \mathbf{E}\cdot\mathbf{k}\right)  ^{2}}{\left(  \omega
_{\mathrm{LO}}+\frac{k^{2}}{2m_{b}}\right)  ^{2}}\left\vert V_{k}\right\vert
^{2}\delta\left(  \Omega-\omega_{\mathrm{LO}}-\frac{k^{2}}{2m_{b}}\right) \\
&  =\frac{2\pi e^{2}}{m_{b}^{2}\Omega^{2}}\frac{V}{\left(  2\pi\right)  ^{3}%
}\int d^{3}k\left(  \frac{\omega_{\mathrm{LO}}}{k}\right)  ^{2}\frac
{4\pi\alpha}{V}\left(  \frac{1}{2m_{b}\omega_{\mathrm{LO}}}\right)  ^{\frac
{1}{2}}\\
&  \times\frac{\left(  \mathbf{E}\cdot\mathbf{k}\right)  ^{2}}{\left(
\omega_{\mathrm{LO}}+\frac{k^{2}}{2m_{b}}\right)  ^{2}}\delta\left(
\Omega-\omega_{\mathrm{LO}}-\frac{k^{2}}{2m_{b}}\right) \\
&  =\frac{e^{2}\alpha E^{2}}{m_{b}^{2}\Omega^{2}\pi}2\pi%
%TCIMACRO{\dint \nolimits_{-1}^{1}}%
%BeginExpansion
{\displaystyle\int\nolimits_{-1}^{1}}
%EndExpansion
dxx^{2}%
%TCIMACRO{\dint \nolimits_{0}^{\infty}}%
%BeginExpansion
{\displaystyle\int\nolimits_{0}^{\infty}}
%EndExpansion
dk\left(  \frac{1}{2m_{b}\omega_{\mathrm{LO}}}\right)  ^{\frac{1}{2}}\\
&  \times\frac{k^{2}}{\left(  1+\frac{k^{2}}{2m_{b}\omega_{\mathrm{LO}}%
}\right)  ^{2}}\frac{1}{\omega_{\mathrm{LO}}}\delta\left(  \frac{\Omega
}{\omega_{\mathrm{LO}}}-1-\frac{k^{2}}{2m_{b}\omega_{\mathrm{LO}}}\right) \\
&  =\frac{8e^{2}\alpha E^{2}}{3m_{b}\Omega^{2}}%
%TCIMACRO{\dint \nolimits_{0}^{\infty}}%
%BeginExpansion
{\displaystyle\int\nolimits_{0}^{\infty}}
%EndExpansion
d\kappa\frac{\varkappa^{2}}{\left(  1+\varkappa^{2}\right)  ^{2}}\delta\left(
\frac{\Omega}{\omega_{\mathrm{LO}}}-1-\varkappa^{2}\right) \\
&  =\frac{4e^{2}\alpha E^{2}}{3m_{b}\Omega^{2}}%
%TCIMACRO{\dint \nolimits_{0}^{\infty}}%
%BeginExpansion
{\displaystyle\int\nolimits_{0}^{\infty}}
%EndExpansion
d\zeta\frac{\sqrt{\zeta}}{\left(  1+\zeta\right)  ^{2}}\delta\left(
\frac{\Omega}{\omega_{\mathrm{LO}}}-1-\zeta\right) \\
&  =\frac{4e^{2}\alpha E^{2}}{3m_{b}\Omega^{2}}\Theta\left(  \frac{\Omega
}{\omega_{\mathrm{LO}}}-1\right)  \frac{\sqrt{\frac{\Omega}{\omega
_{\mathrm{LO}}}-1}}{\left(  \frac{\Omega}{\omega_{\mathrm{LO}}}\right)  ^{2}%
}=\frac{4e^{2}\alpha E^{2}\omega_{\mathrm{LO}}^{2}}{3m_{b}\Omega^{4}}%
\sqrt{\frac{\Omega}{\omega_{\mathrm{LO}}}-1}\quad\Theta\left(  \frac{\Omega
}{\omega_{\mathrm{LO}}}-1\right)  ,
\end{align*}
where%
\[
\Theta\left(  \frac{\Omega}{\omega_{\mathrm{LO}}}-1\right)  =\left\{
\begin{array}
[c]{cc}%
1 & \text{if\qquad}\frac{\Omega}{\omega_{\mathrm{LO}}}>1\\
0 & \text{if\qquad}\frac{\Omega}{\omega_{\mathrm{LO}}}<1
\end{array}
\right.  .
\]

The absorption coefficient (\ref{DHL1}) for absorption by\ free polarons for
$\alpha\longrightarrow0$ finally takes the form%
\begin{equation}
\Gamma_{p}(\Omega)=\frac{1}{\varepsilon_{0}cn}\frac{2Ne^{2}\alpha
\omega_{\mathrm{LO}}^{2}}{3m_{b}\Omega^{3}}\sqrt{\frac{\Omega}{\omega
_{\mathrm{LO}}}-1}\quad\Theta\left(  \frac{\Omega}{\omega_{\mathrm{LO}}%
}-1\right)  . \label{DHL24}%
\end{equation}

The behaviuor of $\Gamma_{p}(\Omega)$ (\ref{DHL24}) as a function of $\Omega$
is as follows. For $\Omega<\omega_{\mathrm{LO}}$ there is no absorption. The
threshold for absorption is at $\Omega=\omega_{\mathrm{LO}}.$From
$\Omega=\omega_{\mathrm{LO}}$ up to $\Omega=\frac{6}{5}\omega_{\mathrm{LO}%
},\Gamma_{p}$ increases to a maximum and for $\Omega>\frac{6}{5}%
\omega_{\mathrm{LO}}$ the absorption coefficient decreases slowly with
increasing $\Omega.$

\begin{figure}[h]
\begin{center}
\includegraphics[height=.3\textheight]{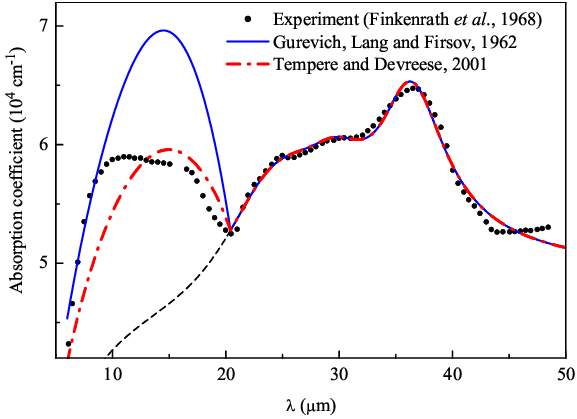}
\end{center}
\caption{Optical absorption spectrum of a CdO-film with the carrier
concentration $n=5.9\times10^{19}$ cm$^{-3}$ at $T=300$ K. The experimental
data (solid dots) of Ref.~\cite{Finkenrath} are compared to different
theoretical results: with (solid curve) and without (dashed line) the
one-polaron contribution of Ref. \cite{GLF62} and for many polarons
(dash-dotted curve) of Ref. \cite{TDPRB01}. }%
\label{fig_absCdO}%
\end{figure}

Experimentally, this one-phonon line has been observed for free polarons in
the infrared absorption spectra of CdO-films, see Fig.\thinspace
\ref{fig_absCdO}. In CdO, which is a weakly polar material with $\alpha
\approx0.74$, the polaron absorption band is observed in the spectral region
between 6 and 20 $\mu$m~(above the LO phonon frequency). The difference
between theory and experiment in the wavelength region where polaron
absorption dominates the spectrum is due to many-polaron effects.

%\newpage

%\newpage

\section{Optical absorption. Strong coupling}

The absorption of light by free large polarons was treated in
Ref.~\cite{KED1969} using the polaron states obtained wihtin the adiabatic
strong-coupling approximation, which was considered above in subsection
\ref{Strong}.

It was argued in Ref.~\cite{KED1969}, that for sufficiently large $\alpha$
($\alpha>3$), the (first) RES of a polaron is a relatively stable state, which
can participate in optical absorption transitions. This idea was necessary to
understand the polaron optical absorption spectrum in the strong-coupling
regime. The following scenario of a transition, which leads to a
\textit{\textquotedblleft zero-phonon\textquotedblright\ peak} in the
absorption by a strong-coupling polaron, can then be suggested. If the
frequency of the incoming photon is equal to
\[
\Omega_{\mathrm{RES}}\equiv{\frac{\text{E}_{\mathrm{RES}}\text{-}E_{0}}{\hbar
}}=0.065\alpha^{2}\omega_{\mathrm{LO}},
\]
then the electron jumps from the ground state (which, at large coupling, is
well-characterized by "$s$"-symmetry for the electron) to an excited state
("$2p$"), while the lattice polarization in the final state is adapted to the
"$2p$" electronic state of the polaron. In Ref. \cite{KED1969} considering the
decay of the RES with emission of one real phonon it is demonstrated, that the
\textquotedblleft zero-phonon\textquotedblright\ peak can be described using
the Wigner-Weisskopf formula valid when the linewidth of that peak is much
smaller than $\hbar\omega_{\mathrm{LO}}.$

For photon energies larger than
\[
\Omega_{\mathrm{RES}}+\omega_{\mathrm{LO}},
\]
a transition of the polaron towards the first scattering state, belonging to
the RES, becomes possible. The final state of the optical absorption process
then consists of a polaron in its lowest RES plus a free phonon. A
\textquotedblleft one-phonon sideband\textquotedblright\ then appears in the
polaron absorption spectrum. This process is called \textit{one-phonon
sideband absorption}.

The one-, two-, ... $K$-, ... phonon sidebands of the zero-phonon peak give
rise to a broad structure in the absorption spectrum. It turns that the
\textit{first moment} of the phonon sidebands corresponds to the FC frequency
$\Omega_{\mathrm{FC}}$:
\[
\Omega_{\mathrm{FC}}\equiv{\frac{E_{\mathrm{FC}}-E_{0}}{\hbar}}=0.141\alpha
^{2}\omega_{\mathrm{LO}}.
\]

To summarize, the polaron optical absorption spectrum at strong coupling is
characterized by the following features ($T=0$):

\begin{enumerate}
\item[a)] An intense absorption peak (``zero-phonon line'') appears, which
corresponds to a transition from the ground state to the first RES at
$\Omega_{\mathrm{RES}}$.

\item[b)] For $\Omega>\Omega_{\mathrm{RES}}+\omega_{\mathrm{LO}}$, a phonon
sideband structure arises. This sideband structure peaks around $\Omega
_{\mathrm{FC}}$.
\end{enumerate}

The qualitative behaviour predicted in Ref.\thinspace\cite{KED1969}, namely,
an intense zero-phonon (RES) line with a broader sideband at the
high-frequency side, was confirmed after an all-coupling expression for the
polaron optical absorption coefficient at $\alpha=5,6,7$ had been studied
\cite{DSG1972}.

In what precedes, the low-frequency end of the polaron absorption spectrum was
discussed; at higher frequencies, transitions to higher RES and their
scattering states can appear. The two-phonon sidebands in the optical
absorption of free polarons in the strong-coupling limit were numerically
studied in Ref. \cite{Goovaerts73}.

The study of the optical absorption of polarons at large coupling is mainly of
formal interest because all reported coupling constants of polar
semiconductors and ionic crystals are smaller than 5 (see Table 1).

\section{Arbitrary coupling}

\subsection{Impedance function of large polarons: An alternative derivation of
FHIP \cite{PD1983}}

\paragraph{Definitions}

We derive here the linear response of the Fr\"{o}hlich polaron, described by
the Hamiltonian%
\begin{equation}
H=\frac{\mathbf{p}^{2}}{2m_{b}}+\sum_{\mathbf{k}}\hbar\omega_{\mathbf{k}%
}a_{\mathbf{k}}^{+}a_{\mathbf{k}}+\sum_{\mathbf{k}}(V_{k}a_{\mathbf{k}%
}e^{i\mathbf{k\cdot r}}+V_{k}^{\ast}a_{\mathbf{k}}^{\dag}e^{-i\mathbf{k\cdot
r}}), \label{IF1}%
\end{equation}
to a spatially uniform, time-varying electric field%
\begin{equation}
\mathbf{E}_{\Omega}(t)=E_{0}\exp\left(  i\Omega t\right)  \mathbf{e}_{x}.
\label{IF2}%
\end{equation}
This field induces a current in the $x$-direction
\begin{equation}
j_{\Omega}(t)=\frac{1}{Z(\Omega)}E_{\Omega}(t). \label{IF3}%
\end{equation}
The complex function $Z(\Omega)$ is called the impedance function. The
frequency-dependent mobility is defined by
\begin{equation}
\mu(\Omega)=\operatorname{Re}\frac{1}{Z(\Omega)}. \label{IF4}%
\end{equation}
For nonzero frequencies (in the case of polarons the frequencies of interest
are in the infrared) one defines the absorption coefficient \cite{DSG1972}%
\begin{equation}
\Gamma(\Omega)=\frac{1}{n\epsilon_{0}c}\operatorname{Re}\frac{1}{Z(\Omega)},
\label{IF5}%
\end{equation}
where $\epsilon_{0}$ is the dielectric constant of the vacuum, $n$ the
refractive index of the crystal, and $c$ the velocity of light. In the
following the amplitude of the electric field $E_{0}$ is taken sufficiently
small so that linear-response theory can be applied.

The impedance function can be expressed via a frequency-dependent conductivity
of a single polaron in a unit volume
\begin{equation}
\frac{1}{Z(\Omega)}=\sigma(\Omega) \label{IFCond}%
\end{equation}
using \textit{the standard Kubo formula} (cf. Eq. (3.8.8) from Ref.
\cite{Mahan}):%
\begin{equation}
\sigma(\Omega)=i\frac{e^{2}}{Vm_{b}\Omega}+\frac{1}{V\hbar\Omega}\int
_{0}^{\infty}e^{i\Omega t}\left\langle \left[  j_{x}(t),j_{x}\right]
\right\rangle dt. \label{IFKubo}%
\end{equation}

In order to introduce a convenient representation of the impedance function,
we give in the next subsection a definition and discuss properties of a scalar
product of two operators [cf. \cite{Forster75}, Chapter 5].

\paragraph{Definition and properties of the scalar product}

For two operators $A$ and $B$ (i.e., elements of the Hilbert space of
operators) the scalar product is defined as
\begin{equation}
\left(  A,B\right)  =%
%TCIMACRO{\dint \nolimits_{0}^{\beta}}%
%BeginExpansion
{\displaystyle\int\nolimits_{0}^{\beta}}
%EndExpansion
d\lambda\left\langle \left(  e^{\lambda\hbar L}A^{\dag}\right)  B\right\rangle
. \label{IF8}%
\end{equation}
The notation $\left(  e^{\lambda\hbar L}A^{\dag}\right)  $ is used in order to
indicate that the operator $e^{\lambda\hbar L}$ acts on the operator $A^{\dag
}$. The time evolution of the operator $A$ is determined by the Liouville
operator $L$:%
\begin{equation}
-i\frac{\partial A}{\partial t}=LA\equiv\frac{1}{\hbar}\left[  H,A\right]
\label{IF9}%
\end{equation}
with a commutator $\left[  H,A\right]  $ , wherefrom%
\begin{equation}
A(t)=e^{iLt}A(0)\equiv e^{iHt/\hbar}A(0)e^{-iHt/\hbar}. \label{IF10}%
\end{equation}
The expectation value in (\ref{IF8}) is taken over the Gibbs' ensemble:%
\begin{equation}
\left\langle A\right\rangle =\mathrm{Tr}(\rho_{0}(H)A) \label{IF10a}%
\end{equation}
with the equilibtium density matrix when the electric field is absent
\begin{equation}
\rho_{0}(H)=e^{-\beta H}/\mathrm{Tr}(e^{-\beta H}). \label{IF10b}%
\end{equation}
One can show that (\ref{IF8}) defines a positive definite scalar product with
the following properties
\begin{align}
\text{(i)}\qquad(A,B)  &  =(B^{\dag},A^{\dag}),\label{IF11a}\\
\text{(ii)}\qquad(A,LB)  &  =(LA,B),\label{IF11b}\\
\text{(iii)}\qquad(A,LB)  &  =\frac{1}{\hbar}\left\langle \left[  A^{\dag
},B\right]  \right\rangle , \label{IF11c}%
\end{align}
and [cf. Eq. (5.11) of \cite{Forster75}]
\begin{equation}
\text{(iv)}\qquad(A,B)^{\ast}=(B,A) \label{IF11d}%
\end{equation}

\subparagraph{Demonstration of the property (\ref{IF11a}).}

Starting from the definition (\ref{IF8}) and using (\ref{IF10}), we obtain%
\begin{equation}
\left(  A,B\right)  =%
%TCIMACRO{\dint \nolimits_{0}^{\beta}}%
%BeginExpansion
{\displaystyle\int\nolimits_{0}^{\beta}}
%EndExpansion
d\lambda\left\langle e^{\lambda H}A^{\dag}e^{-\lambda H}B\right\rangle .
\label{IF11e}%
\end{equation}
Substituting here (\ref{IF10a}) with (\ref{IF10b}), one finds%
\[
\left(  A,B\right)  =%
%TCIMACRO{\dint \nolimits_{0}^{\beta}}%
%BeginExpansion
{\displaystyle\int\nolimits_{0}^{\beta}}
%EndExpansion
d\lambda\mathrm{Tr}\left[  e^{-(\beta-\lambda)H}A^{\dag}e^{-\lambda
H}B\right]  /\mathrm{Tr}(e^{-\beta H}).
\]
Change of the variable $\lambda^{\prime}=\beta-\lambda$ allows us to represnt
this integral as%
\[
\left(  A,B\right)  =%
%TCIMACRO{\dint \nolimits_{0}^{\beta}}%
%BeginExpansion
{\displaystyle\int\nolimits_{0}^{\beta}}
%EndExpansion
d\lambda^{\prime}\mathrm{Tr}\left[  e^{-\lambda^{\prime}H}A^{\dag}%
e^{-(\beta-\lambda^{\prime})H}B\right]  /\mathrm{Tr}(e^{-\beta H}).
\]
Further, a cyclic permutation of the operators under the trace $\mathrm{Tr}$
sign gives
\begin{align*}
\left(  A,B\right)   &  =%
%TCIMACRO{\dint \nolimits_{0}^{\beta}}%
%BeginExpansion
{\displaystyle\int\nolimits_{0}^{\beta}}
%EndExpansion
d\lambda\mathrm{Tr}\left[  e^{-(\beta-\lambda)H}Be^{-\lambda H}A^{\dag
}\right]  /\mathrm{Tr}(e^{-\beta H})\\
&  =%
%TCIMACRO{\dint \nolimits_{0}^{\beta}}%
%BeginExpansion
{\displaystyle\int\nolimits_{0}^{\beta}}
%EndExpansion
d\lambda\left\langle e^{\lambda H}Be^{-\lambda H}A^{\dag}\right\rangle =%
%TCIMACRO{\dint \nolimits_{0}^{\beta}}%
%BeginExpansion
{\displaystyle\int\nolimits_{0}^{\beta}}
%EndExpansion
d\lambda\left\langle \left(  e^{\lambda\hbar L}B\right)  A^{\dag}\right\rangle
\\
&  =%
%TCIMACRO{\dint \nolimits_{0}^{\beta}}%
%BeginExpansion
{\displaystyle\int\nolimits_{0}^{\beta}}
%EndExpansion
d\lambda\left\langle \left[  e^{\lambda\hbar L}\left(  B^{\dag}\right)
^{\dag}\right]  A^{\dag}\right\rangle .
\end{align*}
According to the definition (\ref{IF8}), this finalizes the demonstration of
(\ref{IF11a}).

\subparagraph{Demonstration of the property (\ref{IF11b}).}

Starting from (\ref{IF11e}) and using (\ref{IF9}), we obtain%
\begin{align*}
\left(  A,LB\right)   &  =%
%TCIMACRO{\dint \nolimits_{0}^{\beta}}%
%BeginExpansion
{\displaystyle\int\nolimits_{0}^{\beta}}
%EndExpansion
d\lambda\left\langle e^{\lambda H}A^{\dag}e^{-\lambda H}LB\right\rangle \\
&  =\frac{1}{\hbar}%
%TCIMACRO{\dint \nolimits_{0}^{\beta}}%
%BeginExpansion
{\displaystyle\int\nolimits_{0}^{\beta}}
%EndExpansion
d\lambda\left\langle e^{\lambda H}A^{\dag}e^{-\lambda H}\left(  HB-BH\right)
\right\rangle .
\end{align*}
A cyclic permutation of the operators under the average $\left\langle
\bullet\right\rangle $ sign gives%
\begin{equation}
\left(  A,LB\right)  =\frac{1}{\hbar}%
%TCIMACRO{\dint \nolimits_{0}^{\beta}}%
%BeginExpansion
{\displaystyle\int\nolimits_{0}^{\beta}}
%EndExpansion
d\lambda\left\langle e^{\lambda H}A^{\dag}e^{-\lambda H}HB-He^{\lambda
H}A^{\dag}e^{-\lambda H}B\right\rangle . \label{IF11f}%
\end{equation}
Using the commutation of $H$ and $e^{\pm\lambda H}$, one finds%
\begin{align}
\left(  A,LB\right)   &  =\frac{1}{\hbar}%
%TCIMACRO{\dint \nolimits_{0}^{\beta}}%
%BeginExpansion
{\displaystyle\int\nolimits_{0}^{\beta}}
%EndExpansion
d\lambda\left\langle e^{\lambda H}A^{\dag}He^{-\lambda H}B-e^{\lambda
H}HA^{\dag}e^{-\lambda H}B\right\rangle \\
&  =\frac{1}{\hbar}%
%TCIMACRO{\dint \nolimits_{0}^{\beta}}%
%BeginExpansion
{\displaystyle\int\nolimits_{0}^{\beta}}
%EndExpansion
d\lambda\left\langle e^{\lambda H}\left(  A^{\dag}H-HA^{\dag}\right)
e^{-\lambda H}B\right\rangle \label{IF11g}\\
&  =\frac{1}{\hbar}%
%TCIMACRO{\dint \nolimits_{0}^{\beta}}%
%BeginExpansion
{\displaystyle\int\nolimits_{0}^{\beta}}
%EndExpansion
d\lambda\left\langle e^{\lambda H}\left(  HA-AH\right)  ^{\dag}e^{-\lambda
H}B\right\rangle \\
&  =%
%TCIMACRO{\dint \nolimits_{0}^{\beta}}%
%BeginExpansion
{\displaystyle\int\nolimits_{0}^{\beta}}
%EndExpansion
d\lambda\left\langle e^{\lambda H}\left(  \frac{1}{\hbar}[H,A]\right)  ^{\dag
}e^{-\lambda H}B\right\rangle .
\end{align}
With the definition (\ref{IF9}), this gives
\[
\left(  A,LB\right)  =%
%TCIMACRO{\dint \nolimits_{0}^{\beta}}%
%BeginExpansion
{\displaystyle\int\nolimits_{0}^{\beta}}
%EndExpansion
d\lambda\left\langle e^{\lambda H}\left(  LA\right)  ^{\dag}e^{-\lambda
H}B\right\rangle =.%
%TCIMACRO{\dint \nolimits_{0}^{\beta}}%
%BeginExpansion
{\displaystyle\int\nolimits_{0}^{\beta}}
%EndExpansion
d\lambda\left\langle \left[  e^{\lambda\hbar L}\left(  LA\right)  ^{\dag
}\right]  B\right\rangle .
\]
According to the definition (\ref{IF8}), this finalizes the demonstration of
(\ref{IF11b}).

\subparagraph{Demonstration of the property (\ref{IF11c}).}

Starting from (\ref{IF11g}) and performing a cyclic permutation of the
operators under the average $\left\langle \bullet\right\rangle ,$ we find
\[
\left(  A,LB\right)  =\frac{1}{\hbar}%
%TCIMACRO{\dint \nolimits_{0}^{\beta}}%
%BeginExpansion
{\displaystyle\int\nolimits_{0}^{\beta}}
%EndExpansion
d\lambda\left\langle e^{\lambda H}\left(  A^{\dag}H-HA^{\dag}\right)
e^{-\lambda H}B\right\rangle
\]
Further we notice that
\[
e^{\lambda H}\left(  A^{\dag}H-HA^{\dag}\right)  e^{-\lambda H}=-\frac
{d\left(  e^{\lambda H}A^{\dag}e^{-\lambda H}\right)  }{d\lambda},
\]
consequently,
\begin{align}
\left(  A,LB\right)   &  =-\frac{1}{\hbar}%
%TCIMACRO{\dint \nolimits_{0}^{\beta}}%
%BeginExpansion
{\displaystyle\int\nolimits_{0}^{\beta}}
%EndExpansion
d\lambda\left\langle \frac{d\left(  e^{\lambda H}A^{\dag}e^{-\lambda
H}\right)  }{d\lambda}B\right\rangle \nonumber\\
&  =-\frac{1}{\hbar}\left\langle
%TCIMACRO{\dint \nolimits_{0}^{\beta}}%
%BeginExpansion
{\displaystyle\int\nolimits_{0}^{\beta}}
%EndExpansion
d\lambda\frac{d\left(  e^{\lambda H}A^{\dag}e^{-\lambda H}\right)  }{d\lambda
}B\right\rangle \nonumber\\
&  =-\frac{1}{\hbar}\left\langle \left.  e^{\lambda H}A^{\dag}e^{-\lambda
H}B\right\vert _{0}^{\beta}\right\rangle =\frac{1}{\hbar}\left\langle \left(
A^{\dag}B-e^{\beta H}A^{\dag}e^{-\beta H}B\right)  \right\rangle \nonumber\\
&  =\frac{1}{\hbar}\mathrm{Tr}\left[  e^{-\beta H}\left(  A^{\dag}B-e^{\beta
H}A^{\dag}e^{-\beta H}B\right)  \right]  /\mathrm{Tr}(e^{-\beta H})\nonumber\\
&  =\frac{1}{\hbar}\mathrm{Tr}\left[  e^{-\beta H}A^{\dag}B-A^{\dag}e^{-\beta
H}B\right]  /\mathrm{Tr}(e^{-\beta H}).
\end{align}
Further, a cyclic permutation of the operators in the second term under the
trace $\mathrm{Tr}$ sign gives%
\begin{align*}
\left(  A,LB\right)   &  =\frac{1}{\hbar}\mathrm{Tr}\left[  e^{-\beta
H}A^{\dag}B-e^{-\beta H}BA^{\dag}\right]  /\mathrm{Tr}(e^{-\beta H})\\
&  =\frac{1}{\hbar}\mathrm{Tr}\left[  e^{-\beta H}\left(  A^{\dag}B-BA^{\dag
}\right)  \right]  /\mathrm{Tr}(e^{-\beta H})\\
&  =\frac{1}{\hbar}\left\langle A^{\dag}B-BA^{\dag}\right\rangle =\frac
{1}{\hbar}\left\langle \left[  A^{\dag},B\right]  \right\rangle .
\end{align*}
Thus, the property (\ref{IF11c}) has been demonstrated.

\subparagraph{Demonstration of the property (\ref{IF11d}).}

We start from the represntation of the scalar product (\ref{IF11e}) and take a
complex conjugate:%
\[
\left(  A,B\right)  ^{\ast}=%
%TCIMACRO{\dint \nolimits_{0}^{\beta}}%
%BeginExpansion
{\displaystyle\int\nolimits_{0}^{\beta}}
%EndExpansion
d\lambda\left\langle e^{\lambda H}A^{\dag}e^{-\lambda H}B\right\rangle ^{\ast
}=%
%TCIMACRO{\dint \nolimits_{0}^{\beta}}%
%BeginExpansion
{\displaystyle\int\nolimits_{0}^{\beta}}
%EndExpansion
d\lambda\left\langle B^{\dag}e^{-\lambda H}Ae^{\lambda H}\right\rangle .
\]
A cyclic permutation of the operators under the average $\left\langle
\bullet\right\rangle $ sign gives then%
\begin{align}
\left(  A,B\right)  ^{\ast}  &  =%
%TCIMACRO{\dint \nolimits_{0}^{\beta}}%
%BeginExpansion
{\displaystyle\int\nolimits_{0}^{\beta}}
%EndExpansion
d\lambda\left\langle e^{\lambda H}B^{\dag}e^{-\lambda H}A\right\rangle =%
%TCIMACRO{\dint \nolimits_{0}^{\beta}}%
%BeginExpansion
{\displaystyle\int\nolimits_{0}^{\beta}}
%EndExpansion
d\lambda\left\langle e^{\lambda H}B^{\dag}e^{-\lambda H}A\right\rangle
\nonumber\\
&  =%
%TCIMACRO{\dint \nolimits_{0}^{\beta}}%
%BeginExpansion
{\displaystyle\int\nolimits_{0}^{\beta}}
%EndExpansion
d\lambda\left\langle \left(  e^{\lambda\hbar L}B^{\dag}\right)  A\right\rangle
.
\end{align}
According to the definition (\ref{IF8}), this finalizes the demonstration of
(\ref{IF11d}).

The above scalar product allows one to represent different dynamical
quantities in a rather simple way. For example, let us consider a scalar
product
\begin{align}
\Phi_{AB}\left(  z\right)   &  =\left(  A,\frac{1}{z-L}B\right) \label{IF12}\\
&  =%
%TCIMACRO{\dint \nolimits_{0}^{\beta}}%
%BeginExpansion
{\displaystyle\int\nolimits_{0}^{\beta}}
%EndExpansion
d\lambda\left\langle \left(  e^{\lambda\hbar L}A^{\dag}\right)  \frac{1}%
{z-L}B\right\rangle \nonumber\\
&  =-i%
%TCIMACRO{\dint \nolimits_{0}^{\beta}}%
%BeginExpansion
{\displaystyle\int\nolimits_{0}^{\beta}}
%EndExpansion
d\lambda\left\langle e^{\lambda\hbar L}A^{\dag}\left[
%TCIMACRO{\dint \nolimits_{0}^{\infty}}%
%BeginExpansion
{\displaystyle\int\nolimits_{0}^{\infty}}
%EndExpansion
dte^{i(z-L)t}\right]  B\right\rangle \label{IF12a}\\
&  =-i%
%TCIMACRO{\dint \nolimits_{0}^{\infty}}%
%BeginExpansion
{\displaystyle\int\nolimits_{0}^{\infty}}
%EndExpansion
dte^{izt}%
%TCIMACRO{\dint \nolimits_{0}^{\beta}}%
%BeginExpansion
{\displaystyle\int\nolimits_{0}^{\beta}}
%EndExpansion
d\lambda\left\langle e^{\lambda\hbar L}A^{\dag}e^{-iLt}B\right\rangle
\nonumber\\
&  =-i%
%TCIMACRO{\dint \nolimits_{0}^{\infty}}%
%BeginExpansion
{\displaystyle\int\nolimits_{0}^{\infty}}
%EndExpansion
dte^{izt}%
%TCIMACRO{\dint \nolimits_{0}^{\beta}}%
%BeginExpansion
{\displaystyle\int\nolimits_{0}^{\beta}}
%EndExpansion
d\lambda\left\langle e^{\lambda H}A^{\dag}e^{-\lambda H}e^{-iHt/\hbar
}Be^{iHt/\hbar}\right\rangle \nonumber\\
&  =-i%
%TCIMACRO{\dint \nolimits_{0}^{\infty}}%
%BeginExpansion
{\displaystyle\int\nolimits_{0}^{\infty}}
%EndExpansion
dte^{izt}%
%TCIMACRO{\dint \nolimits_{0}^{\beta}}%
%BeginExpansion
{\displaystyle\int\nolimits_{0}^{\beta}}
%EndExpansion
d\lambda\left\langle e^{iHt/\hbar+\lambda H}A^{\dag}e^{-iHt/\hbar-\lambda
H}B\right\rangle \nonumber\\
&  =-i%
%TCIMACRO{\dint \nolimits_{0}^{\infty}}%
%BeginExpansion
{\displaystyle\int\nolimits_{0}^{\infty}}
%EndExpansion
dte^{izt}%
%TCIMACRO{\dint \nolimits_{0}^{\beta}}%
%BeginExpansion
{\displaystyle\int\nolimits_{0}^{\beta}}
%EndExpansion
d\lambda\left\langle e^{iH\left(  t-i\lambda\hbar\right)  /\hbar}A^{\dag
}e^{-iH\left(  t-i\lambda\hbar\right)  /\hbar}B\right\rangle \nonumber\\
&  =-i%
%TCIMACRO{\dint \nolimits_{0}^{\infty}}%
%BeginExpansion
{\displaystyle\int\nolimits_{0}^{\infty}}
%EndExpansion
dte^{izt}%
%TCIMACRO{\dint \nolimits_{0}^{\beta}}%
%BeginExpansion
{\displaystyle\int\nolimits_{0}^{\beta}}
%EndExpansion
d\lambda\left\langle e^{iL\left(  t-i\lambda\hbar\right)  }A^{\dag
}B\right\rangle \label{IF12b}\\
&  =-i%
%TCIMACRO{\dint \nolimits_{0}^{\infty}}%
%BeginExpansion
{\displaystyle\int\nolimits_{0}^{\infty}}
%EndExpansion
dte^{izt}%
%TCIMACRO{\dint \nolimits_{0}^{\beta}}%
%BeginExpansion
{\displaystyle\int\nolimits_{0}^{\beta}}
%EndExpansion
d\lambda\left\langle A^{\dag}(t-i\hbar\lambda)B(0)\right\rangle . \label{IF13}%
\end{align}

\paragraph{Representation of the impedance function in terms of the relaxation
function}

The impedance function is related to the relaxation function%
\begin{equation}
\Phi\left(  z\right)  \equiv\Phi_{\dot{x}\dot{x}}\left(  z\right)  =\left(
\dot{x},\frac{1}{z-L}\dot{x}\right)  , \label{IF7}%
\end{equation}
where $\dot{x}$ is the velocity operator, by the following expression:
\begin{equation}
\frac{1}{Z(\Omega)}=ie^{2}\underset{\epsilon\rightarrow0}{\lim}\Phi\left(
\Omega+i\epsilon\right)  \label{IF6}%
\end{equation}
($z=\Omega+i\epsilon,\epsilon>0).$

\subparagraph{Demonstration of the representation (\ref{IF7}).\textit{ }}

Apply (\ref{IF12b}) to the relaxation function entering (\ref{IF6}):%
\[
\Phi\left(  z\right)  =-i\int_{0}^{\beta}d\lambda\int_{0}^{\infty}%
dte^{izt}\left\langle \left(  e^{i\left(  t-i\hbar\lambda\right)  L}\dot
{x}\right)  \dot{x}\right\rangle
\]
and perform the integration by parts using the formula%
\begin{align*}
\int_{0}^{\infty}e^{izt}f\left(  t\right)  dt  &  =\left.  -\frac{i}%
{z}f\left(  t\right)  e^{izt}\right\vert _{t=0}^{\infty}+\frac{i}{z}\int
_{0}^{\infty}\frac{\partial f\left(  t\right)  }{\partial t}e^{izt}dt\\
&  =\frac{i}{z}f\left(  0\right)  +\frac{i}{z}\int_{0}^{\infty}\frac{\partial
f\left(  t\right)  }{\partial t}e^{izt}dt:
\end{align*}%
\begin{align*}
\int_{0}^{\infty}e^{izt}\left\langle \left(  e^{i\left(  t-i\hbar
\lambda\right)  L}\dot{x}\right)  \dot{x}\right\rangle dt  &  =\frac{i}%
{z}\left\langle \left(  e^{\hbar\lambda L}\dot{x}\right)  \dot{x}\right\rangle
+\frac{i}{z}\int_{0}^{\infty}e^{izt}\frac{\partial}{\partial t}\left\langle
\left(  e^{i\left(  t-i\hbar\lambda\right)  L}\dot{x}\right)  \dot
{x}\right\rangle dt\\
&  =\frac{i}{z}\left\langle \left(  e^{\hbar\lambda L}\dot{x}\right)  \dot
{x}\right\rangle -\frac{1}{z}\int_{0}^{\infty}e^{izt}\left\langle L\left(
e^{i\left(  t-i\hbar\lambda\right)  L}\dot{x}\right)  \dot{x}\right\rangle dt.
\end{align*}
This allows us to represent the relaxation function in the form%
\begin{align*}
\Phi\left(  z\right)   &  =-i\int_{0}^{\beta}d\lambda\left(  \frac{i}%
{z}\left\langle \left(  e^{\hbar\lambda L}\dot{x}\right)  \dot{x}\right\rangle
-\frac{1}{z}\int_{0}^{\infty}e^{izt}\left\langle L\left(  e^{i\left(
t-i\hbar\lambda\right)  L}\dot{x}\right)  \dot{x}\right\rangle dt\right) \\
&  =\frac{1}{z}\int_{0}^{\beta}d\lambda\left\langle \left(  e^{\hbar\lambda
L}\dot{x}\right)  \dot{x}\right\rangle +\frac{i}{z}\int_{0}^{\beta}%
d\lambda\int_{0}^{\infty}e^{izt}\left\langle L\left(  e^{i\left(
t-i\hbar\lambda\right)  L}\dot{x}\right)  \dot{x}\right\rangle dt\\
&  =\frac{1}{m_{b}z}+\frac{i}{z}\int_{0}^{\beta}d\lambda\int_{0}^{\infty
}e^{izt}\left\langle L\left(  e^{i\left(  t-i\hbar\lambda\right)  L}\dot
{x}\right)  \dot{x}\right\rangle dt,
\end{align*}
where the expression (\ref{IF15a}) is inserted in the first term. Further on,
the integral over $\lambda$ is taken as follows:%
\begin{align*}
\int_{0}^{\beta}d\lambda\left\langle L\left(  e^{i\left(  t-i\hbar
\lambda\right)  L}\dot{x}\right)  \dot{x}\right\rangle  &  =\left\langle
\left(  L\int_{0}^{\beta}e^{i\left(  t-i\hbar\lambda\right)  L}d\lambda\dot
{x}\right)  \dot{x}\right\rangle \\
&  =\frac{1}{\hbar}\left\langle \left(  e^{iLt}\left(  e^{\hbar\beta
L}-1\right)  \dot{x}\right)  \dot{x}\right\rangle \\
&  =\frac{1}{\hbar}\left\langle e^{iLt}\left(  e^{\hbar\beta L}\dot{x}-\dot
{x}\right)  \dot{x}\right\rangle \\
&  =\frac{1}{\hbar}\left\langle \left(  e^{\hbar\beta L}\dot{x}\left(
t\right)  -\dot{x}\left(  t\right)  \right)  \dot{x}\right\rangle \\
&  =\frac{1}{\hbar}\left\langle \left(  e^{\beta H}\dot{x}\left(  t\right)
e^{-\beta H}-\dot{x}\left(  t\right)  \right)  \dot{x}\right\rangle \\
&  =\frac{1}{\hbar\mathrm{Tr}e^{-\beta H}}\mathrm{Tr}\left[  e^{-\beta
H}\left(  e^{\beta H}\dot{x}\left(  t\right)  e^{-\beta H}-\dot{x}\left(
t\right)  \right)  \dot{x}\right] \\
&  =\frac{1}{\hbar\mathrm{Tr}e^{-\beta H}}\mathrm{Tr}\left[  \left(  \dot
{x}\left(  t\right)  e^{-\beta H}-e^{-\beta H}\dot{x}\left(  t\right)
\right)  \dot{x}\right] \\
&  =\frac{1}{\hbar\mathrm{Tr}e^{-\beta H}}\mathrm{Tr}\left[  e^{-\beta
H}\left(  \dot{x}\dot{x}\left(  t\right)  -\dot{x}\left(  t\right)  \right)
\dot{x}\right] \\
&  =\frac{1}{\hbar}\left\langle \dot{x}\dot{x}\left(  t\right)  -\dot
{x}\left(  t\right)  \dot{x}\right\rangle =-\frac{1}{\hbar}\left\langle
\dot{x}\left(  t\right)  ,\dot{x}\right\rangle .
\end{align*}
Hence,
\begin{equation}
\Phi\left(  z\right)  =\frac{1}{m_{b}z}-\frac{i}{\hbar z}\int_{0}^{\infty
}dte^{izt}\left\langle \left[  \dot{x}\left(  t\right)  ,\dot{x}\right]
\right\rangle . \label{IF13a}%
\end{equation}
When setting $z=\Omega+i\varepsilon$ with $\varepsilon\rightarrow+0$, we have%
\begin{equation}
\Phi\left(  \Omega+i\varepsilon\right)  =\frac{1}{m_{b}}\frac{1}%
{\Omega+i\varepsilon}-\frac{i}{\hbar\left(  \Omega+i\varepsilon\right)  }%
\int_{0}^{\infty}e^{i\left(  \Omega+i\varepsilon\right)  t}\left\langle
\left[  \dot{x}\left(  t\right)  ,\dot{x}\right]  \right\rangle dt. \label{K}%
\end{equation}
For $\Omega\neq0$, we can set%
\begin{equation}
\Phi\left(  \Omega+i\varepsilon\right)  =\frac{1}{m_{b}}\frac{1}{\Omega}%
-\frac{i}{\hbar\Omega}\int_{0}^{\infty}e^{i\left(  \Omega+i\varepsilon\right)
t}\left\langle \left[  \dot{x}\left(  t\right)  ,\dot{x}\right]  \right\rangle
dt.
\end{equation}
Multiplying $\Phi\left(  \Omega+i\varepsilon\right)  $ by $ie^{2}$, we find
that%
\begin{align}
ie^{2}\Phi\left(  \Omega+i\varepsilon\right)   &  =ie^{2}\left(  \frac
{1}{m_{b}}\frac{1}{\Omega}-\frac{i}{\hbar\Omega}\int_{0}^{\infty}e^{i\left(
\Omega+i\varepsilon\right)  t}\left\langle \left[  \dot{x}\left(  t\right)
,\dot{x}\right]  \right\rangle dt\right) \nonumber\\
&  =i\frac{e^{2}}{m_{b}\Omega}+\frac{e^{2}}{\hbar\Omega}\int_{0}^{\infty
}e^{i\left(  \Omega+i\varepsilon\right)  t}\left\langle \left[  \dot{x}\left(
t\right)  ,\dot{x}\right]  \right\rangle dt\nonumber\\
&  =i\frac{e^{2}}{m_{b}\Omega}+\frac{1}{\hbar\Omega}\int_{0}^{\infty
}e^{i\left(  \Omega+i\varepsilon\right)  t}\left\langle \left[  j_{x}\left(
t\right)  ,j_{x}\right]  \right\rangle dt, \label{IF13c}%
\end{align}
where the electric current density is
\[
j_{x}=-e\dot{x}.
\]
Substituting further (\ref{IF13c}) in (\ref{IF6}), we arrive at%
\[
\frac{1}{Z(\Omega)}=i\frac{e^{2}}{m_{b}\Omega}+\underset{\epsilon\rightarrow
0}{\lim}\frac{1}{\hbar\Omega}\int_{0}^{\infty}e^{i\left(  \Omega
+i\varepsilon\right)  t}\left\langle \left[  j_{x}\left(  t\right)
,j_{x}\right]  \right\rangle dt,
\]
what coincides with the expression of the impedance function (\ref{IFCond})
through a frequency-dependent conductivity given by the Kubo formula
(\ref{IFKubo}), q.e.d.

\paragraph{Application of the projection operator technique}

Using the Mori-Zwanzig projection operator technique (cf. \cite{Forster75},
Chapter 5), the relaxation function (\ref{IF7})
\[
\Phi\left(  z\right)  =\left(  \dot{x},\frac{1}{z-L}\dot{x}\right)
\]
can be represented in a form, which is especially convenient for the
application in the theory of the optical absorption of polarons.

The projection operator $P$ ($Q=1-P$) is defined as
\begin{equation}
PA=\frac{\dot{x}\left(  \dot{x},A\right)  }{\chi} \label{IF14d}%
\end{equation}
with $A$ an operator and%
\begin{equation}
\chi=\left(  \dot{x},\dot{x}\right)  . \label{IF14a}%
\end{equation}
The projection operator $Q=1-P$ projects an operator onto the space orthogonal
to the space containing $\dot{x}.$ Here we give some examples of the action of
the projection operators:%
\begin{align}
P\dot{x}  &  =\dot{x},\qquad Q\dot{x}=(1-P)\dot{x}=0;\label{IF14e}\\
Px  &  =\frac{\dot{x}\left(  \dot{x},x\right)  }{\chi}=\frac{\dot{x}\left(
iLx,x\right)  }{\chi}=-\frac{i\dot{x}\left(  x,Lx\right)  }{\chi}=-\frac
{i\dot{x}}{\chi}\left\langle \left[  x,x\right]  \right\rangle
=0,\label{IF14f}\\
Qx  &  =(1-P)x=x.\label{IF14ff}\\
Pa_{\mathbf{k}}  &  =\frac{\dot{x}\left(  \dot{x},a_{\mathbf{k}}\right)
}{\chi}=\frac{\dot{x}\left(  iLx,a_{\mathbf{k}}\right)  }{\chi}=-\frac
{i\dot{x}\left(  a_{\mathbf{k}},Lx\right)  }{\chi}=-\frac{i\dot{x}}{\chi
}\left\langle \left[  a_{\mathbf{k}}^{\dag},x\right]  \right\rangle
=0,\label{IF14g}\\
Qa_{\mathbf{k}}  &  =(1-P)a_{\mathbf{k}}=a_{\mathbf{k}} \label{IF14gg}%
\end{align}
The projection operators $P$\ and $Q$ are idempotent:
\begin{align*}
P^{2}A  &  =\frac{\dot{x}\left(  \dot{x},\frac{\dot{x}\left(  \dot
{x},A\right)  }{\chi}\right)  }{\chi}=\frac{\dot{x}\left(  \dot{x},A\right)
}{\chi}=PA;\\
Q^{2}  &  =(1-P)^{2}=1-2P+P^{2}=1-P=Q.
\end{align*}

The Liouville operator can be identiaclly represented as $L=LP+LQ$. Then the
operator $\frac{1}{z-L}$in the relaxation function (\ref{IF7}) can be
represented as follows:%
\[
\frac{1}{z-L}=\frac{1}{z-LQ-LP}.
\]
We use the algebraic operator identity:%
\[
\frac{1}{x+y}=\frac{1}{x}-\frac{1}{x}y\frac{1}{x+y}%
\]
with $x=z-LQ$ and $y=-LP:$%
\[
\frac{1}{z-L}=\frac{1}{z-LQ}+\frac{1}{z-LQ}LP\frac{1}{z-L}.
\]
Consequently, the relaxation function (\ref{IF7}) takes the form%
\begin{equation}
\Phi\left(  z\right)  =\left(  \dot{x},\frac{1}{z-LQ}\dot{x}\right)  +\left(
\dot{x},\frac{1}{z-LQ}LP\frac{1}{z-L}\dot{x}\right)  . \label{IFMori1}%
\end{equation}
The first term in the r.h.s. of (\ref{IFMori1}) simplifies as follows:%
\[
\left(  \dot{x},\frac{1}{z-LQ}\dot{x}\right)  =\left(  \dot{x},\left[
\frac{1}{z}+\frac{1}{z^{2}}LQ+\frac{1}{z^{3}}LQLQ+...\right]  \dot{x}\right)
=\left(  \dot{x},\frac{1}{z}\dot{x}\right)
\]
because $Q\dot{x}=0.$ Using the quantity (\ref{IF14a}) we obtain:
\begin{equation}
\left(  \dot{x},\frac{1}{z-LQ}\dot{x}\right)  =\frac{\chi}{z}. \label{IFMori2}%
\end{equation}
The second term in the r.h.s. of (\ref{IFMori1}) contains the operator%
\[
P\frac{1}{z-L}\dot{x}%
\]
which according to the definition of the projection operator $P$ (\ref{IF14d})
can be transformed as%
\begin{equation}
P\frac{1}{z-L}\dot{x}=\frac{\dot{x}}{\chi}\left(  \dot{x},\frac{1}{z-L}\dot
{x}\right)  =\frac{\dot{x}}{\chi}\Phi\left(  z\right)  . \label{IFMori3}%
\end{equation}
It is remarkable that this term is exactly expressed in terms of the sought
relaxation function (\ref{IF7}). Substituting (\ref{IFMori2}) and
(\ref{IFMori3}) in (\ref{IFMori1}), we find%

\begin{align}
\Phi\left(  z\right)   &  =\frac{\chi}{z}+\left(  \dot{x},\frac{1}{z-LQ}%
L\frac{\dot{x}}{\chi}\right)  \Phi\left(  z\right)  \Rightarrow
\label{IFMori3a}\\
z\Phi\left(  z\right)   &  =\chi+\left(  \dot{x},\frac{z}{z-LQ}L\frac{\dot{x}%
}{\chi}\right)  \Phi\left(  z\right) \nonumber\\
&  =\chi+\left(  \dot{x},\frac{z-LQ+LQ}{z-LQ}L\frac{\dot{x}}{\chi}\right)
\Phi\left(  z\right) \nonumber\\
&  =\chi+\left(  \dot{x},\left[  1+\frac{LQ}{z-LQ}\right]  L\frac{\dot{x}%
}{\chi}\right)  \Phi\left(  z\right)  \Rightarrow\nonumber\\
z\Phi\left(  z\right)   &  =\chi+\left[  \frac{\left(  \dot{x},L\dot
{x}\right)  }{\chi}+\frac{1}{\chi}\left(  \dot{x},\frac{LQ}{z-LQ}L\dot
{x}\right)  \right]  \Phi\left(  z\right)  .\nonumber
\end{align}
Introducring the quantity%
\begin{equation}
O=\frac{\left(  \dot{x},L\dot{x}\right)  }{\chi} \label{IF14b}%
\end{equation}
and the function called the \textit{memory function }%
\begin{equation}
\Sigma(z)=\frac{1}{\chi}\left(  \dot{x},LQ\frac{1}{z-LQ}L\dot{x}\right)  ,
\label{IFMori4}%
\end{equation}
we represent (\ref{IFMori3a}) in the form of the equation%
\[
\left[  z-O-\Sigma(z)\right]  \Phi\left(  z\right)  =\chi.
\]
A solition of this equation gives the relaxation function $\Phi(z)$
represented within the Mori-Zwanzig projection operator technique:
\begin{equation}
\Phi(z)=\frac{\chi}{z-O-\Sigma(z)}. \label{IF14}%
\end{equation}
The memory function (\ref{IFMori4}) can be still transformed to another useful
form.First of all, we apply the property of a scalar product (\ref{IF11b}):%

\begin{align}
\Sigma(z)  &  =\frac{1}{\chi}\left(  L\dot{x},Q\frac{1}{z-LQ}L\dot{x}\right)
\label{IFMori5}\\
&  =\frac{1}{\chi}\left(  (P+Q)L\dot{x},Q\frac{1}{z-LQ}L\dot{x}\right) \\
&  =\frac{1}{\chi}\left(  PL\dot{x},Q\frac{1}{z-LQ}L\dot{x}\right)  +\frac
{1}{\chi}\left(  QL\dot{x},Q\frac{1}{z-LQ}L\dot{x}\right)  .
\end{align}
For any two operators $A$ and $B$
\begin{align*}
(PA,QB)  &  =\left(  \frac{\dot{x}\left(  \dot{x},A\right)  }{\chi},\left[
B-\frac{\dot{x}\left(  \dot{x},B\right)  }{\chi}\right]  \right) \\
&  =\frac{\left(  \dot{x},A\right)  }{\chi}\left[  \left(  \dot{x},B\right)
-\frac{\left(  \dot{x},\dot{x}\right)  \left(  \dot{x},B\right)  }{\chi
}\right] \\
&  =\frac{\left(  \dot{x},A\right)  }{\chi}\left[  \left(  \dot{x},B\right)
-\left(  \dot{x},B\right)  \right]  =0,
\end{align*}
therefore the first term on the r.h.s. in (\ref{IFMori5}) vanishes, and we
obtain%
\begin{equation}
\Sigma(z)=\frac{1}{\chi}\left(  QL\dot{x},Q\frac{1}{z-LQ}L\dot{x}\right)  .
\label{IFMori6}%
\end{equation}
In this expression, the operator $Q\frac{1}{z-LQ}$ can be represented in the
following form, using the fact that $Q$ is the idempotent operator:%
\begin{align}
Q\frac{1}{z-LQ}  &  =Q\left[  \frac{1}{z}+\frac{1}{z^{2}}LQ+\frac{1}{z^{3}%
}LQLQ+...\right] \nonumber\\
&  =\frac{1}{z}Q+\frac{1}{z^{2}}QLQ+\frac{1}{z^{3}}QLQLQ+...\nonumber\\
&  =\frac{1}{z}Q+\frac{1}{z^{2}}QLQ^{2}+\frac{1}{z^{3}}QLQ^{2}LQ^{2}%
+...\nonumber\\
&  =\left[  \frac{1}{z}+\frac{1}{z^{2}}QLQ+\frac{1}{z^{3}}QLQQLQ+...\right]
Q\Rightarrow\nonumber\\
Q\frac{1}{z-LQ}  &  =\frac{1}{z-QLQ}Q. \label{IFMori7}%
\end{align}
A new Liouville operator can be defined, $\mathcal{L}=QLQ$, which describes
the time evolution in the Hilbert space of operators, which is orthogonal
complement of $\dot{x}.$ Substituting then (\ref{IFMori7}) with the operator
$\mathcal{L}$ into (\ref{IFMori6}), we bring it to the form, which will be
used in what follows.%
\begin{equation}
\Sigma(z)=\frac{1}{\chi}\left(  QL\dot{x},\frac{1}{z-\mathcal{L}}QL\dot
{x}\right)  . \label{IF14c}%
\end{equation}

For the Hamiltonian (\ref{eq_1a}) we obtain the following quantities:%
\[
\chi=\left(  \dot{x},\dot{x}\right)  =\left(  \frac{p_{x}}{m_{b}},iLx\right)
=\frac{i}{m_{b}\hbar}\left(  p_{x},Lx\right)
\]
Using (\ref{IF11c}), we find%
\begin{equation}
\chi=\frac{i}{m_{b}\hbar}\left\langle \left[  p_{x},x\right]  \right\rangle
=\frac{i}{m_{b}\hbar}\left\langle \left(  -i\hbar\right)  \right\rangle
=\frac{1}{m_{b}} \label{IF15a}%
\end{equation}
and%
\begin{equation}
O=\frac{\left(  \dot{x},L\dot{x}\right)  }{\chi}=m_{b}\frac{1}{\hbar
}\left\langle \left[  \dot{x},\dot{x}\right]  \right\rangle =0. \label{IF15b}%
\end{equation}
Substituting (\ref{IF15a}) and (\ref{IF15b}) in (\ref{IF14}), one obtains%
\begin{equation}
\Phi(z)=\frac{1}{m_{b}}\frac{1}{z-\Sigma(z)}. \label{IF15c}%
\end{equation}
The operator
\begin{align}
L\dot{x}  &  =L\frac{p_{x}}{m_{b}}=\frac{1}{m_{b}\hbar}\left[  H,p_{x}\right]
=\nonumber\\
&  =-\frac{1}{m_{b}\hbar}\left[  p_{x},\sum_{\mathbf{k}}(V_{k}a_{\mathbf{k}%
}e^{i\mathbf{k\cdot r}}+V_{k}^{\ast}a_{\mathbf{k}}^{\dag}e^{-i\mathbf{k\cdot
r}})\right] \nonumber\\
&  =\frac{i}{m_{b}}\sum_{\mathbf{k}}ik_{x}(V_{k}a_{\mathbf{k}}%
e^{i\mathbf{k\cdot r}}-V_{k}^{\ast}a_{\mathbf{k}}^{\dag}e^{-i\mathbf{k\cdot
r}})\Rightarrow\nonumber\\
L\dot{x}  &  =-\frac{1}{m_{b}}\sum_{\mathbf{k}}k_{x}(V_{k}a_{\mathbf{k}%
}e^{i\mathbf{k\cdot r}}-V_{k}^{\ast}a_{\mathbf{k}}^{\dag}e^{-i\mathbf{k\cdot
r}}) \label{IF15d}%
\end{align}
does not depend on the velocities. Therefore, multiplying both parts of
(\ref{IF15d}) with $Q$ and taking into account (\ref{IF14ff}) and
(\ref{IF14gg}), we obtain%
\[
QL\dot{x}=-\frac{1}{m_{b}}\sum_{\mathbf{k}}k_{x}(V_{k}a_{\mathbf{k}%
}e^{i\mathbf{k\cdot r}}-V_{k}^{\ast}a_{\mathbf{k}}^{\dag}e^{-i\mathbf{k\cdot
r}}),
\]
what allows us to represent the memory function in the form%

\begin{align}
\Sigma(z)  &  =\frac{1}{\chi}\left(  QL\dot{x},\frac{1}{z-\mathcal{L}}%
QL\dot{x}\right) \nonumber\\
&  =\frac{1}{m_{b}}\sum_{\mathbf{k}}\sum_{\mathbf{k}^{\prime}}\left(
\begin{array}
[c]{c}%
k_{x}(V_{k}a_{\mathbf{k}}e^{i\mathbf{k\cdot r}}-V_{k}^{\ast}a_{\mathbf{k}%
}^{\dag}e^{-i\mathbf{k\cdot r}}),\\
\frac{1}{z-\mathcal{L}}k_{x}^{\prime}(V_{k^{\prime}}a_{\mathbf{k}^{\prime}%
}e^{i\mathbf{k}^{\prime}\mathbf{\cdot r}}-V_{k^{\prime}}^{\ast}a_{\mathbf{k}%
^{\prime}}^{\dag}e^{-i\mathbf{k}^{\prime}\mathbf{\cdot r}})
\end{array}
\right) \nonumber\\
&  =\frac{1}{m_{b}}\sum_{\mathbf{k}}\sum_{\mathbf{k}^{\prime}}k_{x}%
k_{x}^{\prime}V_{k}V_{k^{\prime}}^{\ast}\left(
\begin{array}
[c]{c}%
(a_{\mathbf{k}}e^{i\mathbf{k\cdot r}}+a_{\mathbf{k}}^{\dag}e^{-i\mathbf{k\cdot
r}}),\\
\frac{1}{z-\mathcal{L}}(a_{\mathbf{k}^{\prime}}e^{i\mathbf{k}^{\prime
}\mathbf{\cdot r}}+a_{\mathbf{k}^{\prime}}^{\dag}e^{-i\mathbf{k}^{\prime
}\mathbf{\cdot r}})
\end{array}
\right)  . \label{IF16}%
\end{align}
In transition to (\ref{IF16}) we have used the property of the amplitude
(\ref{eq_1b}): $V_{k}^{\ast}=-V_{k}$ and taken into account that according to
the definition (\ref{IF8}), the first operator enters a scalar product in the
hermitian conjugate form. Introducing the operators
\[
b_{\mathbf{k}}=a_{\mathbf{k}}e^{i\mathbf{k\cdot r}};b_{\mathbf{k}}^{\dag
}=a_{\mathbf{k}}^{\dag}e^{-i\mathbf{k\cdot r}},
\]
we represent the memory function as%
\begin{equation}
\Sigma(z)=\frac{1}{m_{b}}\sum_{\mathbf{k}}\sum_{\mathbf{k}^{\prime}}k_{x}%
k_{x}^{\prime}V_{k}V_{k^{\prime}}^{\ast}\left(  (b_{\mathbf{k}}+b_{\mathbf{k}%
}^{\dag}),\frac{1}{z-\mathcal{L}}(b_{\mathbf{k}^{\prime}}+b_{\mathbf{k}%
^{\prime}}^{\dag})\right)  . \label{IF17}%
\end{equation}
We notice that $Qb_{\mathbf{k}}=Q(a_{\mathbf{k}}e^{i\mathbf{k\cdot r}%
})=a_{\mathbf{k}}e^{i\mathbf{k\cdot r}}=b_{\mathbf{k}}$. It will be
represented through the four relaxation functions:%
\begin{align}
\Sigma(z)  &  =\frac{1}{m_{b}}\sum_{\mathbf{k}}\sum_{\mathbf{k}^{\prime}}%
k_{x}k_{x}^{\prime}V_{k}V_{k^{\prime}}^{\ast}\left[  \Phi_{\mathbf{kk}%
^{\prime}}^{+\,+}(z)+\Phi_{\mathbf{kk}^{\prime}}^{-\,-}(z)+\Phi_{\mathbf{kk}%
^{\prime}}^{+\,-}(z)+\Phi_{\mathbf{kk}^{\prime}}^{-\,+}(z)\right]
,\label{IF18a}\\
\Phi_{\mathbf{kk}^{\prime}}^{+\,+}(z)  &  =\left(  b_{\mathbf{k}}^{\dag}%
,\frac{1}{z-\mathcal{L}}b_{\mathbf{k}^{\prime}}^{\dag}\right)  ,
\label{IF18b}\\
\Phi_{\mathbf{kk}^{\prime}}^{-\,-}(z)  &  =\left(  b_{\mathbf{k}},\frac
{1}{z-\mathcal{L}}b_{\mathbf{k}^{\prime}}\right)  ,\label{IF18c}\\
\Phi_{\mathbf{kk}^{\prime}}^{+\,-}(z)  &  =\left(  b_{\mathbf{k}}^{\dag}%
,\frac{1}{z-\mathcal{L}}b_{\mathbf{k}^{\prime}}\right)  ,\label{IF18d}\\
\Phi_{\mathbf{kk}^{\prime}}^{-\,+}(z)  &  =\left(  b_{\mathbf{k}},\frac
{1}{z-\mathcal{L}}b_{\mathbf{k}^{\prime}}^{\dag}\right)  . \label{IF18e}%
\end{align}
There exist relations between the above relaxation functions. For example, the
relaxation function (\ref{IF18c}), takes the form%
\begin{align*}
\Phi_{\mathbf{kk}^{\prime}}^{-\,-}(z)  &  =\left(  b_{\mathbf{k}},\frac
{1}{z-\mathcal{L}}b_{\mathbf{k}^{\prime}}\right) \\
&  =-i%
%TCIMACRO{\dint \nolimits_{0}^{\infty}}%
%BeginExpansion
{\displaystyle\int\nolimits_{0}^{\infty}}
%EndExpansion
dte^{izt}\left(  e^{i\mathcal{L}t}b_{\mathbf{k}}(0),b_{\mathbf{k}^{\prime}%
}(0)\right)  .
\end{align*}
Then the complex conjugate of this relaxation function:
\begin{align*}
\left[  \Phi_{\mathbf{kk}^{\prime}}^{-\,-}(z)\right]  ^{\ast}  &  =i%
%TCIMACRO{\dint \nolimits_{0}^{\infty}}%
%BeginExpansion
{\displaystyle\int\nolimits_{0}^{\infty}}
%EndExpansion
dte^{-iz^{\ast}t}\left(  e^{i\mathcal{L}t}b_{\mathbf{k}}(0),b_{\mathbf{k}%
^{\prime}}(0)\right)  ^{\ast}\\
&  =i%
%TCIMACRO{\dint \nolimits_{0}^{\infty}}%
%BeginExpansion
{\displaystyle\int\nolimits_{0}^{\infty}}
%EndExpansion
dte^{-iz^{\ast}t}\left(  b_{\mathbf{k}^{\prime}}(0),e^{i\mathcal{L}%
t}b_{\mathbf{k}}(0)\right)  ,
\end{align*}
where the property (\ref{IF11d}) has been used. The property (\ref{IF11a})
gives
\begin{align*}
\left[  \Phi_{\mathbf{kk}^{\prime}}^{-\,-}(z)\right]  ^{\ast}  &  =i%
%TCIMACRO{\dint \nolimits_{0}^{\infty}}%
%BeginExpansion
{\displaystyle\int\nolimits_{0}^{\infty}}
%EndExpansion
dte^{-iz^{\ast}t}\left(  b_{\mathbf{k}^{\prime}}(0),e^{i\mathcal{H}t/\hbar
}b_{\mathbf{k}}(0)e^{-i\mathcal{H}t/\hbar}\right) \\
&  =i%
%TCIMACRO{\dint \nolimits_{0}^{\infty}}%
%BeginExpansion
{\displaystyle\int\nolimits_{0}^{\infty}}
%EndExpansion
dte^{-iz^{\ast}t}\left(  e^{i\mathcal{H}t/\hbar}b_{\mathbf{k}}^{\dag
}(0)e^{i\mathcal{H}t/\hbar},b_{\mathbf{k}^{\prime}}^{\dag}(0)\right) \\
&  =i%
%TCIMACRO{\dint \nolimits_{0}^{\infty}}%
%BeginExpansion
{\displaystyle\int\nolimits_{0}^{\infty}}
%EndExpansion
dte^{-iz^{\ast}t}\left(  e^{i\mathcal{L}t}b_{\mathbf{k}}^{\dag}%
(0),b_{\mathbf{k}^{\prime}}^{\dag}(0)\right) \\
&  =i%
%TCIMACRO{\dint \nolimits_{0}^{\infty}}%
%BeginExpansion
{\displaystyle\int\nolimits_{0}^{\infty}}
%EndExpansion
dte^{-iz^{\ast}t}\left(  b_{\mathbf{k}}^{\dag}(t),b_{\mathbf{k}^{\prime}%
}^{\dag}(0)\right)  =-\Phi_{\mathbf{kk}^{\prime}}^{+\,+}(-z^{\ast}),
\end{align*}
wherefrom it follows that
\begin{equation}
\Phi_{\mathbf{kk}^{\prime}}^{-\,-}(z)=-\left[  \Phi_{\mathbf{kk}^{\prime}%
}^{+\,+}(-z^{\ast})\right]  ^{\ast}. \label{IF19c}%
\end{equation}
Similarly, the relation%

\begin{equation}
\Phi_{\mathbf{kk}^{\prime}}^{-\,+}(z)=-\left[  \Phi_{\mathbf{kk}^{\prime}%
}^{+\,-}(-z^{\ast})\right]  ^{\ast} \label{IF19d}%
\end{equation}
is proven.

\paragraph{Memory function}

In this subsection we indicate which approximations must be made in the
calculation of the relaxation functions in order to obtain the FHIP results
for the impedance function. Consider the relaxation function (\ref{IF18b}):%
\begin{align*}
\Phi_{\mathbf{kk}^{\prime}}^{+\,+}(z)  &  =\left(  b_{\mathbf{k}}^{\dag}%
,\frac{1}{z-\mathcal{L}}b_{\mathbf{k}^{\prime}}^{\dag}\right) \\
&  =-i%
%TCIMACRO{\dint \nolimits_{0}^{\infty}}%
%BeginExpansion
{\displaystyle\int\nolimits_{0}^{\infty}}
%EndExpansion
dte^{izt}\left(  e^{i\mathcal{L}t}b_{\mathbf{k}}^{\dag}(0),b_{\mathbf{k}%
^{\prime}}^{\dag}(0)\right)  =-i%
%TCIMACRO{\dint \nolimits_{0}^{\infty}}%
%BeginExpansion
{\displaystyle\int\nolimits_{0}^{\infty}}
%EndExpansion
dte^{izt}\left(  b_{\mathbf{k}}^{\dag}(t),b_{\mathbf{k}^{\prime}}^{\dag
}(0)\right)  ,
\end{align*}
where $b_{\mathbf{k}}^{\dag}(t)=e^{i\mathcal{L}t}b_{\mathbf{k}}^{\dag}(0),$
and perform a partial integration:%
\begin{align}
\Phi_{\mathbf{kk}^{\prime}}^{+\,+}(z)  &  =-\frac{1}{z}%
%TCIMACRO{\dint \nolimits_{0}^{\infty}}%
%BeginExpansion
{\displaystyle\int\nolimits_{0}^{\infty}}
%EndExpansion
d\left(  e^{izt}\right)  \left(  b_{\mathbf{k}}^{\dag}(t),b_{\mathbf{k}%
^{\prime}}^{\dag}(0)\right) \nonumber\\
&  =\left.  -\frac{1}{z}e^{izt}\left(  b_{\mathbf{k}}^{\dag}(t)b_{\mathbf{k}%
^{\prime}}^{\dag}(0)\right)  \right\vert _{0}^{\infty}+\frac{1}{z}%
%TCIMACRO{\dint \nolimits_{0}^{\infty}}%
%BeginExpansion
{\displaystyle\int\nolimits_{0}^{\infty}}
%EndExpansion
dte^{izt}\left(  \frac{db_{\mathbf{k}}^{\dag}(t)}{dt},b_{\mathbf{k}^{\prime}%
}^{\dag}(0)\right) \nonumber\\
&  =\frac{1}{z}\left(  b_{\mathbf{k}}^{\dag}(0),b_{\mathbf{k}^{\prime}}^{\dag
}(0)\right)  +\frac{1}{z}%
%TCIMACRO{\dint \nolimits_{0}^{\infty}}%
%BeginExpansion
{\displaystyle\int\nolimits_{0}^{\infty}}
%EndExpansion
dte^{izt}\left(  i\mathcal{L}b_{\mathbf{k}}^{\dag}(t),b_{\mathbf{k}^{\prime}%
}^{\dag}(0)\right) \label{IF19}\\
&  =\frac{1}{z}\left(  b_{\mathbf{k}}^{\dag}(0),b_{\mathbf{k}^{\prime}}^{\dag
}(0)\right)  -\frac{i}{z}%
%TCIMACRO{\dint \nolimits_{0}^{\infty}}%
%BeginExpansion
{\displaystyle\int\nolimits_{0}^{\infty}}
%EndExpansion
dte^{izt}\left(  \mathcal{L}b_{\mathbf{k}}^{\dag}(t),b_{\mathbf{k}^{\prime}%
}^{\dag}(0)\right)  .\nonumber
\end{align}
Here we supposed that
\[
\underset{t\longrightarrow\infty}{\lim}e^{izt}\left(  b_{\mathbf{k}}^{\dag
}(t),b_{\mathbf{k}^{\prime}}^{\dag}(0)\right)  =0.
\]
In the second term in (\ref{IF19}),%
\begin{align*}
\left(  \mathcal{L}b_{\mathbf{k}}^{\dag}(t),b_{\mathbf{k}^{\prime}}^{\dag
}(0)\right)   &  =\left(  \mathcal{L}e^{i\mathcal{L}t}b_{\mathbf{k}}^{\dag
},b_{\mathbf{k}^{\prime}}^{\dag}\right)  =\left(  QLQe^{iQLQt}b_{\mathbf{k}%
}^{\dag},b_{\mathbf{k}^{\prime}}^{\dag}(0)\right) \\
&  =\left(  QLQe^{iQHQt/\hbar}b_{\mathbf{k}}^{\dag}e^{-iQHQt/\hbar
},b_{\mathbf{k}^{\prime}}^{\dag}\right) \\
&  =\left(  QLe^{iQHQt/\hbar}b_{\mathbf{k}}^{\dag}e^{-iQHQt/\hbar
},b_{\mathbf{k}^{\prime}}^{\dag}\right)
\end{align*}
because $Qb_{\mathbf{k}}^{\dag}=b_{\mathbf{k}}^{\dag}$ and $Q^{2}=Q.$ Further
on, we have%
\begin{align*}
\left(  \mathcal{L}b_{\mathbf{k}}^{\dag}(t),b_{\mathbf{k}^{\prime}}^{\dag
}(0)\right)   &  =%
%TCIMACRO{\dint \nolimits_{0}^{\beta}}%
%BeginExpansion
{\displaystyle\int\nolimits_{0}^{\beta}}
%EndExpansion
d\lambda\left\langle e^{\lambda\hbar L}e^{iQHQt/\hbar}b_{\mathbf{k}%
}e^{-iQHQt/\hbar}LQb_{\mathbf{k}^{\prime}}^{\dag}\right\rangle \\
&  =%
%TCIMACRO{\dint \nolimits_{0}^{\beta}}%
%BeginExpansion
{\displaystyle\int\nolimits_{0}^{\beta}}
%EndExpansion
d\lambda\left\langle e^{\lambda\hbar L}e^{iQHQt/\hbar}b_{\mathbf{k}%
}e^{-iQHQt/\hbar}Lb_{\mathbf{k}^{\prime}}^{\dag}\right\rangle \\
&  =\left(  Le^{iQHQt/\hbar}b_{\mathbf{k}}^{\dag}e^{-iQHQt/\hbar
},b_{\mathbf{k}^{\prime}}^{\dag}\right) \\
&  =\left(  Lb_{\mathbf{k}}^{\dag}(t),b_{\mathbf{k}^{\prime}}^{\dag
}(0)\right)  .
\end{align*}
So, we find from (\ref{IF19})%
\begin{align}
\Phi_{\mathbf{kk}^{\prime}}^{+\,+}(z)  &  =\frac{1}{z}\left(  b_{\mathbf{k}%
}^{\dag}(0),b_{\mathbf{k}^{\prime}}^{\dag}(0)\right)  -\frac{i}{z}%
%TCIMACRO{\dint \nolimits_{0}^{\infty}}%
%BeginExpansion
{\displaystyle\int\nolimits_{0}^{\infty}}
%EndExpansion
dte^{izt}\left(  Lb_{\mathbf{k}}^{\dag}(t),b_{\mathbf{k}^{\prime}}^{\dag
}(0)\right) \nonumber\\
&  =\frac{1}{z}\left(  b_{\mathbf{k}}^{\dag}(0),b_{\mathbf{k}^{\prime}}^{\dag
}(0)\right)  -\frac{i}{z}%
%TCIMACRO{\dint \nolimits_{0}^{\infty}}%
%BeginExpansion
{\displaystyle\int\nolimits_{0}^{\infty}}
%EndExpansion
dte^{izt}\left(  b_{\mathbf{k}}^{\dag}(t),Lb_{\mathbf{k}^{\prime}}^{\dag
}(0)\right) \nonumber\\
&  =\frac{1}{z}\left(  b_{\mathbf{k}}^{\dag}(0),b_{\mathbf{k}^{\prime}}^{\dag
}(0)\right)  -\frac{i}{z\hbar}%
%TCIMACRO{\dint \nolimits_{0}^{\infty}}%
%BeginExpansion
{\displaystyle\int\nolimits_{0}^{\infty}}
%EndExpansion
dte^{izt}\left\langle \left[  b_{\mathbf{k}}(t),b_{\mathbf{k}^{\prime}}^{\dag
}(0)\right]  \right\rangle . \label{IF19a}%
\end{align}
The first term in the r.h.s. of this expression can be represented as
follows:
\begin{align*}
\frac{1}{z}\left(  b_{\mathbf{k}}^{\dag}(0),b_{\mathbf{k}^{\prime}}^{\dag
}(0)\right)   &  =\frac{1}{z}%
%TCIMACRO{\dint \nolimits_{0}^{\beta}}%
%BeginExpansion
{\displaystyle\int\nolimits_{0}^{\beta}}
%EndExpansion
d\lambda\left\langle \left(  e^{\lambda\hbar L}b_{\mathbf{k}}\right)
b_{\mathbf{k}^{\prime}}^{\dag}\right\rangle =\frac{1}{z}\left\langle
%TCIMACRO{\dint \nolimits_{0}^{\beta}}%
%BeginExpansion
{\displaystyle\int\nolimits_{0}^{\beta}}
%EndExpansion
d\lambda\left(  e^{\lambda\hbar L}b_{\mathbf{k}}\right)  b_{\mathbf{k}%
^{\prime}}^{\dag}\right\rangle \\
&  =\frac{1}{z}\left\langle \left(  \frac{e^{\beta\hbar L}-1}{\hbar
L}b_{\mathbf{k}}\right)  b_{\mathbf{k}^{\prime}}^{\dag}\right\rangle =\frac
{1}{z\hbar}\left\langle e^{\beta H}\frac{1}{L}b_{\mathbf{k}}e^{-\beta
H}b_{\mathbf{k}^{\prime}}^{\dag}-\frac{1}{L}b_{\mathbf{k}}b_{\mathbf{k}%
^{\prime}}^{\dag}\right\rangle \\
&  =\frac{1}{z\hbar}\frac{1}{\mathrm{Tr}e^{-\beta H}}\mathrm{Tr}\left\{
e^{-\beta H}\left[  e^{\beta H}\frac{1}{L}b_{\mathbf{k}}e^{-\beta
H}b_{\mathbf{k}^{\prime}}^{\dag}-\frac{1}{L}b_{\mathbf{k}}b_{\mathbf{k}%
^{\prime}}^{\dag}\right]  \right\} \\
&  =\frac{1}{z\hbar}\frac{1}{\mathrm{Tr}e^{-\beta H}}\mathrm{Tr}\left\{
b_{\mathbf{k}^{\prime}}^{\dag}\frac{1}{L}b_{\mathbf{k}}e^{-\beta H}-e^{-\beta
H}\frac{1}{L}b_{\mathbf{k}}b_{\mathbf{k}^{\prime}}^{\dag}\right\} \\
&  =\frac{1}{z\hbar}\left\langle b_{\mathbf{k}^{\prime}}^{\dag}\frac{1}%
{L}b_{\mathbf{k}}-\frac{1}{L}b_{\mathbf{k}}b_{\mathbf{k}^{\prime}}^{\dag
}\right\rangle \\
&  =\frac{i}{z\hbar}\left\langle b_{\mathbf{k}^{\prime}}^{\dag}\frac{1}%
{iL}b_{\mathbf{k}}-\frac{1}{iL}b_{\mathbf{k}}b_{\mathbf{k}^{\prime}}^{\dag
}\right\rangle \\
&  =\frac{i}{z\hbar}\left\langle
%TCIMACRO{\dint \nolimits_{0}^{\infty}}%
%BeginExpansion
{\displaystyle\int\nolimits_{0}^{\infty}}
%EndExpansion
dte^{iLt}b_{\mathbf{k}}b_{\mathbf{k}^{\prime}}^{\dag}-%
%TCIMACRO{\dint \nolimits_{0}^{\infty}}%
%BeginExpansion
{\displaystyle\int\nolimits_{0}^{\infty}}
%EndExpansion
dtb_{\mathbf{k}^{\prime}}^{\dag}e^{iLt}b_{\mathbf{k}}\right\rangle \\
&  =\frac{i}{z\hbar}\left\langle
%TCIMACRO{\dint \nolimits_{0}^{\infty}}%
%BeginExpansion
{\displaystyle\int\nolimits_{0}^{\infty}}
%EndExpansion
dtb_{\mathbf{k}}(t)b_{\mathbf{k}^{\prime}}^{\dag}-%
%TCIMACRO{\dint \nolimits_{0}^{\infty}}%
%BeginExpansion
{\displaystyle\int\nolimits_{0}^{\infty}}
%EndExpansion
dtb_{\mathbf{k}^{\prime}}^{\dag}b_{\mathbf{k}}(t)\right\rangle \\
&  =\frac{i}{z\hbar}%
%TCIMACRO{\dint \nolimits_{0}^{\infty}}%
%BeginExpansion
{\displaystyle\int\nolimits_{0}^{\infty}}
%EndExpansion
dt\left\langle \left[  b_{\mathbf{k}}(t),b_{\mathbf{k}^{\prime}}^{\dag
}(0)\right]  \right\rangle .
\end{align*}
Substituting it in the r.h.s. of (\ref{IF19a}), we find%
\begin{align}
\Phi_{\mathbf{kk}^{\prime}}^{+\,+}(z)  &  =\frac{i}{z\hbar}%
%TCIMACRO{\dint \nolimits_{0}^{\infty}}%
%BeginExpansion
{\displaystyle\int\nolimits_{0}^{\infty}}
%EndExpansion
dt\left\langle \left[  b_{\mathbf{k}}(t),b_{\mathbf{k}^{\prime}}^{\dag
}(0)\right]  \right\rangle -\frac{i}{z\hbar}%
%TCIMACRO{\dint \nolimits_{0}^{\infty}}%
%BeginExpansion
{\displaystyle\int\nolimits_{0}^{\infty}}
%EndExpansion
dte^{izt}\left\langle \left[  b_{\mathbf{k}}(t),b_{\mathbf{k}^{\prime}}^{\dag
}(0)\right]  \right\rangle \nonumber\\
&  =\frac{i}{z\hbar}%
%TCIMACRO{\dint \nolimits_{0}^{\infty}}%
%BeginExpansion
{\displaystyle\int\nolimits_{0}^{\infty}}
%EndExpansion
dt\left(  1-e^{izt}\right)  \left\langle \left[  b_{\mathbf{k}}%
(t),b_{\mathbf{k}^{\prime}}^{\dag}(0)\right]  \right\rangle . \label{IF19b}%
\end{align}
In a similar way one obtains%

\begin{equation}
\Phi_{\mathbf{kk}^{\prime}}^{+\,-}(z)=\frac{i}{z\hbar}%
%TCIMACRO{\dint \nolimits_{0}^{\infty}}%
%BeginExpansion
{\displaystyle\int\nolimits_{0}^{\infty}}
%EndExpansion
dt\left(  1-e^{izt}\right)  \left\langle \left[  b_{\mathbf{k}}%
(t),b_{\mathbf{k}^{\prime}}(0)\right]  \right\rangle . \label{IF19e}%
\end{equation}

Inserting the relaxation functions (\ref{IF19b}), (\ref{IF19e}), (\ref{IF19c})
and (\ref{IF19d}), we find the memory function (\ref{IF18a})%
\begin{align*}
\Sigma(z)  &  =\frac{1}{m_{b}}\sum_{\mathbf{k}}\sum_{\mathbf{k}^{\prime}}%
k_{x}k_{x}^{\prime}V_{k}V_{k^{\prime}}^{\ast}\frac{i}{z\hbar}%
%TCIMACRO{\dint \nolimits_{0}^{\infty}}%
%BeginExpansion
{\displaystyle\int\nolimits_{0}^{\infty}}
%EndExpansion
dt\left(  1-e^{izt}\right) \\
&  \times\left[
\begin{array}
[c]{c}%
\left\langle \left[  b_{\mathbf{k}}(t),b_{\mathbf{k}^{\prime}}^{\dag
}(0)\right]  \right\rangle +\left\langle \left[  b_{\mathbf{k}}%
(t),b_{\mathbf{k}^{\prime}}(0)\right]  \right\rangle \\
-\left\langle \left[  b_{\mathbf{k}}(t),b_{\mathbf{k}^{\prime}}^{\dag
}(0)\right]  \right\rangle ^{\ast}-\left\langle \left[  b_{\mathbf{k}%
}(t),b_{\mathbf{k}^{\prime}}(0)\right]  \right\rangle ^{\ast}%
\end{array}
\right] \\
&  =-\frac{1}{m_{b}}\sum_{\mathbf{k}}\sum_{\mathbf{k}^{\prime}}k_{x}%
k_{x}^{\prime}V_{k}V_{k^{\prime}}^{\ast}\frac{2}{z\hbar}%
%TCIMACRO{\dint \nolimits_{0}^{\infty}}%
%BeginExpansion
{\displaystyle\int\nolimits_{0}^{\infty}}
%EndExpansion
dt\left(  1-e^{izt}\right) \\
&  \times\operatorname{Im}\left[  \left\langle \left[  b_{\mathbf{k}%
}(t),b_{\mathbf{k}^{\prime}}^{\dag}(0)\right]  \right\rangle +\left\langle
\left[  b_{\mathbf{k}}(t),b_{\mathbf{k}^{\prime}}(0)\right]  \right\rangle
\right]  ,
\end{align*}
wherefrom
\begin{equation}
\Sigma(z)=\frac{1}{z}%
%TCIMACRO{\dint \nolimits_{0}^{\infty}}%
%BeginExpansion
{\displaystyle\int\nolimits_{0}^{\infty}}
%EndExpansion
dt\left(  1-e^{izt}\right)  \operatorname{Im}F(t) \label{IF20a}%
\end{equation}
with
\begin{equation}
F(t)=-\frac{2}{m_{b}\hbar}\sum_{\mathbf{k}}\sum_{\mathbf{k}^{\prime}}%
k_{x}k_{x}^{\prime}V_{k}V_{k^{\prime}}^{\ast}\left\{  \left\langle \left[
b_{\mathbf{k}}(t),b_{\mathbf{k}^{\prime}}^{\dag}(0)\right]  \right\rangle
+\left\langle \left[  b_{\mathbf{k}}(t),b_{\mathbf{k}^{\prime}}(0)\right]
\right\rangle \right\}  . \label{IF20b}%
\end{equation}

\paragraph{Derivation of the memory function}

To calculate the expectation values in Eq. (\ref{IF20b}), we shall make the
following approximations (cf. Ref. \cite{PD1981}). The Liouville operator
$\mathcal{L}$, which determines the time evolution of the operator
$b_{\mathbf{k}}^{\dag}(t)=e^{i\mathcal{L}t}b_{\mathbf{k}}^{\dag}(0)$, is
replaced by $L_{ph}+L_{F},$ where $L_{ph}$ is the Liouville operator for free
phonons and $L_{F}$ is the Liouville operator for the Feynman polaron model
\cite{Feynman}. The Fr\"{o}hlich Hamiltonian appearing in the statistical
average $\left\langle \bullet\right\rangle $ is imilarly replaced by
$H_{ph}+H_{F},$with $H_{ph}$ the Hamiltonian of free phonons and $H_{F}$ the
Hamiltonian of the Feynman polaron model. With this approximation, e.g., the
average
\begin{equation}
\left\langle b_{\mathbf{k}}(t)b_{\mathbf{k}^{\prime}}^{\dag}(0)\right\rangle
=\left\langle a_{\mathbf{k}}(t)a_{\mathbf{k}^{\prime}}^{\dag}(0)\right\rangle
\left\langle e^{i\mathbf{k\cdot r(}t\mathbf{)}}e^{-i\mathbf{k}^{\prime
}\mathbf{\cdot r}}\right\rangle =\delta_{\mathbf{k},\mathbf{k}^{\prime}%
}\left\langle a_{\mathbf{k}}(t)a_{\mathbf{k}}^{\dag}(0)\right\rangle
\left\langle e^{i\mathbf{k\cdot r(}t\mathbf{)}}e^{-i\mathbf{k\cdot r}%
}\right\rangle . \label{IF21}%
\end{equation}

The time evolution of the free-phonon annihilation operator (\ref{IF10}),%
\[
a_{\mathbf{k}}(t)=e^{iH_{ph}t/\hbar}a_{\mathbf{k}}e^{-iH_{ph}t/\hbar}%
=\exp\left(  i\omega_{\mathbf{k}}a_{\mathbf{k}}^{+}a_{\mathbf{k}}t\right)
a_{\mathbf{k}}\exp\left(  -i\omega_{\mathbf{k}}a_{\mathbf{k}}^{+}%
a_{\mathbf{k}}t\right)
\]
accorting to (\ref{IF9}) is
\begin{align*}
-i\frac{da_{\mathbf{k}}(t)}{dt}  &  =\omega_{\mathbf{k}}\exp\left(
i\omega_{\mathbf{k}}a_{\mathbf{k}}^{+}a_{\mathbf{k}}t\right)  \left[
a_{\mathbf{k}}^{+}a_{\mathbf{k}},a_{\mathbf{k}}\right]  \exp\left(
-i\omega_{\mathbf{k}}a_{\mathbf{k}}^{+}a_{\mathbf{k}}t\right) \\
&  =\omega_{\mathbf{k}}\exp\left(  i\omega_{\mathbf{k}}a_{\mathbf{k}}%
^{+}a_{\mathbf{k}}t\right)  \left[  a_{\mathbf{k}}^{+},a_{\mathbf{k}}\right]
a_{\mathbf{k}}\exp\left(  -i\omega_{\mathbf{k}}a_{\mathbf{k}}^{+}%
a_{\mathbf{k}}t\right) \\
&  =-\omega_{\mathbf{k}}\exp\left(  i\omega_{\mathbf{k}}a_{\mathbf{k}}%
^{+}a_{\mathbf{k}}t\right)  a_{\mathbf{k}}\exp\left(  -i\omega_{\mathbf{k}%
}a_{\mathbf{k}}^{+}a_{\mathbf{k}}t\right)  \Rightarrow\\
a_{\mathbf{k}}(t)  &  =\exp\left(  -i\omega_{\mathbf{k}}t\right)
a_{\mathbf{k}}.
\end{align*}
Similarly,
\[
a_{\mathbf{k}}^{\dag}(t)=\exp\left(  i\omega_{\mathbf{k}}t\right)
a_{\mathbf{k}}^{\dag}.
\]
Hence, we have
\[
\left\langle a_{\mathbf{k}}(t)a_{\mathbf{k}}^{\dag}\right\rangle =\exp\left(
-i\omega_{\mathbf{k}}t\right)  \left\langle a_{\mathbf{k}}a_{\mathbf{k}}%
^{\dag}\right\rangle =\exp\left(  -i\omega_{\mathbf{k}}t\right)  \left\langle
1+a_{\mathbf{k}}^{\dag}a_{\mathbf{k}}\right\rangle =\exp\left(  -i\omega
_{\mathbf{k}}t\right)  \left[  1+n(\omega_{\mathbf{k}})\right]  ,
\]
where $n(\omega_{\mathbf{k}})=\left[  \exp(\beta\hbar\omega_{\mathbf{k}%
})-1\right]  ^{-1}$ is the average number of phonons with energy $\hbar
\omega_{\mathbf{k}}$.

The calculation of the Fourier component of the electron density-density
correlation function $\left\langle e^{i\mathbf{k\cdot r(}t\mathbf{)}%
}e^{-i\mathbf{k\cdot r}}\right\rangle $ in Eq. (\ref{IF21}) for an electron
described by the Feynman polaron model\ is given below following the approach
of Ref. \cite{PD1981}.

We calculate the correlation function
\begin{equation}
\left\langle e^{i\mathbf{k\cdot r}\left(  t\right)  }e^{-i\mathbf{k\cdot
r}\left(  \tau\right)  }\right\rangle =\frac{\mathrm{Tr}\left(  e^{-\beta
H_{F}}e^{i\mathbf{k\cdot r}\left(  t\right)  }e^{i\mathbf{k\cdot r}\left(
\tau\right)  }\right)  }{\mathrm{Tr}\left(  e^{-\beta H_{F}}\right)  }
\label{TF-eq0}%
\end{equation}
with the Feynman trial Hamiltonian%
\begin{equation}
H_{F}=\frac{\mathbf{p}^{2}}{2m}+\frac{\mathbf{p}_{f}^{2}}{2m_{f}}+\frac{1}%
{2}\chi\left(  \mathbf{r}-\mathbf{r}_{f}\right)  ^{2}. \label{TF-eq1}%
\end{equation}
Here, $\mathbf{r}\left(  t\right)  $ denotes the operator in the Heisenberg
representation%
\begin{equation}
\mathbf{r}\left(  t\right)  =e^{\frac{it}{\hbar}H_{F}}\mathbf{r}e^{-\frac
{it}{\hbar}H_{F}}. \label{TF-Heis}%
\end{equation}

We show that the correlation function $\left\langle e^{i\mathbf{k\cdot
r}\left(  t\right)  }e^{-i\mathbf{k\cdot r}\left(  \tau\right)  }\right\rangle
$ depends on $\left(  \tau-t\right)  $ rather than on $t$ and $\tau$
independently:%
\begin{align}
\left\langle e^{i\mathbf{k\cdot r}\left(  t\right)  }e^{-i\mathbf{k\cdot
r}\left(  \tau\right)  }\right\rangle  &  =\frac{\mathrm{Tr}\left(  e^{-\beta
H_{F}}e^{\frac{it}{\hbar}H_{F}}e^{i\mathbf{k\cdot r}}e^{-\frac{it}{\hbar}%
H_{F}}e^{\frac{i\tau}{\hbar}H_{F}}e^{i\mathbf{k\cdot r}}e^{-\frac{i\tau}%
{\hbar}H_{F}}\right)  }{\mathrm{Tr}\left(  e^{-\beta H_{F}}\right)
}\nonumber\\
&  =\frac{\mathrm{Tr}\left(  e^{-\beta H_{F}}e^{i\mathbf{k\cdot r}}%
e^{\frac{i\left(  \tau-t\right)  }{\hbar}H_{F}}e^{i\mathbf{k\cdot r}}%
e^{-\frac{i\left(  \tau-t\right)  }{\hbar}H_{F}}\right)  }{\mathrm{Tr}\left(
e^{-\beta H_{F}}\right)  }\nonumber\\
&  =\left\langle e^{i\mathbf{k\cdot r}}e^{-i\mathbf{k\cdot r}\left(
\tau-t\right)  }\right\rangle =\left\langle e^{i\mathbf{k\cdot r}%
}e^{-i\mathbf{k\cdot r}\left(  \sigma\right)  }\right\rangle , \label{TF-time}%
\end{align}
where $\sigma=\tau-t$.

The normal coordinates are the center-of-mass vector $\mathbf{R}$ and the
vector of the relative motion $\mathbf{\rho}$:%
\begin{equation}
\left\{
\begin{array}
[c]{c}%
\mathbf{R}=\frac{m\mathbf{r}+m_{f}\mathbf{r}_{f}}{m+m_{f}}\\
\mathbf{\rho}=\mathbf{r}-\mathbf{r}_{f}%
\end{array}
\right.  \label{TF-eq2}%
\end{equation}
The inverse transformation is:%
\begin{equation}
\left\{
\begin{array}
[c]{c}%
\mathbf{r}=\mathbf{R}+\frac{m_{f}}{m+m_{f}}\mathbf{\rho}\\
\mathbf{r}_{f}=\mathbf{R}-\frac{m}{m+m_{f}}\mathbf{\rho}%
\end{array}
\right.  \label{TF-eq3}%
\end{equation}
The same transformation as (\ref{TF-eq2}) takes place for velocities:%
\begin{equation}
\left\{
\begin{array}
[c]{c}%
\mathbf{\dot{r}}=\mathbf{\dot{R}}+\frac{m_{f}}{m+m_{f}}\mathbf{\dot{\rho}}\\
\mathbf{\dot{r}}_{f}=\mathbf{\dot{R}}-\frac{m}{m+m_{f}}\mathbf{\dot{\rho}}%
\end{array}
\right.
\end{equation}
From (\ref{TF-eq2}) we derive the transformation for moments%
\[
\left\{
\begin{array}
[c]{c}%
\frac{\mathbf{p}}{m}=\frac{\mathbf{P}}{m+m_{f}}+\frac{m_{f}}{m+m_{f}}%
\frac{\mathbf{p}_{\rho}}{\frac{mm_{f}}{m+m_{f}}}\\
\frac{\mathbf{p}_{f}}{m_{f}}=\frac{\mathbf{P}}{m+m_{f}}-\frac{m}{m+m_{f}}%
\frac{\mathbf{p}_{\rho}}{\frac{mm_{f}}{m+m_{f}}}%
\end{array}
\right.
\]%
\[
\Downarrow
\]%
\begin{equation}
\left\{
\begin{array}
[c]{c}%
\mathbf{p}=\frac{m}{m+m_{f}}\mathbf{P}+\mathbf{p}_{\rho}\\
\mathbf{p}_{f}=\frac{m_{f}}{m+m_{f}}\mathbf{P}-\mathbf{p}_{\rho}%
\end{array}
\right.  \label{TF-eq4}%
\end{equation}
The Hamiltonian (\ref{TF-eq1}) then takes the form%
\begin{equation}
H_{F}=\frac{\mathbf{P}^{2}}{2M}+\frac{\mathbf{p}_{\rho}^{2}}{2\mu}+\frac{1}%
{2}\mu\bar{\Omega}^{2}\mathbf{\rho}^{2} \label{TF-eq5}%
\end{equation}
with the masses%
\begin{equation}
M=m+m_{f},\; \mu=\frac{mm_{f}}{m+m_{f}} \label{TF-eq6}%
\end{equation}
and with the frequency%
\begin{equation}
\bar{\Omega}=\sqrt{\frac{\chi}{\mu}}. \label{TF-eq7}%
\end{equation}
The Cartesian coordinates and moments corresponding to the relative motion can
be in the standard way expressed in terms of the second quantization
operators:%
\begin{align}
\rho_{j}  &  =\left(  \frac{\hbar}{2\mu\bar{\Omega}}\right)  ^{1/2}\left(
C_{j}+C_{j}^{\dag}\right)  ,\nonumber\\
p_{\rho,j}  &  =-i\left(  \frac{\mu\hbar\bar{\Omega}}{2}\right)  ^{1/2}\left(
C_{j}-C_{j}^{\dag}\right)  .\label{TF-eq8}\\
&  \left(  j=1,2,3\right) \nonumber
\end{align}
In these notations, the Hamiltonian (\ref{TF-eq5}) takes the form%
\begin{equation}
H_{F}=\frac{\mathbf{P}^{2}}{2M}+\sum_{j=1}^{3}\hbar\bar{\Omega}\left(
C_{j}^{\dag}C_{j}+\frac{1}{2}\right)  . \label{TF-eq9}%
\end{equation}

Using (\ref{TF-eq9}), we find the operators in the Heisenberg representation,
(i) for the center-of mass coordinates%
\begin{align*}
X_{j}\left(  \sigma\right)   &  =e^{\frac{i\sigma}{\hbar}H_{F}}X_{j}%
e^{-\frac{i\sigma}{\hbar}H_{F}}=e^{i\frac{\sigma}{2M\hbar}P_{j}^{2}}%
X_{j}e^{-i\frac{\sigma}{2M\hbar}P_{j}^{2}}\\
&  =X_{j}+i\frac{\sigma}{2M\hbar}\left[  P_{j}^{2},X_{j}\right] \\
&  =X_{j}+i\frac{\sigma}{2M\hbar}\left(  P_{j}^{2}X_{j}-P_{j}X_{j}P_{j}%
+P_{j}X_{j}P_{j}-X_{j}P_{j}^{2}\right) \\
&  =X_{j}+i\frac{\sigma}{2M\hbar}\left(  P_{j}\left[  P_{j},X_{j}\right]
+\left[  P_{j},X_{j}\right]  P_{j}\right) \\
&  =X_{j}+i\frac{\sigma}{2M\hbar}P_{j}\left(  -2i\hbar\right)  =X_{j}%
+\frac{\sigma}{M}P_{j},
\end{align*}%
\begin{equation}
X_{j}\left(  \sigma\right)  =X_{j}+\frac{\sigma}{M}P_{j}, \label{TF-eq10}%
\end{equation}
(ii) for the operators $C_{j}$ and $C_{j}^{\dag}$%
\begin{equation}
C_{j}\left(  \sigma\right)  =C_{j}e^{-i\bar{\Omega}\sigma},\;C_{j}^{\dag
}\left(  \sigma\right)  =C_{j}^{\dag}e^{i\bar{\Omega}\sigma}. \label{TF-eq11}%
\end{equation}
Using the first formula of (\ref{TF-eq3}) we find%
\[
\mathbf{r}\left(  \sigma\right)  =\mathbf{R}\left(  \sigma\right)
+\frac{m_{f}}{m+m_{f}}\mathbf{\rho}\left(  \sigma\right)
\]%
\[
\Downarrow
\]%
\begin{equation}
\mathbf{r}\left(  \sigma\right)  =\mathbf{R}+\frac{\sigma}{M}\mathbf{P}%
+\frac{m_{f}}{M}\left(  \frac{\hbar}{2\mu\bar{\Omega}}\right)  ^{1/2}\left(
\mathbf{C}e^{-i\bar{\Omega}\sigma}+\mathbf{C}^{\dag}e^{i\bar{\Omega}\sigma
}\right)  . \label{TF-eq12}%
\end{equation}
We denote%
\begin{align*}
a  &  \equiv\frac{m_{f}}{M}\left(  \frac{\hbar}{2\mu\bar{\Omega}}\right)
^{1/2}=\left(  \frac{\hbar m_{f}^{2}}{2M^{2}\mu\bar{\Omega}}\right)  ^{1/2}\\
&  =\left(  \frac{\hbar m_{f}^{2}}{2\left(  m+m_{f}\right)  ^{2}\frac{mm_{f}%
}{m+m_{f}}\bar{\Omega}}\right)  ^{1/2}=\left(  \frac{\hbar m_{f}}%
{2mM\bar{\Omega}}\right)  ^{1/2}%
\end{align*}%
\begin{equation}
\mathbf{r}\left(  \sigma\right)  =\mathbf{R}+\frac{\sigma}{M}\mathbf{P}%
+a\left(  \mathbf{C}e^{-i\bar{\Omega}\sigma}+\mathbf{C}^{\dag}e^{i\bar{\Omega
}\sigma}\right)  . \label{TF-eq13}%
\end{equation}
Therefore, we obtain%
\begin{align*}
e^{i\mathbf{k\cdot r}}  &  =\exp\left(  i\mathbf{k\cdot R}\right)  \exp\left(
ia\mathbf{k\cdot C}+ia\mathbf{k\cdot C}^{\dag}\right) \\
&  =\prod_{j=1}^{3}\exp\left(  ik_{j}X_{j}\right)  \exp\left(  iak_{j}%
C_{j}-iak_{j}C_{j}^{\dag}\right)
\end{align*}

\begin{align}
e^{-i\mathbf{k\cdot r}\left(  \sigma\right)  }  &  =\exp\left[
-i\mathbf{k\cdot R}-i\frac{\sigma}{M}\mathbf{k\cdot P}-ia\mathbf{k\cdot
}\left(  \mathbf{C}e^{-i\bar{\Omega}\sigma}+\mathbf{C}^{\dag}e^{i\bar{\Omega
}\sigma}\right)  \right] \nonumber\\
&  =\exp\left(  -i\mathbf{k\cdot R}-i\frac{\sigma}{M}\mathbf{k\cdot P}\right)
\exp\left(  -ia\mathbf{k\cdot C}e^{-i\bar{\Omega}\sigma}-ia\mathbf{k\cdot
C}^{\dag}e^{i\bar{\Omega}\sigma}\right) \nonumber\\
&  =\prod_{j=1}^{3}\exp\left(  -ik_{j}X_{j}-i\frac{\sigma}{M}k_{j}%
P_{j}\right)  \exp\left(  -iak_{j}C_{j}e^{-i\bar{\Omega}\sigma}-iak_{j}%
C_{j}^{\dag}e^{i\bar{\Omega}\sigma}\right)  . \label{TF-eq14}%
\end{align}
The disentangling of the exponents is performed using the formula%
\begin{equation}
e^{A+B}=e^{A}T\exp\left(  \int_{0}^{1}d\lambda e^{-\lambda A}Be^{\lambda
A}\right)  . \label{TF-eq15}%
\end{equation}
In the case when $\left[  A,B\right]  $ commutes with both $A$ and $B$, this
formula is reduced to%
\begin{equation}
e^{A+B}=e^{A}e^{B}e^{-\frac{1}{2}\left[  A,B\right]  }. \label{TF-eq16}%
\end{equation}
We perform the necessary commutations:%
\[
-\frac{1}{2}\left[  -ik_{j}X_{j},-i\frac{\sigma}{M}k_{j}P_{j}\right]
=\frac{1}{2}k_{j}^{2}\frac{\sigma}{M}\left[  X_{j},P_{j}\right]  =i\frac{\hbar
k_{j}^{2}}{2M}\sigma,
\]%
\[
-\frac{1}{2}\left[  iak_{j}C_{j}^{\dag},iak_{j}C_{j}\right]  =\frac{1}{2}%
a^{2}k_{j}^{2}\left[  C_{j}^{\dag},C_{j}\right]  =-\frac{1}{2}a^{2}k_{j}^{2}%
\]%
\[
-\frac{1}{2}\left[  -iak_{j}C_{j}^{\dag}e^{i\Omega\sigma},-iak_{j}%
C_{j}e^{-i\Omega\sigma}\right]  =\frac{1}{2}a^{2}k_{j}^{2}\left[  C_{j}^{\dag
},C_{j}\right]  =-\frac{1}{2}a^{2}k_{j}^{2}%
\]%
\[
\Downarrow
\]%
\begin{align*}
e^{i\mathbf{k\cdot r}}  &  =e^{i\mathbf{k\cdot R}}e^{ia\mathbf{k\cdot C}%
^{\dag}}e^{ia\mathbf{k\cdot C}}e^{-\frac{1}{2}a^{2}k^{2}},\\
e^{-i\mathbf{k\cdot r}\left(  \sigma\right)  }  &  =e^{-i\mathbf{k\cdot R}%
}e^{-i\frac{\sigma}{M}\mathbf{k\cdot P}}e^{-ia\mathbf{k\cdot C}^{\dag}%
e^{i\bar{\Omega}\sigma}}e^{-ia\mathbf{k\cdot C}e^{-i\bar{\Omega}\sigma}%
}e^{i\frac{\hbar k^{2}}{2M}\sigma-\frac{1}{2}a^{2}k^{2}}%
\end{align*}%
\[
\Downarrow
\]%
\[
e^{i\mathbf{k\cdot r}}e^{-i\mathbf{k\cdot r}\left(  \sigma\right)
}=e^{-i\frac{\sigma}{M}\mathbf{k\cdot P}}e^{ia\mathbf{k\cdot C}^{\dag}%
}e^{ia\mathbf{k\cdot C}}e^{-ia\mathbf{k\cdot C}^{\dag}e^{i\bar{\Omega}\sigma}%
}e^{-ia\mathbf{k\cdot C}e^{-i\bar{\Omega}\sigma}}e^{i\frac{\hbar k^{2}}%
{2M}\sigma-a^{2}k^{2}}.
\]

It follows from Eq. (\ref{TF-eq16}) that when $\left[  A,B\right]  $ commutes
with both $A$ and $B$,%
\begin{equation}
e^{A}e^{B}=e^{B}e^{A}e^{\left[  A,B\right]  }. \label{TF-eq17}%
\end{equation}
Using (\ref{TF-eq17}), we find%
\begin{align*}
e^{iak_{j}\cdot C_{j}}e^{-iak_{j}\cdot C_{j}^{\dag}e^{i\bar{\Omega}\sigma}}
&  =e^{-iak_{j}\cdot C_{j}^{\dag}e^{i\bar{\Omega}\sigma}}e^{iak_{j}\cdot
C_{j}}e^{\left[  iak_{j}\cdot C_{j},-iak_{j}\cdot C_{j}^{\dag}e^{i\bar{\Omega
}\sigma}\right]  }\\
&  =e^{-iak_{j}\cdot C_{j}^{\dag}e^{i\bar{\Omega}\sigma}}e^{iak_{j}\cdot
C_{j}}e^{a^{2}k_{j}^{2}e^{i\bar{\Omega}\sigma}}.
\end{align*}
Herefrom, we find
\begin{equation}
e^{i\mathbf{k\cdot r}}e^{-i\mathbf{k\cdot r}\left(  \sigma\right)
}=e^{-i\frac{\sigma}{M}\mathbf{k\cdot P}}e^{ia\mathbf{k\cdot C}^{\dag}\left(
1-e^{i\bar{\Omega}\sigma}\right)  }e^{ia\mathbf{k\cdot C}\left(
1-e^{-i\bar{\Omega}\sigma}\right)  }e^{i\frac{\hbar k^{2}}{2M}\sigma
-a^{2}k^{2}\left(  1-e^{i\bar{\Omega}\sigma}\right)  }. \label{TF-eq18}%
\end{equation}
The correlation function then is%
\begin{equation}
\left\langle e^{i\mathbf{k\cdot r}}e^{-i\mathbf{k\cdot r}\left(
\sigma\right)  }\right\rangle =\left\langle e^{-i\frac{\sigma}{M}%
\mathbf{k\cdot P}}\right\rangle \left\langle e^{ia\mathbf{k\cdot C}^{\dag
}\left(  1-e^{i\bar{\Omega}\sigma}\right)  }e^{ia\mathbf{k\cdot C}\left(
1-e^{-i\bar{\Omega}\sigma}\right)  }\right\rangle e^{i\frac{\hbar k^{2}}%
{2M}\sigma-a^{2}k^{2}\left(  1-e^{i\bar{\Omega}\sigma}\right)  },
\label{TF-eq20}%
\end{equation}
since the variables of the center-of mass motion and of the relative motion
are averaged independently.%
\begin{align*}
\left\langle e^{ia\mathbf{k\cdot C}^{\dag}\left(  1-e^{i\bar{\Omega}\sigma
}\right)  }e^{ia\mathbf{k\cdot C}\left(  1-e^{-i\bar{\Omega}\sigma}\right)
}\right\rangle  &  =\left\langle e^{i\mathbf{Q\cdot C}^{\dag}}e^{i\mathbf{Q}%
^{\ast}\mathbf{\cdot C}}\right\rangle \\
&  =\frac{\mathrm{Tr}\left(  e^{-\beta\hbar\bar{\Omega}\mathbf{C}^{\dag}%
\cdot\mathbf{C}}e^{i\mathbf{Q\cdot C}^{\dag}}e^{i\mathbf{Q}^{\ast
}\mathbf{\cdot C}}\right)  }{\mathrm{Tr}\left(  e^{-\beta\hbar\bar{\Omega
}\mathbf{C}^{\dag}\cdot\mathbf{C}}\right)  }%
\end{align*}
with%
\[
\mathbf{Q}=a\left(  1-e^{i\bar{\Omega}\sigma}\right)  \mathbf{k}.
\]

Let us consider the auxiliary expectation value%
\begin{align*}
\left\langle e^{iQC^{\dag}}e^{iQ^{\ast}C}\right\rangle  &  \equiv
\frac{\mathrm{Tr}\left(  e^{-\beta\hbar\bar{\Omega}C^{\dag}C}e^{iQC^{\dag}%
}e^{iQ^{\ast}C}\right)  }{\mathrm{Tr}\left(  e^{-\beta\hbar\bar{\Omega}%
C^{\dag}C}\right)  }\\
&  =\frac{1}{\mathrm{Tr}\left(  e^{-\beta\hbar\bar{\Omega}C^{\dag}C}\right)
}\sum_{n=0}^{\infty}\frac{\left(  iQ\right)  ^{n}}{n!}\sum_{m=0}^{\infty}%
\frac{\left(  iQ^{\ast}\right)  ^{m}}{m!}\mathrm{Tr}\left(  e^{-\beta\hbar
\bar{\Omega}C^{\dag}C}\left(  C^{\dag}\right)  ^{n}C^{m}\right) \\
&  =\frac{1}{\mathrm{Tr}\left(  e^{-\beta\hbar\bar{\Omega}C^{\dag}C}\right)
}\sum_{n=0}^{\infty}\frac{\left(  -1\right)  ^{n}\left\vert Q\right\vert
^{2n}}{\left(  n!\right)  ^{2}}\mathrm{Tr}\left(  e^{-\beta\hbar\bar{\Omega
}C^{\dag}C}\left(  C^{\dag}\right)  ^{n}C^{n}\right)  .
\end{align*}%
\begin{align*}
\mathrm{Tr}\left(  e^{-\beta\hbar\bar{\Omega}C^{\dag}C}\left(  C^{\dag
}\right)  ^{n}C^{n}\right)   &  =\sum_{m=0}^{\infty}\left\langle m\left\vert
e^{-\beta\hbar\bar{\Omega}C^{\dag}C}\left(  C^{\dag}\right)  ^{n}%
C^{n}\right\vert m\right\rangle \\
&  =\sum_{m=0}^{\infty}e^{-\beta\hbar\bar{\Omega}m}\left\langle m\left\vert
\left(  C^{\dag}\right)  ^{n}C^{n}\right\vert m\right\rangle ,
\end{align*}
where $\left\vert m\right\rangle $ are the eigenstates of $\left(  C^{\dag
}C\right)  $. The operators $C$ act on these states as follows:%
\begin{align*}
C\left\vert m\right\rangle  &  =\sqrt{m}\left\vert m-1\right\rangle ,\\
C\left\vert 0\right\rangle  &  =0.
\end{align*}
Therefore, we find%
\begin{align*}
\left\langle m\left\vert \left(  C^{\dag}\right)  ^{n}C^{n}\right\vert
m\right\rangle  &  =m\left(  m-1\right)  \ldots\left(  m-n+1\right)
=\frac{m!}{\left(  m-n\right)  !}\; \text{for }n\leq m,\\
\left\langle m\left\vert \left(  C^{\dag}\right)  ^{n}C^{n}\right\vert
m\right\rangle  &  =0\; \text{for }n>m.
\end{align*}%
\[
\Downarrow
\]%
\[
Tr\left(  e^{-\beta\hbar\bar{\Omega}C^{\dag}C}\left(  C^{\dag}\right)
^{n}C^{n}\right)  =\sum_{m=n}^{\infty}e^{-\beta\hbar\Omega m}\frac{m!}{\left(
m-n\right)  !},
\]
and%
\begin{align*}
\mathrm{Tr}\left(  e^{-\beta\hbar\Omega C^{\dag}C}e^{iQC^{\dag}}e^{iQ^{\ast}%
C}\right)   &  =\sum_{n=0}^{\infty}\frac{\left(  -1\right)  ^{n}\left\vert
Q\right\vert ^{2n}}{\left(  n!\right)  ^{2}}\sum_{m=n}^{\infty}e^{-\beta
\hbar\bar{\Omega}m}\frac{m!}{\left(  m-n\right)  !}\\
&  =\sum_{n=0}^{\infty}\frac{\left(  -1\right)  ^{n}\left\vert Q\right\vert
^{2n}}{n!}\sum_{m=n}^{\infty}e^{-\beta\hbar\bar{\Omega}m}\binom{m}{n}\\
&  =\sum_{n=0}^{\infty}\frac{\left(  -1\right)  ^{n}\left\vert Q\right\vert
^{2n}}{n!}e^{-\beta\hbar\bar{\Omega}n}\sum_{k=0}^{\infty}e^{-\beta\hbar
\bar{\Omega}k}\binom{k+n}{n}\\
&  =\sum_{n=0}^{\infty}\frac{\left(  -1\right)  ^{n}\left\vert Q\right\vert
^{2n}}{n!}e^{-\beta\hbar\bar{\Omega}n}\frac{1}{\left(  1-e^{-\beta\hbar
\bar{\Omega}}\right)  ^{n+1}}\\
&  =\frac{1}{1-e^{-\beta\hbar\bar{\Omega}}}\sum_{n=0}^{\infty}\frac{\left(
-1\right)  ^{n}\left\vert Q\right\vert ^{2n}}{n!}\frac{1}{\left(
e^{\beta\hbar\bar{\Omega}}-1\right)  ^{n}}\\
&  =\frac{1}{1-e^{-\beta\hbar\bar{\Omega}}}\exp\left(  -\frac{\left\vert
Q\right\vert ^{2}}{e^{\beta\hbar\bar{\Omega}}-1}\right)  .
\end{align*}
In particular, for $Q=Q^{\ast}=0,$ we have%
\begin{equation}
Tr\left(  e^{-\beta\hbar\Omega C^{\dag}C}\right)  =\frac{1}{1-e^{-\beta
\hbar\Omega}}.
\end{equation}
As a result, the expectation value $\left\langle e^{iQC^{\dag}}e^{iQ^{\ast}%
C}\right\rangle $ is%
\begin{equation}
\left\langle e^{iQC^{\dag}}e^{iQ^{\ast}C}\right\rangle =\exp\left[  -n\left(
\Omega\right)  \left\vert Q\right\vert ^{2}\right]  , \label{TF-eq19}%
\end{equation}
with%
\[
n\left(  \Omega\right)  \equiv\frac{1}{e^{\beta\hbar\Omega}-1}.
\]
Using this result, we obtain the expression%
\begin{align*}
\left\langle e^{i\mathbf{Q\cdot C}^{\dag}}e^{i\mathbf{Q}^{\ast}\mathbf{\cdot
C}}\right\rangle  &  =\exp\left[  -n\left(  \Omega\right)  \mathbf{Q\cdot
Q}^{\ast}\right] \\
&  =\exp\left[  -n\left(  \Omega\right)  a^{2}k^{2}\left(  1-e^{i\bar{\Omega
}\sigma}\right)  \left(  1-e^{-i\bar{\Omega}\sigma}\right)  \right] \\
&  =\exp\left[  -4n\left(  \Omega\right)  a^{2}k^{2}\sin^{2}\left(  \frac
{1}{2}\Omega\sigma\right)  \right]  .
\end{align*}
The expectation value $\left\langle e^{-i\frac{\sigma}{M}\mathbf{k\cdot P}%
}\right\rangle $ is%
\begin{align*}
\left\langle e^{-i\frac{\sigma}{M}\mathbf{k\cdot P}}\right\rangle  &
=\frac{\int d\mathbf{P}\exp\left(  -\beta\frac{P^{2}}{2M}-i\frac{\sigma}%
{M}\mathbf{k\cdot P}\right)  }{\int d\mathbf{P}\exp\left(  -\beta\frac{P^{2}%
}{2M}\right)  }\\
&  =\frac{\int d\mathbf{P}\exp\left(  \frac{\left(  -i\mathbf{P}%
\beta+\mathbf{k}\sigma\right)  ^{2}}{2M\beta}-\frac{k^{2}\sigma^{2}}{2M\beta
}\right)  }{\int d\mathbf{P}\exp\left(  -\beta\frac{P^{2}}{2M}\right)  }\\
&  =\exp\left(  -\frac{k^{2}\sigma^{2}}{2M\beta}\right)  .
\end{align*}
Collecting all factors in Eq. (\ref{TF-eq20}) together, we find%
\[
\left\langle e^{i\mathbf{k\cdot r}}e^{-i\mathbf{k\cdot r}\left(
\sigma\right)  }\right\rangle =\exp\left[  i\frac{\hbar k^{2}}{2M}\sigma
-\frac{k^{2}\sigma^{2}}{2M\beta}-4n\left(  \bar{\Omega}\right)  a^{2}k^{2}%
\sin^{2}\left(  \frac{1}{2}\bar{\Omega}\sigma\right)  -a^{2}k^{2}\left(
1-e^{i\bar{\Omega}\sigma}\right)  \right]  \Rightarrow.
\]%
\begin{equation}
\left\langle e^{i\mathbf{k\cdot r(}t\mathbf{)}}e^{-i\mathbf{k\cdot r}\left(
\sigma+t\right)  }\right\rangle =e^{-k^{2}D\left(  \sigma\right)  }
\label{TF-eq21}%
\end{equation}
with the function%
\begin{equation}
D(t)=\frac{\hbar}{2M}\left(  -it+\frac{t^{2}}{\beta\hbar}\right)
+a^{2}\left[  1-\exp(i\bar{\Omega}t)+4n\left(  \bar{\Omega}\right)  \sin
^{2}\left(  \frac{\bar{\Omega}t}{2}\right)  \right]  , \label{IF22a}%
\end{equation}
where
\begin{equation}
M=\left(  \frac{v}{w}\right)  ^{2}m_{b},\bar{\Omega}=v\omega_{\mathrm{LO}%
},a^{2}=\frac{\hbar}{2m_{b}\omega_{\mathrm{LO}}}\frac{v^{2}-w^{2}}{v^{3}}.
\label{IF22b}%
\end{equation}
According to (\ref{TF-time}),
\[
\left\langle e^{i\mathbf{k\cdot r(}t\mathbf{)}}e^{-i\mathbf{k\cdot r}\left(
\sigma+t\right)  }\right\rangle =\left\langle e^{i\mathbf{k\cdot r}%
}e^{-i\mathbf{k\cdot r}\left(  \sigma\right)  }\right\rangle .
\]
Taking $\sigma=-t$ in (\ref{TF-eq21}) we finally find the Fourier component of
the electron density-density correlation function $\left\langle
e^{i\mathbf{k\cdot r(}t\mathbf{)}}e^{-i\mathbf{k\cdot r}}\right\rangle $ which
enters Eq. (\ref{IF21}):%

\begin{equation}
\left\langle e^{i\mathbf{k\cdot r(}t\mathbf{)}}e^{-i\mathbf{k\cdot r}%
}\right\rangle =\exp\left[  -k^{2}D(-t)\right]  \label{IF22}%
\end{equation}

Finally, the correlation functions in (\ref{IF20b}) reduce to%
\begin{align*}
\left\langle b_{\mathbf{k}}(t)b_{\mathbf{k}}^{\dag}(0)\right\rangle  &
=\left[  1+n(\omega_{\mathbf{k}})\right]  \exp\left(  -i\omega_{\mathbf{k}%
}t\right)  \exp\left[  -k^{2}D(-t)\right]  ,\\
\left\langle b_{\mathbf{k}}^{\dag}(0)b_{\mathbf{k}}(t)\right\rangle  &
=n(\omega_{\mathbf{k}})\exp\left(  -i\omega_{\mathbf{k}}t\right)  \exp\left[
-k^{2}D(t)\right]  ,\\
\left\langle b_{\mathbf{k}}(t)b_{\mathbf{k}}(0)\right\rangle  &  =0,\\
\left\langle b_{\mathbf{k}}(0)b_{\mathbf{k}}(t)\right\rangle  &  =0.
\end{align*}
Inserting these equations into Eqs. (\ref{IF20a}) and (\ref{IF20b}), one
obtains%
\begin{equation}
\Sigma(z)=\frac{1}{z}%
%TCIMACRO{\dint \nolimits_{0}^{\infty}}%
%BeginExpansion
{\displaystyle\int\nolimits_{0}^{\infty}}
%EndExpansion
dt\left(  1-e^{izt}\right)  \operatorname{Im}S(t) \label{IF23a}%
\end{equation}
with%
\[
\operatorname{Im}S(t)=\frac{2}{m_{b}\hbar}\sum_{\mathbf{k}}k_{x}^{2}\left\vert
V_{k}\right\vert ^{2}\operatorname{Im}\left\{
\begin{array}
[c]{c}%
-\left[  1+n(\omega_{\mathbf{k}})\right]  \exp\left(  -i\omega_{\mathbf{k}%
}t\right)  \exp\left[  -k^{2}D(-t)\right] \\
+n(\omega_{\mathbf{k}})\exp\left(  -i\omega_{\mathbf{k}}t\right)  \exp\left[
-k^{2}D(t)\right]
\end{array}
\right\}  .
\]
Using the property $D(-t)^{\ast}=^{2}D(t)$ for real vaues of $t,$one obtains%
\begin{align*}
-\operatorname{Im}\exp\left(  -i\omega_{\mathbf{k}}t\right)  \exp\left[
-k^{2}D(-t)\right]  \}  &  =\operatorname{Im}\exp\left(  i\omega_{\mathbf{k}%
}t\right)  \exp\left[  -k^{2}D(-t)^{\ast}\right]  \}\\
&  =\operatorname{Im}\exp\left(  i\omega_{\mathbf{k}}t\right)  \exp\left[
-k^{2}D(t)\right]  \}
\end{align*}
and consequently%
\begin{equation}
S(t)=\frac{2}{m_{b}\hbar}\sum_{\mathbf{k}}k_{x}^{2}\left\vert V_{k}\right\vert
^{2}\exp\left[  -k^{2}D(t)\right]  \left\{  \left[  1+n(\omega_{\mathbf{k}%
})\right]  \exp\left(  i\omega_{\mathbf{k}}t\right)  +n(\omega_{\mathbf{k}%
})\exp\left(  -i\omega_{\mathbf{k}}t\right)  \right\}  . \label{IF23b}%
\end{equation}
Owing to the rotational invariance of $\left\vert V_{k}\right\vert ^{2}$ and
$\omega_{\mathbf{k}},$we can substitute in (\ref{IF23b})%
\[
k_{x}^{2}\rightarrow\frac{k^{2}}{3}.
\]
The resulting expression for
\begin{equation}
S(t)=\frac{2}{3m_{b}\hbar}\sum_{\mathbf{k}}k^{2}\left\vert V_{k}\right\vert
^{2}\exp\left[  -k^{2}D(t)\right]  \left\{  \left[  1+n(\omega_{\mathbf{k}%
})\right]  \exp\left(  i\omega_{\mathbf{k}}t\right)  +n(\omega_{\mathbf{k}%
})\exp\left(  -i\omega_{\mathbf{k}}t\right)  \right\}  \label{IF23c}%
\end{equation}
is identical with Eq. (35) of FHIP \cite{FHIP}. In the case of Fr\"{o}hlich
polarons, taking into account (\ref{eq_1b}), Eq. (\ref{IF23c}) simplifies to
\begin{equation}
S(t)=\left(  \frac{\hbar\omega_{\mathrm{LO}}}{m_{b}}\right)  ^{3/2}%
\frac{\alpha}{3\sqrt{2\pi}}\left\{
\begin{array}
[c]{c}%
\left[  1+n(\omega_{\mathrm{LO}})\right]  \exp\left(  i\omega_{\mathrm{LO}%
}t\right) \\
+n(\omega_{\mathrm{LO}})\exp\left(  -i\omega_{\mathrm{LO}}t\right)
\end{array}
\right\}  \left[  D(t)\right]  ^{-3/2}. \label{IF23d}%
\end{equation}

\subsection{Calculation of the memory function (Devreese \textit{et. al.}
\cite{DSG1972})}

Upon substituion of (\ref{IF6}) and (\ref{IF15c}) into Eq. (\ref{IF5}), we
find the absorption coefficient%
\begin{align*}
\Gamma(\Omega)  &  =\frac{1}{n\epsilon_{0}c}\operatorname{Re}\left[
\frac{ie^{2}}{m_{b}}\frac{1}{\Omega-\Sigma(\Omega)}\right]  =-\frac
{1}{n\epsilon_{0}c}\frac{e^{2}}{m_{b}}\operatorname{Im}\left[  \frac{1}%
{\Omega-\Sigma(\Omega)}\right] \\
&  =-\frac{1}{n\epsilon_{0}c}\frac{e^{2}}{m_{b}}\operatorname{Im}\left\{
\frac{\Omega-\Sigma^{\ast}(\Omega)}{\left[  \Omega-\operatorname{Re}%
\Sigma(\Omega)\right]  ^{2}+\left[  \operatorname{Im}\Sigma(\Omega)\right]
^{2}}\right\} \\
&  =\frac{1}{n\epsilon_{0}c}\frac{e^{2}}{m_{b}}\frac{\operatorname{Im}%
\Sigma^{\ast}(\Omega)}{\left[  \Omega-\operatorname{Re}\Sigma(\Omega)\right]
^{2}+\left[  \operatorname{Im}\Sigma(\Omega)\right]  ^{2}}\Rightarrow
\end{align*}
{\normalsize
\begin{equation}
\Gamma(\Omega)=-\frac{1}{n\epsilon_{0}c}\frac{e^{2}}{m_{b}}\frac
{\operatorname{Im}\Sigma(\Omega)}{\left[  \Omega-\operatorname{Re}%
\Sigma(\Omega)\right]  ^{2}+\left[  \operatorname{Im}\Sigma(\Omega)\right]
^{2}}\ . \label{eq:P24-2}%
\end{equation}
}This general expression was the strating point for a derivation of the
theoretical optical absorption spectrum of a single large polaron, at
\textit{all electron-phonon coupling strengths} by Devreese et al. in
Ref.\thinspace\cite{DSG1972}. The memory function $\Sigma(\Omega)$ as given by
Eq. (\ref{IF23a}) with (\ref{IF23d}) contains the dynamics of the polaron and
depends on $\alpha$, temperature and $\Omega.$Following the notation,
introduced in Ref. \cite{FHIP},
\begin{equation}
\Sigma(\Omega)=\frac{\chi^{\ast}(\Omega)}{\Omega} \label{Chi0}%
\end{equation}
we reresent Eq. (\ref{eq:P24-2}) in the form used in Ref.\thinspace
\cite{DSG1972}:{\normalsize
\begin{equation}
\Gamma(\Omega)=\frac{1}{n\epsilon_{0}c}\frac{e^{2}}{m_{b}}\frac{\Omega
\operatorname{Im}\chi(\Omega)}{\left[  \Omega^{2}-\operatorname{Re}\chi
(\Omega)\right]  ^{2}+\left[  \operatorname{Im}\chi(\Omega)\right]  ^{2}}.
\label{eq:P24-2a}%
\end{equation}
}According to (\ref{IF23a}) and (\ref{Chi0}),%
\begin{equation}
\operatorname{Im}\chi(\Omega)=\operatorname{Im}%
%TCIMACRO{\dint \nolimits_{0}^{\infty}}%
%BeginExpansion
{\displaystyle\int\nolimits_{0}^{\infty}}
%EndExpansion
dt\sin(\Omega t)S(t),\quad\operatorname{Re}\chi(\Omega)=\operatorname{Im}%
%TCIMACRO{\dint \nolimits_{0}^{\infty}}%
%BeginExpansion
{\displaystyle\int\nolimits_{0}^{\infty}}
%EndExpansion
dt\left[  1-\cos(\Omega t)\right]  S(t). \label{IF24}%
\end{equation}
In the present Notes we limit our attention to the case $T=0$ $(\beta
\rightarrow\infty).$ It was demonstrated in Ref.\cite{DSG1972} that the exact
zero-temperature limit arises if the limit $\beta\rightarrow\infty$ is taken
directly in the expressions (\ref{IF24}) (see Appendices A and B of
Ref.\cite{DSG1972}). As follows from (\ref{IF23d}),
\begin{equation}
S(t)=\left(  \frac{\hbar\omega_{\mathrm{LO}}}{m_{b}}\right)  ^{3/2}%
\frac{\alpha}{3\sqrt{2}\pi}\exp\left(  i\omega_{\mathrm{LO}}t\right)  \left[
D(t)\right]  ^{-3/2}\qquad(\beta\rightarrow\infty). \label{IF25}%
\end{equation}
Accorting to (\ref{IF22a})%
\[
D(t)=-i\frac{\hbar t}{2M}+a^{2}\left[  1-\exp(i\bar{\Omega}t)\right]
\qquad(\beta\rightarrow\infty).
\]
Using the Feynman units (where $\hbar=1,\omega_{\mathrm{LO}}=1$ and $m_{b}%
=1$), we obtain from (\ref{IF22b}):%
\[
M=\left(  \frac{v}{w}\right)  ^{2},\bar{\Omega}=v,a^{2}=\frac{1}{2}\frac
{v^{2}-w^{2}}{v^{3}},
\]
and consequently%
\begin{equation}
D(t)=\frac{1}{2}\frac{v^{2}-w^{2}}{v^{3}}(1-e^{ivt})-i\frac{1}{2}\left(
\frac{w}{v}\right)  ^{2}t=\frac{1}{2}\left(  \frac{w}{v}\right)  ^{2}\left\{
R(1-e^{ivt})-it\right\}  \qquad(\beta\rightarrow\infty)
\end{equation}
with
\[
R=\frac{v^{2}-w^{2}}{w^{2}v}.
\]
and according to (\ref{IF25})
\begin{equation}
S(t)=\frac{2\alpha}{3\sqrt{\pi}}\left(  \frac{v}{w}\right)  ^{3}e^{it}\left[
R(1-e^{ivt})-it\right]  ^{-3/2}\qquad(\beta\rightarrow\infty).
\end{equation}
From (\ref{IF24}) one obtains immediately%
\begin{align}
\operatorname{Im}\chi(\Omega)  &  =\frac{2\alpha}{3\sqrt{\pi}}\left(  \frac
{v}{w}\right)  ^{3}\operatorname{Im}%
%TCIMACRO{\dint \nolimits_{0}^{\infty}}%
%BeginExpansion
{\displaystyle\int\nolimits_{0}^{\infty}}
%EndExpansion
dt\frac{\sin(\Omega t)e^{it}}{\left[  R(1-e^{ivt})-it\right]  ^{3/2}%
},\label{IF26a}\\
\operatorname{Re}\chi(\Omega)  &  =\frac{2\alpha}{3\sqrt{\pi}}\left(  \frac
{v}{w}\right)  ^{3}\operatorname{Im}%
%TCIMACRO{\dint \nolimits_{0}^{\infty}}%
%BeginExpansion
{\displaystyle\int\nolimits_{0}^{\infty}}
%EndExpansion
dt\frac{\left[  1-\cos(\Omega t)\right]  e^{it}}{\left[  R(1-e^{ivt}%
)-it\right]  ^{3/2}}. \label{IF26b}%
\end{align}
In the limit $\beta\rightarrow\infty$ the function $\operatorname{Im}%
\chi(\Omega)$ was calculated by FHIP \cite{FHIP}. However, to study the
optical absorption to the same approximation as FHIP's treatment of the
impedance, we have also to calculate $\operatorname{Re}\chi(\Omega)$ and use
this result in (\ref{eq:P24-2a}). The calculation of $\operatorname{Re}%
\chi(\Omega),$ which is a Kramers-Kronig-type transform of $\operatorname{Im}%
\chi(\Omega)$, is a key ingredient in Ref.\thinspace\cite{DSG1972}. The
details of those calculations are presented in the Appendices A, B and C to
Ref.\thinspace\cite{DSG1972}.

Developing the denominator of both integrals on the right-hand side of
(\ref{IF26a}) and (\ref{IF26b}), the calculations are reduced to the
evaluation of a sum of integrals of the type%
\begin{equation}
\operatorname{Im}%
%TCIMACRO{\dint \nolimits_{0}^{\infty}}%
%BeginExpansion
{\displaystyle\int\nolimits_{0}^{\infty}}
%EndExpansion
dt\frac{\sin(\Omega t)e^{i(1+nv)t}}{\left(  R-it\right)  ^{3/2+n}}%
,\qquad\operatorname{Im}%
%TCIMACRO{\dint \nolimits_{0}^{\infty}}%
%BeginExpansion
{\displaystyle\int\nolimits_{0}^{\infty}}
%EndExpansion
dt\frac{\cos(\Omega t)e^{i(1+nv)t}}{\left(  R-it\right)  ^{3/2+n}}.
\label{IF26c}%
\end{equation}

In Appendix B\ to Ref.\thinspace\cite{DSG1972} it is shown how such integrals
are evaluated using a recurrence formula. For $\operatorname{Im}\chi(\Omega)$
a very convenient result was found in \cite{DSG1972}:%
\begin{align}
\operatorname{Im}\chi(\Omega)  &  =\frac{2\alpha}{3}\left(  \frac{v}%
{w}\right)  ^{3}%
%TCIMACRO{\dsum \limits_{n=0}^{\infty}}%
%BeginExpansion
{\displaystyle\sum\limits_{n=0}^{\infty}}
%EndExpansion
C_{-3/2}^{n}(-1)^{n}\frac{R^{n}2^{n}}{(2n+1)...3\cdot1}\nonumber\\
&  \times\left\vert \Omega-1-nv\right\vert ^{n+1/2}e^{-\left\vert
\Omega-1-nv\right\vert R}\frac{1+\mathrm{sgn}(\Omega-1-nv)}{2}. \label{IF27}%
\end{align}
This expression is a finite sum and not and infinite series. FHIP gave the
first two terms of (\ref{IF27}) explicitly.

Using the same recurrence relation it is seen the analytical expression (see
Appendix B\ to Ref.\thinspace\cite{DSG1972}), which was found for
$\operatorname{Re}\chi(\Omega)$ is far more complicated. To circumwent the
difficulty with the numerical treatment of $\operatorname{Re}\chi(\Omega)$,
the corresponding integrals in (\ref{IF26c})\ have been transformed in
\cite{DSG1972} to integrals with rapildy convergent integrands:%
\begin{align}
&  \operatorname{Im}%
%TCIMACRO{\dint \nolimits_{0}^{\infty}}%
%BeginExpansion
{\displaystyle\int\nolimits_{0}^{\infty}}
%EndExpansion
dt\frac{\left[  1-\cos(\Omega t)\right]  e^{i(1+nv)t}}{\left(  R-it\right)
^{3/2+n}}\nonumber\\
&  =-\frac{1}{\Gamma(n+\frac{3}{2})}%
%TCIMACRO{\dint \nolimits_{0}^{\infty}}%
%BeginExpansion
{\displaystyle\int\nolimits_{0}^{\infty}}
%EndExpansion
dx\left[  (n+\frac{1}{2})x^{n-1/2}e^{-Rx}-Rx^{n+1/2}e^{-Rx}\right] \nonumber\\
&  \times\ln\left\vert \left(  \frac{\left(  1+nv+x\right)  ^{2}}{\Omega
^{2}-\left(  1+nv+x\right)  ^{2}}\right)  \right\vert ^{1/2}. \label{IF28}%
\end{align}
The integral on the right-hand side of (\ref{IF28}) is adequate for computer
calculations. In Appendix C\ to Ref.\thinspace\cite{DSG1972} some
supplementary details of the computation of (\ref{IF28}) are given. Another
analytical representation for the memory function (\ref{IF23a}) was derived in
Ref. \cite{PD1983}.

\subsection{Discussion of optical absorption of polarons at arbitrary
coupling}

\begin{figure}[h]
\begin{center}
\includegraphics[width=0.7\textwidth]{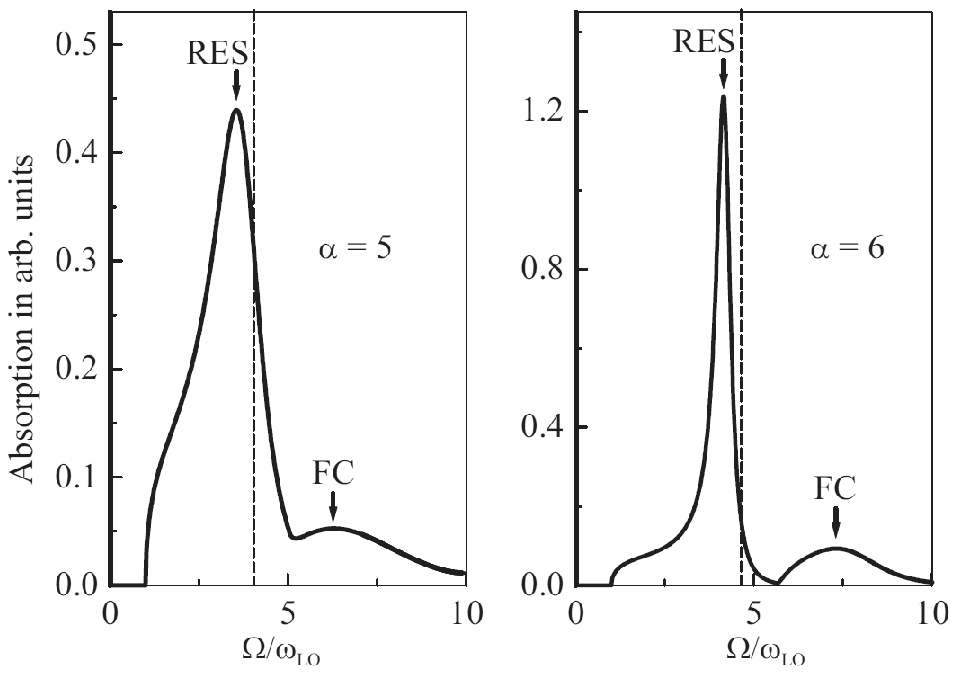}
\end{center}
\caption{Optical absorption spectrum of {a polaron} calculated by Devreese
\textit{et al.} \cite{DSG1972} $\alpha=5$ and $6$. The RES peak is very
intense compared with the FC peak. The frequency $\Omega/\omega_{\mathrm{LO}%
}=v$ is indicated by the dashed lines.)}%
\label{fig_3}%
\end{figure}

At weak coupling, the optical absorption spectrum (\ref{eq:P24-2}) of the
polaron is determined by the absorption of radiation energy, which is
reemitted in the form of LO phonons. For $\alpha\gtrsim5.9$, the polaron can
undergo transitions toward a relatively stable RES (see Fig.~\ref{fig_3}). The
RES peak in the optical absorption spectrum also has a phonon
sideband-structure, whose average transition frequency can be related to a
FC-type transition. Furthermore, at zero temperature, the optical absorption
spectrum of one polaron exhibits also a zero-frequency \textquotedblleft
central peak\textquotedblright\ [$\sim\delta(\Omega)$]. For non-zero
temperature, this \textquotedblleft central peak\textquotedblright\ smears out
and gives rise to an \textquotedblleft anomalous\textquotedblright\ Drude-type
low-frequency component of the optical absorption spectrum.

For example, in Fig. \ref{fig_3} from Ref. \cite{DSG1972}, the main peak of
the polaron optical absorption for $\alpha=$ 5 at $\Omega=3.51\omega
_{\mathrm{LO}}$ is interpreted as due to transitions to a RES. A
\textquotedblleft shoulder\textquotedblright\ at the low-frequency side of the
main peak is attributed to one-phonon transitions to polaron-\textquotedblleft
scattering states\textquotedblright. The broad structure centered at about
$\Omega=6.3\omega_{\mathrm{LO}}$ is interpreted as a FC band. As seen from
Fig. \ref{fig_3}, when increasing the electron-phonon coupling constant to
$\alpha$=6, the RES peak at $\Omega=4.3\omega_{\mathrm{LO}}$ stabilizes. It is
in Ref. \cite{DSG1972} that the all-coupling optical absorption spectrum of a
Fr\"{o}hlich polaron, together with the role of RES-states, FC-states and
scattering states, was first presented.

%\newpage

%\newpage

\begin{figure}[ptbh]
\begin{center}
\includegraphics[width=0.8\textwidth]{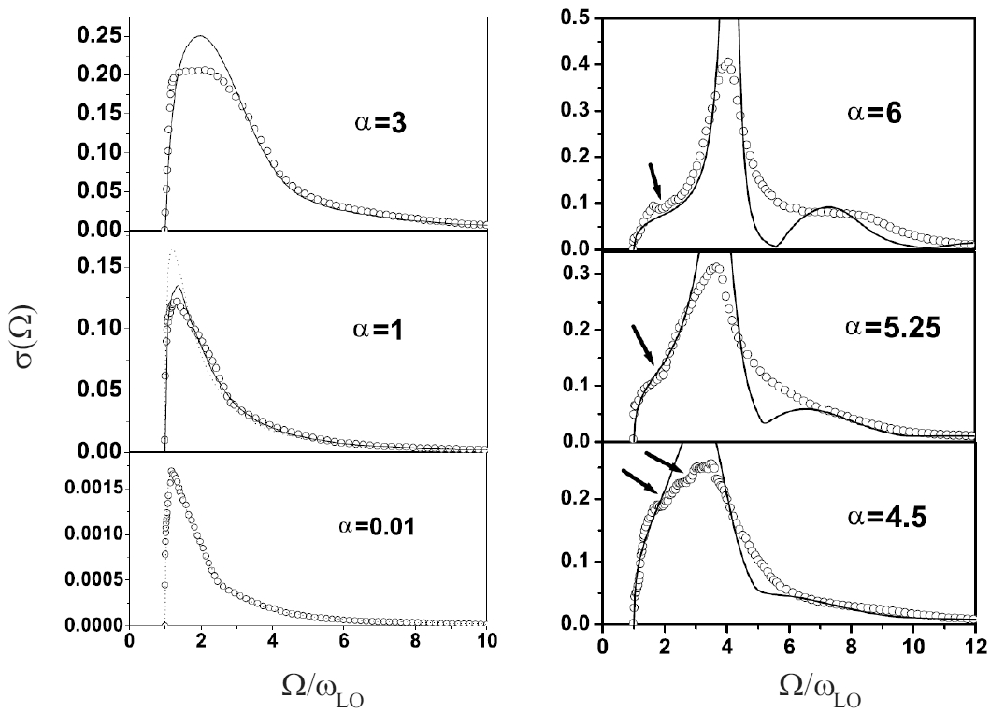}
\end{center}
\caption{{\emph{Left-hand panel}}: Monte Carlo optical conductivity spectra of
one polaron for the weak-coupling regime (open circles) compared to the
second-order perturbation theory (dotted lines) for $\alpha=0.01$ and
$\alpha=1$ and to the analytical DSG calculations \cite{DSG1972} (solid
lines). {\emph{Right-hand panel}}: Monte Carlo optical conductivity spectra
for the intermediate coupling regime (open circles) compared to the analytical
DSG approach \cite{DSG1972} (solid lines). Arrows point to the two- and
three-phonon thresholds. (From Ref.\thinspace\cite{Mishchenko2003}.)}%
\label{fig_4}%
\end{figure}

Recent interesting numerical calculations of the optical conductivity for the
Fr\"{o}hlich polaron performed within the diagrammatic Quantum Monte Carlo
method \cite{Mishchenko2003}, see Fig. \ref{fig_4}, fully confirm the
essential analytical results derived by Devreese et al. in Ref. \cite{DSG1972}
for $\alpha\lesssim3.$ In the intermediate coupling regime $3<\alpha<6,$ the
low-energy behavior and the position of the RES-peak in the optical
conductivity spectrum of Ref. \cite{Mishchenko2003} follow closely the
prediction of Ref. \cite{DSG1972}. There are some minor qualitative
differences between the two approaches in the intermediate coupling regime: in
Ref. \cite{Mishchenko2003}, the dominant (\textquotedblleft
RES\textquotedblright) peak is less intense in the Monte-Carlo numerical
simulations and the second (\textquotedblleft FC\textquotedblright) peak
develops less prominently. There are the following qualitative differences
between the two approaches in the strong coupling regime: in
Ref.\cite{Mishchenko2003}, the dominant peak broadens and the second peak does
not develop, giving instead rise to a flat shoulder in the optical
conductivity spectrum at $\alpha=6.$ This behavior has been tentatively
attributed to the optical processes with participation of two
\cite{Goovaerts73} or more phonons. The above differences can arise also due
to the fact that, within the Feynman polaron model, one-phonon processes are
assigned more oscillator strength and the RES tends to be more stable as
compared to the Monte-Carlo result. The nature of the excited states of a
polaron needs further study. An independent numerical simulation might be
called for.

\newpage

\newpage

\begin{figure}[ptbh]
\begin{center}
\includegraphics[width=0.7\textwidth]{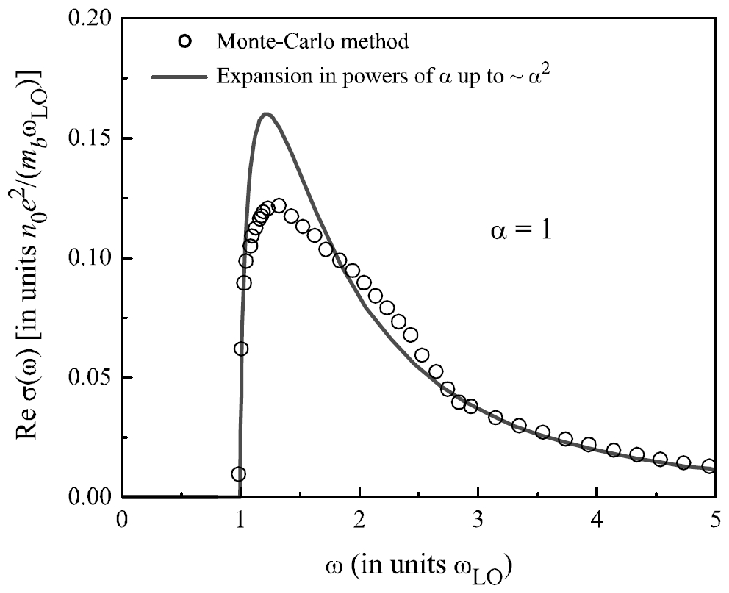}
\end{center}
\caption{One-polaron optical conductivity $\operatorname{Re}\sigma\left(
\omega\right)  $ for $\alpha=1$ calculated within the Monte Carlo approach
\cite{Mishchenko2003} (open circles) and derived using the expansion in powers
of $\alpha$ up to $\alpha^{2}$ \cite{Huybrechts1973} (solid curve).}%
\label{fig_4a}%
\end{figure}

In Fig. \ref{fig_4a}, Monte-Carlo optical conductivity spectrum of one polaron
for $\alpha=1$ compares well with that obtained in Ref. \cite{Huybrechts1973}
within the canonical-transformation formalism taking into account correlation
in processes involving two LO phonons. The difference between the results of
these two approaches becomes less pronounced when decreasing the value of
$\alpha=1$ and might be indicative of a possible precision loss, which
requires an independent check.

%\newpage

%\newpage

The coupling constant $\alpha$ of the known ionic crystals is too small
($\alpha<5$) to allow for the experimental detection of sharp RES peaks, and
the resonance condition $\Omega=Re\Sigma(\Omega)$ cannot be satisfied for
$\alpha\lesssim5.9$ as shown in Ref, \cite{DSG1972}. Nevertheless, for
$3\lesssim\alpha\lesssim5.9$ the development of RES is already reflected in a
broad optical absorption peak. Such a peak, predicted in Ref. \cite{DSG1972},
was identified, e.~g., in the optical absorption of Pr$_{2}$NiO$_{4.22}$ in
Ref. \cite{Eagles1995}. Also, the resonance condition can be fulfilled if an
external magnetic field is applied; the magnetic field stabilizes the RES,
which then can be detected in a cyclotron resonance peak.

\subsubsection{Sum rules for the optical conductivity spectra}

In this section, we analyze the sum rules for the optical conductivity spectra
obtained within the DSG approach \cite{DSG1972} with those obtained within the
diagrammatic Monte Carlo calculation \cite{Mishchenko2003}. The values of the
polaron effective mass for the Monte Carlo approach are taken from Ref.
\cite{Mishchenko2000}. In Tables 3 and 4, we represent the polaron
ground-state $E_{0}$ and the following parameters calculated using the optical
conductivity spectra:%
\begin{align}
M_{0}  &  \equiv\int_{1}^{\omega_{\max}}\operatorname{Re}\sigma\left(
\omega\right)  d\omega,\label{1}\\
M_{1}  &  \equiv\int_{1}^{\omega_{\max}}\omega\operatorname{Re}\sigma\left(
\omega\right)  d\omega, \label{2}%
\end{align}
where $\omega_{\max}$ is the upper value of the frequency available from Ref.
\cite{Mishchenko2003},%
\begin{equation}
\tilde{M}_{0}\equiv\frac{\pi}{2m^{\ast}}+\int_{1}^{\omega_{\max}%
}\operatorname{Re}\sigma\left(  \omega\right)  d\omega, \label{zm}%
\end{equation}
where $m^{\ast}$ is the polaron mass, the optical conductivity is calculated
in units $\frac{n_{0}e^{2}}{m_{b}\omega_{\mathrm{LO}}},$ $m^{\ast}$ is
measured in units of the band mass $m_{b}$, and the frequency is measured in
units of $\omega_{\mathrm{LO}}$. The values of $\omega_{\max}$ are:
$\omega_{\max}=10$ for $\alpha=0.01,$ 1 and 3, $\omega_{\max}=12$ for
$\alpha=4.5,$ 5.25 and 6, $\omega_{\max}=18$ for $\alpha=6.5,$ 7 and 8.

\bigskip

\begin{center}
\textbf{Table 3. Polaron parameters obtained within the diagrammatic Monte
Carlo approach}

\medskip%

\begin{tabular}
[c]{|c|c|c|c|}\hline
$\alpha$ & $M_{0}^{\left(  \mathrm{MC}\right)  }$ & $m^{\ast\left(
\mathrm{MC}\right)  }$ & $\tilde{M}_{0}^{\left(  \mathrm{MC}\right)  }%
$\\\hline
$0.01$ & $0.00249$ & $1.0017$ & $1.5706$\\\hline
$1$ & $0.24179$ & $1.1865$ & $1.5657$\\\hline
$3$ & $0.67743$ & $1.8467$ & $1.5280$\\\hline
$4.5$ & $0.97540$ & $2.8742$ & $1.5219$\\\hline
$5.25$ & $1.0904$ & $3.8148$ & $1.5022$\\\hline
$6$ & $1.1994$ & $5.3708$ & $1.4919$\\\hline
$6.5$ & $1.30$ & $6.4989$ & $1.5417$\\\hline
$7$ & $1.3558$ & $9.7158$ & $1.5175$\\\hline
$8$ & $1.4195$ & $19.991$ & $1.4981$\\\hline
\end{tabular}%
\begin{tabular}
[c]{|c|c|}\hline
$M_{1}^{\left(  \mathrm{MC}\right)  }/\alpha$ & $E_{0}^{\left(  \mathrm{MC}%
\right)  }$\\\hline
$0.634$ & $-0.010$\\\hline
$0.65789$ & $-1.013$\\\hline
$0.73123$ & $-3.18$\\\hline
$0.862$ & $-4.97$\\\hline
$0.90181$ & $-5.68$\\\hline
$0.98248$ & $-6.79$\\\hline
$1.1356$ & $-7.44$\\\hline
$1.2163$ & $-8.31$\\\hline
$1.3774$ & $-9.85$\\\hline
\end{tabular}

\bigskip

\textbf{Table 4. Polaron parameters obtained within the path-integral
approach}

\medskip%

\begin{tabular}
[c]{|c|c|c|c|}\hline
$\alpha$ & $M_{0}^{\left(  \mathrm{DSG}\right)  }$ & $m^{\ast\left(
\mathrm{Feynman}\right)  }$ & $\tilde{M}_{0}^{\left(  \mathrm{DSG}\right)  }%
$\\\hline
$0.01$ & $0.00248$ & $1.0017$ & $1.5706$\\\hline
$1$ & $0.24318$ & $1.1957$ & $1.5569$\\\hline
$3$ & $0.69696$ & $1.8912$ & $1.5275$\\\hline
$4.5$ & $1.0162$ & $3.1202$ & $1.5196$\\\hline
$5.25$ & $1.1504$ & $4.3969$ & $1.5077$\\\hline
$6$ & $1.2608$ & $6.8367$ & $1.4906$\\\hline
$6.5$ & $1.3657$ & $9.7449$ & $1.5269$\\\hline
$7$ & $1.4278$ & $14.395$ & $1.5369$\\\hline
$8$ & $1.4741$ & $31.569$ & $1.5239$\\\hline
\end{tabular}%
\begin{tabular}
[c]{|c|c|}\hline
$M_{1}^{\left(  \mathrm{DSG}\right)  }/\alpha$ & $E_{0}^{\left(
\mathrm{Feynman}\right)  }$\\\hline
$0.633$ & $-0.010$\\\hline
$0.65468$ & $-1.0130$\\\hline
$0.71572$ & $-3.1333$\\\hline
$0.83184$ & $-4.8394$\\\hline
$0.88595$ & $-5.7482$\\\hline
$0.95384$ & $-6.7108$\\\hline
$1.1192$ & $-7.3920$\\\hline
$1.2170$ & $-8.1127$\\\hline
$1.4340$ & $-9.6953$\\\hline
\end{tabular}

\bigskip
\end{center}

The parameters corresponding to the Monte Carlo calculation are obtained using
the numerical data kindly provided by A. Mishchenko. The comparison of the
zero frequency moments $\tilde{M}_{0}^{\left(  \mathrm{MC}\right)  }$ and
$\tilde{M}_{0}^{\left(  \mathrm{DSG}\right)  }$ with each other and with the
value $\pi/2$ corresponding to the sum rule \cite{DLR1977}%
\begin{equation}
\frac{\pi}{2m^{\ast}}+\int_{1}^{\infty}\operatorname{Re}\sigma\left(
\omega\right)  d\omega=\frac{\pi}{2} \label{sr}%
\end{equation}
shows that $\left\vert \tilde{M}_{0}^{\left(  \mathrm{MC}\right)  }-\tilde
{M}_{0}^{\left(  \mathrm{DSG}\right)  }\right\vert $ is smaller than each of
the differences $\frac{\pi}{2}-\tilde{M}_{0}^{\left(  \mathrm{MC}\right)  }$,
$\frac{\pi}{2}-\tilde{M}_{0}^{\left(  \mathrm{DSG}\right)  },$ which appear
due to a finite interval of the integration in (\ref{1}), (\ref{2}).

\begin{figure}[h]
%\begin{center}
\includegraphics[
height=3in]{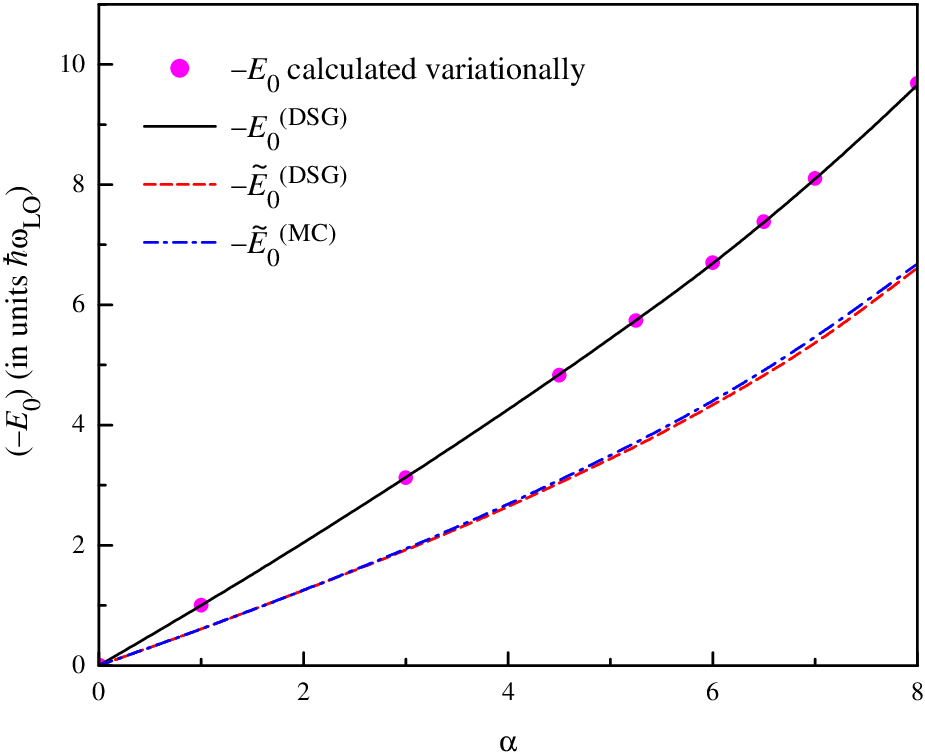}
%
%\end{center}
\caption{The ground-state theorem for a polaron using different data for the
optical conductivity spectra, DSG from Ref. \cite{DSG1972} and MC from Ref.
\cite{Mishchenko2003}. The notations are explained in the text.}%
\label{Moments2-f1}%
\end{figure}

We analyze also the fulfilment of the ground-state theorem \cite{LSD}%
\begin{equation}
E_{0}\left(  \alpha\right)  -E_{0}\left(  0\right)  =-\frac{3}{\pi}\int
_{0}^{\alpha}\frac{d\alpha^{\prime}}{\alpha^{\prime}}\int_{0}^{\infty}%
\omega\operatorname{Re}\sigma\left(  \omega,\alpha^{\prime}\right)
d\omega\label{gst}%
\end{equation}
using the first-frequency moments $M_{1}^{\left(  \mathrm{MC}\right)  }$ and
$M_{1}^{\left(  \mathrm{DSG}\right)  }$. The results of this comparison are
presented in Fig. \ref{Moments2-f1}. The dots indicate the polaron
ground-state energy calculated using the Feynman variational principle. The
solid curve is the value of $E_{0}\left(  \alpha\right)  $ calculated
numerically using the optical conductivity spectra and the ground-state
theorem with the DSG optical conductivity \cite{DSG1972} for a polaron,%
\begin{equation}
E_{0}^{\left(  \mathrm{DSG}\right)  }\left(  \alpha\right)  \equiv-\frac
{3}{\pi}\int_{0}^{\alpha}\frac{d\alpha^{\prime}}{\alpha^{\prime}}\int
_{0}^{\infty}\omega\operatorname{Re}\sigma^{\left(  \mathrm{DSG}\right)
}\left(  \omega,\alpha^{\prime}\right)  d\omega. \label{e1R}%
\end{equation}
The dashed and the dot-dashed curves are the values obtained using
$M_{1}^{\left(  \mathrm{DSG}\right)  }\left(  \alpha\right)  $ and
$M_{1}^{\left(  \mathrm{MC}\right)  }\left(  \alpha\right)  $, respectively:%
\begin{align}
\tilde{E}_{0}^{\left(  \mathrm{DSG}\right)  }\left(  \alpha\right)   &
\equiv-\frac{3}{\pi}\int_{0}^{\alpha}\frac{d\alpha^{\prime}}{\alpha^{\prime}%
}\int_{0}^{\omega_{\max}}\omega\operatorname{Re}\sigma^{\left(  \mathrm{DSG}%
\right)  }\left(  \omega,\alpha^{\prime}\right)  d\omega=-\frac{3}{\pi}%
\int_{0}^{\alpha}d\alpha^{\prime}\frac{M_{1}^{\left(  \mathrm{DSG}\right)
}\left(  \alpha^{\prime}\right)  }{\alpha^{\prime}},\label{e2R}\\
\tilde{E}_{0}^{\left(  \mathrm{MC}\right)  }\left(  \alpha\right)   &
\equiv-\frac{3}{\pi}\int_{0}^{\alpha}\frac{d\alpha^{\prime}}{\alpha^{\prime}%
}\int_{0}^{\omega_{\max}}\omega\operatorname{Re}\sigma^{\left(  \mathrm{MC}%
\right)  }\left(  \omega,\alpha^{\prime}\right)  d\omega=-\frac{3}{\pi}%
\int_{0}^{\alpha}d\alpha^{\prime}\frac{M_{1}^{\left(  \mathrm{MC}\right)
}\left(  \alpha^{\prime}\right)  }{\alpha^{\prime}}. \label{e3}%
\end{align}

As seen from the figure, $E_{0}^{\left(  \mathrm{DSG}\right)  }\left(
\alpha\right)  $ to a high degree of accuracy coincides with the variational
polaron ground-state energy. Both $\tilde{E}_{0}^{\left(  \mathrm{DSG}\right)
}\left(  \alpha\right)  $ and $\tilde{E}_{0}^{\left(  \mathrm{MC}\right)
}\left(  \alpha\right)  $ differ from $E_{0}^{\left(  \mathrm{DSG}\right)
}\left(  \alpha\right)  $ due to the integration over a finite interval of
frequencies. However, $\tilde{E}_{0}^{\left(  \mathrm{DSG}\right)  }\left(
\alpha\right)  $ and $\tilde{E}_{0}^{\left(  \mathrm{MC}\right)  }\left(
\alpha\right)  $ are very close to each other. Herefrom, a conclusion follows
that for integrals over the finite frequency region characteristic for the
polaron optical absorption (i. e., except the \textquotedblleft
tails\textquotedblright), the function $\tilde{E}_{0}^{\left(  \mathrm{MC}%
\right)  }\left(  \alpha\right)  $ (\ref{e3}) reproduces very well the
function $\tilde{E}_{0}^{\left(  \mathrm{DSG}\right)  }\left(  \alpha\right)
$.

%\newpage

%\newpage

\subsection{Scaling relations}

\subsubsection{Derivation of the scaling relations}

The form of the Fr\"{o}hlich Hamiltonian in $n$ dimensions is the same as in
3D,%
\begin{equation}
H=\frac{\mathbf{p}^{2}}{2m_{b}}+\sum_{\mathbf{k}}\hbar\omega_{\mathbf{k}%
}a_{\mathbf{k}}^{\dag}a_{\mathbf{k}}+\sum_{\mathbf{k}}\left(  V_{\mathbf{k}%
}a_{\mathbf{k}}e^{i\mathbf{k}\cdot\mathbf{r}}+V_{\mathbf{k}}^{\ast
}a_{\mathbf{k}}^{\dag}e^{-i\mathbf{k}\cdot\mathbf{r}}\right)  , \label{sr1}%
\end{equation}
except that now all vectors are $n$-dimensional. In this subsection,
dispersionless longitudinal phonons are considered, i.e., $\omega_{\mathbf{k}%
}=\omega_{\text{LO}}$, and units are chosen such that $\hbar=m_{b}%
=\omega_{\text{LO}}=1$.

The electron-phonon interaction is a representation in second quantization of
the electron interaction with the lattice polarization, which in 3D is
essentially the Coulomb potential $1/r.$ $\left\vert V_{\mathbf{k}}\right\vert
^{2}$ is proportional to the Fourier transform of this potential, and as a
consequence we have in $n$ dimensions%
\begin{equation}
\left\vert V_{\mathbf{k}}\right\vert ^{2}=\frac{A_{n}}{L_{n}k^{n-1}},
\label{sr2}%
\end{equation}
where $L_{n}$ is the volume of the $n$-dimensional crystal. Note that
$\left\vert V_{\mathbf{\tilde{k}}}\right\vert ^{2}$, where $\mathbf{\tilde{k}}
$ is an ($n-1$)-dimensional vector, can be obtained from $\left\vert
V_{\mathbf{k}}\right\vert ^{2}$, where $\mathbf{k}=$ $\left(  \mathbf{\tilde
{k}},k_{n}\right)  $ is an $n$-dimensional vector, by summing out one of the
dimensions explicitly:%
\begin{equation}
\left\vert V_{\mathbf{\tilde{k}}}\right\vert ^{2}=\sum_{k_{n}}\left\vert
V_{\mathbf{k}}\right\vert ^{2}. \label{sr3}%
\end{equation}
Inserting Eq. (\ref{sr2}) into Eq. (\ref{sr3}), we have%
\begin{equation}
\frac{A_{n-1}}{L_{n-1}\tilde{k}^{n-2}}=\sum_{k_{n}}\frac{A_{n}}{L_{n}\left(
\tilde{k}^{2}+k_{n}^{2}\right)  ^{\left(  n-1\right)  /2}}. \label{sr4}%
\end{equation}
Replacing the sum in Eq. (\ref{sr4}) by an integral, i.e.,%
\begin{equation}
\frac{L_{n-1}}{L_{n}}\sum_{k_{n}}\longrightarrow\frac{1}{2\pi}\int dk_{n},
\label{sr4a}%
\end{equation}
we obtain%

\begin{equation}
\frac{A_{n-1}}{\tilde{k}^{n-2}}=\frac{A_{n}}{2\pi}\int_{-\infty}^{\infty}%
\frac{dk_{n}}{\left(  \tilde{k}^{2}+k_{n}^{2}\right)  ^{\left(  n-1\right)
/2}}. \label{sr5}%
\end{equation}
Since
\begin{equation}
\int_{-\infty}^{\infty}\frac{dx}{\left(  z^{\mu}+x^{\mu}\right)  ^{\rho}%
}=z^{1-\mu\rho}\frac{\Gamma\left(  \frac{1}{\mu}\right)  \Gamma\left(
\rho-\frac{1}{\mu}\right)  }{\Gamma\left(  \rho\right)  },
\end{equation}
we have%
\begin{equation}
\int_{-\infty}^{\infty}\frac{dk_{n}}{\left(  \tilde{k}^{2}+k_{n}^{2}\right)
^{\left(  n-1\right)  /2}}=\frac{1}{\tilde{k}^{n-2}}\frac{\Gamma\left(
\frac{1}{2}\right)  \Gamma\left(  \frac{n-2}{2}\right)  }{\Gamma\left(
\frac{n-1}{2}\right)  }=\frac{\sqrt{\pi}}{\tilde{k}^{n-2}}\frac{\Gamma\left(
\frac{n-2}{2}\right)  }{\Gamma\left(  \frac{n-1}{2}\right)  }, \label{sr6}%
\end{equation}
where $\Gamma\left(  x\right)  $ is eh $\Gamma$ function. Inserting Eq.
(\ref{sr6}) into Eq. (\ref{sr5}), we obtain%
\begin{equation}
A_{n}=\frac{2\sqrt{\pi}\Gamma\left(  \frac{n-1}{2}\right)  }{\Gamma\left(
\frac{n-2}{2}\right)  }A_{n-1}. \label{sr7}%
\end{equation}

In 3D the interaction coefficient is well known, $\left\vert V_{\mathbf{k}%
}\right\vert ^{2}=2\sqrt{2}\pi\alpha/L_{3}k^{2},$ so that
\begin{equation}
A_{3}=2\sqrt{2}\pi\alpha. \label{sr8}%
\end{equation}
Inserting Eq. (\ref{sr8}) into Eq. (\ref{sr7}), we immediately obtain%
\begin{equation}
A_{2}=2\sqrt{2}\pi\alpha\frac{\Gamma\left(  \frac{1}{2}\right)  }{2\sqrt{\pi
}\Gamma\left(  1\right)  }=\sqrt{2}\pi\alpha. \label{sr9}%
\end{equation}
Applying Eq. (\ref{sr7}) $n-2$ times, we further obtain for $n>3$%
\begin{align}
A_{n}  &  =\frac{\left(  2\sqrt{\pi}\right)  ^{n-2}\prod\limits_{j=2}%
^{n-1}\Gamma\left(  \frac{j}{2}\right)  }{\prod\limits_{j=1}^{n-2}%
\Gamma\left(  \frac{j}{2}\right)  }A_{2}=\frac{\left(  2\sqrt{\pi}\right)
^{n-2}\left(  \frac{n-1}{2}\right)  }{\Gamma\left(  \frac{1}{2}\right)  }%
A_{2}=2^{n-2}\pi^{\left(  n-3\right)  /2}\Gamma\left(  \frac{n-1}{2}\right)
A_{2}\nonumber\\
&  =2^{n-3/2}\pi^{\left(  n-1\right)  /2}\Gamma\left(  \frac{n-1}{2}\right)
\alpha. \label{sr10}%
\end{align}

So, the interaction coefficient in $n$ dimensions becomes \cite{PRB33-3926}
\begin{equation}
\left\vert V_{\mathbf{k}}\right\vert ^{2}=\frac{2^{n-3/2}\pi^{\left(
n-1\right)  /2}\Gamma\left(  \frac{n-1}{2}\right)  \alpha}{L_{n}k^{n-1}}.
\label{srv}%
\end{equation}

Following the Feynman approach \cite{Feynman}, the upper bound for the polaron
ground-state energy can be written down as%
\begin{equation}
E=E_{0}-\lim_{\beta\rightarrow\infty}\frac{1}{\beta}\left\langle
S-S_{0}\right\rangle _{0}, \label{sr11}%
\end{equation}
where $S$ is the exact action functional of the polaron problem, while $S_{0}$
is the trial action functional, which corresponds to a model system where an
electron is coupled by an elastic force to a fictitious particle (i.e., the
model system describes a harmonic oscillator). $E_{0}$ is the ground-state
energy of the above model system, and%
\begin{equation}
\left\langle F\right\rangle _{0}\equiv\frac{\int Fe^{S_{0}}\mathcal{D}%
\mathbf{r}\left(  t\right)  }{\int e^{S_{0}}\mathcal{D}\mathbf{r}\left(
t\right)  }. \label{sr12}%
\end{equation}

As indicated above, the Fr\"{o}hlich Hamiltonian in $n$ dimensions is the same
as in 3D, except that now all vectors are $n$-dimensional [and the coupling
coefficient $\left\vert V_{\mathbf{k}}\right\vert ^{2}$ is modified in
accordance with Eq. (\ref{srv})]. Similarly, the only difference of the model
system in $n$ dimensions from the model system in 3D is that now one deals
with an $n$-dimensional harmonic oscillator. So, directly following
\cite{Feynman}, one can represent $\lim_{\beta\rightarrow\infty}\left\langle
S-S_{0}\right\rangle _{0}/\beta$ as
\begin{equation}
\lim_{\beta\rightarrow\infty}\frac{1}{\beta}\left\langle S-S_{0}\right\rangle
_{0}=A+B, \label{sr13}%
\end{equation}
where%
\begin{equation}
A=\sum_{\mathbf{k}}\left\vert V_{\mathbf{k}}\right\vert ^{2}\int_{0}^{\infty
}\left\langle e^{i\mathbf{k}\cdot\left[  \mathbf{r}\left(  t\right)
-\mathbf{r}\left(  0\right)  \right]  }\right\rangle _{0}e^{-t}dt,
\label{sr14}%
\end{equation}%
\begin{equation}
B=\frac{w\left(  v^{2}-w^{2}\right)  }{4}\int_{0}^{\infty}\left\langle \left[
\mathbf{r}\left(  t\right)  -\mathbf{r}\left(  0\right)  \right]
^{2}\right\rangle _{0}e^{-wt}dt, \label{sr15}%
\end{equation}
$w$ and $v$ are variational parameters, which should be determined by
minimizing $E$ of Eq. (\ref{sr11}). Since the averaging $\left\langle
...\right\rangle _{0}$ in Eq. (\ref{sr14}) is performed with the trial action,
which corresponds to a harmonic oscillator, components of the electron
coordinates, $r_{j}$ ($j=1,...,n$), in $\left\langle e^{i\mathbf{k}%
\cdot\left[  \mathbf{r}\left(  t\right)  -\mathbf{r}\left(  0\right)  \right]
}\right\rangle _{0}$ separate \cite{Feynman}:%
\begin{equation}
\left\langle e^{i\mathbf{k}\cdot\left[  \mathbf{r}\left(  t\right)
-\mathbf{r}\left(  0\right)  \right]  }\right\rangle _{0}=\prod\limits_{j=1}%
^{n}\left\langle e^{ik_{j}\left[  r_{j}\left(  t\right)  -r_{j}\left(
0\right)  \right]  }\right\rangle _{0}. \label{sr16}%
\end{equation}
For the average $\left\langle e^{ik_{j}\left[  r_{j}\left(  t\right)
-r_{j}\left(  0\right)  \right]  }\right\rangle _{0}$, Feynman obtained
\cite{Feynman}%
\begin{equation}
\left\langle e^{ik_{j}\left[  r_{j}\left(  t\right)  -r_{j}\left(  0\right)
\right]  }\right\rangle _{0}=e^{-k_{j}^{2}D_{0}\left(  t\right)  },
\label{sr17}%
\end{equation}
where%
\begin{equation}
D_{0}\left(  t\right)  =\frac{w^{2}}{2v^{2}}t+\frac{v^{2}-w^{2}}{2v^{3}%
}\left(  1-e^{-vt}\right)  . \label{sr18}%
\end{equation}
Inserting Eq. (\ref{sr16}) with Eq. (\ref{sr17}) into Eq. (\ref{sr14}), we
obtain%
\begin{equation}
A=\int_{0}^{\infty}e^{-t}dt\sum_{\mathbf{k}}\left\vert V_{\mathbf{k}%
}\right\vert ^{2}e^{-k^{2}D_{0}\left(  t\right)  }. \label{sr19}%
\end{equation}
Inserting expression (\ref{srv}) for $\left\vert V_{\mathbf{k}}\right\vert
^{2}$ into Eq. (\ref{sr19}) and replacing the sum over $\mathbf{k}$ by an
integral [see (\ref{sr4a})], we have
\begin{align}
A  &  =2^{n-3/2}\pi^{\left(  n-1\right)  /2}\Gamma\left(  \frac{n-1}%
{2}\right)  \alpha\int_{0}^{\infty}e^{-t}dt\int\frac{e^{-k^{2}D_{0}\left(
t\right)  }}{k^{n-1}}\frac{d\mathbf{k}}{\left(  2\pi\right)  ^{n}}\nonumber\\
&  =2^{n-3/2}\pi^{\left(  n-1\right)  /2}\Gamma\left(  \frac{n-1}{2}\right)
\alpha\int_{0}^{\infty}e^{-t}dt\int d\Omega_{n}\int_{0}^{\infty}%
\frac{e^{-k^{2}D_{0}\left(  t\right)  }}{k^{n-1}}\frac{k^{n-1}dk}{\left(
2\pi\right)  ^{n}}, \label{sr20}%
\end{align}
where $d\Omega_{n}$ is the elemental solid angle in $n$ dimensions. Since the
integrand in Eq. (\ref{sr20}) depends only on the modulus $k$ of $\mathbf{k}$,
one have simply $\int d\Omega_{n}=\Omega_{n}$ with%
\begin{equation}
\Omega_{n}=\frac{2\pi^{n/2}}{\Gamma\left(  \frac{n}{2}\right)  }. \label{sr21}%
\end{equation}
So, we obtain for $A$ the result%
\begin{align}
A  &  =\frac{2^{-1/2}\pi^{-1/2}\Gamma\left(  \frac{n-1}{2}\right)  \alpha
}{\Gamma\left(  \frac{n}{2}\right)  }\int_{0}^{\infty}e^{-t}dt\int_{0}%
^{\infty}e^{-k^{2}D_{0}\left(  t\right)  }dk=\frac{2^{-1/2}\pi^{-1/2}%
\Gamma\left(  \frac{n-1}{2}\right)  \alpha}{\Gamma\left(  \frac{n}{2}\right)
}\int_{0}^{\infty}\frac{\sqrt{\pi}e^{-t}}{2\sqrt{D_{0}\left(  t\right)  }%
}dt\nonumber\\
&  =\frac{2^{-3/2}\Gamma\left(  \frac{n-1}{2}\right)  \alpha}{\Gamma\left(
\frac{n}{2}\right)  }\int_{0}^{\infty}\frac{e^{-t}}{\sqrt{D_{0}\left(
t\right)  }}dt. \label{sr22}%
\end{align}

Like in Ref. \cite{Feynman}, $B$ can be easily calculated by noticing that
\begin{align}
\left\langle \left[  \mathbf{r}\left(  t\right)  -\mathbf{r}\left(  0\right)
\right]  ^{2}\right\rangle _{0}  &  =\sum_{j=1}^{n}\left\langle \left[
r_{j}\left(  t\right)  -r_{j}\left(  0\right)  \right]  ^{2}\right\rangle
_{0}=\sum_{j=1}^{n}\left.  \left[  -\frac{\partial^{2}}{\partial k_{j}^{2}%
}\left\langle e^{i\mathbf{k}\cdot\left[  \mathbf{r}\left(  t\right)
-\mathbf{r}\left(  0\right)  \right]  }\right\rangle _{0}\right]  \right\vert
_{\mathbf{k}=0}\nonumber\\
&  =\sum_{j=1}^{n}2D_{0}\left(  t\right)  =2nD_{0}\left(  t\right)  ,
\label{ssr1}%
\end{align}
so that%
\begin{align}
B  &  =\frac{nw\left(  v^{2}-w^{2}\right)  }{2}\int_{0}^{\infty}D_{0}\left(
t\right)  e^{-wt}dt\nonumber\\
&  =\frac{nw\left(  v^{2}-w^{2}\right)  }{2}\int_{0}^{\infty}\left[
\frac{w^{2}}{2v^{2}}te^{-wt}+\frac{v^{2}-w^{2}}{2v^{3}}\left(  e^{-wt}%
-e^{-\left(  v+w\right)  t}\right)  \right]  dt\nonumber\\
&  =\frac{nw\left(  v^{2}-w^{2}\right)  }{2}\left[  \frac{w^{2}}{2v^{2}}%
\frac{1}{v^{2}}+\frac{v^{2}-w^{2}}{2v^{3}}\left(  \frac{1}{w}-\frac{1}%
{v+w}\right)  \right] \nonumber\\
&  =\frac{n\left(  v^{2}-w^{2}\right)  }{4v}. \label{ssr3}%
\end{align}

Inserting Eq. (\ref{srv}) with $A$ and $B$, given by Eqs. (\ref{sr22}) and
(\ref{ssr3}), together with the ground-state energy of the model system
\cite{Feynman} (an isotropic $n$-dimensional harmonic oscillator),%
\begin{equation}
E_{0}=\frac{n\left(  v-w\right)  }{2},
\end{equation}
into Eq. (\ref{sr11}), we obtain
\begin{align}
E  &  =\frac{n\left(  v-w\right)  }{2}-\frac{n\left(  v^{2}-w^{2}\right)
}{4v}-\frac{2^{-3/2}\Gamma\left(  \frac{n-1}{2}\right)  \alpha}{\Gamma\left(
\frac{n}{2}\right)  }\int_{0}^{\infty}\frac{e^{-t}}{\sqrt{D_{0}\left(
t\right)  }}dt\nonumber\\
&  =\frac{n\left(  v-w\right)  ^{2}}{4v}-\frac{\Gamma\left(  \frac{n-1}%
{2}\right)  \alpha}{2\sqrt{2}\Gamma\left(  \frac{n}{2}\right)  }\int
_{0}^{\infty}\frac{e^{-t}}{\sqrt{D_{0}\left(  t\right)  }}dt. \label{ssr5}%
\end{align}
In order to make easier a comparison of $E$ for $n$ dimensions with the
Feynman result \cite{Feynman} for 3D,%
\begin{equation}
E_{3\text{D}}\left(  \alpha\right)  =\frac{3\left(  v-w\right)  ^{2}}%
{4v}-\frac{1}{\sqrt{2\pi}}\alpha\int_{0}^{\infty}\frac{e^{-t}}{\sqrt
{D_{0}\left(  t\right)  }}dt, \label{ssr6}%
\end{equation}
it is convenient to rewrite Eq. (\ref{ssr5}) in the form%
\begin{equation}
E_{n\text{D}}\left(  \alpha\right)  =\frac{n}{3}\left[  \frac{3\left(
v-w\right)  ^{2}}{4v}-\frac{1}{\sqrt{2\pi}}\frac{3\sqrt{\pi}\Gamma\left(
\frac{n-1}{2}\right)  }{2n\Gamma\left(  \frac{n}{2}\right)  }\alpha\int
_{0}^{\infty}\frac{e^{-t}}{\sqrt{D_{0}\left(  t\right)  }}dt\right]  .
\label{ssr7}%
\end{equation}
It is worth recalling that the parameters $w$ and $v$ must be determined by
minimizing $E$. Thus, in the case of Eq. (\ref{ssr7}) one has to minimize the
expression in the square brackets. The only difference of this expression from
the r.h.s. of Eq. (\ref{ssr6}) is that $\alpha$ is multiplied by the factor
\begin{equation}
a_{n}=\frac{3\sqrt{\pi}\Gamma\left(  \frac{n-1}{2}\right)  }{2n\Gamma\left(
\frac{n}{2}\right)  }. \label{sran}%
\end{equation}
This means that the minimizing parameters $w$ and $v$ in $n$D at a given
$\alpha$ will be exactly the same as those calculated in 3D for the
Fr\"{o}hlich constant as large as $a_{n}\alpha$:
\begin{equation}
v_{n\text{D}}\left(  \alpha\right)  =v_{3\text{D}}\left(  a_{n}\alpha\right)
,\text{ }w_{n\text{D}}\left(  \alpha\right)  =w_{3\text{D}}\left(  a_{n}%
\alpha\right)  . \label{srvw}%
\end{equation}
Therefore, comparing Eq. (\ref{ssr7}) to Eq. (\ref{ssr6}), we obtain the
scaling relation \cite{PRB33-3926, prb31-3420, prb36-4442}%
\begin{equation}
E_{n\text{D}}\left(  \alpha\right)  =\frac{n}{3}E_{3\text{D}}\left(
a_{n}\alpha\right)  , \label{ssr8}%
\end{equation}
where $a_{n}$ is given by Eq. (\ref{sran}). As discussed in Ref.
\cite{PRB33-3926}, the above scaling relation is not an exact relation. It is
valid for the Feynman polaron energy and also for the ground-state energy to
order $\alpha$. The next-order term (i.e., $\alpha^{2}$) no longer satisfies
Eq. (\ref{ssr8}). The reason is that in the exact calculation (to order
$\alpha^{2}$) the electron motion in the different space directions is coupled
by the electron-phonon interaction. No such a coupling appears in the Feynman
polaron model [see, e.g., Eq.\ (\ref{sr16})]; and this is the underlying
reason for the validity of the scaling relation for the Feynman approximation.

In Refs. \cite{PRB33-3926, prb36-4442}, scaling relations are obtained also
for the impedance function, the effective mass and the mobility of a polaron.
The inverse of the impedance function $Z\left(  \omega\right)  $ is given by
\begin{equation}
\frac{1}{Z\left(  \omega\right)  }=\frac{i}{\omega-\Sigma\left(
\omega\right)  }, \label{ssr9}%
\end{equation}
where the memory function $\Sigma\left(  \omega\right)  $ can be expressed as
\cite{prb28-6051}%
\begin{equation}
\Sigma\left(  z\right)  =\frac{1}{z}\int_{0}^{\infty}dt\left(  1-e^{izt}%
\right)  \text{Im}S\left(  t\right)  , \label{ssr10}%
\end{equation}
with $z=\omega+i0^{+}$ and%
\begin{equation}
S\left(  t\right)  =\sum_{\mathbf{k}}2k_{1}^{2}\left\vert V_{\mathbf{k}%
}\right\vert ^{2}e^{-k^{2}D\left(  t\right)  }T\left(  t\right)  ,
\label{ssr11}%
\end{equation}%
\begin{equation}
T\left(  t\right)  =\left[  1+n\left(  1\right)  \right]  e^{t}+n\left(
1\right)  e^{-it},
\end{equation}%
\begin{equation}
D\left(  t\right)  =\frac{w^{2}}{2v^{2}}\left(  -it+\frac{t^{2}}{\beta
}\right)  +\frac{v^{2}-w^{2}}{2v^{3}}\left[  1-e^{-ivt}+4n\left(  v\right)
\sin^{2}\left(  \frac{vt}{2}\right)  \right]  .
\end{equation}
Here, $\beta$ is the inverse temperature and $n\left(  \omega\right)  $ is the
occupation number of phonons with frequency $\omega$ (recall that in our units
$\omega_{\text{LO}}=1$).

As implied from Eqs. (\ref{ssr9}) and (\ref{ssr10}), scaling of $\Sigma\left(
\omega\right)  $ and $Z\left(  \omega\right)  $ is determined by scaling of
$S\left(  t\right)  $. For an isotropic crystal, since $\left\vert
V_{\mathbf{k}}\right\vert ^{2}$, $D\left(  t\right)  $ and $T\left(  t\right)
$ do not depend on the direction of $\mathbf{k}$, one can write $\sum
_{\mathbf{k}}k_{1}^{2}\left\vert V_{\mathbf{k}}\right\vert ^{2}e^{-k^{2}%
D\left(  t\right)  }T\left(  t\right)  =\sum_{\mathbf{k}}k_{2}^{2}\left\vert
V_{\mathbf{k}}\right\vert ^{2}e^{-k^{2}D\left(  t\right)  }T\left(  t\right)
=...=\sum_{\mathbf{k}}k_{n}^{2}\left\vert V_{\mathbf{k}}\right\vert
^{2}e^{-k^{2}D\left(  t\right)  }T\left(  t\right)  $ , so that%
\begin{equation}
S\left(  t\right)  =\frac{2}{n}\sum_{\mathbf{k}}k^{2}\left\vert V_{\mathbf{k}%
}\right\vert ^{2}e^{-k^{2}D\left(  t\right)  }T\left(  t\right)  .
\end{equation}
Inserting expression (\ref{srv}) for $\left\vert V_{\mathbf{k}}\right\vert
^{2}$ and replacing the sum over $\mathbf{k}$ by an integral, we have%
\begin{align}
S\left(  t\right)   &  =\frac{2}{n}\frac{2^{n-3/2}\pi^{\left(  n-1\right)
/2}\Gamma\left(  \frac{n-1}{2}\right)  \alpha}{\left(  2\pi\right)  ^{n}}\int
d\Omega_{n}\int_{0}^{\infty}k^{2}e^{-k^{2}D\left(  t\right)  }T\left(
t\right)  dk\nonumber\\
&  =\frac{2}{n}\frac{2^{n-3/2}\pi^{\left(  n-1\right)  /2}\Gamma\left(
\frac{n-1}{2}\right)  \alpha}{\left(  2\pi\right)  ^{n}}\frac{2\pi^{n/2}%
}{\Gamma\left(  \frac{n}{2}\right)  }\int_{0}^{\infty}k^{2}e^{-k^{2}D\left(
t\right)  }T\left(  t\right)  dk\nonumber\\
&  =\sqrt{\frac{2}{\pi}}\frac{\Gamma\left(  \frac{n-1}{2}\right)  \alpha
}{n\Gamma\left(  \frac{n}{2}\right)  }\int_{0}^{\infty}k^{2}e^{-k^{2}D\left(
t\right)  }T\left(  t\right)  dk. \label{ssr13}%
\end{align}
In particular, for 3D one has from Eq. (\ref{ssr13})%
\begin{equation}
S_{3\text{D}}\left(  \alpha;t\right)  =\frac{2\sqrt{2}}{3\pi}\alpha\int
_{0}^{\infty}k^{2}e^{-k^{2}D\left(  t\right)  }T\left(  t\right)  dk.
\label{ssr14}%
\end{equation}
For $n$D, Eq. (\ref{ssr13}) can be rewritten is the form%

\begin{align}
S_{n\text{D}}\left(  \alpha;t\right)   &  =\frac{2\sqrt{2}}{3\pi}\frac
{3\sqrt{\pi}\Gamma\left(  \frac{n-1}{2}\right)  }{2n\Gamma\left(  \frac{n}%
{2}\right)  }\alpha\int_{0}^{\infty}k^{2}e^{-k^{2}D\left(  t\right)  }T\left(
t\right)  dk\nonumber\\
&  =\frac{2\sqrt{2}}{3\pi}a_{n}\alpha\int_{0}^{\infty}k^{2}e^{-k^{2}D\left(
t\right)  }T\left(  t\right)  dk. \label{ssr15}%
\end{align}
So, the only difference of the expression for $S_{3\text{D}}\left(  t\right)
$ from $S_{3\text{D}}\left(  t\right)  $ is that $\alpha$ is multiplied by
$a_{n}$. Since for the minimizing parameters $w$ and $v$, which enter
$D\left(  t\right)  $, scaling is determined by the same product $\alpha$ with
$a_{n}$ [see Eq. (\ref{srvw})], we can write
\begin{equation}
S_{n\text{D}}\left(  \alpha;t\right)  =S_{3\text{D}}\left(  a_{n}%
\alpha;t\right)  ,
\end{equation}
so that \cite{prb36-4442}%
\begin{equation}
\Sigma_{n\text{D}}\left(  \alpha;\omega\right)  =\Sigma_{3\text{D}}\left(
a_{n}\alpha;\omega\right)  , \label{ssr16}%
\end{equation}
and%
\begin{equation}
Z_{n\text{D}}\left(  \alpha;\omega\right)  =Z_{3\text{D}}\left(  a_{n}%
\alpha;\omega\right)  . \label{ssr17}%
\end{equation}

The polaron mass at zero temperature can be obtained from the impedance
function as \cite{prb28-6051,prb2-1212}%
\begin{equation}
\frac{m^{\ast}}{m_{b}}=1-\lim_{\omega\rightarrow0}\frac{\text{Re}\Sigma\left(
\omega\right)  }{\omega},
\end{equation}
so that from the scaling relation (\ref{ssr16}) for the memory function we
also have a scaling relation for the polaron mass \cite{prb36-4442}:%
\begin{equation}
\frac{m_{n\text{D}}^{\ast}\left(  \alpha\right)  }{\left(  m_{b}\right)
_{n\text{D}}}=\frac{m_{3\text{D}}^{\ast}\left(  a_{n}\alpha\right)  }{\left(
m_{b}\right)  _{3\text{D}}}. \label{ssr18}%
\end{equation}
Since the mobility can be obtained from the memory function as \cite{SSP38-81}%
\begin{equation}
\frac{1}{\mu}=-\frac{m_{b}}{e}\lim_{\omega\rightarrow0}\frac{\text{Im}%
\Sigma\left(  \omega\right)  }{\omega},
\end{equation}
fulfilment of the scaling relation (\ref{ssr16}) implies also a scaling
relation for the mobility \cite{prb36-4442}:%
\begin{equation}
\mu_{n\text{D}}\left(  \alpha\right)  =\mu_{3\text{D}}\left(  a_{n}%
\alpha\right)  .
\end{equation}

In the important particular case of 2D, the above scaling relations take the
form \cite{PRB33-3926, prb31-3420, prb36-4442}:%
\begin{equation}
E_{2\text{D}}\left(  \alpha\right)  =\frac{2}{3}E_{3\text{D}}\left(
\frac{3\pi}{4}\alpha\right)  ,
\end{equation}%
\begin{equation}
Z_{2\text{D}}\left(  \alpha;\omega\right)  =Z_{3\text{D}}\left(  \frac{3\pi
}{4}\alpha;\omega\right)  ,
\end{equation}%
\begin{equation}
\frac{m_{2\text{D}}^{\ast}\left(  \alpha\right)  }{\left(  m_{b}\right)
_{n\text{D}}}=\frac{m_{3\text{D}}^{\ast}\left(  \frac{3\pi}{4}\alpha\right)
}{\left(  m_{b}\right)  _{3\text{D}}},
\end{equation}%
\begin{equation}
\mu_{2\text{D}}\left(  \alpha\right)  =\mu_{3\text{D}}\left(  \frac{3\pi}%
{4}\alpha\right)  .
\end{equation}
\bigskip

\subsubsection{Check of the scaling relation for the path integral Monte Carlo
result for the polaron free energy}

The fulfilment of the PD-scaling relation \cite{prb36-4442} is checked for the
path integral Monte Carlo results \cite{Ciuchi} for the polaron free energy.

\begin{figure}
[h]
\begin{center}
\includegraphics[
height=12.2462cm,
width=7.0907cm
]%
{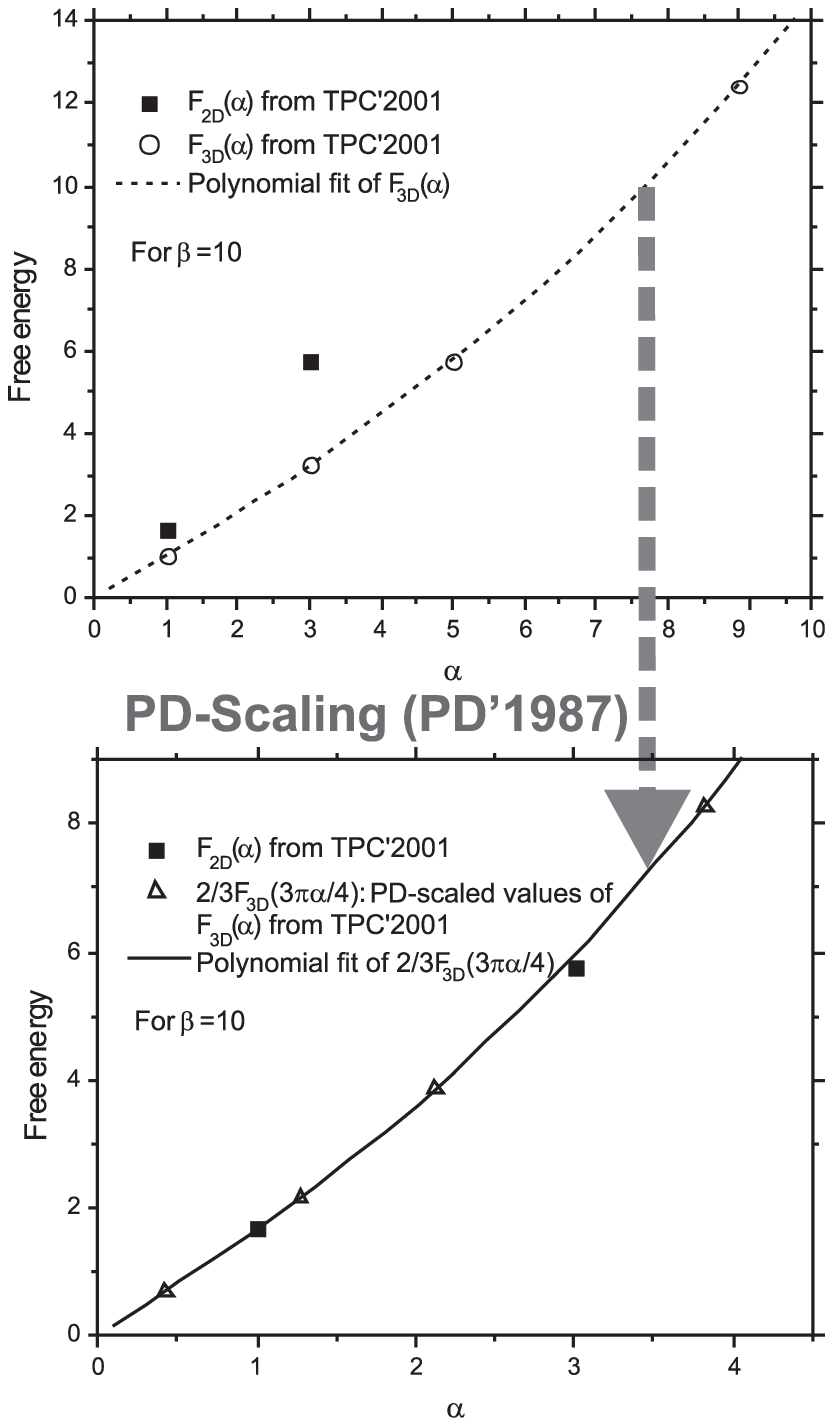}%
\caption{\textit{Upper panel: }The values of the polaron free energy in 3D
(squares) and 2D (open circles) obtained by Ciuchi'2001 \cite{Ciuchi} for
$\beta=10$. The data for $F_{3D}\left(  \alpha\right)  $ are
interpolated\ using a polynomial fit to the available four points (dotted
line). \textit{Lower panel:} Demonstration of the PD-scaling cf. PD'1987: the
values of the polaron free energy in 2D obtained by Ciuchi'2001 \cite{Ciuchi}
for $\beta=10$ (squares) are very close to the \textbf{PD-scaled} according to
PD'1987 \cite{prb36-4442} values of the polaron free energy in 3D from
Ciuchi'2001 for $\beta=10$ (open triangles). The data for $\frac{2}{3}%
F_{3D}\left(  \frac{3\pi\alpha}{4}\right)  $ are interpolated\ using a
polynomial fit to the available four points (solid line).}%
\label{ScComp}%
\end{center}
\end{figure}

The path integral Monte Carlo results of Ref.\cite{Ciuchi} for the polaron
free energy in 3D and in 2D are given for a few values of temperature and for
some selected values of $\alpha.$ For a check of the scaling relation, the
values of the polaron free energy at $\beta=10$ are taken from Ref.
\cite{Ciuchi} in 3D (Table I, for 4 values of $\alpha$) and in 2D (Table II,
for 2 values of $\alpha$) and plotted in Fig. \ref{ScComp}, upper panel, with
squares and open circles, correspondingly.

The PD-scaling relation for the polaron ground-state energy as derived in Ref.
\cite{prb36-4442} reads:%
\begin{equation}
E_{2D}\left(  \alpha\right)  \equiv\frac{2}{3}E_{3D}\left(  \frac{3\pi\alpha
}{4}\right)  . \label{E}%
\end{equation}
In Fig. \ref{ScComp}, lower panel, the available data for the free energy from
Ref \cite{Ciuchi} are plotted in the following form \textit{inspired by the
l.h.s. and the r.h.s parts of Eq. (1)}: $F_{2D}\left(  \alpha\right)  $
(squares) and $\frac{2}{3}F_{3D}\left(  \frac{3\pi\alpha}{4}\right)  $(open
triangles). As follows from the figure, t\textit{he path integral Monte Carlo
results for the polaron free energy in 2D and 3D very closely follow the
PD-scaling relation of the form given by Eq. (\ref{E}):}%
\begin{equation}
F_{2D}\left(  \alpha\right)  \equiv\frac{2}{3}F_{3D}\left(  \frac{3\pi\alpha
}{4}\right)  . \label{F}%
\end{equation}

%\newpage%
%

%TCIMACRO{\FRAME{fhFU}{7.0907cm}{12.2462cm}{0pt}{\Qcb{\QTR{it}{Upper panel:
%}The values of the polaron free energy in 3D (squares) and 2D (open circles)
%obtained by Ciuchi'2001 \cite{Ciuchi} for $\beta=10$. The data for
%$F_{3D}\left(  \alpha\right)  $ are interpolated\ using a polynomial fit to
%the available four points (dotted line). \QTR{it}{Lower panel:} Demonstration
%of the PD-scaling cf. PD'1987: the values of the polaron free energy in 2D
%obtained by Ciuchi'2001 \cite{Ciuchi} for $\beta=10$ (squares) are very close
%to the \QTR{bf}{PD-scaled} according to PD'1987 \cite{prb36-4442} values of
%the polaron free energy in 3D from Ciuchi'2001 for $\beta=10$ (open
%triangles). The data for $\frac{2}{3}F_{3D}\left(  \frac{3\pi\alpha}%
%{4}\right)  $ are interpolated\ using a polynomial fit to the available four
%points (solid line).}}{\Qlb{ScComp}}{part1fig11.eps}%
%maintain-aspect-ratio TRUE;  display "ICON";  valid_file "F";
%width 7.0907cm;  height 12.2462cm;  depth 0pt;  original-width 13.9068cm;
%original-height 24.0157cm;  cropleft "0";  croptop "1";  cropright "1";
%cropbottom "0";  filename 'Part1Fig11.eps';file-properties "XNPEU";}} }%
%BeginExpansion
%EndExpansion

\newpage

\subsection*{Appendix 1. Weak coupling: LLP approach}

Inspired by the work of Tomonaga on quantum electrodynamics (Q. E. D.), Lee,
Low and Pines (L.L.P.) \cite{LLP} derived (\ref{eq_5a}) and $m^{\ast}%
=m_{b}(1+\alpha/6)$ from a canonical transformation formulation, which
establishes (\ref{eq_5a}) as a variational upper bound for the ground-state energy.

The wave equation corresponding to the {Fr\"{o}hlich Hamiltonian (\ref{eq_1a})
}is%

\begin{equation}
H\Phi=E\Phi. \label{LLP1}%
\end{equation}
We shall take advantage of the fact that the total momentum of the system%

\begin{equation}
\mathbf{P}_{op}=\sum_{\mathbf{k}}\hbar\mathbf{k}a_{\mathbf{k}}^{\dag
}a_{\mathbf{k}}+\mathbf{p} \label{LP2}%
\end{equation}
(where $\mathbf{p=-}i\hbar\nabla$\textbf{ }is the momentum of the electron) is
a constant of motion because it commutes with the Hamiltonian (\ref{eq_1a})

Indeed,%
\begin{align*}
\left[  \mathbf{p},H\right]   &  =[\mathbf{p},\sum_{\mathbf{k}}(V_{k}%
a_{\mathbf{k}}e^{i\mathbf{k\cdot r}}+V_{k}^{\ast}a_{\mathbf{k}}^{\dag
}e^{-i\mathbf{k\cdot r}})]=\sum_{\mathbf{k}}(V_{k}a_{\mathbf{k}}\left[
\mathbf{p},e^{i\mathbf{k\cdot r}}\right]  +V_{k}^{\ast}a_{\mathbf{k}}^{\dag
}\left[  \mathbf{p},e^{-i\mathbf{k\cdot r}}\right]  )\\
&  =\sum_{\mathbf{k}}\hbar\mathbf{k}(V_{k}a_{\mathbf{k}}e^{i\mathbf{k\cdot r}%
}-V_{k}^{\ast}a_{\mathbf{k}}^{\dag}e^{-i\mathbf{k\cdot r}});
\end{align*}

\begin{align*}
\left[  \sum_{\mathbf{k}}\hbar\mathbf{k}a_{\mathbf{k}}^{\dag}a_{\mathbf{k}%
},H\right]   &  =\left[  \sum_{\mathbf{k}}\hbar\mathbf{k}a_{\mathbf{k}}^{\dag
}a_{\mathbf{k}},\sum_{\mathbf{k}^{\prime}}(V_{k^{\prime}}a_{\mathbf{k}%
^{\prime}}e^{i\mathbf{k}^{\prime}\mathbf{\cdot r}}+V_{k^{\prime}}^{\ast
}a_{\mathbf{k}^{\prime}}^{\dag}e^{-i\mathbf{k}^{\prime}\mathbf{\cdot r}%
})\right]  =\\
&  =\sum_{\mathbf{k}}\hbar\mathbf{k}\left[  a_{\mathbf{k}}^{\dag}%
a_{\mathbf{k}},(V_{k}a_{\mathbf{k}}e^{i\mathbf{k\cdot r}}+V_{k}^{\ast
}a_{\mathbf{k}}^{\dag}e^{-i\mathbf{k\cdot r}})\right]  =\\
&  =\sum_{\mathbf{k}}\hbar\mathbf{k}\left\{  V_{k}\left[  a_{\mathbf{k}}%
^{\dag}a_{\mathbf{k}},a_{\mathbf{k}}\right]  e^{i\mathbf{k\cdot r}}%
+V_{k}^{\ast}\left[  a_{\mathbf{k}}^{\dag}a_{\mathbf{k}},a_{\mathbf{k}}^{\dag
}\right]  e^{-i\mathbf{k\cdot r}}\right\}  =\\
&  =-\sum_{\mathbf{k}}\hbar\mathbf{k}\left(  V_{k}a_{\mathbf{k}}%
e^{i\mathbf{k\cdot r}}-V_{k}^{\ast}a_{\mathbf{k}}^{\dag}e^{-i\mathbf{k\cdot
r}}\right)  ;
\end{align*}
\bigskip%

\begin{equation}
\left[  \mathbf{P}_{op},H\right]  =\left[  \sum_{\mathbf{k}}\hbar
\mathbf{k}a_{\mathbf{k}}^{\dag}a_{\mathbf{k}}+\mathbf{p},H\right]  =0.
\label{PHCOMM}%
\end{equation}
Because of the commutation (\ref{PHCOMM}), the operators $H$ and
$\mathbf{P}_{op}$ have a common set of basis functions: $H\Phi=E\Phi$ and
$\mathbf{P}_{op}\Phi=\mathbf{P}\Phi.$

It is possible to transform to a representation in which $\mathbf{P}_{op}$
becomes a \textquotedblleft c\textquotedblright\ number $\mathbf{P}$, and in
which the Hamiltonian no longer contains the electron coordinates. The unitary
(canonical) transformation required is $\Phi=S_{1}\psi$, where%

\begin{equation}
S_{1}=\exp\left[  \frac{i}{\hbar}(\mathbf{P}-\sum_{\mathbf{k}}\hbar
\mathbf{k}a_{\mathbf{k}}^{\dagger}a_{\mathbf{k}})\mathbf{\cdot r}\right]  .
\label{eq_6a}%
\end{equation}

Derivation of the transformations of the operators.%

\begin{align}
\mathbf{p}  &  \longrightarrow S_{1}^{-1}\mathbf{p}S_{1}=\nonumber\\
&  =\exp\left[  -\frac{i}{\hbar}(\mathbf{P}-\sum_{\mathbf{k}}\hbar
\mathbf{k}a_{\mathbf{k}}^{\dagger}a_{\mathbf{k}})\mathbf{\cdot r}\right]
\mathbf{p}\exp\left[  \frac{i}{\hbar}(\mathbf{P}-\sum_{\mathbf{k}}%
\hbar\mathbf{k}a_{\mathbf{k}}^{\dagger}a_{\mathbf{k}})\mathbf{\cdot r}\right]
\nonumber\\
&  =\exp\left[  -\frac{i}{\hbar}(\mathbf{P}-\sum_{\mathbf{k}}\hbar
\mathbf{k}a_{\mathbf{k}}^{\dagger}a_{\mathbf{k}})\mathbf{\cdot r}\right]
\left(  -i\hbar\nabla\right)  \exp\left[  \frac{i}{\hbar}(\mathbf{P}%
-\sum_{\mathbf{k}}\hbar\mathbf{k}a_{\mathbf{k}}^{\dagger}a_{\mathbf{k}%
})\mathbf{\cdot r}\right] \nonumber\\
&  =\exp\left[  -\frac{i}{\hbar}(\mathbf{P}-\sum_{\mathbf{k}}\hbar
\mathbf{k}a_{\mathbf{k}}^{\dagger}a_{\mathbf{k}})\mathbf{\cdot r}\right]
\left\{
\begin{array}
[c]{c}%
(\mathbf{P}-\sum_{\mathbf{k}}\hbar\mathbf{k}a_{\mathbf{k}}^{\dagger
}a_{\mathbf{k}})\exp\left[  \frac{i}{\hbar}(\mathbf{P}-\sum_{\mathbf{k}}%
\hbar\mathbf{k}a_{\mathbf{k}}^{\dagger}a_{\mathbf{k}})\mathbf{\cdot r}\right]
\\
+\exp\left[  \frac{i}{\hbar}(\mathbf{P}-\sum_{\mathbf{k}}\hbar\mathbf{k}%
a_{\mathbf{k}}^{\dagger}a_{\mathbf{k}})\mathbf{\cdot r}\right]  \left(
-i\hbar\nabla\right)
\end{array}
\right\} \nonumber\\
&  =\mathbf{P}-\sum_{\mathbf{k}}\hbar\mathbf{k}a_{\mathbf{k}}^{\dagger
}a_{\mathbf{k}}+\mathbf{p,} \label{LLP3a}%
\end{align}

\begin{align}
\mathbf{P}_{op}  &  \longrightarrow S_{1}^{-1}\mathbf{P}_{op}S_{1}=\nonumber\\
&  =\exp\left[  -\frac{i}{\hbar}(\mathbf{P}-\sum_{\mathbf{k}}\hbar
\mathbf{k}a_{\mathbf{k}}^{\dagger}a_{\mathbf{k}})\mathbf{\cdot r}\right]
\left(  \sum_{\mathbf{k}}\hbar\mathbf{k}a_{\mathbf{k}}^{\dag}a_{\mathbf{k}%
}+\mathbf{p}\right)  \exp\left[  \frac{i}{\hbar}(\mathbf{P}-\sum_{\mathbf{k}%
}\hbar\mathbf{k}a_{\mathbf{k}}^{\dagger}a_{\mathbf{k}})\mathbf{\cdot r}\right]
\nonumber\\
&  =\exp\left[  \frac{i}{\hbar}\sum_{\mathbf{k}}\hbar\mathbf{k}a_{\mathbf{k}%
}^{\dagger}a_{\mathbf{k}}\mathbf{\cdot r}\right]  \sum_{\mathbf{k}}%
\hbar\mathbf{k}a_{\mathbf{k}}^{\dag}a_{\mathbf{k}}\exp\left[  -\frac{i}{\hbar
}\sum_{\mathbf{k}}\hbar\mathbf{k}a_{\mathbf{k}}^{\dagger}a_{\mathbf{k}%
}\mathbf{\cdot r}\right]  +S_{1}^{-1}\mathbf{p}S_{1}\nonumber\\
&  =\sum_{\mathbf{k}}\hbar\mathbf{k}a_{\mathbf{k}}^{\dag}a_{\mathbf{k}%
}+\mathbf{P}-\sum_{\mathbf{k}}\hbar\mathbf{k}a_{\mathbf{k}}^{\dagger
}a_{\mathbf{k}}+\mathbf{p=P}+\mathbf{p,} \label{LLP3b}%
\end{align}

\begin{align}
a_{\mathbf{k}}  &  \longrightarrow S_{1}^{-1}a_{\mathbf{k}}S_{1}=\nonumber\\
&  =\exp\left[  -\frac{i}{\hbar}(\mathbf{P}-\sum_{\mathbf{k}}\hbar
\mathbf{k}a_{\mathbf{k}}^{\dagger}a_{\mathbf{k}})\mathbf{\cdot r}\right]
a_{\mathbf{k}}\exp\left[  \frac{i}{\hbar}(\mathbf{P}-\sum_{\mathbf{k}}%
\hbar\mathbf{k}a_{\mathbf{k}}^{\dagger}a_{\mathbf{k}})\mathbf{\cdot r}\right]
\nonumber\\
&  =\exp\left[  \frac{i}{\hbar}\sum_{\mathbf{k}}\hbar\mathbf{k}a_{\mathbf{k}%
}^{\dagger}a_{\mathbf{k}}\mathbf{\cdot r}\right]  a_{\mathbf{k}}\exp\left[
-\frac{i}{\hbar}\sum_{\mathbf{k}}\hbar\mathbf{k}a_{\mathbf{k}}^{\dagger
}a_{\mathbf{k}}\mathbf{\cdot r}\right] \nonumber\\
&  =\exp\left[  i\mathbf{k}a_{\mathbf{k}}^{\dagger}a_{\mathbf{k}}\mathbf{\cdot
r}\right]  a_{\mathbf{k}}\exp\left[  -i\mathbf{k}a_{\mathbf{k}}^{\dagger
}a_{\mathbf{k}}\mathbf{\cdot r}\right] \nonumber\\
&  =\exp\left[  i\mathbf{k}a_{\mathbf{k}}^{\dagger}a_{\mathbf{k}}\mathbf{\cdot
r}\right]  a_{\mathbf{k}}%
%TCIMACRO{\dsum \nolimits_{n=0}^{\infty}}%
%BeginExpansion
{\displaystyle\sum\nolimits_{n=0}^{\infty}}
%EndExpansion
\frac{1}{n!}\left(  -i\mathbf{k}a_{\mathbf{k}}^{\dagger}a_{\mathbf{k}%
}\mathbf{\cdot r}\right)  ^{n}\nonumber\\
&  =\exp\left[  i\mathbf{k}a_{\mathbf{k}}^{\dagger}a_{\mathbf{k}}\mathbf{\cdot
r}\right]
%TCIMACRO{\dsum \nolimits_{n=0}^{\infty}}%
%BeginExpansion
{\displaystyle\sum\nolimits_{n=0}^{\infty}}
%EndExpansion
\frac{1}{n!}a_{\mathbf{k}}\left(  -i\mathbf{k}a_{\mathbf{k}}^{\dagger
}a_{\mathbf{k}}\mathbf{\cdot r}\right)  ^{n}\nonumber\\
&  =\exp\left[  i\mathbf{k\cdot r}a_{\mathbf{k}}^{\dagger}a_{\mathbf{k}%
}\right]
%TCIMACRO{\dsum \nolimits_{n=0}^{\infty}}%
%BeginExpansion
{\displaystyle\sum\nolimits_{n=0}^{\infty}}
%EndExpansion
\frac{1}{n!}\left(  -i\mathbf{k\cdot r}\right)  ^{n}a_{\mathbf{k}}\left(
a_{\mathbf{k}}^{\dagger}a_{\mathbf{k}}\right)  ^{n}\overset{\text{see}(\ast
)}{=}\nonumber\\
&  =\exp\left[  i\mathbf{k\cdot r}a_{\mathbf{k}}^{\dagger}a_{\mathbf{k}%
}\right]
%TCIMACRO{\dsum \nolimits_{n=0}^{\infty}}%
%BeginExpansion
{\displaystyle\sum\nolimits_{n=0}^{\infty}}
%EndExpansion
\frac{1}{n!}[-i\mathbf{k\cdot r}\left(  a_{\mathbf{k}}^{\dagger}a_{\mathbf{k}%
}+1\right)  ]^{n}a_{\mathbf{k}}\nonumber\\
&  =\exp\left[  i\mathbf{k\cdot r}a_{\mathbf{k}}^{\dagger}a_{\mathbf{k}%
}\right]  \exp\left[  -i\mathbf{k\cdot r(}a_{\mathbf{k}}^{\dagger
}a_{\mathbf{k}}+1)\right]  a_{\mathbf{k}}\nonumber\\
&  =a_{\mathbf{k}}\exp\left(  -i\mathbf{k\cdot r}\right)  . \label{LLP3c1}%
\end{align}

Here the property was used:%

\begin{equation}
a_{\mathbf{k}}\left(  a_{\mathbf{k}}^{\dagger}a_{\mathbf{k}}\right)
^{n}=\left(  a_{\mathbf{k}}^{\dagger}a_{\mathbf{k}}+1\right)  ^{n}%
a_{\mathbf{k}}. \tag{*}\label{*}%
\end{equation}
It is evident for $n=0.$For $n=1$ it is demonstrated as follows:%

\[
a_{\mathbf{k}}a_{\mathbf{k}}^{\dagger}=a_{\mathbf{k}}^{\dagger}a_{\mathbf{k}%
}+1\Longrightarrow a_{\mathbf{k}}a_{\mathbf{k}}^{\dagger}a_{\mathbf{k}%
}=(a_{\mathbf{k}}^{\dagger}a_{\mathbf{k}}+1)a_{\mathbf{k}};
\]
then for $n\geq2$ the validity of (\ref{*}) is straightforwardly demonstrated
by induction.

Finally,%

\begin{equation}
a_{\mathbf{k}}^{\dag}\longrightarrow S_{1}^{-1}a_{\mathbf{k}}^{\dag}%
S_{1}=[S_{1}^{-1}a_{\mathbf{k}}S_{1}]^{\dag}=a_{\mathbf{k}}^{\dag}\exp\left(
i\mathbf{k\cdot r}\right)  . \label{LLP3d}%
\end{equation}
Using (\ref{LLP3a}), (\ref{LLP3b}), (\ref{LLP3c1}) and (\ref{LLP3d}), one
arrives at%

\begin{equation}
H\longrightarrow\mathcal{H}=S_{1}^{-1}HS_{1}=\frac{\left(  \mathbf{P}%
-\sum_{\mathbf{k}}\hbar\mathbf{k}a_{\mathbf{k}}^{\dag}a_{\mathbf{k}}\right)
^{2}}{2m_{b}}+\sum_{\mathbf{k}}\hbar\omega_{\mathrm{LO}}a_{\mathbf{k}}^{\dag
}a_{\mathbf{k}}+\sum_{\mathbf{k}}(V_{k}a_{\mathbf{k}}+V_{k}^{\ast
}a_{\mathbf{k}}^{\dag}), \label{LLP3e}%
\end{equation}
where $\mathbf{p}$ is set $0$. \footnote{Transformation of the equation
$\mathbf{P}_{op}\Phi=\mathbf{P}\Phi$ leads to $S_{1}^{-1}\mathbf{P}_{op}%
S_{1}\psi=\mathbf{P}\psi.$At the same time, applying Eq. (\ref{LLP3b}), we
obtain $S_{1}^{-1}\mathbf{P}_{op}S_{1}=\mathbf{P}+\mathbf{p}.$Setting the
gauge $\mathbf{p}\psi=0$ eliminates the operator $\mathbf{p}$ from the
problem.} The wave equation (\ref{LLP1}) takes the form%

\begin{equation}
HS_{1}\psi=ES_{1}\psi\Longrightarrow\mathcal{H}\psi=E\psi. \label{LLP4}%
\end{equation}

Our aim is to calculate for a given momentum $\mathbf{P}$ the lowest
eigenvalue $E(P)$ of the Hamiltonian (\ref{LLP3e}). For the low-lying energy
levels of the electron $E(P)$ may be well represented by the first two terms
of a power series expansion in $P^{2}:E(P)=E_{0}+P^{2}/2m_{p}+O(P^{4}),$where
$m_{p}$ is the effective mass of the polaron.

The canonical transformation (\ref{eq_6a}) formally eliminates the electron
operators from the Hamiltonian. LLP use further a variational method of
calculation. The trial wave function is chosen as%

\begin{equation}
\psi=S_{2}\psi_{0} \label{LLP5}%
\end{equation}
where $\psi_{0}$ is the eigenstate of the unperturbed Hamiltonian with no
phonons present (vacuum state). Specifically, $\psi_{0}$ is defined by%

\begin{equation}
a_{\mathbf{k}}\psi_{0}=0,\qquad\left(  \psi_{0},\psi_{0}\right)  =1
\label{LLP6}%
\end{equation}
and the second canonical transformation:
\begin{equation}
S_{2}=\exp\left[  \sum_{\mathbf{k}}(a_{\mathbf{k}}^{\dagger}f_{\mathbf{k}%
}-a_{\mathbf{k}}f_{\mathbf{k}}^{\ast})\right]  , \label{eq_6b}%
\end{equation}
where $f_{\mathbf{k}}$ are treated as variational functions and will be chosen
to minimize the energy. The physical significance of Eq.\thinspace
(\ref{eq_6b}) is that it \textquotedblleft dresses\textquotedblright\ the
electron with the virtual phonon field, which describes the polarization.
Viewed as a unitary transformation, $S_{2}$ is a \textit{displacement}
operator on $a_{\mathbf{k}}$ and $a_{\mathbf{k}}^{\dagger}:$%

\begin{align}
a_{\mathbf{k}}  &  \longrightarrow S_{2}^{-1}a_{\mathbf{k}}S_{2}=\nonumber\\
&  =\exp\left[  -\sum_{\mathbf{k}}(a_{\mathbf{k}}^{\dagger}f_{\mathbf{k}%
}-a_{\mathbf{k}}f_{\mathbf{k}}^{\ast})\right]  a_{\mathbf{k}}\exp\left[
\sum_{\mathbf{k}}(a_{\mathbf{k}}^{\dagger}f_{\mathbf{k}}-a_{\mathbf{k}%
}f_{\mathbf{k}}^{\ast})\right] \nonumber\\
&  =\exp\left[  -(a_{\mathbf{k}}^{\dagger}f_{\mathbf{k}}-a_{\mathbf{k}%
}f_{\mathbf{k}}^{\ast})\right]  a_{\mathbf{k}}\exp\left[  (a_{\mathbf{k}%
}^{\dagger}f_{\mathbf{k}}-a_{\mathbf{k}}f_{\mathbf{k}}^{\ast})\right]
\overset{\text{see}(\ast\ast)}{=}\nonumber\\
&  =a_{\mathbf{k}}+\left[  a_{\mathbf{k}},(a_{\mathbf{k}}^{\dagger
}f_{\mathbf{k}}-a_{\mathbf{k}}f_{\mathbf{k}}^{\ast})\right]  +\frac{1}%
{2}\left[  \left[  a_{\mathbf{k}},(a_{\mathbf{k}}^{\dagger}f_{\mathbf{k}%
}-a_{\mathbf{k}}f_{\mathbf{k}}^{\ast})\right]  ,(a_{\mathbf{k}}^{\dagger
}f_{\mathbf{k}}-a_{\mathbf{k}}f_{\mathbf{k}}^{\ast})\right]  +...\nonumber\\
&  =a_{\mathbf{k}}+f_{\mathbf{k}}, \label{LLP6a}%
\end{align}

\begin{equation}
a_{\mathbf{k}}^{\dag}\longrightarrow S_{2}^{-1}a\dag_{\mathbf{k}}%
S_{2}=a_{\mathbf{k}}^{\dag}+f_{\mathbf{k}}^{\ast}. \label{LLP6b}%
\end{equation}
Here the relation was used%
\begin{equation}
\exp\left[  -V\right]  a\exp\left[  V\right]  =a+\left[  a,V\right]  +\frac
{1}{2}\left[  \left[  a,V\right]  ,V\right]  +\frac{1}{3!}\left[  \left[
\left[  a,V\right]  ,V\right]  ,V\right]  +... \tag{**}\label{**}%
\end{equation}

Further we seek to minimize the expression for the energy,%

\begin{equation}
E=\left(  \psi,H\psi\right)  =\left(  \psi_{0},S_{2}^{-1}\mathcal{H}S_{2}%
\psi_{0}\right)  . \label{LLP7}%
\end{equation}

In virtue of (\ref{LLP6a}) and (\ref{LLP6b}), we obtain:%

\begin{align*}
S_{2}^{-1}\mathcal{H}S_{2}  &  =\frac{\left[  \mathbf{P}-\sum_{\mathbf{k}%
}\hbar\mathbf{k}\left(  a_{\mathbf{k}}^{\dag}+f_{\mathbf{k}}^{\ast}\right)
\left(  a_{\mathbf{k}}+f_{\mathbf{k}}\right)  \right]  ^{2}}{2m_{b}}\\
&  +\sum_{\mathbf{k}}\hbar\omega_{\mathrm{LO}}\left(  a_{\mathbf{k}}^{\dag
}+f_{\mathbf{k}}^{\ast}\right)  \left(  a_{\mathbf{k}}+f_{\mathbf{k}}\right)
+\sum_{\mathbf{k}}\left[  V_{k}\left(  a_{\mathbf{k}}+f_{\mathbf{k}}\right)
+V_{k}^{\ast}\left(  a_{\mathbf{k}}^{\dag}+f_{\mathbf{k}}^{\ast}\right)
\right] \\
&  =\frac{{\left[  (\mathbf{P}-\sum_{\mathbf{k}}\hbar\mathbf{k}a_{\mathbf{k}%
}^{\dag}a_{\mathbf{k}})-\sum_{\mathbf{k}}\hbar\mathbf{k}\left\vert
f_{\mathbf{k}}\right\vert ^{2}-\sum_{\mathbf{k}}\hbar\mathbf{k}\left(
a_{\mathbf{k}}^{\dag}f_{\mathbf{k}}+a_{\mathbf{k}}f_{\mathbf{k}}^{\ast
}\right)  \right]  ^{2}}}{2m_{b}}\\
&  +\sum_{\mathbf{k}}\hbar\omega_{\mathrm{LO}}\left(  a_{\mathbf{k}}^{\dag
}a_{\mathbf{k}}+\left\vert f_{\mathbf{k}}\right\vert ^{2}+a_{\mathbf{k}}%
^{\dag}f_{\mathbf{k}}+a_{\mathbf{k}}f_{\mathbf{k}}^{\ast}\right) \\
&  +\sum_{\mathbf{k}}\left[  V_{k}\left(  a_{\mathbf{k}}+f_{\mathbf{k}%
}\right)  +V_{k}^{\ast}\left(  a_{\mathbf{k}}^{\dag}+f_{\mathbf{k}}^{\ast
}\right)  \right] \\
&  =H_{0}+H_{1},
\end{align*}
where%
\begin{align}
H_{0}  &  =\frac{{\left[  (\mathbf{P}-\sum_{\mathbf{k}}\hbar\mathbf{k}%
a_{\mathbf{k}}^{\dag}a_{\mathbf{k}})\right]  ^{2}+\left[  \sum_{\mathbf{k}%
}\hbar\mathbf{k}\left\vert f_{\mathbf{k}}\right\vert ^{2}\right]  ^{2}}%
}{2m_{b}}+\sum_{\mathbf{k}}\left[  V_{k}f_{\mathbf{k}}+V_{k}^{\ast
}f_{\mathbf{k}}^{\ast}\right] \nonumber\\
&  +\sum_{\mathbf{k}}\left\vert f_{\mathbf{k}}\right\vert ^{2}\left\{
\hbar\omega_{\mathrm{LO}}-\frac{\hbar\mathbf{k\cdot P}}{m_{b}}+\frac{\hbar
^{2}k^{2}}{2m_{b}}\right\}  +\frac{\hbar^{2}}{m_{b}}\sum_{\mathbf{k}%
}\mathbf{k}a_{\mathbf{k}}^{\dag}a_{\mathbf{k}}\cdot\sum_{\mathbf{k}^{\prime}%
}\mathbf{k}^{\prime}\left\vert f_{\mathbf{k}^{\prime}}\right\vert
^{2}\nonumber\\
&  +\sum_{\mathbf{k}}a_{\mathbf{k}}\left\{  V_{k}+f_{\mathbf{k}}^{\ast}\left[
\hbar\omega_{\mathrm{LO}}-\frac{\hbar\mathbf{k\cdot P}}{m_{b}}+\frac{\hbar
^{2}k^{2}}{2m_{b}}+\frac{\hbar^{2}\mathbf{k}}{m_{b}}\cdot\sum_{\mathbf{k}%
^{\prime}}\mathbf{k}^{\prime}\left\vert f_{\mathbf{k}^{\prime}}\right\vert
^{2}\right]  \right\} \nonumber\\
&  +\sum_{\mathbf{k}}a_{\mathbf{k}}^{\dag}\left\{  V_{k}^{\ast}+f_{\mathbf{k}%
}\left[  \hbar\omega_{\mathrm{LO}}-\frac{\hbar\mathbf{k\cdot P}}{m_{b}}%
+\frac{\hbar^{2}k^{2}}{2m_{b}}+\frac{\hbar^{2}\mathbf{k}}{m_{b}}\cdot
\sum_{\mathbf{k}^{\prime}}\mathbf{k}^{\prime}\left\vert f_{\mathbf{k}^{\prime
}}\right\vert ^{2}\right]  \right\} \nonumber\\
&  +\sum_{\mathbf{k}}\hbar\omega_{\mathrm{LO}}a_{\mathbf{k}}^{\dag
}a_{\mathbf{k}}; \label{LLP7a}%
\end{align}

\begin{align*}
H_{1}  &  =\sum_{\mathbf{k,k}^{\prime}}\frac{\hbar^{2}\mathbf{k\cdot
k}^{\prime}}{2m_{b}}\left\{  a_{\mathbf{k}}a_{\mathbf{k}^{\prime}%
}f_{\mathbf{k}}^{\ast}f_{\mathbf{k}^{\prime}}^{\ast}+2a_{\mathbf{k}}^{\dag
}a_{\mathbf{k}^{\prime}}f_{\mathbf{k}}f_{\mathbf{k}^{\prime}}^{\ast
}+a_{\mathbf{k}}^{\dag}a_{\mathbf{k}^{\prime}}^{\dag}f_{\mathbf{k}%
}f_{\mathbf{k}^{\prime}}\right\}  +\\
&  +\sum_{\mathbf{k,k}^{\prime}}\frac{\hbar^{2}\mathbf{k\cdot k}^{\prime}%
}{2m_{b}}\left\{  a_{\mathbf{k}}^{\dag}a_{\mathbf{k}}a_{\mathbf{k}^{\prime}%
}f_{\mathbf{k}^{\prime}}^{\ast}+a_{\mathbf{k}^{\prime}}^{\dag}a_{\mathbf{k}%
}^{\dag}a_{\mathbf{k}}f_{\mathbf{k}^{\prime}}\right\}  .
\end{align*}
Using (\ref{LLP6}), we obtain from (\ref{LLP7}) that%
\begin{align}
E  &  =H_{0}=\frac{P{^{2}+\left[  \sum_{\mathbf{k}}\hbar\mathbf{k}\left\vert
f_{\mathbf{k}}\right\vert ^{2}\right]  ^{2}}}{2m_{b}}+\sum_{\mathbf{k}}\left[
V_{k}f_{\mathbf{k}}+V_{k}^{\ast}f_{\mathbf{k}}^{\ast}\right] \nonumber\\
&  +\sum_{\mathbf{k}}\left\vert f_{\mathbf{k}}\right\vert ^{2}\left\{
\hbar\omega_{\mathrm{LO}}-\frac{\hbar\mathbf{k\cdot P}}{m_{b}}+\frac{\hbar
^{2}k^{2}}{2m_{b}}\right\}  . \label{LLP8}%
\end{align}
We minimize (\ref{LLP8}) by imposing%

\[
\frac{\delta E}{\delta f_{\mathbf{k}}}=0,\qquad\frac{\delta E}{\delta
f_{\mathbf{k}}^{\dag}}=0.
\]
This results in
\begin{equation}
V_{k}+f_{\mathbf{k}}^{\ast}\left\{  \hbar\omega_{\mathrm{LO}}-\frac
{\hbar\mathbf{k\cdot P}}{m_{b}}+\frac{\hbar^{2}k^{2}}{2m_{b}}+\frac{\hbar^{2}%
}{m_{b}}{\left[  \sum_{\mathbf{k}^{\prime}}\hbar\mathbf{k}^{\prime}\left\vert
f_{\mathbf{k}^{\prime}}\right\vert ^{2}\right]  \cdot}\mathbf{k}\right\}  =0
\label{LLP9}%
\end{equation}
and the appropriate complex conjugate equation for $f_{\mathbf{k}}$. Upon
comparing (\ref{LLP9}) and (\ref{LLP7a}), we see that the linear terms in
$a_{\mathbf{k}}^{\dag}$ and $a_{\mathbf{k}}$ are identically zero if
(\ref{LLP9}) is satisfied, and hence that $H_{0}$ is diagonal in a
representation in which $a_{\mathbf{k}}^{\dag}a_{\mathbf{k}}$ is diagonal. So,
the variational calculation by LLP is equivalent to the use of (\ref{LLP7a})
as the total Hamiltonian provided $f_{\mathbf{k}}^{\ast}$ satisfies
(\ref{LLP9}). An estimate of the accuracy of the LLP variational procedure was
obtained by an estimate of the effect of $H_{1}$ using a perturbation theory.

Now we evaluate the energy of the ground state of the system, which is given
by Eq. (\ref{LLP8}) with $f_{\mathbf{k}}^{\ast}$ satisfying Eq. (\ref{LLP9}).
The only preferred direction in this problem is $\mathbf{P}$. Therefore one
may introduce the parameter $\eta$ defined as%
\begin{equation}
\eta\mathbf{P=}\sum_{\mathbf{k}^{{}}}\hbar\mathbf{k}\left\vert f_{\mathbf{k}%
}\right\vert ^{2}. \label{etaP}%
\end{equation}
Then Eq. (\ref{LLP9}) leads to
\begin{equation}
f_{\mathbf{k}}^{\ast}=-V_{k}\left/  \left[  \hbar\omega_{\mathrm{LO}}%
-\frac{\hbar\mathbf{k\cdot P}}{m_{b}}(1-\eta)+\frac{\hbar^{2}k^{2}}{2m_{b}%
}\right]  \right.  , \label{fk}%
\end{equation}
and we obtain the following implicit equation for $\eta$:%
\begin{align*}
\eta\mathbf{P}  &  \mathbf{=}\sum_{\mathbf{k}^{{}}}\hbar\mathbf{k}\left\vert
V_{k}\right\vert ^{2}\left/  \left[  \hbar\omega_{\mathrm{LO}}-\frac
{\hbar\mathbf{k\cdot P}}{m_{b}}(1-\eta)+\frac{\hbar^{2}k^{2}}{2m_{b}}\right]
^{2}\right. \\
&  =\frac{V}{\left(  2\pi\right)  ^{3}}\int d^{3}k\hbar\mathbf{k}\left(
\frac{\hbar\omega_{\mathrm{LO}}}{k}\right)  ^{2}\frac{4\pi\alpha}{V}\left(
\frac{\hbar}{2m_{b}\omega_{\mathrm{LO}}}\right)  ^{\frac{1}{2}}\left/  \left[
\hbar\omega_{\mathrm{LO}}-\frac{\hbar\mathbf{k\cdot P}}{m_{b}}(1-\eta
)+\frac{\hbar^{2}k^{2}}{2m_{b}}\right]  ^{2}\right.  .
\end{align*}
Let us introduce spherical coordinates with a polar axis along $\mathbf{P}$
and denote $x=\cos(\mathbf{k}^{\symbol{94}}\mathbf{P})$:%
\begin{align*}
\eta P  &  =\frac{\alpha\hbar^{3}\omega_{\mathrm{LO}}^{2}}{2\pi^{2}}\left(
\frac{\hbar}{2m_{b}\omega_{\mathrm{LO}}}\right)  ^{\frac{1}{2}}2\pi%
%TCIMACRO{\dint \nolimits_{-1}^{1}}%
%BeginExpansion
{\displaystyle\int\nolimits_{-1}^{1}}
%EndExpansion
dxx%
%TCIMACRO{\dint \nolimits_{0}^{\infty}}%
%BeginExpansion
{\displaystyle\int\nolimits_{0}^{\infty}}
%EndExpansion
dkk\left/  \left[  \hbar\omega_{\mathrm{LO}}-\frac{\hbar kPx}{m_{b}}%
(1-\eta)+\frac{\hbar^{2}k^{2}}{2m_{b}}\right]  ^{2}\right. \\
&  =\frac{\alpha\hbar}{2\pi^{2}}\left(  \frac{\hbar}{2m_{b}\omega
_{\mathrm{LO}}}\right)  ^{\frac{1}{2}}2\pi%
%TCIMACRO{\dint \nolimits_{-1}^{1}}%
%BeginExpansion
{\displaystyle\int\nolimits_{-1}^{1}}
%EndExpansion
dxx%
%TCIMACRO{\dint \nolimits_{0}^{\infty}}%
%BeginExpansion
{\displaystyle\int\nolimits_{0}^{\infty}}
%EndExpansion
dkk\left/  \left[  1-2\frac{\hbar kPx}{2m_{b}\hbar\omega_{\mathrm{LO}}}%
(1-\eta)+\frac{\hbar^{2}k^{2}}{2m_{b}\hbar\omega_{\mathrm{LO}}}\right]
^{2}\right.  .
\end{align*}
Further, we introduce the parameter
\begin{equation}
q=\frac{P}{\left(  2m_{b}\hbar\omega_{\mathrm{LO}}\right)  ^{1/2}}(1-\eta)
\label{q}%
\end{equation}
and a new variable%
\[
\kappa=\frac{\hbar k}{\left(  2m_{b}\hbar\omega_{\mathrm{LO}}\right)  ^{1/2}%
}.
\]
This gives%
\begin{align*}
\eta &  =\frac{\alpha\hbar}{\pi}\left(  \frac{\hbar}{2m_{b}\omega
_{\mathrm{LO}}}\right)  ^{\frac{1}{2}}\frac{2m_{b}\hbar\omega_{\mathrm{LO}}%
}{\hbar^{2}P}%
%TCIMACRO{\dint \nolimits_{-1}^{1}}%
%BeginExpansion
{\displaystyle\int\nolimits_{-1}^{1}}
%EndExpansion
dxx%
%TCIMACRO{\dint \nolimits_{0}^{\infty}}%
%BeginExpansion
{\displaystyle\int\nolimits_{0}^{\infty}}
%EndExpansion
d\kappa\kappa\left/  \left[  1-2q\kappa x+\kappa^{2}\right]  ^{2}\right. \\
&  =\frac{\alpha}{\pi}\frac{\left(  2m_{b}\hbar\omega_{\mathrm{LO}}\right)
^{1/2}}{P}%
%TCIMACRO{\dint \nolimits_{-1}^{1}}%
%BeginExpansion
{\displaystyle\int\nolimits_{-1}^{1}}
%EndExpansion
dxx%
%TCIMACRO{\dint \nolimits_{0}^{\infty}}%
%BeginExpansion
{\displaystyle\int\nolimits_{0}^{\infty}}
%EndExpansion
d\kappa\kappa\left/  \left[  (\kappa-qx)^{2}+(1-q^{2}x^{2})\right]
^{2}\right. \\
&  =\frac{\alpha}{\pi}\frac{\left(  2m_{b}\hbar\omega_{\mathrm{LO}}\right)
^{1/2}}{P}%
%TCIMACRO{\dint \nolimits_{-1}^{1}}%
%BeginExpansion
{\displaystyle\int\nolimits_{-1}^{1}}
%EndExpansion
dxx%
%TCIMACRO{\dint \nolimits_{-qx}^{\infty}}%
%BeginExpansion
{\displaystyle\int\nolimits_{-qx}^{\infty}}
%EndExpansion
d\kappa\left(  \kappa+qx\right)  \left/  \left[  \kappa^{2}+(1-q^{2}%
x^{2})\right]  ^{2}\right. \\
&  =\frac{\alpha}{\pi}\frac{\left(  2m_{b}\hbar\omega_{\mathrm{LO}}\right)
^{1/2}}{P}%
%TCIMACRO{\dint \nolimits_{-1}^{1}}%
%BeginExpansion
{\displaystyle\int\nolimits_{-1}^{1}}
%EndExpansion
dxx\left\{
\begin{array}
[c]{c}%
-\frac{1}{2\left[  \kappa^{2}+(1-q^{2}x^{2})\right]  }\\
+qx\left[  \frac{\kappa}{2(1-q^{2}x^{2})\left[  \kappa^{2}+(1-q^{2}%
x^{2})\right]  }+\frac{1}{2(1-q^{2}x^{2})^{3/2}}\arctan\left(  \frac{\kappa
}{\left[  1-q^{2}x^{2}\right]  ^{1/2}}\right)  \right]
\end{array}
\right\}  _{-qx}^{\infty}\\
&  =\frac{\alpha}{\pi}\frac{\left(  2m_{b}\hbar\omega_{\mathrm{LO}}\right)
^{1/2}}{P}%
%TCIMACRO{\dint \nolimits_{-1}^{1}}%
%BeginExpansion
{\displaystyle\int\nolimits_{-1}^{1}}
%EndExpansion
dxx\left\{  \frac{1}{2}+\frac{qx\pi}{4(1-q^{2}x^{2})^{3/2}}+\frac{q^{2}x^{2}%
}{2(1-q^{2}x^{2})}+\frac{qx}{2(1-q^{2}x^{2})^{3/2}}\arcsin\left(  qx\right)
\right\}  .
\end{align*}%
\[
\Downarrow
\]%
\begin{align*}
\eta &  =\frac{\alpha}{\pi}\frac{\left(  2m_{b}\hbar\omega_{\mathrm{LO}%
}\right)  ^{1/2}}{P}\frac{q\pi}{4}%
%TCIMACRO{\dint \nolimits_{-1}^{1}}%
%BeginExpansion
{\displaystyle\int\nolimits_{-1}^{1}}
%EndExpansion
\frac{x^{2}}{(1-q^{2}x^{2})^{3/2}}dx\\
&  =\frac{\alpha}{4}\left(  1-\eta\right)
%TCIMACRO{\dint \nolimits_{-1}^{1}}%
%BeginExpansion
{\displaystyle\int\nolimits_{-1}^{1}}
%EndExpansion
\frac{x^{2}}{(1-q^{2}x^{2})^{3/2}}dx\\
&  =\frac{\alpha}{2}\left(  1-\eta\right)  \frac{q-\sqrt{1-q^{2}}%
\arcsin\left(  q\right)  }{q^{3}\sqrt{1-q^{2}}}\\
&  =\frac{\alpha\left(  1-\eta\right)  }{2\left(  \frac{P}{\left(  2m_{b}%
\hbar\omega_{\mathrm{LO}}\right)  ^{1/2}}(1-\eta)\right)  ^{3}}\left(
\frac{q}{\sqrt{1-q^{2}}}-\arcsin\left(  q\right)  \right) \\
&  =\frac{\alpha}{2\left(  1-\eta\right)  ^{2}}\left(  \frac{2m_{b}\hbar
\omega_{\mathrm{LO}}}{P^{2}}\right)  ^{3/2}\left(  \frac{q}{\sqrt{1-q^{2}}%
}-\arcsin\left(  q\right)  \right)
\end{align*}
So, we have arrived at the equation%
\begin{equation}
\eta\left(  1-\eta\right)  ^{2}=\frac{\alpha}{2}\left(  \frac{2m_{b}%
\hbar\omega_{\mathrm{LO}}}{P^{2}}\right)  ^{3/2}\left(  \frac{q}{\sqrt
{1-q^{2}}}-\arcsin\left(  q\right)  \right)  , \label{LLP27-r}%
\end{equation}
or equivalently, using the definition (\ref{q}),%
\begin{equation}
\frac{\eta}{1-\eta}=\frac{\alpha}{2q^{3}}\left(  \frac{q}{\sqrt{1-q^{2}}%
}-\arcsin\left(  q\right)  \right)  . \label{LLP27-a}%
\end{equation}

Using Eqs. (\ref{etaP}) and (\ref{fk}), we can simplify the energy
(\ref{LLP8}) as follows:%
\begin{align*}
E  &  =\frac{P{^{2}+}\left(  {\eta\mathbf{P}}\right)  {^{2}}}{2m_{b}}%
-2\sum_{\mathbf{k}}\frac{\left\vert V_{k}\right\vert ^{2}}{\hbar
\omega_{\mathrm{LO}}-\frac{\hbar\mathbf{k\cdot P}}{m_{b}}(1-\eta)+\frac
{\hbar^{2}k^{2}}{2m_{b}}}\\
&  +\sum_{\mathbf{k}}\frac{\left\vert V_{k}\right\vert ^{2}}{\left(
\hbar\omega_{\mathrm{LO}}-\frac{\hbar\mathbf{k\cdot P}}{m_{b}}(1-\eta
)+\frac{\hbar^{2}k^{2}}{2m_{b}}\right)  ^{2}}\left(  \hbar\omega_{\mathrm{LO}%
}-\frac{\hbar\mathbf{k\cdot P}}{m_{b}}+\frac{\hbar^{2}k^{2}}{2m_{b}}\right)
\end{align*}%
\begin{align*}
&  =\frac{\left(  1+\eta^{2}\right)  P{^{2}}}{2m_{b}}-2\sum_{\mathbf{k}}%
\frac{\left\vert V_{k}\right\vert ^{2}}{\hbar\omega_{\mathrm{LO}}-\frac
{\hbar\mathbf{k\cdot P}}{m_{b}}(1-\eta)+\frac{\hbar^{2}k^{2}}{2m_{b}}}\\
&  +\sum_{\mathbf{k}}\frac{\left\vert V_{k}\right\vert ^{2}}{\left(
\hbar\omega_{\mathrm{LO}}-\frac{\hbar\mathbf{k\cdot P}}{m_{b}}(1-\eta
)+\frac{\hbar^{2}k^{2}}{2m_{b}}\right)  ^{2}}\left(  \hbar\omega_{\mathrm{LO}%
}-\frac{\hbar\mathbf{k\cdot P}}{m_{b}}\left(  1-\eta+\eta\right)  +\frac
{\hbar^{2}k^{2}}{2m_{b}}\right)
\end{align*}%
\begin{align*}
&  =\frac{\left(  1+\eta^{2}\right)  P{^{2}}}{2m_{b}}-2\sum_{\mathbf{k}}%
\frac{\left\vert V_{k}\right\vert ^{2}}{\hbar\omega_{\mathrm{LO}}-\frac
{\hbar\mathbf{k\cdot P}}{m_{b}}(1-\eta)+\frac{\hbar^{2}k^{2}}{2m_{b}}}\\
&  +\sum_{\mathbf{k}}\frac{\left\vert V_{k}\right\vert ^{2}}{\left(
\hbar\omega_{\mathrm{LO}}-\frac{\hbar\mathbf{k\cdot P}}{m_{b}}(1-\eta
)+\frac{\hbar^{2}k^{2}}{2m_{b}}\right)  ^{2}}\left(  \hbar\omega_{\mathrm{LO}%
}-\frac{\hbar\mathbf{k\cdot P}}{m_{b}}\left(  1-\eta\right)  +\frac{\hbar
^{2}k^{2}}{2m_{b}}\right) \\
&  -\sum_{\mathbf{k}}\frac{\left\vert V_{k}\right\vert ^{2}}{\left(
\hbar\omega_{\mathrm{LO}}-\frac{\hbar\mathbf{k\cdot P}}{m_{b}}(1-\eta
)+\frac{\hbar^{2}k^{2}}{2m_{b}}\right)  ^{2}}\left(  \frac{\hbar\mathbf{k\cdot
P}}{m_{b}}\eta\right)
\end{align*}%
\begin{align*}
&  =\frac{\left(  1+\eta^{2}\right)  P{^{2}}}{2m_{b}}-2\sum_{\mathbf{k}}%
\frac{\left\vert V_{k}\right\vert ^{2}}{\hbar\omega_{\mathrm{LO}}-\frac
{\hbar\mathbf{k\cdot P}}{m_{b}}(1-\eta)+\frac{\hbar^{2}k^{2}}{2m_{b}}}\\
&  +\sum_{\mathbf{k}}\frac{\left\vert V_{k}\right\vert ^{2}}{\hbar
\omega_{\mathrm{LO}}-\frac{\hbar\mathbf{k\cdot P}}{m_{b}}(1-\eta)+\frac
{\hbar^{2}k^{2}}{2m_{b}}}\\
&  -\left(  \frac{\mathbf{P}}{m_{b}}\eta\right)  \cdot\sum_{\mathbf{k}}%
\frac{\hbar\mathbf{k}\left\vert V_{k}\right\vert ^{2}}{\left(  \hbar
\omega_{\mathrm{LO}}-\frac{\hbar\mathbf{k\cdot P}}{m_{b}}(1-\eta)+\frac
{\hbar^{2}k^{2}}{2m_{b}}\right)  ^{2}}%
\end{align*}%
\[
=\frac{\left(  1+\eta^{2}\right)  P{^{2}}}{2m_{b}}-\sum_{\mathbf{k}}%
\frac{\left\vert V_{k}\right\vert ^{2}}{\hbar\omega_{\mathrm{LO}}-\frac
{\hbar\mathbf{k\cdot P}}{m_{b}}(1-\eta)+\frac{\hbar^{2}k^{2}}{2m_{b}}}-\left(
\frac{\mathbf{P}}{m_{b}}\eta\right)  \cdot\sum_{\mathbf{k}}\hbar
\mathbf{k}\left\vert f_{\mathbf{k}}\right\vert ^{2}%
\]%
\[
=\frac{\left(  1+\eta^{2}\right)  P{^{2}}}{2m_{b}}-\left(  \frac{\mathbf{P}%
}{m_{b}}\eta\right)  \cdot\eta\mathbf{P}-\sum_{\mathbf{k}}\frac{\left\vert
V_{k}\right\vert ^{2}}{\hbar\omega_{\mathrm{LO}}-\frac{\hbar\mathbf{k\cdot P}%
}{m_{b}}(1-\eta)+\frac{\hbar^{2}k^{2}}{2m_{b}}}%
\]%
\[
\Downarrow
\]%
\begin{equation}
E=\frac{P^{2}}{2m_{b}}\left(  1-\eta^{2}\right)  -\sum_{\mathbf{k}}%
\frac{\left\vert V_{k}\right\vert ^{2}}{\hbar\omega_{\mathrm{LO}}-\frac
{\hbar\mathbf{k\cdot P}}{m_{b}}(1-\eta)+\frac{\hbar^{2}k^{2}}{2m_{b}}}.
\label{EN}%
\end{equation}
The sum over $\mathbf{k}$ in Eq. (\ref{EN}) is calculated as follows:%
\begin{align*}
&  \sum_{\mathbf{k}}\frac{\left\vert V_{k}\right\vert ^{2}}{\hbar
\omega_{\mathrm{LO}}-\frac{\hbar\mathbf{k\cdot P}}{m_{b}}(1-\eta)+\frac
{\hbar^{2}k^{2}}{2m_{b}}}\\
&  =\frac{V}{\left(  2\pi\right)  ^{3}}\int d\mathbf{k}\frac{\left(
\frac{\hbar\omega_{\mathrm{LO}}}{k}\left(  \frac{4\pi\alpha}{V}\right)
^{\frac{1}{2}}\left(  \frac{\hbar}{2m_{b}\omega_{\mathrm{LO}}}\right)
^{\frac{1}{4}}\right)  ^{2}}{\hbar\omega_{\mathrm{LO}}-\frac{\hbar
\mathbf{k\cdot P}}{m_{b}}(1-\eta)+\frac{\hbar^{2}k^{2}}{2m_{b}}}\\
&  =\frac{V}{\left(  2\pi\right)  ^{3}}\hbar^{2}\omega_{\mathrm{LO}}%
^{2}\left(  \frac{4\pi\alpha}{V}\right)  \left(  \frac{\hbar}{2m_{b}%
\omega_{\mathrm{LO}}}\right)  ^{\frac{1}{2}}\int d\mathbf{k}\frac{1}%
{k^{2}\left[  \hbar\omega_{\mathrm{LO}}-\frac{\hbar\mathbf{k\cdot P}}{m_{b}%
}(1-\eta)+\frac{\hbar^{2}k^{2}}{2m_{b}}\right]  }\\
&  =\frac{m_{b}\omega_{\mathrm{LO}}^{2}\alpha}{\pi^{2}}\left(  \frac{\hbar
}{2m_{b}\omega_{\mathrm{LO}}}\right)  ^{\frac{1}{2}}\int d\mathbf{k}\frac
{1}{k^{2}\left(  k^{2}-2\frac{\mathbf{k\cdot P}}{\hbar}(1-\eta)+\frac
{2m_{b}\omega_{\mathrm{LO}}}{\hbar}\right)  }.
\end{align*}
For the calculation of this integral, we can use the auxiliary identity%
\begin{equation}
\frac{1}{ab}=\int_{0}^{1}\frac{1}{\left[  ax+b\left(  1-x\right)  \right]
^{2}}. \label{id}%
\end{equation}
Setting%
\begin{align*}
a  &  =k^{2}-2\frac{\mathbf{k\cdot P}}{\hbar}(1-\eta)+\frac{2m_{b}%
\omega_{\mathrm{LO}}}{\hbar},\\
b  &  =k^{2},
\end{align*}
we find%
\begin{align*}
&  \sum_{\mathbf{k}}\frac{\left\vert V_{k}\right\vert ^{2}}{\hbar
\omega_{\mathrm{LO}}-\frac{\hbar\mathbf{k\cdot P}}{m_{b}}(1-\eta)+\frac
{\hbar^{2}k^{2}}{2m_{b}}}\\
&  =\frac{m_{b}\omega_{\mathrm{LO}}^{2}\alpha}{\pi^{2}}\left(  \frac{\hbar
}{2m_{b}\omega_{\mathrm{LO}}}\right)  ^{\frac{1}{2}}\int_{0}^{1}dx\int
d\mathbf{k}\frac{1}{\left[  x\left(  k^{2}-2\frac{\mathbf{k\cdot P}}{\hbar
}(1-\eta)+\frac{2m_{b}\omega_{\mathrm{LO}}}{\hbar}\right)  +\left(
1-x\right)  k^{2}\right]  ^{2}}\\
&  =\frac{m_{b}\omega_{\mathrm{LO}}^{2}\alpha}{\pi^{2}}\left(  \frac{\hbar
}{2m_{b}\omega_{\mathrm{LO}}}\right)  ^{\frac{1}{2}}\int_{0}^{1}dx\int
d\mathbf{k}\frac{1}{\left(  k^{2}-2\frac{\mathbf{k\cdot P}}{\hbar}%
(1-\eta)x+\frac{2m_{b}\omega_{\mathrm{LO}}}{\hbar}x\right)  ^{2}}\\
&  =\frac{m_{b}\omega_{\mathrm{LO}}^{2}\alpha}{\pi^{2}}\left(  \frac{\hbar
}{2m_{b}\omega_{\mathrm{LO}}}\right)  ^{\frac{1}{2}}\int_{0}^{1}dx\int
d\mathbf{k}\frac{1}{\left(  \left(  \mathbf{k-}\frac{\mathbf{P}}{\hbar}%
(1-\eta)x\right)  ^{2}+\frac{2m_{b}\omega_{\mathrm{LO}}}{\hbar}x-\frac{P^{2}%
}{\hbar^{2}}(1-\eta)^{2}x^{2}\right)  ^{2}}\\
&  =\frac{m_{b}\omega_{\mathrm{LO}}^{2}\alpha}{\pi^{2}}\left(  \frac{\hbar
}{2m_{b}\omega_{\mathrm{LO}}}\right)  ^{\frac{1}{2}}\int_{0}^{1}dx\int
d\mathbf{k}\frac{1}{\left(  k^{2}+\frac{2m_{b}\omega_{\mathrm{LO}}}{\hbar
}x-\frac{P^{2}}{\hbar^{2}}(1-\eta)^{2}x^{2}\right)  ^{2}}%
\end{align*}
As long as $P^{2}/\left(  2m_{b}\right)  $ is sufficiently small so that no
spontaneous emission can occur (roughly $P^{2}/\left(  2m_{b}\right)
\lesssim\hbar\omega_{\mathrm{LO}}$), the quantity
\[
A\equiv\frac{2m_{b}\omega_{\mathrm{LO}}}{\hbar}x-\frac{P^{2}}{\hbar^{2}%
}(1-\eta)^{2}x^{2}%
\]
is supposed to be positive for $0<x<1$. Therefore, we can use the integral%
\[
\int\frac{1}{\left(  k^{2}+A\right)  ^{2}}d\mathbf{k=}\frac{\pi^{2}}{\sqrt{A}%
},
\]
what gives%
\begin{align*}
&  \sum_{\mathbf{k}}\frac{\left\vert V_{k}\right\vert ^{2}}{\hbar
\omega_{\mathrm{LO}}-\frac{\hbar\mathbf{k\cdot P}}{m_{b}}(1-\eta)+\frac
{\hbar^{2}k^{2}}{2m_{b}}}\\
&  =\frac{m_{b}\omega_{\mathrm{LO}}^{2}\alpha}{\pi^{2}}\left(  \frac{\hbar
}{2m_{b}\omega_{\mathrm{LO}}}\right)  ^{\frac{1}{2}}\int_{0}^{1}dx\frac
{\pi^{2}}{\sqrt{\frac{2m_{b}\omega_{\mathrm{LO}}}{\hbar}x-\frac{P^{2}}%
{\hbar^{2}}(1-\eta)^{2}x^{2}}}\\
&  =\frac{1}{2}\alpha\hbar\omega_{\mathrm{LO}}\int_{0}^{1}dx\frac{1}%
{\sqrt{x-\frac{(1-\eta)^{2}P^{2}}{2m_{b}\hbar\omega_{\mathrm{LO}}}x^{2}}}\\
&  =\frac{1}{2}\alpha\hbar\omega_{\mathrm{LO}}\int_{0}^{1}dx\frac{1}%
{\sqrt{x-q^{2}x^{2}}}.
\end{align*}
We change the variable $x=t^{2},$ what gives%
\[
\int_{0}^{1}\frac{1}{\sqrt{x-q^{2}x^{2}}}dx=2\int_{0}^{1}\frac{1}%
{\sqrt{1-q^{2}t^{2}}}dt=\frac{2}{q}\arcsin q,
\]
and hence%
\begin{equation}
\sum_{\mathbf{k}}\frac{\left\vert V_{k}\right\vert ^{2}}{\hbar\omega
_{\mathrm{LO}}-\frac{\hbar\mathbf{k\cdot P}}{m_{b}}(1-\eta)+\frac{\hbar
^{2}k^{2}}{2m_{b}}}=\frac{\alpha\hbar\omega_{\mathrm{LO}}}{q}\arcsin q.
\end{equation}
As a result, the energy (\ref{EN}) is expressed in a closed form%
\begin{equation}
E=\frac{P^{2}}{2m_{b}}\left(  1-\eta^{2}\right)  -\frac{\alpha\hbar
\omega_{\mathrm{LO}}}{q}\arcsin q. \label{LLP29}%
\end{equation}

Further, we expand the r.h.s. of Eq. (\ref{LLP27-a}) to the second order in
powers of $P$ (or, what is the same, in powers of $q$) using the relation%
\begin{equation}
\frac{q}{\sqrt{1-q^{2}}}-\arcsin\left(  q\right)  =\frac{1}{3}q^{3}+O\left(
q^{5}\right)  \label{expand}%
\end{equation}
what results in%
\[
\frac{\eta}{1-\eta}=\frac{\alpha}{2q^{3}}\left[  \frac{1}{3}q^{3}+O\left(
q^{5}\right)  \right]  =\frac{\alpha}{6}+O\left(  q^{2}\right)  .
\]
Solving this equation for $\eta$, we find%
\begin{equation}
\eta=\frac{\alpha/6}{1+\alpha/6}+O\left(  \frac{P^{2}}{2m_{b}\hbar
\omega_{\mathrm{LO}}}\right)  . \label{eta-res}%
\end{equation}
We also apply the expansion in powers of $q$ up to $\sim q^{2}$ to the energy
(\ref{LLP29}):%
\begin{align*}
E  &  =\frac{P^{2}}{2m_{b}}\left(  1-\eta^{2}\right)  -\frac{\alpha\hbar
\omega_{\mathrm{LO}}}{q}\left(  q+\frac{1}{6}q^{3}+O\left(  q^{5}\right)
\right) \\
&  =\frac{P^{2}}{2m_{b}}\left(  1-\eta^{2}\right)  -\alpha\hbar\omega
_{\mathrm{LO}}-\frac{1}{6}\alpha\hbar\omega_{\mathrm{LO}}q^{2}+\hbar
\omega_{\mathrm{LO}}O\left(  q^{4}\right) \\
&  =-\alpha\hbar\omega_{\mathrm{LO}}+\frac{P^{2}}{2m_{b}}\left(  1-\eta
^{2}\right)  -\frac{\alpha P^{2}\left(  1-\eta\right)  ^{2}}{12m_{b}}%
+\hbar\omega_{\mathrm{LO}}O\left(  q^{4}\right) \\
&  =-\alpha\hbar\omega_{\mathrm{LO}}+\frac{P^{2}}{2m_{b}}\left(  1-\eta
^{2}\right)  -\frac{\alpha P^{2}\left(  1-\eta\right)  ^{2}}{12m_{b}}%
+\hbar\omega_{\mathrm{LO}}O\left(  q^{4}\right) \\
&  =-\alpha\hbar\omega_{\mathrm{LO}}+\frac{P^{2}}{12m_{b}}\left(
1-\eta\right)  \left(  \left(  6+\alpha\right)  \eta-\alpha+6\right)
+\hbar\omega_{\mathrm{LO}}O\left(  q^{4}\right) \\
&  =-\alpha\hbar\omega_{\mathrm{LO}}+\frac{P^{2}}{12m_{b}}\left(
1-\frac{\alpha/6}{1+\alpha/6}\right)  \left(  \left(  6+\alpha\right)
\frac{\alpha/6}{1+\alpha/6}-\alpha+6\right)  +\hbar\omega_{\mathrm{LO}%
}O\left(  q^{4}\right) \\
&  =-\alpha\hbar\omega_{\mathrm{LO}}+\frac{P^{2}}{2m_{b}\left(  1+\alpha
/6\right)  }+\hbar\omega_{\mathrm{LO}}O\left(  q^{4}\right)  .
\end{align*}
Finally, we have arrived at the LLP polaron energy%
\begin{equation}
E=-\alpha\hbar\omega_{\mathrm{LO}}+\frac{P^{2}}{2m_{b}\left(  1+\alpha
/6\right)  }+\hbar\omega_{\mathrm{LO}}O\left(  \left(  \frac{P^{2}}%
{2m_{b}\hbar\omega_{\mathrm{LO}}}\right)  ^{2}\right)  . \label{LLP31}%
\end{equation}

\newpage

\subsection*{Appendix 2. Expansion in Stieltjes continuous fractions
\cite{DEK1975}}

In this derivation it is shown that the approximation used in the evaluation
of the function, which determines the polaron mass [see Eqs. (40) and (B1)
from Ref. \cite{DEK1975}]%

\begin{equation}
g(k,z)=\int_{-\infty}^{0}d\tau e^{iz\tau}\exp\left[  -k^{2}C(0)\right]
\exp\left[  k^{2}C(\tau)\right]  \label{B1}%
\end{equation}
with%

\begin{equation}
C(\tau)=\frac{1}{3}%
%TCIMACRO{\dsum \limits_{k^{\prime}}}%
%BeginExpansion
{\displaystyle\sum\limits_{k^{\prime}}}
%EndExpansion
\frac{k^{\prime2}}{m^{2}}\left\vert f_{k^{\prime}}\right\vert ^{2}%
\frac{e^{i\gamma_{k^{\prime}}}\tau}{\gamma_{k^{\prime}}^{2}} \label{B2}%
\end{equation}
is equivalent to an expansion in a continued fraction limited to the first
step. Moreover, it is proved that the choice of the coefficients of the
continued fraction can be justified by a variational principle, at least when
$z$ is real and positive.

Expanding the last exponential of Eq. (\ref{B1}) in a power series leads to%

\begin{align}
g(k,z)  &  =\exp\left[  -k^{2}C(0)\right]  \int_{-\infty}^{0}d\tau e^{iz\tau}%
%TCIMACRO{\dsum \limits_{n=0}^{\infty}}%
%BeginExpansion
{\displaystyle\sum\limits_{n=0}^{\infty}}
%EndExpansion
\frac{1}{n!}\left(  \frac{1}{3m^{2}}\right)  ^{n}\nonumber\\
&  \times%
%TCIMACRO{\dsum \limits_{\vec{k}_{1},...,\vec{k}_{n}}}%
%BeginExpansion
{\displaystyle\sum\limits_{\vec{k}_{1},...,\vec{k}_{n}}}
%EndExpansion
\frac{k_{1}^{2}k_{2}^{2}...k_{n}^{2}\left\vert f_{k_{1}}\right\vert
^{2}\left\vert f_{k_{2}}\right\vert ^{2}...\left\vert f_{k_{n}}\right\vert
^{2}}{\gamma_{k_{1}}^{2}\gamma_{k_{2}}^{2}...\gamma_{k_{n}}^{2}}\nonumber\\
&  \times\exp\left[  i(\gamma_{k_{1}}+\gamma_{k_{2}}+...\gamma_{k_{n}}%
^{2})\tau\right] \nonumber\\
&  =-i\exp\left[  -k^{2}C(0)\right]
%TCIMACRO{\dsum \limits_{n=0}^{\infty}}%
%BeginExpansion
{\displaystyle\sum\limits_{n=0}^{\infty}}
%EndExpansion
\frac{(3m^{2})^{-n}}{n!}\label{B3}\\
&  \times%
%TCIMACRO{\dsum \limits_{\vec{k}_{1},...,\vec{k}_{n}}}%
%BeginExpansion
{\displaystyle\sum\limits_{\vec{k}_{1},...,\vec{k}_{n}}}
%EndExpansion
\frac{k_{1}^{2}k_{2}^{2}...k_{n}^{2}\left\vert f_{k_{1}}\right\vert
^{2}\left\vert f_{k_{2}}\right\vert ^{2}...\left\vert f_{k_{n}}\right\vert
^{2}}{\gamma_{k_{1}}^{2}\gamma_{k_{2}}^{2}...\gamma_{k_{n}}^{2}}\frac
{1}{\gamma_{k_{1}}+\gamma_{k_{2}}+...+\gamma_{k_{n}}+z}.\nonumber
\end{align}
The multiple sum over the $k$'s is in fact an integral with 3n variables. It
is possible to change the variables in that one of the new variables is%

\begin{equation}
x_{n}=\gamma_{k_{1}}+\gamma_{k_{2}}+...+\gamma_{k_{n}}. \label{B4}%
\end{equation}

Then the multiple sum which appears in the last term of Eq. (\ref{B3}) is of
the following type:%

\begin{equation}
J(z)=%
%TCIMACRO{\dint \limits_{n\omega}^{\infty}}%
%BeginExpansion
{\displaystyle\int\limits_{n\omega}^{\infty}}
%EndExpansion
\frac{L(x_{n})}{x_{n}+z}dx_{n}, \label{B5}%
\end{equation}
where%

\[
L(x_{n})=0
\]
is the result of the integration over the $n-1$ other variables. An expansion
of integrals of the type (\ref{B5}) into Stieltjes continued fractions is
known to give good results when $z$ is real and not located on the cut of
$J(z)$, i.e., when%

\begin{equation}
z>-n\omega. \label{B6}%
\end{equation}

The first nontrivial step in the continued fraction expansion is%

\begin{equation}
J(z)=\frac{a_{0}}{a_{1}+z} \label{B7}%
\end{equation}
with%

\begin{equation}
a_{0}=%
%TCIMACRO{\dint \limits_{n\omega}^{\infty}}%
%BeginExpansion
{\displaystyle\int\limits_{n\omega}^{\infty}}
%EndExpansion
L(x_{n})dx_{n}, \label{B8}%
\end{equation}

\begin{equation}
a_{1}=\frac{%
%TCIMACRO{\dint \limits_{n\omega}^{\infty}}%
%BeginExpansion
{\displaystyle\int\limits_{n\omega}^{\infty}}
%EndExpansion
x_{n}L(x_{n})dx_{n}}{%
%TCIMACRO{\dint \limits_{n\omega}^{\infty}}%
%BeginExpansion
{\displaystyle\int\limits_{n\omega}^{\infty}}
%EndExpansion
L(x_{n})dx_{n}}. \label{B9}%
\end{equation}

A \textit{variational principle} can be established, which gives a rather
strong argument in favour of the approximation (\ref{B7}). Let us introduce a
variational parameter $\bar{x}$ writing%

\begin{equation}
J(z)=%
%TCIMACRO{\dint \limits_{n\omega}^{\infty}}%
%BeginExpansion
{\displaystyle\int\limits_{n\omega}^{\infty}}
%EndExpansion
\frac{L(x_{n})}{(x_{n}-\bar{x})+(z+\bar{x})}dx_{n}. \label{B10}%
\end{equation}

Performing two steps of the division, this relation becomes%

\begin{align}
J(z)  &  =\frac{1}{z+\bar{x}}%
%TCIMACRO{\dint \limits_{n\omega}^{\infty}}%
%BeginExpansion
{\displaystyle\int\limits_{n\omega}^{\infty}}
%EndExpansion
L(x_{n})dx_{n}\nonumber\\
&  -\frac{1}{(z+\bar{x})^{2}}%
%TCIMACRO{\dint \limits_{n\omega}^{\infty}}%
%BeginExpansion
{\displaystyle\int\limits_{n\omega}^{\infty}}
%EndExpansion
(x_{n}-\bar{x})L(x_{n})dx_{n}+K(z,\bar{x}) \label{B11}%
\end{align}
with%

\begin{equation}
K(z,\bar{x})=\frac{1}{(z+\bar{x})^{2}}%
%TCIMACRO{\dint \limits_{n\omega}^{\infty}}%
%BeginExpansion
{\displaystyle\int\limits_{n\omega}^{\infty}}
%EndExpansion
\frac{(x_{n}-\bar{x})^{2}L(x_{n})}{x_{n}+z}dx_{n}. \label{B12}%
\end{equation}

The approximation consists of neglecting the term $K(z,\bar{x})$ in Eq.
(\ref{B11}). As this term is positive [cf. (\ref{B6})], the best approximation
is obtained when it is minimum. Therefore let us use the freedom in the choice
of $\bar{x}$ to minimize the expression (\ref{B12}),%

\begin{align}
\frac{\partial K(z,\bar{x})}{\partial\bar{x}}  &  =-2\frac{K(z,\bar{x}%
)}{z+\bar{x}}\nonumber\\
-2\frac{1}{(z+\bar{x})^{2}}%
%TCIMACRO{\dint \limits_{n\omega}^{\infty}}%
%BeginExpansion
{\displaystyle\int\limits_{n\omega}^{\infty}}
%EndExpansion
\frac{(x_{n}-\bar{x})^{2}L(x_{n})}{x_{n}+z}dx_{n}  &  =0, \label{B13a}%
\end{align}
which gives%

\begin{equation}
-2\frac{1}{(z+\bar{x})^{2}}%
%TCIMACRO{\dint \limits_{n\omega}^{\infty}}%
%BeginExpansion
{\displaystyle\int\limits_{n\omega}^{\infty}}
%EndExpansion
\left(  \frac{x_{n}-\bar{x}}{z+\bar{x}}+1\right)  \frac{x_{n}-\bar{x}}%
{x_{n}+z}L(x_{n})dx_{n}=0 \label{B14}%
\end{equation}
or%

\begin{equation}
-2\frac{1}{(z+\bar{x})^{3}}%
%TCIMACRO{\dint \limits_{n\omega}^{\infty}}%
%BeginExpansion
{\displaystyle\int\limits_{n\omega}^{\infty}}
%EndExpansion
(x_{n}-\bar{x})L(x_{n})dx_{n}=0. \label{B15a}%
\end{equation}
This provides us with the best value of the variational parameter%

\begin{equation}
\bar{x}=\frac{%
%TCIMACRO{\dint \limits_{n\omega}^{\infty}}%
%BeginExpansion
{\displaystyle\int\limits_{n\omega}^{\infty}}
%EndExpansion
x_{n}L(x_{n})dx_{n}}{%
%TCIMACRO{\dint \limits_{n\omega}^{\infty}}%
%BeginExpansion
{\displaystyle\int\limits_{n\omega}^{\infty}}
%EndExpansion
L(x_{n})dx_{n}}, \label{B16}%
\end{equation}
which is $a_{1\text{ }}$[cf. Eq. (\ref{B9})].

With this value of $\bar{x}$ and neglecting $K(z,\bar{x})$, the expression
(\ref{B11}) of the calculated quantity $J(z)$ becomes%

\begin{equation}
J(z)=\frac{1}{z+\bar{x}}%
%TCIMACRO{\dint \limits_{n\omega}^{\infty}}%
%BeginExpansion
{\displaystyle\int\limits_{n\omega}^{\infty}}
%EndExpansion
L(x_{n})dx_{n}=J(z)=\frac{a_{0}}{a_{1}+z}, \label{B17}%
\end{equation}
which is the first step (\ref{B7}) of a Stieltjes continued fraction.

To prove that this value of $\bar{x}$ gives a minimum of $K(z,\bar{x})$, let
us calculate the second derivative%

\begin{align}
\frac{\partial^{2}K(z,\bar{x})}{\partial\bar{x}^{2}}  &  =\frac{6}{(z+\bar
{x})^{3}}%
%TCIMACRO{\dint \limits_{n\omega}^{\infty}}%
%BeginExpansion
{\displaystyle\int\limits_{n\omega}^{\infty}}
%EndExpansion
(x_{n}-\bar{x})L(x_{n})dx_{n}\nonumber\\
&  +\frac{2}{(z+\bar{x})^{3}}%
%TCIMACRO{\dint \limits_{n\omega}^{\infty}}%
%BeginExpansion
{\displaystyle\int\limits_{n\omega}^{\infty}}
%EndExpansion
L(x_{n})dx_{n}. \label{B18}%
\end{align}
Now the parameter $\bar{x}$ is replaced by its expression (\ref{B16}). The
relation (\ref{B18}) becomes%

\begin{equation}
\frac{\partial^{2}K(z,\bar{x})}{\partial\bar{x}^{2}}=\frac{2}{(z+a_{1})^{3}}%
%TCIMACRO{\dint \limits_{n\omega}^{\infty}}%
%BeginExpansion
{\displaystyle\int\limits_{n\omega}^{\infty}}
%EndExpansion
L(x_{n})dx_{n}, \label{B18'}%
\end{equation}
which is positive of $z\geqslant-n\omega$, since it follows from relation
(\ref{B16}) that $a_{1}>n\omega.$

Our approximation is related to that used by Feynman which is based on the
following inequality:%

\begin{equation}
\left\langle e^{-sx}\right\rangle \geqslant e^{-s\left\langle x\right\rangle
}, \label{B19}%
\end{equation}
where the brackets denote the expectation value of the random variable $x$.
For instance,%

\begin{equation}
\left\langle e^{-sx}\right\rangle =\frac{%
%TCIMACRO{\dint \nolimits_{a}^{\infty}}%
%BeginExpansion
{\displaystyle\int\nolimits_{a}^{\infty}}
%EndExpansion
L(x)e^{-sx}dx}{%
%TCIMACRO{\dint \nolimits_{a}^{\infty}}%
%BeginExpansion
{\displaystyle\int\nolimits_{a}^{\infty}}
%EndExpansion
L(x)ds}, \label{B20}%
\end{equation}
where L(x) is the non-normalized probability density of $x$. The Laplace
transform of Eq. (\ref{B19}) gives%

\begin{equation}%
%TCIMACRO{\dint \nolimits_{0}^{\infty}}%
%BeginExpansion
{\displaystyle\int\nolimits_{0}^{\infty}}
%EndExpansion
e^{-sz}\left\langle e^{-sx}\right\rangle ds\geqslant%
%TCIMACRO{\dint \nolimits_{0}^{\infty}}%
%BeginExpansion
{\displaystyle\int\nolimits_{0}^{\infty}}
%EndExpansion
e^{-sz}e^{-s\left\langle x\right\rangle }ds, \label{B21}%
\end{equation}
which after integration becomes%

\begin{equation}%
%TCIMACRO{\dint \nolimits_{a}^{\infty}}%
%BeginExpansion
{\displaystyle\int\nolimits_{a}^{\infty}}
%EndExpansion
\frac{L(x)}{x+z}dx\geqslant\frac{%
%TCIMACRO{\dint \nolimits_{a}^{\infty}}%
%BeginExpansion
{\displaystyle\int\nolimits_{a}^{\infty}}
%EndExpansion
L(x)ds}{\left\langle x\right\rangle +z}=\frac{a_{0}}{a_{1}+z}. \label{B22}%
\end{equation}
The last inequality shows the relation with our procedure.

\newpage

\part{Many polarons}

\section{Optical conductivity of an interacting many-polaron gas}

\subsection{Kubo formula for the optical conductivity of the many-polaron gas}

The derivations in the present section are based on Ref. \cite{TDPRB01}. The
Hamiltonian of a system of $N$ interacting continuum polarons is given by:%

\begin{align}
H_{0}  &  =\sum_{j=1}^{N}\frac{p_{j}^{2}}{2m_{b}}+\sum_{\mathbf{q}}\hbar
\omega_{\text{$\mathrm{LO}$}}b_{\mathbf{q}}^{+}b_{\mathbf{q}}\nonumber\\
&  +\sum_{\mathbf{q}}\sum_{j=1}^{N}\left(  e^{i\mathbf{q}\cdot\mathbf{r}_{j}%
}b_{\mathbf{q}}V_{\mathbf{q}}+e^{-i\mathbf{q}\cdot\mathbf{r}_{j}}%
b_{\mathbf{q}}^{+}V_{\mathbf{q}}^{\ast}\right)  +\frac{e^{2}}{2\varepsilon
_{\infty}}\sum_{j=1}^{N}\sum_{\ell(\neq j)=1}^{N}\frac{1}{|\mathbf{r}%
_{i}-\mathbf{r}_{j}|}, \label{hmpol}%
\end{align}
where $\mathbf{r}_{j},\mathbf{p}_{j}$ represent the position and momentum of
the $N$ constituent electrons (or holes) with band mass $m_{b}$;
$b_{\mathbf{q}}^{+},b_{\mathbf{q}}$ denote the creation and annihilation
operators for longitudinal optical (LO) phonons with wave vector $\mathbf{q}$
and frequency $\omega_{\text{\textrm{LO}}}$; $V_{\mathbf{q}}$ describes the
amplitude of the interaction between the electrons and the phonons; and $e$ is
the elementary electron charge. This Hamiltonian can be subdivided into the
following parts:
\begin{equation}
H=H_{e}+H_{e-e}+H_{ph}+H_{e-ph} \label{H}%
\end{equation}
where%
\begin{equation}
H_{e}=\sum_{j=1}^{N}\frac{p_{j}^{2}}{2m_{b}} \label{He}%
\end{equation}
is the kinetic energy of electrons,%
\begin{equation}
H_{e-e}=\frac{e^{2}}{2\varepsilon_{\infty}}\sum_{j=1}^{N}\sum_{\ell(\neq
j)=1}^{N}\frac{1}{|\mathbf{r}_{i}-\mathbf{r}_{j}|} \label{Hee}%
\end{equation}
is the potential energy of the Coulomb electron-electron interaction,%
\begin{equation}
H_{ph}=\sum_{\mathbf{q}}\hbar\omega_{\text{$\mathrm{LO}$}}b_{\mathbf{q}}%
^{+}b_{\mathbf{q}} \label{Hph}%
\end{equation}
is the Hamlitonian of phonons, and%
\begin{equation}
H_{e-ph}=\sum_{\mathbf{q}}\sum_{j=1}^{N}\left(  e^{i\mathbf{q}\cdot
\mathbf{r}_{j}}b_{\mathbf{q}}V_{\mathbf{q}}+e^{-i\mathbf{q}\cdot\mathbf{r}%
_{j}}b_{\mathbf{q}}^{+}V_{\mathbf{q}}^{\ast}\right)  \label{Heph}%
\end{equation}
is the Hamiltonian of the electron-phonon interaction. Further on, we use the
second quantization formalism for electrons, in which the terms $H_{e}$,
$H_{e-e}$ and $H_{e-ph}$ are%
\begin{align}
H_{e}  &  =\sum_{\mathbf{k},\sigma}\frac{\hbar^{2}k^{2}}{2m_{b}}%
a_{\mathbf{k},\sigma}^{+}a_{\mathbf{k},\sigma},\label{Q2}\\
H_{e-e}  &  =\frac{1}{2}\sum_{\mathbf{q}\neq0}v_{\mathbf{q}}\sum_{%
%TCIMACRO{\QATOP{\QTR{bf}{k},\sigma}{\QTR{bf}{k}^{\prime},\sigma^{\prime}}}%
%BeginExpansion
\genfrac{}{}{0pt}{}{\mathbf{k},\sigma}{\mathbf{k}^{\prime},\sigma^{\prime}}%
%EndExpansion
}a_{\mathbf{k}+\mathbf{q},\sigma}^{+}a_{\mathbf{k}^{\prime}-\mathbf{q}%
,\sigma^{\prime}}^{+}a_{\mathbf{k}^{\prime},\sigma^{\prime}}a_{\mathbf{k}%
,\sigma}=\frac{1}{2}\sum_{\mathbf{q}\neq0}v_{\mathbf{q}}:\rho_{\mathbf{q}}%
\rho_{-\mathbf{q}}:,\label{3}\\
H_{e-ph}  &  =\sum_{\mathbf{q}}\left(  V_{\mathbf{q}}b_{\mathbf{q}}%
\rho_{\mathbf{q}}+V_{\mathbf{q}}^{\ast}b_{\mathbf{q}}^{+}\rho_{-\mathbf{q}%
}\right)  , \label{5}%
\end{align}
where $:...:$ is the symbol of the normal product of operators,%
\begin{equation}
v_{\mathbf{q}}=\frac{4\pi e^{2}}{\varepsilon_{\infty}q^{2}V} \label{vq}%
\end{equation}
is the Fourier component of the Coulomb potential, and%
\begin{equation}
\rho_{\mathbf{q}}=\sum_{j=1}^{N}e^{i\mathbf{q\cdot r}_{j}}=\sum_{\mathbf{k}%
,\sigma}a_{\mathbf{k+q},\sigma}^{+}a_{\mathbf{k},\sigma} \label{rho}%
\end{equation}
is the Fourier component of the electron density.

The ground state energy of the many-polaron Hamiltonian (\ref{hmpol}) has been
studied by L. Lemmens, J. T. Devreese and F. Brosens (LDB) \cite{LDB77}, for
weak and intermediate strength of the electron-phonon coupling. They introduce
a variational wave function:
\begin{equation}
\left\vert \psi_{\text{LDB}}\right\rangle =U\left\vert \phi\right\rangle
\left\vert \psi_{el}^{\left(  0\right)  }\right\rangle , \label{psiLDB}%
\end{equation}
where $\left\vert \psi_{el}^{\left(  0\right)  }\right\rangle $ represents the
ground-state many-body wave function for the electron (or hole) system and
$\left\vert \phi\right\rangle $ is the phonon vacuum, $U$ is a many-body
unitary operator which determines a canonical transformation for a fermion gas
interacting with a boson field:
\begin{equation}
U=\exp\left\{  \sum_{j=1}^{N}\sum_{\mathbf{q}}\left(  f_{\mathbf{q}%
}a_{\mathbf{q}}e^{i\mathbf{q}\cdot\mathbf{r}_{j}}-f_{\mathbf{q}}^{\ast
}a_{\mathbf{q}}^{+}e^{-i\mathbf{q}\cdot\mathbf{r}_{j}}\right)  \right\}  .
\label{U}%
\end{equation}
In Ref. \cite{LDB77}, this canonical transformation was used to establish a
many-fermion theory. The $f_{\mathbf{q}}$ were determined variationally
\cite{LDB77} resulting in
\begin{equation}
f_{\mathbf{q}}=\dfrac{V_{\mathbf{q}}}{\hbar\omega_{\mathrm{LO}}+\dfrac
{\hbar^{2}q^{2}}{2m_{b}S(\mathbf{q})}}, \label{Qfk}%
\end{equation}
for a system with total momentum $\mathbf{P=}\sum_{j}\mathbf{p}_{j}=0$. In
this expression, $S(\mathbf{q})$ represents the static structure factor of the
constituent interacting many electron or hole system :
\begin{equation}
NS(\mathbf{q})=\left\langle \sum_{j=1}^{N}\sum_{j^{\prime}=1}^{N}%
e^{i\mathbf{q}\cdot(\mathbf{r}_{j}-\mathbf{r}_{j^{\prime}})}\right\rangle .
\end{equation}
The angular brackets $\left\langle \text{...}\right\rangle $\ represent the
expectation value with respect to the ground state.

The many-polaron optical conductivity is the response of the current-density,
in the system described by the Hamiltonian (\ref{hmpol}), to an applied
electric field (along the $x$-axis) with frequency $\omega$. This applied
electric field introduces a perturbation term in the Hamiltonian
(\ref{hmpol}), which couples the vector potential of the incident
electromagnetic field to the current-density. Within linear response theory,
the optical conductivity can be expressed through the Kubo formula as a
current-current correlation function:
\begin{equation}
\sigma(\omega)=i\frac{Ne^{2}}{\text{$V$}m_{b}\omega}+\frac{1}{\text{$V$}%
\hbar\omega}\int_{0}^{\infty}e^{i\omega t}\left\langle \left[  J_{x}%
(t),J_{x}(0)\right]  \right\rangle dt.
\end{equation}
In this expression, $V$ is the volume of the system, and $J_{x}$ is the
$x$-component of the current operator $\mathbf{J},$ which is related to the
momentum operators of the charge carriers:
\begin{equation}
\mathbf{J}=\frac{q}{m_{b}}\sum_{j=1}^{N}\mathbf{p}_{j}=\frac{q}{m_{b}%
}\mathbf{P,}%
\end{equation}
with $q$ the charge of the charge carriers ($+e$ for holes, $-e$ for
electrons) and $\mathbf{P}$ the total momentum operator of the charge
carriers. The real part of the optical conductivity at temperature zero, which
is proportional to the optical absorption coefficient, can be written as a
function of the total momentum operator of the charge carriers as follows :
\begin{equation}
\operatorname{Re}\sigma(\omega)=\frac{1}{\text{$V$}\hbar\omega}\frac{e^{2}%
}{m_{b}^{2}}\operatorname{Re}\left\{  \int_{0}^{\infty}e^{i\omega
t}\left\langle \left[  P_{x}(t),P_{x}(0)\right]  \right\rangle dt\right\}  .
\label{K1}%
\end{equation}

\subsection{Force-force correlation function}

Let us integrate over time in (\ref{K1}) twice by parts as follows:
\begin{align*}
&  \int_{0}^{\infty}dt\left\langle \left[  P_{x}\left(  t\right)
,P_{x}\right]  \right\rangle e^{i\omega t-\delta t}\\
&  =\frac{1}{i\omega-\delta}\left\{  \left.  \left\langle \left[  P_{x}\left(
t\right)  ,P_{x}\right]  \right\rangle e^{i\omega t-\delta t}\right\vert
_{t=0}^{\infty}-\int_{0}^{\infty}dt\left\langle \left[  \frac{d}{dt}%
P_{x}\left(  t\right)  ,P_{x}\right]  \right\rangle e^{i\omega t-\delta
t}\right\} \\
&  =-\frac{1}{i\omega-\delta}\int_{0}^{\infty}dt\left\langle \left[  \frac
{d}{dt}\left(  e^{\frac{it}{\hbar}H}P_{x}e^{-\frac{it}{\hbar}H}\right)
,P_{x}\right]  \right\rangle e^{i\omega t-\delta t}\\
&  =-\frac{1}{i\omega-\delta}\int_{0}^{\infty}dt\left\langle \left[  \left(
e^{\frac{it}{\hbar}H}\frac{i}{\hbar}\left[  H,P_{x}\right]  e^{-\frac
{it}{\hbar}H}\right)  ,P_{x}\right]  \right\rangle e^{i\omega t-\delta t}\\
&  =-\frac{1}{i\omega-\delta}\int_{0}^{\infty}dt\left\langle \left[  \frac
{i}{\hbar}\left[  H,P_{x}\right]  ,e^{-\frac{it}{\hbar}H}P_{x}e^{\frac
{it}{\hbar}H}\right]  \right\rangle e^{i\omega t-\delta t}\\
&  =-\frac{1}{i\omega-\delta}\int_{0}^{\infty}dt\left\langle \left[
F_{x}\left(  0\right)  ,e^{-\frac{it}{\hbar}H}P_{x}e^{\frac{it}{\hbar}%
H}\right]  \right\rangle e^{i\omega t-\delta t}\\
&  =-\left(  \frac{1}{i\omega-\delta}\right)  ^{2}\left\{  \left.
\left\langle \left[  F_{x}\left(  0\right)  ,e^{-\frac{it}{\hbar}H}%
P_{x}e^{\frac{it}{\hbar}H}\right]  \right\rangle e^{i\omega t-\delta
t}\right\vert _{t=0}^{\infty}\right. \\
&  \left.  -\int_{0}^{\infty}dt\left\langle \left[  F_{x}\left(  0\right)
,\frac{d}{dt}e^{-\frac{it}{\hbar}H}P_{x}e^{\frac{it}{\hbar}H}\right]
\right\rangle e^{i\omega t-\delta t}\right\}
\end{align*}
\begin{align*}
&  =-\left(  \frac{1}{i\omega-\delta}\right)  ^{2}\left\{  -\left\langle
\left[  F_{x},P_{x}\right]  \right\rangle +\int_{0}^{\infty}dt\left\langle
\left[  F_{x}\left(  0\right)  ,e^{-\frac{it}{\hbar}H}\left(  \frac{i}{\hbar
}\left[  H,P_{x}\right]  \right)  e^{\frac{it}{\hbar}H}\right]  \right\rangle
e^{i\omega t-\delta t}\right\} \\
&  =-\left(  \frac{1}{i\omega-\delta}\right)  ^{2}\left\{  -\left\langle
\left[  F_{x},P_{x}\right]  \right\rangle +\int_{0}^{\infty}dt\left\langle
\left[  F_{x}\left(  0\right)  ,e^{-\frac{it}{\hbar}H}F_{x}\left(  0\right)
e^{\frac{it}{\hbar}H}\right]  \right\rangle e^{i\omega t-\delta t}\right\} \\
&  =-\left(  \frac{1}{i\omega-\delta}\right)  ^{2}\left\{  -\left\langle
\left[  F_{x},P_{x}\right]  \right\rangle +\int_{0}^{\infty}dt\left\langle
\left[  e^{\frac{it}{\hbar}H}F_{x}\left(  0\right)  e^{-\frac{it}{\hbar}%
H},F_{x}\left(  0\right)  \right]  \right\rangle e^{i\omega t-\delta
t}\right\} \\
&  =\frac{1}{\left(  \omega+i\delta\right)  ^{2}}\left\{  -\left\langle
\left[  F_{x},P_{x}\right]  \right\rangle +\int_{0}^{\infty}dt\left\langle
\left[  F_{x}\left(  t\right)  ,F_{x}\left(  0\right)  \right]  \right\rangle
e^{i\omega t-\delta t}\right\}  ,
\end{align*}
where $\mathbf{F}\equiv\frac{i}{\hbar}\left[  H,\mathbf{P}\right]  $ is the
operator of the force applied to the center of mass of the electrons.

Since both $F_{x}$ and $P_{x}$ are hermitian operators, the average
$\left\langle \left[  F_{x},P_{x}\right]  \right\rangle $ is imaginary. Hence,
for $\omega\neq0,$ this term does not give a contribution into
$\operatorname{Re}\sigma\left(  \omega\right)  .$ As a result, integrating by
parts twice, the real part of the optical conductivity of the many-polaron
system is written with a force-force correlation function:
\begin{equation}
\operatorname{Re}\sigma(\omega)=\frac{1}{\text{$V$}\hbar\omega^{3}}\frac
{e^{2}}{m_{b}^{2}}\operatorname{Re}\left\{  \int_{0}^{\infty}e^{i\omega
t}\left\langle \left[  F_{x}(t),F_{x}(0)\right]  \right\rangle dt\right\}  .
\label{FF}%
\end{equation}

The force operator is determined as
\[
F_{x}=\frac{i}{\hbar}\left[  H,P_{x}\right]  =\frac{i}{\hbar}\left[
H_{e}+H_{e-e}+H_{ph}+H_{e-ph},P_{x}\right]  .
\]

Further, we use the commutators:
\begin{gather*}
\left[  a_{\mathbf{k+q},\sigma}^{+}a_{\mathbf{k},\sigma},P_{x}\right]
=\sum_{\mathbf{k}^{\prime}}\hbar k_{x}^{\prime}\left[  a_{\mathbf{k+q},\sigma
}^{+}a_{\mathbf{k},\sigma},a_{\mathbf{k}^{\prime},\sigma}^{+}a_{\mathbf{k}%
^{\prime},\sigma}\right] \\
=\sum_{\mathbf{k}^{\prime}}\hbar k_{x}^{\prime}\left(
\begin{array}
[c]{c}%
a_{\mathbf{k+q},\sigma}^{+}a_{\mathbf{k},\sigma}a_{\mathbf{k}^{\prime},\sigma
}^{+}a_{\mathbf{k}^{\prime},\sigma}+a_{\mathbf{k+q},\sigma}^{+}a_{\mathbf{k}%
^{\prime},\sigma}^{+}a_{\mathbf{k},\sigma}a_{\mathbf{k}^{\prime},\sigma}\\
-a_{\mathbf{k}^{\prime},\sigma}^{+}a_{\mathbf{k+q},\sigma}^{+}a_{\mathbf{k}%
^{\prime},\sigma}a_{\mathbf{k},\sigma}-a_{\mathbf{k}^{\prime},\sigma}%
^{+}a_{\mathbf{k}^{\prime},\sigma}a_{\mathbf{k+q},\sigma}^{+}a_{\mathbf{k}%
,\sigma}%
\end{array}
\right) \\
=\sum_{\mathbf{k}^{\prime}}\hbar k_{x}^{\prime}\left(  \delta_{\mathbf{kk}%
^{\prime}}a_{\mathbf{k+q},\sigma}^{+}a_{\mathbf{k}^{\prime},\sigma}%
-\delta_{\mathbf{k}^{\prime},\mathbf{k+q}}a_{\mathbf{k}^{\prime},\sigma}%
^{+}a_{\mathbf{k},\sigma}\right) \\
=\sum_{\mathbf{k}^{\prime}}\hbar k_{x}^{\prime}\left(  \delta_{\mathbf{kk}%
^{\prime}}a_{\mathbf{k+q},\sigma}^{+}a_{\mathbf{k},\sigma}-\delta
_{\mathbf{k}^{\prime},\mathbf{k+q}}a_{\mathbf{k+q},\sigma}^{+}a_{\mathbf{k}%
,\sigma}\right) \\
=a_{\mathbf{k+q},\sigma}^{+}a_{\mathbf{k},\sigma}\sum_{\mathbf{k}^{\prime}%
}\hbar k_{x}^{\prime}\left(  \delta_{\mathbf{kk}^{\prime}}-\delta
_{\mathbf{k}^{\prime},\mathbf{k+q}}\right)  =-\hbar q_{x}a_{\mathbf{k+q}%
,\sigma}^{+}a_{\mathbf{k},\sigma},
\end{gather*}

\[
\left[  \rho_{\mathbf{q}},P_{x}\right]  =-\hbar q_{x}\rho_{\mathbf{q}}.
\]
Hence, $\left[  H_{e},P_{x}\right]  =0,$ $\left[  H_{e-e},P_{x}\right]  =0,$%
\begin{align*}
\left[  H_{e-ph},P_{x}\right]   &  =\sum_{\mathbf{q}}\left(  V_{\mathbf{q}%
}b_{\mathbf{q}}\left[  \rho_{\mathbf{q}},P_{x}\right]  +V_{\mathbf{q}}^{\ast
}b_{\mathbf{q}}^{+}\left[  \rho_{-\mathbf{q}},P_{x}\right]  \right) \\
&  =-\hbar\sum_{\mathbf{q}}q_{x}\left(  V_{\mathbf{q}}b_{\mathbf{q}}%
\rho_{\mathbf{q}}-V_{\mathbf{q}}^{\ast}b_{\mathbf{q}}^{+}\rho_{-\mathbf{q}%
}\right)  ,
\end{align*}
So, the commutator of the Hamiltonian (\ref{hmpol}) with the total momentum
operator of the charge carriers leads to the expression for the force%
\begin{equation}
\mathbf{F}=-i\sum_{\mathbf{q}}\mathbf{q}\left(  V_{\mathbf{q}}b_{\mathbf{q}%
}\rho_{\mathbf{q}}-V_{\mathbf{q}}^{\ast}b_{\mathbf{q}}^{+}\rho_{-\mathbf{q}%
}\right)  . \label{force}%
\end{equation}
This result for the force operator clarifies the significance of using the
force-force correlation function rather than the momentum-momentum correlation
function. The operator product $F_{x}(t)F_{x}(0)$\ is proportional to
$|V_{\mathbf{k}}|^{2}$, the charge carrier - phonon interaction strength. This
will be a distinct advantage for any expansion of the final result in the
charge carrier - phonon interaction strength, since one power of
$|V_{\mathbf{k}}|^{2}$\ is factored out beforehand. Substituting (\ref{force})
into (\ref{FF}), the real part of the optical conductivity then takes the
form:
\begin{gather}
\operatorname{Re}\sigma\left(  \omega\right)  =\frac{1}{V\hbar\omega^{3}}%
\frac{e^{2}}{m_{b}^{2}}\operatorname{Re}\int_{0}^{\infty}dte^{i\omega t-\delta
t}\sum_{\mathbf{q,q}^{\prime}}q_{x}q_{x}^{\prime}\nonumber\\
\times\left\langle \left[  \left[  V_{\mathbf{q}}b_{\mathbf{q}}\left(
t\right)  +V_{-\mathbf{q}}^{\ast}b_{-\mathbf{q}}^{+}\left(  t\right)  \right]
\rho_{\mathbf{q}}\left(  t\right)  ,\left(  V_{-\mathbf{q}^{\prime}%
}b_{-\mathbf{q}^{\prime}}+V_{\mathbf{q}^{\prime}}^{\ast}b_{\mathbf{q}^{\prime
}}^{+}\right)  \rho_{-\mathbf{q}^{\prime}}\right]  \right\rangle . \label{FF2}%
\end{gather}
Up to this point, no approximations other than \textit{linear response theory}
have been made.

\subsection{Canonical transformation}

The expectation value appearing in the right hand side of expression
(\ref{FF2}) for the real part of the optical conductivity is calculated now
with respect to the LDB many-polaron wave function (\ref{psiLDB}). The unitary
operator (\ref{U}) can be written as
\begin{equation}
U=\exp\sum_{\mathbf{q}}A_{\mathbf{q}}\rho_{\mathbf{q}},\quad A_{\mathbf{q}%
}=f_{\mathbf{q}}b_{\mathbf{q}}-f_{-\mathbf{q}}^{\ast}b_{-\mathbf{q}}^{+},
\label{unit}%
\end{equation}
The transformed Hamiltonian (\ref{H}) is denoted as%
\begin{equation}
\tilde{H}=U^{-1}HU. \label{Unit1}%
\end{equation}
The momentum operator of an electron $\mathbf{p}_{j},$ the operator of the
total momentum of electrons $\mathbf{P}$ and the phonon creation and
annihilation operators are transformed by the unitary transformation
(\ref{unit}) as follows:
\begin{align}
U^{-1}\mathbf{p}_{j}U  &  =\mathbf{p}_{j}+\sum_{\mathbf{q}}\hbar
\mathbf{q}A_{\mathbf{q}}e^{i\mathbf{q\cdot r}_{j}},\label{P1}\\
U^{-1}\mathbf{P}U  &  =\mathbf{P}+\sum_{\mathbf{q}}\hbar\mathbf{q}%
A_{\mathbf{q}}\rho_{\mathbf{q}},\\
U^{-1}b_{\mathbf{q}}U  &  =b_{\mathbf{q}}-f_{\mathbf{q}}^{\ast}\rho
_{-\mathbf{q}},\quad U^{-1}b_{\mathbf{q}}^{+}U=b_{\mathbf{q}}^{+}%
-f_{\mathbf{q}}\rho_{\mathbf{q}}.
\end{align}
As a result, after the transformation (\ref{unit}), the Hamiltonian takes the
form (see Ref. \cite{LDB77}):
\begin{equation}
\tilde{H}=H_{e}+\tilde{H}_{e-e}+H_{ph}+\tilde{H}_{e-ph}+H_{N}+H_{ppe},
\label{H1}%
\end{equation}
where the terms are%
\begin{equation}
\tilde{H}_{e-e}=\frac{1}{2}\sum_{\mathbf{q}\neq0}\tilde{v}_{\mathbf{q}}%
:\rho_{\mathbf{q}}\rho_{-\mathbf{q}}:,\quad\tilde{v}_{\mathbf{q}%
}=v_{\mathbf{q}}+2\left(  \hbar\omega_{\mathrm{LO}}\left\vert f_{\mathbf{q}%
}\right\vert ^{2}-V_{\mathbf{q}}f_{\mathbf{q}}^{\ast}-V_{\mathbf{q}}^{\ast
}f_{\mathbf{q}}\right)  , \label{H1a}%
\end{equation}%
\begin{align}
\tilde{H}_{e-ph}  &  =\sum_{\mathbf{q}}\left[  \left(  V_{\mathbf{q}}%
-\hbar\omega_{\mathrm{LO}}f_{\mathbf{q}}\right)  b_{\mathbf{q}}\rho
_{\mathbf{q}}+\left(  V_{\mathbf{q}}^{\ast}-\hbar\omega_{\mathrm{LO}%
}f_{\mathbf{q}}^{\ast}\right)  b_{\mathbf{q}}^{+}\rho_{-\mathbf{q}}\right]
\nonumber\\
&  +\frac{\hbar^{2}}{2m_{b}}\sum_{\mathbf{q}}A_{\mathbf{q}}\sum_{\mathbf{k}%
,\sigma}\left(  \mathbf{q}^{2}+2\mathbf{k\cdot q}\right)  a_{\mathbf{k+q}%
,\sigma}^{+}a_{\mathbf{k},\sigma}, \label{H1b}%
\end{align}%
\begin{equation}
H_{N}=\hat{N}\sum_{\mathbf{q}}\left(  \hbar\omega_{\mathrm{LO}}\left\vert
f_{\mathbf{q}}\right\vert ^{2}-V_{\mathbf{q}}f_{\mathbf{q}}^{\ast
}-V_{\mathbf{q}}^{\ast}f_{\mathbf{q}}\right)  ,\quad\left(  \hat{N}\equiv
\sum_{\mathbf{k},\sigma}a_{\mathbf{k},\sigma}^{+}a_{\mathbf{k},\sigma}\right)
, \label{HN}%
\end{equation}%
\begin{equation}
H_{ppe}=\frac{\hbar^{2}}{2m_{b}}\sum_{\mathbf{qq}^{\prime}}\mathbf{q\cdot
q}^{\prime}A_{\mathbf{q}}A_{\mathbf{q}^{\prime}}\rho_{\mathbf{q}%
+\mathbf{q}^{\prime}}. \label{Hppe}%
\end{equation}
The exact expression for the real part of the conductivity (\ref{FF2}) after
the replacement of $\left\vert \Psi_{0}\right\rangle $ by $\left\vert
\Psi_{LDB}\right\rangle =U\left\vert \phi\right\rangle \left\vert \psi
_{el}^{\left(  0\right)  }\right\rangle $ is transformed to the approximate
one
\begin{align*}
&  \operatorname{Re}\sigma\left(  \omega\right) \\
&  =\frac{1}{V\hbar\omega^{3}}\frac{e^{2}}{m_{b}^{2}}\operatorname{Re}\int
_{0}^{\infty}dte^{i\omega t-\delta t}\sum_{\mathbf{q,q}^{\prime}}q_{x}%
q_{x}^{\prime}\\
&  \times\left\langle \psi_{el}^{\left(  0\right)  }\left\vert \left\langle
\phi\left\vert U^{-1}\left[
\begin{array}
[c]{c}%
e^{\frac{it}{\hbar}H}\left[  V_{\mathbf{q}}b_{\mathbf{q}}+V_{-\mathbf{q}%
}^{\ast}b_{-\mathbf{q}}^{+}\right]  \rho_{\mathbf{q}}e^{-\frac{it}{\hbar}H},\\
\left(  V_{-\mathbf{q}^{\prime}}b_{-\mathbf{q}^{\prime}}+V_{\mathbf{q}%
^{\prime}}^{\ast}b_{\mathbf{q}^{\prime}}^{+}\right)  \rho_{-\mathbf{q}%
^{\prime}}%
\end{array}
\right]  U\right\vert \phi\right\rangle \right\vert \psi_{el}^{\left(
0\right)  }\right\rangle
\end{align*}%
\begin{align*}
&  =\frac{1}{V\hbar\omega^{3}}\frac{e^{2}}{m_{b}^{2}}\operatorname{Re}\int
_{0}^{\infty}dte^{i\omega t-\delta t}\sum_{\mathbf{q,q}^{\prime}}q_{x}%
q_{x}^{\prime}\\
&  \times\left\langle \psi_{el}^{\left(  0\right)  }\left\vert \left\langle
\phi\left\vert \left[
\begin{array}
[c]{c}%
e^{\frac{it}{\hbar}\tilde{H}}U^{-1}\left[  V_{\mathbf{q}}b_{\mathbf{q}%
}+V_{-\mathbf{q}}^{\ast}b_{-\mathbf{q}}^{+}\right]  U\rho_{\mathbf{q}%
}e^{-\frac{it}{\hbar}\tilde{H}},\\
U^{-1}\left(  V_{-\mathbf{q}^{\prime}}b_{-\mathbf{q}^{\prime}}+V_{\mathbf{q}%
^{\prime}}^{\ast}b_{\mathbf{q}^{\prime}}^{+}\right)  U\rho_{-\mathbf{q}%
^{\prime}}%
\end{array}
\right]  \right\vert \phi\right\rangle \right\vert \psi_{el}^{\left(
0\right)  }\right\rangle
\end{align*}%
\begin{align*}
&  =\frac{1}{V\hbar\omega^{3}}\frac{e^{2}}{m_{b}^{2}}\operatorname{Re}\int
_{0}^{\infty}dte^{i\omega t-\delta t}\sum_{\mathbf{q,q}^{\prime}}q_{x}%
q_{x}^{\prime}\\
&  \times\left\langle \psi_{el}^{\left(  0\right)  }\left\vert \left\langle
\phi\left\vert \left[  e^{\frac{it}{\hbar}\tilde{H}}\left(  V_{\mathbf{q}%
}\left(  b_{\mathbf{q}}-f_{\mathbf{q}}^{\ast}\rho_{-\mathbf{q}}\right)
+V_{-\mathbf{q}}^{\ast}\left(  b_{-\mathbf{q}}^{+}-f_{-\mathbf{q}}%
\rho_{-\mathbf{q}}\right)  \right)  \rho_{\mathbf{q}}e^{-\frac{it}{\hbar
}\tilde{H}},\right.  \right.  \right.  \right.  \right. \\
&  \left.  \left.  \left.  \left.  \left.  \left(  V_{-\mathbf{q}^{\prime}%
}\left(  b_{-\mathbf{q}^{\prime}}-f_{-\mathbf{q}^{\prime}}^{\ast}%
\rho_{\mathbf{q}^{\prime}}\right)  +V_{\mathbf{q}^{\prime}}^{\ast}\left(
b_{\mathbf{q}^{\prime}}^{+}-f_{\mathbf{q}^{\prime}}\rho_{\mathbf{q}^{\prime}%
}\right)  \right)  \rho_{-\mathbf{q}^{\prime}}\right]  \right\vert
\phi\right\rangle \right\vert \psi_{el}^{\left(  0\right)  }\right\rangle .
\end{align*}
So, we have arrived at the expression%
\begin{gather*}
\operatorname{Re}\sigma\left(  \omega\right)  =\frac{1}{V\hbar\omega^{3}}%
\frac{e^{2}}{m_{b}^{2}}\operatorname{Re}\int_{0}^{\infty}dte^{i\omega t-\delta
t}\sum_{\mathbf{q,q}^{\prime}}q_{x}q_{x}^{\prime}\\
\times\left\langle \psi_{el}^{\left(  0\right)  }\left\vert \left\langle
\phi\left\vert \left[  e^{\frac{it}{\hbar}\tilde{H}}\left(  V_{\mathbf{q}%
}\left(  b_{\mathbf{q}}-f_{\mathbf{q}}^{\ast}\rho_{-\mathbf{q}}\right)
+V_{-\mathbf{q}}^{\ast}\left(  b_{-\mathbf{q}}^{+}-f_{-\mathbf{q}}%
\rho_{-\mathbf{q}}\right)  \right)  \rho_{\mathbf{q}}e^{-\frac{it}{\hbar
}\tilde{H}},\right.  \right.  \right.  \right.  \right. \\
\left.  \left.  \left.  \left.  \left.  \left(  V_{-\mathbf{q}^{\prime}%
}\left(  b_{-\mathbf{q}^{\prime}}-f_{-\mathbf{q}^{\prime}}^{\ast}%
\rho_{\mathbf{q}^{\prime}}\right)  +V_{\mathbf{q}^{\prime}}^{\ast}\left(
b_{\mathbf{q}^{\prime}}^{+}-f_{\mathbf{q}^{\prime}}\rho_{\mathbf{q}^{\prime}%
}\right)  \right)  \rho_{-\mathbf{q}^{\prime}}\right]  \right\vert
\phi\right\rangle \right\vert \psi_{el}^{\left(  0\right)  }\right\rangle .
\end{gather*}
Since $\rho_{\mathbf{q}}\rho_{-\mathbf{q}}=\rho_{-\mathbf{q}}\rho_{\mathbf{q}%
},$ and $V_{\mathbf{q}}f_{\mathbf{q}}^{\ast}=V_{-\mathbf{q}}f_{-\mathbf{q}%
}^{\ast},$ the terms proportional to $\rho_{-\mathbf{q}}\rho_{\mathbf{q}}$
vanish after the summation over $\mathbf{q}$:
\begin{equation}
\sum_{\mathbf{q}}q_{x}V_{\mathbf{q}}f_{\mathbf{q}}^{\ast}\rho_{-\mathbf{q}%
}\rho_{\mathbf{q}}\overset{\mathbf{q}\rightarrow-\mathbf{q}}{=}-\sum
_{\mathbf{q}}q_{x}V_{\mathbf{q}}f_{\mathbf{q}}^{\ast}\rho_{-\mathbf{q}}%
\rho_{\mathbf{q}}=0.
\end{equation}
Hence we obtain the real part of the optical conductivity in the form%
\begin{align}
\operatorname{Re}\sigma\left(  \omega\right)   &  =\frac{1}{V\hbar\omega^{3}%
}\frac{e^{2}}{m_{b}^{2}}\operatorname{Re}\int_{0}^{\infty}dte^{i\omega
t-\delta t}\sum_{\mathbf{q,q}^{\prime}}q_{x}q_{x}^{\prime}\nonumber\\
&  \times\left\langle \psi_{el}^{\left(  0\right)  }\left\vert \left\langle
\phi\left\vert \left[  e^{\frac{it}{\hbar}\tilde{H}}\left(  V_{\mathbf{q}%
}b_{\mathbf{q}}+V_{-\mathbf{q}}^{\ast}b_{-\mathbf{q}}^{+}\right)
\rho_{\mathbf{q}}e^{-\frac{it}{\hbar}\tilde{H}},\right.  \right.  \right.
\right.  \right. \nonumber\\
&  \left.  \left.  \left.  \left.  \left.  \left(  V_{-\mathbf{q}^{\prime}%
}b_{-\mathbf{q}^{\prime}}+V_{\mathbf{q}^{\prime}}^{\ast}b_{\mathbf{q}^{\prime
}}^{+}\right)  \rho_{-\mathbf{q}^{\prime}}\right]  \right\vert \phi
\right\rangle \right\vert \psi_{el}^{\left(  0\right)  }\right\rangle .
\label{ReS}%
\end{align}
Introducing the factor%
\begin{align}
\mathcal{J}(\mathbf{q},\mathbf{q}^{\prime})  &  =\left\langle \psi
_{el}^{\left(  0\right)  }\left\vert \left\langle \phi\left\vert \left[
e^{\frac{it}{\hbar}\tilde{H}}\left(  V_{\mathbf{q}}b_{\mathbf{q}%
}+V_{-\mathbf{q}}^{\ast}b_{-\mathbf{q}}^{+}\right)  \rho_{\mathbf{q}}%
e^{-\frac{it}{\hbar}\tilde{H}},\right.  \right.  \right.  \right.  \right.
\nonumber\\
&  \left.  \left.  \left.  \left.  \left.  \left(  V_{-\mathbf{q}^{\prime}%
}b_{-\mathbf{q}^{\prime}}+V_{\mathbf{q}^{\prime}}^{\ast}b_{\mathbf{q}^{\prime
}}^{+}\right)  \rho_{-\mathbf{q}^{\prime}}\right]  \right\vert \phi
\right\rangle \right\vert \psi_{el}^{\left(  0\right)  }\right\rangle ,
\end{align}
the optical conductivity can be written as%
\begin{equation}
\operatorname{Re}\sigma\left(  \omega\right)  =\frac{1}{V\hbar\omega^{3}}%
\frac{e^{2}}{m_{b}^{2}}\operatorname{Re}\int_{0}^{\infty}dte^{i\omega t-\delta
t}\sum_{\mathbf{q,q}^{\prime}}q_{x}q_{x}^{\prime}\mathcal{J}(\mathbf{q}%
,\mathbf{q}^{\prime}). \label{ReS1}%
\end{equation}

In the case of a weak electron-phonon coupling, we can neglect in the exponent
$e^{-\frac{it}{\hbar}\tilde{H}}$ of (\ref{ReS}) the terms $\tilde{H}_{e-ph}$
and $H_{ppe}$ [i. e., the renormalized Hamiltonian of the electron-phonon
interaction (\ref{H1a}) and (\ref{Hppe})]. Namely, we replace $\tilde{H}$ in
Eq. (\ref{ReS}) by the Hamiltonian
\begin{equation}
\tilde{H}_{0}=H_{e}+\tilde{H}_{e-e}+H_{ph}+H_{N}.
\end{equation}
In this case, we find%
\begin{align*}
\mathcal{J}(\mathbf{q},\mathbf{q}^{\prime})  &  =\left\langle \psi
_{el}^{\left(  0\right)  }\left\vert \left\langle \phi\left\vert \left[
e^{\frac{it}{\hbar}\tilde{H}_{0}}\left(  V_{\mathbf{q}}b_{\mathbf{q}%
}+V_{-\mathbf{q}}^{\ast}b_{-\mathbf{q}}^{+}\right)  \rho_{\mathbf{q}}%
e^{-\frac{it}{\hbar}\tilde{H}_{0}},\right.  \right.  \right.  \right.  \right.
\\
&  \left.  \left.  \left.  \left.  \left.  \left(  V_{-\mathbf{q}^{\prime}%
}b_{-\mathbf{q}^{\prime}}+V_{\mathbf{q}^{\prime}}^{\ast}b_{\mathbf{q}^{\prime
}}^{+}\right)  \rho_{-\mathbf{q}^{\prime}}\right]  \right\vert \phi
\right\rangle \right\vert \psi_{el}^{\left(  0\right)  }\right\rangle
\end{align*}%
\begin{align*}
&  =|V_{\mathbf{q}}|^{2}\delta_{\mathbf{qq}^{\prime}}\left\langle \psi
_{el}^{\left(  0\right)  }\left\vert \left\langle \phi\left\vert e^{i\tilde
{H}_{0}t/\hbar}\rho_{\mathbf{q}}b_{\mathbf{q}}e^{-i\tilde{H}_{0}t/\hbar}%
\rho_{-\mathbf{q}}b_{\mathbf{q}}^{+}\right.  \right.  \right.  \right. \\
&  \left.  \left.  \left.  \left.  -\rho_{\mathbf{q}}b_{\mathbf{q}}%
e^{i\tilde{H}_{0}t/\hbar}\rho_{-\mathbf{q}}b_{\mathbf{q}}^{+}e^{-i\tilde
{H}_{0}t/\hbar}\right\vert \phi\right\rangle \right\vert \psi_{el}^{\left(
0\right)  }\right\rangle \\
&  =2i|V_{\mathbf{q}}|^{2}\delta_{\mathbf{qq}^{\prime}}\operatorname{Im}%
\left[  \left\langle \psi_{el}^{\left(  0\right)  }\left\vert \left\langle
\phi\left\vert e^{i\tilde{H}_{0}t/\hbar}\rho_{\mathbf{q}}b_{\mathbf{q}%
}e^{-i\tilde{H}_{0}t/\hbar}\rho_{-\mathbf{q}}b_{\mathbf{q}}^{+}\right\vert
\phi\right\rangle \right\vert \psi_{el}^{\left(  0\right)  }\right\rangle
\right]  .
\end{align*}
The time-dependent phonon operators are%
\[
e^{i\tilde{H}_{0}t/\hbar}b_{\mathbf{q}}e^{-i\tilde{H}_{0}t/\hbar
}=b_{\mathbf{q}}e^{-i\omega_{\text{$\mathrm{LO}$}}t},
\]
so that we have%
\begin{align*}
\mathcal{J}(\mathbf{q},\mathbf{q}^{\prime})  &  =2i|V_{\mathbf{q}}|^{2}%
\delta_{\mathbf{qq}^{\prime}}\operatorname{Im}\left[  e^{-i\omega
_{\text{$\mathrm{LO}$}}t}\left\langle \psi_{el}^{\left(  0\right)  }\left\vert
\left\langle \phi\left\vert e^{i\tilde{H}_{0}t/\hbar}\rho_{\mathbf{q}%
}e^{-i\tilde{H}_{0}t/\hbar}\rho_{-\mathbf{q}}b_{\mathbf{q}}b_{\mathbf{q}}%
^{+}\right\vert \phi\right\rangle \right\vert \psi_{el}^{\left(  0\right)
}\right\rangle \right] \\
&  =2i|V_{\mathbf{q}}|^{2}\delta_{\mathbf{qq}^{\prime}}\operatorname{Im}%
\left[  \left\langle \psi_{el}^{\left(  0\right)  }\left\vert e^{i\tilde
{H}_{e}t/\hbar}\rho_{\mathbf{q}}e^{-i\tilde{H}_{e}t/\hbar}\rho_{-\mathbf{q}%
}\right\vert \psi_{el}^{\left(  0\right)  }\right\rangle \left\langle
\phi\left\vert b_{\mathbf{q}}b_{\mathbf{q}}^{+}\right\vert \phi\right\rangle
\right]  ,
\end{align*}
where $\tilde{H}_{e}=H_{e}+\tilde{H}_{e-e}+H_{N}$.

Taking the expectation value with respect to the phonon vacuum, we find
\begin{equation}
\mathcal{J}(\mathbf{q},\mathbf{q}^{\prime})=2i|V_{\mathbf{q}}|^{2}%
\delta_{\mathbf{qq}^{\prime}}\operatorname{Im}\left[  e^{-i\omega
_{\text{$\mathrm{LO}$}}t}\left\langle \psi_{el}^{\left(  0\right)  }\left\vert
e^{i\tilde{H}_{e}t/\hbar}\rho_{\mathbf{q}}e^{-i\tilde{H}_{e}t/\hbar}%
\rho_{-\mathbf{q}}\right\vert \psi_{el}^{\left(  0\right)  }\right\rangle
\right]  .
\end{equation}
The optical conductivity (\ref{ReS}) then takes the form:%
\begin{align}
\operatorname{Re}\sigma\left(  \omega\right)   &  =-\frac{2e^{2}}{V\hbar
m_{b}^{2}\omega^{3}}\operatorname{Im}\int_{0}^{\infty}dte^{i\omega t-\delta
t}\sum_{\mathbf{q}}q_{x}^{2}|V_{\mathbf{q}}|^{2}\nonumber\\
&  \times\operatorname{Im}\left[  e^{-i\omega_{\text{$\mathrm{LO}$}}%
t}\left\langle \psi_{el}^{\left(  0\right)  }\left\vert e^{i\tilde{H}%
_{e}t/\hbar}\rho_{\mathbf{q}}e^{-i\tilde{H}_{e}t/\hbar}\rho_{-\mathbf{q}%
}\right\vert \psi_{el}^{\left(  0\right)  }\right\rangle \right]  \label{ReS2}%
\end{align}
For an isotropic electron-phonon system, $q_{x}^{2}$ in 3D can be replaced by
$\frac{1}{3}\left(  q_{x}^{2}+q_{y}^{2}+q_{z}^{2}\right)  =\frac{1}{3}q^{2},$
what gives us the result%
\begin{equation}
\operatorname{Re}\sigma_{\mathrm{3D}}\left(  \omega\right)  =-\frac{2}%
{3V\hbar\omega^{3}}\frac{e^{2}}{m_{b}^{2}}\sum_{\mathbf{q}}q^{2}\left\vert
V_{\mathbf{q}}\right\vert ^{2}\operatorname{Im}\int_{0}^{\infty}dte^{i\omega
t-\delta t}\operatorname{Im}\left[  e^{-i\omega_{\mathrm{LO}}t}F\left(
\mathbf{q},t\right)  \right]  , \label{resig1}%
\end{equation}
where the two-point correlation function is
\begin{equation}
F\left(  \mathbf{q},t\right)  =\left\langle \psi_{el}^{\left(  0\right)
}\left\vert e^{\frac{it}{\hbar}\tilde{H}_{e}}\rho_{\mathbf{q}}e^{-\frac
{it}{\hbar}\tilde{H}_{e}}\rho_{-\mathbf{q}}\right\vert \psi_{el}^{\left(
0\right)  }\right\rangle . \label{corfun}%
\end{equation}
The same derivation for the 2D case, provides the expression%
\begin{equation}
\operatorname{Re}\sigma_{\mathrm{2D}}\left(  \omega\right)  =-\frac{1}%
{A\hbar\omega^{3}}\frac{e^{2}}{m_{b}^{2}}\sum_{\mathbf{q}}q^{2}\left\vert
V_{\mathbf{q}}\right\vert ^{2}\operatorname{Im}\int_{0}^{\infty}dte^{i\omega
t-\delta t}\operatorname{Im}\left[  e^{-i\omega_{\mathrm{LO}}t}F\left(
\mathbf{q},t\right)  \right]  , \label{resig1-2d}%
\end{equation}
where $A$ is the surface of the 2D system.

\subsection{Dynamic structure factor}

To find the formula for the real part of the optical conductivity in its final
form, we introduce the standard expression for the dynamic structure factor of
the system of charge carriers interacting through a Coulomb potential,%
\begin{equation}
S(\mathbf{q},\omega)=\dfrac{1}{2N}%
%TCIMACRO{\dint \limits_{-\infty}^{\infty}}%
%BeginExpansion
{\displaystyle\int\limits_{-\infty}^{\infty}}
%EndExpansion
\left\langle \psi_{el}^{\left(  0\right)  }\left\vert
%TCIMACRO{\dsum \limits_{j,\ell}}%
%BeginExpansion
{\displaystyle\sum\limits_{j,\ell}}
%EndExpansion
e^{i\mathbf{q}.(\mathbf{r}_{j}(t)-\mathbf{r}_{\ell}(0))}\right\vert \psi
_{el}^{\left(  0\right)  }\right\rangle e^{i\omega t}dt. \label{QS}%
\end{equation}
The dynamic structure factor is expressed in terms of the two-point
correlation function as follows:%
\begin{align*}
S(\mathbf{q},\omega)  &  =\dfrac{1}{2N}%
%TCIMACRO{\dint \limits_{-\infty}^{\infty}}%
%BeginExpansion
{\displaystyle\int\limits_{-\infty}^{\infty}}
%EndExpansion
\left\langle \psi_{el}^{\left(  0\right)  }\left\vert e^{\frac{it}{\hbar
}\tilde{H}_{e}}\rho_{\mathbf{q}}e^{-\frac{it}{\hbar}\tilde{H}_{e}}%
\rho_{-\mathbf{q}}\right\vert \psi_{el}^{\left(  0\right)  }\right\rangle
e^{i\omega t}dt\\
&  =\dfrac{1}{2N}%
%TCIMACRO{\dint \limits_{-\infty}^{\infty}}%
%BeginExpansion
{\displaystyle\int\limits_{-\infty}^{\infty}}
%EndExpansion
F\left(  \mathbf{q},t\right)  e^{i\omega t}dt=\frac{1}{2N}F\left(
\mathbf{q},\omega\right)
\end{align*}%
\[
\Downarrow
\]%
\begin{equation}
S(\mathbf{q},\omega)=\frac{1}{2N}F\left(  \mathbf{q},\omega\right)  ,
\label{SthroughF}%
\end{equation}
where $F\left(  \mathbf{q},\omega\right)  $ is the Fourier image of $F\left(
\mathbf{q},t\right)  $:%
\begin{equation}
F\left(  \mathbf{q},\omega\right)  =%
%TCIMACRO{\dint \limits_{-\infty}^{\infty}}%
%BeginExpansion
{\displaystyle\int\limits_{-\infty}^{\infty}}
%EndExpansion
F\left(  \mathbf{q},t\right)  e^{i\omega t}dt. \label{Fqw}%
\end{equation}
The function $F\left(  \mathbf{q},t\right)  $ obeys the following property:%
\begin{align*}
F^{\ast}\left(  \mathbf{q},t\right)   &  =\left\langle \psi_{el}^{\left(
0\right)  }\left\vert \rho_{-\mathbf{q}}^{+}e^{\frac{it}{\hbar}\tilde{H}_{e}%
}\rho_{\mathbf{q}}^{+}e^{-\frac{it}{\hbar}\tilde{H}_{e}}\right\vert \psi
_{el}^{\left(  0\right)  }\right\rangle \\
&  =\left\langle \psi_{el}^{\left(  0\right)  }\left\vert \rho_{\mathbf{q}%
}e^{\frac{it}{\hbar}\tilde{H}_{e}}\rho_{-\mathbf{q}}e^{-\frac{it}{\hbar}%
\tilde{H}_{e}}\right\vert \psi_{el}^{\left(  0\right)  }\right\rangle \\
&  =\left\langle \psi_{el}^{\left(  0\right)  }\left\vert \rho_{\mathbf{q}%
}e^{\frac{it}{\hbar}\tilde{H}_{e}}\rho_{-\mathbf{q}}\right\vert \psi
_{el}^{\left(  0\right)  }\right\rangle e^{-\frac{it}{\hbar}\tilde{E}_{0}},
\end{align*}
where $\tilde{E}_{0}$ is the eigenvalue of the Hamiltonian $\tilde{H}_{e}$:%
\[
\tilde{H}_{e}\left\vert \psi_{el}^{\left(  0\right)  }\right\rangle =\tilde
{E}_{0}\left\vert \psi_{el}^{\left(  0\right)  }\right\rangle .
\]
Herefrom, we find that%
\begin{align}
F^{\ast}\left(  \mathbf{q},t\right)   &  =e^{-\frac{it}{\hbar}\tilde{E}_{0}%
}\left\langle \psi_{el}^{\left(  0\right)  }\left\vert \rho_{\mathbf{q}%
}e^{\frac{it}{\hbar}\tilde{H}_{e}}\rho_{-\mathbf{q}}\right\vert \psi
_{el}^{\left(  0\right)  }\right\rangle \nonumber\\
&  =\left\langle \psi_{el}^{\left(  0\right)  }\left\vert e^{-\frac{it}{\hbar
}\tilde{H}_{e}}\rho_{\mathbf{q}}e^{\frac{it}{\hbar}\tilde{H}_{e}}%
\rho_{-\mathbf{q}}\right\vert \psi_{el}^{\left(  0\right)  }\right\rangle
=F\left(  \mathbf{q},-t\right)  . \label{sp}%
\end{align}
From (\ref{sp}), for the function%
\begin{equation}
B\left(  \mathbf{q},t\right)  \equiv\operatorname{Im}\left[  e^{-i\omega
_{\mathrm{LO}}t}F\left(  \mathbf{q},t\right)  \right]  \label{B}%
\end{equation}
the following equality is derived:%
\begin{align*}
B\left(  \mathbf{q},-t\right)   &  =\operatorname{Im}\left[  e^{i\omega
_{\mathrm{LO}}t}F\left(  \mathbf{q},-t\right)  \right] \\
&  =\operatorname{Im}\left[  e^{i\omega_{\mathrm{LO}}t}F^{\ast}\left(
\mathbf{q},t\right)  \right] \\
&  =-\operatorname{Im}\left[  e^{-i\omega_{\mathrm{LO}}t}F\left(
\mathbf{q},t\right)  \right]  =-B\left(  \mathbf{q},t\right)  ,
\end{align*}%
\begin{equation}
B\left(  \mathbf{q},-t\right)  =-B\left(  \mathbf{q},t\right)  . \label{QB1}%
\end{equation}
The integral in Eq. (\ref{resig1})
\[
\operatorname{Im}\int_{0}^{\infty}dte^{i\omega t-\delta t}\operatorname{Im}%
\left[  e^{-i\omega_{\mathrm{LO}}t}F\left(  \mathbf{q},t\right)  \right]
=\operatorname{Im}\int_{0}^{\infty}dte^{i\omega t-\delta t}B\left(
\mathbf{q},t\right)
\]
is then transformed as follows:%
\begin{align*}
\operatorname{Im}\int_{0}^{\infty}dte^{i\omega t-\delta t}B\left(
\mathbf{q},t\right)   &  =\frac{1}{2i}\left[  \int_{0}^{\infty}dte^{i\omega
t-\delta t}B\left(  \mathbf{q},t\right)  -\int_{0}^{\infty}dte^{-i\omega
t-\delta t}B\left(  \mathbf{q},t\right)  \right] \\
&  =\frac{1}{2i}\left[  \int_{0}^{\infty}dte^{i\omega t-\delta t}B\left(
\mathbf{q},t\right)  -\int_{-\infty}^{0}dte^{i\omega t+\delta t}B\left(
\mathbf{q},-t\right)  \right] \\
&  =\frac{1}{2i}\left[  \int_{0}^{\infty}dte^{i\omega t-\delta t}B\left(
\mathbf{q},t\right)  +\int_{-\infty}^{0}dte^{i\omega t+\delta t}B\left(
\mathbf{q},t\right)  \right] \\
&  =\frac{1}{2i}\int_{-\infty}^{\infty}dte^{i\omega t-\delta\left\vert
t\right\vert }B\left(  \mathbf{q},t\right) \\
&  =\frac{1}{2i}\int_{-\infty}^{\infty}dte^{i\omega t-\delta\left\vert
t\right\vert }\frac{1}{2i}\left[  e^{-i\omega_{\mathrm{LO}}t}F\left(
\mathbf{q},t\right)  -e^{i\omega_{\mathrm{LO}}t}F^{\ast}\left(  \mathbf{q}%
,t\right)  \right] \\
&  =-\frac{1}{4}\int_{-\infty}^{\infty}dte^{i\omega t-\delta\left\vert
t\right\vert }\left[  e^{-i\omega_{\mathrm{LO}}t}F\left(  \mathbf{q},t\right)
-e^{i\omega_{\mathrm{LO}}t}F\left(  \mathbf{q},-t\right)  \right]  .
\end{align*}
We can show that, as far as $\left\vert \psi_{el}^{\left(  0\right)
}\right\rangle $ is the \emph{ground} state, the integral $\int_{-\infty
}^{\infty}dte^{i\omega t-\delta\left\vert t\right\vert }F\left(
\mathbf{q},-t\right)  $ for positive $\omega$ is equal to zero. Let $\left\{
\left\vert \psi_{el}^{\left(  n\right)  }\right\rangle \right\}  $ is the
total basis set of the eigenfunctions of the Hamiltonian $\tilde{H}_{e}.$
Using these functions we expand $F\left(  \mathbf{q},t\right)  $:%
\begin{align*}
F\left(  \mathbf{q},t\right)   &  =\sum_{n}\left\langle \psi_{el}^{\left(
0\right)  }\left\vert e^{\frac{it}{\hbar}\tilde{H}_{e}}\rho_{\mathbf{q}%
}e^{-\frac{it}{\hbar}\tilde{H}_{e}}\right\vert \psi_{el}^{\left(  n\right)
}\right\rangle \left\langle \psi_{el}^{\left(  n\right)  }\left\vert
\rho_{-\mathbf{q}}\right\vert \psi_{el}^{\left(  0\right)  }\right\rangle \\
&  =\sum_{n}\left\vert \left\langle \psi_{el}^{\left(  n\right)  }\left\vert
\rho_{-\mathbf{q}}\right\vert \psi_{el}^{\left(  0\right)  }\right\rangle
\right\vert ^{2}e^{\frac{it}{\hbar}\left(  \tilde{E}_{0}-\tilde{E}_{n}\right)
},
\end{align*}%
\begin{align*}
\int_{-\infty}^{\infty}dte^{i\omega t-\delta\left\vert t\right\vert }F\left(
\mathbf{q},-t\right)   &  =\sum_{n}\left\vert \left\langle \psi_{el}^{\left(
n\right)  }\left\vert \rho_{-\mathbf{q}}\right\vert \psi_{el}^{\left(
0\right)  }\right\rangle \right\vert ^{2}\int_{-\infty}^{\infty}dte^{i\omega
t+\frac{i}{\hbar}\left(  \tilde{E}_{n}-\tilde{E}_{0}\right)  t-\delta
\left\vert t\right\vert }\\
&  =\sum_{n}\left\vert \left\langle \psi_{el}^{\left(  n\right)  }\left\vert
\rho_{-\mathbf{q}}\right\vert \psi_{el}^{\left(  0\right)  }\right\rangle
\right\vert ^{2}2\pi\delta\left(  \omega+\frac{\tilde{E}_{n}-\tilde{E}_{0}%
}{\hbar}\right)  =0,
\end{align*}
because for $\omega>0,$ $\omega+\frac{\tilde{E}_{n}-\tilde{E}_{0}}{\hbar}$ is
never equal to zero.

So, rewriting expression (\ref{resig1}) with the dynamic structure factor of
the electron (or hole) gas results in:%
\begin{align*}
\operatorname{Re}\sigma_{\mathrm{3D}}\left(  \omega\right)   &  =\frac
{1}{6V\hbar\omega^{3}}\frac{e^{2}}{m_{b}^{2}}\sum_{\mathbf{q}}q^{2}\left\vert
V_{\mathbf{q}}\right\vert ^{2}\int_{-\infty}^{\infty}dte^{i\left(
\omega-\omega_{\mathrm{LO}}\right)  t-\delta\left\vert t\right\vert }F\left(
\mathbf{q},t\right)  ,\\
\operatorname{Re}\sigma_{\mathrm{2D}}\left(  \omega\right)   &  =\frac
{1}{4A\hbar\omega^{3}}\frac{e^{2}}{m_{b}^{2}}\sum_{\mathbf{q}}q^{2}\left\vert
V_{\mathbf{q}}\right\vert ^{2}\int_{-\infty}^{\infty}dte^{i\left(
\omega-\omega_{\mathrm{LO}}\right)  t-\delta\left\vert t\right\vert }F\left(
\mathbf{q},t\right)  ,
\end{align*}
and we obtain%
\begin{subequations}
\begin{align}
\operatorname{Re}\sigma_{\mathrm{3D}}(\omega)  &  =\frac{n_{0}}{3\hbar
\omega^{3}}\frac{e^{2}}{m_{b}^{2}}\sum_{\mathbf{q}}q^{2}|V_{\mathbf{q}}%
|^{2}S(\mathbf{q},\omega-\omega_{\mathrm{LO}}),\label{opticabs}\\
\operatorname{Re}\sigma_{\mathrm{2D}}(\omega)  &  =\frac{n_{0}}{2\hbar
\omega^{3}}\frac{e^{2}}{m_{b}^{2}}\sum_{\mathbf{q}}q^{2}|V_{\mathbf{q}}%
|^{2}S(\mathbf{q},\omega-\omega_{\mathrm{LO}}),
\end{align}
where
\end{subequations}
\[
n_{0}=\left\{
\begin{array}
[c]{c}%
N/V\; \text{in 3D,}\\
N/A\; \text{in 2D}%
\end{array}
\right.
\]
is the density of charge carriers.

For an isotropic medium, the dynamic structure factor does not depend on the
direction of $\mathbf{q}$, so that $S(\mathbf{q},\omega)=S(q,\omega)$, where
$q=\left\vert \mathbf{q}\right\vert $. Let us simplify the expression
(\ref{opticabs}) using explicitly the amplitudes of the Fr\"{o}hlich
electron-phonon interaction. The modulus squared of the Fr\"{o}hlich
electron-phonon interaction amplitude is given by
\begin{equation}
|V_{\mathbf{q}}|^{2}=\left\{
\begin{array}
[c]{l}%
\dfrac{(\hbar\omega_{\mathrm{LO}})^{2}}{q^{2}}\dfrac{4\pi\alpha}{V}\left(
\dfrac{\hbar}{2m_{b}\omega_{\mathrm{LO}}}\right)  ^{1/2}\text{ in 3D}\\
\dfrac{(\hbar\omega_{\mathrm{LO}})^{2}}{q}\dfrac{2\pi\alpha}{A}\left(
\dfrac{\hbar}{2m_{b}\omega_{\mathrm{LO}}}\right)  ^{1/2}\, \text{in 2D,}%
\end{array}
\right.
\end{equation}
where $\alpha$ is the (dimensionless) Fr\"{o}hlich coupling constant
determining the coupling strength between the charge carriers and the
longitudinal optical phonons \cite{prb31-3420,prb36-4442}. In 3D and 2D,
respectively, the sums over $\mathbf{q}$ is transformed to the integrals as
follows:%
\begin{align*}
\text{3D}  &  \text{:}\; \sum_{\mathbf{q}}\ldots=\frac{V}{\left(  2\pi\right)
^{3}}\int d\mathbf{q}\ldots\\
\text{2D}  &  \text{:}\; \sum_{\mathbf{q}}\ldots=\frac{A}{\left(  2\pi\right)
^{2}}\int d\mathbf{q}\ldots
\end{align*}%
\[
\Downarrow
\]%
\begin{subequations}
\[
\operatorname{Re}\sigma_{3\mathrm{D}}(\omega)=\frac{n_{0}}{3\hbar\omega^{3}%
}\frac{e^{2}}{m_{b}^{2}}\frac{V}{\left(  2\pi\right)  ^{3}}\int d\mathbf{q}%
q^{2}\left\vert \frac{\hbar\omega_{\mathrm{LO}}}{iq}\left(  \frac{4\pi\alpha
}{V}\right)  ^{1/2}\left(  \frac{\hbar}{2m_{b}\omega_{\mathrm{LO}}}\right)
^{1/4}\right\vert ^{2}S(q,\omega-\omega_{\mathrm{LO}})
\]%
\end{subequations}
\[
\Downarrow
\]%
\begin{equation}
\operatorname{Re}\sigma_{3\mathrm{D}}(\omega)=\frac{n_{0}e^{2}}{m_{b}^{2}%
}\frac{2\alpha}{3\pi}\frac{\hbar\omega_{\mathrm{LO}}^{2}}{\omega^{3}}\left(
\frac{\hbar}{2m_{b}\omega_{\mathrm{LO}}}\right)  ^{1/2}\int_{0}^{\infty}%
q^{2}S(q,\omega-\omega_{\mathrm{LO}})dq. \label{Resigmaf}%
\end{equation}
In the same way, we transform $\operatorname{Re}\sigma_{2\mathrm{D}}(\omega
)$:
\[
\operatorname{Re}\sigma_{2\mathrm{D}}(\omega)=\frac{n_{0}e^{2}}{m_{b}^{2}%
}\frac{\alpha}{2}\frac{\hbar\omega_{\mathrm{LO}}^{2}}{\omega^{3}}\left(
\dfrac{\hbar}{2m_{b}\omega_{\mathrm{LO}}}\right)  ^{1/2}\int_{0}^{\infty}%
q^{2}S(q,\omega-\omega_{\mathrm{LO}})dq.
\]
Using the Feynman units ($\hbar=1$, $m_{b}=1$, $\omega_{\mathrm{LO}}=1$),
$\operatorname{Re}\sigma(\omega)$ is%
\begin{align}
\operatorname{Re}\sigma_{3\mathrm{D}}(\omega)  &  =n_{0}e^{2}\frac{\sqrt
{2}\alpha}{3\pi}\frac{1}{\omega^{3}}\int_{0}^{\infty}q^{2}S(q,\omega
-1)dq,\label{Resigmaff1}\\
\operatorname{Re}\sigma_{2\mathrm{D}}(\omega)  &  =n_{0}e^{2}\frac{\alpha
}{2\sqrt{2}}\frac{1}{\omega^{3}}\int_{0}^{\infty}q^{2}S(q,\omega-1)dq.
\label{Resigmaff2}%
\end{align}
From these expressions, it is clear that the scaling relation
\begin{equation}
\operatorname{Re}\sigma_{\mathrm{2D}}(\omega,\alpha)=\operatorname{Re}%
\sigma_{\mathrm{3D}}(\omega,\frac{3\pi}{4}\alpha)
\end{equation}
which holds for the one-polaron case introduced in ref.
\cite{prb31-3420,prb36-4442}, is also valid for the many-polaron case if the
corresponding 2D or 3D dynamic structure factor is used.

\subsubsection{Calculation of the dynamic structure factor using the retarded
Green's functions}

The dynamic structure factor $S\left(  \mathbf{q},\omega\right)  $ is
expressed through the two-point correlation function by Eq. (\ref{SthroughF}).
The correlation function $F\left(  \mathbf{q},\omega\right)  $ can be found
using the retarded Green's function of the density operators
\begin{equation}
G^{R}\left(  \mathbf{q},t\right)  =-i\Theta\left(  t\right)  \left\langle
\psi_{el}^{\left(  0\right)  }\left\vert \left[  e^{\frac{it}{\hbar}\tilde
{H}_{0}}\rho_{\mathbf{q}}e^{-\frac{it}{\hbar}\tilde{H}_{0}},\rho_{-\mathbf{q}%
}\right]  \right\vert \psi_{el}^{\left(  0\right)  }\right\rangle ,
\end{equation}
where $\Theta\left(  t\right)  $ is the step function. Let us consider the
more general case of a finite temperature,%
\begin{equation}
G^{R}\left(  \mathbf{q},t\right)  =-i\Theta\left(  t\right)  \left\langle
\left[  e^{\frac{it}{\hbar}\tilde{H}_{0}}\rho_{\mathbf{q}}e^{-\frac{it}{\hbar
}\tilde{H}_{0}},\rho_{-\mathbf{q}}\right]  \right\rangle ,
\end{equation}
where the average is%
\begin{equation}
\left\langle \ldots\right\rangle \equiv\frac{Tr\left(  e^{-\beta\tilde{H}_{0}%
}\ldots\right)  }{Tr\left(  e^{-\beta\tilde{H}_{0}}\right)  },\; \beta
=\frac{1}{k_{B}T}. \label{avd}%
\end{equation}
The Fourier image $G^{R}\left(  \mathbf{q},\omega\right)  $ of the retarded
Green's function $G^{R}\left(  \mathbf{q},t\right)  $ is%
\begin{align*}
G^{R}\left(  \mathbf{q},\omega\right)   &  =\int_{-\infty}^{\infty}%
G^{R}\left(  \mathbf{q},t\right)  e^{i\omega t}dt\\
&  =-i\int_{0}^{\infty}\left\langle \left[  e^{\frac{it}{\hbar}\tilde{H}_{0}%
}\rho_{\mathbf{q}}e^{-\frac{it}{\hbar}\tilde{H}_{0}},\rho_{-\mathbf{q}%
}\right]  \right\rangle e^{i\omega t}dt\\
&  =-i\int_{0}^{\infty}\left(  \left\langle e^{\frac{it}{\hbar}\tilde{H}_{0}%
}\rho_{\mathbf{q}}e^{-\frac{it}{\hbar}\tilde{H}_{0}}\rho_{-\mathbf{q}%
}\right\rangle -\left\langle \rho_{-\mathbf{q}}e^{\frac{it}{\hbar}\tilde
{H}_{0}}\rho_{\mathbf{q}}e^{-\frac{it}{\hbar}\tilde{H}_{0}}\right\rangle
\right)  e^{i\omega t}dt
\end{align*}
The imaginary part of $G^{R}\left(  \mathbf{q},\omega\right)  $ then is%
\begin{align*}
\operatorname{Im}G^{R}\left(  \mathbf{q},\omega\right)   &
=-\operatorname{Re}\int_{0}^{\infty}\left(  \left\langle e^{\frac{it}{\hbar
}\tilde{H}_{0}}\rho_{\mathbf{q}}e^{-\frac{it}{\hbar}\tilde{H}_{0}}%
\rho_{-\mathbf{q}}\right\rangle -\left\langle \rho_{-\mathbf{q}}e^{\frac
{it}{\hbar}\tilde{H}_{0}}\rho_{\mathbf{q}}e^{-\frac{it}{\hbar}\tilde{H}_{0}%
}\right\rangle \right)  e^{i\omega t}dt\\
&  =-\frac{1}{2}\int_{0}^{\infty}\left(  \left\langle e^{\frac{it}{\hbar
}\tilde{H}_{0}}\rho_{\mathbf{q}}e^{-\frac{it}{\hbar}\tilde{H}_{0}}%
\rho_{-\mathbf{q}}\right\rangle -\left\langle \rho_{-\mathbf{q}}e^{\frac
{it}{\hbar}\tilde{H}_{0}}\rho_{\mathbf{q}}e^{-\frac{it}{\hbar}\tilde{H}_{0}%
}\right\rangle \right)  e^{i\omega t}dt\\
&  -\frac{1}{2}\int_{0}^{\infty}\left(  \left\langle \rho_{\mathbf{q}}%
e^{\frac{it}{\hbar}\tilde{H}_{0}}\rho_{-\mathbf{q}}e^{-\frac{it}{\hbar}%
\tilde{H}_{0}}\right\rangle -\left\langle e^{\frac{it}{\hbar}\tilde{H}_{0}%
}\rho_{-\mathbf{q}}e^{-\frac{it}{\hbar}\tilde{H}_{0}}\rho_{\mathbf{q}%
}\right\rangle \right)  e^{-i\omega t}dt\\
&  =-\frac{1}{2}\int_{-\infty}^{\infty}\left\langle e^{\frac{it}{\hbar}%
\tilde{H}_{0}}\rho_{\mathbf{q}}e^{-\frac{it}{\hbar}\tilde{H}_{0}}%
\rho_{-\mathbf{q}}\right\rangle e^{i\omega t}dt+\frac{1}{2}\int_{-\infty
}^{\infty}\left\langle \rho_{-\mathbf{q}}e^{\frac{it}{\hbar}\tilde{H}_{0}}%
\rho_{\mathbf{q}}e^{-\frac{it}{\hbar}\tilde{H}_{0}}\right\rangle e^{i\omega
t}dt.
\end{align*}
In the second integral here, we replace $t$ by $\left(  t^{\prime}+i\hbar
\beta\right)  $:%
\begin{align*}
&  \int_{-\infty}^{\infty}\frac{Tr\left(  e^{-\beta\tilde{H}_{0}}%
\rho_{-\mathbf{q}}e^{\frac{it}{\hbar}\tilde{H}_{0}}\rho_{\mathbf{q}}%
e^{-\frac{it}{\hbar}\tilde{H}_{0}}\right)  }{Tr\left(  e^{-\beta\tilde{H}_{0}%
}\right)  }e^{i\omega t}dt\\
&  =\int_{-\infty-i\hbar\beta}^{\infty-i\hbar\beta}\frac{Tr\left(
e^{-\beta\tilde{H}_{0}}\rho_{-\mathbf{q}}e^{\frac{it^{\prime}}{\hbar}\tilde
{H}_{0}-\beta\tilde{H}_{0}}\rho_{\mathbf{q}}e^{-\frac{it^{\prime}}{\hbar
}\tilde{H}_{0}+\beta\tilde{H}_{0}}\right)  }{Tr\left(  e^{-\beta\tilde{H}_{0}%
}\right)  }e^{i\omega\left(  t^{\prime}+i\hbar\beta\right)  }dt^{\prime}\\
&  =e^{-\beta\hbar\omega}\int_{-\infty-i\hbar\beta}^{\infty-i\hbar\beta}%
\frac{Tr\left(  e^{\frac{it^{\prime}}{\hbar}\tilde{H}_{0}-\beta\tilde{H}_{0}%
}\rho_{\mathbf{q}}e^{-\frac{it^{\prime}}{\hbar}\tilde{H}_{0}}\rho
_{-\mathbf{q}}\right)  }{Tr\left(  e^{-\beta\tilde{H}_{0}}\right)  }e^{i\omega
t^{\prime}}dt^{\prime}.
\end{align*}
As far as the integral over $t$ converges (i. e., $\left\langle e^{\frac
{it}{\hbar}\tilde{H}_{0}}\rho_{\mathbf{q}}e^{-\frac{it}{\hbar}\tilde{H}_{0}%
}\rho_{-\mathbf{q}}\right\rangle $ tends to zero at $\left\vert t\right\vert
\rightarrow\infty$), we can shift the integration contour to the real axis,
what gives us the result%
\begin{align*}
&  \int_{-\infty}^{\infty}\frac{Tr\left(  e^{-\beta\tilde{H}_{0}}%
\rho_{-\mathbf{q}}e^{\frac{it}{\hbar}\tilde{H}_{0}}\rho_{\mathbf{q}}%
e^{-\frac{it}{\hbar}\tilde{H}_{0}}\right)  }{Tr\left(  e^{-\beta\tilde{H}_{0}%
}\right)  }e^{i\omega t}dt\\
&  =e^{-\beta\hbar\omega}\int_{-\infty}^{\infty}\frac{Tr\left(  e^{\frac
{it}{\hbar}\tilde{H}_{0}-\beta\tilde{H}_{0}}\rho_{\mathbf{q}}e^{-\frac
{it}{\hbar}\tilde{H}_{0}}\rho_{-\mathbf{q}}\right)  }{Tr\left(  e^{-\beta
\tilde{H}_{0}}\right)  }e^{i\omega t}dt,\\
&  \Updownarrow\\
&  \int_{-\infty}^{\infty}\left\langle \rho_{-\mathbf{q}}e^{\frac{it}{\hbar
}\tilde{H}_{0}}\rho_{\mathbf{q}}e^{-\frac{it}{\hbar}\tilde{H}_{0}%
}\right\rangle e^{i\omega t}dt\\
&  =e^{-\beta\hbar\omega}\int_{-\infty}^{\infty}\left\langle e^{\frac
{it}{\hbar}\tilde{H}_{0}}\rho_{\mathbf{q}}e^{-\frac{it}{\hbar}\tilde{H}_{0}%
}\rho_{-\mathbf{q}}\right\rangle e^{i\omega t}dt.
\end{align*}
Herefrom, we find that%
\[
\operatorname{Im}G^{R}\left(  \mathbf{q},\omega\right)  =-\frac{1}{2}\left(
1-e^{-\beta\hbar\omega}\right)  \int_{-\infty}^{\infty}\left\langle
e^{\frac{it}{\hbar}\tilde{H}_{0}}\rho_{\mathbf{q}}e^{-\frac{it}{\hbar}%
\tilde{H}_{0}}\rho_{-\mathbf{q}}\right\rangle e^{i\omega t}dt
\]%
\[
\Downarrow
\]%
\[
\operatorname{Im}G^{R}\left(  \mathbf{q},\omega\right)  =-\frac{1}{2}\left(
1-e^{-\beta\hbar\omega}\right)  F\left(  \mathbf{q},\omega\right)  .
\]
So, the equation follows from the analytical properties of the Green's
functions:
\begin{equation}
F\left(  \mathbf{q},\omega\right)  =-\frac{2\operatorname{Im}G^{R}\left(
\mathbf{q},\omega\right)  }{1-e^{-\beta\hbar\omega}}. \label{rel1}%
\end{equation}

The formula (\ref{rel1}) is related to arbitrary temperatures. In the
zero-temperature limit ($\beta\rightarrow\infty$), the factor $\left(
1-e^{-\beta\hbar\omega}\right)  ^{-1}$ (\ref{rel1}) turns into the Heavicide
step function $\Theta\left(  \omega\right)  $, what leads to the formula%
\begin{equation}
\left.  F\left(  \mathbf{q},\omega\right)  \right\vert _{T=0}=-2\Theta\left(
\omega\right)  \operatorname{Im}\left.  G^{R}\left(  \mathbf{q},\omega\right)
\right\vert _{T=0}%
\end{equation}%
\[
\Downarrow
\]%
\begin{equation}
\left.  S\left(  \mathbf{q},\omega\right)  \right\vert _{T=0}=-\frac{1}%
{N}\Theta\left(  \omega\right)  \operatorname{Im}\left.  G^{R}\left(
\mathbf{q},\omega\right)  \right\vert _{T=0} \label{SG}%
\end{equation}
The retarded Green function is related to the dielectric function of the
electron gas by the following equation:
\begin{equation}
G^{R}\left(  \mathbf{q},\omega\right)  =\frac{1}{v_{\mathbf{q}}}\left[
\frac{1}{\varepsilon\left(  \mathbf{q},\omega\right)  }-1\right]  .
\label{diel}%
\end{equation}
Within the random phase approximation (RPA), following \cite{Mahan}, the
expression for $G^{R}\left(  \mathbf{q},\omega\right)  $ is
\begin{equation}
G^{R}\left(  \mathbf{q,}\omega\right)  =\left[  1-v_{\mathbf{q}}P\left(
\mathbf{q},\omega\right)  \right]  ^{-1}\hbar P\left(  \mathbf{q}%
,\omega\right)  , \label{DR1}%
\end{equation}
where the polarization function $P\left(  \mathbf{q},\omega\right)  $ is (see,
e. g., p. 434 of \cite{Mahan})
\begin{equation}
P\left(  \mathbf{q},\omega\right)  =\frac{1}{\hbar}\sum_{\mathbf{k},\sigma
}\frac{f_{\mathbf{k+q},\sigma}-f_{\mathbf{k},\sigma}}{\omega-\frac
{\hbar\mathbf{k}^{2}}{2m_{b}}+\frac{\hbar\left(  \mathbf{k}+\mathbf{q}\right)
^{2}}{2m_{b}}+i\delta},\; \delta\rightarrow+0 \label{P}%
\end{equation}
with the Fermi distribution function of electrons $f_{\mathbf{k},\sigma}$.

For a finite temperature, the explicit analytic expression for the imaginary
part of the structure factor $P_{\mathrm{3D}}\left(  \mathbf{q},\omega\right)
$ is obtained (see \cite{Mahan}),
\begin{align}
\operatorname{Im}P_{\mathrm{3D}}\left(  \mathbf{q},\omega\right)   &
=\frac{Vm_{b}^{2}}{2\pi\hbar^{4}\beta q}\ln\frac{1+\exp\left\{  \beta\left[
\mu-E^{\left(  +\right)  }\left(  q,\omega\right)  \right]  \right\}  }%
{1+\exp\left\{  \beta\left[  \mu-E^{\left(  -\right)  }\left(  q,\omega
\right)  \right]  \right\}  },\nonumber\\
E^{\left(  \pm\right)  }\left(  q,\omega\right)   &  \equiv\frac{\left(
\hbar\omega\pm\frac{\hbar^{2}q^{2}}{2m_{b}}\right)  ^{2}}{4\frac{\hbar
^{2}q^{2}}{2m_{b}}},
\end{align}
with the chemical potential $\mu$. The real part of the structure factor is
obtained using the Kramers-Kronig dispersion relation:
\begin{equation}
\operatorname{Re}P\left(  \mathbf{q},\omega\right)  =\frac{1}{\pi}%
\int_{-\infty}^{\infty}\mathcal{P}\left(  \frac{1}{\omega^{\prime}-\omega
}\right)  \operatorname{Im}P\left(  \mathbf{q},\omega^{\prime}\right)
d\omega^{\prime}.
\end{equation}
Analytical expressions for both real and imaginary parts of $P\left(
\mathbf{q},\omega\right)  $ can be written down for the zero temperature (see
\cite{Mahan}),
\begin{align}
\operatorname{Re}P_{\mathrm{3D}}\left(  q,\omega\right)   &  =-\frac{Vm_{b}%
}{4\pi^{2}\hbar^{2}q}\left\{
\begin{array}
[c]{c}%
\left[  k_{F}^{2}-\frac{m_{b}^{2}}{\hbar^{2}q^{2}}\left(  \omega-\frac{\hbar
q^{2}}{2m_{b}}\right)  ^{2}\right]  \ln\left\vert \frac{\omega-\frac{\hbar
q^{2}}{2m_{b}}-\frac{\hbar k_{F}q}{m_{b}}}{\omega-\frac{\hbar q^{2}}{2m_{b}%
}+\frac{\hbar k_{F}q}{m_{b}}}\right\vert \\
+\left[  k_{F}^{2}-\frac{m_{b}^{2}}{\hbar^{2}q^{2}}\left(  \omega+\frac{\hbar
q^{2}}{2m_{b}}\right)  ^{2}\right]  \ln\left\vert \frac{\omega+\frac{\hbar
q^{2}}{2m_{b}}+\frac{\hbar k_{F}q}{m_{b}}}{\omega+\frac{\hbar q^{2}}{2m_{b}%
}-\frac{\hbar k_{F}q}{m_{b}}}\right\vert \\
+2k_{F}q
\end{array}
\right\}  ,\nonumber\\
\operatorname{Im}P_{\mathrm{3D}}\left(  q,\omega\right)   &  =-\frac{Vm_{b}%
}{4\pi\hbar^{2}q}\left\{
\begin{array}
[c]{c}%
\left[  k_{F}^{2}-\frac{m_{b}^{2}}{\hbar^{2}q^{2}}\left(  \omega-\frac{\hbar
q^{2}}{2m_{b}}\right)  ^{2}\right]  \Theta\left(  k_{F}^{2}-\frac{m_{b}^{2}%
}{\hbar^{2}q^{2}}\left(  \omega-\frac{\hbar q^{2}}{2m_{b}}\right)  ^{2}\right)
\\
-\left[  k_{F}^{2}-\frac{m_{b}^{2}}{\hbar^{2}q^{2}}\left(  \omega+\frac{\hbar
q^{2}}{2m_{b}}\right)  ^{2}\right]  \Theta\left(  k_{F}^{2}-\frac{m_{b}^{2}%
}{\hbar^{2}q^{2}}\left(  \omega+\frac{\hbar q^{2}}{2m_{b}}\right)
^{2}\right)
\end{array}
\right\}  , \label{PRI}%
\end{align}
where $k_{F}=\left(  3\pi^{2}N/V\right)  ^{1/3}$ is the Fermi wave number.

After substituting into Eq. (\ref{rel1}) the retarded Green's function
(\ref{DR1}) in terms of the polarization function we arrive at the formula%
\[
F\left(  \mathbf{q},\omega\right)  =-\frac{2\hbar}{1-e^{-\beta\hbar\omega}%
}\operatorname{Im}\frac{P\left(  \mathbf{q},\omega\right)  }{1-v_{\mathbf{q}%
}P\left(  \mathbf{q},\omega\right)  }%
\]%
\[
\Downarrow
\]%
\begin{align}
S\left(  \mathbf{q},\omega\right)   &  =-\frac{\hbar}{N\left(  1-e^{-\beta
\hbar\omega}\right)  }\frac{\operatorname{Im}P\left(  \mathbf{q,}%
\omega\right)  }{\left[  1-v_{\mathbf{q}}\operatorname{Re}P\left(
\mathbf{q},\omega\right)  \right]  ^{2}+\left[  v_{\mathbf{q}}%
\operatorname{Im}P\left(  \mathbf{q},\omega\right)  \right]  ^{2}}%
,\label{Sqw}\\
\left.  S\left(  \mathbf{q},\omega\right)  \right\vert _{T=0}  &
=-\frac{\hbar}{N}\Theta\left(  \omega\right)  \frac{\operatorname{Im}P\left(
\mathbf{q,}\omega\right)  }{\left[  1-v_{\mathbf{q}}\operatorname{Re}P\left(
\mathbf{q},\omega\right)  \right]  ^{2}+\left[  v_{\mathbf{q}}%
\operatorname{Im}P\left(  \mathbf{q},\omega\right)  \right]  ^{2}}.
\end{align}
With this dynamic structure factor, the optical conductivity (\ref{Resigmaff1}%
) (in the Feynman units) takes the form%
\begin{align}
\operatorname{Re}\sigma_{\mathrm{3D}}(\omega)  &  =-e^{2}\frac{\sqrt{2}\alpha
}{3\pi V}\frac{1}{\omega^{3}}\Theta\left(  \omega-1\right) \nonumber\\
&  \times\int_{0}^{\infty}\frac{\operatorname{Im}P_{\mathrm{3D}}\left(
q\mathbf{,}\omega-1\right)  }{\left[  1-v_{q}\operatorname{Re}P_{\mathrm{3D}%
}\left(  q,\omega-1\right)  \right]  ^{2}+\left[  v_{q}\operatorname{Im}%
P_{\mathrm{3D}}\left(  q,\omega-1\right)  \right]  ^{2}}q^{2}dq. \label{rs3d}%
\end{align}
Correspondingly, in the 2D case we obtain the expression%
\begin{align}
\operatorname{Re}\sigma_{\mathrm{2D}}(\omega)  &  =-e^{2}\frac{\alpha}%
{2\sqrt{2}V}\frac{1}{\omega^{3}}\Theta\left(  \omega-1\right) \nonumber\\
&  \times\int_{0}^{\infty}\frac{\operatorname{Im}P_{\mathrm{2D}}\left(
q\mathbf{,}\omega-1\right)  }{\left[  1-v_{q}\operatorname{Re}P_{\mathrm{2D}%
}\left(  q,\omega-1\right)  \right]  ^{2}+\left[  v_{q}\operatorname{Im}%
P_{\mathrm{2D}}\left(  q,\omega-1\right)  \right]  ^{2}}q^{2}dq. \label{rs2d}%
\end{align}

\subsubsection{Plasmon-phonon contribution}

The RPA dynamic structure factor for the electron (or hole) system can be
separated in two parts, one related to continuum excitations of the electrons
(or holes) $S_{\mathrm{cont}}$, and one related to the undamped plasmon
branch:
\begin{equation}
S_{\mathrm{RPA}}(q,\omega)=A_{\mathrm{pl}}(q)\delta\left(  \omega
-\omega_{\mathrm{pl}}\left(  q\right)  \right)  +S_{\mathrm{cont}}(q,\omega),
\label{DLR1977PA}%
\end{equation}
where $\omega_{\mathrm{pl}}\left(  q\right)  $\ is the wave number dependent
plasmon frequency and $A_{\mathrm{pl}}$\ is the strength of the undamped
plasmon branch.

In Eqs. (\ref{rs3d}), (\ref{rs2d}), the contribution of the continuum
excitations corresponds to the region $\left(  q,\omega\right)  $ where
$\operatorname{Im}P\left(  q\mathbf{,}\omega\right)  \neq0$. The contribution
related to the undamped plasmons is provided by a region of $\left(
q\mathbf{,}\omega\right)  ,$ where the equations
\begin{equation}
\left\{
\begin{array}
[c]{c}%
\operatorname{Im}P\left(  q\mathbf{,}\omega\right)  =0\\
1-v_{q}\operatorname{Re}P\left(  q,\omega\right)  =0
\end{array}
\right.  . \label{c}%
\end{equation}
are fulfilled simultaneously. Using (\ref{diel}), we find that Eqs. (\ref{b})
are equivalent to
\begin{equation}
\operatorname{Im}\frac{1}{\varepsilon\left(  q,\omega\right)  }=0,\; \;
\operatorname{Re}\frac{1}{\varepsilon\left(  q,\omega\right)  }=0. \label{b}%
\end{equation}

In the region where $\operatorname{Im}P\left(  \mathbf{q,}\omega\right)  =0,$
the expression $\frac{\operatorname{Im}P\left(  q\mathbf{,}\omega\right)
}{\left[  1-v_{q}\operatorname{Re}P\left(  q,\omega\right)  \right]
^{2}+\left[  v_{q}\operatorname{Im}P\left(  q,\omega\right)  \right]  ^{2}}$
is proportional to the delta function, which gives a finite contribution to
the memory function after the integration over $q$:
\begin{align}
&  \left.  \frac{\operatorname{Im}P\left(  q\mathbf{,}\omega\right)  }{\left[
1-v_{q}\operatorname{Re}P\left(  q,\omega\right)  \right]  ^{2}+\left[
v_{q}\operatorname{Im}P\left(  q,\omega\right)  \right]  ^{2}}\right\vert
_{\operatorname{Im}P\left(  q\mathbf{,}\omega\right)  =0}\nonumber\\
&  =\frac{1}{\pi v_{q}}\delta\left(  1-v_{q}\operatorname{Re}P\left(
q,\omega\right)  \right)  . \label{vv}%
\end{align}

Using Eq. (\ref{vv}), the coefficients $A_{\mathrm{pl}}(q)$ in Eq.
(\ref{DLR1977PA}) can be expressed in terms of the polarization function
$P\left(  q,\omega\right)  $ as follows:%
\begin{align*}
&  \left.  \frac{\operatorname{Im}P\left(  q\mathbf{,}\omega\right)  }{\left[
1-v_{q}\operatorname{Re}P\left(  q,\omega\right)  \right]  ^{2}+\left[
v_{q}\operatorname{Im}P\left(  q,\omega\right)  \right]  ^{2}}\right\vert
_{\operatorname{Im}P\left(  q\mathbf{,}\omega\right)  =0}\\
&  =\left.  \frac{1}{\pi v_{q}^{2}\left\vert \frac{\partial}{\partial\omega
}\operatorname{Re}P\left(  q,\omega\right)  \right\vert }\right\vert
_{\omega=\omega_{\mathrm{pl}}\left(  q\right)  }\delta\left(  \omega
-\omega_{\mathrm{pl}}\left(  q\right)  \right)
\end{align*}%
\[
\Downarrow
\]%
\begin{equation}
A_{\mathrm{pl}}\left(  q\right)  =\left.  \frac{1}{\pi v_{q}^{2}\left\vert
\frac{\partial}{\partial\omega}\operatorname{Re}P\left(  q,\omega\right)
\right\vert }\right\vert _{\omega=\omega_{\mathrm{pl}}\left(  q\right)  }.
\label{Apl}%
\end{equation}

The contribution derived from the undamped plasmon branch $A_{\mathrm{pl}%
}(q)\delta\left(  \omega-\omega_{\mathrm{pl}}\left(  q\right)  \right)  $\ is
denoted in Ref. \cite{TDPRB01} as the \textit{`plasmon-phonon' contribution}.
The physical process related to this contribution is the emission of both a
phonon and a plasmon in the scattering process.

\subsection{Comparison to the infrared spectrum of Nd$_{2-x}$Ce$_{x}%
$CuO$_{2-y}$}

Calvani and collaborators have performed doping-dependent measurements of the
infrared absorption spectra of the high-T$_{c}$ material Nd$_{2-x}$Ce$_{x}%
$CuO$_{2-y}$ (NCCO). The region of the spectrum examined by these authors
(50-10000 cm$^{-1}$) is very rich in absorption features: they observe is a
\textquotedblleft Drude-like\textquotedblright\ component at the lowest
frequencies, and a set of sharp absorption peaks related to phonons and
infrared active modes (up to about 1000 cm$^{-1}$) possibly associated to
small (Holstein) polarons. Three distinct absorption bands can be
distinguished: the `\emph{d}-band' (around 1000 cm$^{-1}$), the Mid-Infrared
band (MIR, around 5000 cm$^{-1}$) and the Charge-Transfer band (around
10$^{4}$ cm$^{-1}$). Of all these features, the \emph{d}-band and, at a higher
temperatures, the Drude-like component have (hypothetically) been associated
with large polaron optical absorption \cite{calva2}.

%\newpage%
%

%TCIMACRO{\FRAME{fhFU}{9.0325cm}{6.5789cm}{0pt}{\Qcb{The infrared absorption of
%Nd$_{2}$CuO$_{2-\delta}$ ($\delta<0.004$) is shown as a function of frequency,
%up to 3000 cm$^{-1}$. The experimental results of Calvani and co-workers
%\cite{calva2} is represented by the thin black curve and by the shaded area.
%The so-called `d-band' rises in intensity around 600 cm$^{-1}$ and increases
%in intensity up to a maximum around 1000 cm$^{-1}$. The dotted curve shows the
%single polaron result. The full black curve represents the theoretical results
%obtained in the present work for the interacting many-polaron gas with
%$n_{0}=1.5\times10^{17}$ cm$^{-3}$, $\alpha=2.1$ and $m_{b}=0.5$ $m_{e}$.
%(From Ref. \cite{TDPRB01}.)}}{\Qlb{P2Fig1}}{part2fig1.eps}%
%maintain-aspect-ratio TRUE;  display "ICON";  valid_file "F";
%width 9.0325cm;  height 6.5789cm;  depth 0pt;  original-width 16.2968cm;
%original-height 11.8332cm;  cropleft "0";  croptop "1";  cropright "1";
%cropbottom "0";  filename 'Part2Fig1.eps';file-properties "XNPEU";}} }%
%BeginExpansion
\begin{figure}
[h]
\begin{center}
\includegraphics[
height=6.5789cm,
width=9.0325cm
]%
{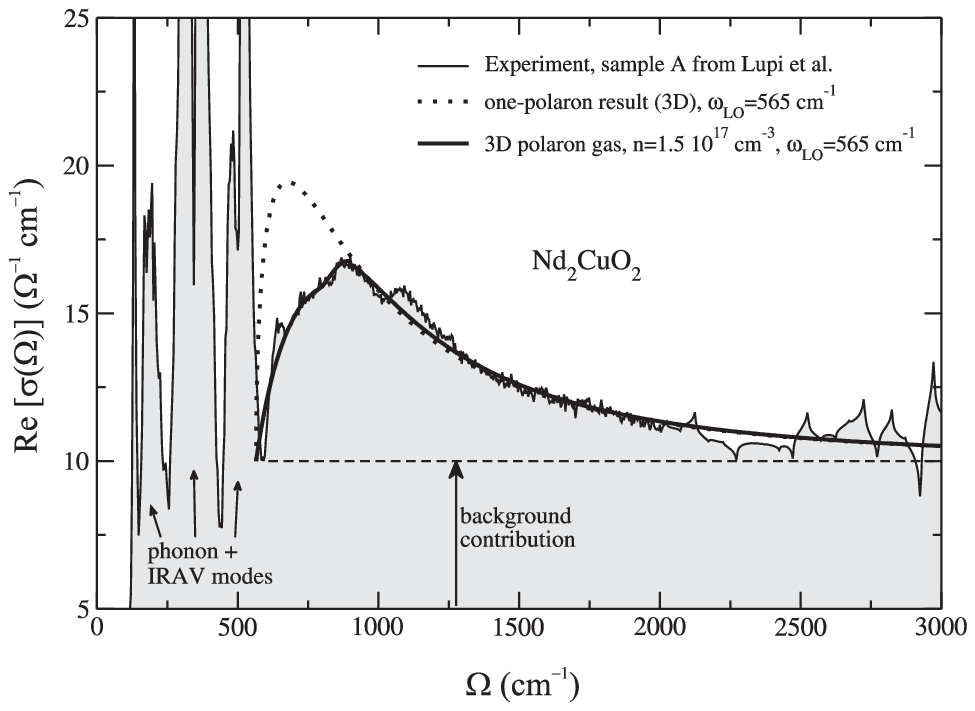}%
\caption{The infrared absorption of Nd$_{2}$CuO$_{2-\delta}$ ($\delta<0.004$)
is shown as a function of frequency, up to 3000 cm$^{-1}$. The experimental
results of Calvani and co-workers \cite{calva2} is represented by the thin
black curve and by the shaded area. The so-called `d-band' rises in intensity
around 600 cm$^{-1}$ and increases in intensity up to a maximum around 1000
cm$^{-1}$. The dotted curve shows the single polaron result. The full black
curve represents the theoretical results obtained in the present work for the
interacting many-polaron gas with $n_{0}=1.5\times10^{17}$ cm$^{-3}$,
$\alpha=2.1$ and $m_{b}=0.5$ $m_{e}$. (From Ref. \cite{TDPRB01}.)}%
\label{P2Fig1}%
\end{center}
\end{figure}
%EndExpansion

%\newpage

For the lowest levels of Ce doping, the \emph{d}-band can be most clearly
distinguished from the other features. The experimental optical absorption
spectrum (up to 3000 cm$^{-1}$) of Nd$_{2}$CuO$_{2-\delta}$ ($\delta<0.004$),
obtained by Calvani and co-workers \cite{calva2}, is shown in Fig.
\ref{P2Fig1} (shaded area) together with the theoretical curve obtained by the
present method (full, bold curve) and, for reference, the one-polaron optical
absorption result (dotted curve). At lower frequencies (600-1000 cm$^{-1}$) a
marked difference between the single polaron optical absorption and the
many-polaron result is manifest. The experimental \emph{d}-band can be clearly
identified, rising in intensity at about 600 cm$^{-1}$, peaking around 1000
cm$^{-1}$, and then decreasing in intensity above that frequency. At a density
of $n_{0}=1.5\times10^{17}$ cm$^{-3}$, we found a remarkable agreement between
our theoretical predictions and the experimental curve.

\subsection{Experimental data on the optical absorption in manganites:
interpretation in terms of a many-polaron response}

\begin{figure}
[h]
\begin{center}
\includegraphics[
height=6.234cm,
width=8.3625cm
]%
{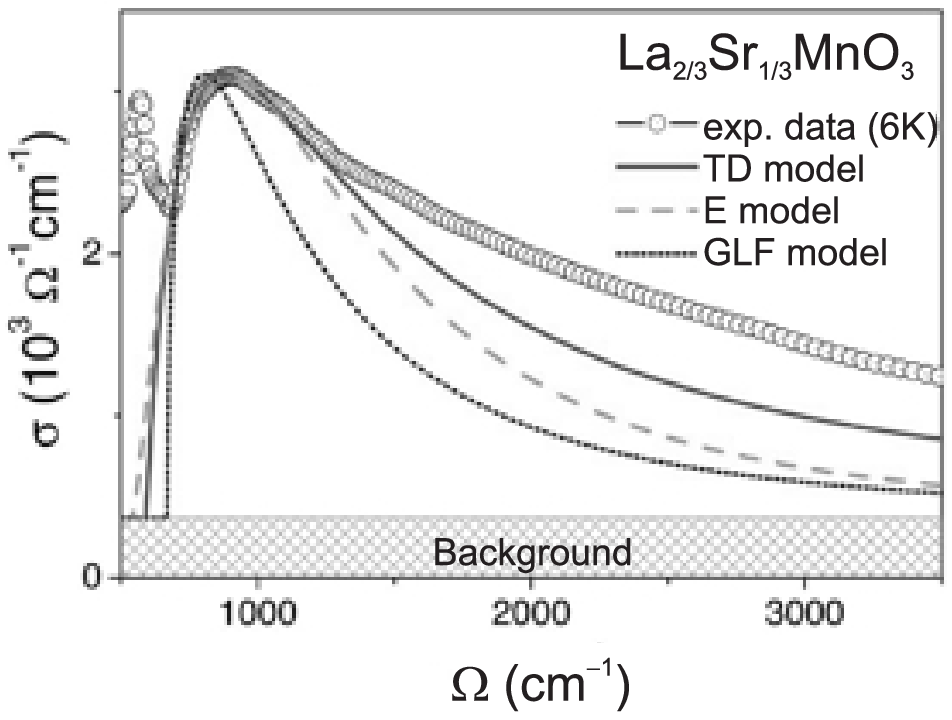}%
\caption{Comparison of the low-temperature MIR optical conductivity to
$\sigma\left(  \omega\right)  $ from various model calculations: the solid
line refers to the weak-coupling approach of Tempere and Devreese
\cite{TDPRB01} modified for an on-site Hubbard interaction, the dashed line is
the result of the phenomenological approach for self-trapped large polarons by
Emin \cite{Emin1993}, the dotted curve is the weak-coupling single-polaron
result \cite{GLF62}. (From Ref. \cite{Hartinger2004}.)}%
\label{P2Fig2}%
\end{center}
\end{figure}

In Refs. \cite{Hartinger2004,HartingerCondMat}, the experimental results on
the optical spectroscopy of La$_{2/3}$Sr$_{1/3}$MnO$_{3}$ (LSMO) and
La$_{2/3}$Ca$_{1/3}$MnO$_{3}$ (LCMO) thin films in the mid-infrared frequency
region are presented. The optical conductivity spectra of LCMO films are
interpreted in \cite{Hartinger2004,HartingerCondMat} in terms of the optical
response of small polarons, while the optical conductivity spectra of LSMO
films are explained using the large-polaron picture (see Fig. \ref{P2Fig2}).

%\newpage%
%

%TCIMACRO{\FRAME{fhFU}{8.3625cm}{6.234cm}{0pt}{\Qcb{Comparison of the
%low-temperature MIR optical conductivity to $\sigma\left(  \omega\right)  $
%from various model calculations: the solid line refers to the weak-coupling
%approach of Tempere and Devreese \cite{TDPRB01} modified for an on-site
%Hubbard interaction, the dashed line is the result of the phenomenological
%approach for self-trapped large polarons by Emin \cite{Emin1993}, the dotted
%curve is the weak-coupling single-polaron result \cite{GLF62}. (From Ref.
%\cite{Hartinger2004}.)}}{\Qlb{P2Fig2}}{part2fig2.eps}%
%maintain-aspect-ratio TRUE;  display "ICON";  valid_file "F";
%width 8.3625cm;  height 6.234cm;  depth 0pt;  original-width 16.2594cm;
%original-height 12.0836cm;  cropleft "0";  croptop "1";  cropright "1";
%cropbottom "0";  filename 'Part2Fig2.eps';file-properties "XNPEU";}} }%
%BeginExpansion
%EndExpansion

%\newpage

The real part of the optical conductivity $\operatorname{Re}\sigma\left(
\omega\right)  $ is expressed in Ref. \cite{HartingerCondMat} by the formula%
\begin{subequations}
\begin{equation}
\operatorname{Re}\sigma(\omega)=\alpha n_{p}\frac{2}{3}\frac{e^{2}}{m^{2}%
}\frac{\left(  \hbar\omega_{0}\right)  ^{2}}{\pi\hbar\omega^{3}}\sqrt
{\frac{\hbar}{2m\omega_{0}}}\int_{0}^{\infty}q^{2}S^{\left(  \mathrm{H}%
\right)  }(q,\omega-\omega_{0})dq, \label{sigma}%
\end{equation}
where $\alpha$ is the electron-phonon coupling constant, $n_{p}$ is the
polaron density, $m$ is the electron band mass, $\omega_{0}$ is the LO-phonon
frequency, and $S^{\left(  \mathrm{H}\right)  }(q,\omega)$ is the dynamic
structure factor, determined in Ref. \cite{HartingerCondMat} through the
dielectric function of an electron gas $\varepsilon\left(  \mathbf{q}%
,\omega\right)  $:%
\end{subequations}
\begin{equation}
S^{\left(  \mathrm{H}\right)  }(\mathbf{q},\omega)=\frac{\hbar}{n_{p}}%
\frac{q^{2}}{4\pi e^{2}}\operatorname{Im}\left[  -\frac{1}{\varepsilon\left(
\mathbf{q},\omega\right)  }\right]  . \label{s1}%
\end{equation}
In Ref. \cite{TDPRB01}, the other definition for the dynamic structure factor
is used, which is equivalent to that given by (\ref{s1}) (see Ref.
\cite{Mahan}) with the factor $N$ (the number of electrons)%
\begin{equation}
S^{\left(  \mathrm{TD}\right)  }(\mathbf{q},\omega)=\dfrac{1}{2}%
%TCIMACRO{\dint \limits_{-\infty}^{+\infty}}%
%BeginExpansion
{\displaystyle\int\limits_{-\infty}^{+\infty}}
%EndExpansion
\left\langle \varphi_{\text{el}}\left\vert \rho_{\mathbf{q}}\left(  t\right)
\rho_{-\mathbf{q}}\left(  0\right)  \right\vert \varphi_{\text{el}%
}\right\rangle e^{i\omega t}dt=NS^{\left(  \mathrm{H}\right)  }(\mathbf{q}%
,\omega). \label{sf}%
\end{equation}
Here, $\varphi_{\text{el}}$ denotes the ground state of the electron subsystem
(without the electron-phonon interaction), $\rho_{\mathbf{q}}=\sum_{j=1}%
^{N}e^{i\mathbf{q\cdot r}_{j}}$ is the Fourier component of the electron density.

In Ref. \cite{TDPRB01}, the dynamic structure factor is calculated within the
random-phase approximation (RPA) taking into account the Coulomb interaction
between electrons with the Fourier component of the Coulomb potential%
\begin{equation}
v_{\mathbf{q}}=\frac{4\pi e^{2}}{q^{2}\varepsilon_{\infty}}, \label{C}%
\end{equation}
where $\varepsilon_{\infty}$ is the high-frequency dielectric constant of the
crystal. In Refs. \cite{Hartinger2004,HartingerCondMat}, the dynamic structure
factor is calculated taking into account the local Hubbard electron-electron
interaction instead of the Coulomb interaction. The local Hubbard interaction
is used in the small-polaron formalism and describes the potential energy of
two electrons on one and the same site (see, e. g., Refs. \cite{Zhang,Hotta}).
In its simplest form the Hubbard interaction is (see Eq. (1) of Ref.
\cite{Zhang})%
\begin{equation}
V_{H}=U\sum_{i}n_{i\uparrow}n_{i\downarrow}, \label{Hub}%
\end{equation}
where $U$ is the coupling constant of the Hubbard interaction, $n_{i\uparrow}$
is the electron occupation number for the $i$-th site.

In Refs. \cite{Hartinger2004,HartingerCondMat}, there are no details of the
calculation using the interaction term (\ref{Hub}). The following procedure
can be supposed. The transition from the summation over the lattice sites to
the integral over the crystal volume $V$ is performed taking into account the
normalization condition%
\[
N_{\uparrow\left(  \downarrow\right)  }=\sum_{i}n_{i\uparrow\left(
\downarrow\right)  }=\int_{V}\tilde{n}_{\uparrow\left(  \downarrow\right)
}\left(  \mathbf{r}\right)  d\mathbf{r},
\]
where the density $\tilde{n}_{\uparrow\left(  \downarrow\right)  }\left(
\mathbf{r}\right)  $ is to be determined through $n_{i\uparrow\left(
\downarrow\right)  }$. As far as the lattice cell volume $\Omega_{0}\ll V$,
the integral $\int_{V}n_{\uparrow\left(  \downarrow\right)  }\left(
\mathbf{r}\right)  d\mathbf{r}$ can be written as the sum over the lattice
sites:%
\[
\int_{V}n_{\uparrow\left(  \downarrow\right)  }\left(  \mathbf{r}\right)
d\mathbf{r}=\Omega_{0}\sum_{i}\tilde{n}_{\uparrow\left(  \downarrow\right)
}\left(  \mathbf{r}_{i}\right)  ,
\]
where $\left\{  \mathbf{r}_{i}\right\}  $ are the vectors of the lattice.
Therefore, from the equality%
\[
\sum_{i}n_{i\uparrow\left(  \downarrow\right)  }=\Omega_{0}\sum_{i}\tilde
{n}_{\uparrow\left(  \downarrow\right)  }\left(  \mathbf{r}_{i}\right)
\]
we find that%
\[
n_{i\uparrow\left(  \downarrow\right)  }=\tilde{n}_{\uparrow\left(
\downarrow\right)  }\left(  \mathbf{r}_{i}\right)  .
\]
The potential (\ref{Hub}) is then transformed from the sum over sites to the
integral:%
\begin{equation}
V_{H}=\frac{U}{\Omega_{0}}\sum_{i}n_{i\uparrow}n_{i\downarrow}\Omega_{0}%
=\int_{V}U\Omega_{0}\delta\left(  \mathbf{r}-\mathbf{r}^{\prime}\right)
\tilde{n}_{\uparrow}\left(  \mathbf{r}\right)  \tilde{n}_{\downarrow}\left(
\mathbf{r}^{\prime}\right)  d\mathbf{r}d\mathbf{\mathbf{r}}^{\prime}.
\label{tran}%
\end{equation}
Consequently, in the continuum approach the Hubbard model is described by the
$\delta$-like interparticle potential $U\Omega_{0}\delta\left(  \mathbf{r}%
-\mathbf{r}^{\prime}\right)  $.

This development of the approach \cite{TDPRB01} performed in Refs.
\cite{Hartinger2004,HartingerCondMat} seems to be contradictory by the
following reason. For a many-polaron system, both the electron-phonon and
electron-electron interactions are provided by the electrostatic potentials.
Therefore, it would be consistent to consider them both within one and the
same approach. Namely, the Coulomb electron-electron interaction with the
potential (\ref{C}) is relevant for large and small polarons with the
Fr\"{o}hlich electron-phonon interaction, while the Hubbard electron-electron
interaction is relevant for small Holstein polarons. Nevertheless, as
recognized in Ref. \cite{Hartinger2004}, this model \textquotedblleft
reproduces the observed shape of the polaron peak quite
convincingly\textquotedblright\ and provides a better agreement with the
experiment \cite{Hartinger2004,HartingerCondMat} than the phenomenological
approach \cite{Emin1993} and the one-polaron theory \cite{GLF62}.

\newpage

\section{Interacting polarons in a quantum dot}

\subsection{The partition function and the free energy of a many-polaron
system}

We consider a system of $N$ electrons with mutual Coulomb repulsion and
interacting with the lattice vibrations following Ref. \cite{MPQD-PRB2004}.
The system is assumed to be confined by a parabolic potential characterized by
the frequency parameter $\Omega_{0}$. The total number of electrons is
represented as $N=\sum_{\sigma}N_{\sigma},$ where $N_{\sigma}$ is the number
of electrons with the spin projection $\sigma=\pm1/2$. The electron 3D (2D)
coordinates are denoted by $\mathbf{x}_{j,\sigma}$ with $j=1,\cdots,N_{\sigma
}.$ The bulk phonons (characterized by 3D wave vectors $\mathbf{k}$ and
frequencies $\omega_{\mathbf{k}}$) are described by the complex coordinates
$Q_{\mathbf{k}},$ which possess the property \cite{Feynman}
\begin{equation}
Q_{\mathbf{k}}^{\ast}=Q_{-\mathbf{k}}. \label{spN}%
\end{equation}
The full set of the electron and phonon coordinates are denoted by
$\mathbf{\bar{x}\equiv}\left\{  \mathbf{x}_{j,\sigma}\right\}  $ and $\bar
{Q}\equiv\left\{  Q_{\mathbf{k}}\right\}  .$

Throughout the present treatment, the Euclidean time variable $\tau=it$ is
used, where $t$ is the real time variable. In this representation the
Lagrangian of the system is
\begin{equation}
L\left(  \mathbf{\dot{\bar{x}}},\dot{\bar{Q}};\mathbf{\bar{x}},\bar{Q}\right)
=L_{e}\left(  \mathbf{\dot{\bar{x}}},\mathbf{\bar{x}}\right)  -V_{C}\left(
\mathbf{\bar{x}}\right)  -U_{b}\left(  \mathbf{\bar{x}}\right)  +L_{ph}\left(
\dot{\bar{Q}},\bar{Q}\right)  +L_{e-ph}\left(  \mathbf{\bar{x}},\bar
{Q}\right)  , \label{L}%
\end{equation}
where $L_{e}\left(  \mathbf{\dot{\bar{x}}},\mathbf{\bar{x}}\right)  $ is the
Lagrangian of an electron with band mass $m_{b}$ in a quantum dot:
\begin{equation}
L_{e}\left(  \mathbf{\dot{\bar{x}}},\mathbf{\bar{x}}\right)  =-\sum
_{\sigma=\pm1/2}\sum_{j=1}^{N_{\sigma}}\frac{m_{b}}{2}\left(  \mathbf{\dot{x}%
}_{j,\sigma}^{2}+\Omega_{0}^{2}\mathbf{x}_{j,\sigma}^{2}\right)
,\qquad\mathbf{\dot{x}\equiv}\frac{d\mathbf{x}}{d\tau}, \label{Le}%
\end{equation}
$\Omega_{0}$ is the confinement frequency, $V_{b}\left(  \mathbf{\bar{x}%
}\right)  $ is the potential of a background charge (supposed to be static and
uniformly distributed with the charge density $en_{b}$ in a sphere of a radius
$R$),%
\begin{align}
U_{b}\left(  \mathbf{\bar{x}}\right)   &  =\sum_{\sigma}\sum_{j=1}^{N}%
V_{b}\left(  r_{j,\sigma}\right)  ,\;r\equiv\left\vert \mathbf{x}\right\vert
,\label{Ub}\\
V_{b}\left(  r\right)   &  =-\frac{4\pi e^{2}n_{b}}{3\varepsilon_{0}}\left[
\Theta\left(  r<R\right)  \frac{3R^{2}-r^{2}}{2}+\Theta\left(  R\leq r\right)
\frac{R^{3}}{r}\right]  , \label{Vb}%
\end{align}
where $\varepsilon_{0}$ is the static dielectric constant of a crystal,
$V_{C}\left(  \mathbf{\bar{x}}\right)  $ is the potential energy of the
electron-electron Coulomb repulsion in the medium with the high-frequency
dielectric constant $\varepsilon_{\infty}$:
\begin{equation}
V_{C}\left(  \mathbf{\bar{x}}\right)  =\underset{\left(  j,\sigma\right)
\neq\left(  l,\sigma^{\prime}\right)  }{\sum_{\sigma,\sigma^{\prime}=\pm
1/2}\sum_{j=1}^{N_{\sigma}}\sum_{l=1}^{N_{\sigma^{\prime}}}\frac{e^{2}%
}{2\varepsilon_{\infty}}}\frac{1}{\left\vert \mathbf{x}_{j,\sigma}%
-\mathbf{x}_{l,\sigma^{\prime}}\right\vert }, \label{Vc}%
\end{equation}
$L_{ph}\left(  \dot{\bar{Q}},\dot{\bar{Q}}^{\ast};\bar{Q},\bar{Q}^{\ast
}\right)  $ is the Lagrangian of free phonons:
\begin{equation}
L_{ph}\left(  \dot{\bar{Q}},\bar{Q}\right)  =-\frac{1}{2}\sum_{\mathbf{k}%
}(\dot{Q}_{\mathbf{k}}^{\ast}\dot{Q}_{\mathbf{k}}+\omega_{\mathbf{k}}%
^{2}Q_{\mathbf{k}}^{\ast}Q_{\mathbf{k}}),\qquad\dot{Q}\mathbf{\equiv}\frac
{dQ}{d\tau}. \label{Lph}%
\end{equation}
Further, $L_{e-ph}\left(  \mathbf{\bar{x}},\bar{Q},\bar{Q}^{\ast}\right)  $ is
the Lagrangian of the electron-phonon interaction:
\begin{equation}
L_{e-ph}\left(  \mathbf{\bar{x}},\bar{Q}\right)  =-\sum_{\mathbf{k}}\left(
\frac{2\omega_{\mathbf{k}}}{\hbar}\right)  ^{1/2}V_{\mathbf{k}}Q_{-\mathbf{k}%
}\rho_{\mathbf{k}}, \label{Leph}%
\end{equation}
where $\rho_{\mathbf{k}}$ is the Fourier transform of the electron density
operator:%
\begin{equation}
\rho_{\mathbf{k}}=\sum_{\sigma=\pm1/2}\sum_{j=1}^{N_{\sigma}}e^{i\mathbf{k}%
\cdot\mathbf{x}_{j,\sigma}}.
\end{equation}
$V_{\mathbf{k}}$ is the amplitude of the electron-phonon interaction. Here, we
consider electrons interacting with the long-wavelength longitudinal optical
(LO) phonons with a dispersionless frequency $\omega_{\mathbf{k}}%
=\omega_{\mathrm{LO}}$, for which the amplitude $V_{\mathbf{k}}$ is
\cite{Devreese72}%
\begin{equation}
V_{\mathbf{k}}=\frac{\hbar\omega_{\mathrm{LO}}}{q}\left(  \frac{2\sqrt{2}%
\pi\alpha}{V}\right)  ^{1/2}\left(  \frac{\hbar}{m_{b}\omega_{\mathrm{LO}}%
}\right)  ^{1/4},
\end{equation}
where $\alpha$ is the electron-phonon coupling constant and $V$ is the volume
of the crystal.

We consider a \emph{canonical} ensemble, where the numbers $N_{\sigma}$ are
fixed. The partition function $Z\left(  \left\{  N_{\sigma}\right\}
,\beta\right)  $ of the system can be expressed as a path integral over the
electron and phonon coordinates:
\begin{equation}
Z\left(  \left\{  N_{\sigma}\right\}  ,\beta\right)  =\sum_{P}\frac{\left(
-1\right)  ^{\mathbf{\xi}_{P}}}{N_{1/2}!N_{-1/2}!}\int d\mathbf{\bar{x}}%
\int_{\mathbf{\bar{x}}}^{P\mathbf{\bar{x}}}D\mathbf{\bar{x}}\left(
\tau\right)  \int d\bar{Q}\int_{\bar{Q}}^{\bar{Q}}D\bar{Q}\left(  \tau\right)
e^{-S\left[  \mathbf{\bar{x}}\left(  \tau\right)  ,\bar{Q}\left(  \tau\right)
\right]  }, \label{Z}%
\end{equation}
where $S\left[  \mathbf{\bar{x}}\left(  \tau\right)  ,\bar{Q}\left(
\tau\right)  \right]  $ is the \textquotedblleft action\textquotedblright%
\ functional:
\begin{equation}
S\left[  \mathbf{\bar{x}}\left(  \tau\right)  ,\bar{Q}\left(  \tau\right)
\right]  =-\frac{1}{\hbar}\int_{0}^{\hbar\beta}L\left(  \mathbf{\dot{\bar{x}}%
},\dot{\bar{Q}};\mathbf{\bar{x}},\bar{Q}\right)  d\tau. \label{SN}%
\end{equation}
The parameter $\beta\equiv1/\left(  k_{B}T\right)  $ is inversely proportional
to the temperature $T$. In order to take the Fermi-Dirac statistics into
account, the integral over the electron paths $\left\{  \mathbf{\bar{x}%
}\left(  \tau\right)  \right\}  $ in Eq. (\ref{Z}) contains a sum over all
permutations $P$ of the electrons with the same spin projection, and
$\mathbf{\xi}_{P}$ denotes the parity of a permutation $P$.

The action functional (\ref{SN}) is quadratic in the phonon coordinates
$\bar{Q}.$ Therefore, \textit{the path integral over the phonon variables} in
$Z\left(  \left\{  N_{\sigma}\right\}  ,\beta\right)  $ \textit{can be
calculated analytically} following Ref. \cite{Feynman}. Let us describe this
path integration in detail. First, we introduce the real phonon coordinates
through the real and imaginary parts of the complex phonon coordinates
$Q_{\mathbf{k}}^{\prime}\equiv\operatorname{Re}Q_{\mathbf{k}}$, $Q_{\mathbf{k}%
}^{\prime\prime}\equiv\operatorname{Im}Q_{\mathbf{k}}$. According to the
symmetry property (\ref{spN}), they obey the equalities%
\begin{equation}
Q_{-\mathbf{k}}^{\prime}=Q_{\mathbf{k}}^{\prime},\;Q_{-\mathbf{k}}%
^{\prime\prime}=-Q_{\mathbf{k}}^{\prime\prime}. \label{sp1}%
\end{equation}%
\begin{equation}
q_{\mathbf{k}}\equiv\left\{
\begin{array}
[c]{c}%
\sqrt{2}Q_{\mathbf{k}}^{\prime},\;k_{x}\geq0,\\
\sqrt{2}Q_{\mathbf{k}}^{\prime\prime},\;k_{x}<0.
\end{array}
\right.  \label{det1}%
\end{equation}
In this representation, the sum over phonon coordinates $\sum_{\mathbf{k}%
}\left\vert Q_{\mathbf{k}}\right\vert ^{2}$ is transformed in the following
way using the symmetry property (\ref{sp1}):%
\begin{align*}
\sum_{\mathbf{k}}\left\vert Q_{\mathbf{k}}\right\vert ^{2}  &  =\sum
_{\mathbf{k}}\left[  \left(  Q_{\mathbf{k}}^{\prime}\right)  ^{2}+\left(
Q_{\mathbf{k}}^{\prime\prime}\right)  ^{2}\right] \\
&  =2\sum_{\substack{\mathbf{k}\\\left(  k_{x}\geq0\right)  }}\left(
Q_{\mathbf{k}}^{\prime}\right)  ^{2}+2\sum_{\substack{\mathbf{k}\\\left(
k_{x}<0\right)  }}\left(  Q_{\mathbf{k}}^{\prime\prime}\right)  ^{2}\\
&  =\sum_{\substack{\mathbf{k}\\\left(  k_{x}\geq0\right)  }}q_{\mathbf{k}%
}^{2}+\sum_{\substack{\mathbf{k}\\\left(  k_{x}<0\right)  }}q_{\mathbf{k}}%
^{2}=\sum_{\mathbf{k}}q_{\mathbf{k}}^{2}.
\end{align*}
Therefore, the phonon Lagrangian (\ref{Lph}) with the real phonon coordinates
is%
\begin{equation}
L_{ph}=-\frac{1}{2}\sum_{\mathbf{k}}(\dot{q}_{\mathbf{k}}^{2}+\omega
_{\mathbf{k}}^{2}q_{\mathbf{k}}^{2}). \label{Lph1}%
\end{equation}
The Lagrangian of the electron-phonon interaction (\ref{Leph}) with the real
phonon coordinates is transformed in the following way using (\ref{sp1}):%
\begin{align*}
L_{e-ph}  &  =-\sum_{\mathbf{k}}\left(  \frac{2\omega_{\mathbf{k}}}{\hbar
}\right)  ^{1/2}V_{\mathbf{k}}\rho_{\mathbf{k}}\left(  Q_{\mathbf{k}}^{\prime
}-iQ_{\mathbf{k}}^{\prime\prime}\right) \\
&  =-\sum_{\substack{\mathbf{k}\\\left(  k_{x}\geq0\right)  }}\left(
\frac{2\omega_{\mathbf{k}}}{\hbar}\right)  ^{1/2}\left(  V_{\mathbf{k}}%
\rho_{\mathbf{k}}+V_{-\mathbf{k}}\rho_{-\mathbf{k}}\right)  Q_{\mathbf{k}%
}^{\prime}\\
&  +i\sum_{\substack{\mathbf{k}\\\left(  k_{x}<0\right)  }}\left(
\frac{2\omega_{\mathbf{k}}}{\hbar}\right)  ^{1/2}\left(  V_{\mathbf{k}}%
\rho_{\mathbf{k}}-V_{-\mathbf{k}}\rho_{-\mathbf{k}}\right)  Q_{\mathbf{k}%
}^{\prime\prime}.
\end{align*}
Let us introduce the real forces:%
\begin{equation}
\gamma_{\mathbf{k}}\equiv\left\{
\begin{array}
[c]{c}%
\frac{1}{\sqrt{2}}\left(  \frac{2\omega_{\mathbf{k}}}{\hbar}\right)
^{1/2}\left(  V_{\mathbf{k}}\rho_{\mathbf{k}}+V_{-\mathbf{k}}\rho
_{-\mathbf{k}}\right)  ,\;k_{x}\geq0,\\
\frac{1}{i\sqrt{2}}\left(  \frac{2\omega_{\mathbf{k}}}{\hbar}\right)
^{1/2}\left(  V_{\mathbf{k}}\rho_{\mathbf{k}}-V_{-\mathbf{k}}\rho
_{-\mathbf{k}}\right)  ,\;k_{x}>0.
\end{array}
\right.  \label{realforces}%
\end{equation}
This gives us the Lagrangian of the electron-phonon interaction in terms of
the real forces and real phonon coordinates:%
\begin{equation}
L_{e-ph}=-\sum_{\mathbf{k}}\gamma_{\mathbf{k}}q_{\mathbf{k}}. \label{Leph1}%
\end{equation}
So, the sum of the Lagrangians of phonons and of the electron-phonon
interaction is expressed through ordinary real oscillator variables:%
\begin{equation}
L_{ph}+L_{e-ph}=-\frac{1}{2}\sum_{\mathbf{k}}(\dot{q}_{\mathbf{k}}^{2}%
+\omega_{\mathbf{k}}^{2}q_{\mathbf{k}}^{2}+\gamma_{\mathbf{k}}q_{\mathbf{k}}).
\label{L2}%
\end{equation}
The path integration for each phonon mode with the coordinate $q_{\mathbf{k}}$
is performed independently as described in Sec. 2 of Ref. \cite{Feynman} and
gives the result%
\begin{align*}
&  \int_{-\infty}^{\infty}dq_{\mathbf{k}}\int_{q_{\mathbf{k}}}^{q_{\mathbf{k}%
}}Dq_{\mathbf{k}}\left(  \tau\right)  \exp\left[  -\frac{1}{\hbar}\int
_{0}^{\hbar\beta}d\tau\frac{1}{2}(\dot{q}_{\mathbf{k}}^{2}+\omega_{\mathbf{k}%
}^{2}q_{\mathbf{k}}^{2}+\gamma_{\mathbf{k}}q_{\mathbf{k}})\right] \\
&  =\frac{1}{2\sinh\left(  \frac{\beta\hbar\omega_{\mathbf{k}}}{2}\right)  }\\
&  \times\exp\left\{  \frac{1}{4\hbar}\int_{0}^{\hbar\beta}d\tau\int
_{0}^{\hbar\beta}d\tau^{\prime}\frac{\cosh\left[  \omega_{\mathbf{k}}\left(
\left\vert \tau-\tau^{\prime}\right\vert -\hbar\beta/2\right)  \right]
}{\omega_{\mathbf{k}}\sinh\left(  \beta\hbar\omega_{\mathbf{k}}/2\right)
}\gamma_{\mathbf{k}}\left(  \tau\right)  \gamma_{\mathbf{k}}\left(
\tau^{\prime}\right)  ,\right\}
\end{align*}
where the exponential is the influence functional of a driven oscillator \{
\cite{Feynman}, Eq. (3.43)\}. Therefore, the path integral over all phonon
modes is%
\begin{align*}
&  \int d\left\{  q_{\mathbf{k}}\right\}  \int_{\left\{  q_{\mathbf{k}%
}\right\}  }^{\left\{  q_{\mathbf{k}}\right\}  }D\left\{  q_{\mathbf{k}%
}\left(  \tau\right)  \right\}  \exp\left[  -\frac{1}{\hbar}\int_{0}%
^{\hbar\beta}d\tau\sum_{\mathbf{k}}\frac{1}{2}(\dot{q}_{\mathbf{k}}^{2}%
+\omega_{\mathbf{k}}^{2}q_{\mathbf{k}}^{2}+\gamma_{\mathbf{k}}q_{\mathbf{k}%
})\right] \\
&  =\left(  \prod_{\mathbf{k}}\frac{1}{2\sinh\left(  \frac{\beta\hbar
\omega_{\mathbf{k}}}{2}\right)  }\right) \\
&  \times\exp\left\{  \frac{1}{4\hbar}\int_{0}^{\hbar\beta}d\tau\int
_{0}^{\hbar\beta}d\tau^{\prime}\sum_{\mathbf{k}}\frac{\cosh\left[
\omega_{\mathbf{k}}\left(  \left\vert \tau-\tau^{\prime}\right\vert
-\hbar\beta/2\right)  \right]  }{\omega_{\mathbf{k}}\sinh\left(  \beta
\hbar\omega_{\mathbf{k}}/2\right)  }\gamma_{\mathbf{k}}\left(  \tau\right)
\gamma_{\mathbf{k}}\left(  \tau^{\prime}\right)  .\right\}
\end{align*}
Here, the product $\prod_{\mathbf{k}}\ldots$ is the partition function of free
phonons, and the exponential is the influence functional of the phonon
subsystem on the electron subsystem. This influence functional results from
the above described \textit{elimination of the phonon coordinates} and is
usually written down as $e^{-\Phi}$, where $\Phi$ is%
\begin{equation}
\Phi=-\frac{1}{4\hbar}\int_{0}^{\hbar\beta}d\tau\int_{0}^{\hbar\beta}%
d\tau^{\prime}\sum_{\mathbf{k}}\frac{\cosh\left[  \omega_{\mathbf{k}}\left(
\left\vert \tau-\tau^{\prime}\right\vert -\hbar\beta/2\right)  \right]
}{\omega_{\mathbf{k}}\sinh\left(  \beta\hbar\omega_{\mathbf{k}}/2\right)
}\gamma_{\mathbf{k}}\left(  \tau\right)  \gamma_{\mathbf{k}}\left(
\tau^{\prime}\right)  . \label{Phi}%
\end{equation}
The sum over the phonon wave vectors $\mathbf{k}$ can be simplified as
follows:%
\begin{align*}
&  \sum_{\mathbf{k}}\frac{\cosh\left[  \omega_{\mathbf{k}}\left(  \left\vert
\tau-\tau^{\prime}\right\vert -\hbar\beta/2\right)  \right]  }{\omega
_{\mathbf{k}}\sinh\left(  \beta\hbar\omega_{\mathbf{k}}/2\right)  }%
\gamma_{\mathbf{k}}\left(  \tau\right)  \gamma_{\mathbf{k}}\left(
\tau^{\prime}\right) \\
&  =\frac{1}{\hbar}\sum_{\substack{\mathbf{k}\\\left(  k_{x}\geq0\right)
}}\frac{\cosh\left[  \omega_{\mathbf{k}}\left(  \left\vert \tau-\tau^{\prime
}\right\vert -\hbar\beta/2\right)  \right]  }{\sinh\left(  \beta\hbar
\omega_{\mathbf{k}}/2\right)  }\\
&  \times\left[  V_{\mathbf{k}}\rho_{\mathbf{k}}\left(  \tau\right)
+V_{-\mathbf{k}}\rho_{-\mathbf{k}}\left(  \tau\right)  \right]  \left[
V_{\mathbf{k}}\rho_{\mathbf{k}}\left(  \tau^{\prime}\right)  +V_{-\mathbf{k}%
}\rho_{-\mathbf{k}}\left(  \tau^{\prime}\right)  \right] \\
&  -\frac{1}{\hbar}\sum_{\substack{\mathbf{k}\\\left(  k_{x}<0\right)  }%
}\frac{\cosh\left[  \omega_{\mathbf{k}}\left(  \left\vert \tau-\tau^{\prime
}\right\vert -\hbar\beta/2\right)  \right]  }{\sinh\left(  \beta\hbar
\omega_{\mathbf{k}}/2\right)  }\\
&  \times\left[  V_{\mathbf{k}}\rho_{\mathbf{k}}\left(  \tau\right)
-V_{-\mathbf{k}}\rho_{-\mathbf{k}}\left(  \tau\right)  \right]  \left[
V_{\mathbf{k}}\rho_{\mathbf{k}}\left(  \tau^{\prime}\right)  -V_{-\mathbf{k}%
}\rho_{-\mathbf{k}}\left(  \tau^{\prime}\right)  \right] \\
&  =\frac{2}{\hbar}\sum_{\substack{\mathbf{k}\\\left(  k_{x}\geq0\right)
}}\frac{\cosh\left[  \omega_{\mathbf{k}}\left(  \left\vert \tau-\tau^{\prime
}\right\vert -\hbar\beta/2\right)  \right]  }{\sinh\left(  \beta\hbar
\omega_{\mathbf{k}}/2\right)  }V_{\mathbf{k}}V_{-\mathbf{k}}\\
&  \times\left[  \rho_{\mathbf{k}}\left(  \tau\right)  \rho_{-\mathbf{k}%
}\left(  \tau^{\prime}\right)  +\rho_{-\mathbf{k}}\left(  \tau\right)
\rho_{\mathbf{k}}\left(  \tau^{\prime}\right)  \right] \\
&  =\frac{2}{\hbar}\sum_{\mathbf{k}}\frac{\cosh\left[  \omega_{\mathbf{k}%
}\left(  \left\vert \tau-\tau^{\prime}\right\vert -\hbar\beta/2\right)
\right]  }{\sinh\left(  \beta\hbar\omega_{\mathbf{k}}/2\right)  }\left\vert
V_{\mathbf{k}}\right\vert ^{2}\rho_{\mathbf{k}}\left(  \tau\right)
\rho_{-\mathbf{k}}\left(  \tau^{\prime}\right)  .
\end{align*}
Herefrom, we find that%
\begin{equation}
\Phi=-\sum_{\mathbf{k}}\frac{\left\vert V_{\mathbf{k}}\right\vert ^{2}}%
{2\hbar^{2}}\int_{0}^{\hbar\beta}d\tau\int_{0}^{\hbar\beta}d\tau^{\prime}%
\frac{\cosh\left[  \omega_{\mathbf{k}}\left(  \left\vert \tau-\tau^{\prime
}\right\vert -\frac{\hbar\beta}{2}\right)  \right]  }{\sinh\left(  \frac
{\beta\hbar\omega_{\mathbf{k}}}{2}\right)  }\rho_{\mathbf{k}}\left(
\tau\right)  \rho_{-\mathbf{k}}\left(  \tau^{\prime}\right)  . \label{Phi1}%
\end{equation}
As a result, the partition function of the electron-phonon system (\ref{Z})
factorizes into a product%

\begin{equation}
Z\left(  \left\{  N_{\sigma}\right\}  ,\beta\right)  =Z_{p}\left(  \left\{
N_{\sigma}\right\}  ,\beta\right)  \prod_{\mathbf{k}}\frac{1}{2\sinh\left(
\beta\hbar\omega_{\mathbf{k}}/2\right)  } \label{Factorized}%
\end{equation}
of the partition function of free phonons with a partition function
$Z_{p}\left(  \left\{  N_{\sigma}\right\}  ,\beta\right)  $ of interacting
polarons, which is a path integral over the electron coordinates only:%
\begin{equation}
Z_{p}\left(  \left\{  N_{\sigma}\right\}  ,\beta\right)  =\sum_{P}%
\frac{\left(  -1\right)  ^{\mathbf{\xi}_{P}}}{N_{1/2}!N_{-1/2}!}\int
d\mathbf{\bar{x}}\int_{\mathbf{\bar{x}}}^{P\mathbf{\bar{x}}}D\mathbf{\bar{x}%
}\left(  \tau\right)  e^{-S_{p}\left[  \mathbf{\bar{x}}\left(  \tau\right)
\right]  }. \label{Zp}%
\end{equation}
The functional
\begin{equation}
S_{p}\left[  \mathbf{\bar{x}}\left(  \tau\right)  \right]  =-\frac{1}{\hbar
}\int_{0}^{\hbar\beta}\left[  L_{e}\left(  \mathbf{\dot{\bar{x}}}\left(
\tau\right)  ,\mathbf{\bar{x}}\left(  \tau\right)  \right)  -V_{C}\left(
\mathbf{\bar{x}}\left(  \tau\right)  \right)  \right]  d\tau+\Phi\left[
\mathbf{\bar{x}}\left(  \tau\right)  \right]  \label{Sp}%
\end{equation}
describes the phonon-induced retarded interaction between the electrons,
including the retarded self-interaction of each electron.

Using (\ref{Le}) and (\ref{Vc}) we write down $S_{p}\left[  \mathbf{\bar{x}%
}\left(  \tau\right)  \right]  $ explicitly:%
\begin{align}
S_{p}\left[  \mathbf{\bar{x}}\left(  \tau\right)  \right]   &  =\frac{1}%
{\hbar}\int_{0}^{\hbar\beta}\left[  \sum_{\sigma}\sum_{j=1}^{N_{\sigma}}%
\frac{m_{b}}{2}\left(  \mathbf{\dot{x}}_{j,\sigma}^{2}+\Omega_{0}%
^{2}\mathbf{x}_{j,\sigma}^{2}\right)  \underset{\left(  j,\sigma\right)
\neq\left(  l,\sigma^{\prime}\right)  }{+\sum_{\sigma,\sigma^{\prime}}%
\sum_{j=1}^{N_{\sigma}}\sum_{l=1}^{N_{\sigma^{\prime}}}}\frac{e^{2}%
}{2\varepsilon_{\infty}\left\vert \mathbf{x}_{j,\sigma}-\mathbf{x}%
_{l,\sigma^{\prime}}\right\vert }\right]  d\tau\nonumber\\
&  -\sum_{\mathbf{q}}\frac{\left\vert V_{\mathbf{q}}\right\vert ^{2}}%
{2\hbar^{2}}\int\limits_{0}^{\hbar\beta}d\tau\int\limits_{0}^{\hbar\beta}%
d\tau^{\prime}\frac{\cosh\left[  \omega_{\mathrm{LO}}\left(  \left\vert
\tau-\tau^{\prime}\right\vert -\hbar\beta/2\right)  \right]  }{\sinh\left(
\beta\hbar\omega_{\mathrm{LO}}/2\right)  }\rho_{\mathbf{q}}\left(
\tau\right)  \rho_{-\mathbf{q}}\left(  \tau^{\prime}\right)  . \label{Sp1}%
\end{align}

The free energy of a system of interacting polarons $F_{p}\left(  \left\{
N_{\sigma}\right\}  ,\beta\right)  $ is related to their partition function
(\ref{Zp}) by the equation:
\begin{equation}
F_{p}\left(  \left\{  N_{\sigma}\right\}  ,\beta\right)  =-\frac{1}{\beta}\ln
Z_{p}\left(  \left\{  N_{\sigma}\right\}  ,\beta\right)  . \label{Fp}%
\end{equation}

At present no method is known to calculate the non-gaussian path integral
(\ref{Zp}) analytically. For \emph{distinguishable} particles, the
Jensen-Feynman variational principle \cite{Feynman} provides a convenient
approximation technique. It yields a lower bound to the partition function,
and hence an upper bound to the free energy.

It can be shown \cite{PRE96} that the path-integral approach to the many-body
problem for a fixed number of identical particles can be formulated as a
Feynman-Kac functional on a state space for $N$ indistinguishable particles,
by imposing an ordering on the configuration space and by the introduction of
a set of boundary conditions at the boundaries of this state space. The
resulting variational inequality for identical particles takes the same form
as the Jensen-Feynman variational principle:
\begin{align}
F_{p}  &  \leqslant F_{var},\label{JF}\\
F_{var}  &  =F_{0}+\frac{1}{\beta}\left\langle S_{p}-S_{0}\right\rangle
_{S_{0}}, \label{Fvar}%
\end{align}
where $S_{0}$ is a model action with the corresponding free energy $F_{0}$.
The angular brackets mean a weighted average over the paths%
\begin{equation}
\left\langle \left(  \bullet\right)  \right\rangle _{S_{0}}=\frac{\sum
_{P}\frac{\left(  -1\right)  ^{\mathbf{\xi}_{P}}}{N_{1/2}!N_{-1/2}!}\int
d\mathbf{\bar{x}}\int_{\mathbf{\bar{x}}}^{P\mathbf{\bar{x}}}D\mathbf{\bar{x}%
}\left(  \tau\right)  \left(  \bullet\right)  e^{-S_{0}\left[  \mathbf{\bar
{x}}\left(  \tau\right)  \right]  }}{\sum_{P}\frac{\left(  -1\right)
^{\mathbf{\xi}_{P}}}{N_{1/2}!N_{-1/2}!}\int d\mathbf{\bar{x}}\int
_{\mathbf{\bar{x}}}^{P\mathbf{\bar{x}}}D\mathbf{\bar{x}}\left(  \tau\right)
e^{-S_{0}\left[  \mathbf{\bar{x}}\left(  \tau\right)  \right]  }}.
\label{Aver}%
\end{equation}

In the zero-temperature limit, the polaron ground-state energy
\begin{equation}
E_{p}^{0}=\lim_{\beta\rightarrow\infty}F_{p}%
\end{equation}
obeys the inequality following from (\ref{JF}) with (\ref{Fvar}):%
\[
E_{p}^{0}\leqslant E_{var}%
\]
with%
\begin{align}
E_{var}  &  =E_{0}^{0}+\lim_{\beta\rightarrow\infty}\left(  \frac{1}{\beta
}\left\langle S_{p}-S_{0}\right\rangle _{S_{0}}\right)  ,\\
E_{0}^{0}  &  =\lim_{\beta\rightarrow\infty}F_{0}.
\end{align}

\subsection{Model system}

We consider a model system consisting of $N$ electrons with coordinates
$\mathbf{\bar{x}\equiv}\left\{  \mathbf{x}_{j,\sigma}\right\}  $ and $N_{f}$
fictitious particles with coordinates $\mathbf{\bar{y}\equiv}\left\{
\mathbf{y}_{j}\right\}  $ in a harmonic confinement potential with elastic
interparticle interactions as studied in Refs. \cite{SSC114-305,MPQD-PRB2004}.
The Lagrangian of this model system takes the form%
\begin{align}
L_{M}\left(  \mathbf{\dot{\bar{x}}},\mathbf{\dot{\bar{y}}};\mathbf{\bar{x}%
},\mathbf{\bar{y}}\right)   &  =-\frac{m_{b}}{2}\sum_{\sigma}\sum
_{j=1}^{N_{\sigma}}\left(  \mathbf{\dot{x}}_{j,\sigma}^{2}+\Omega
^{2}\mathbf{x}_{j,\sigma}^{2}\right)  +\frac{m_{b}\omega^{2}}{4}\sum
_{\sigma,\sigma^{\prime}}\sum_{j=1}^{N_{\sigma}}\sum_{l=1}^{N_{\sigma^{\prime
}}}\left(  \mathbf{x}_{j,\sigma}-\mathbf{x}_{l,\sigma^{\prime}}\right)
^{2}\nonumber\\
&  -\frac{m_{f}}{2}\sum_{j=1}^{N_{f}}\left(  \mathbf{\dot{y}}_{j}^{2}%
+\Omega_{f}^{2}\mathbf{y}_{j}^{2}\right)  -\frac{k}{2}\sum_{\sigma}\sum
_{j=1}^{N_{\sigma}}\sum_{l=1}^{N_{f}}\left(  \mathbf{x}_{j,\sigma}%
-\mathbf{y}_{l}\right)  ^{2}. \label{LM}%
\end{align}
The frequencies $\Omega,$ $\omega,$ $\Omega_{f},$ the mass of a fictitious
particle $m_{f},$ and the force constant $k$ are variational parameters.
Clearly, this Lagrangian is symmetric with respect to electron permutations.
Performing the path integral over the coordinates of the fictitious particles
in the same way as described above for phonons, the partition function
$Z_{0}\left(  \left\{  N_{\sigma}\right\}  ,\beta\right)  $ of the model
system of interacting polarons becomes a path integral over the electron
coordinates:%
\begin{equation}
Z_{0}\left(  \left\{  N_{\sigma}\right\}  ,\beta\right)  =\sum_{P}%
\frac{\left(  -1\right)  ^{\mathbf{\xi}_{P}}}{N_{1/2}!N_{-1/2}!}\int
d\mathbf{\bar{x}}\int_{\mathbf{\bar{x}}}^{P\mathbf{\bar{x}}}D\mathbf{\bar{x}%
}\left(  \tau\right)  e^{-S_{0}\left[  \mathbf{\bar{x}}\left(  \tau\right)
\right]  }, \label{Z0}%
\end{equation}
with the action functional $S_{0}\left[  \mathbf{\bar{x}}\left(  \tau\right)
\right]  $ given by%
\begin{align}
S_{0}\left[  \mathbf{\bar{x}}\left(  \tau\right)  \right]   &  =\frac{1}%
{\hbar}\int_{0}^{\hbar\beta}\sum_{\sigma}\sum_{j=1}^{N_{\sigma}}\frac{m_{b}%
}{2}\left[  \mathbf{\dot{x}}_{j,\sigma}^{2}\left(  \tau\right)  +\Omega
^{2}\mathbf{x}_{j,\sigma}^{2}\left(  \tau\right)  \right]  d\tau\nonumber\\
&  -\frac{1}{\hbar}\int_{0}^{\hbar\beta}\sum_{\sigma,\sigma^{\prime}}%
\sum_{j=1}^{N_{\sigma}}\sum_{l=1}^{N_{\sigma^{\prime}}}\frac{m_{b}\omega^{2}%
}{4}\left[  \mathbf{x}_{j,\sigma}\left(  \tau\right)  -\mathbf{x}%
_{l,\sigma^{\prime}}\left(  \tau\right)  \right]  ^{2}d\tau\nonumber\\
&  -\frac{k^{2}N^{2}N_{f}}{4m_{f}\hbar\Omega_{f}}\int\limits_{0}^{\hbar\beta
}d\tau\int\limits_{0}^{\hbar\beta}d\tau^{\prime}\frac{\cosh\left[  \Omega
_{f}\left(  \left\vert \tau-\tau^{\prime}\right\vert -\hbar\beta/2\right)
\right]  }{\sinh\left(  \beta\hbar\Omega_{f}/2\right)  }\mathbf{X}\left(
\tau\right)  \cdot\mathbf{X}\left(  \tau^{\prime}\right)  , \label{S0}%
\end{align}
where $\mathbf{X}$ is the center-of-mass coordinate of the electrons,%
\begin{equation}
\mathbf{X=}\frac{1}{N}\sum_{\sigma}\sum_{j=1}^{N_{\sigma}}\mathbf{x}%
_{j,\sigma}. \label{X}%
\end{equation}

\subsubsection{Analytical calculation of the model partition function}

The partition function $Z_{0}\left(  \left\{  N_{\sigma}\right\}
,\beta\right)  $ [Eq. (\ref{Z0})] for the model system of interacting polarons
can be expressed in terms of the partition function $Z_{M}\left(  \left\{
N_{\sigma}\right\}  ,N_{f},\beta\right)  $ of the model system of interacting
electrons and fictitious particles with the Lagrangian $L_{M}$ [Eq.
(\ref{LM})] as follows:%
\begin{equation}
Z_{0}\left(  \left\{  N_{\sigma}\right\}  ,\beta\right)  =\frac{Z_{M}\left(
\left\{  N_{\sigma}\right\}  ,N_{f},\beta\right)  }{Z_{f}\left(  N_{f}%
,w_{f},\beta\right)  }, \label{r1}%
\end{equation}
where $Z_{f}\left(  N_{f},w_{f},\beta\right)  $ is the partition function of
fictitious particles,%
\begin{equation}
Z_{f}\left(  N_{f},\beta\right)  =\frac{1}{\left(  2\sinh\frac{1}{2}\beta\hbar
w_{f}\right)  ^{DN_{f}}}, \label{Zf}%
\end{equation}
with the frequency%
\begin{equation}
w_{f}=\sqrt{\Omega_{f}^{2}+kN/m_{f}} \label{wf}%
\end{equation}
and D=3(2) for 3D(2D) systems. The partition function $Z_{M}\left(  \left\{
N_{\sigma}\right\}  ,N_{f},\beta\right)  $ is the path integral for both the
electrons and the fictitious particles:%
\begin{align}
Z_{M}\left(  \left\{  N_{\sigma}\right\}  ,N_{f},\beta\right)   &  =\sum
_{P}\frac{\left(  -1\right)  ^{\mathbf{\xi}_{P}}}{N_{1/2}!N_{-1/2}%
!}\nonumber\\
&  \int d\mathbf{\bar{x}}\int_{\mathbf{\bar{x}}}^{P\mathbf{\bar{x}}%
}D\mathbf{\bar{x}}\left(  \tau\right)  \int d\mathbf{\bar{y}}\int
_{\mathbf{\bar{y}}}^{\mathbf{\bar{y}}}D\mathbf{\bar{y}}\left(  \tau\right)
e^{-S_{M}\left[  \mathbf{\bar{x}}\left(  \tau\right)  ,\mathbf{\bar{y}}\left(
\tau\right)  \right]  } \label{ZM}%
\end{align}
with the \textquotedblleft action\textquotedblright\ functional
\begin{equation}
S_{M}\left[  \mathbf{\bar{x}}\left(  \tau\right)  ,\mathbf{\bar{y}}\left(
\tau\right)  \right]  =-\frac{1}{\hbar}\int_{0}^{\hbar\beta}L_{M}\left(
\mathbf{\dot{\bar{x}}},\mathbf{\dot{\bar{y}}};\mathbf{\bar{x}},\mathbf{\bar
{y}}\right)  d\tau, \label{SM}%
\end{equation}
where the Lagrangian is given by Eq. (\ref{LM}).

Let us consider an auxiliary \textquotedblleft ghost\textquotedblright%
\ subsystem with the Lagrangian
\begin{equation}
L_{g}\left(  \mathbf{\dot{X}}_{g},\mathbf{\dot{Y}}_{g},\mathbf{X}%
_{g},\mathbf{Y}_{g}\right)  =-\frac{m_{b}N}{2}\left(  \mathbf{\dot{X}}_{g}%
^{2}+w^{2}\mathbf{X}_{g}^{2}\right)  -\frac{m_{f}N_{f}}{2}\left(
\mathbf{\dot{Y}}_{g}^{2}+w_{f}^{2}\mathbf{Y}_{g}^{2}\right)  \label{ghost}%
\end{equation}
with two frequencies $w$ and $w_{f},$ where $w$ is given by%
\begin{equation}
w=\sqrt{\Omega^{2}-N\omega^{2}+kN_{f}/m_{b}}. \label{wab}%
\end{equation}
The partition function $Z_{g}$ of this subsystem
\begin{equation}
Z_{g}=\int d\mathbf{X}_{g}\int d\mathbf{Y}_{g}\int\limits_{\mathbf{X}_{g}%
}^{\mathbf{X}_{g}}D\mathbf{X}_{g}\left(  \tau\right)  \int\limits_{\mathbf{Y}%
_{g}}^{\mathbf{Y}_{g}}D\mathbf{Y}_{g}\left(  \tau\right)  \exp\left\{
-S_{g}\left[  \mathbf{X}_{g}\left(  \tau\right)  ,\mathbf{Y}_{g}\left(
\tau\right)  \right]  \right\}  , \label{Zg1}%
\end{equation}
with the \textquotedblleft action\textquotedblright\ functional%
\begin{equation}
S_{g}\left[  \mathbf{X}_{g}\left(  \tau\right)  ,\mathbf{Y}_{g}\left(
\tau\right)  \right]  =-\frac{1}{\hbar}\int\limits_{0}^{\hbar\beta}%
L_{g}\left(  \mathbf{\dot{X}}_{g},\mathbf{X}_{g},\mathbf{\dot{Y}}%
_{g},\mathbf{Y}_{g}\right)  \,d\tau\label{Sg}%
\end{equation}
is calculated in the standard way, because its Lagrangian (\ref{ghost}) has a
simple oscillator form. Consequently, the partition function $Z_{g}$ is%
\begin{equation}
Z_{g}=\frac{1}{\left[  2\sinh\left(  \frac{\beta\hbar w}{2}\right)  \right]
^{D}}\frac{1}{\left[  2\sinh\left(  \frac{\beta\hbar w_{f}}{2}\right)
\right]  ^{D}}. \label{Zg2}%
\end{equation}

The product $Z_{g}Z_{M}$ of the two partition functions $Z_{g}$ and
$Z_{M}\left(  \left\{  N_{\sigma}\right\}  ,N_{f},\beta\right)  $ is a path
integral in the state space of $N$ electrons, $N_{f}$ fictitious particles and
two \textquotedblleft ghost\textquotedblright\ particles with the coordinate
vectors $\mathbf{X}_{g}$ and $\mathbf{Y}_{g}.$ The Lagrangian $\tilde{L}_{M}$
of this system is a sum of $L_{M}$ and $L_{g},$%
\begin{equation}
\tilde{L}_{M}\left(  \mathbf{\dot{\bar{x}}},\mathbf{\dot{\bar{y}},\dot{X}}%
_{g},\mathbf{\dot{Y}}_{g};\mathbf{\bar{x}},\mathbf{\bar{y},X}_{g}%
,\mathbf{Y}_{g}\right)  \equiv L_{M}\left(  \mathbf{\dot{\bar{x}}%
},\mathbf{\dot{\bar{y}}};\mathbf{\bar{x}},\mathbf{\bar{y}}\right)
+L_{g}\left(  \mathbf{\dot{X}}_{g},\mathbf{\dot{Y}}_{g},\mathbf{X}%
_{g},\mathbf{Y}_{g}\right)  . \label{LMt}%
\end{equation}
The \textquotedblleft ghost\textquotedblright\ subsystem is introduced because
the center-of-mass coordinates in $\tilde{L}_{M}$ can be explicitly separated
much more transparently than in $L_{M}$. This separation is realized by the
linear transformation of coordinates,%
\begin{equation}
\left\{
\begin{array}
[c]{l}%
\mathbf{x}_{j,\sigma}=\mathbf{x}_{j,\sigma}^{\prime}+\mathbf{X}-\mathbf{X}%
_{g},\\
\mathbf{y}_{j\sigma}=\mathbf{y}_{j\sigma}^{\prime}+\mathbf{Y}-\mathbf{Y}_{g},
\end{array}
\right.  \label{Trans}%
\end{equation}
where $\mathbf{X}$\ and $\mathbf{Y}$ are the center-of-mass coordinate vectors
of the electrons and of the fictitious particles, correspondingly:%
\begin{equation}
\mathbf{X=}\frac{1}{N}\sum_{\sigma}\sum_{j=1}^{N_{\sigma}}\mathbf{x}%
_{j,\sigma},\quad\mathbf{Y=}\frac{1}{N_{f}}\sum_{j=1}^{N_{f}}\mathbf{y}_{j}.
\label{XY}%
\end{equation}
Before the transformation (\ref{Trans}), the independent variables are
$\left(  \mathbf{\bar{x}},\mathbf{\bar{y},X}_{g},\mathbf{Y}_{g}\right)  ,$
with the center-of-mass coordinates $\mathbf{X}$ and $\mathbf{Y}$ determined
by Eq. (\ref{XY}). When applying the transformation (\ref{Trans}) to the
centers of mass (\ref{XY}), we find that%
\begin{align}
\mathbf{X}  &  =\frac{1}{N}\sum_{\sigma}\sum_{j=1}^{N_{\sigma}}\left(
\mathbf{x}_{j,\sigma}^{\prime}+\mathbf{X}-\mathbf{X}_{g}\right)  =\frac{1}%
{N}\sum_{\sigma}\sum_{j=1}^{N_{\sigma}}\mathbf{x}_{j,\sigma}^{\prime
}+\mathbf{X}-\mathbf{X}_{g},\label{t1}\\
\mathbf{Y}  &  \mathbf{=}\frac{1}{N_{f}}\sum_{j=1}^{N_{f}}\left(
\mathbf{y}_{j\sigma}^{\prime}+\mathbf{Y}-\mathbf{Y}_{g}\right)  =\frac
{1}{N_{f}}\sum_{j=1}^{N_{f}}\mathbf{y}_{j\sigma}^{\prime}+\mathbf{Y}%
-\mathbf{Y}_{g}. \label{t2}%
\end{align}
As seen from Eqs. (\ref{t1}), (\ref{t2}), after the transformation
(\ref{Trans}) the independent variables are $\left(  \mathbf{\bar{x}}^{\prime
},\mathbf{\bar{y}}^{\prime},\mathbf{X,Y}\right)  ,$ while the coordinates
$\left(  \mathbf{X}_{g},\mathbf{Y}_{g}\right)  $ obey the equations%
\begin{equation}
\mathbf{X}_{g}\mathbf{=}\frac{1}{N}\sum_{\sigma}\sum_{j=1}^{N_{\sigma}%
}\mathbf{x}_{j,\sigma}^{\prime},\quad\mathbf{Y}_{g}\mathbf{=}\frac{1}{N_{f}%
}\sum_{j=1}^{N_{f}}\mathbf{y}_{j}^{\prime}. \label{XgYg}%
\end{equation}

In order to find the explicit form of the Lagrangian (\ref{LMt}) after the
transformation (\ref{Trans}), we use the following relations for the quadratic
sums of coordinates:
\begin{gather}
\sum_{\sigma}\sum_{j=1}^{N_{\sigma}}\mathbf{x}_{j,\sigma}^{2}=\sum_{\sigma
}\sum_{j=1}^{N_{\sigma}}\left(  \mathbf{x}_{j,\sigma}^{\prime}\right)
^{2}+N\left(  \mathbf{X}^{2}-\mathbf{X}_{g}^{2}\right)  ,\quad\sum
_{j=1}^{N_{f}}\mathbf{y}_{j}^{2}=\sum_{j=1}^{N_{f}}\left(  \mathbf{y}%
_{j}^{\prime}\right)  ^{2}+N_{f}\left(  \mathbf{Y}^{2}-\mathbf{Y}_{g}%
^{2}\right)  ,\nonumber\\
\sum_{\sigma,\sigma^{\prime}}\sum_{j=1}^{N_{\sigma}}\sum_{l=1}^{N_{\sigma
^{\prime}}}\left(  \mathbf{x}_{j,\sigma}-\mathbf{x}_{l,\sigma^{\prime}%
}\right)  ^{2}=2N\sum_{\sigma}\sum_{j=1}^{N_{\sigma}}\left(  \mathbf{x}%
_{j,\sigma}^{\prime}\right)  ^{2}-2N_{f}^{2}\mathbf{X}_{g}^{2},\nonumber\\
\sum_{j=1}^{N_{f}}\sum_{l=1}^{N_{f}}\left(  \mathbf{y}_{j}-\mathbf{y}%
_{l}\right)  ^{2}=2N_{f}\sum_{j=1}^{N_{f}}\left(  \mathbf{y}_{j}^{\prime
}\right)  ^{2}-2N_{f}^{2}\mathbf{Y}_{g}^{2},\nonumber\\
\sum_{\sigma}\sum_{j=1}^{N_{\sigma}}\sum_{l=1}^{N_{f}}\left(  \mathbf{x}%
_{j,\sigma}-\mathbf{y}_{l}\right)  ^{2}=N_{f}\sum_{\sigma}\sum_{j=1}%
^{N_{\sigma}}\left(  \mathbf{x}_{j,\sigma}^{\prime}\right)  ^{2}+N\sum
_{j=1}^{N_{f}}\left(  \mathbf{y}_{j}^{\prime}\right)  ^{2}+\nonumber\\
NN_{f}\left(  \mathbf{X}^{2}+\mathbf{Y}^{2}-2\mathbf{X}\cdot\mathbf{Y}%
-\mathbf{X}_{g}^{2}-\mathbf{Y}_{g}^{2}\right)  . \label{tosum}%
\end{gather}

The substitution of Eq. (\ref{XY}) into Eq. (\ref{LMt}) then results in the
following 3 terms:%
\begin{equation}
\tilde{L}_{M}\left(  \mathbf{\dot{\bar{x}}}^{\prime},\mathbf{\dot{\bar{y}}%
}^{\prime},\mathbf{\dot{X},\dot{Y};\bar{x}}^{\prime},\mathbf{\bar{y}}^{\prime
},\mathbf{X,Y}\right)  =L_{w}\left(  \mathbf{\dot{\bar{x}}}^{\prime
},\mathbf{\bar{x}}^{\prime}\right)  +L_{w_{f}}\left(  \mathbf{\dot{\bar{y}}%
}^{\prime},\mathbf{\bar{y}}^{\prime}\right)  +L_{C}\left(  \mathbf{\dot{X}%
,X};\mathbf{\dot{Y},Y}\right)  , \label{LM1}%
\end{equation}
where $L_{w}\left(  \mathbf{\dot{\bar{x}}}^{\prime},\mathbf{\bar{x}}^{\prime
}\right)  $ and $L_{w_{f}}\left(  \mathbf{\dot{\bar{y}}}^{\prime}%
,\mathbf{\bar{y}}^{\prime}\right)  $ are Lagrangians of non-interacting
identical oscillators with the frequencies $w$ and $w_{f},$ respectively,%
\begin{align}
L_{w}\left(  \mathbf{\dot{\bar{x}}}^{\prime},\mathbf{\bar{x}}^{\prime}\right)
&  =-\frac{m_{b}}{2}\sum_{\sigma=\pm1/2}\sum_{j=1}^{N_{\sigma}}\left[  \left(
\mathbf{\dot{x}}_{j,\sigma}^{\prime}\right)  ^{2}+w^{2}\left(  \mathbf{x}%
_{j,\sigma}^{\prime}\right)  ^{2}\right]  ,\label{Lwa}\\
L_{w_{f}}\left(  \mathbf{\dot{\bar{y}}}^{\prime},\mathbf{\bar{y}}^{\prime
}\right)   &  =-\frac{m_{f}}{2}\sum_{j=1}^{N_{f}}\left[  \left(
\mathbf{\dot{y}}_{j,\sigma}^{\prime}\right)  ^{2}+w_{f}^{2}\left(
\mathbf{y}_{j,\sigma}^{\prime}\right)  ^{2}\right]  . \label{Lwb}%
\end{align}
The Lagrangian $L_{C}\left(  \mathbf{\dot{X},X};\mathbf{\dot{Y},Y}\right)  $
describes the combined motion of the centers-of-mass of the electrons and of
the fictitious particles,%
\begin{equation}
L_{C}\left(  \mathbf{\dot{X},X};\mathbf{\dot{Y},Y}\right)  =-\frac{m_{b}N}%
{2}\left(  \mathbf{\dot{X}}^{2}+\tilde{\Omega}^{2}\mathbf{X}^{2}\right)
-\frac{m_{f}N_{f}}{2}\left(  \mathbf{\dot{Y}}^{2}+w_{f}^{2}\mathbf{Y}%
^{2}\right)  +kNN_{f}\mathbf{X\cdot Y,} \label{LC}%
\end{equation}
with%
\begin{equation}
\tilde{\Omega}=\sqrt{\Omega^{2}+kN_{f}/m_{b}}. \label{Ot}%
\end{equation}

The Lagrangian (\ref{LC}) is reduced to a diagonal quadratic form in the
coordinates and the velocities by a unitary transformation for two interacting
oscillators using the following replacement of variables:
\begin{align}
\mathbf{X}  &  =\frac{1}{\sqrt{m_{b}N}}\left(  a_{1}\mathbf{r}+a_{2}%
\mathbf{R}\right)  ,\nonumber\\
\mathbf{Y}  &  =\frac{1}{\sqrt{m_{f}N_{f}}}\left(  -a_{2}\mathbf{r}%
+a_{1}\mathbf{R}\right)
\end{align}
with the coefficients
\begin{align}
a_{1}  &  =\left[  \frac{1+\chi}{2}\right]  ^{1/2},\quad a_{2}=\left[
\frac{1-\chi}{2}\right]  ^{1/2},\label{a12}\\
\chi &  \equiv\frac{\tilde{\Omega}^{2}-\tilde{\Omega}_{f}^{2}}{\left[  \left(
\tilde{\Omega}^{2}-\tilde{\Omega}_{f}^{2}\right)  ^{2}+4\gamma^{2}\right]
^{1/2}},\quad\gamma\equiv k\sqrt{\frac{NN_{f}}{m_{b}m_{f}}}. \label{CH}%
\end{align}
The eigenfrequencies of the center-of-mass subsystem are then given by the
expression
\begin{equation}
\left\{
\begin{array}
[c]{c}%
\Omega_{1}=\sqrt{\frac{1}{2}\left[  \tilde{\Omega}^{2}+\tilde{\Omega}_{f}%
^{2}+\sqrt{\left(  \tilde{\Omega}^{2}-\tilde{\Omega}_{f}^{2}\right)
^{2}+4\gamma^{2}}\right]  },\\
\Omega_{2}=\sqrt{\frac{1}{2}\left[  \tilde{\Omega}^{2}+\tilde{\Omega}_{f}%
^{2}-\sqrt{\left(  \tilde{\Omega}^{2}-\tilde{\Omega}_{f}^{2}\right)
^{2}+4\gamma^{2}}\right]  }.
\end{array}
\right.  \label{O12}%
\end{equation}
As a result, four independent frequencies $\Omega_{1},$ $\Omega_{2},$ $w$ and
$w_{f}$ appear in the problem. Three of them ($\Omega_{1},$ $\Omega_{2},$ $w$)
are the eigenfrequencies of the model system. $\Omega_{1}$\ is the frequency
of the relative motion of the center of mass of the electrons with respect to
the center of mass of the fictitious particles; $\Omega_{2}$\ is the frequency
related to the center of mass of the model system as a whole; $w$\ is the
frequency of the relative motion of the electrons with respect to their center
of mass. The parameter $w_{f}$ is an analog of the second variational
parameter $w$ of the one-polaron Feynman model. Further, the Lagrangian
(\ref{LC}) takes the form
\begin{equation}
L_{C}=-\frac{1}{2}\left(  \mathbf{\dot{r}}^{2}+\Omega_{1}^{2}\mathbf{r}%
^{2}\right)  -\frac{1}{2}\left(  \mathbf{\dot{R}}^{2}+\Omega_{2}^{2}%
\mathbf{R}^{2}\right)  , \label{QF}%
\end{equation}
leading to the partition function corresponding to the combined motion of the
centers-of-mass of the electrons and of the fictitious particles
\begin{equation}
Z_{C}=\frac{1}{\left[  2\sinh\left(  \frac{\beta\hbar\Omega_{1}}{2}\right)
\right]  ^{D}}\frac{1}{\left[  2\sinh\left(  \frac{\beta\hbar\Omega_{2}}%
{2}\right)  \right]  ^{D}}. \label{ZC}%
\end{equation}

Taking into account Eqs. (\ref{Zg2}) and (\ref{ZC}), we obtain finally the
partition function of the model system for interacting polarons
\begin{equation}
Z_{0}\left(  \left\{  N_{\sigma}\right\}  ,\beta\right)  =\left[  \frac
{\sinh\left(  \frac{\beta\hbar w}{2}\right)  \sinh\left(  \frac{\beta\hbar
w_{f}}{2}\right)  }{\sinh\left(  \frac{\beta\hbar\Omega_{1}}{2}\right)
\sinh\left(  \frac{\beta\hbar\Omega_{2}}{2}\right)  }\right]  ^{D}%
\mathbb{\tilde{Z}}_{F}\left(  \left\{  N_{\sigma}\right\}  ,w,\beta\right)  .
\label{Z02}%
\end{equation}
Here%
\begin{equation}
\mathbb{\tilde{Z}}_{F}\left(  \left\{  N_{\sigma}\right\}  ,w,\beta\right)
=\mathbb{Z}_{F}\left(  N_{1/2},w,\beta\right)  \mathbb{Z}_{F}\left(
N_{-1/2},w,\beta\right)  \label{sup}%
\end{equation}
is the partition function of $N=N_{1/2}+N_{-1/2}$ non-interacting fermions in
a parabolic confinement potential with the frequency $w.$ The analytical
expressions for the partition function of $N_{\sigma}$ spin-polarized fermions
$\mathbb{Z}_{F}\left(  N_{\sigma},w,\beta\right)  $ were derived in Ref.
\cite{PRE97}.

\subsection{Variational functional}

In order to obtain an upper bound to the free energy $E_{var}$, we substitute
the model action functional (\ref{S0}) into the right-hand side of the
variational inequality (\ref{JF}) and consider the limit $\beta\rightarrow
\infty$:%
\begin{align}
&  E_{var}\left(  \left\{  N_{\sigma}\right\}  \right) \nonumber\\
&  =\mathbb{E}_{F}\left(  \left\{  N_{\sigma}\right\}  ,w\right)  +\frac
{m_{b}}{2}\left(  \Omega_{0}^{2}-\Omega^{2}+N\omega^{2}\right)  \left\langle
\sum_{j=1}^{N}\mathbf{x}_{j}^{2}\left(  0\right)  \right\rangle _{S_{0}%
}\nonumber\\
&  -\frac{m_{b}\omega^{2}N^{2}}{2}\left\langle \mathbf{X}^{2}\left(  0\right)
\right\rangle _{S_{0}}+\left\langle U_{b}\left(  \mathbf{\bar{x}}\right)
\right\rangle _{S_{0}}+\sum_{\mathbf{q}\neq0}\frac{2\pi e^{2}}{V\varepsilon
_{\infty}q^{2}}\left[  \mathcal{G}\left(  \mathbf{q},0|\left\{  N_{\sigma
}\right\}  ,\beta\rightarrow\infty\right)  -N\right] \nonumber\\
&  +\lim_{\beta\rightarrow\infty}\frac{k^{2}N^{2}N_{f}}{4m_{f}\beta\hbar
\Omega_{f}}\int\limits_{0}^{\hbar\beta}d\tau\int\limits_{0}^{\hbar\beta}%
d\tau^{\prime}\frac{\cosh\left[  \Omega_{f}\left(  \left\vert \tau
-\tau^{\prime}\right\vert -\hbar\beta/2\right)  \right]  }{\sinh\left(
\beta\hbar\Omega_{f}/2\right)  }\left\langle \mathbf{X}\left(  \tau\right)
\cdot\mathbf{X}\left(  \tau^{\prime}\right)  \right\rangle _{S_{0}}\nonumber\\
&  -\lim_{\beta\rightarrow\infty}\sum_{\mathbf{q}}\frac{\left\vert
V_{\mathbf{q}}\right\vert ^{2}}{2\hbar^{2}\beta}\int\limits_{0}^{\hbar\beta
}d\tau\int\limits_{0}^{\hbar\beta}d\tau^{\prime}\frac{\cosh\left[
\omega_{\mathrm{LO}}\left(  \left\vert \tau-\tau^{\prime}\right\vert
-\hbar\beta/2\right)  \right]  }{\sinh\left(  \beta\hbar\omega_{\mathrm{LO}%
}/2\right)  }\mathcal{G}\left(  \mathbf{q},\tau-\tau^{\prime}|\left\{
N_{\sigma}\right\}  ,\beta\right)  . \label{varfun}%
\end{align}
Here, $\mathbb{E}_{F}\left(  N,w\right)  $ is the energy of $N$
non-interacting fermions in a parabolic confinement potential with the
confinement frequency $w$,%
\begin{align}
\mathbb{E}_{F}\left(  \left\{  N_{\sigma}\right\}  ,w\right)   &  =\hbar
w\sum_{\sigma=\pm1/2}\left\{  \sum_{n=0}^{L_{\sigma}-1}\left(  n+\frac{3}%
{2}\right)  g\left(  n\right)  \right. \nonumber\\
&  \left.  +\left(  N_{\sigma}-N_{L_{\sigma}}\right)  \left(  L_{\sigma}%
+\frac{3}{2}\right)  \right\}  , \label{Ew}%
\end{align}
where $\sigma$ is the spin of an electron, $L_{\sigma}$ is the lower partly
filled or empty level for $N_{\sigma}$ electrons with the spin projection
$\sigma$. The first term in the curly brackets of Eq. (\ref{2N}) (the upper
line) is the number of electrons at fully filled energy levels, while the
second term (square brackets) is the number of electrons at the next upper
level (which can be empty or filled partially). The energy levels of a 3D
oscillator are degenerate, so that
\begin{equation}
g\left(  n\right)  =\frac{1}{2}\left(  n+1\right)  \left(  n+2\right)
\end{equation}
is the degeneracy of the $n$-th energy level. The parameter
\begin{equation}
N_{L_{\sigma}}=\frac{1}{6}L_{\sigma}\left(  L_{\sigma}+1\right)  \left(
L_{\sigma}+2\right)
\end{equation}
is the number of electrons at all fully filled levels. The summation in Eq.
(\ref{Ew}) is performed explicitly, what gives us the result%
\begin{equation}
\mathbb{E}_{F}\left(  \left\{  N_{\sigma}\right\}  ,w\right)  =\hbar
w\sum_{\sigma}\left[  \frac{1}{8}L_{\sigma}\left(  L_{\sigma}+1\right)
^{2}\left(  L_{\sigma}+2\right)  +\left(  N_{\sigma}-N_{L_{\sigma}}\right)
\left(  L_{\sigma}+\frac{3}{2}\right)  \right]  .
\end{equation}

In Eq. (\ref{varfun}), $\mathcal{G}\left(  \mathbf{q},\tau-\tau^{\prime
}|\left\{  N_{\sigma}\right\}  ,\beta\right)  $ is the two-point correlation
function for the electron density operators:%
\begin{equation}
\mathcal{G}\left(  \mathbf{q},\tau|\left\{  N_{\sigma}\right\}  ,\beta\right)
=\left\langle \rho_{\mathbf{q}}\left(  \tau\right)  \rho_{-\mathbf{q}}\left(
0\right)  \right\rangle _{S_{0}}. \label{CiuchiF}%
\end{equation}

The averages $\left\langle \mathbf{X}\left(  \tau\right)  \cdot\mathbf{X}%
\left(  \tau^{\prime}\right)  \right\rangle _{S_{0}}$ are calculated using the
generating function method:
\begin{equation}
\left\langle X_{k}\left(  \tau\right)  X_{k}\left(  \tau^{\prime}\right)
\right\rangle _{S_{0}}=\left.  -\frac{\partial^{2}}{\partial\xi_{k}%
\partial\eta_{k}}\left\langle \exp\left[  i\left(  \mathbf{\xi}\cdot
\mathbf{X}\left(  \tau\right)  +\mathbf{\eta}\cdot\mathbf{X}\left(
\tau^{\prime}\right)  \right)  \right]  \right\rangle _{S_{0}}\right\vert
_{\substack{\mathbf{\xi}=0,\\\mathbf{\eta}=0}}, \label{b2aa}%
\end{equation}%
\begin{equation}
\Longrightarrow\left\langle \mathbf{X}\left(  \tau\right)  \cdot
\mathbf{X}\left(  \tau^{\prime}\right)  \right\rangle _{S_{0}}=\frac{3\hbar
}{2mN}\sum_{i=1}^{2}\frac{a_{i}^{2}\cosh\left[  \Omega_{i}\left(  \left\vert
\tau-\sigma\right\vert -\hbar\beta/2\right)  \right]  }{\Omega_{i}\sinh\left(
\hbar\beta\Omega_{i}/2\right)  }. \label{XtXs}%
\end{equation}
Substituting this expression into Eq. (\ref{varfun}) and performing
integrations over $\tau$ and $\sigma$ analytically, we obtain the result%
\begin{align}
&  \frac{k^{2}N^{2}N_{f}}{4m_{f}\beta\hbar\Omega_{f}}\int\limits_{0}%
^{\hbar\beta}d\tau\int\limits_{0}^{\hbar\beta}d\tau^{\prime}\frac{\cosh\left[
\Omega_{f}\left(  \left\vert \tau-\tau^{\prime}\right\vert -\hbar
\beta/2\right)  \right]  }{\sinh\left(  \beta\hbar\Omega_{f}/2\right)
}\left\langle \mathbf{X}\left(  \tau\right)  \cdot\mathbf{X}\left(
\tau^{\prime}\right)  \right\rangle _{S_{0}}\nonumber\\
&  =\frac{3\hbar\gamma}{4}\sum_{i=1}^{2}\frac{a_{i}^{2}}{\Omega_{f}^{2}%
-\Omega_{i}^{2}}\left[  \frac{\coth\left(  \beta\Omega_{i}/2\right)  }%
{\Omega_{i}}-\frac{\coth\left(  \beta\Omega_{f}/2\right)  }{\Omega_{f}%
}\right]  ,
\end{align}
and in the zero-temperature limit we have%
\begin{align}
&  \lim_{\beta\rightarrow\infty}\frac{k^{2}N^{2}N_{f}}{4m_{f}\beta\hbar
\Omega_{f}}\int\limits_{0}^{\hbar\beta}d\tau\int\limits_{0}^{\hbar\beta}%
d\tau^{\prime}\frac{\cosh\left[  \Omega_{f}\left(  \left\vert \tau
-\tau^{\prime}\right\vert -\hbar\beta/2\right)  \right]  }{\sinh\left(
\beta\hbar\Omega_{f}/2\right)  }\left\langle \mathbf{X}\left(  \tau\right)
\cdot\mathbf{X}\left(  \tau^{\prime}\right)  \right\rangle _{S_{0}}\nonumber\\
&  =\frac{3\hbar\gamma}{4}\sum_{i=1}^{2}\frac{a_{i}^{2}}{\Omega_{f}^{2}%
-\Omega_{i}^{2}}\left(  \frac{1}{\Omega_{i}}-\frac{1}{\Omega_{f}}\right)  .
\end{align}

The average $\left\langle \sum\limits_{j=1}^{N}\mathbf{x}_{j}^{2}\right\rangle
_{S_{0}}$ is transformed, using the described above operations with the
\textquotedblleft ghost\textquotedblright\ subsystem,
\begin{equation}
\mathbf{x}_{j}=\mathbf{x}_{j}^{\prime}+\mathbf{X}-\mathbf{X}_{g}, \label{b4}%
\end{equation}
and taking into account the first of equations (\ref{tosum})
\begin{equation}
\sum_{j=1}^{N}\mathbf{x}_{j}^{2}=\sum_{j=1}^{N}\left(  \mathbf{x}_{j}^{\prime
}\right)  ^{2}+N\left(  \mathbf{X}^{2}-\mathbf{X}_{g}^{2}\right)  . \label{b5}%
\end{equation}
Consequently, averaging the left-hand side of Eq. (\ref{b5}) on the model
action functional $S_{0},$ one obtains
\begin{align}
\left\langle \sum\limits_{j=1}^{N}\mathbf{x}_{j}^{2}\right\rangle _{S_{0}}  &
=\left\langle \sum\limits_{j=1}^{N}\mathbf{x}_{j}^{2}\right\rangle _{S_{M}%
}=\left\langle \sum\limits_{j=1}^{N}\mathbf{x}_{j}^{2}\right\rangle
_{S_{M}+S_{g}}\nonumber\\
&  =\left\langle \sum\limits_{j=1}^{N}\mathbf{x}_{j}^{2}\right\rangle _{S_{w}%
}+N\left(  \left\langle \mathbf{X}^{2}\right\rangle _{S_{C}}-\left\langle
\mathbf{X}_{g}^{2}\right\rangle _{S_{g}}\right)  . \label{Trr}%
\end{align}
The term $\left\langle \sum\limits_{j=1}^{N}\mathbf{x}_{j}^{2}\right\rangle
_{S_{w}}$ is expressed using the virial theorem\ through the ground-state
energy $\mathbb{E}_{F}\left(  N,w\right)  $ of $N$ independent 3D fermion
oscillators with the frequency $w$ and with the mass $m_{b}$,
\begin{equation}
\left\langle \sum\limits_{j=1}^{N}\mathbf{x}_{j}^{2}\right\rangle _{S_{w}%
}=\frac{\mathbb{E}_{F}\left(  N,w\right)  }{m_{b}w^{2}}=-\frac{1}{m_{b}w^{2}%
}\frac{\partial}{\partial\lambda}\ln\mathbb{Z}_{I}\left(  N\right)  ,
\label{1term}%
\end{equation}
Two other terms in Eq. (\ref{Trr}) are [cf. Eq. (\ref{XtXs})]:
\begin{align}
\left\langle \mathbf{X}^{2}\right\rangle _{S_{C}}  &  =\frac{3\hbar}{2m_{b}%
N}\sum_{i=1}^{2}\frac{a_{i}^{2}\coth\left(  \beta\Omega_{i}/2\right)  }%
{\Omega_{i}},\nonumber\\
\left\langle \mathbf{X}_{g}^{2}\right\rangle _{S_{g}}  &  =\frac{3\hbar
}{2m_{b}N}\frac{\coth\left(  \beta w/2\right)  }{w}. \label{a5}%
\end{align}
So, we obtain
\begin{equation}
\left\langle \sum\limits_{j=1}^{N}\mathbf{x}_{j}^{2}\right\rangle _{S_{0}%
}=\frac{\mathbb{E}_{F}\left(  \left\{  N_{\sigma}\right\}  ,w\right)  }%
{m_{b}w^{2}}+\frac{3\hbar}{2m_{b}}\left(  \sum_{i=1}^{2}\frac{a_{i}^{2}%
}{\Omega_{i}}-\frac{1}{w}\right)  . \label{vir}%
\end{equation}

The averaging of the background-charge potential gives us the result%
\begin{align}
\left\langle U_{b}\left(  \mathbf{\bar{x}}\right)  \right\rangle _{S_{0}}  &
=\frac{3\sqrt{2}\alpha\eta}{\pi\left(  1-\eta\right)  }\sum_{\sigma}\sum
_{n=0}^{\infty}\left.  f_{1}\left(  n,\sigma|\beta,N_{\sigma}\right)
\right\vert _{\beta\rightarrow\infty}\sum_{k=0}^{n}\frac{\left(  -1\right)
^{k}}{k!}\binom{n+2}{n-k}\left(  \frac{1}{2w}\right)  ^{k}\nonumber\\
&  \times\left\{  \frac{\Gamma\left(  k-\frac{1}{2}\right)  }{A^{k-1/2}%
}\left[  _{1}F_{1}\left(  k-\frac{1}{2};\frac{1}{2};-\frac{R^{2}}{4A}\right)
-\,_{1}F_{1}\left(  k-\frac{1}{2};\frac{3}{2};-\frac{R^{2}}{4A}\right)
\right]  \right\}  ,\\
\eta &  \equiv\varepsilon_{\infty}/\varepsilon_{0},\; \;A\equiv\frac{\hbar
}{4m_{b}N}\left(  \sum\limits_{i=1}^{2}\frac{a_{i}^{2}}{\Omega_{i}}+\frac
{N-1}{w}\right)  ,\nonumber
\end{align}
where $f_{1}\left(  n,\sigma|\beta,N_{\sigma}\right)  $ is the one-particle
distribution function of fermions (the distribution functions are considered
in more details in the next subsection).

Collecting all terms together, we arrive at the variational functional%
\begin{gather}
E_{var}\left(  \Omega_{1},\Omega_{2},w,\Omega_{f}\right)  =\hbar\left\{
\frac{\Omega_{0}^{2}+w^{2}}{2w^{2}}\left[  \frac{\tilde{E}\left(  w,N\right)
}{\hbar}-\frac{3}{2}w\right]  +\frac{3}{2}\left(  \Omega_{1}+\Omega_{2}%
-\Omega_{f}\right)  \right. \nonumber\\
+\left.  \frac{3}{4}\left(  \Omega_{0}^{2}-\Omega_{1}^{2}-\Omega_{2}%
^{2}+\Omega_{f}^{2}\right)  \sum_{i=1}^{2}\frac{a_{i}^{2}}{\Omega_{i}}%
+\frac{3\gamma^{2}}{4\Omega_{f}}\sum_{i=1}^{2}\frac{a_{i}^{2}}{\Omega
_{i}\left(  \Omega_{i}+\Omega_{f}\right)  }\right\} \nonumber\\
+\left\langle U_{b}\left(  \mathbf{\bar{x}}\right)  \right\rangle _{S_{0}%
}+E_{C}+E_{e-ph}, \label{Evar}%
\end{gather}
where $E_{C}$ and $E_{e-ph}$ are the Coulomb and polaron contributions, respectively:%

\begin{equation}
E_{C}=\frac{e^{2}}{4\pi^{2}\varepsilon_{\infty}}\int d\mathbf{q}\frac{1}%
{q^{2}}\left[  \left.  \mathcal{G}\left(  \mathbf{q},0|\left\{  N_{\sigma
}\right\}  ,\beta\right)  \right\vert _{\beta\rightarrow\infty}-N\right]  ,
\label{EC}%
\end{equation}%
\begin{equation}
E_{e-ph}=-\frac{\sqrt{2}\alpha}{4\pi^{2}\hbar}\int d\mathbf{q}\frac{1}{q^{2}%
}\int\limits_{0}^{\infty}d\tau\exp\left(  -\omega_{\mathrm{LO}}\tau\right)
\left.  \mathcal{G}\left(  \mathbf{q},\tau|\left\{  N_{\sigma}\right\}
,\beta\right)  \right\vert _{\beta\rightarrow\infty}. \label{Epol}%
\end{equation}

The correlation function (\ref{CiuchiF}) is calculated analytically in the
next subsection. With this correlation function, the variational ground-state
energy is calculated and minimized numerically.

\subsection{Two-point correlation functions}

The two-point correlation function (\ref{CiuchiF}) is represented as the
following path integral:%
\begin{align}
\mathcal{G}\left(  \mathbf{q},\tau|\left\{  N_{\sigma}\right\}  ,\beta\right)
&  =\frac{1}{Z_{0}\left(  \left\{  N_{\sigma}\right\}  ,\beta\right)  }%
\sum_{P}\frac{\left(  -1\right)  ^{\mathbf{\xi}_{P}}}{N_{1/2}!N_{-1/2}%
!}\nonumber\\
&  \times\int d\mathbf{\bar{x}}\int_{\mathbf{\bar{x}}}^{P\mathbf{\bar{x}}%
}D\mathbf{\bar{x}}\left(  \tau\right)  e^{-S_{0}\left[  \mathbf{\bar{x}%
}\left(  \tau\right)  \right]  }\rho_{\mathbf{q}}\left(  \tau\right)
\rho_{-\mathbf{q}}\left(  0\right)  . \label{g1}%
\end{align}
We observe that $\mathcal{G}\left(  \mathbf{q},\tau|\left\{  N_{\sigma
}\right\}  ,\beta\right)  $ can be rewritten as an average within the model
\textquotedblleft action\textquotedblright\ $S_{M}\left[  \mathbf{\bar{x}%
}\left(  \tau\right)  ,\mathbf{\bar{y}}\left(  \tau\right)  \right]  $ of
interacting electrons and fictitious particles:%
\begin{align}
\mathcal{G}\left(  \mathbf{q},\tau|\left\{  N_{\sigma}\right\}  ,\beta\right)
&  =\frac{1}{Z_{M}\left(  \left\{  N_{\sigma}\right\}  ,N_{f},\beta\right)
}\sum_{P}\frac{\left(  -1\right)  ^{\mathbf{\xi}_{P}}}{N_{1/2}!N_{-1/2}%
!}\nonumber\\
&  \times\int d\mathbf{\bar{x}}\int_{\mathbf{\bar{x}}}^{P\mathbf{\bar{x}}%
}D\mathbf{\bar{x}}\left(  \tau\right)  \int d\mathbf{\bar{y}}\int
_{\mathbf{\bar{y}}}^{\mathbf{\bar{y}}}D\mathbf{\bar{y}}\left(  \tau\right)
e^{-S_{M}\left[  \mathbf{\bar{x}}\left(  \tau\right)  ,\mathbf{\bar{y}}\left(
\tau\right)  \right]  }\nonumber\\
&  \times\rho_{\mathbf{q}}\left(  \tau\right)  \rho_{-\mathbf{q}}\left(
0\right)  . \label{g2}%
\end{align}
Indeed, one readily derives that the elimination of the fictitious particles
in (\ref{g2}) leads to (\ref{g1}). The representation (\ref{g2}) allows one to
calculate the correlation function $\mathcal{G}\left(  \mathbf{q}%
,\tau|\left\{  N_{\sigma}\right\}  ,\beta\right)  $ in a much simpler way than
through Eq. (\ref{g1}), using the separation of the coordinates of the centers
of mass of the electrons and of the fictitious particles. This separation is
performed for the two-point correlation function (\ref{g2}) by the same method
as it has been done for the partition function (\ref{ZM}). As a result, one
obtains%
\begin{equation}
\mathcal{G}\left(  \mathbf{q},\tau|\left\{  N_{\sigma}\right\}  ,\beta\right)
=\tilde{g}\left(  \mathbf{q},\tau|\left\{  N_{\sigma}\right\}  ,\beta\right)
\frac{\left\langle \exp\left[  i\mathbf{q\cdot}\left(  \mathbf{X}\left(
\tau\right)  -\mathbf{X}\left(  \sigma\right)  \right)  \right]  \right\rangle
_{S_{C}}}{\left\langle \exp\left[  i\mathbf{q\cdot}\left(  \mathbf{X}%
_{g}\left(  \tau\right)  -\mathbf{X}_{g}\left(  \sigma\right)  \right)
\right]  \right\rangle _{S_{g}}}, \label{Fact2}%
\end{equation}
where $\tilde{g}\left(  \mathbf{q},\tau|\left\{  N_{\sigma}\right\}
,\beta\right)  $ is the time-dependent correlation function of $N$
non-interacting electrons in a parabolic confinement potential with the
frequency $w$,
\begin{equation}
\tilde{g}\left(  \mathbf{q},\tau|\left\{  N_{\sigma}\right\}  ,\beta\right)
=\left\langle \rho_{\mathbf{q}}\left(  \tau\right)  \rho_{-\mathbf{q}}\left(
0\right)  \right\rangle _{S_{w}}. \label{Gqtild}%
\end{equation}
The action functional $S_{w}\left[  \mathbf{\bar{x}}_{\tau}\right]  $ is
related to the Lagrangian $L_{w}\left(  \mathbf{\dot{\bar{x}}},\mathbf{\bar
{x}}\right)  $ [Eq. (\ref{Lwa})]%
\begin{equation}
S_{w}\left[  \mathbf{\bar{x}}_{\tau}\right]  =\frac{1}{\hbar}\int
\limits_{0}^{\hbar\beta}L_{w}\left(  \mathbf{\dot{\bar{x}}},\mathbf{\bar{x}%
}\right)  \,d\tau. \label{Sw}%
\end{equation}

The averages in (\ref{Fact2}) are calculated using Feynman's method of
generating functions \cite{Feynman}. Namely, according to \cite{Feynman}, the
average
\begin{equation}
G\left[  f\left(  \tau\right)  \right]  \equiv\left\langle \exp\left\{
\frac{i}{\hbar}\int\limits_{0}^{\beta}f\left(  \tau\right)  x_{\tau}%
\,d\tau\right\}  \right\rangle _{S_{\omega}}, \label{Gosc}%
\end{equation}
where $S_{\omega}$ is the action functional of a one-dimensional harmonic
oscillator with the frequency $\omega$ and with the mass $m$, results in
\begin{equation}
G\left[  f\left(  \tau\right)  \right]  =\exp\left\{  -\frac{1}{4m\hbar\omega
}\int\limits_{0}^{\beta}d\tau\int\limits_{0}^{\beta}d\sigma\frac{\cosh\left[
\omega\left(  \left\vert \tau-\sigma\right\vert -\beta/2\right)  \right]
}{\sinh\left(  \beta\omega/2\right)  }f\left(  \tau\right)  f\left(
\sigma\right)  \right\}  . \label{Gosc1}%
\end{equation}
The diagonalization procedure for the Lagrangian $L_{C}$ (\ref{LC}) allows us
to represent that Lagrangian as a sum of Lagrangians of independent harmonic
oscillators, what gives the following explicit expressions for averages in Eq.
(\ref{Fact2}):%
\begin{align*}
&  \left\langle \exp\left[  i\mathbf{q\cdot}\left(  \mathbf{X}\left(
\tau\right)  -\mathbf{X}\left(  \sigma\right)  \right)  \right]  \right\rangle
_{S_{C}}\\
&  =\exp\left\{  -\frac{\hbar q^{2}}{Nm_{b}}\left[  \sum_{i=1}^{2}a_{i}%
^{2}\frac{\sinh\left(  \frac{\Omega_{i}\left\vert \tau-\sigma\right\vert }%
{2}\right)  \sinh\left(  \frac{\Omega_{i}\left(  \hbar\beta-\left\vert
\tau-\sigma\right\vert \right)  }{2}\right)  }{\Omega_{i}\sinh\left(
\frac{\beta\hbar\Omega_{i}}{2}\right)  }\right]  \right\}  ,\\
&  \left\langle \exp\left[  i\mathbf{q\cdot}\left(  \mathbf{X}_{g}\left(
\tau\right)  -\mathbf{X}_{g}\left(  \sigma\right)  \right)  \right]
\right\rangle _{S_{g}}\\
&  =\exp\left[  -\frac{\hbar q^{2}}{Nm_{b}}\frac{\sinh\left(  \frac
{w\left\vert \tau-\sigma\right\vert }{2}\right)  \sinh\left(  \frac{w\left(
\hbar\beta-\left\vert \tau-\sigma\right\vert \right)  }{2}\right)  }%
{w\sinh\left(  \frac{\beta\hbar w}{2}\right)  }\right]  .
\end{align*}

\subsubsection{The correlation function $\tilde{g}\left(  \mathbf{q}%
,\tau|\left\{  N_{\sigma}\right\}  ,\beta\right)  $}

As seen from the formula (\ref{Gqtild}), $\tilde{g}\left(  \mathbf{q}%
,\tau|\left\{  N_{\sigma}\right\}  ,\beta\right)  $ is the time-dependent
correlation function of $N$ non-interacting fermions in a parabolic
confinement potential with the frequency $w$. Let us consider first of all a
system of $N$ identical spin-polarized oscillators with the Lagrangian
\begin{equation}
L=\frac{m}{2}\sum_{j=1}^{N}\left(  \mathbf{\dot{x}}_{j}^{2}-\omega
^{2}\mathbf{x}_{j}^{2}\right)  . \label{Q1}%
\end{equation}
The corresponding Hamiltonian is
\begin{equation}
\hat{H}=\sum_{j=1}^{N}\left(  \frac{\mathbf{\hat{p}}_{j}^{2}}{2m}%
+\frac{m\omega^{2}\mathbf{\hat{x}}_{j}^{2}}{2}\right)  , \label{2N}%
\end{equation}

\begin{equation}
\hat{H}=\sum_{j=1}^{N}\hat{h}_{j},\quad\hat{h}\equiv\frac{\mathbf{\hat{p}}%
^{2}}{2m}+\frac{m\omega^{2}\mathbf{\hat{x}}^{2}}{2}. \label{3N}%
\end{equation}
A set of eigenfunctions of the one-particle Hamiltonian $\hat{h}$ is
determined as follows:
\begin{equation}
\hat{h}\psi_{\mathbf{n}}\left(  \mathbf{x}\right)  =\varepsilon_{n}%
\psi_{\mathbf{n}}\left(  \mathbf{x}\right)  , \label{4}%
\end{equation}
where
\begin{align}
\mathbf{n}  &  \equiv\left(  n_{1},n_{2},n_{3}\right)  ,\quad n\equiv
n_{1}+n_{2}+n_{3},\nonumber\\
\varepsilon_{\mathbf{n}}  &  =\varepsilon_{n}=\hbar\omega\left(  n+\frac{3}%
{2}\right)  ,\quad\psi_{\mathbf{n}}\left(  \mathbf{x}\right)  =\varphi_{n_{1}%
}\left(  x_{1}\right)  \varphi_{n_{2}}\left(  x_{2}\right)  \varphi_{n_{3}%
}\left(  x_{3}\right)  , \label{5N}%
\end{align}
$\varphi_{n}\left(  x\right)  $ is the $n$-th eigenfunction of a
one-dimensional oscillator with the frequency $\omega.$

The Hamiltonian (\ref{2N}) can be written down in terms of the annihilation
$\left(  \hat{a}_{\mathbf{n}}\right)  $ and creation $\left(  \hat
{a}_{\mathbf{n}}^{+}\right)  $ operators:
\begin{equation}
\hat{H}=\sum_{\mathbf{n}}\varepsilon_{\mathbf{n}}\hat{a}_{\mathbf{n}}^{+}%
\hat{a}_{\mathbf{n}}=\sum_{\mathbf{n}}\varepsilon_{\mathbf{n}}\hat
{N}_{\mathbf{n}},\quad\hat{N}_{\mathbf{n}}\equiv\hat{a}_{\mathbf{n}}^{+}%
\hat{a}_{\mathbf{n}}. \label{6}%
\end{equation}

The many-particle quantum states in the representation of ``occupation
numbers'' are written down as $\left|  \dots N_{\mathbf{n}}\dots\right\rangle
,$ where $N_{\mathbf{n}}$ is the number of particles in the $\mathbf{n}$-th
one-particle quantum state. The states $\left|  \dots N_{\mathbf{n}}%
\dots\right\rangle $ are defined as the eigenstates of the operator of the
number of particles in the $\mathbf{n}$-th state $\hat{N}_{\mathbf{n}}$:
\begin{equation}
\hat{N}_{\mathbf{n}}\left|  \dots N_{\mathbf{n}}\dots\right\rangle
=N_{\mathbf{n}}\left|  \dots N_{\mathbf{n}}\dots\right\rangle . \label{7}%
\end{equation}
Let us determine a set of quantum states with a \emph{finite} total number of
particles
\begin{equation}
\sum_{\mathbf{n}}N_{\mathbf{n}}=N \label{Restriction}%
\end{equation}
as follows:
\begin{equation}
\left.  \left|  \dots N_{\mathbf{n}}\dots\right\rangle \right|  _{\sum
_{\mathbf{n}}N_{\mathbf{n}}=N}\equiv\left|  \Psi_{N,\left\{  N_{\mathbf{n}%
}\right\}  }\right\rangle . \label{8}%
\end{equation}
Further on, we use the basis set of quantum states (\ref{8}) for the
derivation of the partition function, of the density function and of the
two-point correlation function.

\begin{quote}
\textbf{Partition function}
\end{quote}

The density matrix of the \emph{canonical} Hibbs ensemble is
\[
\hat{\rho}=\exp\left(  -\beta\hat{H}\right)  ,\quad\beta\equiv\frac{1}{k_{B}%
T}.
\]

The partition function of this ensemble is the trace of the density matrix on
the set of quantum states (\ref{8}):
\begin{align}
\mathbb{Z}_{I}\left(  \beta|N\right)   &  =\sum_{\left\{  N_{\mathbf{n}%
}\right\}  }\left\langle \Psi_{N,\left\{  N_{\mathbf{n}}\right\}  }\left|
\exp\left(  -\beta\hat{H}\right)  \right|  \Psi_{N,\left\{  N_{\mathbf{n}%
}\right\}  }\right\rangle \nonumber\\
&  =\left[  \sum_{\left\{  N_{\mathbf{n}}\right\}  }\exp\left(  -\beta
\sum_{\mathbf{n}}\varepsilon_{\mathbf{n}}N_{\mathbf{n}}\right)  \right]
_{\sum_{\mathbf{n}}N_{\mathbf{n}}=N}. \label{9}%
\end{align}
This expression can be written down also in the form
\begin{equation}
\mathbb{Z}_{I}\left(  \beta|N\right)  =\sum_{\left\{  N_{\mathbf{n}}\right\}
}\exp\left(  -\beta\sum_{\mathbf{n}}\varepsilon_{\mathbf{n}}N_{\mathbf{n}%
}\right)  \delta_{N,\sum_{\mathbf{n}}N_{\mathbf{n}}}, \label{10}%
\end{equation}
where
\[
\delta_{j,k}=\left\{
\begin{array}
[c]{c}%
1,\quad j=k\\
0,\quad j\neq k
\end{array}
\right.
\]
is the delta symbol.

Let us introduce the generating function for the partition function in the
same way as in Ref. \cite{PRE97}:
\begin{align*}
\Xi\left(  \beta,u\right)   &  =\sum_{N=0}^{\infty}u^{N}\mathbb{Z}_{I}\left(
\beta|N\right)  =\sum_{N=0}^{\infty}u^{N}\sum_{\left\{  N_{\mathbf{n}%
}\right\}  }\exp\left(  -\beta\sum_{\mathbf{n}}\varepsilon_{\mathbf{n}%
}N_{\mathbf{n}}\right)  \delta_{N,\sum_{\mathbf{n}}N_{\mathbf{n}}}\\
&  =\sum_{\left\{  N_{\mathbf{n}}\right\}  }\exp\left(  -\beta\sum
_{\mathbf{n}}\varepsilon_{\mathbf{n}}N_{\mathbf{n}}\right)  \sum_{N=0}%
^{\infty}u^{\sum_{\mathbf{n}}N_{\mathbf{n}}}\delta_{N,\sum_{\mathbf{n}%
}N_{\mathbf{n}}}\\
&  =\sum_{\left\{  N_{\mathbf{n}}\right\}  }\exp\left(  -\beta\sum
_{\mathbf{n}}\varepsilon_{\mathbf{n}}N_{\mathbf{n}}\right)  u^{\sum
_{\mathbf{n}}N_{\mathbf{n}}}\quad\Longrightarrow
\end{align*}
\begin{equation}
\Xi\left(  \beta,u\right)  =\prod_{\mathbf{n}}\left\{  \sum_{N_{\mathbf{n}}%
}\left[  u\exp\left(  -\beta\varepsilon_{\mathbf{n}}\right)  \right]
^{N_{\mathbf{n}}}\right\}  . \label{GF1}%
\end{equation}

\begin{quote}
\textbf{Fermions}
\end{quote}

For fermions, the number $N_{\mathbf{n}}$ can take only values $N_{\mathbf{n}%
}=0$ and $N_{\mathbf{n}}=1.$ Hence, for fermions (denoted by the index $F$),
we obtain:
\[
\Xi_{F}\left(  \beta,u\right)  =\prod_{\mathbf{n}}\left[  1+u\exp\left(
-\beta\varepsilon_{\mathbf{n}}\right)  \right]  .
\]
Since the $n$-th level of a 3D oscillator is degenerate with the degeneracy
\[
g\left(  n\right)  =\frac{\left(  n+1\right)  \left(  n+2\right)  }{2},
\]
we find that the generating function $\Xi_{F}\left(  \beta,u\right)  $ is
given by%
\begin{equation}
\Xi_{F}\left(  \beta,u\right)  =\prod_{n=0}^{\infty}\left[  1+u\exp\left(
-\beta\varepsilon_{n}\right)  \right]  ^{g\left(  n\right)  }. \label{Equiv1}%
\end{equation}

\begin{quote}
\textbf{Bosons}
\end{quote}

For bosons (denoted by the index $B$), $N_{\mathbf{n}}=0,1,\dots,\infty.$ The
summations over $\left\{  N_{\mathbf{n}}\right\}  $ in Eq. (\ref{GF1}) gives:
\begin{equation}
\Xi_{B}\left(  \beta,u\right)  =\prod_{n=0}^{\infty}\left[  \frac{1}%
{1-u\exp\left(  -\beta\varepsilon_{n}\right)  }\right]  ^{g\left(  n\right)
}. \label{GF1a}%
\end{equation}

The results (\ref{GF1}) and (\ref{GF1a}) prove (for the partition function)
the equivalence of the path-integral approach for identical particles
\cite{PRE97} and of the second-quantization method.

\begin{quote}
\textbf{Integral representation}
\end{quote}

Let us use the Fourier representation for the delta symbol:
\begin{equation}
\delta_{N,\sum_{\mathbf{n}}N_{\mathbf{n}}}=\frac{1}{2\pi}\int\limits_{0}%
^{2\pi}\exp\left[  i\left(  \sum_{\mathbf{n}}N_{\mathbf{n}}-N\right)  \left(
\theta-i\zeta\right)  \right]  d\theta, \label{delta}%
\end{equation}
where $\zeta$ is an arbitrary constant. Substituting Eq. (\ref{delta}) into
Eq. (\ref{10}) we obtain
\begin{align*}
\mathbb{Z}_{I}\left(  \beta|N\right)   &  =\sum_{\left\{  N_{\mathbf{n}%
}\right\}  }\exp\left(  -\beta\sum_{\mathbf{n}}\varepsilon_{\mathbf{n}%
}N_{\mathbf{n}}\right)  \frac{1}{2\pi}\int\limits_{0}^{2\pi}\exp\left[
i\left(  \sum_{\mathbf{n}}N_{\mathbf{n}}-N\right)  \left(  \theta
-i\zeta\right)  \right]  d\theta\\
&  =\frac{1}{2\pi}\int\limits_{0}^{2\pi}d\theta\exp\left[  -iN\left(
\theta-i\zeta\right)  \right]  \sum_{\left\{  N_{\mathbf{n}}\right\}  }%
\exp\left(  -\beta\sum_{\mathbf{n}}\varepsilon_{\mathbf{n}}N_{\mathbf{n}%
}+i\sum_{\mathbf{n}}N_{\mathbf{n}}\left(  \theta-i\zeta\right)  \right) \\
&  =\frac{1}{2\pi}\int\limits_{0}^{2\pi}d\theta\exp\left(  -iN\theta
-N\zeta\right)  \Xi\left(  \beta,e^{i\theta+\zeta}\right)  \quad
\Longrightarrow
\end{align*}%
\begin{equation}
\mathbb{Z}_{I}\left(  \beta|N\right)  =\frac{1}{2\pi}\int\limits_{0}^{2\pi
}d\theta\exp\left[  \ln\Xi\left(  \beta,e^{i\theta+\zeta}\right)
-N\zeta-iN\theta\right]  . \label{qq}%
\end{equation}

The partition function for a finite number of particles can be obtained from
the generation function also by the inversion formula \cite{SSC99}
\begin{align}
\mathbb{Z}_{I}\left(  \beta|N\right)   &  =\frac{1}{2\pi i}\oint\frac
{\Xi\left(  \beta,z\right)  }{z^{N+1}}dz\label{210}\\
&  =\frac{1}{2\pi}\int_{0}^{2\pi}e^{\left[  \ln\Xi\left(  \beta,ue^{i\theta
}\right)  -N\ln u\right]  }e^{-iN\theta}d\theta. \label{211}%
\end{align}
Let us denote in Eq. (\ref{qq}):
\begin{equation}
\zeta\equiv\ln u. \label{zu}%
\end{equation}
In these notations, Eqs. (\ref{qq}) and (\ref{211}) are \emph{identical}. For
the numerical calculation, it is more convenient to choose in Eq.
(\ref{delta}) the interval of the integration over $\theta$ as $\left[
-\pi,\pi\right]  $ instead of $\left[  0,2\pi\right]  ,$ what gives:
\begin{equation}
\mathbb{Z}_{I}\left(  \beta|N\right)  =\frac{1}{2\pi}\int\limits_{-\pi}^{\pi
}\Phi_{N}\left(  \theta\right)  d\theta, \label{Zi5}%
\end{equation}
with the function
\begin{equation}
\Phi_{N}\left(  \theta\right)  =\exp\left[  \ln\Xi\left(  \beta,ue^{i\theta
}\right)  -N\ln u-iN\theta\right]  . \label{FiN}%
\end{equation}

The aforesaid method of derivation of the partition function [Eqs.
(\ref{delta}) to (\ref{qq})] is heuristically useful, because it allows a
simple generalization to spin-mixed systems with various polarization distributions.

The two-point density-density correlation function in the operator formalism
is%
\begin{equation}
\tilde{g}\left(  \mathbf{q},\tau|\left\{  N_{\sigma}\right\}  ,\beta\right)
=\left\langle \hat{\rho}_{\mathbf{q}}\left(  \tau\right)  \hat{\rho
}_{-\mathbf{q}}\left(  0\right)  \right\rangle , \label{g}%
\end{equation}
where $\hat{\rho}_{\mathbf{q}}\left(  t\right)  $ is the density operator in
the Heisenberg representation:
\begin{equation}
\hat{\rho}_{\mathbf{q}}\left(  \tau\right)  =\exp\left(  \frac{\tau}{\hbar
}\hat{H}\right)  \rho_{\mathbf{q}}\exp\left(  -\frac{\tau}{\hbar}\hat
{H}\right)  . \label{Heis}%
\end{equation}
In the \textquotedblleft second-quantization\textquotedblright%
\ representation, $\hat{\rho}_{\mathbf{q}}\left(  t\right)  $ is
\begin{align}
\hat{\rho}_{\mathbf{q}}\left(  \tau\right)   &  =\sum_{\mathbf{n}%
,\mathbf{n}^{\prime}}\left(  e^{i\mathbf{q\cdot\hat{x}}}\right)
_{\mathbf{nn}^{\prime}}\hat{a}_{\mathbf{n}}^{+}\left(  \tau\right)  \hat
{a}_{\mathbf{n}^{\prime}}\left(  \tau\right) \nonumber\\
&  =\sum_{\mathbf{n},\mathbf{n}^{\prime}}\left(  e^{i\mathbf{q\cdot\hat{x}}%
}\right)  _{\mathbf{nn}^{\prime}}\hat{a}_{\mathbf{n}}^{+}\hat{a}%
_{\mathbf{n}^{\prime}}\exp\left[  \frac{\tau}{\hbar}\left(  \varepsilon
_{\mathbf{n}}-\varepsilon_{\mathbf{n}^{\prime}}\right)  \right]  .
\label{Rhoqt}%
\end{align}

After substituting Eq. (\ref{Rhoqt}) into (\ref{g}), we find that
\begin{equation}
\tilde{g}\left(  \mathbf{q},\tau|\left\{  N_{\sigma}\right\}  ,\beta\right)
=\sum_{\mathbf{n},\mathbf{n}^{\prime}}\sum_{\mathbf{m},\mathbf{m}^{\prime}%
}\left(  e^{i\mathbf{q\cdot\hat{x}}}\right)  _{\mathbf{nn}^{\prime}}\left(
e^{-i\mathbf{q\cdot\hat{x}}}\right)  _{\mathbf{mm}^{\prime}}\exp\left[
\frac{\tau}{\hbar}\left(  \varepsilon_{\mathbf{n}}-\varepsilon_{\mathbf{n}%
^{\prime}}\right)  \right]  \left\langle \hat{a}_{\mathbf{n}}^{+}\hat
{a}_{\mathbf{n}^{\prime}}\hat{a}_{\mathbf{m}}^{+}\hat{a}_{\mathbf{m}^{\prime}%
}\right\rangle . \label{gt1}%
\end{equation}

The operator $\hat{a}_{\mathbf{n}}^{+}\hat{a}_{\mathbf{n}^{\prime}}\hat
{a}_{\mathbf{m}}^{+}\hat{a}_{\mathbf{m}^{\prime}}$ has non-zero diagonal
matrix elements in the basis of quantum states $\left\vert \Psi_{N,\left\{
N_{\mathbf{n}}\right\}  }\right\rangle $ only in the cases
\begin{equation}
\left\{
\begin{array}
[c]{c}%
\mathbf{n}=\mathbf{n}^{\prime}\\
\mathbf{m}=\mathbf{m}^{\prime}%
\end{array}
\right.  \quad\mathrm{or}\quad\left\{
\begin{array}
[c]{c}%
\mathbf{n}=\mathbf{m}^{\prime}\\
\mathbf{m}=\mathbf{n}^{\prime}%
\end{array}
\right.  . \label{cond}%
\end{equation}
Hence, the average
\begin{equation}
\left\langle \hat{a}_{\mathbf{n}}^{+}\hat{a}_{\mathbf{n}^{\prime}}\hat
{a}_{\mathbf{m}}^{+}\hat{a}_{\mathbf{m}^{\prime}}\right\rangle =\frac
{1}{\mathbb{Z}_{I}\left(  \beta|N\right)  }\sum_{\left\{  N_{\mathbf{n}%
}\right\}  }\left\langle \Psi_{N,\left\{  N_{\mathbf{n}}\right\}  }\left\vert
\exp\left(  -\beta\hat{H}\right)  \hat{a}_{\mathbf{n}}^{+}\hat{a}%
_{\mathbf{n}^{\prime}}\hat{a}_{\mathbf{m}}^{+}\hat{a}_{\mathbf{m}^{\prime}%
}\right\vert \Psi_{N,\left\{  N_{\mathbf{n}}\right\}  }\right\rangle
\label{Av2}%
\end{equation}
is not equal to zero only when the condition (\ref{cond}) is fulfilled. This
allows us to write down the average (\ref{Av2}) as
\begin{align}
\left\langle \hat{a}_{\mathbf{n}}^{+}\hat{a}_{\mathbf{n}^{\prime}}\hat
{a}_{\mathbf{m}}^{+}\hat{a}_{\mathbf{m}^{\prime}}\right\rangle  &
=\delta_{\mathbf{n}^{\prime}\mathbf{n}}\delta_{\mathbf{m}^{\prime}\mathbf{m}%
}\left(  1-\delta_{\mathbf{mn}}\right)  \left\langle \hat{a}_{\mathbf{n}}%
^{+}\hat{a}_{\mathbf{n}}\hat{a}_{\mathbf{m}}^{+}\hat{a}_{\mathbf{m}%
}\right\rangle +\delta_{\mathbf{m}^{\prime}\mathbf{n}}\delta_{\mathbf{n}%
^{\prime}\mathbf{m}}\left(  1-\delta_{\mathbf{mn}}\right)  \left\langle
\hat{a}_{\mathbf{n}}^{+}\hat{a}_{\mathbf{m}}\hat{a}_{\mathbf{m}}^{+}\hat
{a}_{\mathbf{n}}\right\rangle \nonumber\\
&  +\delta_{\mathbf{n}^{\prime}\mathbf{n}}\delta_{\mathbf{m}^{\prime
}\mathbf{m}}\delta_{\mathbf{mn}}\left\langle \hat{a}_{\mathbf{n}}^{+}\hat
{a}_{\mathbf{n}}\hat{a}_{\mathbf{n}}^{+}\hat{a}_{\mathbf{n}}\right\rangle
\nonumber\\
&  =\delta_{\mathbf{n}^{\prime}\mathbf{n}}\delta_{\mathbf{m}^{\prime
}\mathbf{m}}\left(  1-\delta_{\mathbf{mn}}\right)  \left\langle \hat
{N}_{\mathbf{n}}\hat{N}_{\mathbf{m}}\right\rangle +\delta_{\mathbf{m}^{\prime
}\mathbf{n}}\delta_{\mathbf{n}^{\prime}\mathbf{m}}\left(  1-\delta
_{\mathbf{mn}}\right)  \left(  \left\langle \hat{N}_{\mathbf{n}}\right\rangle
-\left\langle \hat{N}_{\mathbf{n}}\hat{N}_{\mathbf{m}}\right\rangle \right)
\nonumber\\
&  +\delta_{\mathbf{n}^{\prime}\mathbf{n}}\delta_{\mathbf{m}^{\prime
}\mathbf{m}}\delta_{\mathbf{mn}}\bar{N}_{\mathbf{n}}\nonumber\\
&  =\delta_{\mathbf{n}^{\prime}\mathbf{n}}\delta_{\mathbf{m}^{\prime
}\mathbf{m}}\left\langle \hat{N}_{\mathbf{n}}\hat{N}_{\mathbf{m}}\right\rangle
+\delta_{\mathbf{m}^{\prime}\mathbf{n}}\delta_{\mathbf{n}^{\prime}\mathbf{m}%
}\left\langle \hat{N}_{\mathbf{n}}\left(  1-\hat{N}_{\mathbf{m}}\right)
\right\rangle . \label{Av3}%
\end{align}

Here, the notation is used for the average occupation number $\bar
{N}_{\mathbf{n}}$:
\begin{equation}
\left\langle \hat{N}_{\mathbf{n}}\right\rangle =\frac{1}{\mathbb{Z}_{I}\left(
\beta|N\right)  }\sum_{\left\{  N_{\mathbf{n}}\right\}  }\left\langle
\Psi_{N,\left\{  N_{\mathbf{n}}\right\}  }\left\vert \exp\left(  -\beta\hat
{H}\right)  \hat{N}_{\mathbf{n}}\right\vert \Psi_{N,\left\{  N_{\mathbf{n}%
}\right\}  }\right\rangle . \label{Nav}%
\end{equation}%
\begin{align}
\mathbb{Z}_{I}\left(  \beta|N\right)   &  =\sum_{\left\{  N_{\mathbf{n}%
}\right\}  }\left\langle \Psi_{N,\left\{  N_{\mathbf{n}}\right\}  }\left\vert
\exp\left(  -\beta\hat{H}\right)  \right\vert \Psi_{N,\left\{  N_{\mathbf{n}%
}\right\}  }\right\rangle \nonumber\\
&  =\left[  \sum_{\left\{  N_{\mathbf{n}}\right\}  }\exp\left(  -\beta
\sum_{\mathbf{n}}\varepsilon_{\mathbf{n}}N_{\mathbf{n}}\right)  \right]
_{\sum_{\mathbf{n}}N_{\mathbf{n}}=N}.
\end{align}
In the same way as Eq. (\ref{9}), the average (\ref{Nav}) can be written down
in the form
\[
\left\langle \hat{N}_{\mathbf{n}}\right\rangle =\frac{1}{\mathbb{Z}_{I}\left(
\beta|N\right)  }\left[  \sum_{\left\{  N_{\mathbf{n}^{\prime}}\right\}
}N_{\mathbf{n}}\exp\left(  -\beta\sum_{\mathbf{n}^{\prime}}\varepsilon
_{\mathbf{n}^{\prime}}N_{\mathbf{n}^{\prime}}\right)  \right]  _{\sum
_{\mathbf{n}^{\prime}}N_{\mathbf{n}^{\prime}}=N}.
\]

\[
\left\langle \hat{N}_{\mathbf{n}}\right\rangle =-\frac{1}{\beta\mathbb{Z}%
_{I}\left(  \beta|N\right)  }\frac{\delta\mathbb{Z}_{I}\left(  \beta|N\right)
}{\delta\varepsilon_{\mathbf{n}}},
\]%
\begin{equation}
\left\langle \hat{N}_{\mathbf{n}}\right\rangle =\frac{1}{2\pi\mathbb{Z}%
_{I}\left(  \beta|N\right)  }\int\limits_{-\pi}^{\pi}\frac{\Phi_{N}\left(
\theta\right)  }{\exp\left(  \beta\varepsilon_{\mathbf{n}}-\zeta
-i\theta\right)  +1}d\theta. \label{Num1}%
\end{equation}
Since $\varepsilon_{\mathbf{n}}=\varepsilon_{n},$ $\left\langle \hat
{N}_{\mathbf{n}}\right\rangle $ depends only on $n$.

Using Eq. (\ref{zu}), we can write $\bar{N}_{n}$ as
\begin{align}
\left\langle \hat{N}_{\mathbf{n}}\right\rangle  &  =\frac{1}{2\pi
\mathbb{Z}_{I}\left(  \beta|N\right)  }\int\limits_{-\pi}^{\pi}\frac{\Phi
_{N}\left(  \theta\right)  }{\frac{1}{u}\exp\left(  \beta\varepsilon
_{n}-i\theta\right)  +1}d\theta\nonumber\\
&  =\frac{1}{2\pi\mathbb{Z}_{I}\left(  \beta|N\right)  }\int\limits_{-\pi
}^{\pi}\frac{\exp\left[  \ln\Xi\left(  \beta,ue^{i\theta}\right)  -N\ln
u-iN\theta\right]  }{\frac{1}{u}\exp\left(  \beta\varepsilon_{n}%
-i\theta\right)  +1}d\theta\nonumber\\
&  =\frac{1}{2\pi\mathbb{Z}_{I}\left(  \beta|N\right)  }\int\limits_{-\pi
}^{\pi}\frac{\Xi\left(  \beta,ue^{i\theta}\right)  }{u^{N-1}}\frac{\exp\left[
-i\theta\left(  N-1\right)  -\beta\varepsilon_{n}\right]  }{1+u\exp\left(
i\theta-\beta\varepsilon_{n}\right)  }d\theta. \label{co}%
\end{align}

The averages $\left\langle \hat{N}_{\mathbf{n}}\hat{N}_{\mathbf{m}%
}\right\rangle $ for $\mathbf{m}\neq\mathbf{n}$ can be also expressed in terms
of the integral representation:
\begin{align*}
\left.  \left\langle \hat{N}_{\mathbf{n}}\hat{N}_{\mathbf{m}}\right\rangle
\right\vert _{\mathbf{m}\neq\mathbf{n}}  &  =\frac{1}{\mathbb{Z}_{I}\left(
\beta|N\right)  }\left.  \sum_{\left\{  N_{\mathbf{n}}\right\}  }\left\langle
\Psi_{N,\left\{  N_{\mathbf{n}}\right\}  }\left\vert \exp\left(  -\beta\hat
{H}\right)  \hat{N}_{\mathbf{n}}\hat{N}_{\mathbf{m}}\right\vert \Psi
_{N,\left\{  N_{\mathbf{n}}\right\}  }\right\rangle \right\vert _{\mathbf{m}%
\neq\mathbf{n}}\\
&  =\frac{1}{\mathbb{Z}_{I}\left(  \beta|N\right)  }\left[  \sum_{\left\{
N_{\mathbf{n}^{\prime}}\right\}  }N_{\mathbf{n}}N_{\mathbf{m}}\exp\left(
-\beta\sum_{\mathbf{n}^{\prime}}\varepsilon_{\mathbf{n}^{\prime}}%
N_{\mathbf{n}^{\prime}}\right)  \right]  _{\sum_{\mathbf{n}^{\prime}%
}N_{\mathbf{n}^{\prime}}=N,\quad\mathbf{m}\neq\mathbf{n}}\\
&  =\frac{1}{\mathbb{Z}_{I}\left(  \beta|N\right)  \beta^{2}}\frac{\delta
^{2}\mathbb{Z}_{I}\left(  \beta|N\right)  }{\delta\varepsilon_{\mathbf{m}%
}\delta\varepsilon_{\mathbf{n}}}\\
&  =\frac{1}{\mathbb{Z}_{I}\left(  \beta|N\right)  \beta^{2}}\frac{\delta^{2}%
}{\delta\varepsilon_{\mathbf{m}}\delta\varepsilon_{\mathbf{n}}}\left(
\frac{1}{2\pi}\int\limits_{0}^{2\pi}d\theta\exp\left[  \ln\Xi\left(
\beta,e^{i\theta+\zeta}\right)  -N\zeta-iN\theta\right]  \right)
\quad\Rightarrow
\end{align*}
We obtain the integral representation for the average of the product of
operators $\hat{N}_{\mathbf{n}}\hat{N}_{\mathbf{m}}$ for $\mathbf{m\neq n}$:
\begin{equation}
\left.  \left\langle \hat{N}_{\mathbf{n}}\hat{N}_{\mathbf{m}}\right\rangle
\right\vert _{\mathbf{m}\neq\mathbf{n}}=\frac{1}{2\pi\mathbb{Z}_{I}\left(
N\right)  }\int\limits_{-\pi}^{\pi}\frac{\Phi_{N}\left(  \theta\right)
}{\left[  \exp\left(  \beta\varepsilon_{\mathbf{n}}-\zeta-i\theta\right)
+1\right]  \left[  \exp\left(  \beta\varepsilon_{\mathbf{m}}-\zeta
-i\theta\right)  +1\right]  }d\theta. \label{Num2}%
\end{equation}

Let us introduce the notation
\begin{equation}
f\left(  \varepsilon,\theta\right)  \equiv\frac{1}{\exp\left(  \beta
\varepsilon-\zeta-i\theta\right)  +1}, \label{QQF}%
\end{equation}
which formally coincides with the Fermi distribution function of the energy
$\varepsilon$ with the \textquotedblleft chemical potential\textquotedblright%
\ $\left(  \zeta+i\theta\right)  /\beta.$ Using this notation, the averages
(\ref{Num1}) and (\ref{Num2}) can be written down in the form
\begin{align}
\left\langle \hat{N}_{\mathbf{n}}\right\rangle  &  =\frac{1}{2\pi
\mathbb{Z}_{I}\left(  \beta|N\right)  }\int\limits_{-\pi}^{\pi}f\left(
\varepsilon_{\mathbf{n}},\theta\right)  \Phi_{N}\left(  \theta\right)
d\theta,\label{Num1a}\\
\left.  \left\langle \hat{N}_{\mathbf{n}}\hat{N}_{\mathbf{m}}\right\rangle
\right\vert _{\mathbf{m}\neq\mathbf{n}}  &  =\frac{1}{2\pi\mathbb{Z}%
_{I}\left(  N\right)  }\int\limits_{-\pi}^{\pi}f\left(  \varepsilon
_{\mathbf{n}},\theta\right)  f\left(  \varepsilon_{\mathbf{m}},\theta\right)
\Phi_{N}\left(  \theta\right)  d\theta. \label{Num2a}%
\end{align}
We can develop the aforesaid procedure for the average of a product of any
number of operators $\hat{N}_{\mathbf{n}_{1}}\hat{N}_{\mathbf{n}_{2}}\dots
\hat{N}_{\mathbf{n}_{K}},$ where all quantum numbers $\mathbf{n}%
_{1},\mathbf{n}_{2},\dots,\mathbf{n}_{K}$ are \emph{different}. The result
is:
\begin{equation}
\left.  \left\langle \hat{N}_{\mathbf{n}_{1}}\hat{N}_{\mathbf{n}_{2}}\dots
\hat{N}_{\mathbf{n}_{K}}\right\rangle \right\vert _{\mathbf{n}_{j}%
\neq\mathbf{n}_{l}}=\frac{1}{2\pi\mathbb{Z}_{I}\left(  N\right)  }%
\int\limits_{-\pi}^{\pi}f\left(  \varepsilon_{\mathbf{n}_{1}},\theta\right)
f\left(  \varepsilon_{\mathbf{n}_{2}},\theta\right)  \dots f\left(
\varepsilon_{\mathbf{n}_{K}},\theta\right)  \Phi_{N}\left(  \theta\right)
d\theta. \label{Gen}%
\end{equation}
It should be emphasized, that all expressions above [including Eq.
(\ref{Gen})] are derived for a \emph{canonical} Hibbs ensemble (i. e., for a
fixed number of particles) and for both closed-shell and open-shell systems.

Let us substitute the average (\ref{Av3}) into the correlation function
$\tilde{g}\left(  \mathbf{q},\tau|\left\{  N_{\sigma}\right\}  ,\beta\right)
$:
\begin{align*}
\tilde{g}\left(  \mathbf{q},\tau|\left\{  N_{\sigma}\right\}  ,\beta\right)
&  =\sum_{\mathbf{n},\mathbf{n}^{\prime}}\sum_{\mathbf{m},\mathbf{m}^{\prime}%
}\left(  e^{i\mathbf{q\cdot\hat{x}}}\right)  _{\mathbf{nn}^{\prime}}\left(
e^{-i\mathbf{q\cdot\hat{x}}}\right)  _{\mathbf{mm}^{\prime}}\exp\left[
\frac{\tau}{\hbar}\left(  \varepsilon_{\mathbf{n}}-\varepsilon_{\mathbf{n}%
^{\prime}}\right)  \right] \\
&  \times\left(  \delta_{\mathbf{n}^{\prime}\mathbf{n}}\delta_{\mathbf{m}%
^{\prime}\mathbf{m}}\left\langle \hat{N}_{\mathbf{n}}\hat{N}_{\mathbf{m}%
}\right\rangle +\delta_{\mathbf{m}^{\prime}\mathbf{n}}\delta_{\mathbf{n}%
^{\prime}\mathbf{m}}\left\langle \hat{N}_{\mathbf{n}}\left(  1-\hat
{N}_{\mathbf{m}}\right)  \right\rangle \right)
\end{align*}
$\Longrightarrow$%
\begin{align}
\tilde{g}\left(  \mathbf{q},\tau|\left\{  N_{\sigma}\right\}  ,\beta\right)
&  =\sum_{\mathbf{n},\mathbf{m}}\left(  e^{i\mathbf{q\cdot\hat{x}}}\right)
_{\mathbf{nn}}\left(  e^{-i\mathbf{q\cdot\hat{x}}}\right)  _{\mathbf{mm}%
}\left\langle \hat{N}_{\mathbf{n}}\hat{N}_{\mathbf{m}}\right\rangle
\nonumber\\
&  +\sum_{\mathbf{n,m}}\left\vert \left(  e^{i\mathbf{q\cdot\hat{x}}}\right)
_{\mathbf{nm}}\right\vert ^{2}\exp\left[  \frac{\tau}{\hbar}\left(
\varepsilon_{\mathbf{n}}-\varepsilon_{\mathbf{m}}\right)  \right]
\left\langle \hat{N}_{\mathbf{n}}\left(  1-\hat{N}_{\mathbf{m}}\right)
\right\rangle . \label{gq}%
\end{align}
The matrix elements $\left(  e^{i\mathbf{q\cdot\hat{x}}}\right)
_{\mathbf{nm}}$ has the following form%

\[
\left(  e^{i\mathbf{q\cdot\hat{x}}}\right)  _{\mathbf{nm}}=\left\langle
m_{1}\left\vert e^{iq_{1}\hat{x}_{1}}\right\vert n_{1}\right\rangle
\left\langle m_{2}\left\vert e^{iq_{2}\hat{x}_{2}}\right\vert n_{2}%
\right\rangle \left\langle m_{3}\left\vert e^{iq_{3}\hat{x}_{3}}\right\vert
n_{3}\right\rangle ,
\]
where $\left\langle m\left\vert e^{iq\hat{x}}\right\vert n\right\rangle $ is
the matrix element of a one-dimensional oscillator with the frequency $w$:%

\begin{align}
\left\langle m\left\vert e^{iq\hat{x}}\right\vert n\right\rangle  &
=\exp\left(  -\frac{\gamma^{2}}{2}\right)  \left(  i\gamma\right)
^{n_{>}-n_{<}}\sqrt{\frac{n_{<}!}{n_{>}!}}L_{n_{<}}^{\left(  n_{>}%
-n_{<}\right)  }\left(  \gamma^{2}\right)  ,\nonumber\\
&  \left\{
\begin{array}
[c]{l}%
n_{>}\equiv\max\left(  n,m\right)  ;\\
n_{<}\equiv\min\left(  n,m\right)  ,
\end{array}
\right.  \label{Mel1}%
\end{align}
$\gamma\equiv q\sqrt{\frac{\hbar}{2mw}}$, $\left\vert n\right\rangle $ are the
quantum states of the one-dimensional oscillator with the frequency $w$,
$L_{n}^{\left(  \alpha\right)  }\left(  z\right)  $ is the generalized
Laguerre polynomial.

\begin{quote}
\textbf{System with mixed spins}
\end{quote}

The correlation functions for a system with mixed spins can be explicitly
derived by the generalization of Eqs. (\ref{Zi5}) and (\ref{Gen}) to the case
of the particles with the non-zero spin. We use the fact, that the derivation
of Eqs. (\ref{Zi5}) and (\ref{Gen}), performed in this section, does not
depend on the concrete form of the energy spectrum $\varepsilon_{\mathbf{n}}.$
Hence, in the formulae, derived above, the replacement should be made:
\begin{equation}
\left\vert \mathbf{n}\right\rangle \rightarrow\left\vert \mathbf{n}%
,\sigma\right\rangle ,\quad\hat{N}_{\mathbf{n}}\rightarrow\hat{N}%
_{\mathbf{n},\sigma}=a_{\mathbf{n},\sigma}^{+}a_{\mathbf{n},\sigma},
\label{repl}%
\end{equation}
where $\sigma$ is the electron spin projection. Consequently, the matrix
elements $\left(  e^{i\mathbf{q\cdot\hat{x}}}\right)  _{\mathbf{mn}}$ are
replaced by
\begin{equation}
\left(  e^{i\mathbf{q\cdot\hat{x}}}\right)  _{\mathbf{mn}}\rightarrow
\left\langle \mathbf{m},\sigma\left\vert e^{i\mathbf{q\cdot\hat{x}}%
}\right\vert \mathbf{n},\sigma^{\prime}\right\rangle =\delta_{\sigma
\sigma^{\prime}}\left(  e^{i\mathbf{q\cdot\hat{x}}}\right)  _{\mathbf{mn}}.
\label{repl1}%
\end{equation}
Taking into account Eqs. (\ref{repl}) and (\ref{repl1}), the two-point
correlation function (\ref{gq}) becomes
\begin{align}
\tilde{g}\left(  \mathbf{q},\tau|\left\{  N_{\sigma}\right\}  ,\beta\right)
&  =\sum_{\mathbf{n},\mathbf{m}}\sum_{\sigma_{1},\sigma_{2}}\left(
e^{i\mathbf{q\cdot\hat{x}}}\right)  _{\mathbf{nn}}\left(  e^{-i\mathbf{q\cdot
\hat{x}}}\right)  _{\mathbf{mm}}\left\langle \hat{N}_{\mathbf{n},\sigma_{1}%
}\hat{N}_{\mathbf{m},\sigma_{2}}\right\rangle \nonumber\\
&  +\sum_{\mathbf{n,m}}\sum_{\sigma}\left\vert \left(  e^{i\mathbf{q\cdot
\hat{x}}}\right)  _{\mathbf{nm}}\right\vert ^{2}\exp\left[  \frac{\tau}{\hbar
}\left(  \varepsilon_{\mathbf{n}}-\varepsilon_{\mathbf{m}}\right)  \right]
\left\langle \hat{N}_{\mathbf{n},\sigma}\left(  1-\hat{N}_{\mathbf{m},\sigma
}\right)  \right\rangle . \label{gq1}%
\end{align}

The averages $\left\langle \hat{N}_{\mathbf{n},\sigma}\right\rangle $ and
$\left\langle \hat{N}_{\mathbf{n},\sigma_{1}}\hat{N}_{\mathbf{m},\sigma_{2}%
}\right\rangle $ are, respectively, one-particle and two-particle distribution
functions,%
\begin{align}
\left\langle \hat{N}_{\mathbf{n},\sigma}\right\rangle  &  \equiv f_{1}\left(
n,\sigma|N_{\sigma},\beta\right)  ,\\
\left\langle \hat{N}_{\mathbf{n},\sigma}\hat{N}_{\mathbf{n}^{\prime}%
,\sigma^{\prime}}\right\rangle  &  \equiv f_{2}\left(  n,\sigma;n^{\prime
},\sigma^{\prime}|\left\{  N_{\sigma}\right\}  ,\beta\right)  .
\end{align}
The one-electron distribution function $f_{1}\left(  n,\sigma|N_{\sigma}%
,\beta\right)  $ is the average number of electrons with the spin projection
$\sigma$ at the $n$-th energy level, while the two-electron distribution
function $f_{2}\left(  n,\sigma;n^{\prime},\sigma^{\prime}|\left\{  N_{\sigma
}\right\}  ,\beta\right)  $ is the average product of the numbers of electrons
with the spin projections $\sigma$ and $\sigma^{\prime}$ at the levels $n$ and
$n^{\prime}.$ These functions are expressed through the following integrals
[see (\ref{Num1a}), (\ref{Num2a})]:%
\begin{align}
f_{1}\left(  n,\sigma|N_{\sigma},\beta\right)   &  =\frac{1}{2\pi
\mathbb{Z}_{F}\left(  N_{\sigma},w,\beta\right)  }\int\limits_{-\pi}^{\pi
}f\left(  \varepsilon_{n},\theta\right)  \Phi\left(  \theta,\beta,N_{\sigma
}\right)  d\theta,\label{Num1a-1}\\
f_{2}\left(  n,\sigma;n^{\prime},\sigma^{\prime}|\left\{  N_{\sigma}\right\}
,\beta\right)   &  =\left\{
\begin{array}
[c]{c}%
\frac{1}{2\pi\mathbb{Z}_{F}\left(  N_{\sigma},w,\beta\right)  }\int
\limits_{-\pi}^{\pi}f\left(  \varepsilon_{n},\theta\right)  f\left(
\varepsilon_{n\mathbf{^{\prime}}},\theta\right)  \Phi\left(  \theta
,\beta,N_{\sigma}\right)  d\theta,\; \text{if }\sigma^{\prime}=\sigma;\\
f_{1}\left(  n,\sigma|N_{\sigma},\beta\right)  f_{1}\left(  n,\sigma^{\prime
}|N_{\sigma^{\prime}},\beta\right)  ,\; \text{if }\sigma^{\prime}\neq\sigma
\end{array}
\right.  \label{Num2a-1}%
\end{align}
with the notations%
\begin{equation}
\Phi\left(  \theta,\beta,N_{\sigma}\right)  =\exp\left[  \sum_{n=0}^{\infty
}\ln\left(  1+e^{i\theta+\mathbf{\xi}-\beta\varepsilon_{n}}\right)
-N_{\sigma}\left(  \mathbf{\xi}+i\theta\right)  \right]  , \label{FiNN}%
\end{equation}%
\begin{equation}
f\left(  \varepsilon,\theta\right)  \equiv\frac{1}{\exp\left(  \beta
\varepsilon-\mathbf{\xi}-i\theta\right)  +1}. \label{ferm}%
\end{equation}
The function $f\left(  \varepsilon,\theta\right)  $ formally coincides with
the Fermi-Dirac distribution function of the energy $\varepsilon$ with the
\textquotedblleft chemical potential\textquotedblright\ $\left(  \mathbf{\xi
}+i\theta\right)  /\beta.$

Here we consider the zero-temperature limit, for which the integrals
(\ref{Num1a-1}) and (\ref{Num2a-1}) can be calculated analytically. The result
for the one-electron distribution function is%
\begin{equation}
\left.  f_{1}\left(  n,\sigma|\beta,N_{\sigma}\right)  \right\vert
_{\beta\rightarrow\infty}=\left\{
\begin{array}
[c]{cc}%
1, & n<L_{\sigma};\\
0, & n>L_{\sigma};\\
\frac{N_{\sigma}-N_{L_{\sigma}}}{g_{L_{\sigma}}}, & n=L_{\sigma}.
\end{array}
\right.  \label{p1}%
\end{equation}
According to (\ref{p1}), $L_{\sigma}$ is the number of the lowest open shell,
and
\[
g_{n}=\left\{
\begin{array}
[c]{ccc}%
\frac{1}{2}\left(  n+1\right)  \left(  n+2\right)  &  & \left(  3D\right)  ,\\
n+1 &  & \left(  2D\right)  .
\end{array}
\right.
\]
is the degeneracy of the $n$-th shell. $N_{L_{\sigma}}$ is the number of
electrons in all the closed shells with the spin projection $\sigma,$%
\begin{equation}
N_{L_{\sigma}}\equiv\sum_{n=0}^{L_{\sigma}-1}g_{n}=\left\{
\begin{array}
[c]{ccc}%
\frac{1}{6}L_{\sigma}\left(  L_{\sigma}+1\right)  \left(  L_{\sigma}+2\right)
&  & \left(  3D\right)  ,\\
\frac{1}{2}L_{\sigma}\left(  L_{\sigma}+1\right)  &  & \left(  2D\right)  .
\end{array}
\right.  \label{NLs}%
\end{equation}
The two-electron distribution function $f_{2}\left(  n,\sigma;n^{\prime
},\sigma^{\prime}|\left\{  N_{\sigma}\right\}  ,\beta\right)  $ at $T=0$ takes
the form%
\begin{align}
&  \left.  f_{2}\left(  n,\sigma;n^{\prime},\sigma^{\prime}|\beta,\left\{
N_{\sigma}\right\}  \right)  \right\vert _{\beta\rightarrow\infty}\nonumber\\
&  =\left\{
\begin{array}
[c]{cc}%
\left.  f_{1}\left(  n,\sigma|\beta,N_{\sigma}\right)  \right\vert
_{\beta\rightarrow\infty}\left.  f_{1}\left(  n^{\prime},\sigma^{\prime}%
|\beta,N_{\sigma^{\prime}}\right)  \right\vert _{\beta\rightarrow\infty}, &
n\neq n^{\prime}\; \mathrm{or}\; \sigma\neq\sigma^{\prime}\\
1, & \sigma=\sigma^{\prime}\; \mathrm{and}\;n=n^{\prime}<L_{\sigma};\\
0, & \sigma=\sigma^{\prime}\; \mathrm{and}\;n=n^{\prime}>L_{\sigma};\\
\frac{N_{\sigma}-N_{L_{\sigma}}}{g_{L_{\sigma}}}\frac{N_{\sigma}-N_{L_{\sigma
}}-1}{g_{L_{\sigma}}-1}, & \sigma=\sigma^{\prime}\; \mathrm{and}\;n=n^{\prime
}=L_{\sigma}.
\end{array}
\right.  \label{p2}%
\end{align}

In summary, we have obtained the following expression for $\tilde{g}\left(
\mathbf{q},\tau|\left\{  N_{\sigma}\right\}  ,\beta\right)  $:%
\begin{align}
\tilde{g}\left(  \mathbf{q},\tau|\left\{  N_{\sigma}\right\}  ,\beta\right)
&  =\sum_{\mathbf{n},\sigma,\mathbf{n}^{\prime},\sigma^{\prime}}\left(
e^{i\mathbf{q\cdot x}}\right)  _{\mathbf{nn}}\left(  e^{-i\mathbf{q\cdot x}%
}\right)  _{\mathbf{n^{\prime}n^{\prime}}}f_{2}\left(  n,\sigma;n^{\prime
},\sigma^{\prime}|\left\{  N_{\sigma}\right\}  ,\beta\right) \nonumber\\
&  +\sum_{\mathbf{n},\mathbf{n}^{\prime},\sigma}\left\vert \left(
e^{i\mathbf{q\cdot x}}\right)  _{\mathbf{nn^{\prime}}}\right\vert ^{2}%
\exp\left[  \frac{\tau}{\hbar}\left(  \varepsilon_{n}-\varepsilon_{n^{\prime}%
}\right)  \right] \nonumber\\
&  \times\left[  f_{1}\left(  n,\sigma|\left\{  N_{\sigma}\right\}
,\beta\right)  -f_{2}\left(  n,\sigma;n^{\prime},\sigma|\left\{  N_{\sigma
}\right\}  ,\beta\right)  \right]  . \label{g4}%
\end{align}
This formula is valid for both closed and open shells. The correlation
functions derived in this subsection are used both for the calculation of the
ground-state energy and, a shown below, for the calculation of the optical
conductivity of an $N$-polaron system in a quantum dot.

\subsection{Many-polaron ground state in a quantum dot: extrapolation to the
homogeneous limit and comparison to the results for a polaron gas in bulk
\cite{KBD05}}

The correlation function given by Eq. (\ref{g4}) can be subdivided as%
\begin{equation}
\tilde{g}\left(  \mathbf{q},\tau|\left\{  N_{\sigma}\right\}  ,\beta\right)
=\tilde{g}_{1}\left(  \mathbf{q},\tau|\left\{  N_{\sigma}\right\}
,\beta\right)  +\tilde{g}_{2}\left(  \mathbf{q},\tau|\left\{  N_{\sigma
}\right\}  ,\beta\right)  , \label{subd}%
\end{equation}
with%
\begin{align}
&  \tilde{g}_{1}\left(  \mathbf{q},\tau|\left\{  N_{\sigma}\right\}
,\beta\right)  \equiv\sum_{\mathbf{n},\mathbf{n}^{\prime},\sigma}\left\vert
\left(  e^{i\mathbf{q\cdot x}}\right)  _{\mathbf{nn^{\prime}}}\right\vert
^{2}\exp\left[  \frac{\tau}{\hbar}\left(  \varepsilon_{n}-\varepsilon
_{n^{\prime}}\right)  \right] \nonumber\\
&  \times\left[  f_{1}\left(  n,\sigma|\left\{  N_{\sigma}\right\}
,\beta\right)  -f_{2}\left(  n,\sigma;n^{\prime},\sigma|\left\{  N_{\sigma
}\right\}  ,\beta\right)  \right]  ,
\end{align}%
\begin{equation}
\tilde{g}_{2}\left(  \mathbf{q},\tau|\left\{  N_{\sigma}\right\}
,\beta\right)  \equiv\sum_{\mathbf{n},\sigma,\mathbf{n}^{\prime}%
,\sigma^{\prime}}\left(  e^{i\mathbf{q\cdot x}}\right)  _{\mathbf{nn}}\left(
e^{-i\mathbf{q\cdot x}}\right)  _{\mathbf{n^{\prime}n^{\prime}}}f_{2}\left(
n,\sigma;n^{\prime},\sigma^{\prime}|\left\{  N_{\sigma}\right\}
,\beta\right)  .
\end{equation}
In accordance with the subdivision (\ref{subd}) of the correlation function,
we subdivide the Coulomb and polaron contributions:%
\begin{align}
E_{C}  &  =E_{C}^{\left(  1\right)  }+E_{C}^{\left(  2\right)  }%
,\label{Ec12}\\
E_{e-ph}  &  =E_{e-ph}^{\left(  1\right)  }+E_{e-ph}^{\left(  2\right)  }.
\label{Epol12}%
\end{align}

\begin{figure}
[h]
\begin{center}
\includegraphics[
height=6.4778cm,
width=8.4416cm
]%
{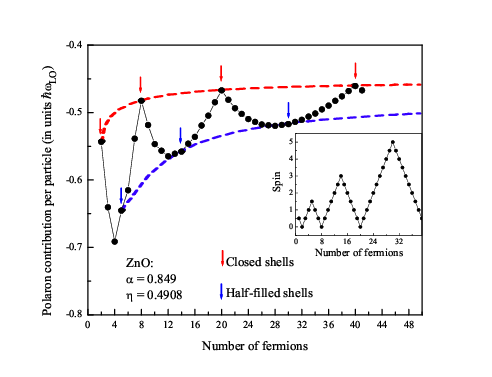}%
\caption{Polaron contribution $E_{e-ph}^{\left(  1\right)  }/N$ as a function
of $N$ for a ZnO quantum dot.The material parameters for ZnO are taken from
Ref. \cite{LDB77}. The value $r_{s}^{\ast}=2$ corresponds to $n_{0}%
=4.34\times10^{19}$ cm$^{-2}$. Inset: the total spin as a function of $N$.}%
\label{PolCon3}%
\end{center}
\end{figure}

\begin{figure}
[h]
\begin{center}
\includegraphics[
height=6.4075cm,
width=8.334cm
]%
{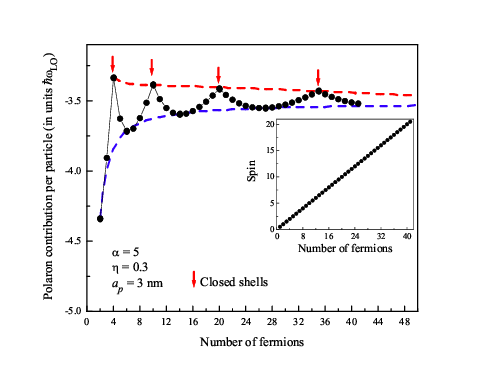}%
\caption{Polaron contribution $E_{e-ph}^{\left(  1\right)  }/N$ as a function
of $N$ for a quantum dot of a polar medium with $\alpha=5$, $\eta=0.3$. The
value $r_{s}^{\ast}=20$ corresponds to $n_{0}=1.14\times10^{18}$ cm$^{-3}.$
Inset: the total spin as a function of $N$.}%
\label{PolCon4}%
\end{center}
\end{figure}

We have numerically checked whether the polaron contribution per particle
$E_{e-ph}^{\left(  1\right)  }/N$ tends to a finite value at $N\rightarrow
\infty$. In Figs. \ref{PolCon3} and \ref{PolCon4}, we have plotted the polaron
contributions $E_{e-ph}^{\left(  1\right)  }/N$ as a function of $N$ for a
quantum dot in ZnO and in a polar medium with $\alpha=5$, $\eta=0.3$,
respectively. \footnote{Since, as discussed above, for a single polaron only
the whole polaron contribution $E_{e-ph}=E_{e-ph}^{\left(  1\right)
}+E_{e-ph}^{\left(  2\right)  }$ has a physical meaning, the plots for
$E_{e-ph}^{\left(  1\right)  }$ in Figs. 3 and 4 start from $N=2$. The total
polaron contribution $E_{e-ph}$ for $N=1$ is plotted below, in Fig. 9.} As
seen from Fig. \ref{PolCon3}, the polaron contribution $E_{e-ph}^{\left(
1\right)  }/N$ in ZnO as a function of $N$ oscillates taking expressed maxima
for $N$ corresponding to the closed shells $N=2,8,20,40,\ldots$. There exist
kinks of $E_{e-ph}^{\left(  1\right)  }/N$ at $N$ corresponding to the
half-filled shells, but these kinks are extremely small. In the case of the
medium with $\alpha=5,$ $\eta=0.3,$ for $r_{s}^{\ast}=20$ (what corresponds to
the density $n_{0}\approx1.14\times10^{18}$ cm$^{-3}$), the polaron
contribution $E_{e-ph}^{\left(  1\right)  }/N$ oscillates taking maximal
values at the numbers of fermions, which correspond to the closed shells for a
spin-polarized system with parallel spins.

In Figs. \ref{PolCon3} and \ref{PolCon4}, the dashed curves are the envelopes
for local maxima (closed shells) and local minima of $E_{e-ph}^{\left(
1\right)  }/N.$We see that when these envelopes are \textit{extrapolated} to
larger number of fermions, the distance between the envelopes decreases.
Therefore, the magnitude of the variations of $E_{e-ph}^{\left(  1\right)
}/N$ related to the shell filling diminishes with increasing $N$, and it is
safe to suppose that in the limit of large $N,$ the envelopes tend to one and
the same value. That value corresponds to the \textit{homogeneous}
(\textquotedblleft bulk\textquotedblright) limit $\lim_{N\rightarrow\infty
}\left(  E_{e-ph}^{\left(  1\right)  }/N\right)  $.

\newpage%

%TCIMACRO{\FRAME{fhFU}{8.4416cm}{6.4778cm}{0pt}{\Qcb{Polaron contribution
%$E_{e-ph}^{\left(  1\right)  }/N$ as a function of $N$ for a ZnO quantum
%dot.The material parameters for ZnO are taken from Ref. \cite{LDB77}. The
%value $r_{s}^{\ast}=2$ corresponds to $n_{0}=4.34\times10^{19}$ cm$^{-2}$.
%Inset: the total spin as a function of $N$.}}{\Qlb{PolCon3}}{part2fig3.eps}%
%maintain-aspect-ratio TRUE;  display "ICON";  valid_file "F";
%width 8.4416cm;  height 6.4778cm;  depth 0pt;  original-width 8.3252cm;
%original-height 6.3724cm;  cropleft "0";  croptop "1";  cropright "1";
%cropbottom "0";  filename 'Part2Fig3.eps';file-properties "XNPEU";}} }%
%BeginExpansion
%EndExpansion
%

%TCIMACRO{\FRAME{fhFU}{8.334cm}{6.4075cm}{0pt}{\Qcb{Polaron contribution
%$E_{e-ph}^{\left(  1\right)  }/N$ as a function of $N$ for a quantum dot of a
%polar medium with $\alpha=5$, $\eta=0.3$. The value $r_{s}^{\ast}=20$
%corresponds to $n_{0}=1.14\times10^{18}$ cm$^{-3}.$ Inset: the total spin as a
%function of $N$.}}{\Qlb{PolCon4}}{part2fig4.eps}%
%maintain-aspect-ratio TRUE;  display "ICON";  valid_file "F";  width 8.334cm;
%height 6.4075cm;  depth 0pt;  original-width 8.2176cm;
%original-height 6.3021cm;  cropleft "0";  croptop "1";  cropright "1";
%cropbottom "0";  filename 'Part2Fig4.eps';file-properties "XNPEU";}} }%
%BeginExpansion
%EndExpansion

%\newpage

\begin{figure}
[h]
\begin{center}
\includegraphics[
height=10.1857cm,
width=8.3691cm
]%
{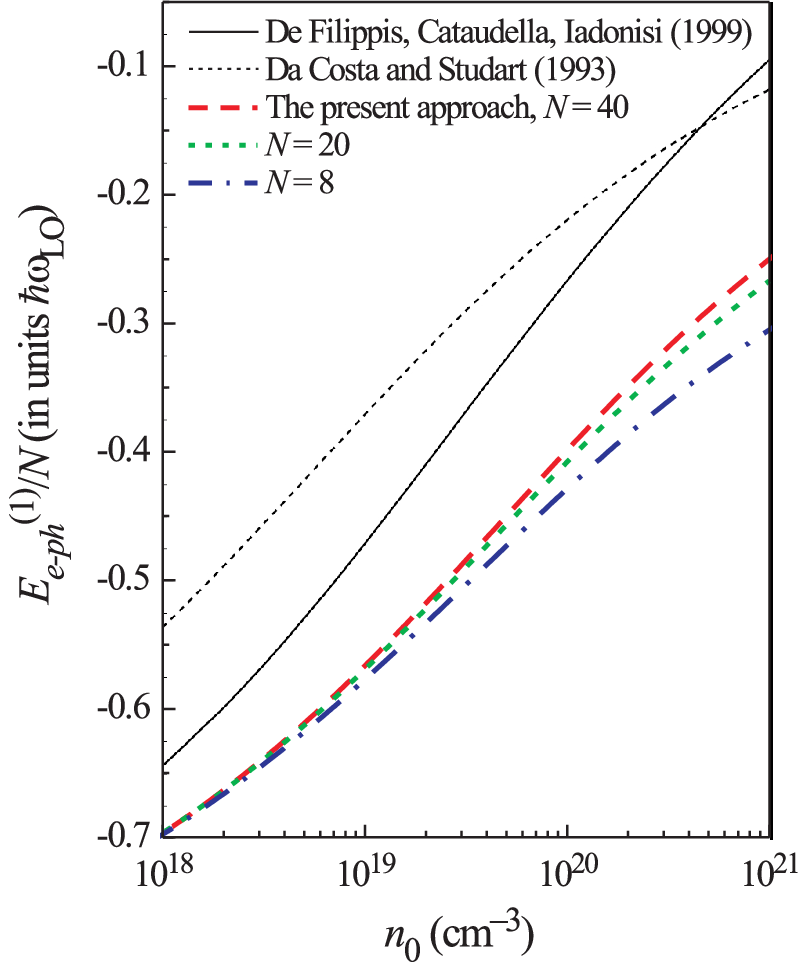}%
\caption{Polaron contribution to the polaron ground-state energy per particle
$E_{e-ph}^{\left(  1\right)  }$ in an $N$-polaron quantum dot as a function of
the effective density. The parameters are taken for ZnO (see Ref.
\cite{LDB77}): $\alpha=0.849$, $\varepsilon_{0}=8.15$, $\varepsilon_{\infty
}=4.0$, $\hbar\omega_{\mathrm{LO}}=73.27$ meV, $m_{b}=0.24m_{e}$, where
$m_{e}$ is the electron mass in the vacuum. This polaron contribution is
compared with the polaron contribution to the ground-state energy of a polaron
gas in bulk calculated in Refs. \cite{FCI1999,CS1993}.}%
\label{PolCon}%
\end{center}
\end{figure}

In Fig. \ref{PolCon}, we compare the polaron contribution $E_{e-ph}^{\left(
1\right)  }/N$ calculated within our variational path-integral method for
different numbers of fermions with the polaron contribution to the
ground-state energy per particle for a polaron gas in bulk, calculated (i) in
Ref. \cite{FCI1999} within an intermediate-coupling approach (the thin solid
curve), (ii) in Ref. \cite{CS1993}, using a variational approach developed
first in Ref. \cite{LDB77}. As seen from this figure, our all-coupling
variational method provides lower values for the polaron contribution than
those obtained in Refs. \cite{FCI1999,CS1993}. The difference between the
polaron contribution calculated within our method and that of Ref.
\cite{FCI1999} is smaller at low densities and increases in magnitude with
increasing density. The difference between the polaron contribution calculated
within our method and that of Ref. \cite{CS1993} very slightly depends on the
density. The result of Ref. \cite{CS1993} becomes closer to our result only at
high densities.

%\newpage%
%

%TCIMACRO{\FRAME{fhFU}{8.3691cm}{10.1857cm}{0pt}{\Qcb{Polaron contribution to
%the polaron ground-state energy per particle $E_{e-ph}^{\left(  1\right)  }$
%in an $N$-polaron quantum dot as a function of the effective density. The
%parameters are taken for ZnO (see Ref. \cite{LDB77}): $\alpha=0.849$,
%$\varepsilon_{0}=8.15$, $\varepsilon_{\infty}=4.0$, $\hbar\omega
%_{\QTR{rm}{LO}}=73.27$ meV, $m_{b}=0.24m_{e}$, where $m_{e}$ is the electron
%mass in the vacuum. This polaron contribution is compared with the polaron
%contribution to the ground-state energy of a polaron gas in bulk calculated in
%Refs. \cite{FCI1999,CS1993}.}}{\Qlb{PolCon}}{part2fig5.eps}%
%maintain-aspect-ratio TRUE;  display "ICON";  valid_file "F";
%width 8.3691cm;  height 10.1857cm;  depth 0pt;  original-width 13.6037cm;
%original-height 16.5823cm;  cropleft "0";  croptop "1";  cropright "1";
%cropbottom "0";  filename 'Part2Fig5.eps';file-properties "XNPEU";}} }%
%BeginExpansion
%EndExpansion

\newpage

\subsection{Optical conductivity}

In Ref. \cite{MPQD-PRB2004} we have extended the memory-function approach to a
system of arbitrary-coupling interacting polarons confined to a parabolic
confinement potential. The optical conductivity relates the current
$\mathbf{J}\left(  t\right)  $ per electron to a time-dependent uniform
electric field $\mathbf{E}\left(  t\right)  $ in the framework of linear
response theory. Further on, the Fourier components of the electric field are
denoted by $\mathbf{E}\left(  \omega\right)  :$
\begin{equation}
\mathbf{E}\left(  t\right)  =\frac{1}{2\pi}\int_{-\infty}^{\infty}%
\mathbf{E}\left(  \omega\right)  e^{-i\omega t}d\omega, \label{Fourier}%
\end{equation}
and the similar denotations are used for other time-dependent quantities. The
electric current per electron $\mathbf{J}\left(  t\right)  $ is related to the
mean electron coordinate response $\mathbf{R}\left(  t\right)  $ by
\begin{equation}
\mathbf{J}\left(  t\right)  =-e\frac{d\mathbf{R}\left(  t\right)  }{dt},
\label{J}%
\end{equation}
and hence%
\begin{equation}
\mathbf{J}\left(  \omega\right)  =ie\omega\mathbf{R}\left(  \omega\right)  .
\label{J1}%
\end{equation}
Within the linear-response theory, both the electric current and the
coordinate response are proportional to $\mathbf{E}\left(  \omega\right)  $:%
\begin{equation}
\mathbf{J}\left(  \omega\right)  =\sigma\left(  \omega\right)  \mathbf{E}%
\left(  \omega\right)  ,\quad\mathbf{R}\left(  \omega\right)  =\frac
{\sigma\left(  \omega\right)  }{ie\omega}\mathbf{E}\left(  \omega\right)  ,
\label{J2}%
\end{equation}
where $\sigma\left(  \omega\right)  $ is the conductivity per electron.
Because we treat an isotropic electron-phonon system, $\sigma\left(
\omega\right)  $ is a scalar function. It is determined from the time
evolution of the center-of-mass coordinate:
\begin{equation}
\mathbf{R}\left(  t\right)  \equiv\frac{1}{N}\left\langle \left\langle
\sum_{j=1}^{N}\mathbf{x}_{j}\left(  t\right)  \right\rangle \right\rangle
_{S}. \label{R}%
\end{equation}
The symbol $\left\langle \left\langle \left(  \bullet\right)  \right\rangle
\right\rangle _{S}$ denotes an average in the \emph{real-time} representation
for a system with action functional $S$:%
\begin{equation}
\left\langle \left\langle \left(  \bullet\right)  \right\rangle \right\rangle
_{S}\equiv\int d\mathbf{\bar{x}}\int d\mathbf{\bar{x}}_{0}\int d\mathbf{\bar
{x}}_{0}^{\prime}\int\limits_{\mathbf{\bar{x}}_{0}}^{\mathbf{\bar{x}}%
}D\mathbf{\bar{x}}\left(  t\right)  \int\limits_{\mathbf{\bar{x}}_{0}^{\prime
}}^{\mathbf{\bar{x}}}D\mathbf{\bar{x}}^{\prime}\left(  t\right)  e^{\frac
{i}{\hbar}S\left[  \mathbf{\bar{x}}\left(  t\right)  ,\mathbf{\bar{x}}%
^{\prime}\left(  t\right)  \right]  }\left(  \bullet\right)  \left.
\left\langle \mathbf{\bar{x}}_{0}\left\vert \hat{\rho}\left(  t_{0}\right)
\right\vert \mathbf{\bar{x}}_{0}^{^{\prime}}\right\rangle \right\vert
_{t_{0}\rightarrow-\infty}, \label{Average}%
\end{equation}
where $\left\langle \mathbf{\bar{x}}_{0}\left\vert \hat{\rho}\left(
t_{0}\right)  \right\vert \mathbf{\bar{x}}_{0}^{^{\prime}}\right\rangle $ is
the density matrix before the onset of the electric field in the infinite past
$\left(  t_{0}\rightarrow-\infty\right)  $. The corresponding action
functional is%
\begin{equation}
S\left[  \mathbf{\bar{x}}\left(  t\right)  ,\mathbf{\bar{x}}^{\prime}\left(
t\right)  \right]  =\int\limits_{-\infty}^{t}\left[  L_{e}\left(
\mathbf{\dot{\bar{x}}}\left(  t\right)  ,\mathbf{\bar{x}}\left(  t\right)
,t\right)  -L_{e}\left(  \mathbf{\dot{\bar{x}}}^{\prime}\left(  t\right)
,\mathbf{\bar{x}}^{\prime}\left(  t\right)  ,t\right)  \right]  dt^{\prime
}-i\hbar\Phi\left[  \mathbf{\bar{x}}\left(  t\right)  ,\mathbf{\bar{x}%
}^{\prime}\left(  t\right)  \right]  , \label{S1}%
\end{equation}
where $L_{e}\left(  \mathbf{\dot{\bar{x}}},\mathbf{\bar{x}},t\right)  $ is the
Lagrangian of $N$ interacting electrons in a time-dependent uniform electric
field $\mathbf{E}\left(  t\right)  $
\begin{equation}
L_{e}\left(  \mathbf{\dot{\bar{x}}},\mathbf{\bar{x}},t\right)  =\sum_{\sigma
}\sum_{j=1}^{N_{\sigma}}\left(  \frac{m_{b}\mathbf{\dot{x}}_{j,\sigma}^{2}}%
{2}-\frac{m_{b}\Omega_{0}^{2}\mathbf{x}_{j,\sigma}^{2}}{2}-e\mathbf{x}%
_{j,\sigma}\cdot\mathbf{E}\left(  t\right)  \right)  -\underset{\left(
j,\sigma\right)  \neq\left(  l,\sigma^{\prime}\right)  }{\sum_{\sigma
,\sigma^{\prime}}\sum_{j=1}^{N_{\sigma}}\sum_{l=1}^{N_{\sigma^{\prime}}}}%
\frac{e^{2}}{2\varepsilon_{\infty}\left\vert \mathbf{x}_{j,\sigma}%
-\mathbf{x}_{l,\sigma^{\prime}}\right\vert }. \label{L1}%
\end{equation}
The influence phase of the phonons
\begin{align}
\Phi\left[  \mathbf{\bar{x}}\left(  s\right)  ,\mathbf{\bar{x}}^{\prime
}\left(  s\right)  \right]   &  =-\sum_{\mathbf{q}}\frac{\left\vert
V_{\mathbf{q}}\right\vert ^{2}}{\hbar^{2}}\int\limits_{-\infty}^{t}%
ds\int\limits_{-\infty}^{s}ds^{\prime}\left[  \rho_{\mathbf{q}}\left(
s\right)  -\rho_{\mathbf{q}}^{\prime}\left(  s\right)  \right] \nonumber\\
&  \times\left[  T_{\omega_{\mathbf{q}}}^{\ast}\left(  s-s^{\prime}\right)
\rho_{\mathbf{q}}\left(  s^{\prime}\right)  -T_{\omega_{\mathbf{q}}}\left(
s-s^{\prime}\right)  \rho_{\mathbf{q}}^{\prime}\left(  s^{\prime}\right)
\right]
\end{align}
describes both a retarded interaction between different electrons and a
retarded self-interaction of each electron due to the elimination of the
phonon coordinates. This functional contains the free-phonon Green's
function:
\begin{equation}
T_{\omega}\left(  t\right)  =\frac{e^{i\omega t}}{1-e^{-\beta\hbar\omega}%
}+\frac{e^{-i\omega t}}{e^{\beta\hbar\omega}-1}. \label{PGF}%
\end{equation}
The equation of motion for $\mathbf{R}\left(  t\right)  $ is%
\begin{equation}
m_{b}\frac{d^{2}\mathbf{R}\left(  t\right)  }{dt^{2}}+m_{b}\Omega_{0}%
^{2}\mathbf{R}\left(  t\right)  +e\mathbf{E}\left(  t\right)  =\mathbf{F}%
_{ph}\left(  t\right)  , \label{EqMotion}%
\end{equation}
where $\mathbf{F}_{ph}\left(  t\right)  $ is the average force due to the
electron-phonon interaction,
\begin{equation}
\mathbf{F}_{ph}\left(  t\right)  =-\operatorname{Re}\sum_{\mathbf{q}}%
\frac{2\left\vert V_{\mathbf{q}}\right\vert ^{2}\mathbf{q}}{N\hbar}%
\int\limits_{-\infty}^{t}ds\;T_{\omega_{\mathrm{LO}}}^{\ast}\left(
t-s\right)  \left\langle \left\langle \rho_{\mathbf{q}}\left(  t\right)
\rho_{-\mathbf{q}}\left(  s\right)  \right\rangle \right\rangle _{S}.
\label{Fph}%
\end{equation}
The two-point correlation function $\langle\left\langle \rho_{\mathbf{q}%
}\left(  t\right)  \rho_{-\mathbf{q}}\left(  s\right)  \right\rangle
\rangle_{S}$ should be calculated from Eq.~(\ref{Average}) using the exact
action (\ref{S1}), but like for the free energy above, this path integral
cannot be calculated analytically. Instead, we perform an approximate
calculation, replacing the two-point correlation function in Eq. (\ref{Fph})
by $\langle\left\langle \rho_{\mathbf{q}}\left(  t\right)  \rho_{-\mathbf{q}%
}\left(  s\right)  \right\rangle \rangle_{S_{0}},$ where $S_{0}\left[
\mathbf{\bar{x}}\left(  t\right)  ,\mathbf{\bar{x}}^{\prime}\left(  t\right)
\right]  $ is the action functional with the optimal values of the variational
parameters for the model system considered in the previous section in the
presence of the electric field $\mathbf{E}\left(  t\right)  $. The functional
$S_{0}\left[  \mathbf{\bar{x}}\left(  t\right)  ,\mathbf{\bar{x}}^{\prime
}\left(  t\right)  \right]  $ is quadratic and describes a system of coupled
harmonic oscillators in the uniform electric field $\mathbf{E}\left(
t\right)  $. This field enters the term $-e\mathbf{E}\left(  t\right)
\cdot\sum_{\sigma}\sum_{j=1}^{N_{\sigma}}\mathbf{x}_{j,\sigma}$ in the
Lagrangian, which only affects the center-of-mass coordinate. Hence, a shift
of variables to the frame of reference with the origin at the center of mass
\begin{equation}
\left\{
\begin{array}
[c]{l}%
\mathbf{x}_{n}\left(  t\right)  =\mathbf{\tilde{x}}_{n}\left(  t\right)
+\mathbf{R}\left(  t\right)  ,\\
\mathbf{x}_{n}^{\prime}\left(  t\right)  =\mathbf{\tilde{x}}_{n}^{\prime
}\left(  t\right)  +\mathbf{R}\left(  t\right)  ,
\end{array}
\right.  \label{Tr1}%
\end{equation}
results in%
\begin{equation}
\left\langle \left\langle \rho_{q}\left(  t\right)  \rho_{-q}\left(  s\right)
\right\rangle \right\rangle _{S_{0}}=\left.  \left\langle \left\langle
\rho_{q}\left(  t\right)  \rho_{-q}\left(  s\right)  \right\rangle
\right\rangle _{S_{0}}\right\vert _{E=0}e^{i\mathbf{q}\cdot\left[
\mathbf{R}\left(  t\right)  -\mathbf{R}\left(  s\right)  \right]  }.
\label{Tr2}%
\end{equation}
This result (\ref{Tr2}) is valid for any \emph{quadratic} model action
$S_{0}.$

The applicability of the parabolic approximation is confirmed by the fact that
a self-induced polaronic potential, created by the polarization cloud around
an electron, is rather well described by a parabolic potential whose
parameters are determined by a variational method. For weak coupling, our
variational method is at least of the same accuracy as the perturbation
theory, which results from our approach at a special choice of the variational
parameters. For strong coupling, an interplay of the electron-phonon
interaction and the Coulomb correlations within a confinement potential can
lead to the assemblage of polarons in multi-polaron systems. Our choice of the
model variational system is reasonable because of this trend, apparently
occurring in a many-polaron system with arbitrary $N$ for a finite confinement strength.

The correlation function $\left.  \left\langle \rho_{\mathbf{q}}\left(
t\right)  \rho_{-\mathbf{q}}\left(  s\right)  \right\rangle _{S_{0}%
}\right\vert _{\mathbf{E}=0}$ corresponds to the model system in the absence
of an electric field. For $t>s,$ this function is related to the
imaginary-time correlation function $\mathcal{G}\left(  \mathbf{q}%
,\tau|\left\{  N_{\sigma}\right\}  ,\beta\right)  ,$ described in the previous
section:%
\begin{equation}
\left.  \left\langle \left\langle \rho_{\mathbf{q}}\left(  t\right)
\rho_{-\mathbf{q}}\left(  s\right)  \right\rangle \right\rangle _{S_{0}%
}\right\vert _{\mathbf{E}=0,t>s}=\mathcal{G}\left(  \mathbf{q},i\left(
t-s\right)  |\left\{  N_{\sigma}\right\}  ,\beta\right)  . \label{Tr3}%
\end{equation}
Using the transformation (\ref{Tr1}) and the relation (\ref{Tr3}), we obtain
from Eq. (\ref{Fph})%
\begin{equation}
\mathbf{F}_{ph}\left(  t\right)  =-\operatorname{Re}\sum_{\mathbf{q}}%
\frac{2\left\vert V_{\mathbf{q}}\right\vert ^{2}\mathbf{q}}{N\hbar}%
\int\limits_{-\infty}^{t}T_{\omega_{\mathrm{LO}}}^{\ast}\left(  t-s\right)
\mathrm{e}^{\mathrm{i}\mathbf{q}\cdot\left[  \mathbf{R}\left(  t\right)
-\mathbf{R}\left(  s\right)  \right]  }\mathcal{G}\left(  \mathbf{q},i\left(
t-s\right)  |\left\{  N_{\sigma}\right\}  ,\beta\right)  ds. \label{F1}%
\end{equation}

Within the linear-response theory, we expand the function $\mathrm{e}%
^{\mathrm{i}\mathbf{q}\cdot\left[  \mathbf{R}\left(  t\right)  -\mathbf{R}%
\left(  s\right)  \right]  }$ in Eq. (\ref{F1}) as a Taylor series in $\left[
\mathbf{R}\left(  t\right)  -\mathbf{R}\left(  s\right)  \right]  $ up to the
first-order term. The zeroth-order term gives no contribution into
$\mathbf{F}_{ph}\left(  t\right)  $ due to the symmetry of $\left\vert
V_{\mathbf{q}}\right\vert ^{2}$ and of $f_{\mathbf{q}}\left(  t-s\right)  $
with respect to the inversion $\mathbf{q}\rightarrow-\mathbf{q}$. In this
approach, the Cartesian coordinates of the force $\left(  j=1,2,3\right)  $
become
\begin{align}
\left(  \mathbf{F}_{ph}\left(  t\right)  \right)  _{j}  &  =\sum_{k=1}^{3}%
\sum_{\mathbf{q}}\frac{2\left\vert V_{\mathbf{q}}\right\vert ^{2}q_{j}q_{k}%
}{N\hbar}\int\limits_{-\infty}^{t}\left[  R_{k}\left(  t\right)  -R_{k}\left(
s\right)  \right] \nonumber\\
&  \times\operatorname{Im}\left[  T_{\omega_{\mathrm{LO}}}^{\ast}\left(
t-s\right)  \mathcal{G}\left(  \mathbf{q},i\left(  t-s\right)  |\left\{
N_{\sigma}\right\}  ,\beta\right)  \right]  ds. \label{aa}%
\end{align}

Further on, we perform the Fourier expansion:%
\begin{equation}
\mathbf{R}\left(  t\right)  =\frac{1}{2\pi}\int_{-\infty}^{\infty}%
\mathbf{R}\left(  \omega\right)  e^{-i\omega t}d\omega. \label{Fourier1}%
\end{equation}
In Eq. (\ref{aa}), we make the replacement%
\[
\tau\equiv t-s,\quad\Longrightarrow\quad s=t-\tau,
\]
what gives%
\begin{align*}
\left(  \mathbf{F}_{ph}\left(  t\right)  \right)  _{j}  &  =\sum_{k=1}^{3}%
\sum_{\mathbf{q}}\frac{2\left\vert V_{\mathbf{q}}\right\vert ^{2}q_{j}q_{k}%
}{N\hbar}\int\limits_{0}^{\infty}d\tau\; \left[  R_{k}\left(  t\right)
-R_{k}\left(  t-\tau\right)  \right]  \operatorname{Im}\left[  T_{\omega
_{\mathrm{LO}}}^{\ast}\left(  \tau\right)  \mathcal{G}\left(  \mathbf{q}%
,it|\left\{  N_{\sigma}\right\}  ,\beta\right)  \right] \\
&  =\sum_{k=1}^{3}\sum_{\mathbf{q}}\frac{2\left\vert V_{\mathbf{q}}\right\vert
^{2}q_{j}q_{k}}{N\hbar}\frac{1}{2\pi}\int_{-\infty}^{\infty}d\omega
R_{k}\left(  \omega\right)  e^{-i\omega t}\int\limits_{0}^{\infty}d\tau\;
\left(  1-e^{i\omega\tau}\right)  \operatorname{Im}\left[  T_{\omega
_{\mathrm{LO}}}^{\ast}\left(  \tau\right)  f_{\mathbf{q}}\left(  \tau\right)
\right] \\
&  =\frac{1}{2\pi}\int_{-\infty}^{\infty}d\omega F_{j}\left(  \omega\right)
e^{-i\omega t},
\end{align*}
where the Fourier component of the force is%
\begin{equation}
\left(  \mathbf{F}_{ph}\left(  \omega\right)  \right)  _{j}=\sum_{k=1}^{3}%
\sum_{\mathbf{q}}\frac{2\left\vert V_{\mathbf{q}}\right\vert ^{2}q_{j}q_{k}%
}{N\hbar}\int\limits_{0}^{\infty}dt\; \left(  1-e^{i\omega t}\right)
\operatorname{Im}\left[  T_{\omega_{\mathrm{LO}}}^{\ast}\left(  \tau\right)
\mathcal{G}\left(  \mathbf{q},it|\left\{  N_{\sigma}\right\}  ,\beta\right)
\right]  R_{k}\left(  \omega\right)  . \label{bb}%
\end{equation}
The expression (\ref{bb}) can be written down as%
\begin{equation}
\left(  \mathbf{F}_{ph}\left(  \omega\right)  \right)  _{j}=-m_{b}\sum
_{k=1}^{3}\chi_{jk}\left(  \omega\right)  R_{k}\left(  \omega\right)  ,
\label{tt}%
\end{equation}
where $\chi_{jk}\left(  \omega\right)  $ are components of the tensor%
\begin{equation}
\chi_{jk}\left(  \omega\right)  =\sum_{\mathbf{q}}\frac{2\left\vert
V_{\mathbf{q}}\right\vert ^{2}q_{j}q_{k}}{N\hbar m_{b}}\int\limits_{0}%
^{\infty}dt\; \left(  e^{i\omega t}-1\right)  \operatorname{Im}\left[
T_{\omega_{\mathrm{LO}}}^{\ast}\left(  t\right)  \mathcal{G}\left(
\mathbf{q},it|\left\{  N_{\sigma}\right\}  ,\beta\right)  \right]  . \label{T}%
\end{equation}
In the abstract tensor form, Eq. (\ref{tt}) is%
\begin{equation}
\mathbf{F}\left(  \omega\right)  =-\overleftrightarrow{\chi}\left(
\omega\right)  \mathbf{R}\left(  \omega\right)  . \label{tt2}%
\end{equation}
In particular, for the isotropic electron-phonon interaction and in the
absence of the magnetic field, the tensor $\overleftrightarrow{\chi}\left(
\omega\right)  $ is proportional to the unity tensor $\mathbb{I}$,%
\begin{equation}
\overleftrightarrow{\chi}\left(  \omega\right)  =\chi\left(  \omega\right)
\mathbb{I}, \label{scalar}%
\end{equation}
where $\chi\left(  \omega\right)  $ is the scalar memory function:%
\begin{equation}
\chi\left(  \omega\right)  =\sum_{\mathbf{q}}\frac{2\left\vert V_{\mathbf{q}%
}\right\vert ^{2}q^{2}}{3N\hbar m_{b}}\int\limits_{0}^{\infty}dt\; \left(
e^{i\omega t}-1\right)  \operatorname{Im}\left[  T_{\omega_{\mathrm{LO}}%
}^{\ast}\left(  t\right)  \mathcal{G}\left(  \mathbf{q},it|\left\{  N_{\sigma
}\right\}  ,\beta\right)  \right]  .
\end{equation}

Let us perform the Fourier transformation of the equation of motion
(\ref{EqMotion}):%
\begin{equation}
m_{b}\left(  \Omega_{0}^{2}-\omega^{2}\right)  \mathbf{R}\left(
\omega\right)  +e\mathbf{E}\left(  \omega\right)  =\mathbf{F}_{ph}\left(
\omega\right)  . \label{eqmot2}%
\end{equation}
With Eq. (\ref{tt2}), this equation takes the form%
\[
m_{b}\left(  \Omega_{0}^{2}-\omega^{2}\right)  \mathbf{R}\left(
\omega\right)  +e\mathbf{E}\left(  \omega\right)  =-m_{b}\overleftrightarrow
{\chi}\left(  \omega\right)  \mathbf{R}\left(  \omega\right)
\]%
\[
\Downarrow
\]%
\begin{equation}
m_{b}\left[  \omega^{2}-\Omega_{0}^{2}-\overleftrightarrow{\chi}\left(
\omega\right)  \right]  \mathbf{R}\left(  \omega\right)  =e\mathbf{E}\left(
\omega\right)  . \label{eqm3}%
\end{equation}

Comparing Eqs. (\ref{J2}) and (\ref{eqm3}) between each other, we find that%
\[
m_{b}\left[  \omega^{2}-\Omega_{0}^{2}-\overleftrightarrow{\chi}\left(
\omega\right)  \right]  \frac{\sigma\left(  \omega\right)  }{ie\omega
}\mathbf{E}\left(  \omega\right)  =e\mathbf{E}\left(  \omega\right)  ,
\]
so that%
\[
\sigma\left(  \omega\right)  =\frac{ie^{2}\omega}{m_{b}}\left[  \omega
^{2}-\Omega_{0}^{2}-\overleftrightarrow{\chi}\left(  \omega\right)  \right]
^{-1}.
\]
In the case when Eq. (\ref{scalar}) is valid, we obtain the conductivity in
the scalar form%
\[
\sigma\left(  \omega\right)  =\frac{ie^{2}\omega}{m_{b}\left[  \omega
^{2}-\Omega_{0}^{2}-\chi\left(  \omega\right)  \right]  }.
\]
The real part of the conductivity is%
\begin{align*}
\operatorname{Re}\sigma\left(  \omega\right)   &  =\operatorname{Re}%
\frac{ie^{2}\omega\left[  \left(  \omega^{2}-\Omega_{0}^{2}\right)
-\operatorname{Re}\chi\left(  \omega\right)  +i\operatorname{Im}\chi\left(
\omega\right)  \right]  }{m_{b}\left\{  \left[  \left(  \omega^{2}-\Omega
_{0}^{2}\right)  -\operatorname{Re}\chi\left(  \omega\right)  \right]
^{2}+\left[  \operatorname{Im}\chi\left(  \omega\right)  \right]
^{2}\right\}  }\\
&  =-\frac{e^{2}\omega}{m_{b}}\frac{\operatorname{Im}\chi\left(
\omega\right)  }{\left[  \left(  \omega^{2}-\Omega_{0}^{2}\right)
-\operatorname{Re}\chi\left(  \omega\right)  \right]  ^{2}+\left[
\operatorname{Im}\chi\left(  \omega\right)  \right]  ^{2}}.
\end{align*}

In summary, the optical conductivity can be expressed in terms of the memory
function $\chi\left(  \omega\right)  $ (cf. Ref. \cite{DSG1972}),%
\begin{equation}
\operatorname{Re}\sigma\left(  \omega\right)  =-\frac{e^{2}}{m_{b}}%
\frac{\omega\operatorname{Im}\chi\left(  \omega\right)  }{\left[  \omega
^{2}-\Omega_{0}^{2}-\operatorname{Re}\chi\left(  \omega\right)  \right]
^{2}+\left[  \operatorname{Im}\chi\left(  \omega\right)  \right]  ^{2}},
\label{Kw}%
\end{equation}
where $\chi\left(  \omega\right)  $ is given by
\begin{equation}
\chi\left(  \omega\right)  =\sum_{\mathbf{q}}\frac{2\left\vert V_{\mathbf{q}%
}\right\vert ^{2}q^{2}}{3N\hbar m_{b}}\int\limits_{0}^{\infty}dt\, \left(
e^{i\omega t}-1\right)  \operatorname{Im}\left[  T_{\omega_{\mathrm{LO}}%
}^{\ast}\left(  t\right)  \mathcal{G}\left(  \mathbf{q},it|\left\{  N_{\sigma
}\right\}  ,\beta\right)  \right]  . \label{Hi}%
\end{equation}
It is worth noting that the optical conductivity (\ref{Kw}) differs from that
for a translationally invariant polaron system both by the explicit form of
$\chi\left(  \omega\right)  $ and by the presence of the term $\Omega_{0}^{2}$
in the denominator. For $\alpha\rightarrow0,$ the optical conductivity tends
to a $\delta$-like peak at $\omega=\Omega_{0},$%
\begin{equation}
\lim_{\alpha\rightarrow0}\operatorname{Re}\sigma\left(  \omega\right)
=\frac{\pi e^{2}}{2m_{b}}\delta\left(  \omega-\Omega_{0}\right)  .
\label{Limit}%
\end{equation}
For a translationally invariant system $\Omega_{0}\rightarrow0$, and this
weak-coupling expression (\ref{Limit}) reproduces the \textquotedblleft
central peak\textquotedblright\ of the polaron optical conductivity
\cite{DLR1977}.

The further simplification of the memory function (\ref{Hi}) is performed in
the following way. With the Fr\"{o}hlich amplitudes of the electron-phonon
interaction, we transform the summation over $\mathbf{q}$ to the integral and
use the Feynman units ($\hbar=1$, $\omega_{\mathrm{LO}}=1$, $m_{b}=1$), in
which $\left\vert V_{\mathbf{q}}\right\vert ^{2}=\frac{2\sqrt{2}\pi\alpha
}{q^{2}V}$. We also use the fact that in an isotropic crystal, $\mathcal{G}%
\left(  \mathbf{q},it|\left\{  N_{\sigma}\right\}  ,\beta\right)
=\mathcal{G}\left(  q,it|\left\{  N_{\sigma}\right\}  ,\beta\right)  $. As a
result, we find
\begin{align*}
\chi\left(  \omega\right)   &  =\frac{V}{\left(  2\pi\right)  ^{3}}\int
_{0}^{\infty}4\pi q^{2}dq\frac{2q^{2}}{3N}\frac{2\sqrt{2}\pi\alpha}{q^{2}V}\\
&  \times\int\limits_{0}^{\infty}dt\left(  e^{i\omega t}-1\right)
\operatorname{Im}\left[  T_{\omega_{\mathrm{LO}}}^{\ast}\left(  t\right)
\mathcal{G}\left(  q,it|\left\{  N_{\sigma}\right\}  ,\beta\right)  \right] \\
&  =\frac{2\sqrt{2}\alpha}{3\pi N}\int_{0}^{\infty}q^{2}dq\int\limits_{0}%
^{\infty}dt\left(  e^{i\omega t}-1\right)  \operatorname{Im}\left[
T_{\omega_{\mathrm{LO}}}^{\ast}\left(  t\right)  \mathcal{G}\left(
q,it|\left\{  N_{\sigma}\right\}  ,\beta\right)  \right]  .
\end{align*}
In the zero-temperature case, $T_{\omega_{\mathrm{LO}}}^{\ast}\left(
t\right)  \rightarrow e^{-it},$ and we arrive at the expression%
\begin{equation}
\chi\left(  \omega\right)  =\frac{2\sqrt{2}\alpha}{3\pi N}\int_{0}^{\infty
}q^{2}dq\int\limits_{0}^{\infty}dt\left(  e^{i\omega t}-1\right)
\operatorname{Im}\left[  e^{-it}\mathcal{G}\left(  q,it|\left\{  N_{\sigma
}\right\}  ,\beta\right)  \right]  . \label{Hi1}%
\end{equation}

Substituting the two-point correlation function $\mathcal{G}\left(
q,it|\left\{  N_{\sigma}\right\}  ,\beta\right)  $ with the one-electron
(\ref{p1}) and the two-electron (\ref{p2}) distribution functions into the
memory function (\ref{Hi1}) and expanding $\mathcal{G}\left(  q,it|\left\{
N_{\sigma}\right\}  ,\beta\right)  $ in powers of $e^{-iwt}$, $e^{-i\Omega
_{1}t}$ and $e^{-i\Omega_{2}t}$, the integrations over $q$ and $t$ in Eq.
(\ref{Hi1}) are performed analytically. The similar transformations are
performed also in the 2D case. As a result, the memory function (\ref{Hi}) is
represented in the unified form for 3D and 2D interacting polarons:
\begin{align}
\chi\left(  \omega\right)   &  =\lim_{\varepsilon\rightarrow+0}\frac{2\alpha
}{3\pi N}\left(  \frac{3\pi}{4}\right)  ^{3-D}\left(  \frac{\omega
_{\mathrm{LO}}}{A}\right)  ^{3/2}\nonumber\\
&  \times\sum_{p_{1}=0}^{\infty}\sum_{p_{2}=0}^{\infty}\sum_{p_{3}=0}^{\infty
}\frac{\left(  -1\right)  ^{p_{3}}}{p_{1}!p_{2}!p_{3}!}\left(  \frac{a_{1}%
^{2}}{N\Omega_{1}A}\right)  ^{p_{1}}\left(  \frac{a_{2}^{2}}{N\Omega_{2}%
A}\right)  ^{p_{2}}\left(  \frac{1}{NwA}\right)  ^{p_{3}}\nonumber\\
&  \times\left\{  \left[  \sum\limits_{m=0}^{\infty}\sum\limits_{n=0}^{\infty
}\sum\limits_{\sigma}\left.  \left[  f_{1}\left(  n,\sigma|\left\{  N_{\sigma
}\right\}  ,\beta\right)  -f_{2}\left(  n,\sigma;m,\sigma|\left\{  N_{\sigma
}\right\}  ,\beta\right)  \right]  \right\vert _{\beta\rightarrow\infty
}\right.  \right. \nonumber
\end{align}%
\begin{align}
&  \times\left(
\begin{array}
[c]{c}%
\frac{1}{\omega-\omega_{\mathrm{LO}}-\left[  p_{1}\Omega_{1}+p_{2}\Omega
_{2}+\left(  p_{3}-m+n\right)  w\right]  +\mathrm{i}\varepsilon}-\frac
{1}{\omega+\omega_{\mathrm{LO}}+p_{1}\Omega_{1}+p_{2}\Omega_{2}+\left(
p_{3}-m+n\right)  w+\mathrm{i}\varepsilon}\\
+\mathcal{P}\left(  \frac{2}{\omega_{\mathrm{LO}}+p_{1}\Omega_{1}+p_{2}%
\Omega_{2}+\left(  p_{3}-m+n\right)  w}\right)
\end{array}
\right) \nonumber\\
&  \times\sum\limits_{l=0}^{m}\sum\limits_{k=n-m+l}^{n}\frac{\left(
-1\right)  ^{n-m+l+k}\Gamma\left(  p_{1}+p_{2}+p_{3}+k+l+\frac{3}{2}\right)
}{k!l!}\left(  \frac{1}{wA}\right)  ^{l+k}\nonumber\\
&  \left.  \times\binom{n+D-1}{n-k}\binom{2k}{k-l-n+m}\right] \nonumber
\end{align}%
\begin{align}
&  +\left[  \left(
\begin{array}
[c]{c}%
\frac{1}{\omega-\omega_{\mathrm{LO}}-\left(  p_{1}\Omega_{1}+p_{2}\Omega
_{2}+p_{3}w\right)  +\mathrm{i}\varepsilon}-\frac{1}{\omega+\omega
_{\mathrm{LO}}+p_{1}\Omega_{1}+p_{2}\Omega_{2}+p_{3}w+\mathrm{i}\varepsilon}\\
+\mathcal{P}\left(  \frac{2}{\omega_{\mathrm{LO}}+p_{1}\Omega_{1}+p_{2}%
\Omega_{2}+p_{3}w}\right)
\end{array}
\right)  \right. \nonumber\\
&  \times\sum\limits_{m=0}^{\infty}\sum\limits_{n=0}^{\infty}\sum
\limits_{\sigma,\sigma^{\prime}}\left.  f_{2}\left(  n,\sigma;m,\sigma
^{\prime}|\left\{  N_{\sigma}\right\}  ,\beta\right)  \right\vert
_{\beta\rightarrow\infty}\nonumber\\
&  \times\sum\limits_{k=0}^{n}\sum\limits_{l=0}^{m}\frac{\left(  -1\right)
^{k+l}\Gamma\left(  p_{1}+p_{2}+p_{3}+k+l+\frac{3}{2}\right)  }{k!l!}\left(
\frac{1}{wA}\right)  ^{k+l}\nonumber\\
&  \left.  \left.  \times\binom{n+D-1}{n-k}\binom{m+D-1}{m-l}\right]
\right\}  , \label{mf}%
\end{align}
where $D=2,3$ is the dimensionality of the space, $\mathcal{P}$ denotes the
principal value, $A$ is defined as $A\equiv\left[  \sum_{i=1}^{2}a_{i}%
^{2}/\Omega_{i}+\left(  N-1\right)  /w\right]  /N$, $\Omega_{1},\Omega_{2},$
and $w$ are the eigenfrequencies of the model system, $a_{1}$ and $a_{2}$ are
the coefficients of the canonical transformation which diagonalizes the model
Lagrangian (\ref{LM}).

\subsubsection{Selected results: the manifestations of the shell filling in
optical conductivity}

\begin{figure}
[h]
\begin{center}
\includegraphics[
height=8.8875cm,
width=9.7289cm
]%
{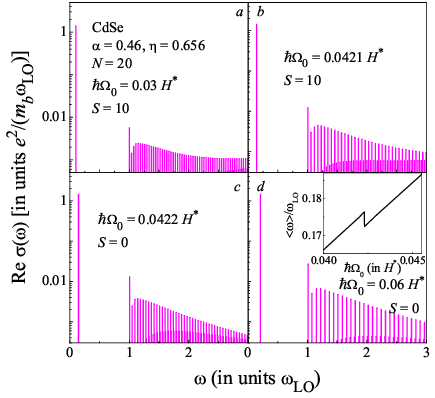}%
\caption{Optical conductivity spectra of $N=20$ interacting polarons in CdSe
quantum dots with $\alpha=0.46$, $\eta=0.656$ for different confinement
energies close to the transition from a spin-polarized ground state to a
ground state obeying Hund$^{\text{'}}$s rule. \emph{Inset}: the first
frequency moment $\left\langle \omega\right\rangle $ of the optical
conductivity as a function of the confinement energy. (From Ref.
\cite{MPQD-PRB2004}.)}%
\label{Spectra}%
\end{center}
\end{figure}

The shell filling schemes for an $N$-polaron system in a quantum dot can
manifest themselves in the spectra of the optical conductivity. In
Fig.\thinspace\ref{Spectra}, optical conductivity spectra for $N=20$ polarons
are presented for a quantum dot with the parameters of CdSe: $\alpha=0.46,$
$\eta=0.656$ \cite{KartheuserGreenbook} and with different values of the
confinement energy $\hbar\Omega_{0}$. \footnote{For the numerical
calculations, we use effective atomic units, where $\hbar,$ the electron band
mass $m_{b}$ and $e/\sqrt{\varepsilon_{\infty}}$ have the numerical value of
1. This means that the unit of length is the effective Bohr radius
$a_{B}^{\ast}=\hbar^{2}\varepsilon_{\infty}/\left(  m_{b}e^{2}\right)  $,
while the unit of energy is the effective Hartree $H^{\ast}=m_{b}e^{4}/\left(
\hbar^{2}\varepsilon_{\infty}^{2}\right)  $.} In this case, the spin-polarized
ground state changes to the ground state satisfying Hund's rule with
increasing $\hbar\Omega_{0}$ in the interval $0.0421H^{\ast}<\hbar\Omega
_{0}<0.0422H^{\ast}$.

%\newpage%
%

%TCIMACRO{\FRAME{fhFU}{9.7289cm}{8.8875cm}{0pt}{\Qcb{Optical conductivity
%spectra of $N=20$ interacting polarons in CdSe quantum dots with $\alpha
%=0.46$, $\eta=0.656$ for different confinement energies close to the
%transition from a spin-polarized ground state to a ground state obeying
%Hund$^{\text{'}}$s rule. \QTR{em}{Inset}: the first frequency moment
%$\left\langle \omega\right\rangle $ of the optical conductivity as a function
%of the confinement energy. (From Ref. \cite{MPQD-PRB2004}.)}}{\Qlb{Spectra}%
%}{part2fig6.eps}{\special{ language "Scientific Word";  type "GRAPHIC";
%maintain-aspect-ratio TRUE;  display "ICON";  valid_file "F";
%width 9.7289cm;  height 8.8875cm;  depth 0pt;  original-width 7.3719cm;
%original-height 6.7283cm;  cropleft "0";  croptop "1";  cropright "1";
%cropbottom "0";  filename 'Part2Fig6.eps';file-properties "XNPEU";}} }%
%BeginExpansion
%EndExpansion

%\newpage

In the inset to Fig.\thinspace\ref{Spectra}, the first frequency moment of the
optical conductivity
\begin{equation}
\left\langle \omega\right\rangle \equiv\frac{\int_{0}^{\infty}\omega
\operatorname{Re}\sigma\left(  \omega\right)  d\omega}{\int_{0}^{\infty
}\operatorname{Re}\sigma\left(  \omega\right)  d\omega}, \label{Moment}%
\end{equation}
as a function of $\hbar\Omega_{0}$ shows a \emph{discontinuity}, at the value
of the confinement energy corresponding to the change of the shell filling
schemes from the spin-polarized ground state to the ground state obeying
Hund's rule. This discontinuity might be observable in optical measurements.

\begin{figure}
[h]
\begin{center}
\includegraphics[
height=8.044cm,
width=6.6272cm
]%
{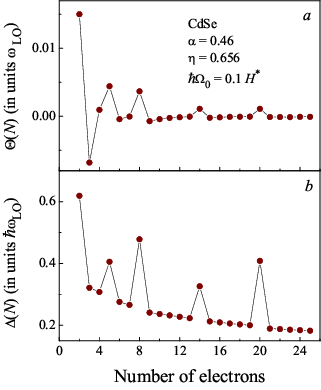}%
\caption{The function $\Theta\left(  N\right)  $ and the addition energy
$\Delta\left(  N\right)  $ for systems of interacting polarons in CdSe quantum
dots with $\alpha=0.46$, $\eta=0.656$ for $\hbar\Omega_{0}=0.1H^{\ast}$. (From
Ref. \cite{MPQD-PRB2004}.)}%
\label{Moments}%
\end{center}
\end{figure}

The shell structure for a system of interacting polarons in a quantum dot is
clearly revealed when analyzing the addition energy and the first frequency
moment of the optical conductivity in parallel. In Fig \ref{Moments}, we show
both the function
\begin{equation}
\Theta\left(  N\right)  \equiv\left.  \left\langle \omega\right\rangle
\right\vert _{N+1}-2\left.  \left\langle \omega\right\rangle \right\vert
_{N}+\left.  \left\langle \omega\right\rangle \right\vert _{N-1},
\label{Theta}%
\end{equation}
and the addition energy%
\begin{equation}
\Delta\left(  N\right)  =E^{0}\left(  N+1\right)  -2E^{0}\left(  N\right)
+E^{0}\left(  N-1\right)  . \label{Add}%
\end{equation}
for interacting polarons in a 3D CdSe quantum dot.

%\newpage%
%

%TCIMACRO{\FRAME{fhFU}{6.6272cm}{8.044cm}{0pt}{\Qcb{The function $\Theta\left(
%N\right)  $ and the addition energy $\Delta\left(  N\right)  $ for systems of
%interacting polarons in CdSe quantum dots with $\alpha=0.46$, $\eta=0.656$ for
%$\hbar\Omega_{0}=0.1H^{\ast}$. (From Ref. \cite{MPQD-PRB2004}.)}%
%}{\Qlb{Moments}}{part2fig7.eps}{\special{ language "Scientific Word";
%type "GRAPHIC";  maintain-aspect-ratio TRUE;  display "ICON";
%valid_file "F";  width 6.6272cm;  height 8.044cm;  depth 0pt;
%original-width 5.4169cm;  original-height 6.5877cm;  cropleft "0";
%filename 'Part2Fig7.eps';file-properties "XNPEU";}} }%
%BeginExpansion
%EndExpansion

%\newpage

As seen from Fig \ref{Moments}, distinct peaks appear in $\Theta\left(
N\right)  $ and $\Delta\left(  N\right)  $ at the \textquotedblleft magic
numbers\textquotedblright\ corresponding to closed-shell configurations at
$N=8,20$ and to half-filled-shell configurations at $N=5,14$. We see that each
of the peaks of $\Theta\left(  N\right)  $ corresponds to a peak of the
addition energy. The filling patterns for a many-polaron system in a quantum
dot can be therefore determined from the analysis of the first moment of the
optical absorption for different numbers of polarons.\newpage

\section{Variational path-integral treatment of a translation invariant
$N$-polaron system}

\subsection{The many-polaron system}

In the present section, the ground-state properties of a translation invariant
$N$-polaron system are theoretically studied in the framework of the
variational path-integral method for identical particles, using a further
development \cite{BKD-PRB2005} of the model introduced in Refs.
\cite{SSC114-305,MPPhysE,MPQD-PRB2004}.

In order to describe a many-polaron system, we start from the translation
invariant $N$-polaron Hamiltonian
\begin{equation}
H=\sum_{j=1}^{N}\frac{\mathbf{p}_{j}^{2}}{2m}+\frac{1}{2}\sum_{j=1}^{N}%
\sum_{l=1,\neq j}^{N}\frac{e^{2}}{\epsilon_{\infty}\left\vert \mathbf{r}%
_{j}\mathbf{-r}_{l}\right\vert }+\sum_{\mathbf{k}}\hbar\omega_{\mathrm{LO}%
}a_{\mathbf{k}}^{\dagger}a_{\mathbf{k}}+\left(  \sum_{j=1}^{N}\sum
_{\mathbf{k}}V_{k}a_{\mathbf{k}}e^{i\mathbf{k\cdot r}_{j}}+H.c.\right)  ,
\end{equation}
where $m$ is the band mass, $e$ is the electron charge, $\omega_{\mathrm{LO}}$
is the longitudinal optical (LO) phonon frequency, and $V_{k}$ are the
amplitudes of the Fr\"{o}hlich electron-LO-phonon interaction%
\begin{equation}
V_{k}=i\frac{\hbar\omega_{\mathrm{LO}}}{k}\left(  \frac{4\pi\alpha}{V}\right)
^{1/2}\left(  \frac{\hbar}{2m\omega_{\mathrm{LO}}}\right)  ^{1/4},\quad
\alpha=\frac{e^{2}}{2\hbar\omega_{\mathrm{LO}}}\left(  \frac{2m\omega
_{\mathrm{LO}}}{\hbar}\right)  ^{1/2}\left(  \frac{1}{\epsilon_{\infty}}%
-\frac{1}{\epsilon_{0}}\right)  ,
\end{equation}
with the electron-phonon coupling constant $\alpha>0,$ the high-frequency
dielectric constant $\epsilon_{\infty}>0$ and the static dielectric constant
$\epsilon_{0}>0,$ and consequently
\begin{equation}
\frac{e^{2}}{\epsilon_{\infty}}>\hbar\left(  \frac{2\hbar\omega_{\mathrm{LO}}%
}{m}\right)  ^{1/2}\alpha\Longleftrightarrow\alpha\sqrt{2}<\left(
\frac{H^{\ast}}{\hbar\omega_{\mathrm{LO}}}\right)  ^{1/2}\equiv U.
\label{units}%
\end{equation}
In the expression (\ref{units}), $H^{\ast}$ is the effective Hartree
\begin{equation}
H^{\ast}=\frac{e^{2}}{\epsilon_{\infty}a_{B}^{\ast}},\quad a_{B}^{\ast}%
=\frac{\hbar^{2}}{me^{2}/\epsilon_{\infty}}%
\end{equation}
where $a_{B}^{\ast}$ is the effective Bohr radius. The partition function of
the system can be expressed as a path integral over all electron and phonon
coordinates. The path integral over the phonon variables can be calculated
analytically \cite{Feynman}. Feynman's phonon elimination technique for this
system is well known and leads to the partition function, which is a path
integral over the electron coordinates only:%
\begin{equation}
Z=\left(  \prod_{\mathbf{k}}\frac{e^{\frac{1}{2}\beta\hbar\omega_{\mathrm{LO}%
}}}{2\sinh\frac{1}{2}\beta\hbar\omega_{\mathrm{LO}}}\right)  \oint
e^{S}\mathcal{D}\mathbf{\bar{r}}%
\end{equation}
where $\mathbf{\bar{r}=}\left\{  \mathbf{r}_{1},\cdots,\mathbf{r}_{N}\right\}
$ denotes the set of electron coordinates, and $\oint\mathcal{D}%
\mathbf{\bar{r}}$ denotes the path integral over all the electron coordinates,
integrated over equal initial and final points, i.e.
\[
\oint e^{S}\mathcal{D}\mathbf{\bar{r}}\equiv\int d\mathbf{\bar{r}}%
\int_{\mathbf{\bar{r}}\left(  0\right)  =\mathbf{\bar{r}}}^{\mathbf{\bar{r}%
}\left(  \beta\right)  =\mathbf{\bar{r}}}e^{S}\mathcal{D}\mathbf{\bar{r}%
}\left(  \tau\right)  .
\]
Throughout this paper, imaginary time variables are used. The effective action
for the $N$-polaron system is retarded and given by%
\begin{align}
S  &  =-\int_{0}^{\beta}\left(  \frac{m}{2}\sum_{j=1}^{N}\left(
\frac{d\mathbf{r}_{j}\left(  \tau\right)  }{d\tau}\right)  ^{2}+\frac{1}%
{2}\sum_{j=1}^{N}\sum_{l=1,\neq j}^{N}\frac{e^{2}}{\epsilon_{\infty}\left\vert
\mathbf{r}_{j}\left(  \tau\right)  \mathbf{-r}_{l}\left(  \tau\right)
\right\vert }\right)  d\tau\nonumber\\
&  +\frac{1}{2}\int_{0}^{\beta}\int_{0}^{\beta}\sum_{j,l=1}^{N}\sum
_{\mathbf{k}}\left\vert V_{k}\right\vert ^{2}e^{i\mathbf{k\cdot}\left(
\mathbf{r}_{j}\left(  \tau\right)  -\mathbf{r}_{l}\left(  \sigma\right)
\right)  }\frac{\cosh\hbar\omega_{\mathrm{LO}}\left(  \frac{1}{2}%
\beta-\left\vert \tau-\sigma\right\vert \right)  }{\sinh\frac{1}{2}\beta
\hbar\omega_{\mathrm{LO}}}d\sigma d\tau. \label{eq:Spol}%
\end{align}
Note that the electrons are fermions. Therefore the path integral for the
electrons with parallel spin has to be interpreted as the required
antisymmetric projection of the propagators for distinguishable particles.

We below use units in which $\hbar=1$, $m=1$, and $\omega_{\mathrm{LO}}=1$.
The units of distance and energy are thus the effective polaron radius
$\left[  \hbar/\left(  m\omega_{\mathrm{LO}}\right)  \right]  ^{1/2}$ and the
LO-phonon energy $\hbar\omega_{\mathrm{LO}}$.

\subsection{Variational principle}

For distinguishable particles, it is well known that the Jensen-Feynman
inequality \cite{Feynman} provides a lower bound on the partition function $Z$
(and consequently an upper bound on the free energy $F$)%
\begin{equation}
Z=\oint e^{S}\mathcal{D}\mathbf{\bar{r}}=\left(  \oint e^{S_{0}}%
\mathcal{D}\mathbf{\bar{r}}\right)  \left\langle e^{S-S_{0}}\right\rangle
_{0}\geq\left(  \oint e^{S_{0}}\mathcal{D}\mathbf{\bar{r}}\right)
e^{\left\langle S-S_{0}\right\rangle _{0}}\text{ with }\left\langle
A\right\rangle _{0}\equiv\frac{\oint A\left(  \mathbf{\bar{r}}\right)
e^{S_{0}}\mathcal{D}\mathbf{\bar{r}}}{\oint e^{S_{0}}\mathcal{D}%
\mathbf{\bar{r}}}, \label{eq:JensenFeynman}%
\end{equation}%
\begin{equation}
e^{-\beta F}\geq e^{-\beta F_{0}}e^{\left\langle S-S_{0}\right\rangle _{0}%
}\Longrightarrow F\leq F_{0}-\frac{\left\langle S-S_{0}\right\rangle _{0}%
}{\beta} \label{VarIneq}%
\end{equation}
for a system with real action $S$ and a real trial action $S_{0}.$The
many-body extension (Ref.\thinspace\cite{PRE96,NoteKleinert}) of the
Jensen-Feynman inequality, requires that the potentials are symmetric with
respect to all particle permutations, and that the exact propagator as well as
the model propagator are defined on the same state space. Within this
interpretation we consider the following generalization of Feynman's trial
action%
\begin{align}
S_{0}  &  =-\int_{0}^{\beta}\left(  \frac{1}{2}\sum_{j=1}^{N}\left(
\frac{d\mathbf{r}_{j}\left(  \tau\right)  }{d\tau}\right)  ^{2}+\frac
{\omega^{2}+w^{2}-v^{2}}{4N}\sum_{j,l=1}^{N}\left(  \mathbf{r}_{j}\left(
\tau\right)  \mathbf{-r}_{l}\left(  \tau\right)  \right)  ^{2}\right)
d\tau\nonumber\\
&  -\frac{w}{8}\frac{v^{2}-w^{2}}{N}\sum_{j,l=1}^{N}\int_{0}^{\beta}\int
_{0}^{\beta}\left(  \mathbf{r}_{j}\left(  \tau\right)  -\mathbf{r}_{l}\left(
\sigma\right)  \right)  ^{2}\frac{\cosh w\left(  \frac{1}{2}\beta-\left\vert
\tau-\sigma\right\vert \right)  }{\sinh\frac{1}{2}\beta w}d\sigma
d\tau\label{eq:S0}%
\end{align}
with the variational frequency parameters $v,w,\omega$.

Using the explicit forms of the exact (\ref{eq:Spol}) and the trial
(\ref{eq:S0}) actions, the variational inequality (\ref{VarIneq}) takes the
form%
\begin{align}
F\left(  \beta|N_{\uparrow},N_{\downarrow}\right)   &  \leq F_{0}\left(
\beta|N_{\uparrow},N_{\downarrow}\right)  +\frac{U}{2\beta}\int_{0}^{\beta
}\left\langle \sum_{j,l=1,\neq j}^{N}\frac{1}{\left\vert \mathbf{r}%
_{j}\mathbf{\left(  \tau\right)  -r}_{l}\mathbf{\left(  \tau\right)
}\right\vert }\right\rangle _{0}d\tau\nonumber\\
&  -\frac{\omega^{2}+w^{2}-v^{2}}{4N\beta}\int_{0}^{\beta}\left\langle
\sum_{j,l=1}^{N}\left(  \mathbf{r}_{j}\left(  \tau\right)  \mathbf{-r}%
_{l}\left(  \tau\right)  \right)  ^{2}\right\rangle _{0}d\tau\nonumber\\
&  -\frac{w}{8}\frac{v^{2}-w^{2}}{N\beta}\int_{0}^{\beta}\int_{0}^{\beta
}\left\langle \sum_{j,l=1}^{N}\left(  \mathbf{r}_{j}\left(  \tau\right)
-\mathbf{r}_{l}\left(  \sigma\right)  \right)  ^{2}\right\rangle _{0}%
\frac{\cosh w\left(  \frac{1}{2}\beta-\left\vert \tau-\sigma\right\vert
\right)  }{\sinh\frac{1}{2}\beta w}d\sigma d\tau\nonumber\\
&  -\frac{1}{2\beta}\int_{0}^{\beta}\int_{0}^{\beta}\sum_{\mathbf{k}%
}\left\vert V_{k}\right\vert ^{2}\left\langle \sum_{j,l=1}^{N}%
e^{i\mathbf{k\cdot}\left(  \mathbf{r}_{j}\left(  \tau\right)  -\mathbf{r}%
_{l}\left(  \sigma\right)  \right)  }\right\rangle _{0}\frac{\cosh
\omega_{\mathrm{LO}}\left(  \frac{1}{2}\beta-\left\vert \tau-\sigma\right\vert
\right)  }{\sinh\frac{1}{2}\beta\omega_{\mathrm{LO}}}d\sigma d\tau.
\label{VI1}%
\end{align}

In the zero-temperature limit ($\beta\rightarrow\infty$), we arrive at the
following upper bound for the ground-state energy $E^{0}\left(  N_{\uparrow
},N_{\downarrow}\right)  $ of a translation invariant $N$-polaron system%
\[
E^{0}\left(  N_{\uparrow},N_{\downarrow}\right)  \leq E_{var}\left(
N_{\uparrow},N_{\downarrow}|v,w,\omega\right)  ,
\]
with%
\begin{align}
E_{var}\left(  N_{\uparrow},N_{\downarrow}|v,w,\omega\right)   &  =\frac{3}%
{4}\frac{\left(  v-w\right)  ^{2}}{v}-\frac{3}{4}\omega+\frac{1}{2}%
\mathbb{E}_{F}\left(  N_{\downarrow}\right)  +\frac{1}{2}\mathbb{E}_{F}\left(
N_{\downarrow}\right) \nonumber\\
&  +E_{C\Vert}\left(  N_{\uparrow}\right)  +E_{C\Vert}\left(  N_{\downarrow
}\right)  +E_{C\uparrow\downarrow}\left(  N_{\uparrow},N_{\downarrow}\right)
\nonumber\\
&  +E_{\alpha\Vert}\left(  N_{\uparrow}\right)  +E_{\alpha\Vert}\left(
N_{\downarrow}\right)  +E_{\alpha\uparrow\downarrow}\left(  N_{\uparrow
},N_{\downarrow}\right)  , \label{Egr}%
\end{align}
where $\mathbb{E}_{F}\left(  N\right)  $ is the energy of $N$ spin-polarized
fermions confined to a parabolic potential with the confinement frequency
$\omega$, $E_{C\Vert}\left(  N_{\uparrow\left(  \downarrow\right)  }\right)  $
is the Coulomb energy of the electrons with parallel spins, $E_{C\uparrow
\downarrow}\left(  N_{\uparrow},N_{\downarrow}\right)  $ is the Coulomb energy
of the electrons with opposite spins, $E_{\alpha\Vert}\left(  N_{\uparrow
\left(  \downarrow\right)  }\right)  $ is the electron-phonon energy of the
electrons with parallel spins, and $E_{\alpha\uparrow\downarrow}\left(
N_{\uparrow},N_{\downarrow}\right)  $ is the electron-phonon energy of the
electrons with opposite spins.

\subsection{Results}

Here, we discuss some results of the numerical minimization of $E_{var}\left(
N_{\uparrow},N_{\downarrow}|v,w,\omega\right)  $ with respect to the three
variational parameters $v$, $w$, and $\omega$. The Fr\"{o}hlich constant
$\alpha$ and the Coulomb parameter
\begin{equation}
\alpha_{0}\equiv\frac{U}{\sqrt{2}}\equiv\frac{\alpha}{1-\eta}\text{ with
}\frac{1}{\eta}=\frac{\varepsilon_{0}}{\varepsilon_{\infty}}%
\end{equation}
characterize the strength of the electron-phonon and of the Coulomb
interaction, obeying the physical condition $\alpha\geq\alpha_{0}$ [see
(\ref{units})]. The optimal values of the variational parameters $v,$$w,$and
$\omega$ are denoted $v_{op},$$w_{op},$and $\omega_{op}$, respectively. The
optimal value of the total spin was always determined by choosing the
combination $\left(  N_{\uparrow},N_{\downarrow}\right)  $ for fixed
$N=N_{\uparrow}+N_{\downarrow}$which corresponds to the lowest value
$E^{0}\left(  N\right)  $ of the variational functional%
\begin{equation}
E^{0}\left(  N\right)  \equiv\min_{N_{\uparrow}}E_{var}\left(  N_{\uparrow
},N-N_{\uparrow}|v_{op},w_{op},\omega_{op}\right)  .
\end{equation}

%\newpage%
%

%TCIMACRO{\FRAME{fhFU}{8.2681cm}{8.345cm}{0pt}{\Qcb{The \textquotedblleft phase
%diagrams\textquotedblright\ of a translation invariant $N$-polaron system. The
%grey area is the non-physical region, for which $\alpha>\alpha_{0}$. The
%stability region for each number of electrons is determined by the equation
%$\alpha_{c}<\alpha<\alpha_{0}$. (From Ref. \cite{BKD-PRB2005}.)}%
%}{\Qlb{trinv-PD}}{part2fig8.eps}{\special{ language "Scientific Word";
%type "GRAPHIC";  maintain-aspect-ratio TRUE;  display "ICON";
%valid_file "F";  width 8.2681cm;  height 8.345cm;  depth 0pt;
%original-width 16.0727cm;  original-height 16.2221cm;  cropleft "0";
%filename 'Part2Fig8.eps';file-properties "XNPEU";}} }%
%BeginExpansion
\begin{figure}
[h]
\begin{center}
\includegraphics[
height=8.345cm,
width=8.2681cm
]%
{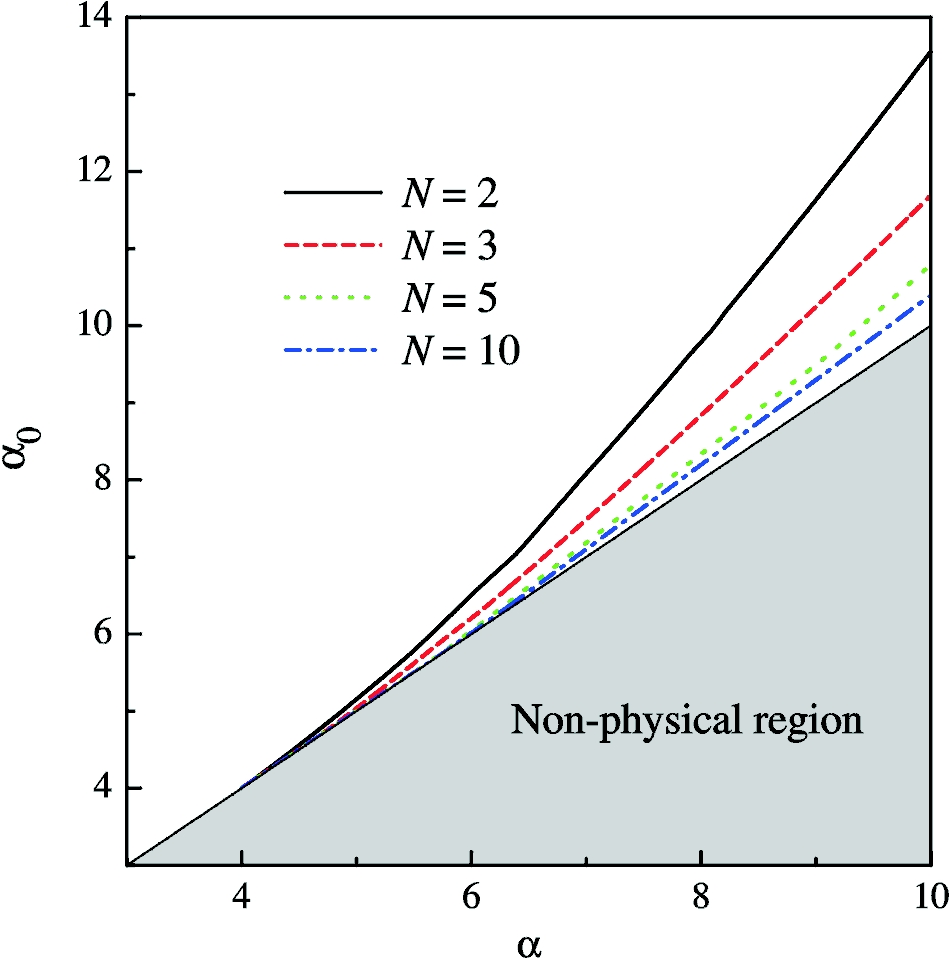}%
\caption{The \textquotedblleft phase diagrams\textquotedblright\ of a
translation invariant $N$-polaron system. The grey area is the non-physical
region, for which $\alpha>\alpha_{0}$. The stability region for each number of
electrons is determined by the equation $\alpha_{c}<\alpha<\alpha_{0}$. (From
Ref. \cite{BKD-PRB2005}.)}%
\label{trinv-PD}%
\end{center}
\end{figure}
%EndExpansion

%\newpage

In Fig. \ref{trinv-PD}, the \textquotedblleft phase diagrams\textquotedblright%
\ analogous to the bipolaron \textquotedblleft phase diagram\textquotedblright%
\ of Ref. \cite{VPD91} are plotted for an $N$-polaron system in bulk with
$N=2,3,5,$ and $10$. The area where $\alpha_{0}\leq\alpha$ is the non-physical
region. For $\alpha>\alpha_{0}$, each sector between a curve corresponding to
a well defined $N$ and the line indicating $\alpha_{0}=\alpha$ shows the
stability region where $\omega_{op}\neq0$, while the white area corresponds to
the regime with $\omega_{op}=0$. When comparing the stability region for $N=2$
from Fig. \ref{trinv-PD} with the bipolaron \textquotedblleft phase
diagram\textquotedblright\ of Ref. \cite{VPD91}, the stability region in the
present work starts from the value $\alpha_{c}\approx4.1$ (instead of
$\alpha_{c}\approx6.9$ in Ref. \cite{VPD91}). The width of the stability
region within the present model is also larger than the width of the stability
region within the model of Ref. \cite{VPD91}. Also, the absolute values of the
ground-state energy of a two-polaron system given by the present model are
smaller than those given by the approach of Ref. \cite{VPD91}.

The difference between the numerical results of the present work and of Ref.
\cite{VPD91} is due to the following distinction between the used model
systems. The model system of Ref. \cite{VPD91} consists of two electrons
interacting with two fictitious particles and with each other through
quadratic interactions. But the trial Hamiltonian given by Eq. (6) of Ref.
\cite{VPD91} is not symmetric with respect to the permutation of the
electrons. It is only symmetric under the permutation of the pairs
\textquotedblleft electron + fictitious particle\textquotedblright. As a
consequence, this trial system is only applicable if the electrons are
distinguishable, i.e. have opposite spin. In contrast to the model of Ref.
\cite{VPD91}, the model used in the present paper is described by the trial
action (9), which is fully symmetric with respect to the permutations of the
electrons, as is required to describe identical particles.

\begin{figure}
[h]
\begin{center}
\includegraphics[
height=12.6855cm,
width=7.3345cm
]%
{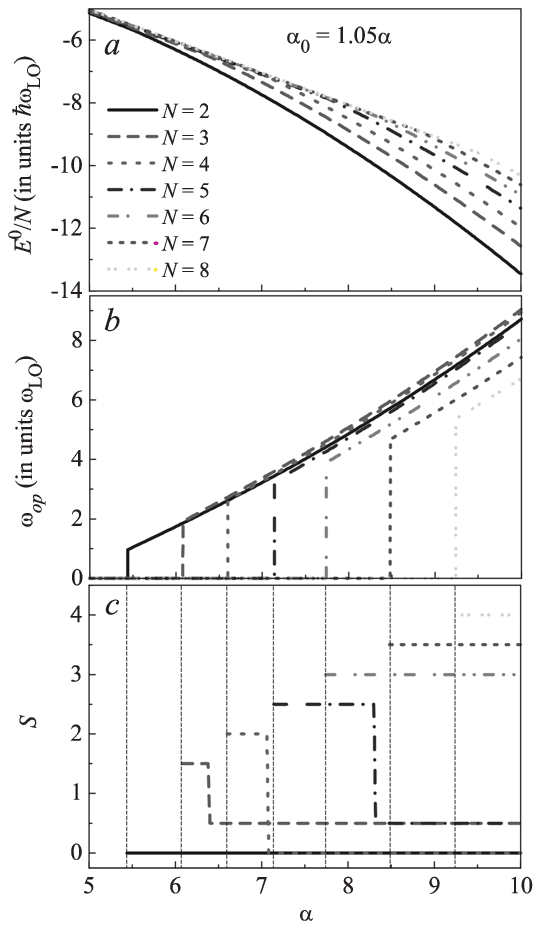}%
\caption{The ground-state energy per particle (\emph{a}), the optimal value
$\omega_{op}$ of the confinement frequency (\emph{b}), and the total spin
(\emph{c}) of a translation invariant $N$-polaron system as a function of the
coupling strength $\alpha$ for $\alpha_{0}/\alpha=0.5$. The vertical dashed
lines in the panel \emph{c} indicate the critical values $\alpha_{c}$
separating the regimes of $\alpha>\alpha_{c}$, where the multipolaron ground
state with $\omega_{op}\neq0$ exists, and $\alpha<\alpha_{c}$, where
$\omega_{op}=0$. (From Ref. \cite{BKD-PRB2005}.)}%
\label{trinv-GSE}%
\end{center}
\end{figure}

The \textquotedblleft phase diagrams\textquotedblright\ for $N>2$ demonstrate
the existence of stable multipolaron states (see Ref. \cite{SVPD1993}). As
distinct from Ref. \cite{SVPD1993}, here the ground state of an $N$-polaron
system is investigated supposing that the electrons are fermions. As seen from
Fig. \ref{trinv-GSE}, for $N>2$, the stability region for a multipolaron state
is narrower than the stability region for $N=2$, and its width decreases with
increasing $N$.

%\newpage%
%

%TCIMACRO{\FRAME{fhFU}{7.3345cm}{12.6855cm}{0pt}{\Qcb{The ground-state energy
%per particle (\QTR{em}{a}), the optimal value $\omega_{op}$ of the confinement
%frequency (\QTR{em}{b}), and the total spin (\QTR{em}{c}) of a translation
%invariant $N$-polaron system as a function of the coupling strength $\alpha$
%for $\alpha_{0}/\alpha=0.5$. The vertical dashed lines in the panel
%\QTR{em}{c} indicate the critical values $\alpha_{c}$ separating the regimes
%of $\alpha>\alpha_{c}$, where the multipolaron ground state with $\omega
%_{op}\neq0$ exists, and $\alpha<\alpha_{c}$, where $\omega_{op}=0$. (From Ref.
%\cite{BKD-PRB2005}.)}}{\Qlb{trinv-GSE}}{part2fig9.eps}%
%maintain-aspect-ratio TRUE;  display "ICON";  valid_file "F";
%width 7.3345cm;  height 12.6855cm;  depth 0pt;  original-width 8.9666cm;
%original-height 15.5719cm;  cropleft "0";  croptop "1";  cropright "1";
%cropbottom "0";  filename 'Part2Fig9.eps';file-properties "XNPEU";}} }%
%BeginExpansion
%EndExpansion

%\newpage

A consequence of the Fermi statistics is the dependence of the polaron
characteristics and of the total spin of an $N$-polaron system on the
parameters ($\alpha,\, \alpha_{0},$$N$). In Fig. \ref{trinv-GSE}, we present
the ground-state energy per particle, the confinement frequency $\omega_{op}$
and the total spin $S$ as a function of the coupling constant $\alpha$ for
$\alpha_{0}/\alpha=1.05$ and for a different numbers of polarons. The
ground-state energy turns out to be a continuous function of $\alpha$, while
$\omega_{op}$and $S$ reveal jumps. For $N=2$ (the case of a bipolaron), we see
from Fig. \ref{trinv-GSE} that the ground state has a total spin $S=0$ for all
values of $\alpha$, i.~e., the ground state of a bipolaron is a singlet. This
result is in agreement with earlier investigations on the large-bipolaron
problem (see, e.~g., \cite{Kashirina2003}).

In summary, using the extension of the Jensen-Feynman variational principle to
the systems of identical particles, we have derived a rigorous upper bound for
the free energy of a translation invariant system of $N$ interacting polarons.
The developed approach is valid for an arbitrary coupling strength. The
resulting ground-state energy is obtained taking into account the Fermi
statistics of electrons.\newpage

\section{Ripplonic polarons in multielectron bubbles}

\subsection{Ripplon-phonon modes of a MEB}

Spherical shells of charged particles appear in a variety of physical systems,
such as fullerenes, metallic nanoshells, charged droplets and neutron stars. A
particularly interesting physical realization of the spherical electron gas is
found in multielectron bubbles (MEBs) in liquid helium-4. These MEBs are 0.1
$\mu$m -- 100 $\mu$m sized cavities inside liquid helium, that contain helium
vapor at vapor pressure and a nanometer-thick electron layer anchored to the
surface of the bubble \cite{VolodinJETP26}. They exist as a result of
equilibrium between the surface tension of liquid helium and the Coulomb
repulsion of the electrons \cite{ShikinJETP27}. Recently proposed experimental
schemes to stabilize MEBs \cite{SilveraBAPS46} have stimulated theoretical
investigation of their properties.

We describe the dynamical modes of an MEB by considering the motion of the
helium surface (\textquotedblleft ripplons\textquotedblright) and the
vibrational modes of the electrons together. In particular, we analyze the
case when the ripplopolarons form a Wigner lattice \cite{TempereEPJ2003}.

First, we derive the Lagrangian of interacting ripplons and phonons within a
continuum approach. The shape of the surface of a bubble is described by the
function $R\left(  \theta,\varphi\right)  =R_{\text{b}}+u\left(
\theta,\varphi\right)  ,$ where $u\left(  \theta,\varphi\right)  $ is the
deformation of the surface from a sphere with radius $R_{\text{b}}.$ The
deformation can be expanded in a series of spherical harmonics $Y_{lm}\left(
\theta,\varphi\right)  $ with amplitudes $Q_{lm},$
\begin{equation}
u\left(  \theta,\varphi\right)  =\sum_{l=1}^{\infty}\sum_{m=-l}^{l}%
Q_{lm}Y_{lm}\left(  \theta,\varphi\right)  . \label{radius}%
\end{equation}
We suppose that the amplitudes are small in such a way that $\sqrt{l\left(
l+1\right)  }\left\vert Q_{lm}\right\vert \ll R_{\text{b}}.$

The ripplon contribution ($T_{\text{r}}$) to the kinetic energy of an MEB, and
the contributions to the potential energy due to the surface tension
($U_{\sigma}$) and due to the pressure ($U_{\text{V}}$) were described in Ref.
\cite{TemperePRL87}:
\begin{equation}%
\begin{array}
[c]{l}%
T_{\text{r}}=\dfrac{\rho}{2}R_{\text{b}}^{3}%
%TCIMACRO{\dsum \limits_{l=1}^{\infty}}%
%BeginExpansion
{\displaystyle\sum\limits_{l=1}^{\infty}}
%EndExpansion%
%TCIMACRO{\dsum \limits_{m=-l}^{l}}%
%BeginExpansion
{\displaystyle\sum\limits_{m=-l}^{l}}
%EndExpansion
\dfrac{1}{l+1}\left|  \dot{Q}_{lm}\right|  ^{2},\\
U_{\sigma}=4\pi\sigma R_{\text{b}}^{2}+\dfrac{\sigma}{2}%
%TCIMACRO{\dsum \limits_{l=1}^{\infty}}%
%BeginExpansion
{\displaystyle\sum\limits_{l=1}^{\infty}}
%EndExpansion%
%TCIMACRO{\dsum \limits_{m=-l}^{l}}%
%BeginExpansion
{\displaystyle\sum\limits_{m=-l}^{l}}
%EndExpansion
\left(  l^{2}+l+2\right)  \left|  Q_{lm}\right|  ^{2},\\
U_{\text{V}}=\dfrac{4\pi}{3}pR_{\text{b}}^{3}+pR_{\text{b}}%
%TCIMACRO{\dsum \limits_{l=1}^{\infty}}%
%BeginExpansion
{\displaystyle\sum\limits_{l=1}^{\infty}}
%EndExpansion%
%TCIMACRO{\dsum \limits_{m=-l}^{l}}%
%BeginExpansion
{\displaystyle\sum\limits_{m=-l}^{l}}
%EndExpansion
\left|  Q_{lm}\right|  ^{2}.
\end{array}
\label{Usp}%
\end{equation}
Here $\rho\approx145$ kg/m$^{3}$ is the density of liquid helium,
$\sigma\approx3.6\times10^{-4}$ J/m$^{2}$ is its surface tension, and $p$ is
the difference of pressures outside and inside the bubble.

Expanding the surface electron density $n\left(  \theta,\varphi\right)  $ in a
series of spherical harmonics with amplitudes $n_{lm},$
\begin{equation}
n\left(  \theta,\varphi\right)  =\sum_{l=0}^{\infty}\sum_{m=-l}^{l}%
n_{lm}Y_{lm}\left(  \theta,\varphi\right)  , \label{n1}%
\end{equation}
the kinetic energy of the motion of electrons can be written as
\begin{equation}
T_{\text{p}}=\frac{1}{2}\sum_{l=1}^{\infty}\sum_{m=-l}^{l}\frac{4\pi
m_{\text{e}}R_{\text{b}}^{6}}{l(l+1)N}\left\vert \dot{n}_{lm}\right\vert ^{2},
\label{Tsd}%
\end{equation}
where $m_{\text{e}}$\ is the bare electron mass and $N$\ is the number of
electrons. Finally, the electrostatic energy ($U_{\text{C}}$) of the deformed
MEB with a non-uniform surface electron density (\ref{n1}) is calculated using
the Maxwell equations and the electrostatic boundary conditions at the
surface. The result is:
\begin{align}
U_{\text{C}}  &  =\frac{e^{2}N^{2}}{2\varepsilon R_{\text{b}}}+2\pi
e^{2}R_{\text{b}}^{3}\sum_{l=1}^{\infty}\sum_{m=-l}^{l}\frac{\left\vert
n_{lm}\right\vert ^{2}}{l+\varepsilon\left(  l+1\right)  }\nonumber\\
&  -\frac{e^{2}N^{2}}{8\pi\varepsilon R_{\text{b}}^{3}}\sum_{l=1}^{\infty}%
\sum_{m=-l}^{l}\frac{l^{2}-\varepsilon\left(  l+1\right)  }{l+\varepsilon
\left(  l+1\right)  }\left\vert Q_{lm}\right\vert ^{2}\nonumber\\
&  -e^{2}N\sum_{l=1}^{\infty}\sum_{m=-l}^{l}\frac{l+1}{l+\varepsilon\left(
l+1\right)  }n_{lm}Q_{lm}^{\ast}, \label{UC}%
\end{align}
with the dielectric constant of liquid helium $\varepsilon\approx1.0572$. The
last term in Eq. (\ref{UC}) describes the ripplon-phonon mixing. Only ripplon
and phonon modes which have the same angular momentum couple to each other.
After the diagonalization of the Lagrangian of this ripplon-phonon system, we
arrive at the eigenfrequencies:
\begin{align}
\Omega_{1,2}\left(  l\right)   &  =\left\{  \frac{1}{2}\left[  \omega
_{\text{p}}^{2}\left(  l\right)  +\omega_{\text{r}}^{2}\left(  l\right)
%TCIMACRO{\QDATOP{{}}{{}}}%
%BeginExpansion
\genfrac{}{}{0pt}{0}{{}}{{}}%
%EndExpansion
_{{}}^{{}}\right.  \right. \nonumber\\
&  \left.  \left.  \pm\sqrt{\left[  \omega_{\text{p}}^{2}\left(  l\right)
-\omega_{\text{r}}^{2}\left(  l\right)  \right]  ^{2}+4\gamma^{2}\left(
l\right)  }\right]  \right\}  ^{1/2}, \label{Fab}%
\end{align}
where $\omega_{\text{r}}\left(  l\right)  $ is the bare ripplon frequency,
\begin{align}
\omega_{\text{r}}\left(  l\right)   &  =\left\{  \frac{l+1}{\rho R_{\text{b}%
}^{3}}\left[  \sigma\left(  l^{2}+l+2\right)  \left.
%TCIMACRO{\QDATOP{{}}{{}}}%
%BeginExpansion
\genfrac{}{}{0pt}{0}{{}}{{}}%
%EndExpansion
_{{}}^{{}}\right.  _{{}}^{{}}\right.  \right. \nonumber\\
&  \left.  \left.  -\frac{e^{2}N^{2}}{4\pi\varepsilon R_{\text{b}}^{3}}%
\frac{l^{2}-\varepsilon\left(  l+1\right)  }{l+\varepsilon\left(  l+1\right)
}+2pR_{\text{b}}\right]  \right\}  ^{1/2}, \label{wr}%
\end{align}
while $\omega_{\text{p}}\left(  l\right)  $ is the bare phonon frequency,
\begin{equation}
\omega_{\text{p}}\left(  l\right)  =\left(  \frac{e^{2}N}{m_{\text{e}%
}R_{\text{b}}^{3}}\frac{l\left(  l+1\right)  }{l+\varepsilon\left(
l+1\right)  }\right)  ^{1/2}, \label{wp}%
\end{equation}
and $\gamma\left(  l\right)  $ describes the ripplon-phonon coupling:
\begin{equation}
\gamma\left(  l\right)  =\frac{e^{2}N}{R_{\text{b}}^{3}}\left(  \frac{Nl}{4\pi
m_{\text{e}}\rho R_{\text{b}}^{3}}\right)  ^{1/2}\frac{\left(  l+1\right)
^{2}}{l+\varepsilon\left(  l+1\right)  }. \label{gRI}%
\end{equation}

\subsection{Electron-ripplon interaction in the MEB}

The interaction energy between the ripplons and the electrons in the
multielectron bubble can be derived from the following considerations: (i) the
distance between the layer electrons and the helium surface is fixed (the
electrons find themselves confined to an effectively 2D surface anchored to
the helium surface) and (ii) the electrons are subjected to a force field,
arising from the electric field of the other electrons. For a spherical
bubble, this electric field lies along the radial direction and equals
\begin{equation}
\mathbf{E}=-{\frac{Ne}{2R_{b}^{2}}}\mathbf{e}_{\mathbf{r}}.
\end{equation}
A bubble shape oscillation will displace the layer of electrons anchored to
the surface. The interaction energy which arises from this, equals the
displacement of the electrons times the force $e\mathbf{E}$ acting on them.
Thus, we get for the interaction Hamiltonian
\begin{equation}
\hat{H}_{int}=\sum_{j}e|\mathbf{E}|\times u(\hat{\Omega}_{j}).
\end{equation}
Here $u(\Omega)$ is the radial displacement of the surface in the direction
given by the spherical angle $\Omega$; and $\hat{\Omega}_{j}$ is the (angular)
position operator for electron $j$. The displacement can be rewritten using
(\ref{radius}) and we find
\begin{equation}
\hat{H}_{int}=\sum_{j}e|\mathbf{E}|\sum_{\ell,m}\hat{Q}_{\ell m}Y_{\ell
m}(\hat{\Omega}_{j}).
\end{equation}
Using the relation
\begin{equation}
\hat{Q}_{\ell,m}=(-1)^{(m-|m|)/2}\sqrt{{\textstyle{\frac{\hbar(\ell+1)}{2\rho
R_{b}^{3}\omega_{\ell}}}}}(\hat{a}_{\ell,m}+\hat{a}_{\ell,-m}^{+}),
\label{Qriplon}%
\end{equation}
the interaction Hamiltonian can be written in the suggestive form
\begin{equation}
\hat{H}_{int}=\sum_{\ell,m}\sum_{j}M_{\ell,m}Y_{\ell,m}(\hat{\Omega}_{j}%
)(\hat{a}_{\ell,m}+\hat{a}_{\ell,-m}^{+}), \label{Hint}%
\end{equation}
with the electron-ripplon coupling amplitude for a MEB given by
\begin{equation}
M_{\ell,m}=(-1)^{(m-|m|)/2}{\displaystyle{\frac{Ne^{2}}{2R_{b}^{2}}}}%
\sqrt{{\displaystyle{\frac{\hbar(\ell+1)}{2\rho R_{b}^{3}\omega_{\ell}}}}}%
\end{equation}

\subsection{Locally flat approximation}

Substituting $M_{\ell,m}$ into (\ref{Hint}), we get
\begin{align}
\lefteqn{\hat{H}_{int}=\sum_{\ell,m}\sum_{j}\frac{Ne^{2}}{2R_{b}^{2}}%
\sqrt{\frac{\hbar(\ell+1)}{2\rho R_{b}^{3}\omega_{\ell}}}}\\
&  \times\left[  (-1)^{(m-|m|)/2}{\frac{Y_{\ell,m}(\hat{\Omega}_{j})}{R_{b}}%
}\right]  (\hat{a}_{\ell,m}+\hat{a}_{\ell,-m}^{+}).\nonumber
\end{align}
In this expression, we consider the limit of a bubble so large that the
surface becomes flat on all length scales of interest. Hence we let
$R_{b}\rightarrow\infty$ but keep $\ell/R_{b}=q$ a constant. This means we
have to let $\ell\rightarrow\infty$ as well. In this limit,
\begin{equation}
\lim_{\ell\rightarrow\infty}Y_{\ell,0}(\theta)={\displaystyle{\frac{i^{\ell}%
}{\pi\sqrt{\sin\theta}}}}\sin[(\ell+1/2)\theta+\pi/4],
\end{equation}
and $Y_{\ell,0}(\theta)$ varies locally as a plane wave with wave vector
$q=\ell/R_{b}$. The wave function $Y_{\ell,m}(\hat{\Omega}_{j})/R_{b}$ is
furthermore normalized with respect to integration over the surface (with
total area $4\pi R_{b}^{2}$). Thus, we get in the locally flat approximation
\begin{equation}
\hat{H}_{int}=\sum_{\mathbf{q}}\sum_{j}{\displaystyle{\frac{Ne^{2}}{2R_{b}%
^{2}}}}\sqrt{{\displaystyle{\frac{\hbar q}{2\rho\omega(q)}}}}e^{i\mathbf{q}%
.\mathbf{\hat{r}}_{j}}(\hat{a}_{\mathbf{q}}+\hat{a}_{-\mathbf{q}}^{+}),
\end{equation}
or
\begin{align}
\hat{H}_{int}  &  =\sum_{\mathbf{q}}\sum_{j}M_{q}e^{i\mathbf{q}.\mathbf{\hat
{r}}_{j}}(\hat{a}_{\mathbf{q}}+\hat{a}_{-\mathbf{q}}^{+}),\nonumber\\
M_{q}  &  =e|\mathbf{E|}\sqrt{{\displaystyle{\frac{\hbar q}{2\rho\omega(q)}}}%
}.
\end{align}
This corresponds in the limit of large bubbles to the interaction Hamiltonian
expected for a flat surface.

\subsection{Ripplopolaron in a Wigner lattice: the mean-field approach}

In their treatment of the electron Wigner lattice embedded in a polarizable
medium such as a semiconductors or an ionic solid, Fratini and Qu\'{e}merais
\cite{FratiniEPJB14} described the effect of the electrons on a particular
electron through a mean-field lattice potential. The (classical) lattice
potential $V_{lat}$ is obtained by approximating all the electrons acting on
one particular electron by a homogenous charge density in which a hole is
punched out; this hole is centered in the lattice point of the particular
electron under investigation and has a radius given by the lattice distance
$d$.

Within this particular mean-field approximation, the lattice potential can be
calculated from classical electrostatics and we find that for a 2D electron
gas it can be expressed in terms of the elliptic functions of first and second
kind, $E\left(  x\right)  $ and $K\left(  x\right)  $,
\begin{align}
V_{lat}\left(  \mathbf{r}\right)   &  =-\frac{2e^{2}}{\pi d^{2}}\left\{
\left\vert d-r\right\vert E\left[  -\frac{4rd}{\left(  d-r\right)  ^{2}%
}\right]  \right. \nonumber\\
&  \left.  +\left(  d+r\right)  \mathop{\rm sgn}\left(  d-r\right)  K\left[
-\frac{4rd}{\left(  d-r\right)  ^{2}}\right]  \right\}  . \label{Potential}%
\end{align}
Here, $\mathbf{r}$ is the position vector measured from the lattice position.
We can expand this potential around the origin to find the small-amplitude
oscillation frequency of the electron lattice:
\begin{equation}
\lim_{r\ll d}V_{lat}\left(  \mathbf{r}\right)  =-\frac{2e^{2}}{d}+\frac{1}%
{2}m_{e}\omega_{lat}^{2}r^{2}+\mathcal{O}\left(  r^{4}\right)  ,
\label{Potlimit}%
\end{equation}
with the confinement frequency
\begin{equation}
\omega_{lat}=\sqrt{\frac{e^{2}}{m_{e}d^{3}}}. \label{phonfreq}%
\end{equation}
In the mean-field approximation, the Hamiltonian for a ripplopolaron in a
lattice on a \textit{locally flat} helium surface is given by
\begin{align}
\hat{H}  &  ={\displaystyle{\frac{\hat{p}^{2}}{2m_{e}}}}+V_{lat}\left(
\mathbf{\hat{r}}\right)  +\sum_{\mathbf{q}}\hbar\omega(q)\hat{a}_{\mathbf{q}%
}^{+}\hat{a}_{\mathbf{q}}\nonumber\\
&  +\sum_{\mathbf{q}}M_{q}e^{-i\mathbf{q.r}}\left(  \hat{a}_{\mathbf{q}}%
+\hat{a}_{-\mathbf{q}}^{+}\right)  , \label{H1RI}%
\end{align}
where $\mathbf{\hat{r}}$ is the electron position operator.

Now that the lattice potential has been introduced, we can move on and include
effects of the bubble geometry. If we restrict our treatment to the case of
large bubbles (with $N>10^{5}$ electrons), then both the ripplopolaron radius
and the inter-electron distance $d$ are much smaller than the radius of the
bubble $R_{b}$. This gives us ground to use the locally flat approximation
using the auxiliary model of a ripplonic polaron in a planar system described
by (\ref{H1RI}), but with a modified ripplon dispersion relation and an
modified pressing field. We find for the modified ripplon dispersion relation
in the MEB:
\begin{equation}
\omega(q)=\sqrt{{\displaystyle{\frac{\sigma}{\rho}}}q^{3}+{\displaystyle{\frac
{p}{\rho R_{b}}}}q}, \label{ripplodisp}%
\end{equation}
where $R_{b}$ is the equilibrium bubble radius which depends on the pressure
and the number of electrons. The bubble radius is found by balancing the
surface tension and the pressure with the Coulomb repulsion. The modified
electron-ripplon interaction amplitude in an MEB is given by
\begin{equation}
M_{\mathbf{q}}=e|\mathbf{E}|\sqrt{{\displaystyle{\frac{\hbar q}{2\rho
\omega(q)}}}}. \label{elripcoupl}%
\end{equation}
The effective electric pressing field pushing the electrons against the helium
surface and determining the strength of the electron-ripplon interaction is
\begin{equation}
\mathbf{E}=-{\displaystyle{\frac{Ne}{2R_{b}^{2}}}}\mathbf{e}_{\mathbf{r}}.
\label{pressfield}%
\end{equation}

\subsection{Ripplopolaron Wigner lattice at finite temperature}

To study the ripplopolaron Wigner lattice at finite temperature and for any
value of the electron-ripplon coupling, we use the variational path-integral
approach \cite{Feynman}. This variational principle distinguishes itself from
Rayleigh-Ritz variation in that it uses a trial action functional $S_{trial}$
instead of a trial wave function.

The action functional of the system described by Hamiltonian (\ref{H1RI}),
becomes, after elimination of the ripplon degrees of freedom,
\begin{align}
S  &  =-{\displaystyle{\frac{1}{\hbar}}}\displaystyle \int\limits_{0}%
^{\hbar\beta}d\tau\left\{  {\displaystyle{\frac{m_{e}}{2}}}\dot{r}^{2}%
(\tau)+V_{lat}[r(\tau)]\right\}  +\sum_{\mathbf{q}}\left\vert M_{q}\right\vert
^{2}\nonumber\\
&  \times\displaystyle \int\limits_{0}^{\hbar\beta}d\tau\displaystyle \int
\limits_{0}^{\hbar\beta}d\sigma G_{\omega(q)}(\tau-\sigma)e^{i\mathbf{q}%
\cdot\lbrack\mathbf{r}(\tau)-\mathbf{r}(\sigma)]}, \label{DLR1977I}%
\end{align}
with
\begin{equation}
G_{\nu}(\tau-\sigma)={\displaystyle{\frac{\cosh[\nu(|\tau-\sigma|-\hbar
\beta/2)]}{\sinh(\beta\hbar\nu/2)}}}.
\end{equation}
In preparation of its customary use in the Jensen-Feynman inequality, the
action functional (\ref{DLR1977I}) is written in imaginary time $t=i\tau$ with
$\beta=1/(k_{B}T)$ where $T$\thinspace is the temperature. We introduce a
quadratic trial action of the form
\begin{align}
S_{trial}  &  =-{\displaystyle{\frac{1}{\hbar}}}\displaystyle \int
\limits_{0}^{\hbar\beta}d\tau\left[  {\displaystyle{\frac{m_{e}}{2}}}\dot
{r}^{2}(\tau)+{\displaystyle{\frac{m_{e}\Omega^{2}}{2}}}r^{2}(\tau)\right]
\nonumber\\
&  -{\displaystyle{\frac{Mw^{2}}{4\hbar}}}\displaystyle \int\limits_{0}%
^{\hbar\beta}d\tau\displaystyle \int\limits_{0}^{\hbar\beta}d\sigma G_{w}%
(\tau-\sigma)\mathbf{r}(\tau)\cdot\mathbf{r}(\sigma). \label{S0RI}%
\end{align}
where $M,w,$ and $\Omega$ are the variationally adjustable parameters. This
trial action corresponds to the Lagrangian
\begin{equation}
\mathcal{L}_{0}={\displaystyle{\frac{m_{e}}{2}}}\dot{r}^{2}%
+{\displaystyle{\frac{M}{2}}}\dot{R}^{2}-{\displaystyle{\frac{\kappa}{2}}%
}r^{2}-{\displaystyle{\frac{K}{2}}}(\mathbf{r}-\mathbf{R})^{2}, \label{L0}%
\end{equation}
from which the degrees of freedom associated with $\mathbf{R}$ have been
integrated out. This Lagrangian can be interpreted as describing an electron
with mass $m_{e}$ at position $\mathbf{r}$, coupled through a spring with
spring constant $\kappa$ to its lattice site, and to which a fictitious mass
$M$ at position $\mathbf{R}$ has been attached with another spring, with
spring constant $K$. The relation between the spring constants in (\ref{L0})
and the variational parameters $w,\Omega$ is given by
\begin{align}
w  &  =\sqrt{K/m_{e}},\\
\Omega &  =\sqrt{(\kappa+K)/m_{e}}.
\end{align}

Based on the trial action $S_{trial}$, Feynman's variational method allows one
to obtain an upper bound for the free energy $F$ of the system (at temperature
$T$) described by the action functional $S$ by minimizing the following function:%

\begin{equation}
F=F_{0}-\frac{1}{\beta}\left\langle S-S_{trial}\right\rangle , \label{JFRI}%
\end{equation}
with respect to the variational parameters of the trial action. In this
expression, $F_{0}$ is the free energy of the trial system characterized by
the Lagrangian $\mathcal{L}_{0}$, $\beta=1/(k_{b}T)$ is the inverse
temperature, and the expectation value $\left\langle S-S_{trial}\right\rangle
$ is to be taken with respect to the ground state of this trial system. The
evaluation of expression (\ref{JFRI}) is straightforward though lengthy. We
find
\begin{align}
\lefteqn{F=\displaystyle{2 \over\beta}\ln\left[  2\sinh\left(  \displaystyle
{\beta\hbar\Omega_{1} \over2}\right)  \right]  +\displaystyle{2 \over\beta}%
\ln\left[  2\sinh\left(  \displaystyle{\beta\hbar\Omega_{2} \over2}\right)
\right]  }\nonumber\\
&  -{\displaystyle{\frac{2}{\beta}}}\ln\left[  2\sinh\left(
{\displaystyle{\frac{\beta\hbar w}{2}}}\right)  \right]  -{\displaystyle{\frac
{\hbar}{2}}}\sum_{i=1}^{2}a_{i}^{2}\Omega_{i}\coth\left(  {\displaystyle{\frac
{\beta\hbar\Omega_{i}}{2}}}\right) \nonumber\\
&  -{\displaystyle{\frac{\sqrt{\pi}e^{2}}{D}}}e^{-d^{2}/(2D)}\left[
I_{0}\left(  {\displaystyle{\frac{d^{2}}{2D}}}\right)  +I_{1}\left(
{\displaystyle{\frac{d^{2}}{2D}}}\right)  \right] \label{FRI}\\
&  -{\displaystyle{\frac{1}{2\pi\hbar\beta}}}\int_{1/R_{b}}^{\infty}%
dqq|M_{q}|^{2}\int_{0}^{\hbar\beta/2}d\tau{\displaystyle{\frac{\cosh
[\omega(q)(\tau-\hbar\beta/2)]}{\sinh[\beta\hbar\omega(q)/2]}}}\nonumber\\
&  \times\exp\left[  -{\textstyle{\frac{\hbar q^{2}}{2m_{e}}}}\sum_{j=1}%
^{2}a_{j}^{2}{\textstyle{\frac{\cosh(\hbar\Omega_{j}\beta/2)-\cosh[\hbar
\Omega_{j}(\tau-\beta/2)]}{\Omega_{j}\sinh(\hbar\Omega_{j}\beta/2)}}}\right]
.\nonumber
\end{align}
In this expression, $I_{0}$ and $I_{1}$ are Bessel functions of imaginary
argument, and
\begin{equation}
D={\displaystyle{\frac{\hbar}{m_{e}}}}\sum_{j=1}^{2}{\displaystyle{\frac
{a_{j}^{2}}{\Omega_{j}}}}\coth\left(  \hbar\Omega_{j}\beta/2\right)  ,
\end{equation}%
\begin{equation}
a_{1}=\sqrt{{\displaystyle{\frac{\Omega_{1}^{2}-w^{2}}{\Omega_{1}^{2}%
-\Omega_{2}^{2}}}}};a_{2}=\sqrt{{\displaystyle{\frac{w^{2}-\Omega_{2}^{2}%
}{\Omega_{1}^{2}-\Omega_{2}^{2}}}}}.
\end{equation}
Finally, $\Omega_{1}$ and $\Omega_{2}$ are the eigenfrequencies of the trial
system, given by
\begin{equation}
\Omega_{1,2}^{2}={\displaystyle{\frac{1}{2}}}\left[  \Omega^{2}+w^{2}\pm
\sqrt{\left(  \Omega^{2}-w^{2}\right)  ^{2}+4K/(Mm_{e})}\right]  .
\end{equation}
Optimal values of the variational parameters are determined by the numerical
minimization of the variational functional $F$ as given by expression
(\ref{FRI}).

\subsection{Melting of the ripplopolaron Wigner lattice}

The Lindemann melting criterion \cite{LindemanZPhys11} states in general that
a crystal lattice of objects (be it atoms, molecules, electrons, or
ripplopolarons) will melt when the average motion of the objects around their
lattice site is larger than a critical fraction $\delta_{0}$ of the lattice
parameter $d$. It would be a strenuous task to calculate from first principles
the exact value of the critical fraction $\delta_{0}$, but for the particular
case of electrons on a helium surface, we can make use of an experimental
determination. Grimes and Adams \cite{GrimesPRL42} found that the Wigner
lattice melts when $\Gamma=137\pm15$, where $\Gamma$ is the ratio of potential
energy to the kinetic energy per electron. At temperature $T$ the average
kinetic energy in a lattice potential $V_{lat}$ is
\begin{equation}
E_{kin}={\displaystyle{\frac{\hbar\omega_{lat}}{2}}}\coth\left(
{\displaystyle{\frac{\hbar\omega_{lat}}{2k_{B}T}}}\right)  ,
\end{equation}
and the average distance that an electron moves out of the lattice site is
determined by
\begin{equation}
\left\langle \mathbf{r}^{2}\right\rangle ={\displaystyle{\frac{\hbar}%
{m_{e}\omega_{lat}}}}\coth\left(  {\displaystyle{\frac{\hbar\omega_{lat}%
}{2k_{B}T}}}\right)  ={\displaystyle{\frac{2E_{kin}}{m_{e}\omega_{lat}^{2}}}}.
\end{equation}
From this we find that for the melting transition in Grimes and Adams'
experiment \cite{GrimesPRL42}, the critical fraction equals $\delta_{0}%
\approx0.13$. This estimate is in agreement with previous (empirical)
estimates yielding $\delta_{0}\approx0.1$ \cite{BedanovPRB49}, and we shall
use it in the rest of this section.

Within the approach of Fratini and Qu\'{e}merais \cite{FratiniEPJB14}, the
Wigner lattice of (ripplo)polarons melts when at least one of the two
following Lindemann criteria are met:
\begin{equation}
\delta_{r}={\displaystyle{\frac{\sqrt{\left\langle \mathbf{R}_{cms}%
^{2}\right\rangle }}{d}}}>\delta_{0}, \label{Lind1}%
\end{equation}%
\begin{equation}
\delta_{\rho}={\displaystyle{\frac{\sqrt{\left\langle \mathbf{\rho}%
^{2}\right\rangle }}{d}}}>\delta_{0}. \label{Lind2}%
\end{equation}
where $\mathbf{\rho}$ and $\mathbf{R}_{cms}$ are, respectively, the relative
coordinate and the center of mass coordinate of the model system (\ref{L0}):
if $\mathbf{r}$ is the electron coordinate and $\mathbf{R}$ is the position
coordinate of the fictitious ripplon mass $M$, this is
\begin{equation}
\mathbf{R}_{cms}={\displaystyle{\frac{m_{e}\mathbf{r}+M\mathbf{R}}{m_{e}+M}}%
};\mathbf{\rho}=\mathbf{r}-\mathbf{R.}%
\end{equation}
The appearance of two Lindemann criteria takes into account the composite
nature of (ripplo)polarons. As follows from the physical sense of the
coordinates $\mathbf{\rho}$ and $\mathbf{R}_{cms}$, the first criterion
(\ref{Lind1}) is related to the melting of the ripplopolaron Wigner lattice
towards a ripplopolaron liquid, where the ripplopolarons move as a whole, the
electron together with its dimple. The second criterion (\ref{Lind2}) is
related to the dissociation of ripplopolarons: the electrons shed their dimple.

\begin{figure}
[h]
\begin{center}
\includegraphics[
height=8.7184cm,
width=9.5817cm
]%
{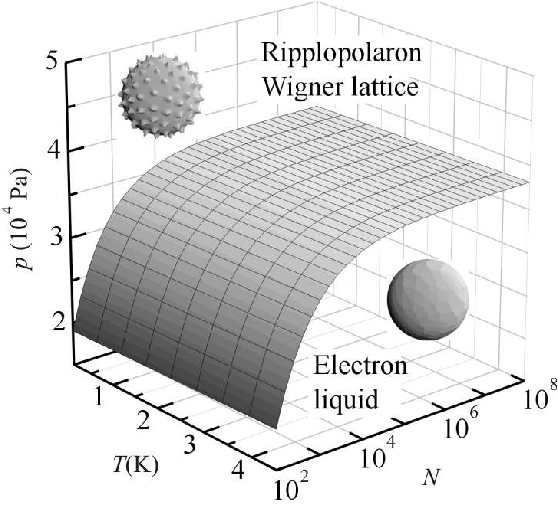}%
\caption{The phase diagram for the spherical 2D layer of electrons in the MEB.
Above a critical pressure, a ripplopolaron solid (a Wigner lattice of
electrons with dimples in the helium surface underneath them) is formed. Below
the critical pressure, the ripplopolaron solid melts into an electron liquid
through dissociation of ripplopolarons. (From Ref. \cite{TempereEPJ2003}.)}%
\label{PhaseD1}%
\end{center}
\end{figure}

The path-integral variational formalism allows us to calculate the expectation
values $\left\langle \mathbf{R}_{cms}^{2}\right\rangle $ and $\left\langle
\mathbf{\rho}^{2}\right\rangle $ with respect to the ground state of the
variationally optimal model system. We find
\begin{align}
\left\langle \mathbf{R}_{cms}^{2}\right\rangle  &  ={\displaystyle{\frac{\hbar
w^{4}}{m_{e}\left[  w^{2}(\Omega_{1}^{2}+\Omega_{2}^{2})-\Omega_{1}^{2}%
\Omega_{2}^{2}\right]  \left(  \Omega_{1}^{2}-\Omega_{2}^{2}\right)  }}%
}\nonumber\\
&  \times\left[  \Omega_{2}^{4}(\Omega_{1}^{2}-w^{2})\coth(\hbar\Omega
_{1}\beta/2)/\Omega_{1}\right. \nonumber\\
&  \left.  +\Omega_{1}^{4}(w^{2}-\Omega_{2}^{2})\coth(\hbar\Omega_{2}%
\beta/2)/\Omega_{2}\right]  , \label{Rcms}%
\end{align}%
\begin{align}
\left\langle \mathbf{\rho}^{2}\right\rangle  &  ={\displaystyle{\frac{\hbar
}{m_{e}\left(  \Omega_{1}^{2}-\Omega_{2}^{2}\right)  \left(  \Omega_{1}%
^{2}-w^{2}\right)  \left(  w^{2}-\Omega_{2}^{2}\right)  }}}\nonumber\\
&  \times\left[  \Omega_{1}^{3}(w^{2}-\Omega_{2}^{2})\coth\left(  \hbar
\Omega_{1}\beta/2\right)  \right. \nonumber\\
&  \left.  +\Omega_{2}^{3}(\Omega_{1}^{2}-w^{2})\coth(\hbar\Omega_{2}%
\beta/2)\right]  . \label{rhoRI}%
\end{align}
Numerical calculation shows that for ripplopolarons in an MEB the inequality
$\Omega_{1}\gg w$ is fulfilled ($w/\Omega_{1}\approx10^{-3}$ to $10^{-2}$) so
that the strong-coupling regime is realized. Owing to this inequality, we find
from Eqs. (\ref{Rcms}),(\ref{rhoRI}) that
\begin{equation}
\left\langle \mathbf{R}_{cms}^{2}\right\rangle \ll\left\langle \mathbf{\rho
}^{2}\right\rangle .
\end{equation}
So, the destruction of the ripplopolaron Wigner lattice in an MEB occurs
through the dissociation of ripplopolarons, since the second criterion
(\ref{Lind2}) will be fulfilled before the first (\ref{Lind1}). The results
for the melting of the ripplopolaron Wigner lattice are summarized in the
phase diagram shown in Fig. \ref{PhaseD1}.

\newpage%

%TCIMACRO{\FRAME{fhFU}{9.5817cm}{8.7184cm}{0pt}{\Qcb{The phase diagram for the
%spherical 2D layer of electrons in the MEB. Above a critical pressure, a
%ripplopolaron solid (a Wigner lattice of electrons with dimples in the helium
%surface underneath them) is formed. Below the critical pressure, the
%ripplopolaron solid melts into an electron liquid through dissociation of
%ripplopolarons. (From Ref. \cite{TempereEPJ2003}.)}}{\Qlb{PhaseD1}%
%}{part2fig10.eps}{\special{ language "Scientific Word";  type "GRAPHIC";
%maintain-aspect-ratio TRUE;  display "ICON";  valid_file "F";
%width 9.5817cm;  height 8.7184cm;  depth 0pt;  original-width 9.4587cm;
%original-height 8.602cm;  cropleft "0";  croptop "1";  cropright "1";
%cropbottom "0";  filename 'Part2Fig10.eps';file-properties "XNPEU";}} }%
%BeginExpansion
%EndExpansion

\newpage

For every value of $N$, pressure $p$ and temperature $T$ in an experimentally
accessible range, this figure shows whether the ripplopolaron Wigner lattice
is present (points above the surface) or molten (points below the surface).
Below a critical pressure (on the order of 10$^{4}$ Pa) the ripplopolaron
solid will melt into an electron liquid. This critical pressure is nearly
independent of the number of electrons (except for the smallest bubbles) and
is weakly temperature dependent, up to the helium critical temperature 5.2 K.
This can be understood since the typical lattice potential well in which the
ripplopolaron resides has frequencies of the order of THz or larger, which
correspond to $\sim10$ K.

The new phase that we predict, the ripplopolaron Wigner lattice, will not be
present for electrons on a flat helium surface. At the values of the pressing
field necessary to obtain a strong enough electron-ripplon coupling, the flat
helium surface is no longer stable against long-wavelength deformations
\cite{GorkovJETP18}. Multielectron bubbles, with their different ripplon
dispersion and the presence of stabilizing factors such as the energy barrier
against fissioning \cite{TemperePRB67}, allow for much larger electric fields
pressing the electrons against the helium surface. The regime of $N$, $p$, $T$
parameters suitable for the creation of a ripplopolaron Wigner lattice lies
within the regime that would be achievable in recently proposed experiments
aimed at stabilizing multielectron bubbles \cite{SilveraBAPS46}. The
ripplopolaron Wigner lattice and its melting transition might be detected by
spectroscopic techniques \cite{GrimesPRL42,FisherPRL42} probing for example
the transverse phonon modes of the lattice \cite{DevillePRL53}.

\begin{acknowledgments}
I thank S. N. Klimin for discussions in the course of the preparation of the
third edition of the Lectures.
\end{acknowledgments}

\appendix

\section{Optical conductivity of a strong-coupling Fr\"{o}hlich polaron
[\emph{S. N. Klimin and J. T. Devreese, Phys. Rev. B 89, 035201 (2014)}]}

\subsection{Introduction \label{sec:intro}}

The optical conductivity of the Fr\"{o}hlich polaron model attracted attention
for years \cite{Devreese2009}. In the regime of weak coupling, the optical
absorption of a polaron was calculated using different methods, e. g., Green's
function method \cite{GLF62}, the Low-Lee-Pines formalism
\cite{DHL1971,Huybrechts1973}, perturbation expansion of the current-current
correlation function \cite{Sernelius1993}. The strong-coupling polaron optical
conductivity was calculated taking into account one-phonon \cite{KED1969} and
two-phonon \cite{Goovaerts73} transitions from the polaron ground state to the
polaron relaxed excited state (RES). In fact the present work finalizes the
project started in Ref. \cite{KED1969}. Using the path integral response
formalism, the impedance function of an all-coupling polaron was calculated by
FHIP \cite{FHIP} on the basis of the Feynman polaron model \cite{Feynman}.
Developing further the FHIP approach, the optical conductivity was calculated
in the path-integral formalism at zero temperature \cite{DSG1972} and at
finite temperatures \cite{PD1983}. In Ref. \cite{DeFilippis2006}, the
extension of the method of Ref. \cite{DSG1972} accounting for the polaron
damping (for the polaron coupling constant $\alpha\lesssim8$) and the
asymptotic strong-coupling approach using the Franck-Condon (FC) picture for
the optical conductivity (for $\alpha\gtrsim8$) have given reasonable results
for the polaron optical conductivity at all values of $\alpha$. The concept of
the RES and FC polaron states played a key role in the understanding of the
mechanism of the polaron optical conductivity
\cite{KED1969,Goovaerts73,DE64,DSG1972,KartheuserGreenbook,PD1983}.

Recently, the Diagrammatic Quantum Monte Carlo (DQMC) numerical method has
been developed \cite{Mishchenko2000,Mishchenko2003}, which provides accurate
results for the polaron characteristics in all coupling regimes. The analytic
treatment \cite{DSG1972} was intended to be valid at all coupling strengths.
However, it is established in \cite{DSG1972,KartheuserGreenbook,Goovaerts73}
that the linewidth of the obtained spectra \cite{DSG1972} is unreliable for
$\alpha\gtrapprox7$. Nevertheless, the position of the peak attributed to RES
in Ref. \cite{DSG1972} is close to the maximum of the polaron optical
conductivity band calculated using DQMC up to very large values of $\alpha$
(see Fig. 1).

An extension of the path-integral approach \cite{DSG1972} performed in Ref.
\cite{DeFilippis2006} gives a good agreement with DQMC for weak and
intermediate coupling strengths. In the strong-coupling limit, in Ref.
\cite{DeFilippis2006} the adiabatic strong-coupling expansion was applied.
That expansion, however, is not exact in the strong-coupling limit because of
a parabolic approximation \cite{LP48} for the adiabatic potential.

In the present work, the strong-coupling approach of Ref.
\cite{DeFilippis2006} is extended in order to obtain the polaron optical
conductivity which is \emph{asymptotically exact in the strong-coupling
limit}. We develop the multiphonon strong-coupling expansion using numerically
accurate in the strong-coupling limit polaron energies and wave functions and
accounting for non-adiabaticity.

\subsection{Optical conductivity}

We consider the electron-phonon system with the Hamiltonian written down in
the Feynman units ($\hbar=1,$ the carrier band mass $m_{b}=1$, and the
LO-phonon frequency $\omega_{\mathrm{LO}}=1$)%
\begin{equation}
H=\frac{\mathbf{p}^{2}}{2}+\sum_{\mathbf{q}}\left(  b_{\mathbf{q}}%
^{+}b_{\mathbf{q}}+\frac{1}{2}\right)  +\frac{1}{\sqrt{V}}\sum_{\mathbf{q}%
}\frac{\sqrt{2\sqrt{2}\pi\alpha}}{q}\left(  b_{\mathbf{q}}+b_{-\mathbf{q}}%
^{+}\right)  e^{i\mathbf{q\cdot r}}.
\end{equation}
where $\mathbf{r},\mathbf{p}$ represent the position and momentum of an
electron, $b_{\mathbf{q}}^{+},b_{\mathbf{q}}$ denote the creation and
annihilation operators for longitudinal optical (LO) phonons with wave vector
$\mathbf{q}$, and $V_{\mathbf{q}}$ describes the amplitude of the interaction
between the electrons and the phonons. For the Fr\"{o}hlich electron-phonon
interaction, the amplitude of the electron -- LO-phonon interaction is%
\begin{equation}
V_{\mathbf{q}}=\frac{1}{\sqrt{V}}\frac{\sqrt{2\sqrt{2}\pi\alpha}}{q}
\label{Vq}%
\end{equation}
with the crystal volume $V$, and the electron-phonon coupling constant
$\alpha$.

The polaron optical conductivity describes the response of the system with the
Hamiltonian (\ref{H}) to an applied electromagnetic field (along the $z$-axis)
with frequency $\omega$. This optical response is expressed using the Kubo
formula with a dipole-dipole correlation function:
\begin{equation}
\operatorname{Re}\sigma\left(  \omega\right)  =\frac{n_{0}\omega}{2}\left(
1-e^{-\beta\omega}\right)  \int_{-\infty}^{\infty}e^{i\omega t}\left\langle
d_{z}\left(  t\right)  d_{z}\right\rangle \,dt, \label{KuboDD}%
\end{equation}
where $\mathbf{d}=-e_{0}\mathbf{r}$ is the electric dipole moment, $e_{0}$ is
the unit charge, $\beta=\frac{1}{k_{B}T}$, $n_{0}$ is the electron density. In
the zero-temperature limit, the optical conductivity (\ref{KuboDD}) measured
in units of $e_{0}^{2}$ becomes%
\begin{equation}
\operatorname{Re}\sigma\left(  \omega\right)  =\frac{\omega}{2}\int_{-\infty
}^{\infty}e^{i\omega t}f_{zz}\left(  t\right)  \,dt, \label{KuboDD0}%
\end{equation}
with the correlation function
\begin{equation}
f_{zz}\left(  t\right)  \equiv\left\langle z\left(  t\right)  z\left(
0\right)  \right\rangle =\left\langle \Psi_{0}\left\vert e^{itH}%
ze^{-itH}z\right\vert \Psi_{0}\right\rangle , \label{fzz}%
\end{equation}
where $\left\vert \Psi_{0}\right\rangle $ is the ground-state wave function of
the electron-phonon system.

Within the strong-coupling approach, the ground-state wave function is chosen
as the product of a trial wave function of an electron $\left\vert \psi
_{0}^{\left(  e\right)  }\right\rangle $ and of a trial wave function of a
phonon subsystem $\left\vert \Phi_{ph}\right\rangle $:%
\begin{equation}
\left\vert \Psi_{0}\right\rangle =\left\vert \psi_{0}^{\left(  e\right)
}\right\rangle \left\vert \Phi_{ph}\right\rangle . \label{ansatz}%
\end{equation}
The phonon trial wave function is written as the strong-coupling unitary
transformation applied to the phonon vacuum
\begin{equation}
\left\vert \Phi_{ph}\right\rangle =U\left\vert 0_{ph}\right\rangle .
\label{Fph1}%
\end{equation}
with the unitary operator%
\begin{equation}
U=e^{\sum_{\mathbf{q}}\left(  f_{\mathbf{q}}b_{\mathbf{q}}-f_{\mathbf{q}%
}^{\ast}b_{\mathbf{q}}^{+}\right)  }, \label{unit2}%
\end{equation}
and the variational parameters $\left\{  f_{\mathbf{q}}\right\}  $. The
transformed Hamiltonian $\tilde{H}\equiv U^{-1}HU$ takes the form%
\begin{equation}
\tilde{H}=\tilde{H}_{0}+W \label{HT2}%
\end{equation}
with the terms%
\begin{align}
\tilde{H}_{0}  &  =\frac{\mathbf{p}^{2}}{2}+\sum_{\mathbf{q}}\left\vert
f_{\mathbf{q}}\right\vert ^{2}+V_{a}\left(  r\right)  +\sum_{\mathbf{q}%
}\left(  b_{\mathbf{q}}^{+}b_{\mathbf{q}}+\frac{1}{2}\right)  ,\label{H0}\\
W  &  =\sum_{\mathbf{q}}\left(  W_{\mathbf{q}}b_{\mathbf{q}}+W_{\mathbf{q}%
}^{\ast}b_{\mathbf{q}}^{+}\right)  . \label{W}%
\end{align}
Here, $W_{\mathbf{q}}$ are the amplitudes of the renormalized electron-phonon
interaction%
\begin{equation}
W_{\mathbf{q}}=\frac{\sqrt{2\sqrt{2}\pi\alpha}}{q\sqrt{V}}\left(
e^{i\mathbf{q\cdot r}}-\rho_{\mathbf{q}}\right)  ,
\end{equation}
where $\rho_{\mathbf{q}}$ is the expectation value of the operator
$e^{i\mathbf{q\cdot r}}$ with the trial electron wave function $\left\vert
\psi_{0}^{\left(  e\right)  }\right\rangle $:%
\begin{equation}
\rho_{\mathbf{q}}=\left\langle \psi_{0}^{\left(  e\right)  }\left\vert
e^{i\mathbf{q\cdot r}}\right\vert \psi_{0}^{\left(  e\right)  }\right\rangle ,
\end{equation}
and $V_{a}\left(  r\right)  $ is the self-consistent potential energy for the
electron,%
\begin{equation}
V_{a}\left(  r\right)  =-\sum_{\mathbf{q}}\frac{4\sqrt{2}\pi\alpha}{q^{2}%
V}\rho_{-\mathbf{q}}e^{i\mathbf{q}\cdot\mathbf{r}}.
\end{equation}

Averaging the Hamiltonian (\ref{HT2}) with the phonon vacuum $\left\vert
0\right\rangle $ and with the trial electron wave function $\left\vert
\psi_{0}\right\rangle $, we arrive at the following variational expression for
the ground-state energy%
\begin{align}
E_{0}  &  =\left\langle \Psi_{0}\left\vert H\right\vert \Psi_{0}\right\rangle
=\left\langle \psi_{0}\left\vert \frac{\mathbf{p}^{2}}{2}\right\vert \psi
_{0}\right\rangle +\sum_{\mathbf{q}}\left\vert f_{\mathbf{q}}\right\vert
^{2}\nonumber\\
&  -\sum_{\mathbf{q}}\left(  V_{\mathbf{q}}f_{\mathbf{q}}^{\ast}%
\rho_{\mathbf{q}}+V_{\mathbf{q}}^{\ast}f_{\mathbf{q}}\rho_{-\mathbf{q}%
}\right)  , \label{E0var}%
\end{align}
After minimization of the polaron ground-state energy (\ref{E0var}), the
parameters $f_{\mathbf{q}}$ acquire their optimal values%
\begin{equation}
f_{\mathbf{q}}=V_{\mathbf{q}}\rho_{\mathbf{q}}. \label{ov}%
\end{equation}
The ground-state energy with $\left\{  f_{\mathbf{q}}\right\}  $ given by Eq.
(\ref{ov}) takes the form%
\begin{equation}
E_{0}=\left\langle \psi_{0}\left\vert \frac{\mathbf{p}^{2}}{2}\right\vert
\psi_{0}\right\rangle -\sum_{\mathbf{q}}\left\vert V_{\mathbf{q}}\right\vert
^{2}\left\vert \rho_{\mathbf{q}}\right\vert ^{2}. \label{E0}%
\end{equation}

With the strong-coupling Ansatz (\ref{ansatz}) for the polaron ground-state
wave function and after the application of the unitary transformation
(\ref{unit2}), the correlation function (\ref{fzz}) takes the form%
\begin{equation}
f_{zz}\left(  t\right)  =\left\langle 0_{ph}\left\vert \left\langle \psi
_{0}\left\vert e^{it\tilde{H}}ze^{-it\tilde{H}}z\right\vert \psi
_{0}\right\rangle \right\vert 0_{ph}\right\rangle . \label{fzz2}%
\end{equation}

This correlation function can be expanded using a complete orthogonal set of
intermediate states $\left\vert j\right\rangle $ and the completeness
property:%
\begin{equation}
\sum_{j}\left\vert j\right\rangle \left\langle j\right\vert =1. \label{compl}%
\end{equation}
In the present work, we use the intermediate basis of the Franck-Condon (FC)
states. The FC states correspond to the equilibrium phonon configuration for
the ground state. Thus the FC wave functions are the exact eigenstates of the
Hamiltonian $\tilde{H}_{0}$. Further on, the FC wave functions are written in
the spherical-wave representation as $\left\vert \psi_{n,l,m}\right\rangle
=R_{n,l}\left(  r\right)  Y_{l,m}\left(  \theta,\varphi\right)  $ where
$R_{n,l}\left(  r\right)  $ are the radial wave functions, and $Y_{l,m}\left(
\theta,\varphi\right)  $ are the spherical harmonics, $l$ is the quantum
number of the angular momentum, $m$ is the $z$-projection of the angular
momentum, and $n$ is the radial quantum number\footnote{In this
classification, the ground-state wave function is $\left\vert \psi
_{0,0,0}\right\rangle \equiv\left\vert \psi_{0}\right\rangle $.}. The energy
levels for the eigenstates of the Hamiltonian $\tilde{H}_{0}$ are denoted
$E_{n,l}$.

Using (\ref{compl}) with that complete and orthogonal basis , we transform
(\ref{fzz2}) to the expression%
\begin{align}
f_{zz}\left(  t\right)   &  =\sum_{\substack{n,l,m,\\n^{\prime},l^{\prime
},m^{\prime},\\n^{\prime\prime},l^{\prime\prime},m^{\prime\prime}%
}}\left\langle \psi_{n,l,m}\left\vert z\right\vert \psi_{n^{\prime\prime
},l^{\prime\prime},m^{\prime\prime}}\right\rangle \left\langle \psi
_{n^{\prime},l^{\prime},m^{\prime}}\left\vert z\right\vert \psi_{0}%
\right\rangle \nonumber\\
&  \times\left\langle 0_{ph}\left\vert \left\langle \psi_{0}\left\vert
e^{it\tilde{H}}\right\vert \psi_{n,l,m}\right\rangle \left\langle
\psi_{n^{\prime\prime},l^{\prime\prime},m^{\prime\prime}}\left\vert
e^{-it\tilde{H}}\right\vert \psi_{n^{\prime},l^{\prime},m^{\prime}%
}\right\rangle \right\vert 0_{ph}\right\rangle . \label{fzz3}%
\end{align}

So far, the only approximation made in (\ref{fzz3}) is the strong-coupling
Ansatz for the polaron ground-state wave function. However, in order to obtain
a numerically tractable expression for the polaron optical conductivity, an
additional approximation valid in the strong-coupling limit must be applied to
the matrix elements of the evolution operator $e^{-it\tilde{H}}$ with the
Hamiltonian of the electron-phonon system $\tilde{H}$ given by formula
(\ref{HT2}). According to Ref. \cite{Allcock1}, in the strong-coupling limit,
the matrix elements of the Hamiltonian of the electron-phonon system between
states corresponding to different energy levels are of order of magnitude
$\alpha^{-4}$. Therefore in the strong-coupling regime these matrix elements
can be neglected; this is called the adiabatic or the Born-Oppenheimer (BO)
approximation \cite{Allcock1}, because of its strict analogy with the
Born-Oppenheimer adiabatic approximation in the theory of molecules and
crystals (\cite{Born}, p. 171). Consequently, in the further treatment we
neglect the matrix elements $\left\langle \psi_{n,l,m}\left\vert
e^{-it\tilde{H}}\right\vert \psi_{n^{\prime},l^{\prime},m^{\prime}%
}\right\rangle $ for the FC states with different energies, $E_{n,l}\neq
E_{n^{\prime},l^{\prime}}$. The same scheme was used in the theory of the
multi-phonon optical processes for bound electrons interacting with phonons
\cite{Pekar,Perlin}.

Strictly speaking, the summation over the excited polaron states in Eq.
(\ref{fzz3}) must involve the transitions to both the discrete and continuous
parts of the polaron spectrum. A transition to the states of the continuous
spectrum means that the electron leaves the polaron potential well. Therefore
these transitions can be attributed to the \textquotedblleft polaron
dissociation\textquotedblright. The transitions to the continuous spectrum are
definitely beyond the adiabatic approximation. As shown in Ref. \cite{Pekar},
the transition probability to the states of the continuous spectrum is very
small compared with the transition probability between the ground and the
first excited state (which belongs to the discrete part of the polaron energy
spectrum). We neglect here the contribution to the polaron optical
conductivity due to the transitions to the continuous spectrum.

The matrix elements neglected within the adiabatic approximation correspond to
the transitions between FC states with different energies due to the
electron-phonon interaction. Hence these transitions can be called
non-adiabatic. The adiabatic approximation is related to the matrix elements
of the evolution operator $e^{-it\tilde{H}}$. On the contrary, the matrix
elements of the transitions between different FC states for the electric
dipole moment are, in general, not equal to zero. Moreover, these transitions
can be accompanied by the emission of phonons. The electron FC wave functions
constitute a complete orthogonal set. However, the corresponding phonon wave
functions can be non-orthogonal because of a different shift of phonon
coordinates for different electron states. This makes multi-phonon transitions
possible \cite{Perlin}. It is important to note that in our treatment we
neglect only the non-adiabatic transitions between the electron states with
\emph{different} energies. On the contrary, the transitions within one and the
same degenerate level can be non-adiabatic. This \emph{internal
non-adiabaticity} (i.~e., the non-adiabaticity of the transitions within one
and the same degenerate level) is taken into account in the subsequent treatment.

It is useful to stress the difference between the strong-coupling Ansatz and
the adiabatic approximation. The strong-coupling Ansatz consists of the choice
of the trial variational ground state wave function for the electron-phonon
system in the factorized form (\ref{ansatz}). The adiabatic approximation
means neglecting the matrix elements of the evolution operator between
internal polaron states with different energies. These two approximations are
not the same, but they both are valid in the strong-coupling regime and
consistent with each other.

The correlation function (\ref{fzz3}) is transformed in the following way. The
exponents $e^{it\tilde{H}}$ and $e^{-it\tilde{H}}$ are disentangled:%
\begin{align}
e^{-it\tilde{H}}  &  =e^{-it\tilde{H}_{0}}\mathrm{T}\exp\left(  -i\int_{0}%
^{t}dsW\left(  s\right)  \right)  ,\\
e^{it\tilde{H}}  &  =e^{it\tilde{H}_{0}}\mathrm{T}\exp\left(  i\int_{0}%
^{t}dsW\left(  -s\right)  \right)
\end{align}
where $W\left(  s\right)  $ is the renormalized electron-phonon interaction
Hamiltonian $W$ in the interaction representation,%
\begin{equation}
W\left(  s\right)  \equiv e^{is\tilde{H}_{0}}We^{-is\tilde{H}_{0}}.
\end{equation}
This gives us the result%
\begin{align}
&  f_{zz}\left(  t\right)  =\sum_{\substack{n,l,m,\\n^{\prime},l^{\prime
},m^{\prime},\\n^{\prime\prime},l^{\prime\prime},m^{\prime\prime}%
}}\left\langle \psi_{n,l,m}\left\vert z\right\vert \psi_{n^{\prime\prime
},l^{\prime\prime},m^{\prime\prime}}\right\rangle \left\langle \psi
_{n^{\prime},l^{\prime},m^{\prime}}\left\vert z\right\vert \psi_{0}%
\right\rangle e^{it\left(  E_{0}-E_{n^{\prime\prime},l^{\prime\prime}}\right)
}\nonumber\\
&  \times\left\langle 0_{ph}\left\vert \left\langle \psi_{0}\left\vert
\mathrm{T}\exp\left(  i\int_{0}^{t}dsW\left(  -s\right)  \right)  \right\vert
\psi_{n,l,m}\right\rangle \right.  \right. \nonumber\\
&  \left.  \left.  \times\left\langle \psi_{n^{\prime\prime},l^{\prime\prime
},m^{\prime\prime}}\left\vert \mathrm{T}\exp\left(  -i\int_{0}^{t}dsW\left(
s\right)  \right)  \right\vert \psi_{n^{\prime},l^{\prime},m^{\prime}%
}\right\rangle \right\vert 0_{ph}\right\rangle . \label{fzz4}%
\end{align}

Within the adiabatic approximation, the optical conductivity is simplified.
The full details of the derivation are described in the Appendix A. First,
using the selection rules for the dipole matrix elements, the spherical
symmetry of the Hamiltonian $\tilde{H}$ and the adiabatic approximation, the
correlation function (\ref{fzz4}) is reduced to the form%
\begin{align}
f_{zz}\left(  t\right)   &  =\sum_{n}D_{n}e^{-i\Omega_{n,0}t}\nonumber\\
&  \times\left\langle \psi_{n,1,0}\left\vert \left\langle 0_{ph}\left\vert
\mathrm{T}\exp\left[  -i\int_{0}^{t}dsW\left(  s\right)  \right]  \right\vert
0_{ph}\right\rangle \right\vert \psi_{n,1,0}\right\rangle \label{fzz4a}%
\end{align}
where $\Omega_{n,0}$ is the FC transition frequency%
\begin{equation}
\Omega_{n,0}\equiv E_{n,1}-E_{0}, \label{Wn0}%
\end{equation}
and $D_{n}$ is the squared modulus of the dipole transition matrix element%
\begin{equation}
D_{n}=\left\vert \left\langle \psi_{0}\left\vert z\right\vert \psi
_{n,1,0}\right\rangle \right\vert ^{2}. \label{Dn}%
\end{equation}

Within the adiabatic approximation, the partial (with the electron wave
functions) averaging of the operator T-exponent in (\ref{fzz4a}) can be
exactly performed (see details in Appendix A). As a result, the optical
conductivity is transformed to the expression%
\begin{align}
\operatorname{Re}\sigma\left(  \omega\right)   &  =\frac{\omega}{6}\sum
_{n}D_{n}\int_{-\infty}^{\infty}e^{i\left(  \omega-\Omega_{n,0}\right)
t}\nonumber\\
&  \times\left\langle 0_{ph}\left\vert \mathrm{Tr}\left(  \mathrm{T}%
\exp\left[  -i\int_{0}^{t}ds\mathbb{W}^{\left(  n\right)  }\left(  s\right)
\right]  \right)  \right\vert 0_{ph}\right\rangle dt. \label{f-ad3}%
\end{align}
The T-exponent in (\ref{f-ad3}) contains the finite-dimensional matrix
$\mathbb{W}^{\left(  n\right)  }\left(  s\right)  $ depending on the phonon
coordinates:%
\begin{equation}
\left(  \mathbb{W}_{k,l,m}^{\left(  n\right)  }\right)  _{m_{1},m_{2}%
}=\left\langle \psi_{n,1,m_{1}}\left\vert W_{k,l,m}\right\vert \psi
_{n,1,m_{2}}\right\rangle \label{W1}%
\end{equation}
where $W_{k,l,m}$ are the amplitudes of the electron-phonon interaction in the
basis of spherical wave functions.

Because the kinetic energy of the phonons is of order $\alpha^{-4}$ compared
to the leading term of the Hamiltonian \cite{Allcock1}, we neglect this
kinetic energy in the present work, because the treatment is related to the
strong-coupling regime. As a result, $Q_{k,l,m}$ commute with the Hamiltonian
$\tilde{H}_{0},$ so that in (\ref{f-ad3}), $\mathbb{W}^{\left(  n\right)
}\left(  s\right)  =\mathbb{W}^{\left(  n\right)  }$. Furthermore, in a
finite-dimensional basis $\left\{  \left\vert \psi_{n,l,m}\right\rangle
\right\}  $ for a given level $\left(  n,l\right)  $, all eigenvalues of the
Hamiltonian $\tilde{H}_{0}$ are the same. Therefore the $\mathrm{T}$-exponent
entering (\ref{f-ad3}) in that finite-dimensional basis turns into a usual
exponent. As a result, the strong-coupling polaron optical conductivity
(\ref{f-ad3}) takes the form%
\begin{equation}
\operatorname{Re}\sigma\left(  \omega\right)  =\frac{\omega}{6}\sum_{n}%
D_{n}\int_{-\infty}^{\infty}e^{i\left(  \omega-\Omega_{n,0}\right)
t}\left\langle 0_{ph}\left\vert \mathrm{Tr}\exp\left(  -i\mathbb{W}^{\left(
n\right)  }t\right)  \right\vert 0_{ph}\right\rangle dt. \label{fzz1}%
\end{equation}

The matrix interaction Hamiltonian (\ref{W1}) depends on the phonon
coordinates, and the matrices $\mathbb{W}_{k,l,m}^{\left(  n\right)  }$ with
different $m$ for one and the same degenerate energy level do not commute with
each other. According to the Jahn -- Teller theorem \cite{JT}, for a
degenerate level there does not exist a unitary transformation which
simultaneously diagonalizes all matrices $\mathbb{W}_{k,l,m}^{\left(
n\right)  }$ in a basis that does not depend on the phonon coordinates. The
manifestations of that theorem are attributed to the Jahn -- Teller effect.
Therefore, because we neglect the non-commutation of the matrices
$\mathbb{W}_{k,l,m}^{\left(  n\right)  }$, the Jahn -- Teller effect is omitted.

In fact, neglecting the Jahn -- Teller effect is not necessary. The averaging
in Eq. (\ref{fzz1}) is performed exactly using the effective phonon modes
similarly to Ref. \cite{Lumin1998} (see the details in Appendix B). As a
result, we arrive at the following expression for the strong-coupling polaron
optical conductivity%
\begin{align}
\operatorname{Re}\sigma\left(  \omega\right)   &  =\frac{\omega}{3\pi^{2}}%
\sum_{n}\frac{D_{n}}{a_{0}^{\left(  n\right)  }}\int_{-\infty}^{\infty}%
dx_{0}\int_{-\infty}^{\infty}dx_{1}\int_{-\infty}^{\infty}dx_{2}\int_{-\infty
}^{\infty}dy_{1}\int_{-\infty}^{\infty}dy_{2}\nonumber\\
&  \times\sum_{j=1}^{3}\exp\left\{  -\frac{1}{2}\left[  x_{0}^{2}+\sum
_{m=1,2}\left(  x_{m}^{2}+y_{m}^{2}\right)  +\frac{\left(  \omega-\Omega
_{n,0}-\frac{a_{2}^{\left(  n\right)  }}{2\sqrt{5\pi}}\lambda_{j}\left(
Q_{2}\right)  \right)  ^{2}}{\left(  a_{0}^{\left(  n\right)  }\right)  ^{2}%
}\right]  \right\}  . \label{Resig}%
\end{align}
Here, $\lambda_{j}\left(  Q_{2}\right)  $ are the eigenvalues for the matrix
interaction Hamiltonian, which are explicitly determined in the Appendix B by
the formula (\ref{eigenv}). The coefficients $a_{0}^{\left(  n\right)  }$ and
$a_{2}^{\left(  n\right)  }$ are given by (\ref{c0n}) and (\ref{c2n}),
respectively. The polaron optical conductivity given by the expression
(\ref{Resig}), is in fact an envelope of the multiphonon polaron optical
conductivity band with the correlation function (\ref{f-ad3}) provided by the
phonon-assisted transitions from the polaron ground state to the polaron RES.
This result is consistent with Ref. \cite{KED1969}, where the same paradigm of
the phonon-assisted transitions to the polaron RES was exploited, but the
calculation was limited to the one-phonon transition.

In order to reveal the significance of the Jahn -- Teller effect for the
polaron, we alternatively calculate $\left\langle 0_{ph}\left\vert
\mathrm{Tr}\exp\left(  -i\mathbb{W}^{\left(  n\right)  }t\right)  \right\vert
0_{ph}\right\rangle $ neglecting the non-commutation of the matrices
$\mathbb{W}_{k,l,m}^{\left(  n\right)  }$, as described in the Appendix B.~2.
The resulting expression for the polaron optical conductivity is much simpler
than formula (\ref{Resig}) and is similar to the expression (3) of Ref.
\cite{DeFilippis2006}:%
\begin{equation}
\operatorname{Re}\sigma\left(  \omega\right)  =\omega\sum_{n}\sqrt{\frac{\pi
}{2\omega_{s}^{\left(  n\right)  }}}D_{n}\exp\left(  -\frac{\left(
\omega-\Omega_{n,0}\right)  ^{2}}{2\omega_{s}^{\left(  n\right)  }}\right)  ,
\label{SJT}%
\end{equation}
with the parameter (often called the Huang-Rhys factor)%
\begin{equation}
\omega_{s}^{\left(  n\right)  }=\frac{1}{2}\left(  a_{0}^{\left(  n\right)
}\right)  ^{2}+\frac{1}{4\pi}\left(  a_{2}^{\left(  n\right)  }\right)  ^{2}.
\label{S1a}%
\end{equation}

The strong-coupling electron energies and wave functions in Eq. (\ref{f-ad3})
can be calculated using different approximations. For example, within the
Landau-Pekar (LP) approximation \cite{LP48}, the trial wave function
$\left\vert \psi_{0}\right\rangle $ is chosen as the ground state of a 3D
oscillator. Within the Pekar approximation \cite{Pekar}, $\left\vert \psi
_{0}\right\rangle $ is chosen in the form
\begin{equation}
\left\vert \psi_{0}\left(  r\right)  \right\rangle =Ce^{-ar}\left(
1+ar+br^{2}\right)  \label{P1a}%
\end{equation}
with the variational parameters $a$ and $b$. Finally, the trial ground state
wave function can be determined numerically exactly following Miyake
\cite{M75} (see also \cite{Kleinert}, Chap. 5.22). Within the LP
approximation, formula (\ref{SJT}) reproduces the polaron optical conductivity
obtained in Ref. \cite{DeFilippis2006}.

In the LP approximation, the matrix elements $\left\langle \psi_{0}\left\vert
z\right\vert \psi_{n,1,0}\right\rangle $ are different from zero only for
$n=1$, i. e. only for the $1s\rightarrow2p$ transition. Beyond the LP
approximation, also the transitions to other excited states are allowed
because of the nonparabolicity of the self-consistent potential $V_{a}\left(
r\right)  $. The use of exact strong-coupling wave functions, instead of the
LP wave functions, may significantly influence the optical conductivity. In
the present treatment we use the numerically exact electron energies and wave
functions of both ground and first excited states according to Ref.
\cite{M75}. The FC transition energies $\Omega_{n,0}$ to leading order of the
strong-coupling approximation are determined according to (\ref{Wn0}). In
order to account for the corrections of the FC energy with accuracy up to
$\alpha^{0}$, we add to $\Omega_{n,0}$ the correction $\Delta\Omega
_{\mathrm{FC}}\approx-3.8$ from Ref. \cite{DeFilippis2006}. Because we use the
numerically accurate strong-coupling wave functions and energies corresponding
to Miyake \cite{M75}, the formula (\ref{fzz4}) \emph{is asymptotically exact
in the strong-coupling limit, at least in its leading term in powers of}
$\alpha^{-2}$.

\subsection{Results and discussion}

In Figs. 2 to 3, we have plotted the polaron optical conductivity spectra
calculated for different values of the coupling constant $\alpha$. The optical
conductivity spectra calculated within the present strong-coupling approach
taking into account the Jahn -- Teller effect are shown by the solid curves.
The optical conductivity derived neglecting the Jahn -- Teller effect is shown
by the dashed curves. It is worth mentioning that there is little difference
in the optical conductivity spectra between those calculated with and without
the Jahn -- Teller effect. The optical conductivity obtained in Ref.
\cite{DeFilippis2006} with the Landau-Pekar (LP) adiabatic approximation is
plotted with dash-dotted curves. The full dots show the numerical Diagrammatic
Quantum Monte Carlo (DQMC) data \cite{Mishchenko2003,DeFilippis2006}. The FC
transition frequency for the transition to the first excited FC state
$\Omega_{1,0}\equiv\Omega_{\mathrm{FC}}$ and the RES transition frequency
$\Omega_{\mathrm{RES}}$ are explicitly indicated in the figures.

The polaron optical conductivity spectra calculated within the present
strong-coupling approach are shifted to lower frequencies with respect to the
optical conductivity spectra calculated within the LP approximation of Ref.
\cite{DeFilippis2006}. This shift is due to the use of the numerically
accurate strong coupling energy levels and wave functions of the internal
polaron states, and of the numerically accurate self-consistent adiabatic
polaron potential.

According to the selection rules for the matrix elements of the
electron-phonon interaction, there is a contribution to the polaron optical
conductivity from the phonon modes with angular momentum $l=0$ ($s$-phonons)
and with angular momentum $l=2$ ($d$-phonons). The $s$-phonons are fully
symmetric, therefore they do not contribute to the Jahn -- Teller effect,
while the $d$-phonons are active in the Jahn -- Teller effect. The
contribution of the $d$-phonons to the optical conductivity spectra is not
small compared to the contribution of the $s$-phonons. However, the
distinction between the optical conductivity spectra calculated with and
without the Jahn -- Teller effect is relatively small.

For $\alpha=8$ and $\alpha=8.5$, the maxima of the polaron optical
conductivity spectra, calculated within the present strong-coupling approach
are positioned to the low frequency side of the maxima of those calculated
using the DQMC method. The agreement between our strong-coupling polaron
optical conductivity spectra and the numerical DQMC data improves with
increasing alpha. This is in accordance with the fact that the present
strong-coupling approach for the polaron optical conductivity is
asymptotically exact in the strong-coupling limit.

The total polaron optical conductivity must satisfy the sum rule
\cite{DLR1977}%
\begin{equation}
\int_{0}^{\infty}\operatorname{Re}\sigma\left(  \omega\right)  d\omega
=\frac{\pi}{2}. \label{sr1a}%
\end{equation}
In the weak- and intermediate-coupling regimes at $T=0$, there are two
contributions to the left-hand side of that sum rule: (1) the contribution
from the polaron optical conductivity for $\omega>\omega_{\mathrm{LO}}$ and
(2) the contribution from the \textquotedblleft central peak\textquotedblright%
\ at $\omega=0$, which is proportional to the inverse polaron mass
\cite{DLR1977}. In the asymptotic strong-coupling regime, the inverse to the
polaron mass is of order $\alpha^{-4}$, and hence the contribution from the
\textquotedblleft central peak\textquotedblright\ to the polaron optical
conductivity is beyond the accuracy of the present approximation (where we
keep the terms $\propto\alpha^{-2}$ and $\propto\alpha^{0}$).

As discussed above, in the present work the transitions from the ground state
to the states of the continuous part of the polaron energy spectrum are
neglected. Therefore the integral over the frequency [the left-hand side of
(\ref{sr1a})] for the optical conductivity calculated within the present
strong-coupling approximation can be (relatively slightly) smaller than
$\pi/2$. The relative contribution of the transitions to the continuous part
of the polaron spectrum, $\Delta_{c}$, can be therefore estimated as%
\begin{equation}
\Delta_{c}\equiv1-\frac{2}{\pi}\int_{0}^{\infty}\operatorname{Re}\sigma\left(
\omega\right)  d\omega, \label{Delta}%
\end{equation}
where the right-hand side is obtained by a numerical integration of
$\operatorname{Re}\sigma\left(  \omega\right)  $ calculated within the present
strong-coupling approach. This numeric estimation shows that for $\alpha>8$,
$\Delta_{c}<0.01$. Moreover, with increasing $\alpha$, the relative
contribution of the transitions to the continuous part of the polaron spectrum
falls down. This confirms the accuracy of the present strong-coupling approach.

In Refs. \cite{Emin1993,Myasnikov2006}, the optical conductivity of a
strong-coupling polaron was calculated assuming that in the strong-coupling
regime the polaron optical response is provided mainly by the transitions to
the continuous part of the spectrum (these transitions are called there
\textquotedblleft the polaron dissociation\textquotedblright). This concept is
in contradiction both with the early estimation by Pekar \cite{Pekar}
discussed above and with the very small weight of those transitions shown in
Fig. 4. The approach of Ref. \cite{Emin1993} in fact takes into account only a
small part of the strong-coupling polaron optical conductivity -- namely, the
high-frequency \textquotedblleft tail\textquotedblright\ of the optical
conductivity spectrum.

When comparing the polaron optical conductivity spectra calculated in the
present work with the DQMC data \cite{Mishchenko2003,DeFilippis2006}, we can
see that the present approach, with respect to DQMC, underestimates the
high-frequency part of the polaron optical conductivity. This difference,
however, gradually diminishes with increasing $\alpha$, in accordance with the
fact that the present method is an asymptotic strong-coupling approximation.

Because the optical conductivity spectra calculated in the present
strong-coupling approximation using the expressions (\ref{Resig}) and
(\ref{SJT}) represent the envelopes of the RES peak with the multi-phonon
satellites, the separate peeks are not explicitly seen in those spectra. The
FC and RES peaks are indicated in the figures by the arrows. The FC transition
frequency $\Omega_{1,0}$ in the strong-coupling case is positioned close to
the maximum of the polaron optical conductivity band (both calculated within
the present approach and within DQMC). The RES transition frequency is
positioned one $\omega_{\mathrm{LO}}$ below the onset of the LO-sidebands.
Note that the strong-coupling polaron optical conductivity derived in Refs.
\cite{Spohn1987} contains only the zero-phonon (RES) line and no phonon
satellites at all. In contrast, in the present calculation, the maximum of the
polaron optical conductivity spectrum shifts to higher frequencies with
increasing $\alpha$, so that the multiphonon processes invoking large number
of phonons become more and more important, in accordance with predictions of
Refs. \cite{KED1969,DSG1972}.

It is worth noting the following important point: the maximum of the polaron
optical conductivity band can be hardly interpreted as a broadened transition
to an FC state on the following reasons.\ Formula (\ref{f-ad3}) describes a
set of multi-phonon peaks. In the simplifying approximation which neglects the
Jahn -- Teller effect (see Ref. \cite{DeFilippis2006}), those peaks are
positioned at the frequencies $\omega=\tilde{\Omega}_{n,0}+k$, where $k$ is
the number of emitted phonons and is the frequency of the zero-phonon line.
The frequencies $\tilde{\Omega}_{n,0}$ do not coincide with the FC transition
frequencies but are determined by%
\begin{equation}
\tilde{\Omega}_{n,0}=\Omega_{n,0}-\omega_{s}^{\left(  n\right)  },
\end{equation}
where the Huang-Rhys factor $\omega_{s}^{\left(  n\right)  }$ describes the
energy shift due to lattice relaxation. The physical meaning of the parameters
$\omega_{s}^{\left(  n\right)  }$ obviously implies that the peaks at
$\omega=\tilde{\Omega}_{n,0}+k$ should be attributed to transitions to the RES
with emission of $k$ phonons. So, the so-called \textquotedblleft FC
transition\textquotedblright\ is realized as the envelope of a series of
phonon sidebands of the polaron RES but not as a transition to the FC state.
The account of the Jahn-Teller effects in general makes the multiphonon peak
series non-equidistant, but it changes nothing in the concept of the internal
polaron states which is discussed above.

\subsection{Conclusions}

We have derived the polaron optical conductivity which is asymptotically exact
in the strong-coupling limit. The strong-coupling polaron optical conductivity
band is provided by the multiphonon transitions from the polaron ground state
to the polaron RES and has the maximum positioned close to the FC transition
frequency. With increasing the electron-phonon coupling constant $\alpha$, the
polaron optical conductivity band shape gradually tends to that provided by
the Diagrammatic Quantum Monte Carlo (DQMC) method. This agreement
demonstrates the importance of the multiphonon processes for the polaron
optical conductivity in the strong-coupling regime.

The obtained polaron optical conductivity with a high accuracy satisfies the
sum rule \cite{DLR1977}, what gives us an evidence of the fact that in the
strong-coupling regime the dominating contribution to the polaron optical
conductivity is due to the transitions to the \emph{internal} polaron states,
while the contribution due to the transitions to the continuum states is
negligibly small.

Accurate numerical results, obtained using DQMC method \cite{Mishchenko2003},
-- modulo the linewidths for sufficiently large $\alpha$ -- and the
analytically exact in the strong-coupling limit polaron optical conductivity
of the present work, as well as the analytical approximation of Ref.
\cite{DeFilippis2006} confirm the essence of the mechanism for the optical
absorption of Fr\"{o}hlich polarons, which were proposed in Refs.
\cite{DSG1972,KartheuserGreenbook}.

\subsection{Appendix 1. Correlation function}

The dipole-dipole correlation function $f_{zz}\left(  t\right)  $ given by
(\ref{fzz4}) is further simplified within the adiabatic approximation and
using the selection rules for the dipole transition matrix elements and the
symmetry properties of the polaron Hamiltonian. First, according to the
selection rules, the matrix element\textrm{ }$\left\langle \psi_{0}\left\vert
z\right\vert \psi_{n,l,m}\right\rangle \ $is
\begin{equation}
\left\langle \psi_{n^{\prime},l^{\prime},m^{\prime}}\left\vert z\right\vert
\psi_{0}\right\rangle =\delta_{l^{\prime},1}\delta_{m^{\prime},0}\left\langle
\psi_{n^{\prime},1,0}\left\vert z\right\vert \psi_{0}\right\rangle
\label{B1aa}%
\end{equation}
\textrm{ }

Second, the interaction Hamiltonian $W$ (and hence, also the evolution
operator which involves $W$) is a scalar of the rotation symmetry group. The
matrix elements $\left\langle \psi_{n,l,m}\left\vert W\left(  s\right)
\right\vert \psi_{n,l^{\prime},m^{\prime}}\right\rangle $ for $l\neq
l^{\prime}$ and $m\neq m^{\prime}$ are then exactly equal to zero. Therefore,
in the adiabatic approximation and due to the symmetry of the Hamiltonian
$\tilde{H}$, we obtain the relations%
\begin{align}
&  \left\langle \psi_{0}\left\vert \mathrm{T}\exp\left(  i\int_{0}%
^{t}dsW\left(  -s\right)  \right)  \right\vert \psi_{n,l,m}\right\rangle
\nonumber\\
&  \approx\delta_{n,0}\delta_{l,0}\delta_{m,0}\left\langle \psi_{0}\left\vert
\mathrm{T}\exp\left(  -i\int_{0}^{t}dsW\left(  s\right)  \right)  \right\vert
\psi_{0}\right\rangle , \label{b2}%
\end{align}%
\begin{align}
&  \left\langle \psi_{n^{\prime\prime},l^{\prime\prime},m^{\prime\prime}%
}\left\vert \mathrm{T}\exp\left(  -i\int_{0}^{t}dsW\left(  s\right)  \right)
\right\vert \psi_{n^{\prime},l^{\prime},m^{\prime}}\right\rangle \nonumber\\
&  \approx\delta_{n^{\prime\prime},n^{\prime}}\delta_{l^{\prime\prime
},l^{\prime}}\left\langle \psi_{n^{\prime},l^{\prime},m^{\prime}}\left\vert
\mathrm{T}\exp\left(  -i\int_{0}^{t}dsW\left(  s\right)  \right)  \right\vert
\psi_{n^{\prime},l^{\prime},m^{\prime}}\right\rangle . \label{b3}%
\end{align}
Furthermore, because the ground state $\psi_{0}$ is non-degenerate, we find
that
\[
\left\langle \psi_{0}\left\vert \mathrm{T}\exp\left(  -i\int_{0}^{t}dsW\left(
s\right)  \right)  \right\vert \psi_{0}\right\rangle \approx1,
\]
because within the adiabatic approximation, for any $n\geq1$ the averages
$\left\langle \psi_{0}\left\vert W^{n}\right\vert \psi_{0}\right\rangle =0$.

The correlation function (\ref{fzz4}) using (\ref{B1aa}) to (\ref{b3}) takes
the form%
\begin{align}
f_{zz}\left(  t\right)   &  =\sum_{n}D_{n}e^{-i\Omega_{n,0}t}\nonumber\\
&  \times\left\langle \psi_{n,1,0}\left\vert \left\langle 0_{ph}\left\vert
\mathrm{T}\exp\left[  -i\int_{0}^{t}dsW\left(  s\right)  \right]  \right\vert
0_{ph}\right\rangle \right\vert \psi_{n,1,0}\right\rangle \label{f-ad}%
\end{align}
with the squared matrix elements of the dipole transitions%
\begin{equation}
D_{n}\equiv\left\vert \left\langle \psi_{n,1,0}\left\vert z\right\vert
\psi_{0}\right\rangle \right\vert ^{2}=\frac{1}{3}\left(  \int_{0}^{\infty
}R_{n,1}\left(  r\right)  R_{0,0}\left(  r\right)  r^{3}dr\right)  ^{2},
\label{dd}%
\end{equation}
and the FC transition frequencies%
\begin{equation}
\Omega_{n,0}\equiv E_{n,1}-E_{0}. \label{FC}%
\end{equation}
Further on, the interaction Hamiltonian is expressed in terms of the complex
phonon coordinates $Q_{\mathbf{k}}$:%
\begin{equation}
W=\sqrt{2}\sum_{\mathbf{k}}W_{\mathbf{k}}Q_{\mathbf{k}},\quad Q_{\mathbf{k}%
}=\frac{b_{\mathbf{k}}+b_{-\mathbf{k}}^{+}}{\sqrt{2}} \label{W3}%
\end{equation}
Here, we use the spherical-wave basis for phonon modes:%
\begin{equation}
\varphi_{k,l,m}\left(  \mathbf{r}\right)  \equiv\left(  -1\right)
^{\frac{m-\left\vert m\right\vert }{2}}\phi_{k,l}\left(  r\right)
Y_{l,m}\left(  \theta,\varphi\right)  ,
\end{equation}
where the radial part of the basis function is expressed through the spherical
Bessel function $j_{l}\left(  kr\right)  $:
\begin{equation}
\phi_{k,l}\left(  r\right)  =\left(  \frac{2}{R}\right)  ^{1/2}k\ j_{l}\left(
kr\right)  ,\;R=\left(  \frac{3V}{4\pi}\right)  ^{1/3}.
\end{equation}
The factor $\left(  -1\right)  ^{\frac{m-\left\vert m\right\vert }{2}}$ is
chosen in order to fulfil the symmetry property%
\[
\varphi_{k,l,m}^{\ast}\left(  \mathbf{r}\right)  =\varphi_{k,l,-m}\left(
\mathbf{r}\right)  .
\]
In the spherical-wave basis, the interaction Hamiltonian is%
\begin{equation}
W=\sqrt{2}\sum_{k,l,m}W_{k,l,m}Q_{k,l,m}, \label{W4}%
\end{equation}
with the complex phonon coordinates%
\begin{equation}
Q_{k,l,m}=\frac{b_{k,l,m}+b_{k,l,-m}^{+}}{\sqrt{2}} \label{Q}%
\end{equation}
and with the interaction amplitudes%
\begin{equation}
W_{k,l,m}=\frac{\sqrt{2\sqrt{2}\pi\alpha}}{k}\left(  \varphi_{k,l,m}\left(
\mathbf{r}\right)  -\rho_{k,l,m}\right)  ,\; \rho_{k,l,m}\equiv\left\langle
\psi_{0}\left\vert \varphi_{k,l,m}\right\vert \psi_{0}\right\rangle .
\end{equation}
The dipole-dipole correlation function (\ref{f-ad}) is then%
\begin{align}
f_{zz}\left(  t\right)   &  =\sum_{n}D_{n}e^{-i\Omega_{n,0}t}\nonumber\\
&  \times\left\langle \psi_{n,1,0}\left\vert \left\langle 0_{ph}\left\vert
\mathrm{T}\exp\left[  -i\sqrt{2}\int_{0}^{t}ds\sum_{k,l,m}W_{k,l,m}\left(
s\right)  Q_{k,l,m}\left(  s\right)  \right]  \right\vert 0_{ph}\right\rangle
\right\vert \psi_{n,1,0}\right\rangle . \label{f-ad1}%
\end{align}
The operators $W_{k,l,m}\left(  s\right)  $ in (\ref{f-ad1}) are equivalent to
the $\left(  2l+1\right)  $-dimensional matrices $\mathbb{W}_{k,l,m}^{\left(
n\right)  }$ determined in the basis of the level $\left(  n,l\right)  $. The
matrix elements of these matrices are%
\begin{equation}
\left(  \mathbb{W}_{k,l,m}^{\left(  n\right)  }\right)  _{m_{1},m_{2}%
}=\left\langle \psi_{n,1,m_{1}}\left\vert W_{k,l,m}\right\vert \psi
_{n,1,m_{2}}\right\rangle . \label{matr}%
\end{equation}
In these notations, $f_{zz}\left(  t\right)  $ given by (\ref{f-ad1}) can be
written down as%
\begin{equation}
f_{zz}\left(  t\right)  =\sum_{n}D_{n}e^{-i\Omega_{n,0}t}\left\langle
0_{ph}\left\vert \left(  \mathrm{T}\exp\left[  -i\int_{0}^{t}ds\mathbb{W}%
^{\left(  n\right)  }\left(  s\right)  \right]  \right)  _{0,0}\right\vert
0_{ph}\right\rangle . \label{f-ad2}%
\end{equation}
where $\mathbb{W}^{\left(  n\right)  }$ is the matrix electron-phonon
interaction Hamiltonian expressed through the phonon complex coordinates in
the spherical-wave representation as follows:%
\begin{equation}
\mathbb{W}^{\left(  n\right)  }=\sqrt{2}\sum_{k,l,m}\mathbb{W}_{k,l,m}%
^{\left(  n\right)  }Q_{k,l,m}. \label{WM}%
\end{equation}
Here, $\mathbb{W}_{k,l,m}^{\left(  n\right)  }$ is a $\left(  3\times3\right)
$ matrix in a basis of a level $\left(  n,l\right)  _{l=1}$ of the Hamiltonian
$\tilde{H}_{0}$.

Because $\mathbb{W}^{\left(  n\right)  }$ is a scalar of the rotation group,
we can replace the diagonal matrix element of the T-exponent in (\ref{f-ad2})
with the trace in the aforesaid-finite-dimensional basis. As a result, we
obtain for the polaron optical conductivity (\ref{KuboDD0}) with (\ref{f-ad2})
the expression%
\begin{align}
\operatorname{Re}\sigma\left(  \omega\right)   &  =\frac{\omega}{6}\sum
_{n}D_{n}\int_{-\infty}^{\infty}e^{i\left(  \omega-\Omega_{n,0}\right)
t}\nonumber\\
&  \times\left\langle 0_{ph}\left\vert \mathrm{Tr}\left(  \mathrm{T}%
\exp\left[  -i\int_{0}^{t}ds\mathbb{W}^{\left(  n\right)  }\left(  s\right)
\right]  \right)  \right\vert 0_{ph}\right\rangle dt. \label{f-ad3a}%
\end{align}

\subsection{Appendix 2. Effective phonon modes}

In order to perform the averaging in Eq. (\ref{fzz1}) analytically, we
introduce the effective phonon modes $Q_{0,0}$ and $Q_{2,m}$ similarly to Ref.
\cite{Lumin1998}. The Hamiltonian $\mathbb{W}^{\left(  n\right)  }$ in terms
of these effective phonon modes is expressed as%
\begin{equation}
\mathbb{W}^{\left(  n\right)  }=\sqrt{2}\sum_{l,m}\mathbb{\tilde{W}}%
_{l,m}^{\left(  n\right)  }Q_{l,m} \label{W2a}%
\end{equation}
where the matrices $\mathbb{\tilde{W}}_{l,m}^{\left(  n\right)  }$ (depending
on the vibration coordinates $Q_{l,m}$) are explicitly given by the
expressions (cf. Ref. \cite{Lumin1998}),%
\begin{equation}
\mathbb{W}^{\left(  n\right)  }=a_{0}^{\left(  n\right)  }\mathbb{I}%
Q_{0,0}+a_{2}^{\left(  n\right)  }\sum_{m=-2}^{2}\mathbb{B}_{m}Q_{2,m}
\label{W2}%
\end{equation}
with the matrices $\mathbb{B}_{j}$
\begin{equation}
\mathbb{B}_{0}=\frac{1}{2\sqrt{5\pi}}\left(
\begin{array}
[c]{ccc}%
-1 & 0 & 0\\
0 & 2 & 0\\
0 & 0 & -1
\end{array}
\right)  , \label{B0}%
\end{equation}%
\begin{equation}
\mathbb{B}_{1}=\mathbb{B}_{-1}^{+}=\frac{1}{2}\sqrt{\frac{3}{5\pi}}\left(
\begin{array}
[c]{ccc}%
0 & 0 & 0\\
-1 & 0 & 0\\
0 & 1 & 0
\end{array}
\right)  , \label{B1a}%
\end{equation}%
\begin{equation}
\mathbb{B}_{2}=\mathbb{B}_{-2}^{+}=\sqrt{\frac{3}{10\pi}}\left(
\begin{array}
[c]{ccc}%
0 & 0 & 0\\
0 & 0 & 0\\
-1 & 0 & 0
\end{array}
\right)  . \label{B2a}%
\end{equation}
The coefficients $a_{0}^{\left(  n\right)  }$ and $a_{2}^{\left(  n\right)  }$
in Eq. (\ref{W2}) are%
\begin{align}
a_{0}^{\left(  n\right)  }  &  =\left(  \sqrt{2}\alpha\sum_{k}\frac{1}{k^{2}%
}\left[  \left\langle \phi_{k,0}\right\rangle _{n,1}-\left\langle \phi
_{k,0}\right\rangle _{0,0}\right]  ^{2}\right)  ^{1/2},\label{c0}\\
a_{2}^{\left(  n\right)  }  &  =\left(  4\sqrt{2}\pi\alpha\sum_{k}\frac
{1}{k^{2}}\left\langle \phi_{k,2}\right\rangle _{n,1}^{2}\right)  ^{1/2}.
\label{c2}%
\end{align}
Here $\phi_{k,l}$ is the radial part of the basis function expressed through
the spherical Bessel function $j_{l}\left(  kr\right)  $:
\begin{equation}
\phi_{k,l}\left(  r\right)  =\left(  \frac{2}{R}\right)  ^{1/2}k\ j_{l}\left(
kr\right)  ,\;R=\left(  \frac{3V}{4\pi}\right)  ^{1/3}, \label{Rad}%
\end{equation}
$V$ is the volume of the crystal, and $\left\langle f\left(  r\right)
\right\rangle _{n,l}$ is the average%
\begin{equation}
\left\langle f\left(  r\right)  \right\rangle _{n,l}=\int_{0}^{\infty}f\left(
r\right)  R_{n,l}^{2}\left(  r\right)  r^{2}dr.
\end{equation}
The normalization of the phonon wave functions corresponds to the condition%
\begin{equation}
\int_{0}^{R}\phi_{k,l}\left(  r\right)  \phi_{k^{\prime},l}\left(  r\right)
r^{2}dr=\delta_{k,k^{\prime}}. \label{norm}%
\end{equation}
After the straightforward calculation using (\ref{norm}), we express the
coefficients $a_{0}^{\left(  n\right)  }$ and $a_{2}^{\left(  n\right)  }$
through the integrals with the radial wave functions:%
\begin{align}
a_{0}^{\left(  n\right)  }  &  =\left(  2\sqrt{2}\alpha\int_{0}^{\infty}%
dr\int_{0}^{r}dr^{\prime}\ r\left(  r^{\prime}\right)  ^{2}\left[  R_{n,1}%
^{2}\left(  r\right)  -R_{0,0}^{2}\left(  r\right)  \right]  \left[
R_{n,1}^{2}\left(  r^{\prime}\right)  -R_{0,0}^{2}\left(  r^{\prime}\right)
\right]  \right)  ^{1/2},\label{c0n}\\
a_{2}^{\left(  n\right)  }  &  =\left(  \frac{8\sqrt{2}\pi\alpha}{5}\int
_{0}^{\infty}dr\int_{0}^{r}dr^{\prime}\frac{\left(  r^{\prime}\right)  ^{4}%
}{r}R_{n,1}^{2}\left(  r\right)  R_{n,1}^{2}\left(  r^{\prime}\right)
\right)  ^{1/2}. \label{c2n}%
\end{align}

\subsubsection{Exact averaging}

Let us substitute the matrix interaction Hamiltonian (\ref{W2}) to the
dipole-dipole correlation function (\ref{fzz1}), what gives us the result%
\begin{equation}
f_{zz}\left(  t\right)  =\frac{1}{3}\sum_{n}D_{n}e^{-i\Omega_{n,0}%
t}\left\langle 0_{ph}\left\vert \exp\left(  -ita_{0}^{\left(  n\right)  }%
Q_{0}\right)  \mathrm{Tr}\exp\left(  -it\frac{a_{2}^{\left(  n\right)  }%
}{2\sqrt{5\pi}}\mathbb{V}\left(  Q_{2}\right)  \right)  \right\vert
0_{ph}\right\rangle . \label{fzz2a}%
\end{equation}
Here, we use the matrix depending on the phonon coordinates,%
\begin{equation}
\mathbb{V}\left(  Q_{2}\right)  \equiv2\sqrt{5\pi}\sum_{m=-2}^{2}%
\mathbb{B}_{m}Q_{2m}, \label{VQ}%
\end{equation}
whose explicit form is%
\begin{equation}
\mathbb{V}\left(  Q_{2}\right)  =\left(
\begin{array}
[c]{ccc}%
-Q_{2,0} & -\sqrt{3}Q_{2,-1} & -\sqrt{6}Q_{2,-2}\\
-\sqrt{3}Q_{2,1} & 2Q_{2,0} & \sqrt{3}Q_{2,-1}\\
-\sqrt{6}Q_{2,2} & \sqrt{3}Q_{2,1} & -Q_{2,0}%
\end{array}
\right)  . \label{MatrA}%
\end{equation}

The matrix $\mathbb{V}\left(  Q_{2}\right)  $ is analytically diagonalized.
The equation for the eigenvectors $\left\vert \chi\left(  Q_{2}\right)
\right\rangle $ and eigenvalues $\lambda\left(  Q_{2}\right)  $ of
$\mathbb{V}\left(  Q_{2}\right)  $ is%
\begin{equation}
\mathbb{V}\left(  Q_{2}\right)  \left\vert \chi\left(  Q_{2}\right)
\right\rangle =\lambda\left(  Q_{2}\right)  \left\vert \chi\left(
Q_{2}\right)  \right\rangle . \label{SL}%
\end{equation}
The eigenvalues are found from the equation%
\begin{equation}
\det\left(  \mathbb{V}\left(  Q_{2}\right)  -\lambda\left(  Q_{2}\right)
\mathbb{I}\right)  =0. \label{SE}%
\end{equation}
We make the transformation to the real phonon coordinates,
\begin{align*}
Q_{2,0}  &  \equiv x_{0},\\
Q_{2,m}  &  \equiv\frac{x_{m}+iy_{m}}{\sqrt{2}},\quad Q_{2,-m}=Q_{2,m}^{\ast
}=\frac{x_{m}-iy_{m}}{\sqrt{2}}.
\end{align*}
Five variables $x_{0},x_{1},x_{2},y_{1},y_{2}$ are the independent real phonon
coordinates. The l.h.s. of Eq. (\ref{SE}) is expressed in terms of these
coordinates as
\begin{equation}
\det\left(  \mathbb{V}\left(  Q_{2}\right)  -\lambda\left(  Q_{2}\right)
\mathbb{I}\right)  =-\lambda^{3}+3p\lambda+2q
\end{equation}
with the coefficients
\begin{align*}
p  &  =x_{0}^{2}+x_{1}^{2}+x_{2}^{2}+y_{1}^{2}+y_{2}^{2},\\
q  &  =x_{0}^{3}+\frac{3}{2}x_{0}\left(  x_{1}^{2}+y_{1}^{2}\right)
+\frac{3\sqrt{3}}{2}x_{2}\left(  x_{1}^{2}-y_{1}^{2}\right)  -3x_{0}\left(
x_{2}^{2}+y_{2}^{2}\right)  +3\sqrt{3}x_{1}y_{1}y_{2}.
\end{align*}
So, we have the cubic equation for $\lambda$:%
\begin{equation}
\lambda^{3}-3p\lambda-2q=0. \label{cub}%
\end{equation}
Because the matrix $\mathbb{V}\left(  Q_{2}\right)  $ is Hermitian, all its
eigenvalues are real. Therefore, $\frac{\left\vert q\right\vert }{p^{3/2}}%
\leq1$ (otherwise, $\sin\left(  3\varphi\right)  $ is not real). Herefrom, we
have three explicit eigenvalues:%
\begin{align}
\lambda_{1}\left(  Q_{2}\right)   &  =2\sqrt{p}\sin\left[  \frac{\pi}{3}%
+\frac{1}{3}\arcsin\left(  \frac{q}{p^{3/2}}\right)  \right]  ,\nonumber\\
\lambda_{2}\left(  Q_{2}\right)   &  =-2\sqrt{p}\sin\left[  \frac{1}{3}%
\arcsin\left(  \frac{q}{p^{3/2}}\right)  \right]  ,\nonumber\\
\lambda_{3}\left(  Q_{2}\right)   &  =-2\sqrt{p}\sin\left[  \frac{\pi}%
{3}-\frac{1}{3}\arcsin\left(  \frac{q}{p^{3/2}}\right)  \right]  .
\label{eigenv}%
\end{align}

The trace in (\ref{fzz2a}) is invariant with respect to the choice of the
basis. Consequently, after the diagonalization $f_{zz}\left(  t\right)  $
takes the form%
\begin{equation}
f_{zz}\left(  t\right)  =\frac{1}{3}\sum_{n}D_{n}e^{-i\Omega_{n,0}t}\sum
_{j=1}^{3}\left\langle 0_{ph}\left\vert \exp\left(  -it\left[  a_{0}^{\left(
n\right)  }Q_{0}+\frac{a_{2}^{\left(  n\right)  }}{2\sqrt{5\pi}}\lambda
_{j}\left(  Q_{2}\right)  \right]  \right)  \right\vert 0_{ph}\right\rangle .
\label{fzz3b}%
\end{equation}
After inserting $f_{zz}\left(  t\right)  $ given by (\ref{fzz3b}) into
(\ref{KuboDD0}), the integration over time gives the delta function multiplied
by $2\pi$, and we arrive at the result%
\begin{equation}
\operatorname{Re}\sigma\left(  \omega\right)  =\frac{\pi\omega}{3}\sum
_{n}D_{n}\sum_{j=1}^{3}\left\langle 0_{ph}\left\vert \delta\left(
\omega-\Omega_{n,0}-a_{0}^{\left(  n\right)  }Q_{0}-\frac{a_{2}^{\left(
n\right)  }}{2\sqrt{5\pi}}\lambda_{j}\left(  Q_{2}\right)  \right)
\right\vert 0_{ph}\right\rangle . \label{fzz5}%
\end{equation}
The ground-state wave function for the effective phonon modes is%
\begin{equation}
\left\vert 0_{ph}\right\rangle \equiv\Phi_{0}\left(  Q\right)  =\Phi
_{0}^{\left(  0\right)  }\left(  Q_{0}\right)  \Phi_{0}^{\left(  2\right)
}\left(  Q_{2}\right)  . \label{0ph}%
\end{equation}
$\Phi_{0}^{\left(  0\right)  }\left(  Q_{0}\right)  $ is the one-oscillator
ground-state wave function:
\begin{equation}
\Phi_{0}^{\left(  0\right)  }\left(  Q_{0}\right)  =\pi^{-1/4}\exp\left(
-\frac{Q_{0}^{2}}{2}\right)  .
\end{equation}
The ground-state wave function of phonons with $l=2$ is:
\begin{equation}
\Phi_{0}^{\left(  2\right)  }\left(  Q_{2}\right)  =\pi^{-5/4}\exp\left[
-\frac{1}{2}\left(  x_{0}^{2}+\sum_{m=1,2}\left(  x_{m}^{2}+y_{m}^{2}\right)
\right)  \right]  .
\end{equation}
The phonon ground-state wave function (\ref{0ph}) is then%
\begin{equation}
\Phi_{0}\left(  Q\right)  =\frac{1}{\pi^{3/2}}\exp\left[  -\frac{1}{2}\left(
x_{0}^{2}+\sum_{m=1,2}\left(  x_{m}^{2}+y_{m}^{2}\right)  +Q_{0}^{2}\right)
\right]  .
\end{equation}
With these phonon wave functions, Eq. (\ref{fzz5}) results in the following
expression for the polaron optical conductivity
\begin{align}
\operatorname{Re}\sigma\left(  \omega\right)   &  =\frac{\omega}{3\pi^{2}}%
\sum_{n}\frac{D_{n}}{a_{0}^{\left(  n\right)  }}\int_{-\infty}^{\infty}%
dx_{0}\int_{-\infty}^{\infty}dx_{1}\int_{-\infty}^{\infty}dx_{2}\int_{-\infty
}^{\infty}dy_{1}\int_{-\infty}^{\infty}dy_{2}\nonumber\\
&  \times\sum_{j=1}^{3}\exp\left\{  -\frac{1}{2}\left[  x_{0}^{2}+\sum
_{m=1,2}\left(  x_{m}^{2}+y_{m}^{2}\right)  +\frac{\left(  \omega-\Omega
_{n,0}-\frac{a_{2}^{\left(  n\right)  }}{2\sqrt{5\pi}}\lambda_{j}\left(
Q_{2}\right)  \right)  ^{2}}{\left(  a_{0}^{\left(  n\right)  }\right)  ^{2}%
}\right]  \right\}  . \label{Resig1}%
\end{align}

\subsubsection{Averaging neglecting the Jahn-Teller effect}

In order to perform the phonon averaging explicitly, we disentangle the
exponent $\exp\left(  -it\sqrt{2}\sum_{l,m}\mathbb{\tilde{W}}_{l,m}^{\left(
n\right)  }Q_{l,m}\right)  $ as follows.%
\begin{align}
&  \exp\left(  -it\sqrt{2}\sum_{l,m}\mathbb{\tilde{W}}_{l,m}^{\left(
n\right)  }Q_{l,m}\right)  =\exp\left(  -it\sum_{l,m}\mathbb{\tilde{W}}%
_{l,-m}^{\left(  n\right)  }b_{l,m}^{+}\right) \nonumber\\
&  \times\mathrm{T}\exp\left(  -i\int_{0}^{t}ds\sum_{l,m}e^{is\sum_{l^{\prime
},m^{\prime}}\mathbb{\tilde{W}}_{l^{\prime},-m^{\prime}}^{\left(  n\right)
}b_{l^{\prime},m^{\prime}}^{+}}\mathbb{\tilde{W}}_{l,m}^{\left(  n\right)
}b_{l,m}e^{-is\sum_{l^{\prime},m^{\prime}}\mathbb{\tilde{W}}_{l^{\prime
},-m^{\prime}}^{\left(  n\right)  }b_{l^{\prime},m^{\prime}}^{+}}\right)  .
\end{align}

Neglecting non-commutation of matrices $\mathbb{\tilde{W}}_{l,m}^{\left(
n\right)  }$ we find that%
\begin{align}
&  \sum_{l,m}e^{is\sum_{l^{\prime},m^{\prime}}\mathbb{\tilde{W}}_{l^{\prime
},-m^{\prime}}^{\left(  n\right)  }b_{l^{\prime},m^{\prime}}^{+}%
}\mathbb{\tilde{W}}_{l,m}^{\left(  n\right)  }b_{l,m}e^{-is\sum_{l^{\prime
},m^{\prime}}\mathbb{\tilde{W}}_{l^{\prime},-m^{\prime}}^{\left(  n\right)
}b_{l^{\prime},m^{\prime}}^{+}}\nonumber\\
&  =\sum_{l,m}\mathbb{\tilde{W}}_{l,m}^{\left(  n\right)  }b_{l,m}%
-is\sum_{l,m}\mathbb{\tilde{W}}_{l,-m}^{\left(  n\right)  }\mathbb{\tilde{W}%
}_{l,m}^{\left(  n\right)  }.
\end{align}

The sum $\sum_{l,m}\mathbb{\tilde{W}}_{l,-m}^{\left(  n\right)  }%
\mathbb{\tilde{W}}_{l,m}^{\left(  n\right)  }$ in the basis ($l,m$) for a
definite $n$ is proportional to the unity matrix. Therefore, $\exp\left(
-it\sqrt{2}\sum_{l,m}\mathbb{\tilde{W}}_{l,m}^{\left(  n\right)  }%
Q_{l,m}\right)  $ is%
\begin{align}
&  e^{-it\sqrt{2}\sum_{l,m}\mathbb{\tilde{W}}_{l,m}^{\left(  n\right)
}Q_{l,m}}\nonumber\\
&  =e^{-it\sum_{l,m}\mathbb{\tilde{W}}_{l,-m}^{\left(  n\right)  }b_{l,m}^{+}%
}e^{-it\sum_{l,m}\mathbb{\tilde{W}}_{l,m}^{\left(  n\right)  }b_{l,m}%
-\frac{t^{2}}{2}\sum_{l,m}\mathbb{\tilde{W}}_{l,-m}^{\left(  n\right)
}\mathbb{\tilde{W}}_{l,m}^{\left(  n\right)  }},
\end{align}
that gives us the result%
\begin{equation}
\left\langle 0_{ph}\left\vert e^{-it\sqrt{2}\sum_{l,m}\mathbb{\tilde{W}}%
_{l,m}^{\left(  n\right)  }Q_{l,m}}\right\vert 0_{ph}\right\rangle
=e^{-\frac{t^{2}}{2}\sum_{l,m}\mathbb{\tilde{W}}_{l,-m}^{\left(  n\right)
}\mathbb{\tilde{W}}_{l,m}^{\left(  n\right)  }}.
\end{equation}
Using the explicit formulae for the matrices $\tilde{W}_{l,m}^{\left(
n\right)  }$, the matrix sum takes the form%
\begin{equation}
\sum_{l,m}\mathbb{\tilde{W}}_{l,-m}^{\left(  n\right)  }\mathbb{\tilde{W}%
}_{l,m}^{\left(  n\right)  }=\omega_{s}^{\left(  n\right)  }\mathbb{I}
\label{exp1}%
\end{equation}
with the parameter%
\begin{equation}
\omega_{s}^{\left(  n\right)  }=\frac{1}{2}\left(  a_{0}^{\left(  n\right)
}\right)  ^{2}+\frac{1}{4\pi}\left(  a_{2}^{\left(  n\right)  }\right)  ^{2}.
\end{equation}
Using (\ref{exp1}), the optical conductivity (\ref{fzz1}) is transformed to
the expression%
\begin{equation}
\operatorname{Re}\sigma\left(  \omega\right)  =\omega\sum_{n}\sqrt{\frac{\pi
}{2S_{n}}}D_{n}\exp\left(  -\frac{\left(  \omega-\Omega_{n,0}\right)  ^{2}%
}{2S_{n}}\right)  . \label{ReS5}%
\end{equation}

\newpage

\begin{quote}
\textbf{\emph{Figures to Appendix A}}
\end{quote}

%

%TCIMACRO{\FRAME{dhFU}{5.0376in}{5.1391in}{0pt}{\Qcb{Frequency of the main peak
%in the optical conductivity spectra calculated within the model of Ref.
%\cite{DSG1972} (red dots) and the main-peak energy extracted from the DQMC
%data \cite{Mishchenko2003,DeFilippis2006} (black squares).}}{}%
%{optabsstrongcouplpolaron-figure1.eps}{\special{ language "Scientific Word";
%type "GRAPHIC";  maintain-aspect-ratio TRUE;  display "ICON";
%valid_file "F";  width 5.0376in;  height 5.1391in;  depth 0pt;
%original-width 5.3309in;  original-height 5.4379in;  cropleft "0";
%filename 'OptAbsStrongCouplPolaron-Figure1.EPS';file-properties "XNPEU";}} }%
%BeginExpansion
\begin{center}
\includegraphics[
height=5.1391in,
width=5.0376in
]%
{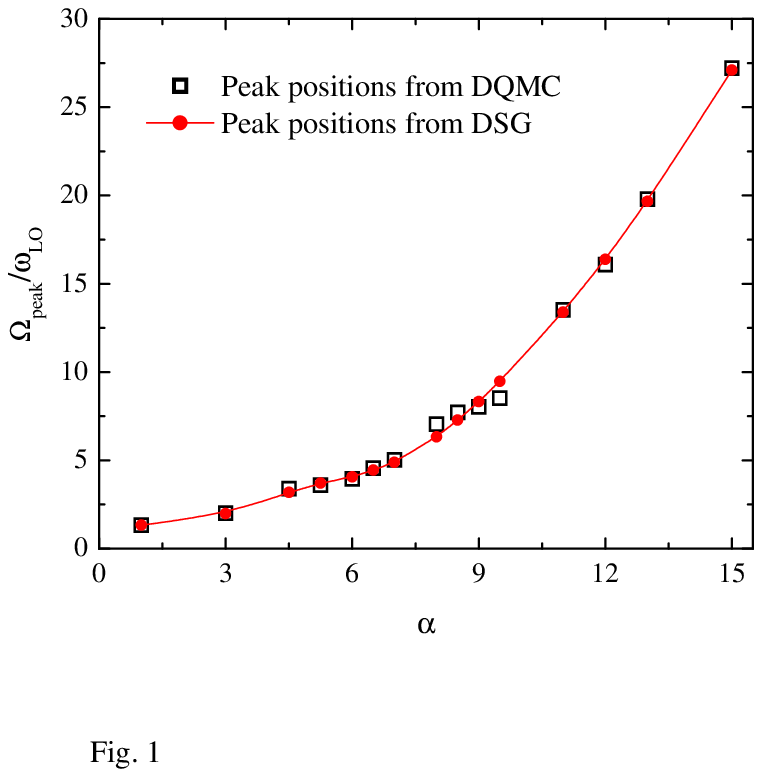}%
\\
Frequency of the main peak in the optical conductivity spectra calculated
within the model of Ref. \cite{DSG1972} (red dots) and the main-peak energy
extracted from the DQMC data \cite{Mishchenko2003,DeFilippis2006} (black
squares).
\end{center}
%EndExpansion

\newpage%

%TCIMACRO{\FRAME{dhFU}{3.2831in}{5.4407in}{0pt}{\Qcb{The strong-coupling
%polaron optical conductivity calculated within the rigorous strong-coupling
%approach of the present work (black solid curves), within the present approach
%but neglecting the dynamic Jahn-Teller effect (red dashed curves), within the
%adiabatic approximation of Ref. \cite{DeFilippis2006} (blue dot-dashed
%curves), and the numerical Diagrammatic Monte Carlo data (full dots) for
%$\alpha=8$ and 8.5.}}{}{optabsstrongcouplpolaron-figure2.eps}%
%maintain-aspect-ratio TRUE;  display "ICON";  valid_file "F";
%width 3.2831in;  height 5.4407in;  depth 0pt;  original-width 4.3421in;
%original-height 7.2174in;  cropleft "0";  croptop "1";  cropright "1";
%filename 'OptAbsStrongCouplPolaron-Figure2.EPS';file-properties "XNPEU";}} }%
%BeginExpansion
\begin{center}
\includegraphics[
height=5.4407in,
width=3.2831in
]%
{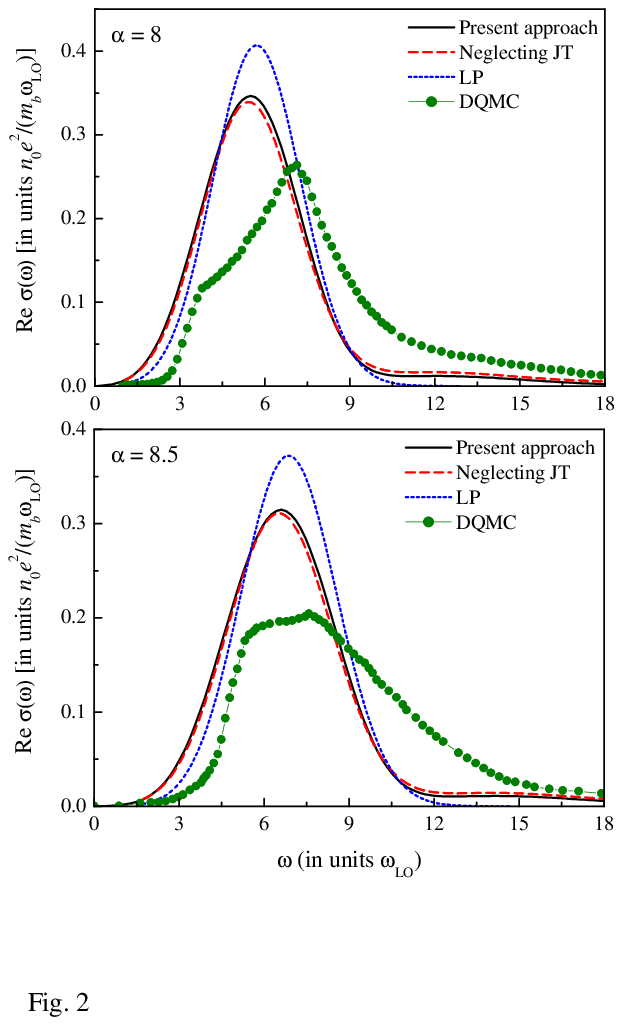}%
\\
The strong-coupling polaron optical conductivity calculated within the
rigorous strong-coupling approach of the present work (black solid curves),
within the present approach but neglecting the dynamic Jahn-Teller effect (red
dashed curves), within the adiabatic approximation of Ref.
\cite{DeFilippis2006} (blue dot-dashed curves), and the numerical Diagrammatic
Monte Carlo data (full dots) for $\alpha=8$ and 8.5.
\end{center}
%EndExpansion

\newpage%

%TCIMACRO{\FRAME{dhFU}{3.3366in}{7.4794in}{0pt}{\Qcb{The strong-coupling
%polaron optical conductivity calculated within the rigorous strong-coupling
%approach of the present work (black solid curves), within the present approach
%but neglecting the dynamic Jahn-Teller effect (red dashed curves), within the
%adiabatic approximation of Ref. \cite{DeFilippis2006} (blue dot-dashed
%curves), and the numerical Diagrammatic Monte Carlo data (full dots) for
%$\alpha=9$, 13 and 15.}}{}{optabsstrongcouplpolaron-figure3.eps}%
%maintain-aspect-ratio TRUE;  display "ICON";  valid_file "F";
%width 3.3366in;  height 7.4794in;  depth 0pt;  original-width 4.4112in;
%original-height 9.935in;  cropleft "0";  croptop "1";  cropright "1";
%filename 'OptAbsStrongCouplPolaron-Figure3.EPS';file-properties "XNPEU";}} }%
%BeginExpansion
\begin{center}
\includegraphics[
height=7.4794in,
width=3.3366in
]%
{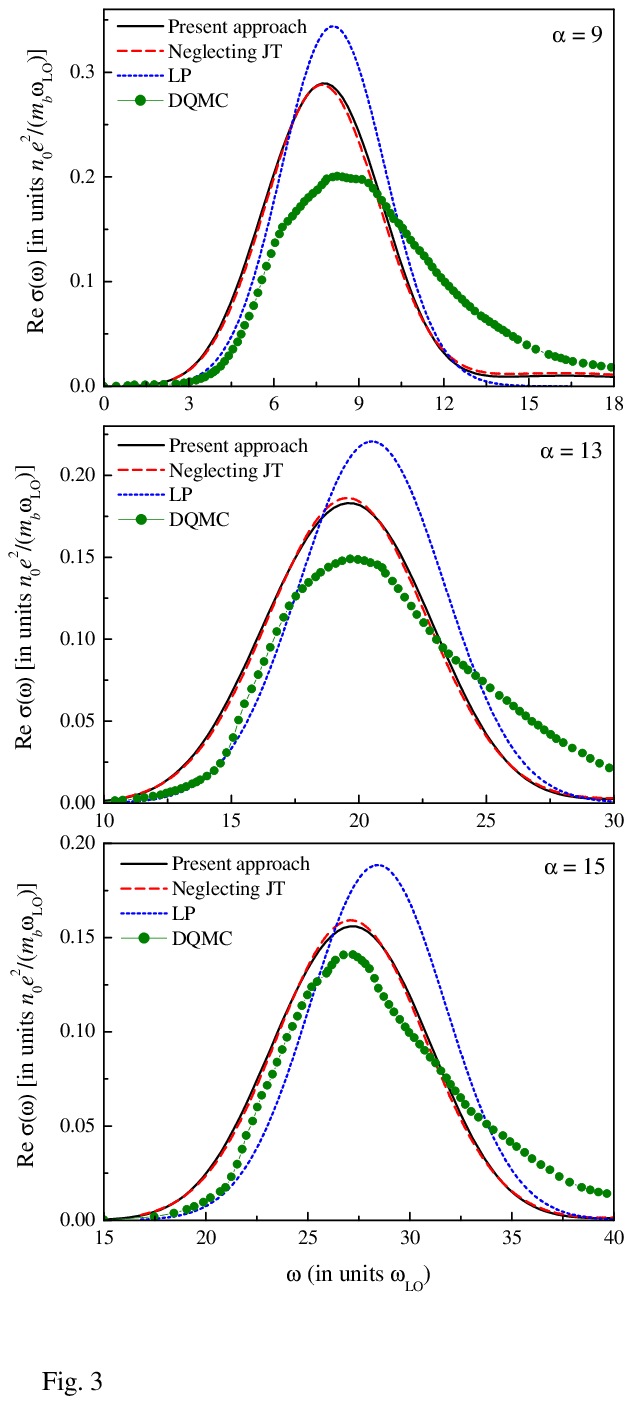}%
\\
The strong-coupling polaron optical conductivity calculated within the
rigorous strong-coupling approach of the present work (black solid curves),
within the present approach but neglecting the dynamic Jahn-Teller effect (red
dashed curves), within the adiabatic approximation of Ref.
\cite{DeFilippis2006} (blue dot-dashed curves), and the numerical Diagrammatic
Monte Carlo data (full dots) for $\alpha=9$, 13 and 15.
\end{center}
%EndExpansion

\newpage%

%TCIMACRO{\FRAME{dhFU}{5.4601in}{4.6502in}{0pt}{\Qcb{Relative contribution of
%the transitions to the continuum polaron states to the zeroth frequency moment
%of the strong-coupling polaron optical conductivity as a function of the
%coupling constant $\alpha$.}}{}{optabsstrongcouplpolaron-figure4.eps}%
%maintain-aspect-ratio TRUE;  display "ICON";  valid_file "F";
%width 5.4601in;  height 4.6502in;  depth 0pt;  original-width 5.7793in;
%original-height 4.9167in;  cropleft "0";  croptop "1";  cropright "1";
%filename 'OptAbsStrongCouplPolaron-Figure4.EPS';file-properties "XNPEU";}} }%
%BeginExpansion
\begin{center}
\includegraphics[
height=4.6502in,
width=5.4601in
]%
{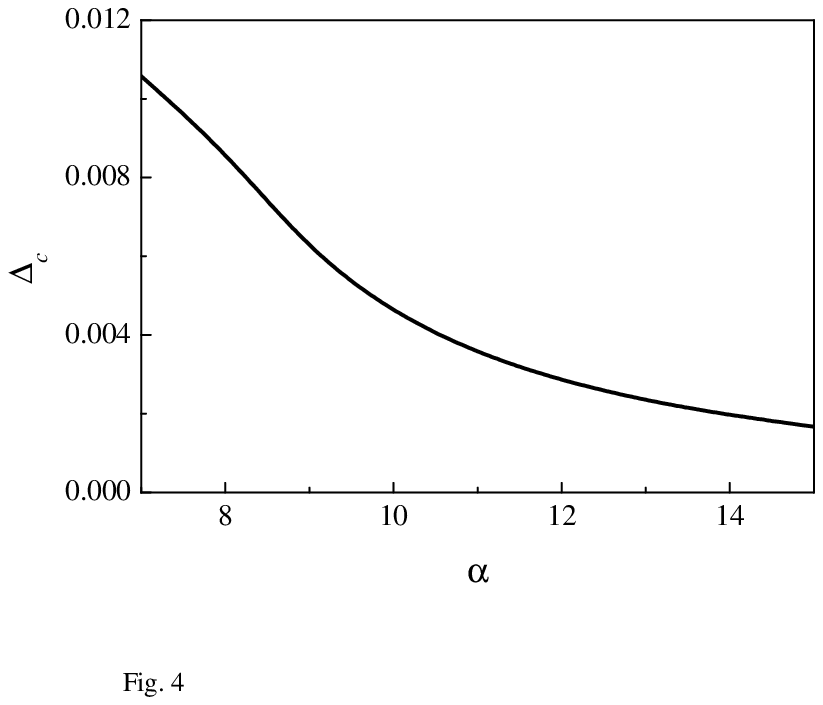}%
\\
Relative contribution of the transitions to the continuum polaron states to
the zeroth frequency moment of the strong-coupling polaron optical
conductivity as a function of the coupling constant $\alpha$.
\end{center}
%EndExpansion

.

\section{Feynman's path-integral polaron treatment approached using
time-ordered operator calculus [\emph{S. N. Klimin and J. T. Devreese, Solid
State Communications 151, 144 (2011)}]}

Several studies have been devoted to the search of a Hamiltonian formalism
equivalent to Feynman's path integral approximation to polaron theory.
Bogolubov \cite{Bogolubov} reproduced the Feynman result for the polaron free
energy \cite{Feynman} using time-ordering T-products . Yamazaki
\cite{Yamazaki} introduced two kinds of auxiliary vector fields to derive
Feynman's ground state polaron energy expression with the operator technique,
however he found no proof of the variational nature of this result. Cataudella
\emph{et al}. \cite{Cataudella} formally re-obtained Feynman's polaron
ground-state energy expression by introducing additional degrees of freedom,
but again their result could not be proved to constitute an upper bound for
the polaron ground state energy.

The study of the excited polaron states is of interest i.~a. for its
application to the polaron response properties. In \cite{FHIP} a path-integral
based response-formalism was introduced that was applied to derive polaron
optical absorption spectra in \cite{KartheuserGreenbook}. The results for the
polaron response obtained in \cite{KartheuserGreenbook} were re-derived with a
Hamiltonian technique (Mori- formalism) in \cite{PD1983}.

To the best of our knowledge, no explicit description of the polaron excited
states has been derived within the \textquotedblleft all
coupling-\textquotedblleft\ Feynman approach. Only for the limiting cases of
weak and strong coupling approximations (and for a 1D-model system) such
excitation spectra were derived \cite{E65,Devreese2009,DE64}.

In principle, the spectrum of the polaron excited states can be derived
indirectly -- using a Laplace transform of the finite-temperature partition
function. However, it is not clear how to realize this program in practice.

The polaron excitation spectrum is interesting by itself. E.g. the existence
and the nature of \textquotedblleft relaxed excited states\textquotedblright,
\textquotedblleft Franck-Condon states\textquotedblright, \textquotedblleft
scattering states\textquotedblright\ is understood from the mathematical
structure of corresponding eigenstates.

In the present letter we first present a re-derivation of the original Feynman
variational path integral polaron model \cite{Feynman} for the ground state,
using a Hamiltonian formalism, and we do provide a proof of the upper bound
nature of the obtained ground state energy. Furthermore, using Feynman's
(Hamiltonian-) time-ordered operator calculus (and an \emph{ad hoc} unitary
transformation) we obtain explicitly -- and for the first time -- the excited
polaron states that correspond to the Feynman polaron model.

The novelty of the present approach consists (a) in the \emph{direct
calculation of the energies and the lifetimes of the excited polaron states}
(within a Hamiltonian all-coupling approach -- developed in this work --
equivalent to the Feynman path integral polaron model) and (b) in the
extension of the Feynman variational technique to non-parabolic trial
potentials. Although the time-ordered operator calculus is formally equivalent
to the path-integral formalism, it is not obvious how to directly calculate
the excited polaron states using path integrals.

The present work, formulated with the (Hamiltonian) time ordered operator
calculus, thus provides an (equivalent) tool complementary with respect to the
Feynman path integral approach to the polaron, to study the polaron problem.
Additionally we directly study the excited polaron states.

Consider an electron-phonon system with the Fr\"{o}hlich Hamiltonian%

\begin{align}
H  &  =\frac{\mathbf{p}^{2}}{2}+H_{ph}+H_{e-ph},\label{FH}\\
H_{ph}  &  =\sum_{\mathbf{q}}\left(  a_{\mathbf{q}}^{+}a_{\mathbf{q}}+\frac
{1}{2}\right)  ,\label{FHp}\\
H_{e-ph}  &  =\frac{1}{\sqrt{V}}\sum_{\mathbf{q}}\frac{\sqrt{2\sqrt{2}%
\pi\alpha}}{q}\left(  a_{\mathbf{q}}+a_{-\mathbf{q}}^{+}\right)
e^{i\mathbf{q\cdot r}}. \label{FHep}%
\end{align}
Here, the Feynman units are used: $\hbar=1,$ the band mass $m_{b}=1,$ the
LO-phonon frequency $\omega_{\mathrm{LO}}=1$.

The polaron partition function after exact averaging over phonon states is%
\begin{equation}
Z_{pol}=\mathtt{Tr}\left[  \mathtt{T}\exp\left(  -\int_{0}^{\beta}%
\frac{\mathbf{p}_{\tau}^{2}}{2}d\tau+\hat{\Phi}\left[  \mathbf{r}_{\tau
}\right]  \right)  \right]  , \label{Fzpol}%
\end{equation}
where $\beta=\frac{1}{k_{B}T}$. The \textquotedblleft influence
phase\textquotedblright\ of the phonons $\hat{\Phi}\left[  \mathbf{r}_{\tau
}\right]  $ in the the time-ordered operator calculus has the same form as in
the path-integral representation. The polaron free energy is determined as%
\begin{equation}
F_{pol}=-\frac{1}{\beta}\ln Z_{pol}. \label{Ffpol}%
\end{equation}

The trial Hamiltonian describes the electron interacting with a fictitious
particle of the mass $m_{f}$ through an attractive potential $V_{f}$:
\begin{equation}
H_{tr}=\frac{\mathbf{p}^{2}}{2}+\frac{\mathbf{p}_{f}^{2}}{2m_{f}}+V_{f}\left(
\mathbf{r}-\mathbf{r}_{f}\right)  . \label{FHtr1}%
\end{equation}
The trial potential $V_{f}$ is, in general, non-parabolic. The parabolic
potential with frequency parameter $w$ corresponds to the Feynman polaron model.

Consider the \textquotedblleft extended\textquotedblright\ partition function
of the electron-phonon system%
\begin{equation}
Z_{ext}=Z_{f}Z_{pol} \label{Faa}%
\end{equation}
where $Z_{f}$ is the partition function of a fictitious particle,%
\begin{equation}
Z_{f}\equiv\mathtt{Tr}\left[  \mathtt{T}\exp\left(  -\int_{0}^{\beta}d\tau
H_{f,\tau}\right)  \right]  , \label{Fzf}%
\end{equation}
with Hamiltonian%
\begin{equation}
H_{f}=\frac{\mathbf{p}_{f}^{2}}{2m_{f}}+V_{f}\left(  \mathbf{r}_{f}\right)  .
\label{FHf}%
\end{equation}

The polaron free energy is expressed as the difference%
\begin{equation}
F_{pol}=F_{ext}-F_{f}, \label{Ffp}%
\end{equation}
where $F_{f}$ is the free energy of the fictitious particle confined to the
potential $V\left(  \mathbf{r}_{f}\right)  $. The free energies $F_{ext}$ and
$F_{f}$ are determined similarly to (\ref{Ffpol}), with corresponding
partition functions. In the zero-temperature limit, the free energies
$F_{pol}$, $F_{ext}$ and $F_{f}$ become, respectively, the ground-state
energies $E_{pol}^{0}$, $E_{ext}^{0}$ and $E_{f}^{0}$.

The key element of the present approach is the unitary transformation%
\begin{equation}
U=e^{-i\mathbf{p}_{f}\cdot\mathbf{r}}. \label{FU}%
\end{equation}
Application of this canonical transformation results in the transformed
\textquotedblleft extended\textquotedblright\ Hamiltonian $H_{ext}^{\prime
}=UH_{ext}U^{-1}$,%
\begin{align}
H_{ext}^{\prime}  &  =\frac{\left(  \mathbf{p}+\mathbf{p}_{f}\right)  ^{2}}%
{2}+\frac{\mathbf{p}_{f}^{2}}{2m_{f}}+V_{f}\left(  \mathbf{r}_{f}%
-\mathbf{r}\right) \nonumber\\
&  +H_{ph}+H_{e-ph}. \label{FHext}%
\end{align}
This Hamiltonian can be represented as a sum of an unperturbed Hamiltonian
\begin{equation}
H_{0}\equiv H_{tr}+H_{ph} \label{FH0}%
\end{equation}
and an interaction term%
\begin{equation}
V\equiv\frac{1}{2}\mathbf{p}_{f}^{2}+\mathbf{p}\cdot\mathbf{p}_{f}+H_{e-ph}.
\label{FV}%
\end{equation}

Further we use the variational principle for the ground-state energy in terms
of the time-ordered operators following Ref. \cite{DB1992}. The exact ground
state $\left\vert 0\right\rangle $ of the system with the Hamiltonian
(\ref{FHext}) can be written in the interaction representation starting from
the unperturbed ground state $\left\vert -\infty\right\rangle $:%
\begin{equation}
\left\vert 0\right\rangle =\mathcal{U}\left(  \infty,-\infty\right)
\left\vert -\infty\right\rangle \label{F0}%
\end{equation}
where $\mathcal{U}\left(  \infty,-\infty\right)  $ is the time-evolution
operator,%
\begin{equation}
\mathcal{U}\left(  t_{2},t_{1}\right)  =\mathcal{T}\exp\left(  -i\int_{t_{1}%
}^{t_{2}}e^{-\delta\left\vert t\right\vert }e^{iH_{0}t}Ve^{-iH_{0}t}\right)  .
\label{FT}%
\end{equation}
Here, $\delta\rightarrow+0$ and $\mathcal{T}$ denotes time ordering.

In the exact expectation value for the ground state energy $E_{ext}^{0}%
\equiv\left\langle 0\left\vert H_{ext}^{\prime}\right\vert 0\right\rangle $,
the phonons are eliminated using the time ordered-operator calculus as in Ref.
\cite{DB1992}. The average of the interaction term becomes then%
\begin{align}
&  \left\langle 0\left\vert H_{e-ph}\right\vert 0\right\rangle \nonumber\\
&  =-i\frac{\sqrt{2}\pi\alpha}{V}\int_{-\infty}^{\infty}dte^{-i\left\vert
t\right\vert -\delta\left\vert t\right\vert }\nonumber\\
&  \times\sum_{\mathbf{q}}\frac{1}{q^{2}}\left\langle \infty\left\vert
\mathcal{T}\left[  \mathcal{U}\left(  \infty,-\infty\right)  e^{i\mathbf{q}%
\cdot\left[  \mathbf{r}\left(  t\right)  -\mathbf{r}\left(  0\right)  \right]
}\right]  \right\vert -\infty\right\rangle . \label{FAv}%
\end{align}
This means that the polaron ground state energy is exactly described using a
retarded potential in the interaction representation, cf. Eq. (2.16) of Ref.
\cite{DB1992}.

The ground state energy satisfies the Ritz variational principle with a trial
state. Choosing the trial state as the ground state of the Hamiltonian
(\ref{FH0}), the variational principle can be written as \cite{DB1992}%
\begin{align}
E_{ext}^{0}  &  \leq E_{tr}^{0}\nonumber\\
&  +\left\langle \infty\left\vert \mathcal{T}\left\{  \mathcal{U}_{tr}\left(
\infty,-\infty\right)  \left[  H_{ext}^{\prime}\left(  0\right)  -H_{0}\left(
0\right)  \right]  \right\}  \right\vert -\infty\right\rangle , \label{FFVP}%
\end{align}
where $\mathcal{U}_{tr}\left(  \infty,-\infty\right)  $ is the time-evolution
operator corresponding to the trial Hamiltonian (\ref{FHtr1}).

The exact polaron ground state energy is denoted here as $E^{0}\left(
\mathbf{k}\right)  $, where $\mathbf{k}$ is the polaron translation momentum.
We find an upper bound for $E^{0}\left(  \mathbf{k}\right)  $ substituting
(\ref{FAv}) in (\ref{FFVP}) and using the exact wave functions and energy
levels of the trial Hamiltonian. The trial Hamiltonian (\ref{FHtr1}) can be
rewritten in terms of the coordinates $\left(  \mathbf{R},\boldsymbol{\rho
}\right)  $ and momenta $\left(  \mathbf{P},\vec{\pi}\right)  $ of the
center-of-mass and relative (internal) motions of the trial system with the
masses $M=1+m_{f}$ and $\mu=m_{f}/\left(  1+m_{f}\right)  $ using the
frequency $v=wM$. The energy spectrum of the trial system is the sum of the
translation- and oscillation contributions,%
\begin{equation}
E_{\mathbf{k},n}=\frac{\mathbf{k}^{2}}{2M}+\varepsilon_{n},\; \varepsilon
_{n}=v\left(  n+\frac{3}{2}\right)  . \label{Fener}%
\end{equation}
The eigenfunctions of the Hamiltonian (\ref{FHtr1}) are products of
translational- and oscillatory wave functions:%
\begin{equation}
\psi_{\mathbf{k};l,n,m}\left(  \mathbf{R},\boldsymbol{\rho}\right)  =\frac
{1}{\sqrt{V}}e^{i\mathbf{k\cdot R}}\varphi_{l,n,m}\left(  \boldsymbol{\rho
}\right)  , \label{Fpsi}%
\end{equation}
where $\varphi_{l,n,m}\left(  \boldsymbol{\rho}\right)  $ is the 3D
harmonic-oscillator wave function with a given angular momentum. The result is%
\begin{align}
E^{0}\left(  \mathbf{k}\right)   &  \leq\mathcal{E}_{p}^{\left(  0,0\right)
}\left(  \mathbf{k}\right)  ,\\
\mathcal{E}_{p}^{\left(  0,0\right)  }\left(  \mathbf{k}\right)   &  =\frac
{3}{4}\frac{\left(  v-w\right)  ^{2}}{v}\nonumber\\
&  +\frac{1}{2}\left(  1-\frac{1}{\left(  1+m_{f}\right)  ^{2}}\right)
\mathbf{k}^{2}-\frac{\sqrt{2}\alpha}{4\pi^{2}}\int\frac{d\mathbf{q}}{q^{2}%
}\nonumber\\
&  \times\sum_{\mathbf{k}^{\prime},l^{\prime},n^{\prime},m^{\prime}}%
\frac{\left\vert \left\langle \psi_{\mathbf{k};0,0,0}\left\vert
e^{i\mathbf{q\cdot r}}\right\vert \psi_{\mathbf{k}^{\prime};l^{\prime
},n^{\prime},m^{\prime}}\right\rangle \right\vert ^{2}}{\frac{1}{2\left(
m_{f}+1\right)  }\left(  \left(  \mathbf{k}^{\prime}\right)  ^{2}%
-\mathbf{k}^{2}\right)  +vn^{\prime}+1}, \label{FE0}%
\end{align}
where $v>w$ are the Feynman variational frequencies. The functional
(\ref{FE0}) can be reduced to the known Feynman result for the polaron
ground-state energy. In the r.h.s. of (\ref{FE0} at the polaron momentum
$\mathbf{k}=0$, we introduce the integral over the Euclidean time:
\begin{equation}
\frac{1}{\frac{\left(  \mathbf{k}^{\prime}\right)  ^{2}}{2\left(
m_{f}+1\right)  }+vn^{\prime}+1}=\int_{0}^{\infty}e^{-\left(  \frac{\left(
\mathbf{k}^{\prime}\right)  ^{2}}{2\left(  m_{f}+1\right)  }+vn^{\prime
}+1\right)  \tau}d\tau. \label{FIntegr}%
\end{equation}
After this, the summations and integrations in (\ref{FEpol}) are performed
analytically, and we arrive at the Feynman variational expression for the
polaron ground-state energy:%
\begin{align}
\left.  E^{0}\left(  \mathbf{k}\right)  \right\vert _{\mathbf{k}=0}  &
\leq\frac{3}{4}\frac{\left(  v-w\right)  ^{2}}{v}\nonumber\\
&  -\frac{\alpha v}{\sqrt{\pi}}\int_{0}^{\infty}\frac{e^{-\tau}}{\sqrt
{w^{2}\tau+\frac{v^{2}-w^{2}}{v}\left(  1-e^{-v\tau}\right)  }}d\tau.
\label{FEF}%
\end{align}

The electron-phonon contribution in (\ref{FE0}) is structurally similar to the
second-order perturbation correction to the polaron ground-state energy due to
the electron-phonon interaction (using states of the Feynman model
$\psi_{\mathbf{k};l,n,m}$ as the zero-order approximation). Therefore we can
estimate the energies of the excited polaron states when averaging the
difference between exact and unperturbed Hamiltonians in (\ref{FFVP}) with an
excited trial state. We then arrive at the following extension for the r.h.s.
of (\ref{FE0}):%
\begin{align}
\mathcal{E}_{p}^{\left(  l,n\right)  }\left(  \mathbf{k}\right)   &
=\frac{v^{2}+w^{2}}{2v}\left(  n+\frac{3}{2}\right)  -\frac{3}{2}w\nonumber\\
&  +\frac{1}{2}\left(  1-\frac{1}{\left(  1+m_{f}\right)  ^{2}}\right)
\mathbf{k}^{2}-\frac{\sqrt{2}\alpha}{4\pi^{2}}\int\frac{d\mathbf{q}}{q^{2}%
}\nonumber\\
&  \times\sum_{\mathbf{k}^{\prime},l^{\prime},n^{\prime},m^{\prime}}%
\frac{\left\vert \left\langle \psi_{\mathbf{k};l,n,m}\left\vert
e^{i\mathbf{q\cdot r}}\right\vert \psi_{\mathbf{k}^{\prime};l^{\prime
},n^{\prime},m^{\prime}}\right\rangle \right\vert ^{2}}{\frac{1}{2\left(
m_{f}+1\right)  }\left(  \left(  \mathbf{k}^{\prime}\right)  ^{2}%
-\mathbf{k}^{2}\right)  +v\left(  n^{\prime}-n\right)  +1}. \label{FEpol}%
\end{align}

In the same approach, we obtain the inverse lifetimes for the excited states
of the polaron:%
\begin{align}
\Gamma_{l,n}\left(  \mathbf{k}\right)   &  =\frac{\sqrt{2}\alpha}{4\pi}%
\sum_{\mathbf{k}^{\prime},l^{\prime},n^{\prime},m^{\prime}}\int d\mathbf{q}%
\frac{1}{q^{2}}\nonumber\\
&  \times\left\vert \left\langle \psi_{\mathbf{k};l,n,m}\left\vert
e^{i\mathbf{q\cdot r}}\right\vert \psi_{\mathbf{k}^{\prime};l^{\prime
},n^{\prime},m^{\prime}}\right\rangle \right\vert ^{2}\nonumber\\
&  \times\delta\left(  \frac{q^{2}}{2\left(  m_{f}+1\right)  }+v\left(
n^{\prime}-n\right)  +1\right)  . \label{Fgamma}%
\end{align}
The broadening of the excited polaron \textquotedblleft
non-scattering\textquotedblright\ states must be taken into account for an
analytical study of the polaron optical conductivity.

Using the above expressions, we determine the transition energies for the
transitions between the ground and the first excited state $\hbar
\Omega_{0\rightarrow1exc}\equiv E_{p}^{\left(  1exc\right)  }-E_{p}^{\left(
0\right)  }$. Let us first consider the transition energies in which
$E_{p}^{\left(  1exc\right)  }$ are calculated using optimal values of the
parameters of the Feynman model obtained from the minimization of the
variational ground-state energy $E_{p}^{\left(  0\right)  }$. This method
formally leads to the Franck-Condon (FC) excited states, with the
\textquotedblleft frozen\textquotedblright\ phonon configuration corresponding
to the ground state of the polaron. Note that the existence of Franck-Condon
states as eigenstates of the Fr\"{o}hlich polaron Hamiltonian has not been
proved: Ref \cite{DE64} suggests their non-existence as eigenstates for a
simplified polaron model. Nevertheless the Franck-Condon concept can be
significant, e.~g. for approximate treatments using a basis of Franck-Condon
states, as indicative for the frequency of the maxima of phonon-sidebands, etc.%

%TCIMACRO{\FRAME{dhFU}{7.8774cm}{6.5512cm}{0pt}{\Qcb{\QTR{bf}{Fig. 1.}
%Franck-Condon transition energies as a function of the coupling constant
%compared to the lowest-energy peak position of the polaron optical
%conductivity from Ref. \cite{KartheuserGreenbook} and the maximum of the
%polaron optical conductivity band from Ref. \cite{Mishchenko2003}.}}%
%{}{kd-hamiltonianpolaron-fig1.eps}{\special{ language "Scientific Word";
%type "GRAPHIC";  maintain-aspect-ratio TRUE;  display "ICON";
%valid_file "F";  width 7.8774cm;  height 6.5512cm;  depth 0pt;
%original-width 8.6086cm;  original-height 7.1478cm;  cropleft "0";
%filename 'KD-HamiltonianPolaron-Fig1.eps';file-properties "XNPEU";}} }%
%BeginExpansion
\begin{center}
\includegraphics[
height=6.5512cm,
width=7.8774cm
]%
{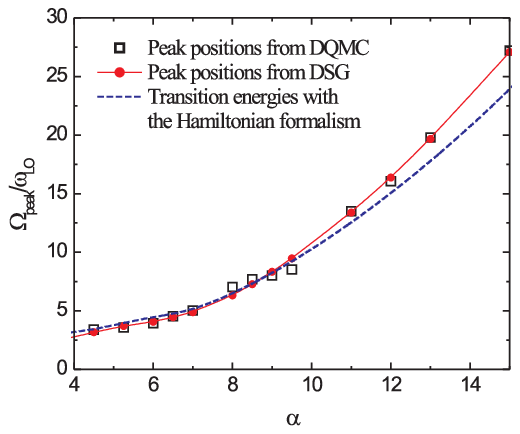}%
\\
\textbf{Fig. 1.} Franck-Condon transition energies as a function of the
coupling constant compared to the lowest-energy peak position of the polaron
optical conductivity from Ref. \cite{KartheuserGreenbook} and the maximum of
the polaron optical conductivity band from Ref. \cite{Mishchenko2003}.
\end{center}
%EndExpansion

In Fig. 1, the FC transition energies calculated with the approach introduced
in the present work for polaron momentum $\mathbf{k}=0$ are plotted as a
function of the coupling constant $\alpha$. They are compared with the peak
energies of the polaron optical conductivity calculated using the diagrammatic
Monte Carlo method (DQMC) \cite{Mishchenko2003,DeFilippis2006} and with the
peak energies attributed to polaron \textquotedblleft relaxed excited
states\textquotedblright\ (RES) in Ref. \cite{KartheuserGreenbook}
(\textquotedblleft DSG\textquotedblright). The DQMC and DSG main-peak energies
are close to each other in the whole range of the coupling strength. In the
range $4\lessapprox\alpha\lessapprox10$, the present result for the transition
energy is close to the DQMC and the DSG peak energies. Furthermore, in this
range of $\alpha$, the non-monotonous behavior of the curvature is remarkably
the same for the DQMC and DSG peak energies and for the present result.

There is a remarkable agreement between the peaks attributed to the RES in
Ref. \cite{KartheuserGreenbook}, the peak positions obtained within the
strong-coupling approach, Eq. (3) of Ref. \cite{DeFilippis2006}, and the
positions of the maximum of the optical conductivity band calculated in Ref.
\cite{Mishchenko2003} using DQMC. It is reasonable that the three aforesaid
peaks must be interpreted in one and the same way. In order to clarify this,
we can refer to Ref. \cite{Mishchenko2003}. In the strong-coupling regime, the
dominant broad peak of the polaron optical conductivity spectrum can be
considered as a \textquotedblleft Franck-Condon sideband\textquotedblright\ of
the \textquotedblleft groundstate to RES-transition\textquotedblright, even if
this latter transition can have a negligible oscillator strength (see also
\cite{KED1969}). The optical conductivity spectra of Ref.
\cite{DeFilippis2006} in the strong-coupling approximation have been
calculated taking into account the polaronic shift of the energy levels. The
polaronic shift in Ref. \cite{DeFilippis2006} has been calculated with the
Franck-Condon wave functions (i. e., with the strong-coupling wave functions
corresponding to the \textquotedblleft frozen\textquotedblright\ lattice
configuration for the ground state). Note that the exact excitation spectrum
of the Fr\"{o}hlich-Hamiltonian might be devoid of Franck-Condon eigenstates,
cf. Ref. \cite{DE64}). It should be remarked that the maxima of the
FC-sideband structures of Ref. \cite{KartheuserGreenbook} are positioned at
the frequency $\Omega=v$, i. e., at the transition frequency for the model
system without the polaron shift.

The Franck-Condon peak energies calculated in the present work also take into
account the polaron shift. As follows from the above analysis, in the
strong-coupling limit they must correspond to the Franck-Condon peak energies
of the strong-coupling expansion of Ref. \cite{DeFilippis2006}. The agreement
of the position of the maxima of these peaks with those attributed to
transitions to the RES in Ref. \cite{KartheuserGreenbook} shows that in the
strong-coupling range of $\alpha$, the latter should be associated to the
Franck-Condon sidebands rather than to the RES.

Another approach, in which the parameters of the first excited state are
determined self-consistently (Ref. \cite{KED1969}), was used i.~a. to
calculate (in the strong-coupling case) the (lowest) energy level of the
relaxed excited state (RES). The transitions from the polaron ground state to
the RES correspond to a zero-phonon peak in the optical conductivity.

For the study of the energies of excited states of the polaron, a variational
approach requires special care, because the excited states of the polaron are
not stable. A variational approach, strictly speaking, is only valid for
excited states when the variational wave function of the excited state is
orthogonal to the exact ground-state wave function.

For the estimation of the energy of the first RES with our present formalism,
we determine a minimum of the expression (\ref{FEpol}) in a physically
reasonable range of the variational parameters. In order to determine that
range, we refer to Ref. \cite{Lepine}, where the energy of the polaron RES is
calculated variationally within the Green's function formalism.

The expression for the RES energy in Ref. \cite{Lepine} contains the
electron-phonon contribution corresponding to the second-order perturbation
formula. It differs, however, from the weak-coupling second-order perturbation
expression by the choice of the unperturbed states: in Ref.\cite{Lepine} those
are variational states rather than free-electron states. There exists some
analogy between our approach and that of Ref. \cite{Lepine}. The latter,
however, does not take into account the translation invariance of the polaron problem.

In Ref. \cite{Lepine}, the energy of the polaron RES is calculated
variationally. The unperturbed wave function of the RES is chosen orthogonal
(due to symmetry) to the unperturbed ground state wave function. In the
present approach, this orthogonality is also exactly satisfied because of symmetry.

The expressions for the polaron RES energy of Ref. \cite{Lepine} contain
singularities, which occur when the energies of the unperturbed ground state
and that of the first excited states are in resonant with the LO-phonon
energy. These singularities are related to the instability of the excited
polaron with respect to the emission of LO-phonons. Using the same reasoning
as in Ref. \cite{Lepine} we search for a local minimum of the polaron RES
energy in the range where the confinement frequency $v$ of the Feynman model
satisfies the inequality $v>1$. The instability of the excited polaron state
is then avoided.

The resulting numerical values of the transition energy to the first RES as a
function of $\alpha$ are shown in Fig. 2. They are compared with the
numerical-DQMC peak energies of the polaron optical conductivity band
\cite{Mishchenko2003,DeFilippis2006}, with the FC transition energies obtained
in the present work, and with the leading term of the strong-coupling
approximation for the RES transition energy from Ref. \cite{KED1969}.%

%TCIMACRO{\FRAME{dhFU}{7.8774cm}{6.5512cm}{0pt}{\Qcb{\QTR{bf}{Fig. 2.} The
%transition energy for the transition from the polaron ground state to the
%first RES(\QTR{em}{solid black curve}) and to the first excited FC state
%(\QTR{em}{dashed red curve}) as a function of $\alpha$ obtained in the present
%work, compared with the maximum of the polaron optical conductivity band from
%numerical DQMC (\QTR{em}{black squares}, Ref. \cite{Mishchenko2003}).
%\QTR{em}{The dashed-dot green curve}: the strong-coupling result for this
%transition energy as given in Ref. \cite{KED1969}.}}{}%
%{kd-hamiltonianpolaron-fig2.eps}{\special{ language "Scientific Word";
%type "GRAPHIC";  maintain-aspect-ratio TRUE;  display "ICON";
%valid_file "F";  width 7.8774cm;  height 6.5512cm;  depth 0pt;
%original-width 9.25cm;  original-height 6.7678cm;  cropleft "0";
%filename 'KD-HamiltonianPolaron-Fig2.eps';file-properties "XNPEU";}} }%
%BeginExpansion
\begin{center}
\includegraphics[
height=6.5512cm,
width=7.8774cm
]%
{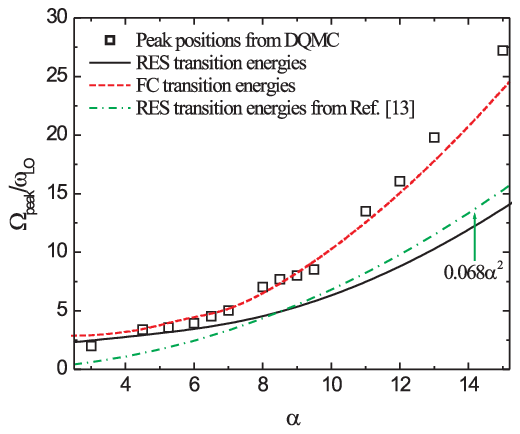}%
\\
\textbf{Fig. 2.} The transition energy for the transition from the polaron
ground state to the first RES(\emph{solid black curve}) and to the first
excited FC state (\emph{dashed red curve}) as a function of $\alpha$ obtained
in the present work, compared with the maximum of the polaron optical
conductivity band from numerical DQMC (\emph{black squares}, Ref.
\cite{Mishchenko2003}). \emph{The dashed-dot green curve}: the strong-coupling
result for this transition energy as given in Ref. \cite{KED1969}.
\end{center}
%EndExpansion

For $\alpha\lesssim2.5$, there exists no minimum of $E_{p}^{\left(
1exc\right)  }$ in the range $v>1$. We can interpret this result as a
manifestation of the fact that for decreasing coupling strength, the RES is
suppressed at sufficiently weak coupling. We see that for sufficiently small
$\alpha$ ($\alpha\lesssim6$), the RES transition energies show good agreement
with the DQMC peak energies, what confirms the concept of RES developed in
Refs. \cite{KartheuserGreenbook,KED1969}. For higher coupling strengths, the
DQMC data appear to be closer to the FC (rather than to RES) transition
energies. This result can be an indication of the fact that with increasing
$\alpha$, the mechanism of the polaron optical absorption changes its nature
as suggested in Ref. \cite{DeFilippis2006}, from a regime with dynamic lattice
relaxation (for which the RES are relevant) at weak and intermediate coupling
to the Franck-Condon (\textquotedblleft LO-phonon sidebands\textquotedblright%
-) regime at strong coupling.

In summary, we have re-formulated the Feynman all-coupling path integral
method for the polaron problem within a Hamiltonian formalism using
time-ordered operator calculus. This reformulation allows us to describe not
only the free energy and the ground state, but also to directly determine --
for the first time -- the excited polaron states that correspond to the
Feynman all-coupling polaron model. A variational procedure for the polaron
RES energy has been developed, within the formalism presented in this work,
which provides results i.a. in agreement with the strong-coupling limit of
Ref. \cite{KED1969}. The present treatment offers the prospect of further
elucidation of the nature of the polaron resonances (\textquotedblleft relaxed
excited states\textquotedblright\ versus \textquotedblleft Franck-Condon
sidebands\textquotedblright\ \cite{DeFilippis2006}) at intermediate coupling.

\newpage

\section{Many-body large polaron optical conductivity in SrTi$_{1-x}$Nb$_{x}%
$O$_{3}$ [\emph{J. T. Devreese, S. N. Klimin, J. L. M. van Mechelen, and D.
van der Marel, Phys. Rev. B \textbf{81}, 125119 (2010)}]}

\subsection{Introduction \label{sec:intro copy(1)}}

The infrared optical absorption of perovskite-type materials, in particular,
{of} copper oxide based high-$T_{c}$ superconductors {and of the manganites}
has been the subject of intensive investigations
\cite{zLupi1999,zcalva0,zfalck1,zcalva1b,zQQ4,zCrawford90,zzhang,zHomes1997,zRonnow,zHartinger}%
. {Insulating SrTiO$_{3}$ has a perovskite structure and manifests a
metal--insulator transition at room temperature around a doping of 0.002\% La
or Nb per unit cell \cite{zCalvani1993}. At low doping concentrations, between
0.003\% and 3\%, strontium titanate reveals a} superconducting phase
transition \cite{zSchooley1964} below 0.7 K. {Various optical experiments
\cite{zGervais93,zCalvani1993,zEagles96,zAng2000,zJPCM2006,zVDM-PRL2008} show
a mid-infrared band in the normal state optical conductivity of doped
SrTiO$_{3}$ which is often explained by polaronic behavior.} In the recently
observed optical conductivity spectra of Ref.~\cite{zVDM-PRL2008}, shown in
Fig.~1, there is a broad mid-infrared optical conductivity band starting at a
photon energy of $\hbar\Omega\sim100$ meV, which is within the range of the
LO-phonon energies of SrTi$_{1-x}$Nb$_{x}$O$_{3}$. The peaks/shoulders of the
experimental optical conductivity band at $\hbar\Omega\sim200$ to 400 meV
resemble the peaks provided by the mixed plasmon-phonon excitations as
described in Ref.~\cite{zTD2001}. Based on the experimental data, the authors
deduce a coupling constant $3<\alpha<4$ and conclude the mid-infrared peaks to
originate from large polaron formation. The high and narrow peaks positioned
at the lower frequencies with respect to the mid-infrared band are attributed
in Ref.~\cite{zVDM-PRL2008} to the optical absorption of the TO-phonons.%

%TCIMACRO{\FRAME{fhFU}{2.6697in}{3.8977in}{0pt}{\Qcb{Optical conductivity of
%SrTi$_{1-x}$Nb$_{x}$O$_{3}$ for 0.1\% (grey curves), 0.2\% (blue curves),
%0.9\% (\QTR{group}{green} curves) and 2\% (pink curves) at 300 K (panel
%\QTR{group}{a}) and 7 K (panel \QTR{group}{b}). For clarity, the mid-infrared
%conductivities of $x=0.1\%$ and 0.2\% are magnified by \QTR{group}{a} factor
%4. (From Ref.~\cite{zVDM-PRL2008}.)}}{}{sto-fig1.eps}%
%maintain-aspect-ratio TRUE;  display "ICON";  valid_file "F";
%width 2.6697in;  height 3.8977in;  depth 0pt;  original-width 1.7495in;
%original-height 2.5624in;  cropleft "0";  croptop "1";  cropright "1";
%cropbottom "0";  filename 'STO-Fig1.eps';file-properties "XNPEU";}} }%
%BeginExpansion
\begin{figure}
[h]
\begin{center}
\includegraphics[
height=3.8977in,
width=2.6697in
]%
{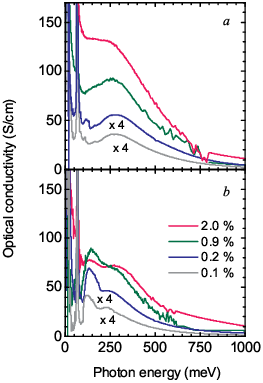}%
\caption{Optical conductivity of SrTi$_{1-x}$Nb$_{x}$O$_{3}$ for 0.1\% (grey
curves), 0.2\% (blue curves), 0.9\% ({green} curves) and 2\% (pink curves) at
300 K (panel {a}) and 7 K (panel {b}). For clarity, the mid-infrared
conductivities of $x=0.1\%$ and 0.2\% are magnified by {a} factor 4. (From
Ref.~\cite{zVDM-PRL2008}.)}%
\end{center}
\end{figure}
%EndExpansion

There are different types of polaron states in solids. In the effective mass
approximation for the electron placed in a continuum polarizable medium, a
so-called large or continuum polaron can exist. Large polaron wave functions
and the corresponding lattice distortions spread over many lattice sites. Due
to the finite phonon frequencies the ion polarizations can follow the polaron
motion if the motion is sufficiently slow. Hence, large polarons with a low
kinetic energy propagate through the lattice as free electrons but with an
enhanced effective mass. When the polaron binding energy is larger than the
half bandwidth of the electron band, all states in the Bloch bands are
`dressed' by phonons. In this strong-coupling regime, the finite electron
bandwidth becomes important, so the continuum approximation cannot be applied.
In this case the carriers are described as \textquotedblleft
small\textquotedblright\ or discrete (lattice) polarons that can hop between
different states localized at lattice sites. A key distinction between large
and small polarons is then the radius of the polaron state. For large
polarons, that radius substantially exceeds the lattice constant, while for
small polarons it is comparable to the lattice constant. A review of the
properties of large and small polarons can be found, e.~g., in Refs.
\cite{zReview2009,zBook}. In the theory of \textquotedblleft
mixed\textquotedblright\ polarons \cite{zE1,zE2,zE3,zEagles85} the states of
the electron-phonon system are composed of a mixture of large and small
polaron states.

Polaron states are formed due to the electron-phonon interaction, which is
different in the cases of large and small polarons. For a large polaron, the
electron-phonon interaction is provided by a macroscopic (continuum)
polarization of the lattice. This interaction is characterized by the coupling
constant $\alpha$ introduced by Fr\"{o}hlich \cite{zFr1954},%
\begin{equation}
\alpha=\frac{1}{2}\left(  \frac{1}{\varepsilon_{\infty}}-\frac{1}%
{\varepsilon_{0}}\right)  \frac{e^{2}}{\hbar\omega_{L}}\left(  \frac
{2m_{b}\omega_{L}}{\hbar}\right)  ^{1/2}, \label{alpha}%
\end{equation}
where $\varepsilon_{\infty}$ and $\varepsilon_{0}$ are, respectively, the
high-frequency and the static dielectric constants, $e$ is the electron
charge, $\omega_{L}$ is the longitudinal optical (LO) phonon frequency in the
Brillouin zone centre, and $m_{b}$ is the band electron (or hole) mass. The
large-polaron coupling constant is thus expressed through macroscopic
observable parameters of a polarizable medium. On the contrary, the
electron-phonon interaction for a small polaron is described through
microscopic parameters.

The nature of the polaron states in SrTi$_{1-x}$Nb$_{x}$O$_{3}$ is not yet
clear. Previous optical measurements on strontium titanate were interpreted in
terms of small polarons \cite{zReik67,zEagles85}. However, that assumption
contradicts the interpretation of transport measurements \cite{zFrederikse},
which rather support the large-polaron picture. Also the heat capacity
measurements \cite{zAmbler66}, provide effective masses similar to those of
large polarons. In Ref.~\cite{zEagles96}, the experimental results of Ref.
\cite{zGervais93} on the temperature-dependent plasma frequencies in
SrTi$_{1-x}$Nb$_{x}$O$_{3}$ were interpreted within the theory of mixed
polarons \cite{zE1,zE2,zE3,zEagles85}. Thermoelectric power measurements
\cite{zFrederikse} have shown that the density-of-states masses increase with
increasing temperature, which can be explained by a theory of mixed polarons
\cite{zE1}. It has been supposed \cite{zJPCM2006} that the polaron optical
conductivity in SrTi$_{1-x}$Nb$_{x}$O$_{3}$ is probably provided by mixed
polarons. A possible coexistence of large and small mass polarons has been
suggested in Ref. \cite{zIadonisi1998}. In Ref. \cite{zEagles95}, coexistence
of small and large polarons in the same solid is invoked to interpret
experimental data on the optical absorption in oxides.

The key question is to determine the type of polarons that provide the
mechanism of the polaron optical conductivity in SrTi$_{1-x}$Nb$_{x}$O$_{3}$.
The optical response of large polarons in various approximations was studied,
e.~g., in Refs. \cite{zGLF,zDHL1971,zKED1969,zDSG}. The same problem for the
small polaron was investigated in \cite{zReik67,zEmin1993}. In the
large-polaron theory, the optical absorption is provided by transitions (with
$0,1,\ldots$ phonon emission) between different continuum electron states. In
the small-polaron theory, the optical absorption occurs when the self-trapped
carrier is induced to transfer from its localized state to a localized state
at an adjacent site, with emission of phonons. Because of the different
physical mechanisms involved, the optical conductivity spectra of large and
small polarons are different from each other. In the large-polaron theory the
polaron optical conductivity behaves at high frequencies $\Omega$ as a power
function $\left(  \propto\Omega^{-5/2}\right)  $. In the small-polaron theory,
the polaron optical conductivity at high frequencies decreases much faster
than for large polarons: as a Gaussian exponent. Therefore the analysis of
optical measurements can shed some light on the aforesaid question on the type
of polarons responsible for the optical conductivity in SrTi$_{1-x}$Nb$_{x}%
$O$_{3}$.

The polaron optical conductivity band of SrTi$_{1-x}$Nb$_{x}$O$_{3}$ occupies
the mid-infrared range of the photon energies $\hbar\Omega\lesssim1$ eV, and
the threshold for interband electron-hole transitions lies at the band gap
energy, which is around 3.3 eV in SrTi$_{1-x}$Nb$_{x}$O$_{3}$
\cite{zVDM-PRL2008}. Therefore interband transitions do not interfere with the
polaron optical conductivity. Other mechanisms of electron intraband
scattering (for example, electron-phonon interaction with acoustic phonons
and/or electron or hole transitions from impurity centers) may be manifested
together with the polaron mechanism in the energy range $\hbar\Omega\lesssim1$
eV. The treatment of those mechanisms is, however, beyond the scope of the
present investigation.

We can make some preliminary suggestions concerning the dominating mechanism
of the mid-infrared optical conductivity in the Nb doped strontium titanate.
The low-frequency edge of the mid-infrared band in SrTi$_{1-x}$Nb$_{x}$O$_{3}$
at a low temperature ($T=7$ K) lies in the range of the LO-phonon energies
obtained in \cite{zGervais93}. The maximum of the mid-infrared band lies
relatively close to this low-frequency edge (the difference in frequency
between the low-frequency edge and the maximum of the mid-infrared band is
comparable to the LO-phonon frequencies in SrTi$_{1-x}$Nb$_{x}$O$_{3}$). This
behavior is characteristic of large-polaron optical conductivity rather than
of small-polaron optical conductivity. Indeed, the maximum of the small
polaron optical conductivity band is expected to be shifted to considerably
higher frequencies with respect to the low-frequency edge of the polaron
optical conductivity band (see, e.g.,~Ref. \cite{zEmin1993}). Also, at
sufficiently high frequencies, the experimental mid-infrared band from Ref.
\cite{zVDM-PRL2008} decreases with increasing $\Omega$ rather slowly, which is
characteristic for large-polaron optical conductivity rather than for
small-polaron optical conductivity. We therefore can suggest that the
large-polaron picture is the most appropriate for the interpretation of the
mid-infrared band of SrTi$_{1-x}$Nb$_{x}$O$_{3}$ observed in Ref.
\cite{zVDM-PRL2008}.

In order to interpret the mid-infrared band of the experimental optical
conductivity spectra of SrTi$_{1-x}$Nb$_{x}$O$_{3}$ \cite{zVDM-PRL2008} in
terms of polarons, we calculate the large-polaron optical conductivity spectra
for SrTi$_{1-x}$Nb$_{x}$O$_{3}$ using the model for the optical conductivity
of a large-polaron gas developed in Ref. \cite{zTD2001}, adapted to take into
account multiple LO-phonon branches \cite{zdraft}. The degeneracy and the
anisotropy of the conduction band in SrTi$_{1-x}$Nb$_{x}$O$_{3}$ are taken
into account.

\subsection{Optical conductivity of a gas of large polarons \label{sec:theory}%
}

The optical absorption spectra of SrTi$_{1-x}$Nb$_{x}$O$_{3}$ are sensitive to
the doping level \cite{zVDM-PRL2008}. Therefore a many-polaron description is
in order. In our context, \textquotedblleft many-polaron
description\textquotedblright\ means an account of many-electron effects on
the optical conductivity of a polaron gas. These effects include the influence
of the electron-electron Coulomb interaction (which leads to screening
effects) and of the Fermi statistics of the polaron gas on the optical
conductivity spectra. In the low-density limit, those many-body effects are
not important, and the optical conductivity of a polaron gas is well described
by the optical conductivity of a single polaron multiplied by the electron
density. The scope of the present study embraces a wide range of electron
densities for which the single-polaron approach is, in general, insufficient.
As shown below, even at the lowest electron density involved in the experiment
\cite{zVDM-PRL2008}, the shape and magnitude of the optical conductivity
spectrum is strongly affected by many-body effects.

We wish to compare the experiments of Ref. \cite{zVDM-PRL2008}, in particular
the observed mid-infrared band, to the theoretical optical conductivity of a
gas of large polarons. For that purpose we use the many-body large polaron
approach of Refs. \cite{zTD2001,zdraft}, which takes into account the
electron-electron interaction and the Fermi statistics of polarons.

Refs. \cite{zTD2001,zdraft} are limited to the study of weak-coupling
polarons. Up to $\alpha\approx3$, which includes the case of SrTi$_{1-x}%
$Nb$_{x}$O$_{3}$, the weak coupling approximation can be expected to describe
the main characteristics of the many-polaron optical response (see, e.g.,
Refs. \cite{zTD2001,zReview2009,zBook}). In Ref. \cite{zdraft} a
generalization of Ref. \cite{zTD2001} is presented that takes into account the
electron-phonon interaction with \emph{multiple LO-phonon branches} as they
exist, e. g., in complex oxides. For a single polaron, effects related to
multiple LO-phonon branches were investigated in Ref. \cite{zFerro}. The
starting point for the treatment of a many-polaron system is the Fr\"{o}hlich
Hamiltonian%
\begin{align}
H  &  =\sum_{\mathbf{k}}\sum_{\sigma=\pm1/2}\frac{\hbar^{2}k^{2}}{2m_{b}%
}c_{\mathbf{k},\sigma}^{+}c_{\mathbf{k},\sigma}+\sum_{\mathbf{q}}\sum
_{j=1}^{n}\hbar\omega_{L,j}a_{\mathbf{q},j}^{+}a_{\mathbf{q},j}+U_{e-e}%
\nonumber\\
&  +\frac{1}{\sqrt{V}}\sum_{\mathbf{q}}\sum_{j=1}^{n}\left(  V_{\mathbf{q}%
,j}a_{\mathbf{q},j}\sum_{\mathbf{k}}\sum_{\sigma=\pm1/2}c_{\mathbf{k}%
+\mathbf{q},\sigma}^{+}c_{\mathbf{k},\sigma}+\mathtt{h.c.}\right)  ,
\label{zH}%
\end{align}
where $c_{\mathbf{k},\sigma}^{+}$ ($c_{\mathbf{k},\sigma}$) are the creation
(annihilation) operators for an electron with momentum $\mathbf{k}$ and with
the spin $z$-projection $\sigma$, $a_{\mathbf{q},j}^{+}$ ($a_{\mathbf{q},j}$)
are the creation (annihilation) operators for a phonon of the $j$-th branch
with the momentum $q$, $\omega_{L,j}$ are the LO-phonon frequencies
(approximated here as non-dispersive), and $V$ is the volume of the crystal.
The polaron interaction amplitude $V_{\mathbf{q},j}$ is \cite{zFerro}%
\begin{equation}
V_{\mathbf{q},j}=\frac{\hbar\omega_{L,j}}{q}\left(  \frac{4\pi\alpha_{j}}%
{V}\right)  ^{1/2}\left(  \frac{\hbar}{2m_{b}\omega_{L,j}}\right)  ^{1/4},
\label{V2}%
\end{equation}
where $\alpha_{j}$ is a dimensionless partial coupling constant characterizing
the interaction between an electron and the $j$-th LO-phonon branch. The
electron-electron interaction is described by the Coulomb potential energy
\begin{equation}
U_{e-e}=\frac{1}{2}\sum_{\mathbf{q}\neq0}\frac{4\pi e^{2}}{\varepsilon
_{\infty}q^{2}}\sum_{\mathbf{k},\mathbf{k}^{\prime},\sigma,\sigma^{\prime}%
}c_{\mathbf{k}+\mathbf{q},\sigma}^{+}c_{\mathbf{k}^{\prime}-\mathbf{q}%
,\sigma^{\prime}}^{+}c_{\mathbf{k}^{\prime},\sigma^{\prime}}c_{\mathbf{k}%
,\sigma}. \label{Uee}%
\end{equation}

Optical phonons in SrTiO$_{3}$ show a considerable dispersion (see, e. g.,
Ref. \cite{zChoudhury} and references therein). The effect of the phonon
dispersion is a broadening of features of the polaron optical conductivity
band. The magnitude of the broadening is characterized by the dispersion
parameter $\Delta\omega$ of the optical phonons, that contribute to the
integrals over $\mathbf{q}$ entering the polaron optical conductivity. In a
polar crystal with a single LO-phonon branch, that range of convergence is
approximately $q_{0}=\left(  m_{b}\omega_{LO}/\hbar\right)  ^{1/2}$. For
SrTiO$_{3}$, taking $\omega_{LO}=\max\left\{  \omega_{L,j}\right\}  $, we
obtain $q_{0}\approx1.02\times10^{9}%
%TCIMACRO{\unit{m}}%
%BeginExpansion
\operatorname{m}%
%EndExpansion
^{-1}$. The boundary of the Brillouin zone $\pi/a_{0}$ in SrTiO$_{3}$ (where
the lattice constant $a_{0}\approx0.3905$ nm) is at $8\times10^{9}%
%TCIMACRO{\unit{m}}%
%BeginExpansion
\operatorname{m}%
%EndExpansion
^{-1}$. Therefore the integration domain for the relevant integrals is one
order smaller than the size of the Brillouin zone. In the region $0<q<q_{0}$,
the dispersion parameter of the LO-phonon frequencies, $\Delta\omega$, is a
few percent of $\omega_{L,j}$. Consequently, $\Delta\omega$ is very small
compared with the characteristic width of the polaron band. Therefore, in the
present treatment, we apply the approximation of non-dispersive phonons.

For a description of a polarizable medium with $n$ optical-phonon branches, we
use the model dielectric function \cite{zT1972,zmmc3}\textrm{ }%
\begin{equation}
\varepsilon\left(  \omega\right)  =\varepsilon_{\infty}\prod_{j=1}^{n}\left(
\frac{\omega^{2}-\omega_{L,j}^{2}}{\omega^{2}-\omega_{T,j}^{2}}\right)  ,
\label{DF}%
\end{equation}
whose zeros (poles) correspond to the LO(TO) phonon frequencies $\omega_{L,j}$
($\omega_{T,j}$). This dielectric function is the result of the
straightforward extension of the Born-Huang approach \cite{zBH1954} to the
case where more than one optical-phonon branch exists in a polar crystal. The
Born-Huang approach and its extension \cite{zT1972} generate expressions for
the macroscopic polarization induced by the polar vibrations, and for the
corresponding electrostatic potential. This electrostatic potential is a basis
element of the Hamiltonian of the electron-phonon interaction. In Ref.
\cite{zT1972}, the Hamiltonian of the electron-phonon interaction has been
explicitly derived with the amplitudes%
\begin{equation}
V_{\mathbf{q}j}=\frac{1}{\sqrt{V}}\frac{e}{iq}\left(  \frac{4\pi\hbar}{\left.
\frac{\partial\varepsilon\left(  \omega\right)  }{\partial\omega}\right\vert
_{\omega=\omega_{L,j}}}\right)  ^{1/2}. \label{rrr3}%
\end{equation}
Using Eqs. (\ref{V2}) and (\ref{rrr3}) with the dielectric function
(\ref{DF}), we arrive at the following set of linear equations for the
coupling constants $\alpha_{j}$ ($j=1,\ldots,n$):%
\begin{equation}
\sum_{k=1}^{n}\hbar\omega_{L,k}^{3}\left(  \frac{\hbar}{2m_{b}\omega_{L,k}%
}\right)  ^{1/2}\frac{\alpha_{k}}{\omega_{L,k}^{2}-\omega_{T,j}^{2}}%
=\frac{e^{2}}{2\varepsilon_{\infty}}. \label{set0}%
\end{equation}
Knowledge of the band mass, of the electronic dielectric constant
$\varepsilon_{\infty}$ and of the LO- and TO-phonon frequencies is sufficient
to determine the coupling constants $\alpha_{j}$ taking into account mixing
between different optical-phonon branches. In the particular case of a single
LO-phonon branch, Eq. (\ref{set0}) is reduced to (\ref{alpha}).

In order to describe the optical conductivity of a polaron gas, we refer to
the work \cite{zPD1983}, where the Mori-Zwanzig projection operator technique
has been used to rederive the path-integral result of Ref. \cite{zFHIP} and
the impedance of Ref. \cite{zDSG}. We repeat the derivations of Ref.
\cite{zPD1983} with the replacement of single-electron functions by their
many-electron analogs. For example, $e^{i\mathbf{q\cdot r}}$ in the
Hamiltonian of the electron-phonon interaction is replaced by the Fourier
component of the electron density for an $N$-electron system,
\begin{equation}
\rho\left(  \mathbf{q}\right)  \equiv\sum_{s=1}^{N}e^{i\mathbf{q\cdot r}_{s}%
}=\sum_{\mathbf{k},\sigma}c_{\mathbf{k}+\mathbf{q},\sigma}^{+}c_{\mathbf{k}%
,\sigma}. \label{rhoq}%
\end{equation}
As a result, we arrive at a formula which is structurally similar to the
single-polaron optical conductivity \cite{zDSG,zPD1983},%
\begin{equation}
\sigma\left(  \Omega\right)  =\frac{e^{2}n_{0}}{m_{b}}\frac{i}{\Omega
-\chi\left(  \Omega\right)  /\Omega}, \label{z4}%
\end{equation}
where $n_{0}=N/V$ is the carrier density, and $\chi\left(  \Omega\right)  $ is
the memory function. The same many-electron derivation as in the present work,
to the best of our knowledge, was first performed for the polaron gas in 2D in
Ref. \cite{zWu1986} in the weak electron-phonon coupling limit.

In Refs. \cite{zDSG,zPD1983} the single-polaron memory function was calculated
starting from the all-coupling Feynman variational principle
\cite{zFeynman1955}. For a many-polaron system, an effective all-coupling
extension of that variational principle has not been worked out yet. In the
present treatment, we restrict ourselves to the weak-coupling approximation
for the electron-phonon interaction to derive the memory function. In this
approximation, the memory function $\chi\left(  \Omega\right)  $ is similar to
that of Ref. \cite{zWu1986}, with two distinctions: (1) the electron gas in
the present treatment is three-dimensional, (2) several LO phonon branches are
taken into account. The resulting form of the memory function is
\begin{align}
\chi\left(  \Omega\right)   &  =\frac{4}{3\hbar m_{b}n_{0}V}\sum
_{\mathbf{q},j}q^{2}\left\vert V_{\mathbf{q},j}\right\vert ^{2}\int
_{0}^{\infty}dt\left(  e^{i\Omega t}-1\right) \nonumber\\
&  \times\operatorname{Im}\left[  \frac{\cos\left[  \omega_{L,j}\left(
t+i\hbar\beta/2\right)  \right]  }{\sinh\left(  \beta\hbar\omega
_{L,j}/2\right)  }S\left(  \mathbf{q},t\right)  \right]  , \label{xi}%
\end{align}
where $\beta=1/\left(  k_{B}T\right)  $. The dynamical structure factor
$S\left(  \mathbf{q},t\right)  $ is proportional to the two-point correlation
function (cf. Ref. \cite{zTD2001}),%
\begin{equation}
S\left(  \mathbf{q},t\right)  \equiv\frac{1}{2}\left\langle \sum_{i,j=1}%
^{N}e^{i\mathbf{q}\cdot\left[  \mathbf{r}_{j}\left(  t\right)  -\mathbf{r}%
_{k}\left(  0\right)  \right]  }\right\rangle =\frac{1}{2}\left\langle
\rho\left(  \mathbf{q,}t\right)  \rho\left(  -\mathbf{q},0\right)
\right\rangle . \label{zFF}%
\end{equation}
To obtain $\chi\left(  \Omega\right)  $ to order $\alpha$ it is sufficient to
perform the averaging in the correlation function (\ref{zFF}) using the
Hamiltonian (\ref{zH}) without the electron-phonon interaction and keeping the
electron-electron interaction term $U_{e-e}$.

We calculate the dynamical structure factor (\ref{zFF}) extending the method
\cite{zTD2001} to nonzero temperatures. In Ref. \cite{zTD2001}, the key
advantage of the many-polaron variational approach \cite{zLDB77} is exploited:
the fact that the many-body effects are entirely contained in the dynamical
structure factor $S\left(  \mathbf{q},t\right)  $. The structure factor can be
calculated using various approximations. Terms of order of $|V_{\mathbf{q}%
,j}|^{2}$ are automatically taken into account in the memory function
(\ref{xi}). Consequently, up to order $\alpha$ for $\sigma\left(
\Omega\right)  $, it is sufficient to calculate $S\left(  \mathbf{q},t\right)
$ without the electron-phonon coupling. In Ref. \cite{zTD2001}, $S\left(
\mathbf{q},t\right)  $ was calculated within two different approximations: (i)
the Hartree-Fock approximation, (ii) the random-phase approximation (RPA). As
shown in Ref. \cite{zTD2001}, the RPA dynamical structure factor, contrary to
the Hartree-Fock approximation, takes into account the effects both of the
Fermi statistics and of the electron-electron interaction on the many-polaron
optical-absorption spectra.

The dynamical structure factor is expressed through the density-density
Green's functions defined as%
\begin{align}
\mathcal{G}\left(  \mathbf{q},\Omega\right)   &  \equiv-i\int_{0}^{\infty
}e^{i\Omega t}\left\langle \rho\left(  \mathbf{q,}t\right)  \rho\left(
-\mathbf{q},0\right)  \right\rangle dt,\label{zGF1}\\
G^{R}\left(  \mathbf{q},\Omega\right)   &  \equiv-i\int_{0}^{\infty}e^{i\Omega
t}\left\langle \left[  \rho\left(  \mathbf{q,}t\right)  ,\rho\left(
-\mathbf{q},0\right)  \right]  \right\rangle dt. \label{GF2}%
\end{align}
In terms of $G\left(  \mathbf{q},\Omega\right)  $ and $G^{R}\left(
\mathbf{q},\Omega\right)  $, the memory function (\ref{xi}) takes the form:%
\begin{align}
\chi\left(  \Omega\right)   &  =\sum_{j}\frac{\alpha_{j}\hbar\omega_{L,j}^{2}%
}{6\pi^{2}Nm_{b}}\left(  \frac{\hbar}{2m_{b}\omega_{L,j}}\right)
^{1/2}\nonumber\\
&  \times\int d\mathbf{q}\left\{  \mathcal{G}\left(  \mathbf{q},\Omega
-\omega_{L,j}\right)  +\mathcal{G}^{\ast}\left(  \mathbf{q},-\Omega
-\omega_{L,j}\right)  -\mathcal{G}\left(  \mathbf{q},-\omega_{L,j}\right)
-\mathcal{G}^{\ast}\left(  \mathbf{q},-\omega_{L,j}\right)  \right.
\nonumber\\
&  +\frac{1}{e^{\beta\hbar\omega_{L,j}}-1}\left[  G^{R}\left(  \mathbf{q}%
,\Omega-\omega_{L,j}\right)  +\left(  G^{R}\left(  \mathbf{q},-\Omega
-\omega_{L,j}\right)  \right)  ^{\ast}\right. \nonumber\\
&  \left.  \left.  -G^{R}\left(  \mathbf{q},-\omega_{L,j}\right)  -\left(
G^{R}\left(  \mathbf{q},-\omega_{L,j}\right)  \right)  ^{\ast}\right]
\right\}  . \label{MF}%
\end{align}
Taking into account the Coulomb electron-electron interaction within RPA, the
retarded Green's function $G^{R}\left(  \mathbf{q},\Omega\right)  $ is given
by%
\begin{equation}
G^{R}\left(  \mathbf{q},\Omega\right)  =\frac{\hbar VP^{\left(  1\right)
}\left(  \mathbf{q},\Omega\right)  }{1-\frac{4\pi e^{2}}{\varepsilon_{\infty
}q^{2}}P^{\left(  1\right)  }\left(  \mathbf{q},\Omega\right)  }, \label{GH}%
\end{equation}
where $P^{\left(  1\right)  }\left(  \mathbf{q},\Omega\right)  $ is the
polarization function of the free electron gas, see, e.g., \cite{zMahan}%
\begin{equation}
P^{\left(  1\right)  }\left(  \mathbf{q},\Omega\right)  =\frac{1}{V}%
\sum_{\mathbf{k},\sigma}\frac{f_{\mathbf{k}+\mathbf{q},\sigma}-f_{\mathbf{k}%
,\sigma}}{\hbar\Omega+\frac{\hbar^{2}\left(  \mathbf{k}+\mathbf{q}\right)
^{2}}{2m_{b}}-\frac{\hbar^{2}k^{2}}{2m_{b}}+i\delta},\quad\delta\rightarrow+0
\label{zP1}%
\end{equation}
with the electron average occupation numbers $f_{\mathbf{k},\sigma}$. The
function $\mathcal{G}\left(  \mathbf{q},\Omega\right)  $ is obtained from
$G^{R}\left(  \mathbf{q},\Omega\right)  $ using the exact analytical relation%
\begin{equation}
\left(  1-e^{-\beta\hbar\Omega}\right)  \operatorname{Im}\mathcal{G}\left(
\mathbf{q},\Omega\right)  =\operatorname{Im}G^{R}\left(  \mathbf{q}%
,\Omega\right)  \label{pr1}%
\end{equation}
and the Kramers-Kronig dispersion relations for $\mathcal{G}\left(
\mathbf{q},\Omega\right)  $.

The above expressions are written for an isotropic conduction band. However,
the conduction band of SrTi$_{1-x}$Nb$_{x}$O$_{3}$ is strongly anisotropic and
triply degenerate. The electrons are doped in three bands: $d_{xy}$, $d_{yz}$
and $d_{xz}$, which all have their minima at $\mathbf{k}=0$. Each of these
bands has light masses along two direction ($x$ and $y$ for $d_{xy}$, etc.)
and a heavy mass along the third direction. While each electron has a strongly
anisotropic mass, the electronic transport remains isotropic due to the fact
that 2 light masses and 1 heavy mass contribute along each crystallographic axis.

The anisotropy of the electronic effective mass {of} the conduction band can
be approximately taken into account in the following way. We use the averaged
inverse band mass
\begin{equation}
\frac{1}{\bar{m}_{b}}=\frac{1}{3}\left(  \frac{1}{m_{x}}+\frac{1}{m_{y}}%
+\frac{1}{m_{z}}\right)  \label{mb1}%
\end{equation}
and the density-of-states band mass
\begin{equation}
m_{D}=\left(  m_{x}m_{y}m_{z}\right)  ^{1/3}. \label{md}%
\end{equation}
{The mass }${m}_{D}$ {appears in the prefactor of the linear term of the
specific heat. Comparing the mass }$m_{D}${ obtained from the experimental
specific heat~\cite{zAmbler66,zphillips} with the mass }$\bar{m}_{b}${
obtained using optical spectral weights \cite{zVDM-PRL2008} reveals the mass
ratio of the heavy and light bands to be about 27.} The expression (\ref{mb1})
replaces the bare mass $m_{b}$ in the optical conductivity (\ref{z4}) and in
the memory function (\ref{MF}). The polarization function of the free electron
gas (\ref{zP1}) is calculated with the density-of-states mass $m_{D}$ instead
of $m_{b}$. The band degeneracy is taken into account through the degeneracy
factor which is equal to 3, both in the polarization function and in the
normalization equation for the chemical potential.\ The reduction of the
polaron optical conductivity band due to screening with band degeneracy turns
out to be less significant than without band degeneracy.

\subsubsection{Theory and experiment \label{sec:results}}

\subsubsection{Material parameters}

Several experimental parameters characterizing SrTi$_{1-x}$Nb$_{x}$O$_{3}$ are
necessary for the calculation of the large-polaron optical conductivity (see,
e.g., Refs. \cite{zJPCM2006,zGervais93}): the LO- and TO-phonon frequencies,
the electron band mass, and the electronic dielectric constant $\varepsilon
_{\infty}$.

The electronic dielectric constant can be obtained using reflectivity spectra
of SrTi$_{1-x}$Nb$_{x}$O$_{3}$. At $T=10$ K, the reflectivity of SrTi$_{1-x}%
$Nb$_{x}$O$_{3}$ is $R\approx0.16$ for $\Omega\approx5000$ cm$^{-1}$. The
electronic dielectric constant can be approximated using the expression%
\begin{equation}
R\left(  \Omega\right)  =\left\vert \frac{\sqrt{\varepsilon\left(
\Omega\right)  }-1}{\sqrt{\varepsilon\left(  \Omega\right)  }+1}\right\vert
^{2} \label{zR}%
\end{equation}
and assuming that $\Omega=5000$ cm$^{-1}$ is a sufficiently high frequency to
characterize the electronic response. From (\ref{zR}) it follows that for
SrTi$_{1-x}$Nb$_{x}$O$_{3}$, $\varepsilon_{\infty}\approx5.44$.

In order to determine the optical-phonon frequencies, we use {the}
experimental data from available sources \cite{zVDM-PRL2008,zGervais93}. In
Ref. \cite{zVDM-PRL2008}, three infrared active phonon modes are observed at
room temperature: at 11.0 meV, 21.8 meV and 67.6 meV. With decreasing
temperature, the lowest-frequency infrared-active phonon mode shows a strong
red shift upon cooling, and saturates at about 2.3 meV at 7 K. Those
infrared-active phonon modes are associated with the polar TO-phonons. The
TO-phonon frequencies determined in Ref. \cite{zGervais93} for SrTi$_{1-x}%
$Nb$_{x}$O$_{3}$ with $x=0.9\%$ at $T=300$ K are 100 cm$^{-1}$, 175 cm$^{-1}$
and 550 cm$^{-1}$. The corresponding TO-phonon energies are 12.4 meV, 21.7 meV
and 68.2 meV.

Refs. \cite{zVDM-PRL2008} and \cite{zGervais93} are used as sources for phonon
parameters. In Ref. \cite{zGervais93}, the TO-phonon frequencies are
calculated on the basis of reflectivity measurements using a model dielectric
function to fit experimental data. In Ref. \cite{zVDM-PRL2008}, the TO-phonon
frequencies are obtained from an analysis of both reflectivity and
transmission spectra, using inversion of the Fresnel equations of reflection
and transmission coefficients and the Kramers-Kronig transformation of the
reflectivity spectra. The TO-phonon energies reported in Refs.
\cite{zVDM-PRL2008} and \cite{zGervais93} are in close agreement. This
confirms the reliability of both experimental data sources
\cite{zVDM-PRL2008,zGervais93}. The values of the TO-phonon frequencies used
in our calculation are taken from the experiment \cite{zVDM-PRL2008} because
they are directly related to the samples of SrTi$_{1-x}$Nb$_{x}$O$_{3}$ for
which the comparison of theory and experiment is made in the present work.

The TO phonon frequencies from Ref. \cite{zVDM-PRL2008} can be used when they
are complemented with corresponding LO phonon frequencies. However,
Ref.~\cite{zVDM-PRL2008} does not contain data of the LO-phonon frequencies.
In the present calculation we use the LO phonon frequencies from
Ref.~\cite{zGervais93}.

The averaged band mass (\ref{mb1}) is taken to be $\bar{m}_{b}=0.81m_{e}$
(where $m_{e}$ is the electron mass in vacuum) according to experimental data
from Ref.~\cite{zComments2}. Using the ratio of the heavy mass ($m_{z}$) to
the light mass ($m_{x}=m_{y}$), $m_{z}/m_{x}=27$, we find the density-of
states band mass $m_{D}\approx1.65m_{e}.$

The TO- and LO- phonon frequencies and the resulting partial coupling
constants calculated using the mass $\bar{m}_{b}$ are presented in Table 1.%

%TCIMACRO{\TeXButton{B}{\begin{table}[h] \centering}}%
%BeginExpansion
\begin{table}[h] \centering
%EndExpansion
\caption{Optical-phonon frequencies and partial coupling constants of doped strontium titanate}%
\begin{tabular}
[c]{|l|l|l|l|l|l|l|l|l|}\hline
$x$ & $x=0.1\%$ & $x=0.1\%$ & $x=0.2\%$ & $x=0.2\%$ & $x=0.9\%$ & $x=0.9\%$ &
$x=2\%$ & $x=2\%$\\
$T$ & $T=7$ K & $T=300$ K & $T=7$ K & $T=300$ K & $T=7$ K & $T=300$ K & $T=7$
K & $T=300$ K\\\hline
$\hbar\omega_{T,1}$ (meV) & 2.27 & 11.5 & 2.63 & 11.5 & 6.01 & 12.1 & 8.51 &
13.0\\\hline
$\hbar\omega_{L,1}$ (meV) & 21.2 & 21.2 & 21.2 & 21.2 & 21.2 & 21.2 & 21.2 &
21.2\\\hline
$\alpha_{1}$ & 0.021 & 0.013 & 0.021 & 0.013 & 0.017 & 0.013 & 0.017 &
0.013\\\hline
$\hbar\omega_{T,2}$ (meV) & 21.2 & 21.8 & 21.2 & 21.8 & 21.2 & 21.8 & 21.2 &
21.8\\\hline
$\hbar\omega_{L,2}$ (meV) & 58.4 & 58.4 & 58.4 & 58.4 & 58.4 & 58.4 & 58.4 &
58.4\\\hline
$\alpha_{2}$ & 0.457 & 0.414 & 0.457 & 0.414 & 0.452 & 0.414 & 0.447 &
0.409\\\hline
$\hbar\omega_{T,3}$ (meV) & 67.6 & 67.1 & 67.6 & 67.1 & 67.6 & 67.1 & 67.6 &
67.1\\\hline
$\hbar\omega_{L,3}$ (meV) & 98.7 & 98.7 & 98.7 & 98.7 & 98.7 & 98.7 & 98.7 &
98.7\\\hline
$\alpha_{3}$ & 1.582 & 1.582 & 1.582 & 1.580 & 1.576 & 1.578 & 1.570 &
1.574\\\hline
$\alpha_{\mathrm{eff}}$ & 2.06 & 2.01 & 2.06 & 2.01 & 2.05 & 2.01 & 2.03 &
2.01\\\hline
\end{tabular}
\label{table1}%
%TCIMACRO{\TeXButton{E}{\end{table}}}%
%BeginExpansion
\end{table}%
%EndExpansion

The effective coupling constant in Table 1 is determined following Ref.
\cite{zFerro}, as a sum of partial coupling constants $\alpha_{j}$,%
\begin{equation}
\alpha_{\mathrm{eff}}\equiv\sum_{j}\alpha_{j} \label{aeff}%
\end{equation}
The result $\alpha_{\mathrm{eff}}\sim2$ shows that the electron-phonon
coupling strength in SrTi$_{1-x}$Nb$_{x}$O$_{3}$ lies in the {intermediate to
weak coupling range}, and the conditions for small polaron formation are not
fulfilled. This analysis indicates that the large-polaron picture -- rather
than the small-polaron description is suitable for the interpretation of the
mid-infrared band of the optical conductivity of SrTi$_{1-x}$Nb$_{x}$O$_{3}$.

We use the actual electron densities for the samples studied in
Ref.~\cite{zVDM-PRL2008} based on the unit cell volume (59.5 cubic angstrom)
and the chemical composition ($x$ is the doping level). These carrier
densities (see Table 2) are confirmed by measurements of the Hall constants.%

%TCIMACRO{\TeXButton{B}{\begin{table}[h] \centering}}%
%BeginExpansion
\begin{table}[h] \centering
%EndExpansion
\caption{Electron densities of SrTi$_{1-x}$Nb$_{x}$O$_3$}
\begin{tabular}
[c]{|l|l|}\hline
$x\,(\%)$ & $n_{0}$ (cm$^{-3}$)\\\hline
0.1 & $1.7\times10^{19}$\\\hline
0.2 & $3.4\times10^{19}$\\\hline
0.9 & $1.5\times10^{20}$\\\hline
2.0 & $3.4\times10^{20}$\\\hline
\end{tabular}
\label{Table2}%
%TCIMACRO{\TeXButton{E}{\end{table}}}%
%BeginExpansion
\end{table}%
%EndExpansion

\subsection{Optical conductivity spectra}

We calculate the large-polaron optical conductivity spectra for SrTi$_{1-x}%
$Nb$_{x}$O$_{3}$ using the approach of Ref. \cite{zTD2001} as adapted in Ref.
\cite{zdraft} to take into account multiple LO-phonon branches. We also
include in the numerical calculation the TO-phonon contribution to the optical
conductivity, described by an oscillatory-like model dielectric function (see,
e.g., Ref. \cite{zGervais93}):%
\begin{equation}
\operatorname{Re}\sigma_{TO}\left(  \Omega\right)  =\sum_{j}\sigma_{0,j}%
\frac{\gamma_{j}^{2}}{\left(  \Omega-\omega_{T,j}\right)  ^{2}+\gamma_{j}^{2}%
}, \label{sto}%
\end{equation}
where the weight coefficients $\sigma_{0,j}$ and the damping parameters
$\gamma_{j}$ for each $j$-th TO-phonon branch are extracted from the
experimental optical conductivity spectra of Ref. \cite{zVDM-PRL2008}. The
polaron-and the TO-phonon optical responses are treated as independent {of}
each other. Consequently the polaron-(\ref{z4}) and TO-phonon (\ref{sto})
contributions enter the optical conductivity additively.

{Following the procedure described above using the material parameters
discussed above, we obtain the theoretical large-polaron optical conductivity
spectra of SrTi$_{1-x}$Nb$_{x}$O$_{3}$ shown in Fig.~2 and Fig.~3 at 7 K and
300 K, respectively. In each graph also the experimental optical conductivity
spectra of Ref.~\cite{zVDM-PRL2008} are shown. It should be emphasized that}
\emph{in the present calculation, there is no fitting of material constants
for the polaron contribution to }$\operatorname{Re}\sigma\left(
\Omega\right)  $\emph{. Even the magnitude of the optical conductivity, which
is often arbitrarily scaled in the literature, follows from first principles}.%

%TCIMACRO{\FRAME{fhFU}{6.0087in}{4.7885in}{0pt}{\Qcb{The many-large-polaron
%optical conductivity compared with the experiment \cite{zVDM-PRL2008} at $T=7$
%K. The doping level is $x=0.1\%$ (\QTR{em}{a}), 0.2\QTR{group}{\%}
%(\QTR{em}{b}), 0.9\% (\QTR{em}{c}) and 2\%(\QTR{em}{d}).}}{}{sto-fig2.eps}%
%maintain-aspect-ratio TRUE;  display "ICON";  valid_file "F";
%width 6.0087in;  height 4.7885in;  depth 0pt;  original-width 6.0148in;
%original-height 4.7876in;  cropleft "0";  croptop "1";  cropright "1";
%cropbottom "0";  filename 'STO-Fig2.eps';file-properties "XNPEU";}} }%
%BeginExpansion
\begin{figure}
[h]
\begin{center}
\includegraphics[
height=4.7885in,
width=6.0087in
]%
{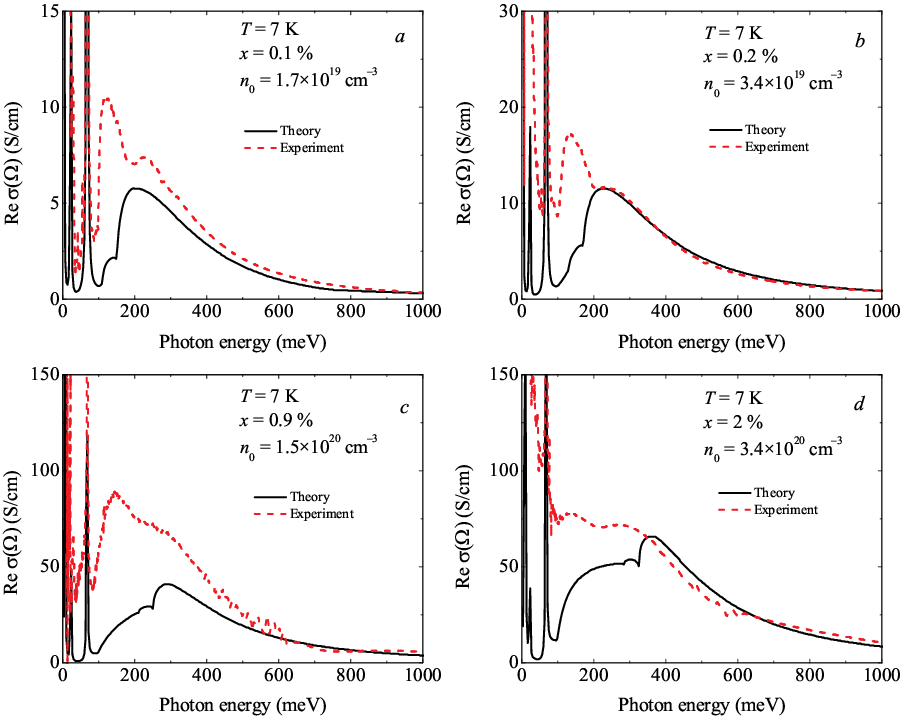}%
\caption{The many-large-polaron optical conductivity compared with the
experiment \cite{zVDM-PRL2008} at $T=7$ K. The doping level is $x=0.1\%$
(\emph{a}), 0.2{\%} (\emph{b}), 0.9\% (\emph{c}) and 2\%(\emph{d}).}%
\end{center}
\end{figure}
%EndExpansion

At 7 K, the calculated optical conductivity based on the Fr\"{o}hlich model
and extended for a gas of large polarons as described in the present paper,
shows convincing agreement with the behavior of the experimental optical
conductivity for the high energy part of the spectra, i.e., $\hbar
\Omega\gtrapprox300$ meV. The experimental polaron optical conductivity of
SrTi$_{1-x}$Nb$_{x}$O$_{3}$ falls down at high frequencies following the power
law (derived in the present work and typical for large polarons) rather than
as a Gaussian exponent that would follow from the small-polaron theory. At
lower photon energies $\hbar\Omega\lessapprox200$ meV, the experiment shows
distinct peaks that are not explained within the polaron theory. They can be
due to other scattering mechanisms as discussed below.

The minor deviations between theoretical and experimental $\operatorname{Re}%
\sigma\left(  \Omega\right)  $ in the frequency range $\hbar\Omega
\gtrapprox300$ meV may be attributed to the difference between the actual
electron densities and the densities calculated on the basis of the unit cell
volume and the chemical composition. However, we prefer not to fit of the density.

The optical conductivity calculated for a single large-polaron
absorption~\cite{zDSG} predicts an intensity 3-4 times larger than the
experimental data for the lowest doping level $x=0.1\%$, and therefore cannot
explain those data. For higher dopings, the overestimation of the magnitude of
the optical conductivity within the single-polaron theory is even larger than
for $x=0.1\%$. Therefore the many-polaron approach, used in the present work,
is essential.

At 300 K, in Fig.~3~(\emph{a},\ \emph{b\ },\emph{d}), the agreement between
theory and experiment is qualitative. Both experimental and theoretical
spectra show a maximum at the room-temperature optical conductivity spectra in
the range $\hbar\Omega\sim250$ meV. For the doping level $x=0.9\%$ the
calculated optical conductivity spectrum underestimates the experimental data,
as also observed at 7 K.

Many-body effects considerably influence the optical conductivity spectra of a
polaron gas. First, features related to the emission of a plasmon together
with a LO phonon \cite{zTD2001} are manifested in the optical conductivity
spectra of the many-polaron gas at $T=7$ K as separate peaks whose positions
shift to higher energies with increasing doping level. At room temperature,
those peaks are strongly broadened and smoothened, and only a broad plasmon
feature is apparent. Second, the mid-infrared optical conductivity (per
particle) in SrTi$_{1-x}$Nb$_{x}$O$_{3}$ is decreasing at higher doping levels
due to the screening of the polar interactions, which is accounted for in the
present approach in which $S\left(  \mathbf{q},t\right)  $ is based on RPA.
The effect of screening can be illustrated by the fact that for $n_{0}%
\sim10^{20}$ cm$^{-3}$, the many-polaron optical conductivity \emph{per
particle} is reduced by about an order of magnitude compared to the
single-polaron optical conductivity. The reduction in intensity of the polaron
optical conductivity band can be interpreted as a decrease of the overall
electron-phonon coupling strength due to many-body effects. Correspondingly,
at high doping levels, the polaron mass $m^{\ast}$, determined by the sum rule
introduced in Ref. \cite{zDLR77}%
\begin{equation}
\frac{\pi e^{2}n_{0}}{2m^{\ast}}+\int_{\omega_{L}}^{\infty}\operatorname{Re}%
\left(  \Omega\right)  d\Omega=\frac{\pi e^{2}n_{0}}{2\bar{m}_{b}} \label{zsr}%
\end{equation}
is reduced, compared to the single-polaron effective mass. As shown in Refs.
\cite{zTD2001,zTD2001a}, the sum rule \cite{zDLR77} remains valid for an
interacting polaron gas.%

%TCIMACRO{\FRAME{fhFU}{6.0087in}{4.7885in}{0pt}{\Qcb{The many-large-polaron
%optical conductivity compared with the experiment \cite{zVDM-PRL2008} at
%$T=300$ K. The doping level is $x=0.1\%$ (\QTR{em}{a}), 0.2\QTR{group}{\%}
%(\QTR{em}{b}), 0.9\% (\QTR{em}{c}) and 2\% (\QTR{em}{d}).}}{}{sto-fig3.eps}%
%maintain-aspect-ratio TRUE;  display "ICON";  valid_file "F";
%width 6.0087in;  height 4.7885in;  depth 0pt;  original-width 6.0148in;
%original-height 4.7876in;  cropleft "0";  croptop "1";  cropright "1";
%cropbottom "0";  filename 'STO-Fig3.eps';file-properties "XNPEU";}} }%
%BeginExpansion
\begin{figure}
[h]
\begin{center}
\includegraphics[
height=4.7885in,
width=6.0087in
]%
{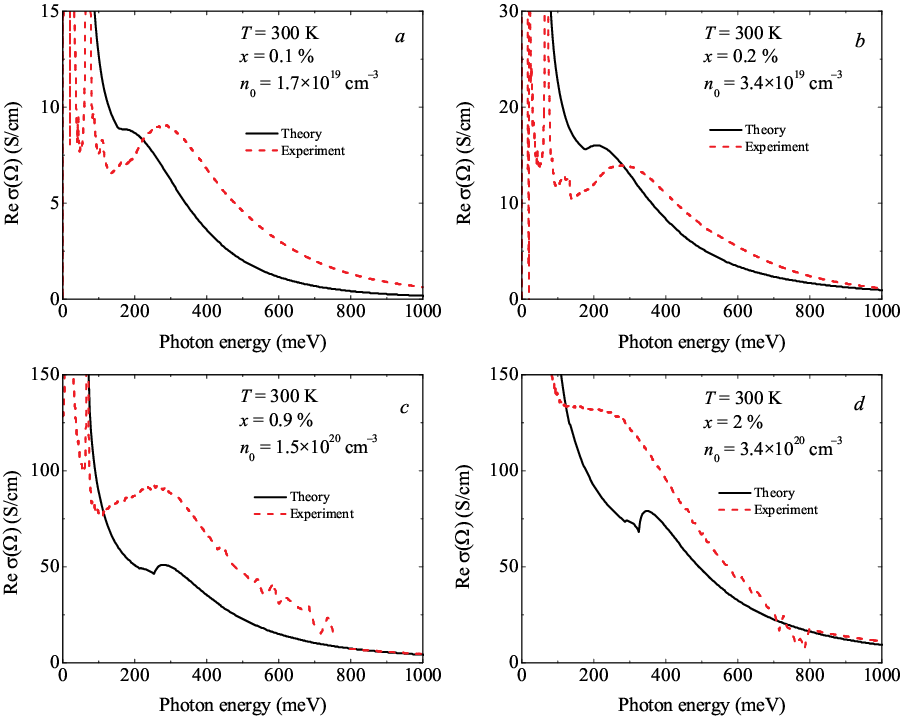}%
\caption{The many-large-polaron optical conductivity compared with the
experiment \cite{zVDM-PRL2008} at $T=300$ K. The doping level is $x=0.1\%$
(\emph{a}), 0.2{\%} (\emph{b}), 0.9\% (\emph{c}) and 2\% (\emph{d}).}%
\end{center}
\end{figure}
%EndExpansion

The large-polaron theory of the optical absorption based on Ref.
\cite{zTD2001} explains without any fitting parameters the main
characteristics and trends of the observed spectra of Ref. \cite{zVDM-PRL2008}
in SrTi$_{1-x}$Nb$_{x}$O$_{3}$, including doping- and temperature dependence.
Nevertheless, some features of the experimental spectra remain to be
explained. In particular, at $T=7$ K, the pronounced peak at $\hbar\Omega
\sim130$ meV in the experimental optical conductivity is not accounted for by
the present theoretical analysis. In the theoretical spectra, peaks of much
smaller intensity appear at about the same frequency. In the large-polaron
theory, those peaks are provided by the interaction between electrons and the
LO-phonon branch with energy $\hbar\omega_{L,2}\approx58.4$ meV, accompanied
by the emission of a plasmon as described in Ref. \cite{zTD2001}.

The intensity of the experimentally observed absorption peak at $\hbar
\Omega\sim130$ meV is considerably higher than described by the large-polaron
theory. In the low density limit, the experimental optical data more rapidly
approach the single polaron limit~\cite{zDSG} than the theoretical predictions
based on Eq.~(\ref{xi}). This absorption peak at $\hbar\Omega\sim130$ meV may
be provided by other mechanisms, not controlled in the present study. E. g.,
electron-phonon interaction with low-frequency non-polar (e. g., acoustic)
phonons may contribute to the optical conductivity. The squared modulus
$\left\vert V_{q}\right\vert ^{2},$ which characterizes the coupling strength,
for the deformation electron-phonon interaction is $\left\vert V_{\mathbf{q}%
}\right\vert ^{2}\propto q$ \cite{zPD1985}, while for the Fr\"{o}hlich
interaction, $\left\vert V_{\mathbf{q}}\right\vert ^{2}\propto q^{-2}$.
Consequently, for the deformation electron-phonon interaction, the
short-wavelength phonons may provide non-negligible contributions to the
optical conductivity. Also, at sufficiently large $q,$ Umklapp scattering
processes with acoustic phonons can play a role. The treatment of
contributions due to acoustic phonons (and other mechanisms) is the subject of
the future work. Another possible explanation of the absorption peak at
$\hbar\Omega\sim130$ meV is weakened screening in the corresponding energy
range due to dynamical-exchange \cite{zBDL}.

\subsection{Conclusions \label{sec:conclusions}}

Many-polaron optical conductivity spectra, calculated (based on Ref.
\cite{zTD2001}) within the large-polaron picture without adjustment of
material constants, explain essential characteristics of the experimental
optical conductivity~\cite{zVDM-PRL2008}. The intensities of the calculated
many-polaron optical conductivity spectra and the intensities of the
experimental mid-infrared bands of the optical conductivity spectra of
SrTi$_{1-x}$Nb$_{x}$O$_{3}$ (from Ref.~\cite{zVDM-PRL2008}) are comparable for
all considered values of the doping parameter. The doping dependence of the
intensity of the mid-infrared band in the theoretical large-polaron spectra is
similar to that of the experimental data of Ref.~\cite{zVDM-PRL2008}. In the
high-frequency range, the theoretical absorption curves describe well the
experimental data (especially at low temperature). A remarkable difference
between the present theoretical approach and experiment is manifested on the
low frequency side of the mid-infrared range, where the experimental optical
conductivity shows a sharp and pronounced peak for $\hbar\Omega\sim130$ meV at
7 K. Although the theoretical curve also shows a feature around the same
frequency, its intensity is clearly underestimated. This peak in the
absorption spectrum at $\hbar\Omega\sim130$ meV remains to be explained. The
value of the effective electron-phonon coupling constant obtained in the
present work ($\alpha_{eff}\approx2$) corresponds to the intermediate coupling
strength of the large-polaron theory.

The alternative small-polaron and mixed-polaron models for the optical
conductivity require several fitting parameters. Furthermore, we find that the
mixed-polaron model would need a major adjustment of the overall intensity in
order to fit experimental spectra.

Contrary to the case of the large polaron, the small-polaron parameters cannot
be extracted from experimental data. Moreover, the small-polaron model, for
any realistic choice of parameters, shows a frequency dependence in the
high-frequency range which is different from that of the experimental optical
conductivity. Both the experimental and the theoretical large-polaron optical
conductivity decrease as a power function at high frequencies, while the
small-polaron optical conductivity falls down exponentially for sufficiently
high $\Omega$.

In summary, the many-body large-polaron model based on the Fr\"{o}hlich
interaction accounts for the essential characteristics (except --
interestingly -- for the intensity of a prominent peak at $\hbar\Omega\sim130$
meV, that constitutes an interesting challenge for theory) of the experimental
mid-infrared optical conductivity band in SrTi$_{1-x}$Nb$_{x}$O$_{3}$ without
any adjustment of material parameters. The large-polaron model gives then a
convincing interpretation of the experimentally observed mid-infrared band of
SrTi$_{1-x}$Nb$_{x}$O$_{3}$.

\newpage

\section{Notes on the polaron mobility}

\begin{quote}
\emph{J. T. Devreese and S. N. Klimin}
\end{quote}

Here, we are focused on some important issues related to the polaron mobility.
First, we discuss the polaron mobility in the weak-coupling regime on the
basis of Ref. \cite{DB1981}. Several theoretical methods have been applied to
study the transport properties of the Fr\"{o}hlich polaron. The polaron
mobility was calculated using various approaches: the calculation of the
scattering amplitude \cite{LP1955}, the kinetic equation \cite{Kadanoff}, the
Green's function technique, the Kubo formula \cite{LK1964,Osaka1961}, the
path-integral formalism \cite{FHIP,TF70,Thornber}.

A challenging difficulty is that, even for weak coupling and in the ohmic
regime, there is a remarkable difference in the mobility as obtained via a
relaxation-time approximation \cite{HS1953,LP1955,ac3,Osaka1961,Kadanoff}, and
as obtained via the path-integral formalism, worked out by Thornber and
Feynman \cite{TF70}, and which is based on the Feynman polaron model
\cite{Feynman}.

At weak coupling and small electric field, the relaxation time result for the
mobility \cite{LK1964} seems more reliable than the Thornber-Feynman result.
This might be partly due to the deviation of the electron velocity
distribution from a drifted Maxwellian as shown analytically \cite{ac10} from
the Boltzmann equation at weak electron-phonon coupling and low temperature in
the steady state regime.

Because the Boltzmann equation is valid at weak electron-phonon coupling, and
because of its intuitively transparent structure, this equation is an
important tool to study transport properties of polarons for weak coupling. In
Ref. \cite{DB1981}, its solution is discussed in the ohmic regime and for the
steady state. The mobility in the zero temperature limit from Ref.
\cite{DB1981} is given by:
\begin{equation}
\left.  \mu\right\vert _{T\rightarrow0}\rightarrow\frac{e}{2\alpha}\bar
{N}^{-1},
\end{equation}
where $\bar{N}$ is the average number of phonons. This is equivalent to the
result from the relaxation-time approximation \cite{Kadanoff}, which therefore
holds in the zero-temperature limit.

\begin{figure}
[h]
\begin{center}
\includegraphics[
height=3.3287in,
width=3.7939in
]%
{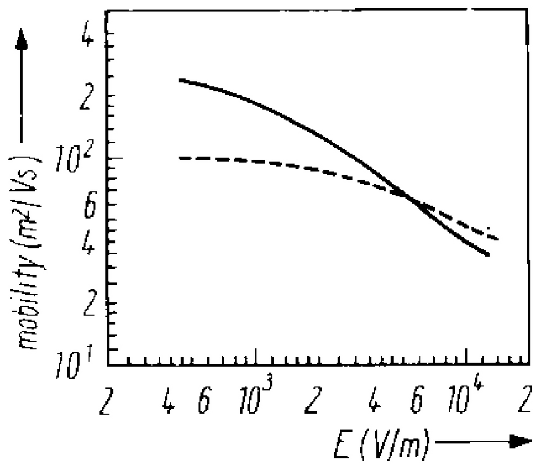}%
\caption{Mobility of weak-coupling polarons, obtained from the exact solution
of the Boltzmann equation \cite{DEK1978} (solid line) and from \cite{TF70}
(dashed line). (After Ref. \cite{DE1978}.)}%
\label{fig:apc1}%
\end{center}
\end{figure}

An analytical solution of the Boltzmann equation at $T=0$ was obtained in Ref.
\cite{DEK1978}. In Fig. \ref{fig:apc1}, the mobility of a polaron in the
weak-coupling regime, calculated using the exact solution of the Boltzmann
equation \cite{DEK1978} is calculated for InSb at $T=77$ K and compared to the
mobility from Ref. \cite{TF70}. For weak electric fields the result of Ref.
\cite{DEK1978} is quite close to that of the polaron theory with a relaxation
time but differs by the factor $\frac{3}{2}\frac{k_{B}T}{\hbar\omega_{0}}%
\frac{1}{2.5}$ from \cite{TF70}. These results seem to confirm the validity of
a relaxation time approach for the electric field $E\rightarrow0$ (at least
for InSb at 77 K). Nevertheless, as pointed out in \cite{DE1978}, a system of
non-interacting polarons is not ergodic and this point should be examined
carefully before definite conclusions can be drawn when $E\rightarrow0$. In
Ref. \cite{Nonlin}, arbitrary temperature and electric field are considered,
and an exact recursion relation is obtained for the time-dependent expansion
coefficients of the electron distribution function in terms of Legendre polynomials.%

%TCIMACRO{\FRAME{fhFU}{3.7939in}{3.3287in}{0pt}{\Qcb{Mobility of weak-coupling
%polarons, obtained from the exact solution of the Boltzmann equation
%\cite{DEK1978} (solid line) and from \cite{TF70} (dashed line). (After Ref.
%\cite{DE1978}.)}}{\Qlb{fig:apc1}}{appc-fig1.eps}%
%maintain-aspect-ratio TRUE;  display "ICON";  valid_file "F";
%width 3.7939in;  height 3.3287in;  depth 0pt;  original-width 3.7664in;
%original-height 3.3015in;  cropleft "0";  croptop "1";  cropright "1";
%cropbottom "0";  filename 'AppC-Fig1.eps';file-properties "XNPEU";}} }%
%BeginExpansion
%EndExpansion

The DC mobility of a polaron in the strong-coupling regime was investigated by
Volovik \emph{et al}. \cite{Volovik1}. They showed that the interaction
Hamiltonian corresponding to scattering of a phonon by a polaron can be
separated in the strong-coupling limit with the aid of the transformations of
Bogoliubov and Tyablikov in conjunction with the LLP canonical transformation.
In the leading order in powers of the inverse coupling constant $\alpha^{-1}$,
the principal role in the scattering is played by two-phonon processes. In the
system of units with $\hbar=1$, the following result for the strong-coupling
polaron mobility has been obtained:%
\begin{equation}
\mu=\frac{\gamma}{m\omega_{\mathrm{LO}}\alpha^{2}}\frac{T}{\omega_{0}%
}e^{\omega_{0}/T}, \label{apc1}%
\end{equation}
with a numerical temperature-independent coefficient $\gamma\sim1$.

The \textquotedblleft$\frac{3}{2}k_{B}T$\textquotedblright\ problem reveals a
key distinction between the polaron mobility obtained in Refs.
\cite{FHIP,TF70,Thornber} and the other approaches. Also the other problem is
worth discussing. The results of the approaches
\cite{Kadanoff,LK1964,Osaka1961} are in agreement with each other. Therefore
during a long time they were recognized well-established. Several works
appeared in which the polaron mobility at low temperatures differs from the
Kadanoff result by the numerical factor 3, e. g., Refs.
\cite{Los1984,Sokolovsky2011}. However, they have not yet attract a proper attention.

In the work by V. F. Los \cite{Los1984}, the polaron mobility was calculated
on the basis of Kubo's formula using a Green's superoperator technique. As
stated in Ref. \cite{Los1984}, the relaxation-time approximation
\cite{Kadanoff} does not take into account the change in the electron velocity
in all the electron-phonon scattering processes allowed by the energy and
momentum conservation laws. The polaron mobility obtained in Ref.
\cite{Los1984} gives the correct temperature dependence of the polaron
mobility but exceeds the expression obtained by Kadanoff exactly by a factor 3.

Recently, this \textquotedblleft factor 3\textquotedblright\ has been again
confirmed using a rigorous derivation. An approach to the polaron mobility has
been proposed by F. Brosens and D. Sels \cite{SB2014}, based on the dynamics
of the Wigner distribution function, using the kinetic equations derived in
Refs. \cite{Sels2013-2,Sels2013-3}.

In the paper \cite{SB2014}, the mobility of the Fr\"{o}hlich polaron is
calculated within approach, based on the dynamics of the Wigner distribution
function. The approach proposed by the authors is based on a path integral
description of the Wigner distribution function. The time evolution of the
electron distribution function $f\left(  \mathbf{p},t\right)  $ in the
electron-phonon system under an external time-dependent uniform electric field
$\mathbf{E}\left(  t\right)  $ is governed by the generalized kinetic
equation
\begin{align}
&  \left(  \frac{\partial}{\partial t}+e\mathbf{E}\left(  t\right)  \cdot
\frac{d}{d\mathbf{p}}\right)  f\left(  \mathbf{p},t\right) \nonumber\\
&  =\sum_{\mathbf{k}}\frac{2\left\vert \gamma\left(  k\right)  \right\vert
^{2}}{\hbar^{2}}\iiint dt^{\prime}d\mathbf{x}^{\prime}d\mathbf{p}^{\prime
}\Theta\left(  t^{\prime}\leq t\right)  f\left(  \mathbf{p}^{\prime}%
,t^{\prime}\right) \nonumber\\
&  \times\left(
\begin{array}
[c]{c}%
\left(
\begin{array}
[c]{c}%
\left(  n_{B}\left(  \omega_{k}\right)  +1\right)  \cos\left(  \mathbf{k}%
\cdot\left(  \mathbf{x}-\mathbf{x}^{\prime}\right)  +\omega_{k}\left(
t-t^{\prime}\right)  \right) \\
+n_{B}\left(  \omega_{k}\right)  \cos\left(  \mathbf{k}\cdot\left(
\mathbf{x}-\mathbf{x}^{\prime}\right)  -\omega_{k}\left(  t-t^{\prime}\right)
\right)
\end{array}
\right) \\
\times\left(  K_{0}\left(  \mathbf{x},\mathbf{p}-\frac{\hbar\mathbf{k}}%
{2},t|\mathbf{x}^{\prime},\mathbf{p}^{\prime}+\frac{\hbar\mathbf{k}}%
{2},t^{\prime}\right)  -K_{0}\left(  \mathbf{x},\mathbf{p}+\frac
{\hbar\mathbf{k}}{2},t|\mathbf{x}^{\prime},\mathbf{p}^{\prime}+\frac
{\hbar\mathbf{k}}{2},t^{\prime}\right)  \right)
\end{array}
\right)  \label{1sb}%
\end{align}
with the propagator%
\begin{align}
K_{0}\left(  \mathbf{x},\mathbf{p},t|\mathbf{x}^{\prime},\mathbf{p}^{\prime
},t^{\prime}\right)   &  =\delta\left(  \mathbf{p}-\mathbf{p}^{\prime}%
-\int_{t^{\prime}}^{t}e\mathbf{E}\left(  \sigma\right)  d\sigma\right)
\nonumber\\
&  \times\delta\left(  \mathbf{x}-\mathbf{x}^{\prime}-\frac{\mathbf{p}%
^{\prime}}{m}\left(  t-t^{\prime}\right)  -\int_{t^{\prime}}^{t}%
\frac{e\mathbf{E}\left(  \sigma\right)  }{m}\left(  t-\sigma\right)
d\sigma\right)  . \label{2sb}%
\end{align}
Here, $\omega_{k}$ is the phonon frequency, $\gamma\left(  k\right)  $ is the
amplitude of the electron-phonon interaction, and $n_{B}\left(  \omega
_{k}\right)  $ is the free-phonon distribution function,%
\begin{equation}
n_{B}\left(  \omega_{k}\right)  =\frac{1}{e^{\beta\hbar\omega_{k}}-1}.
\label{2asb}%
\end{equation}

The key result of Ref. \cite{SB2014} is the mobility of a weak-coupling
polaron given by formula (IV.4):%
\begin{equation}
\sigma_{DC}=\frac{3e^{2}}{2\alpha m\omega_{LO}}e^{\beta\hbar\omega_{LO}}.
\label{3sb}%
\end{equation}
This expression differs by the factor 3 from the result by Kadanoff and Osaka
\cite{Kadanoff,LK1964,Osaka1961}. Also the critical re-derivation of the
polaron optical conductivity on the basis of the Feynman polaron model results
in the optical conductivity
\begin{equation}
\sigma_{DC}=\frac{3e^{2}}{2\alpha m^{\ast}\omega_{LO}}e^{\beta\hbar\omega
_{LO}} \label{4sb}%
\end{equation}
that differs by the factor $\frac{2\hbar\omega_{LO}}{k_{B}T}$ from the FHIP
polaron mobility.

V. F. Los derived the polaron mobility \cite{Los1984} on the basis of the Kubo
formula and the Bogoliubov technique of an exact elimination of the phonon
operators. The mobility was obtained in Ref. \cite{Los1984} both in the
weak-coupling approximation and using the Feynman polaron model. The
weak-coupling expression given in Ref. \cite{Los1984} by formula (18) is the
same as (\ref{1sb}), i. e. three times larger than the Kadanoff polaron mobility.

The polaron mobility obtained in Ref. \cite{Los1984} using the Feynman polaron
model, formula (36),%
\begin{equation}
\mu^{F}=\frac{3e}{2\alpha}e^{\beta}\frac{e^{m_{f}/v}}{\left(  m_{f}+1\right)
^{3/2}} \label{5sb}%
\end{equation}
differs from (\ref{4sb}). However, the factor 3 with respect to the Kadanoff
result is definitely present in (\ref{5sb}).

The key distinction between the derivation of the polaron mobility in Refs.
\cite{Los1984,SB2014}, on the one hand, and in Refs.
\cite{Kadanoff,LK1964,Osaka1961} is the relaxation-time approximation (RTA)
(see Ref. \cite{PD1984}). The RTA consists in disregarding the contribution of
the so called \textquotedblleft re-population term\textquotedblright\ (using
terminology of Ref. \cite{PD1984})\ in the kinetic equation. RTA is used in
Refs. \cite{Kadanoff,LK1964,Osaka1961}. On the contrary, in Refs.
\cite{Los1984,SB2014} there is no relaxation-time approximation. In Ref.
\cite{Los1984}, the parameter attributed to the relaxation time appears in a
natural way quite rigorously. Moreover, in Ref. \cite{SB2014}, the relaxation
time does not appear at all.

The difference of the results by Los from the theory by Kadanoff and Osaka
appears already at an intermediate stage. Los derived the evolution equation
for the correlation function velocity-velocity $\left\langle v_{\nu}\left(
0\right)  v_{\mu}\left(  \tau\right)  \right\rangle $ [formula (9)], with the
kernel function%
\begin{align}
\Gamma_{\nu}\left(  \mathbf{p}\right)   &  =2\pi\sum_{\mathbf{k}}\left\vert
V_{\mathbf{k}}\right\vert ^{2}\left(  1-\frac{v_{\nu}\left(  \mathbf{p}%
-\mathbf{k}\right)  }{v_{\nu}\left(  \mathbf{p}\right)  }\right) \nonumber\\
&  \times\left\{  \left(  1+N_{\mathbf{k}}\right)  \delta\left(  T\left(
\mathbf{p}\right)  -T\left(  \mathbf{p}-\mathbf{k}\right)  -\omega
_{\mathbf{k}}\right)  +N_{\mathbf{k}}\delta\left(  T\left(  \mathbf{p}\right)
-T\left(  \mathbf{p}-\mathbf{k}\right)  +\omega_{\mathbf{k}}\right)  \right\}
\label{7sb}%
\end{align}
(the notations are in Ref. \cite{Los1984}).

The factor $\left[  1-v_{\nu}\left(  \mathbf{p}-\mathbf{k}\right)  /v_{\nu
}\left(  \mathbf{p}\right)  \right]  $ in the kernel function is an essential
difference between the theory by Los and the theory by Osaka/Kadanoff. Without
this factor, as checked in Ref. \cite{Los1984}, the theory by Los would give
the same result as the Kadanoff theory. This factor describes the change in
the electron velocity in the electron-phonon scattering processes allowed by
the energy and momentum conservation laws. It is important to note that in
relaxation times of the kinetic equations corresponding to elastic scattering
mechanisms (e. g., impurity scattering) or approximately elastic mechanisms
(e. g., acoustic phonons), the factor $\left(  1-v_{\nu}^{\prime}/v_{\nu
}\right)  $ is always present. In those kinetic equations, the second
(subtracted) part, $v_{\nu}^{\prime}/v_{\nu}$, comes from the aforesaid
\textquotedblleft re-population\textquotedblright\ term. This
\textquotedblleft re-population\textquotedblright\ term was neglected in Refs.
\cite{Kadanoff,LK1964,Osaka1961}. It appears that the \textquotedblleft
re-population\textquotedblright\ term is in fact non-negligible. The analogous
reasoning is developed in Ref. \cite{SB2014}, where the \textquotedblleft
re-population\textquotedblright\ contribution to the kinetic equation is taken
into account. It is shown in Ref. \cite{SB2014} that the neglect of the
re-population term in the kinetic equation is an unwarranted approximation,
because it \emph{violates the particle number conservation}. The conclusions
of Refs. \cite{Los1984,SB2014} are contrary to the assumption made in Refs.
\cite{Kadanoff,LK1964,Osaka1961}, where that term is neglected.

In Ref. \cite{Los1984}, the expression (\ref{7sb}) for $\Gamma_{\nu}\left(
\mathbf{p}\right)  $ arises here from the rigorous microscopic treatment. It
is important to note that such a relaxation time was introduced
phenomenologically in 1939 in \cite{FM}, but in subsequent studies the
expression without the factor describing the change in the velocity was obtained.

In Ref. \cite{FM}, the relaxation time has been phenomenologically defined by:%
\begin{equation}
\frac{1}{\tau}=-\sum_{\mathbf{q}}\left(  \frac{\Delta k}{k}\right)  _{x}%
\phi_{\mathbf{q}}\left(  \mathbf{k}\right)  \label{tau}%
\end{equation}
where $\phi_{\mathbf{q}}\left(  \mathbf{k}\right)  $ is the probability per
unit time that an electron with the wave number $\mathbf{k}$ makes a collision
with a lattice wave of wave number $\mathbf{q},$ and $\Delta k$ is the average
change of the $x$-component of the wave number $k_{x}$ on each collision.

The factor
\[
-\left(  \frac{\Delta k}{k}\right)  _{x}=\frac{k_{x}-k_{x}^{\prime}}{k_{x}%
}=1-\frac{k_{x}^{\prime}}{k_{x}}%
\]
entering (\ref{tau}) has exactly the same meaning as the factor $\left(
1-\frac{v_{\nu}\left(  \mathbf{p}-\mathbf{k}\right)  }{v_{\nu}\left(
\mathbf{p}\right)  }\right)  $ which appears in the work by Los \cite{Los1984}
and ensures the particle number conservation. Without this factor, the formula%
\begin{equation}
\frac{1}{\tau}=\sum_{\mathbf{q}}W_{\mathbf{q}}\left(  \mathbf{k}\right)
\label{tau2}%
\end{equation}
gives the result of Refs. \cite{Kadanoff,LK1964,Osaka1961}. With this factor,
the derivation reproduced in Ref. \cite{Anselm} gives the same relaxation time
and mobility as in Ref. \cite{Los1984}.

In the paper by B. I. Davydov and I. M. Shmushkevich \cite{Davydov}, the
derivation of the mean free path and the electron mobility in ionic crystal is
performed using the parameters of the medium which are not immediately
measurable. Later, Born and Huang in their \emph{Dynamical Theory of Crystal
Lattices} \cite{Born} introduced the description of optical phonons and the
electron-LO phonon interaction using only observable parameters, such as
high-frequency and static dielectric constants. At present, these notations
are of common use. In the monograph by A. Anselm \cite{Anselm}, the theory by
Davydov and Shmushkevich has been reproduced using these contemporary
notations. The physics of the approach by Davydov and Shmushkevich is
described in Ref. \cite{Anselm} in the following way.

\textquotedblleft At low temperatures the scattering is inelastic, and,
therefore, general the relaxation time cannot be introduced with the aid of
Boltzmann equation ... However, as was demonstrated by B. I. Davydov and I. M.
Shmushkevich in 1940, in the low-temperature case as well the relaxation time
can be introduced, provided a correct calculation procedure is followed.

Qualitatively this can be explained as follows. At low temperatures, when
$k_{B}T\ll\hbar\omega_{0}$ the absolute majority of the electrons are able
only to absorb the phonons. Such absorption of a phonon results in the
electron going over to the energy interval from $\hbar\omega_{0}$ to
$2\hbar\omega_{0}$. Such an electron will immediately emit a phonon, because
the ratio of the emission probability to the absorption probability is equal,
according to (6.1), to $\frac{N_{q}+1}{N_{q}}\approx\exp\left(  \frac
{\hbar\omega_{0}}{k_{B}T}\right)  \gg1$. The variation of the electron energy
in the result of such an absorption and an almost immediate emission of a
phonon will be very small (only at the expense of the $\omega_{0}$ vs $q$
dependence), but the variation of wave vector will be substantial. This makes
it possible to regard the electron scattering in a definite sense as elastic
and to introduce the relaxation time.\textquotedblright\

Remarkably, the phenomenological definition (\ref{tau}) coincides with the
Davydov-Shmushkevich formula for the inverse relaxation time. The resulting
low-temperature relaxation time within the approach by Davydov and
Shmushkevich is given by:%
\begin{equation}
\tau=\frac{3\sqrt{2}}{2}\frac{\hbar^{2}\varepsilon^{\ast}}{e^{2}m_{b}%
^{1/2}\left(  \hbar\omega_{0}\right)  ^{1/2}}\exp\left(  \frac{\hbar\omega
_{0}}{k_{B}T}\right)  .
\end{equation}
with $\varepsilon^{\ast}$ defined through the high-frequency and static
dielectric constants:%
\begin{equation}
\frac{1}{\varepsilon^{\ast}}=\frac{1}{\varepsilon_{\infty}}-\frac
{1}{\varepsilon_{0}}.
\end{equation}
The mobility is expressed through the relaxation time in the standard way:%
\begin{equation}
\mu=\frac{e}{m_{b}}\tau.
\end{equation}
Hence the mobility is:%
\begin{equation}
\mu=\frac{3\sqrt{2}}{2}\frac{\hbar^{2}\varepsilon^{\ast}}{em_{b}^{3/2}\left(
\hbar\omega_{0}\right)  ^{1/2}}\exp\left(  \frac{\hbar\omega_{0}}{k_{B}%
T}\right)  .
\end{equation}
Using the Fr\"{o}hlich coupling constant $\alpha$,%
\begin{equation}
\alpha=\frac{1}{2\varepsilon^{\ast}}\frac{e^{2}}{\hbar\omega_{0}}\left(
\frac{2m_{b}\omega_{0}}{\hbar}\right)  ^{1/2},
\end{equation}
the mobility is transformed to the expression%
\begin{equation}
\mu=\frac{3e}{2m_{b}\alpha\omega_{0}}\exp\left(  \frac{\hbar\omega_{0}}%
{k_{B}T}\right)  .
\end{equation}
This result is three times larger than the mobility obtained by many authors,
e. g., Kadanoff.

In the recent paper \cite{DF2014}, the alternative representation has been
found for the optical conductivity described by the Kubo formula. The
treatment is based on the expression for the optical conductivity:%
\begin{equation}
\sigma\left(  z\right)  =\frac{i}{zV}\left[  \Pi\left(  z\right)  -e^{2}%
\Gamma\right]  \qquad\left(  z=\Omega+i\delta,\;\delta\rightarrow+0\right)
\label{s}%
\end{equation}
where $V$ is the system volume, $e$ is the electronic charge, $\Pi\left(
z\right)  $ is the current-current correlation function,
\begin{equation}
\Pi\left(  z\right)  =-i\int_{0}^{\infty}dt~e^{izt}\left\langle \left[
J\left(  t\right)  ,J\left(  0\right)  \right]  \right\rangle
\end{equation}
and the coefficient $\Gamma$ is determined through the correlation function in
the Euclidean time:%
\begin{equation}
e^{2}\Gamma=-\int_{0}^{\beta}d\tau\left\langle J\left(  \tau\right)  J\left(
0\right)  \right\rangle ,\qquad\beta=\frac{1}{k_{B}T}. \label{G}%
\end{equation}
Here, the current operator is%
\begin{equation}
J=-ev_{x}=-\frac{e}{m_{b}}p_{x}.
\end{equation}
In the known expressions for the Kubo formula, $\Gamma$ is given by explicit
constants:
\begin{equation}
e^{2}\Gamma=-\frac{e^{2}}{m_{b}} \label{eqv}%
\end{equation}
for a single electron with the band mass $m_{b}$ (see, e. g., Ref.
\cite{DSG1972}).

In the memory-function representation, the polaron optical conductivity is
given by formula (7) of Ref. \cite{DF2014}:%
\begin{equation}
\sigma\left(  z\right)  =-\frac{i}{V}\frac{e^{2}\Gamma}{z+iM\left(  z\right)
}.
\end{equation}
with the memory function $M\left(  z\right)  $. The equivalence relation
(\ref{eqv}) is important for the sum rule \cite{DLR1977} due to the following
reasons. On the one hand, it is easily checked by hand that the expression
(\ref{s1}) explicitly satisfies the sum rule given by formula (6) of Ref.
\cite{DF2014}:%
\begin{equation}
\int_{-\infty}^{\infty}\operatorname{Re}\sigma\left(  \Omega+i\delta\right)
d\Omega=-\frac{\pi e^{2}\Gamma}{V}.
\end{equation}
On the other hand, the polaron optical conductivity must satisfy the $f$-sum
rule \cite{DLR1977}:
\begin{equation}
\int_{-\infty}^{\infty}\operatorname{Re}\sigma\left(  \Omega+i\delta\right)
d\Omega=\frac{1}{V}\frac{\pi e^{2}}{m_{b}}.
\end{equation}
Thus the relation (\ref{eqv}) ensures the fulfilment of the $f$-sum rule for
the polaron optical conductivity. When using exact polaron states, the
integral in (\ref{G}) gives analytically $e^{2}/m_{b}$. However, any
approximation for the polaron states may violate (\ref{eqv}) and consequently
violate the $f$-sum rule.

The DC mobility of a Fr\"{o}hlich polaron is obtained in Ref. \cite{DF2014} in
the weak-coupling regime at low temperatures as $\mu=\frac{10}{3}\mu_{FHIP}$,
i.e., the mobility differs by a numerical factor 10/3 from the result of FHIP
\cite{FHIP} and by $5k_{B}T/\left(  \hbar\omega_{0}\right)  $ from the value
obtained by Kadanoff. Accounting for the above discussion, the fulfilment of
the $f$-sum rule within the theory \cite{DF2014} and, consequently, the DC
mobility need further verification.

In summary, the most reliable results for the mobility of a Fr\"{o}hlich
polaron are obtained in Refs. \cite{Los1984,SB2014}. It is proven in those
works that the \textquotedblleft re-population\textquotedblright\ term in the
kinetic equation cannot be neglected, that leads to a significant change of
the polaron mobility. Consequently, the results obtained in Refs.
\cite{Los1984,SB2014} bring an important correction to the theory of the
polaron response.

\newpage

\section{Selected publications on polarons in high-rating journals (Nature,
Science, Physical Review Letters -- 2005-2015)}

\begin{enumerate}
\item \emph{Method for Analyzing Second-Order Phase Transitions: Application
to the Ferromagnetic Transition of a Polaronic System}, J. A. Souza, Yi-Kuo
Yu, J. J. Neumeier, H. Terashita, and R. F. Jardim, Phys. Rev. Lett.
\textbf{94}, 207209 (2005).\newline{\small \textbf{Abstract}}\newline{\small A
new method for analyzing second-order phase transitions is presented and
applied to the polaronic system La}$_{0.7}${\small Ca}$_{0.3}${\small MnO}%
$_{3}${\small . It utilizes heat capacity and thermal expansion data
simultaneously to correctly predict the critical temperature's pressure
dependence. Analysis of the critical phenomena reveals second-order behavior
and an unusually large heat capacity exponent.}

\item \emph{Validity of the Franck-Condon Principle in the Optical
Spectroscopy: Optical Conductivity of the Fr\"{o}hlich Polaron}, G. De
Filippis, V. Cataudella, A. S. Mishchenko, C. A. Perroni, and J. T. Devreese,
Phys. Rev. Lett. \textbf{96}, 136405 (2006).\newline{\small \textbf{Abstract}%
}\newline{\small The optical absorption of the Fr\"{o}hlich polaron model is
obtained by an approximation-free diagrammatic Monte Carlo method and compared
with two new approximate approaches that treat lattice relaxation effects in
different ways. We show that: (i) a strong coupling expansion, based on the
Franck-Condon principle, well describes the optical conductivity for large
coupling strengths (}$\alpha>10${\small ); (ii) a memory function formalism
with phonon broadened levels reproduces the optical response for weak coupling
strengths (}$\alpha<6${\small ) taking the dynamic lattice relaxation into
account. In the coupling regime }$6<\alpha<10${\small , the optical
conductivity is a rapidly changing superposition of both Franck-Condon and
dynamic contributions.}

\item \emph{Remanent Zero Field Spin Splitting of Self-Assembled Quantum Dots
in a Paramagnetic Host}, C. Gould, A. Slobodskyy, D. Supp, T. Slobodskyy, P.
Grabs, P. Hawrylak, F. Qu, G. Schmidt, and L. W. Molenkamp, Phys. Rev. Lett.
\textbf{97}, 017202 (2006).

\item \emph{Quantum Transport of Slow Charge Carriers in Quasicrystals and
Correlated Systems}, Guy Trambly de Laissardi\`{e}re, Jean-Pierre Julien, and
Didier Mayou, Phys. Rev. Lett. \textbf{97}, 026601 (2006).\newline%
{\small \textbf{Abstract}}\newline{\small We show that the semiclassical model
of conduction breaks down if the mean free path of charge carriers is smaller
than a typical extension of their wave function. This situation is realized
for sufficiently slow charge carriers and leads to a transition from a
metalliclike to an insulatinglike regime when scattering by defects increases.
This explains the unconventional conduction properties of quasicrystals and
related alloys. The conduction properties of some heavy fermions or polaronic
systems, where charge carriers are also slow, present a deep analogy.}

\item \emph{Occurrence of Intersubband Polaronic Repellons in a
Two-Dimensional Electron Ga}s, Stefan Butscher and Andreas Knorr, Phys. Rev.
Lett. \textbf{97}, 197401 (2006).

\item \emph{Subsecond Spin Relaxation Times in Quantum Dots at Zero Applied
Magnetic Field Due to a Strong Electron-Nuclear Interaction}, R. Oulton, A.
Greilich, S. Yu. Verbin, R. V. Cherbunin, T. Auer, D. R. Yakovlev, M. Bayer,
I. A. Merkulov, V. Stavarache, D. Reuter, and A. D. Wieck, Phys. Rev. Lett.
\textbf{98}, 107401 (2007).

\item \emph{Exciton Dephasing in Quantum Dots due to LO-Phonon Coupling: An
Exactly Solvable Model}, E. A. Muljarov and R. Zimmermann, Phys. Rev. Lett.
\textbf{98}, 187401 (2007)

\item \emph{Electron-Phonon Interaction and Charge Carrier Mass Enhancement in
SrTiO}$_{3}$, J. L. M. van Mechelen, D. van der Marel, C. Grimaldi, A. B.
Kuzmenko, N. P. Armitage, N. Reyren, H. Hagemann, and I. I. Mazin, Phys. Rev.
Lett. \textbf{100}, 226403 (2008).\newline{\small \textbf{Abstract}}%
\newline{\small We report a comprehensive THz, infrared and optical study of
Nb-doped SrTiO}$_{3}$ {\small as well as dc conductivity and Hall effect
measurements. Our THz spectra at 7 K show the presence of an unusually narrow
(}$<2${\small meV) Drude peak. For all carrier concentrations the Drude
spectral weight shows a factor of three mass enhancement relative to the
effective mass in the local density approximation, whereas the spectral weight
contained in the incoherent midinfrared response indicates that the mass
enhancement is at least a factor two. We find no evidence of a particularly
large electron-phonon coupling that would result in small polaron formation.}

\item \emph{Orbital and Charge-Resolved Polaron States in CdSe Dots and Rods
Probed by Scanning Tunneling Spectroscopy}, Zhixiang Sun, Ingmar Swart,
Christophe Delerue, Dani\"{e}l Vanmaekelbergh, and Peter Liljeroth, Phys. Rev.
Lett. \textbf{102}, 196401 (2009).

\item \emph{Dynamical Response and Confinement of the Electrons at the
LaAlO}$_{3}$\emph{/SrTiO}$_{3}$\emph{ Interface}, A. Dubroka, M. R\"{o}ssle,
K. W. Kim, V. K. Malik, L. Schultz, S. Thiel, C. W. Schneider, J. Mannhart, G.
Herranz, O. Copie, M. Bibes, A. Barth\'{e}l\'{e}my, and C. Bernhard, Phys.
Rev. Lett. \textbf{104}, 156807 (2010).\newline{\small \textbf{Abstract}%
}\newline{\small With infrared ellipsometry and transport measurements we
investigated the electrons at the interface between LaAlO}$_{3}${\small and
SrTiO}$_{3}${\small . We obtained a sheet carrier concentration of }%
$N_{s}\approx5-9\times10^{13}$ {\small cm}$^{-2}${\small , an effective mass
of }$m^{\ast}=3.2\pm0.4m_{e}${\small , and a strongly frequency dependent
mobility. The latter are similar as in bulk SrTi}$_{1-x}${\small Nb}$_{x}%
${\small O}$_{3}$ {\small and therefore suggestive of polaronic correlations.
We also determined the vertical concentration profile which has a strongly
asymmetric shape with a rapid initial decay over the first 2 nm and a
pronounced tail that extends to about 11 nm.}

\item \emph{Bipolaron and N-Polaron Binding Energies}, Rupert L. Frank,
Elliott H. Lieb, Robert Seiringer, and Lawrence E. Thomas, Phys. Rev. Lett.
\textbf{104}, 210402 (2010).\newline{\small \textbf{Abstract}}\newline%
{\small The binding of polarons, or its absence, is an old and subtle topic.
Here we prove two things rigorously. First, the transition from many-body
collapse to the existence of a thermodynamic limit for N polarons occurs
precisely at }$U=2\alpha${\small , where }$U$ {\small is the electronic
Coulomb repulsion and }$\alpha${\small is the polaron coupling constant.
Second, if }$U$ {\small is large enough, there is no multipolaron binding of
any kind. Considering the known fact that there is binding for some
}$U>2\alpha${\small , these conclusions are not obvious and their proof has
been an open problem for some time.}

\item \emph{Polaronic Conductivity in the Photoinduced Phase of 1T-TaS}$_{2}$,
N. Dean, J. C. Petersen, D. Fausti, R. I. Tobey, S. Kaiser, L. V. Gasparov, H.
Berger, and A. Cavalleri, Phys. Rev. Lett. \textbf{106}, 016401 (2011).

\item \emph{Spectroscopy of Single Donors at ZnO(0001) Surfaces}, Hao Zheng,
J\"{o}rg Kr\"{o}ger, and Richard Berndt, Phys. Rev. Lett. \textbf{108}, 076801 (2012)

\item \emph{Polarons in Suspended Carbon Nanotubes}, I. Snyman, and Yu. V.
Nazarov, Phys. Rev. Lett. \textbf{108}, 076805 (2012)

\item \emph{Two-Dimensional Polaronic Behavior in the Binary Oxides
m-HfO}$_{\emph{2}}$\emph{ and m-ZrO}$_{\emph{2}}$, K. P. McKenna, M. J. Wolf,
A. L. Shluger, S. Lany, and A. Zunger, Phys. Rev. Lett. \textbf{108}, 116403 (2012)

\item \emph{Polaron-to-Polaron Transitions in the Radio-Frequency Spectrum of
a Quasi-Two-Dimensional Fermi Gas}, Y. Zhang, W. Ong, I. Arakelyan, and J. E.
Thomas, Phys. Rev. Lett. \textbf{108}, 235302 (2012)\newline%
{\small \textbf{Abstract}}\newline{\small We measure radio-frequency spectra
for a two-component mixture of a }$^{6}${\small Li atomic Fermi gas in a
quasi-two-dimensional regime with the Fermi energy comparable to the energy
level spacing in the tightly confining potential. Near the Feshbach resonance,
we find that the observed resonances do not correspond to transitions between
confinement-induced dimers. The spectral shifts can be fit by assuming
transitions between noninteracting polaron states in two dimensions.}

\item \emph{Model of the Electron-Phonon Interaction and Optical Conductivity
of Ba}$_{1-x}$\emph{K}$_{x}$\emph{BiO}$_{3}$, R. Nourafkan, F. Marsiglio, and
G. Kotliar, Phys. Rev. Lett. \textbf{109}, 017001 (2012)

\item \emph{p-Wave Polaron}, Jesper Levinsen, Pietro Massignan,
Fr\'{e}d\'{e}ric Chevy, and Carlos Lobo, Phys. Rev. Lett. \textbf{109}, 075302 (2012)

\item \emph{Effect of Electron-Phonon Interaction Range for a Half-Filled Band
in One Dimension}, Martin Hohenadler, Fakher F. Assaad, and Holger Fehske,
Phys. Rev. Lett. \textbf{109}, 116407 (2012)

\item \emph{Digital Quantum Simulation of the Holstein Model in Trapped Ions},
A. Mezzacapo, J. Casanova, L. Lamata, and E. Solano, Phys. Rev. Lett.
\textbf{109}, 200501 (2012)

\item \emph{Bilayers of Rydberg Atoms as a Quantum Simulator for
Unconventional Superconductors}, J. P. Hague and C. MacCormick, Phys. Rev.
Lett. \textbf{109}, 223001 (2012)

\item \emph{Relaxation Dynamics of the Holstein Polaron}, Denis Gole\v{z},
Janez Bon\v{c}a, Lev Vidmar, and Stuart A. Trugman, Phys. Rev. Lett.
\textbf{109}, 236402 (2012)

\item \emph{Quantum Simulation of Small-Polaron Formation with Trapped Ions},
Vladimir M. Stojanovi\'{c}, Tao Shi, C. Bruder, and J. Ignacio Cirac, Phys.
Rev. Lett. \textbf{109}, 250501 (2012)

\item \emph{Condensed-matter physics: Repulsive polarons found}, P. Hannaford,
Nature \textbf{485}, 588 (2012)\newline{\small \textbf{Abstract}\newline
Quasiparticles known as repulsive polarons are predicted to occur when
'impurity' fermionic particles interact repulsively with a fermionic
environment. They have now been detected in two widely differing systems. See
Letters p.615 \& p.619}

\item \emph{Quantum Breathing of an Impurity in a One-Dimensional Bath of
Interacting Bosons}, Sebastiano Peotta, Davide Rossini, Marco Polini,
Francesco Minardi, and Rosario Fazio, Phys. Rev. Lett. \textbf{110}, 015302
(2013)\newline{\small \textbf{Abstract}}\newline{\small By means of the
time-dependent density-matrix renormalization-group (TDMRG) method we are able
to follow the real-time dynamics of a single impurity embedded in a
one-dimensional bath of interacting bosons. We focus on the impurity breathing
mode, which is found to be well described by a single oscillation frequency
and a damping rate. If the impurity is very weakly coupled to the bath, a
Luttinger-liquid description is valid and the impurity suffers an
Abraham-Lorentz radiation-reaction friction. For a large portion of the
explored parameter space, the TDMRG results fall well beyond the
Luttinger-liquid paradigm.}

\item \emph{Measurement of Coherent Polarons in the Strongly Coupled
Antiferromagnetically Ordered Iron-Chalcogenide Fe}$_{1.02}$\emph{Te using
Angle-Resolved Photoemission Spectroscopy}, Z. K. Liu, R.-H. He, D. H. Lu, M.
Yi, Y. L. Chen, M. Hashimoto, R. G. Moore, S.-K. Mo, E. A. Nowadnick, J. Hu,
T. J. Liu, Z. Q. Mao, T. P. Devereaux, Z. Hussain, and Z.-X. Shen, Phys. Rev.
Lett. \textbf{110}, 037003 (2013)

\item \emph{Decoherence of a Single-Ion Qubit Immersed in a Spin-Polarized
Atomic Bath}, L. Ratschbacher, C. Sias, L. Carcagni, J. M. Silver, C. Zipkes,
and M. K\"{o}hl, Phys. Rev. Lett. \textbf{110}, 160402 (2013)

\item \emph{Tunable Polaronic Conduction in Anatase TiO}$_{\emph{2}}$, S.
Moser, L. Moreschini, J. Ja\'{c}imovi\'{c}, O. S. Bari\v{s}i\'{c}, H. Berger,
A. Magrez, Y. J. Chang, K. S. Kim, A. Bostwick, E. Rotenberg, L. Forr\'{o},
and M. Grioni, Phys. Rev. Lett. \textbf{110}, 196403 (2013)

\item \emph{Investigating Polaron Transitions with Polar Molecules}, Felipe
Herrera, Kirk W. Madison, Roman V. Krems, and Mona Berciu, Phys. Rev. Lett.
\textbf{110}, 223002 (2013)\newline{\small \textbf{Abstract}}\newline%
{\small We determine the phase diagram of a polaron model with mixed
breathing-mode and Su-Schrieffer-Heeger couplings and show that it has two
sharp transitions, in contrast to pure models which exhibit one (for
Su-Schrieffer-Heeger coupling) or no (for breathing-mode coupling) transition.
We then show that ultracold molecules trapped in optical lattices can be used
as a quantum simulator to study precisely this mixed Hamiltonian, and that the
relative contributions of the two couplings can be tuned with external
electric fields. The parameters of current experiments place them in the
region where one of the transitions occurs. We also propose a scheme to
measure the polaron dispersion using stimulated Raman spectroscopy.}

\item \emph{Electronic Instability in a Zero-Gap Semiconductor: The
Charge-Density Wave in (TaSe}$_{4}$\emph{)}$_{2}$, C. Tournier-Colletta, L.
Moreschini, G. Aut\`{e}s, S. Moser, A. Crepaldi, H. Berger, A. L. Walter, K.
S. Kim, A. Bostwick, P. Monceau, E. Rotenberg, O. V. Yazyev, and M. Grioni,
Phys. Rev. Lett. \textbf{110}, 236401 (2013).

\item \emph{Itinerant Ferromagnetism in a Polarized Two-Component Fermi Gas},
Pietro Massignan, Zhenhua Yu, and Georg M. Bruun, Phys. Rev. Lett.
\textbf{110}, 230401 (2013).

\item \emph{Suppression of the Hanle Effect in Organic Spintronic Devices}, Z.
G. Yu, Phys. Rev. Lett. \textbf{111}, 016601 (2013).

\item \emph{Energy and Contact of the One-Dimensional Fermi Polaron at Zero
and Finite Temperature}, E. V. H. Doggen and J. J. Kinnunen, Phys. Rev. Lett.
\textbf{111}, 025302 (2013).

\item \emph{Measurement of the Femtosecond Optical Absorption of
LaAlO}$_{\emph{3}}$\emph{/SrTiO}$_{\emph{3}}$ \emph{Heterostructures: Evidence
for an Extremely Slow Electron Relaxation at the Interface}, Yasuhiro Yamada,
Hiroki K. Sato, Yasuyuki Hikita, Harold Y. Hwang, and Yoshihiko Kanemitsu,
Phys. Rev. Lett. \textbf{111}, 047403 (2013)\newline{\small \textbf{Abstract}%
}\newline{\small The photocarrier relaxation dynamics of an n-type LaAlO}%
$_{3}${\small /SrTiO}$_{3}$ {\small heterointerface is investigated using
femtosecond transient absorption (TA) spectroscopy at low temperatures. In
both LaAlO}$_{3}${\small /SrTiO}$_{3}$ {\small heterostructures and
electron-doped SrTiO}$_{3}$ {\small bulk crystals, the TA spectrum shows a
Drude-like free carrier absorption immediately after excitation. In addition,
a broad absorption band gradually appears within 40 ps, which corresponds to
the energy relaxation of photoexcited free electrons into self-trapped polaron
states. We reveal that the polaron formation time is enhanced considerably at
the LaAlO}$_{3}${\small /SrTiO}$_{3}$ {\small heterointerface as compared to
bulk crystals. Further, we discuss the interface effects on the electron
relaxation dynamics in conjunction with the splitting of the }${\small t}%
_{{\small 2g}}$ {\small subbands due to the interface potential.}

\item Pauli Spin Blockade and the Ultrasmall Magnetic Field Effect, Jeroen
Danon, Xuhui Wang, and Aur\'{e}lien Manchon, Phys. Rev. Lett. \textbf{111},
066802 (2013)

\item \emph{Tkachenko Polarons in Vortex Lattices}, M. A. Caracanhas, V. S.
Bagnato, and R. G. Pereira, Phys. Rev. Lett. \textbf{111}, 115304 (2013).

\item \emph{Impurity Problem in a Bilayer System of Dipoles}, N. Matveeva and
S. Giorgini, Phys. Rev. Lett. \textbf{111}, 220405 (2013).

\item \emph{Single-Polariton Optomechanics}, Juan Restrepo, Cristiano Ciuti,
and Ivan Favero, Phys. Rev. Lett. \textbf{112}, 013601 (2014)

\item \emph{Ferromagnetism of a Repulsive Atomic Fermi Gas in an Optical
Lattice: A Quantum Monte Carlo Study}, S. Pilati, I. Zintchenko, and M.
Troyer, Phys. Rev. Lett. \textbf{112}, 015301 (2014)

\item \emph{Ultrafast Photoemission Spectroscopy of the Uranium Dioxide UO2
Mott Insulator: Evidence for a Robust Energy Gap Structure}, Steve M.
Gilbertson, Tomasz Durakiewicz, Georgi L. Dakovski, Yinwan Li, Jian-Xin Zhu,
Steven D. Conradson, Stuart A. Trugman, and George Rodriguez, Phys. Rev. Lett.
\textbf{112}, 087402 (2014).

\item \emph{Direct View at Excess Electrons in TiO}$_{2}$\emph{ Rutile and
Anatase}, Martin Setvin, Cesare Franchini, Xianfeng Hao, Michael Schmid,
Anderson Janotti, Merzuk Kaltak, Chris G. Van de Walle, Georg Kresse, and
Ulrike Diebold, Phys. Rev. Lett. \textbf{113}, 086402 (2014) \newline%
{\small \textbf{Abstract} }\newline{\small A combination of scanning tunneling
microscopy and spectroscopy and density functional theory is used to
characterize excess electrons in TiO2 rutile and anatase, two prototypical
materials with identical chemical composition but different crystal lattices.
In rutile, excess electrons can localize at any lattice Ti atom, forming a
small polaron, which can easily hop to neighboring sites. In contrast,
electrons in anatase prefer a free-carrier state, and can only be trapped near
oxygen vacancies or form shallow donor states bound to Nb dopants. The present
study conclusively explains the differences between the two polymorphs and
indicates that even small structural variations in the crystal lattice can
lead to a very different behavior.}

\item \emph{Diagrammatic Monte Carlo Method for Many-Polaron Problems}, Andrey
S. Mishchenko, Naoto Nagaosa, and Nikolay Prokof'ev, Phys. Rev. Lett.
\textbf{113}, 166402 (2014) \newline{\small \textbf{Abstract}}\newline%
{\small We introduce the first bold diagrammatic Monte Carlo approach to deal
with polaron problems at a finite electron density nonperturbatively, i.e., by
including vertex corrections to high orders. Using the Holstein model on a
square lattice as a prototypical example, we demonstrate that our method is
capable of providing accurate results in the thermodynamic limit in all
regimes from a renormalized Fermi liquid to a single polaron, across the
nonadiabatic region where Fermi and Debye energies are of the same order of
magnitude. By accounting for vertex corrections, the accuracy of the
theoretical description is increased by orders of magnitude relative to the
lowest-order self-consistent Born approximation employed in most studies. We
also find that for the electron-phonon coupling typical for real materials,
the quasiparticle effective mass increases and the quasiparticle residue
decreases with increasing the electron density at constant electron-phonon
coupling strength.}

\item \emph{Polaron spin current transport in organic semiconductors}, Shun
Watanabe, Kazuya Ando, Keehoon Kang, Sebastian Mooser, Yana Vaynzof, Hidekazu
Kurebayashi, Eiji Saitoh, and Henning Sirringhaus, Nature Physics \textbf{10},
308 (2014)

\item \emph{Real Space Imaging of Spin Polarons in Zn-Doped SrCu}$_{2}%
$\emph{(BO}$_{3}$\emph{)}$_{2}$, M. Yoshida, H. Kobayashi, I. Yamauchi et al.,
Phys. Rev. Lett. \textbf{114}, 056402 (2015)

\item \emph{Crossover from Super- to Subdiffusive Motion and Memory Effects in
Crystalline Organic Semiconductors}, G. De Filippis, V. Cataudella,
A.\thinspace S. Mishchenko, N. Nagaosa, A. Fierro, and A. de Candia, Phys.
Rev. Lett. \textbf{114}, 086601 (2015)\newline{\small \textbf{Abstract}%
\newline The transport properties at finite temperature of crystalline organic
semiconductors are investigated, within the Su-Schrieffer-Heeger model, by
combining an exact diagonalization technique, Monte Carlo approaches, and a
maximum entropy method. The temperature-dependent mobility data measured in
single crystals of rubrene are successfully reproduced: a crossover from
super-to subdiffusive motion occurs in the range }$150<T<200${\small K, where
the mean free path becomes of the order of the lattice parameter and strong
memory effects start to appear. We provide an effective model, which can
successfully explain features of the absorption spectra at low frequencies.
The observed response to slowly varying electric field is interpreted by means
of a simple model where the interaction between the charge carrier and lattice
polarization modes is simulated by a harmonic interaction between a fictitious
particle and an electron embedded in a viscous fluid.}

\item \emph{Mobility of Holstein Polaron at Finite Temperature: An Unbiased
Approach}, A.\thinspace S. Mishchenko, N. Nagaosa, G. De Filippis, A. de
Candia, and V. Cataudella, Phys. Rev. Lett. \textbf{114}, 146401
(2015).\newline{\small \textbf{Abstract}\newline We present the first unbiased
results for the mobility }$\mu${\small of a one-dimensional Holstein polaron
obtained by numerical analytic continuation combined with diagrammatic and
worldline Monte Carlo methods in the thermodynamic limit. We have identified
for the first time several distinct regimes in the }$\lambda-T${\small plane
including a band conduction region, incoherent metallic region, an activated
hopping region, and a high-temperature saturation region. We observe that
although mobilities and mean free paths at different values of }$\lambda
${\small differ by many orders of magnitude at small temperatures, their
values at }$T${\small larger than the bandwidth become very close to each
other.}

\item \emph{Band Structures of Plasmonic Polarons}, F. Caruso, H. Lambert, and
F. Giustino, Phys. Rev. Lett. \textbf{114}, 146404 (2015)\newline%
{\small \textbf{Abstract}\newline Using state-of-the-art many-body
calculations based on the \textquotedblleft GW plus cumulant\textquotedblright%
\ approach, we show that electron-plasmon interactions lead to the emergence
of plasmonic polaron bands in the band structures of common semiconductors.
Using silicon and group IV transition-metal dichalcogenide monolayers (AX(2)
with A = Mo, W and X = S, Se) as prototypical examples, we demonstrate that
these new bands are a general feature of systems characterized by well-defined
plasmon resonances. We find that the energy versus momentum dispersion
relations of these plasmonic structures closely follow the standard valence
bands, although they appear broadened and blueshifted by the plasmon energy.
Based on our results, we identify general criteria for observing plasmonic
polaron bands in the angle-resolved photoelectron spectra of solids.}

\item \emph{Long-lived photoinduced polaron formation in conjugated
polyelectrolyte-fullerene assemblies}, R. C. Huber, A. S. Ferreira, R.
Thompson \emph{et al}., Science \textbf{348}, 1340 (2015)

\item \emph{Electron-Phonon Interactions, Metal-Insulator Transitions, and
Holographic Massive Gravity}, M. Baggioli and O. Pujolas, Phys. Rev. Lett.
\textbf{114}, 251602 (2015)\newline{\small \textbf{Abstract}\newline Massive
gravity is holographically dual to \textquotedblleft
realistic\textquotedblright\ materials with momentum relaxation. The dual
graviton potential encodes the phonon dynamics, and it allows for a much
broader diversity than considered so far. We construct a simple family of
isotropic and homogeneous materials that exhibit an interaction-driven
metal-insulator transition. The transition relates to the formation of
polarons -- phonon-electron quasibound states that dominate the
conductivities, shifting the spectral weight above a mass gap. We characterize
the polaron gap, width, and dispersion.}

\item \emph{Electron-Phonon Coupling in the Bulk of Anatase TiO}$_{2}$\emph{
Measured by Resonant Inelastic X-Ray Spectroscopy}, S. Moser, S. Fatale, P.
Krueger \emph{et al}., Phys. Rev. Lett. \textbf{115}, 096404 (2015).\newline%
{\small \textbf{Abstract}\newline We investigate the polaronic ground state of
anatase TiO2 by bulk-sensitive resonant inelastic x-ray spectroscopy (RIXS) at
the Ti L-3 edge. We find that the formation of the polaron cloud involves a
single 95 meV phonon along the c axis, in addition to the 108 meV ab-plane
mode previously identified by photoemission. The coupling strength to both
modes is the same within error bars, and it is unaffected by the carrier
density. These data establish RIXS as a directional bulk-sensitive probe of
electron-phonon coupling in solids.}

\item \emph{Impurity in a Bose-Einstein Condensate and the Efimov Effect}, J.
Levinsen, M. M. Parish, and G. M. Bruun, Phys. Rev. Lett. \textbf{115}, 125302 (2015).

\item \emph{Decoherence of Impurities in a Fermi Sea of Ultracold Atoms}, M.
Cetina, M. Jag, R. S. Lous, et al., Phys. Rev. Lett. \textbf{115}, 135302 (2015).

\item \emph{Impurities in Bose-Einstein Condensates: From Polaron to Soliton},
S. Shadkhoo and R., Shahriar, Phys. Rev. Lett. \textbf{115}, 135305
(2015).\newline{\small \textbf{Abstract}\newline We propose that impurities in
a Bose-Einstein condensate which is coupled to a transversely laser-pumped
multimode cavity form an experimentally accessible and analytically tractable
model system for the study of impurities solvated in correlated liquids and
the breakdown of linear-response theory. As the strength of the coupling
constant between the impurity and the Bose-Einstein condensate is increased,
which is possible through Feshbach resonance methods, the impurity passes from
a large to a small polaron state, and then to an impurity-soliton state. This
last transition marks the breakdown of linear-response theory.}

\item \emph{Quasiparticle Properties of a Mobile Impurity in a Bose-Einstein
Condensate}, R. S. Christensen, J. Levinsen, and G. M. Bruun, Phys. Rev. Lett.
\textbf{115}, 160401 (2015).\newline{\small \textbf{Abstract}\newline We
develop a systematic perturbation theory for the quasiparticle properties of a
single impurity immersed in a Bose-Einstein condensate. Analytical results are
derived for the impurity energy, effective mass, and residue to third order in
the impurity-boson scattering length. The energy is shown to depend
logarithmically on the scattering length to third order, whereas the residue
and the effective mass are given by analytical power series. When the
boson-boson scattering length equals the boson-impurity scattering length, the
energy has the same structure as that of a weakly interacting Bose gas,
including terms of the Lee-Huang-Yang and fourth order logarithmic form. Our
results, which cannot be obtained within the canonical Frohlich model of an
impurity interacting with phonons, provide valuable benchmarks for many-body
theories and for experiments.}
\end{enumerate}

\end{document}